\documentclass[11pt,a4paper]{JHEP3}
\usepackage{amsmath,epsfig}
\usepackage{amssymb,amsfonts}
\usepackage{latexsym}
\usepackage[latin1]{inputenc}
\usepackage{slashed}
\usepackage{empheq}
\numberwithin{equation}{section}
\usepackage{cancel}
\usepackage{overpic}
\usepackage{subcaption}
\usepackage{subeqnarray}
\usepackage{xcolor}
\let\normalcolor\relax
\usepackage{graphicx}
\usepackage{longtable}
\usepackage{color}
\usepackage{bbm}
\usepackage{bm}
\usepackage{ulem}
\usepackage{array} 
\usepackage{tikz}
\usepackage{subcaption}
\usepackage{placeins} 
\usepackage{diagbox} 
\usepackage{makecell} 

\newcolumntype{C}{>{$}c<{$}}  

\newcommand{\exclude}[1]{}
\def\nn{\nonumber}

\def\L{\mathcal{L}}

\def\<{\langle}
\def\>{\rangle}

\def\d{\alphaathrm{d}}

\def\a#1{\alpha_{#1}}

\def\tr{\alphaathrm{tr}}

\def\beq{\begin{equation}}
\def\eeq{\end{equation}}
\def\be{\begin{equation}}
\def\ee{\end{equation}}
\def\bea{\begin{eqnarray}}
\def\eea{\end{eqnarray}}
\def\bal{\begin{align}}
\def\eal{\end{align}}

\def\2b2[#1,#2][#3,#4]{\left( \begin{array}{cc} #1 & #2 \\ #3 & #4 \end{array}
\right)}
\def\3b3[#1,#2,#3][#4,#5,#6][#7,#8,#9]{\left( \begin{array}{ccc} #1 & #2 #3 \\
#4 & #5 & #6\\#7&#8&#9\end{array} \right)}

\newcommand{\g}{\gamma}

\newcommand\fverb{\setbox\pippobox=\hbox\bgroup\verb}
\newcommand\fverbdo{\egroup\alphaedskip\noindent%
                        \fbox{\unhbox\pippobox}\ }
\newcommand\fverbit{\egroup\item[\fbox{\unhbox\pippobox}]}

\newcommand{\bear}{\begin{eqnarray}}

\newcommand{\eear}{\end{eqnarray}}
\newcommand{\de}{\partial}
\newcommand{\bsea}{\begin{subeqnarray}}
\newcommand{\esea}{\end{subeqnarray}}
\newbox\pippobox

\def\f{\varphi}
\def\d{\delta}
\def\g{\gamma}
\def\G{\Gamma}

\def\6{\partial}
\def\a{\alpha}
\def\nn{\nonumber}

\def\pa{\partial}

\def\e{\epsilon}
\def\m{\mu}
\def\n{\nu}

\def\s{\sigma}
\def\t{\theta}
\def\sp{\;\;\;,\;\;\;}

\def\z{\zeta}
\def\sq
\def\a{\alpha}
\def\b{\beta}
\def\l{\lambda}

\def\tr{{\rm Tr}}
\def\k{\chi}
\def\hrj#1#2{\href{www.doi.org/#1}{#2}}
\def\hri#1#2{\href{http://arxiv.org/abs/#1}{[ArXiv:#1]#2}}
\def\hre#1#2{\href{http://arxiv.org/abs/#1/#2}{[ArXiv:#1/#2]}}

\def\hree#1#2{\href{https://doi.org/#1}{#2}}

\def\e{\epsilon}

\def\d{\delta}

\def\L{\Lambda}

\usepackage{tensor}

\def\k{{\kappa}}

\def\tbe{\tilde{\b}_\text{eff}}

\def\ar{{~~~\Rightarrow~~~}}

\title{Quantum (in)stability of maximally symmetric space-times}

\author{Jewel K.~Ghosh$^{\star,\S}$, Elias Kiritsis$^\natural$$^\flat$, Francesco Nitti$^\natural$, Valentin Nourry$^\natural$
~\\
$^\star$  \href{http://www.iub.edu.bd/}{Independent University Bangladesh (IUB)}, Bashundhara RA, Dhaka 1229, Bangladesh
~\\
$^\S$ \href{https://ccds.ai/compas/} {Center for Computational and Data Sciences}, Independent University, Bangladesh, Bashundhara RA, Dhaka 1229, Bangladesh
~\\
$\natural$  \href{http://www.apc.univ-paris7.fr}{Universit\'e Paris Cit\' e, CNRS, Astroparticule et Cosmologie}, F-75013 Paris, France
~\\
$^\flat$ \href{http://hep.physics.uoc.gr}{Crete Center for Theoretical Physics}, Institute for Theoretical and Computational Physics,
Department of Physics, Voutes University Campus,\\
GR-70013, Vasilika Vouton, Heraklion, Greece
}

\preprint{CCTP-2023-3\\
ITCP-IPP 2023/3}

\abstract{Classical gravity coupled to a CFT$_4$ (matter) is considered. The effect of the quantum dynamics of matter on gravity is studied around maximally symmetric spaces (flat, de Sitter and Anti de Sitter).
The structure of the graviton propagator is modified and non-trivial poles appear due to matter quantum effects.
The position and residues of such poles are mapped as a function of the relevant parameters, the central charge of the CFT$_4$, the two $R^2$ couplings of gravity as well as the curvature of the background space-time.
The instabilities induced are determined. Such instabilities can be important in cosmology as they trigger the departure from de Sitter space and in some regions of parameters
are more important than the well-known  scalar instabilities.
It is also determined when the presence of such instabilities is unreliable if the associated scales are larger than the ``species" cutoff of the gravitational theory.
}

\dedicated{\bf Dedicated to the memory of Stavros Katsanevas,\\ a universal scientist and a visionary.}

\begin{document}
\maketitle

\section{Introduction}

The interplay between semiclassical gravity and the quantum effects of Quantum Field Theory (QFT) is a topic of research that has been in the spotlight for several decades.
The most important area of applicability motivating these issues arises from cosmology. In cosmology,  we treat gravity as semiclassical\footnote{Exceptions exist where gravity  is treated in perturbation theory, \cite{TW} or where the curvature is very strong and a more fundamental theory (an example is string theory) can take over.}
and we couple it with QFTs. The class of semiclassical metrics relevant in this case is cosmological metrics, with the most prominent example being the maximally symmetric cosmological metric, i.e., de Sitter.

Other contexts also are relevant, like semiclassical effective actions of string theory with asymptotically AdS or flat asymptotics.  In this context, quantum effects in gravity and matter seem to go hand in hand as they are controlled by the same underlying parameter, the string coupling constant.  However, we understand that there are two types of quantum effects associated with string loops or the $\a'$ expansion, although their separation is ``duality-frame" dependent.
There are however limits, (known as double scaling limits) in which gravitational quantum effects can be made subleading to ``matter" quantum effects.
In the context of holography, relevant for AdS or asymptotically AdS spaces typically (bulk) gravity and matter are treated semiclassically, but subleading corrections in N
involve quantum effects in both sectors.

The case of QFTs on de Sitter space is an especially hot issue, as we believe that our universe has been near de Sitter at least twice during its history.
Defining a QFT on de Sitter space and, in particular,
answering questions about QFT backreaction on the geometry is a
subtle issue. Perturbative field quantization (and
renormalization) on fixed classical curved space-times is text-book material \cite{BD}.
However, answering concrete questions about the observable effects of quantum
fields backreacting on classical geometry is not
straightforward.  This concerns particularly theories which are gapless
in the infrared, due to the presence of infrared divergences (and in most cases strong IR dynamics).

It is well known that in classical GR, both de Sitter space and Flat Minkowski space are non-linearly stable\footnote{ Famously, AdS is not non-linearly stable, \cite{BR} , (and citations to it) even at the level of classical gravity. The instability is towards the formation of black holes.}, \cite{S1}-\cite{S5}.
However, in the presence of quantum effects from matter, instabilities can appear. For the case of flat space this was established in \cite{Q1}-\cite{Q10}.

In (quasi) de Sitter space, there are several other issues arising when one considers QFTs.
In  \cite{AIT}-\cite{TW2} a divergence of scalar correlators was observed at large times.
This was addressed in \cite{Staro,STW,GS} using a stochastic approach. The case of interacting massive scalar fields has been treated thoroughly more recently in \cite{Marolf,Hollands2}.  A systematic approach to compute corrections in the massless case is lacking and the problem remains still open.

The accumulation of long wavelength fluctuations in an expanding universe
is another issue that has been studied,  starting with references \cite{Mukha,AWoo}. Their analysis was extended further in \cite{unruh}.
Another issue concerned the fact that the two-point function of a massless scalar in de Sitter space had to break de Sitter symmetry, due to the presence of a zero mode, \cite{MM,BAllen}. This issue is of a different nature and is more similar to the fact that in two dimensions, a massless scalar is IR singular.
The resolution of this issue may be therefore similar: massless scalars are not good acceptable fields on de Sitter space\footnote{We shall later conclude in this paper that in all the theories we examine and which are all gapless, there is no breaking of de Sitter invariance, and the de Sitter invariant vacuum is chosen.} as argued in \cite{Hollands1}.

The expectation that QFTs in de Sitter space render the
manifold unstable at the non-linear level has been entertained for a long time \cite{Mottola}-\cite{dvali}. In particular, destabilizing effects were most important from massless particles, and a gravity 2-loop computation in \cite{TW} suggested such an instability. Similar calculations with massless scalars implied similar effects, \cite{Mukha,AWoo}, see however \cite{1702.05694}.

The topic of quantum effects has been revived after cosmological (CMB) data became precise, and in \cite{Weinberg,Cha,Sloth} it was argued that large time-dependent logs from quantum effects of quantum  fields, could give large corrections to inflationary observables. A different approach in \cite{Sena},  provided different results.
Therefore, the question of  the consequences  of the
secular terms (growing with time) which arise in perturbation theory
of a massless scalar field in the cosmological patch of de Sitter remains controversial. Do
these contributions  indicate an instability of de
Sitter space against quantum perturbations? Or is this conclusion an artefact of
 finite orders in perturbation theory, which is expected to disappear  once an
 appropriate resummation is performed (as is the case for infra-red
 effects in thermal perturbation theory)?
A review of these developments and additional references can be found in \cite{Seery}. Furthermore, the persistent difficulty of constructing de Sitter vacua in string theory, see \cite{1804.01120} for a review, has led to the conjecture that de Sitter space cannot be attained in a weakly curved/coupled  quantum theory of gravity \cite{Obied}.

There is another issue where de Sitter instabilities induced by quantum dynamics can be important: they trigger an exit from the inflationary regime that is an important ingredient of any inflationary model. Indeed, scalar instabilities triggered by coupling gravity to a CFT$_4$, provided an exit from  inflation, \cite{St,VilenkinStaro,BB}
induced by the conformal anomaly, \cite{duff}.
However, this exit was not good enough and the model was modified to what is now called the Starobinsky model, where inflation is triggered by a (rather large) $R^2$ term. This is a very successful model when compared to current data \cite{Planck:2018inflation}.  In that model also the exit from inflation is triggered by the unstable scalar mode, which in this context is the scalaron of $R^2$ gravity.

So far the quantum effects studied and the backreaction  on de Sitter and Minkowski spaces used weakly-coupled QFTs.
The holographic gauge/gravity duality provides a way to tackle non-perturbative (large-$N$) four-dimensional quantum field theories by mapping them to higher-dimensional semiclassical General Relativity. Moreover, one controls the dynamics of holographic QFTs even when the manifold they are defined upon is curved.
Holography has been applied to cosmological issues already in several works, \cite{anomalyinflation0,anomalyinflation1,BB}, \cite{H0}-\cite{Pen}.
One important arena where this technique can provide important information is when   {\it four-dimensional} gravity (described by Einstein General Relativity or its higher-derivative extensions) is coupled to a strongly coupled QFT. Among the issues that may arise are for example the stability under small (metric and matter) perturbations as well as the non-perturbative stability of cosmological backgrounds.

The holographic approach allows for recasting these questions in terms of a classical higher-dimensional gravity theory, which via holography captures the effects of the QFT, coupled to a classical four-dimensional gravitational theory. To be more  specific, the holographic setup consists of two coupled sectors:
\begin{enumerate}
\item The holographic sector describes the strongly coupled (holographic)  CFT, whose dual is living in a higher-dimensional space-time ({\it the bulk}) with metric $\mathcal{G}_{ab}$.
\item The four-dimensional gravity sector is defined in terms of a metric $g^{(0)}_{\omega\s}$. This metric plays the role of a boundary condition for  $\mathcal{G}_{ab}$, and it has no bulk dynamics. From the CFT point of view, it corresponds to the source of the field theory stress tensor.
\end{enumerate}
The action describing classical 4d gravity coupled to the holographic QFT has the form:
\be \label{intro1}
S = S_\text{grav}[g^{(0)}_{\omega\s}]  + S_{bulk}[\mathcal{G}_{ab},\ldots]
\ee
The first term can be taken to be the usual Einstein-Hilbert action plus eventually higher curvature terms, and it has the effect of making $g^{(0)}_{\omega\s}$  dynamical. The second term describes  the  higher-dimensional holographic dual of the QFT, and   the dots represent  bulk fields other than the metric. Both terms are treated classically, but the bulk action encodes holographically the full quantum dynamics of the dual field theory. The two sectors are coupled by the requirement that  on the conformal boundary $\mathcal{G}_{\omega\s}$ asymptotes  to  $g^{(0)}_{\omega\s}$.  ``Integrating out'' the CFT consists in evaluating   $S_{bulk}[\mathcal{G}_{ab},\ldots] $ on shell. This  results in an effective gravitational action for $g^{(0)}_{\omega\s}$ alone,
\be \label{intro2}
S_\text{eff}[g^{(0)}_{\omega\s}]  = S_\text{grav}[g^{(0)}_{\omega\s}]  +  S_{bulk}^{on-shell}[g^{(0)}_{\omega\s},\ldots]
\ee
The second term in (\ref{intro2}) is now a functional of the boundary value $g^{(0)}_{\omega\s}$, i.e. the four-dimensional metric.  Varying the effective action results in the semi-classical Einstein equation :
\be \label{intro3}
E_{\omega\s}  =  \< T_{\omega\s} \>_{CFT}
\ee
where $E_{\omega\s}$ is the variation of the first term in  (\ref{intro1}) (i.e. the Einstein tensor if  $S_\text{grav}$ is purely GR) and $ \< T_{\omega\s} \>_{CFT}$ is obtained by varying $ S_{bulk}^{on-shell}$ with respect to $g^{(0)}_{\omega\s}$.

In a previous work \cite{GKNW}, some of the authors have used the  setup described above to address the issue of {non-perturbative existence} of 4d de Sitter space coupled to  a (gap-less) holographic field theory.  For this,  the attention was limited to boundary metrics $g^{(0)}_{\omega\s}$ of maximal symmetry. In this case, the effective action (\ref{intro2})  takes the form of an effective $f(R)$ theory,
\be \label{intro4}
S_\text{eff}[g^{(0)}_{\omega\s}] = \int d^4x \sqrt{g^{(0)}} f(R),
\ee
where $R$ is the Ricci scalar of $g^{(0)}_{\omega\s}$. It was shown in \cite{GKNW} that  such  a theory generically still admits de Sitter solutions (albeit with a smaller cosmological constant than that of the ``bare'' 4d gravity) after the full quantum effects of the field theory are taken into account.

The results of \cite{GKNW} suggest that, at least in the holographic context, IR effects from a QFT do not always  destroy de Sitter space-time. However, that analysis could only be applied   to {\it constant curvature} 4d space-times.  Therefore, it says nothing about the stability of the solutions under non-homogeneous perturbations. The  form  (\ref{intro4})  of the effective action is only good for obtaining maximally symmetric solutions, and it would be misleading to expand it in perturbations around one such background. Rather, to study small perturbations of a holographic QFT coupled to gravity one has to go back to the original theory (\ref{intro1}) and study its perturbation spectrum.
This has been considered already in \cite{anomalyinflation1}, albeit in a slightly different context (Randall-Sundrum cosmology), and more recently in \cite{Chesler} for the case of a two derivative gravitational action coupled to a holographic CFT around de Sitter space.
In \cite{Chesler}, beyond the scalar instabilities that have been known for several decades, \cite{VilenkinStaro},  a spin-2 instability was found for small enough de Sitter curvatures.

Our goal in this paper is to extend previous results on the stability of maximally symmetric spacetimes due to quantum effects in several ways:

\begin{itemize}
\item We consider not only de Sitter but also flat space and Anti-de Sitter space.

\item We consider a classical gravitational theory with all couplings necessary for renormalization which are relevant at low energies. This implies that we have a cosmological constant and Einstein terms as well as the two independent $R^2$ terms with (finite) dimensionless renormalized couplings $\a,\b$.

\item In our case, the only non-trivial (i.e. in which the CFT degrees of freedom participate) quadratic action for the fluctuations is that in the spin-2 sector.  We analyze not only the possible tachyonic poles that are responsible for the instabilities but also the presence of negative residues that signal the presence of ghosts

\item Moreover, we investigate when such tachyons or ghosts are below the effective UV cutoff for the classical gravitational theory, which is given by the so-called species cutoff, \cite{dvali2}.

\item Although we use holographic techniques to solve our problems, {\it our results apply to all four-dimensional CFTs}. The reason is that {\it our results depend only on the  two-point function of the energy-momentum tensor of the CFT.  For a CFT around a maximally symmetric space, and  given a coordinate system and an invariant state, such a two-point function is universal up to an overall multiplicative constant that is the central charge $c$ of the CFT.}

\item Our results do not apply to non-conformal theories. However, as in the case of reference \cite{GKNW},  if we consider a QFT$_4$  as a flow between CFT$_{UV}$ and $CFT_{IR}$, then our  results relevant to CFT$_{UV}$, $CFT_{IR}$, provide bounds on the results relevant to QFT$_4$. This is expected to be valid for generic QFT$_4$     although special cases may need further analysis. The analysis for general QFTs will be done in a future publication.

\item Our results on ghost instabilities are independent of the choice of coordinates on the maximally symmetric space in which the CFT lives.

\item In the case of de Sitter, the presence of tachyonic instabilities  may depend on the choice of coordinates. We have performed the analysis in global, cosmological and static coordinates, and found that the conditions for the absence of tachyonic instabilities are the same in all three cases.

\end{itemize}

Concretely, we consider  general perturbations around maximally symmetric four-dimensional space-times  in a holographic setup, given by (\ref{intro1}),  where the matter content is a four-dimensional holographic conformal field theory. The coupling to gravity is entirely described by the exact one-point correlation function of the CFT stress-energy tensor, as in equation (\ref{intro3}). This stress-tensor is obtained on a boundary of $AdS_5$ from a holographic calculation and has the correct conformal Weyl anomaly (see \cite{duff} for a review of the conformal quantum anomaly, and \cite{HSS} for the holographic CFT stress-tensor).

The perturbation analysis is obtained by fluctuating the bulk and boundary metrics and then writing the  linearized version of the effective Einstein equation (\ref{intro3}) for the four-dimensional metric $g^{(0)}_{\omega\s}$. We perform this analysis  around maximally symmetric 4d space-times of positive curvature (dS), negative curvature\footnote{The case of negative curvature is special. It corresponds to foliating AdS$_5$ by AdS$_4$ slices, which leads to a geometry with two {connected}  boundaries, which is dual  to two copies of the CFTs with {an interface between them }(see e.g. \cite{Janus} and the recent discussion in \cite{AdS1}). One has then different options on how to couple dynamical gravity to the system, the most general case being a bi-gravity theory with each metric coupled to one of the CFTs. Here, we discuss the special cases in which only one metric is dynamical.}  (AdS) and vanishing curvature (Minkowski).

In this work, we pursue two main objectives:
\begin{enumerate}
\item   We first  obtain analytic  spectral equations  (in terms of transcendental functions)  for the boundary metric fluctuations  of the 4d gravity+ holographic CFT system, around a general maximally symmetric background;
\item We then perform a full numerical analysis of  the spectrum in momentum space, (defined by the eigenvalue of the Laplacian on the corresponding maximally symmetric space-time) and determine criteria for the presence of  instabilities  of both tachyonic and ghostlike type.
 This way, we obtain a detailed  map of  the stable and unstable regions of parameter space.
\end{enumerate}

Our approach is similar in spirit to other works in the context of weakly coupled field theory: a similar analysis was performed with a matter content given by quantum corrections of free massless scalar CFT \cite{VilenkinStaro}, for {\it homogeneous}  time-dependent metric perturbations around de Sitter.  A similar perturbation analysis around flat space was carried out for a free scalar coupled to higher-derivative gravity in \cite{Q7a}.

Here,  we use the holographic setup to perform a full parameter-space analysis of the gravity+CFT system, and we establish stable and unstable regions of parameter space around background solutions with zero, negative and positive constant curvature. Specifically, the parameters of the model are:
\begin{itemize}
\item The  boundary gravity renormalized couplings $\Lambda, G, \alpha, \beta$, corresponding to local covariant functions of $g^{(0)}_{\omega\s}$ of dimension up to four, which we define as follows:
\be \label{intro4.1}
{\Lambda\over 8\pi G} \sqrt{g^{(0)}},  \quad  {1\over 16\pi G} \sqrt{g^{(0)}} R, \quad {\alpha \over 384 \pi}  \sqrt{g^{(0)}} R^2, \quad {\beta \over 64\pi} \sqrt{g^{(0)}} \left[ R_{\omega\s}R^{\omega\s} - {1\over 3}R^2 \right]
\ee
Here $\Lambda$ is the 4d cosmological constant, $G$ is  4d Newton's constant,    $R_{\omega\s}$ is the Ricci tensor of  $g^{(0)}_{\omega\s}$ and $R$ the corresponding Ricci scalar and $\alpha,\beta$ two dimensionless parameters.

\item The parameter $N$, counting the degrees of freedom of the CFT. This can be traded for the central charge of the CFT.

\item  The value  $R$ of the curvature  of the background solution. This is not an independent parameter, as it is determined by the other parameters via the background solution. However, it is convenient to use this as an independent parameter instead of e.g. the 4d cosmological constant.

\item An extra parameter is the {\it renormalization scale} $\mu$ of the CFT, which arises from the conformal anomaly. Since this is universal, its   only effect is  to shift the parameter $\beta$ so that it enters only in the combination:
\be \label{intro4.2}
\beta_\text{eff} = \b - {N^2\over \pi} \log \left(4 \mu^2 G N^2 \right).
\ee
\end{itemize}

 Although the  QFT we couple to gravity is a holographic CFT$_4$, it is important to stress that  our result holds for {\it any} generic CFT coupled to gravity: as we shall observe  below, the spectral properties of the system are determined by the stress-tensor 2-point function, which for a CFT  in any conformally flat space-time is completely fixed by the central charge\footnote{In general such a two-point function depends on initial conditions or equivalently, the initial state. We assume that we use the maximal symmetry invariant state. In de Sitter, this is  the Bunch-Davies vacuum.}. Therefore, although the method we use to compute the fluctuation spectra is specific to a large-$N$ holographic CFT, to obtain the result for a generic CFT it is enough to trade the parameter $N$ with the appropriate central charge.

\subsection{Cutoffs in effective gravity theories}

In order to have semiclassical 4d gravity, we require that the 4d curvature $R$ is small compared to the cutoff of the theory.
Typically, this is assumed to be the Planck scale or, in the case of string theory,  the string scale, but it may be different from both in other possible realizations.

  However, as argued in \cite{dvali2}, if the matter theory has many degrees of freedom, the  perturbativity condition imposes a lower cutoff (the so-called ``species scale'') than the Planck scale. For example if, like in our case, the matter theory has ${\cal O}(N^2)$ degrees of freedom\footnote{For a general CFT$_4$, a measure of the number of degrees of freedom is the central charge $a$. One can always replace everywhere in the paper $N^2\to 4 a$ as explained later on in section \ref{aa}.}, and the original cutoff was the Planck scale, then  the species cutoff $M_\text{species} $ is
\be
M_\text{species} \equiv {M_\text{Planck} \over N} \sim {1\over \sqrt{G} N}.
\label{sp-a}
\ee
One reason for this is that the bare Planck scale receives corrections that are ${\cal O}(N^2\Lambda^2)$ where $\Lambda$ is the overall cutoff of the gravity+QFT.
For $N\gg 1$,  $M_P\sim N\Lambda$ and then solving for $\Lambda$ we obtain (\ref{sp-a}).
Another reason may be that one would like to keep perturbativity, which  implies that the QFT corrections to $M_P$  (which are of order $N\Lambda$) should be small:  demanding $N\Lambda \ll M_P$  again implies that $\Lambda \ll {M_P\over N}$.

However, both arguments have loopholes.

$\bullet$ The corrections to the Planck mass, being quadratic in the cutoff are very difficult to calculate in a theory, preserving diffeomorphism invariance.
The $\e$-expansion is not well suited for this calculation as was extensively argued in \cite{Pl0}. The only reliable calculations of the coefficients of the  ${\cal O}(N^2\Lambda^2)$ corrections to the Planck mass have been done in  many ground-states of string theory at the one-loop level,
\cite{Pl1}-\cite{Pl5}. In such cases, the coefficients can be either positive or negative depending on the type of matter fields that are integrated out.
Moreover,the duality frame matters, in the sense that on-shell equivalent states give different one-loop corrections to the Planck scale.
In short, generic coefficients are of order one, but fine-tuning is possible (like in susy theories where such corrections may vanish depending on the amount of supersymmetry.

$\bullet$ The original argument for the species cutoff has another loophole:  the cutoff $\Lambda$ should be a physical scale, ie. a scale associated with a relevant coupling constant of the theory.
If for example, we have gravity and a CFT as in our examples, the CFT has no intrinsic cutoff.
We need to introduce one to renormalize the combined CFT+gravity theory, but we can then remove it by keeping the renormalized couplings finite.
In that case, the renormalized Planck scale is completely independent of $N$.

$\bullet$ There is no reason why one should require that matter corrections are perturbative, especially if such corrections can be computed beyond perturbations theory, a fact that may happen in holographic or supersymmetric theories.

The conclusion of the above discussion is that although assuming the validity of the species cutoff is a rather conservative approach, there may be many cases where the cutoff may be different and in particular much higher than the species cutoff. Our results will be presented at all scales, and one can then apply the relevant cutoff before concluding.

Generically, the effective action of gravity may contain higher derivative terms before adding the effects of the CFT. These terms can be schematically written as
\be
S_{grav}=\bar M^2_\text{Planck}  R+\bar\a R^2+\sum_{n=1}^{\infty}\bar a_n R^{n+2}
\label{nn1}\ee
The bar over the coefficients indicates they are bare coefficients that arose from integrating out the high-energy degrees of freedom of the gravitational sector.
However, these bare coefficients will be renormalized by the CFT$_4$.
To do this we must introduce an arbitrary cutoff $\Lambda$ in the CFT$_4$, perform our quantum calculations and write the effective corrections to the bare gravitational couplings.
All such corrections will be controlled by the number of degrees of freedom ($N^2$) of the CFT$_4$ as well as the cutoff scale.
\be
\delta \bar M^2_\text{Planck} \sim N^2 \Lambda ^2\sp \delta\bar\a \sim N^2 \log{\Lambda^2}\sp \delta \bar a_n\sim N^2 \Lambda ^{-2n}
\ee 
We now add counterterms to remove the divergent contributions as $\Lambda\to\infty$, and we then take $\Lambda\to \infty$ obtaining  
\be
S= M^2_\text{Planck}  R+\a R^2+\sum_{n=1}^{\infty}\bar a_n R^{n+2}+W_{CFT}(R)
\label{nn2}\ee
where now $M^2_{\rm Planck}$, $\a$ are renormalized couplings (independent of $N$), $a_{n\geq 1}$ are unaffected by renormalization, and $W_{CFT}(R)$ are the renormalized CFT$_4$ contributions to the gravitational action.

Typically the size of $a_n$ is controlled by the original bare $\bar M^2_{\rm Planck}$ as
\be
a_n\sim {1\over \bar M^{2n}_{\rm Planck}}
\ee 
As long as $M_{\rm Planck }\ll \bar M_{\rm Planck}$ the higher curvature terms beyond 
the quadratic ones can be neglected, and this is what we shall assume in this paper.

Moreover, as $W_{CFT}\sim N^2$, it is clear that our results will not depend separately on $M^2_{\rm Planck}$ and $N^2$ but only on the combination ${M^2_{\rm Planck}\over N^2}$ which is the species scale.

The same considerations about the cut-off apply to  the analysis of fluctuations and in particular  of instabilities.
In effective field theory, any mode with a mass of the order or above the cutoff is outside the reach of the theory.
In particular,  it is only when the ghost or tachyon mass is well below the EFT cut-off that one can unambiguously conclude that the theory truly has stability issues.

Indeed, as it is well known (and as we review in a simple example in Appendix \ref{app:free scalars}), an EFT originating from an otherwise healthy UV theory may display some unstable modes as an artefact of the low-energy expansion. In  this case, the unstable modes will have masses of the order of the EFT cut-off (the scale of the fields that have been integrated out). Turning this around, we can say that  one cannot conclude anything about the actual stability or instability  of an EFT based on  the occurrence of ghosts or tachyons whose mass scale is at or above the EFT cut-off: one would have to know the UV completion to reach a definite conclusion. For the same reason, within EFT one cannot reach any definite conclusion about the stability of space-times whose curvature scale ($H$ or $\chi$) is at or larger than the cut-off.

Throughout this work, we shall encounter many instabilities (tensor and scalar ghosts and tachyons).
Whether they are physical, depends on whether they are above or below the cutoff of the effective theory.
We may take the attitude that the cut-off is the one in (\ref{sp-a}) but other cutoffs can be in principle envisaged as we can freely choose our renormalized parameters.

An important point is the following. It seems that our previous argument indicates that the condition on tachyons and ghosts depends on the combination $M_{\rm species}={M_{\rm PLanck}\over N}$. This in turn implies that for scales well below this cutoff, the CFT$_4$ corrections should be negligible.
 This argument is correct for local contributions to the dynamics of fluctuations.
However, as we shall later see, most of the relevant contributions are non-local due to massless modes being integrated out. Such non-local contributions, due to logs can
become large, and therefore the question of instabilities is non-trivial.

We devote the rest of this introduction to an extended summary of these techniques as well as a discussion of  the results obtained in this work. Although they stem from a holographic calculation, we insist that they are valid for a generic CFT, and  describe the results mostly in the language  of the field theory side. We leave the details of the holographic approach to the rest of the paper.

\subsection{Summary and results} \label{sec:intro-sum}

The setup studied  here consists of higher curvature gravity coupled to a Conformal Field Theory (CFT). For the gravitational part, we  consider the Einstein-Hilbert action with a cosmological constant plus quadratic curvature term:
\be \label{res-0}
S_\text{grav} = S_{EH} + S_2
\ee
where
\be \label{res-1}
S_{EH} =- \frac{1}{16 \pi G}\int d^4x\sqrt{-g^{(0)}}(R - 2\Lambda)
\ee
and
\begin{equation} \label{res-2}
S_{2} = \frac{\a}{384 \pi}\int d^4x\sqrt{-g^{(0)}}R^2 + \frac{\b}{64\pi} \int d^4x\sqrt{-g^{(0)}} \left( R_{\omega\s}R^{\omega\s}-\frac{1}{3}R^2 \right)
\end{equation}
 Here, $g^{(0)}_{\mu\nu}$ is the 4d  space-time metric, $g^{(0)}$ its determinant and $R$ its curvature. The parameters $G,\Lambda,\alpha,\beta$ are the (finite) renormalized parameters which already contain the contributions of the CFT\footnote{These  parameters are defined so that they are finite in an appropriate scaling limit after removing the UV cut-off. The detailed procedure is described in section \ref{section:action}.  }

Coupling to a large-$N$ CFT is implemented via holography: we identify the 4d space-time with the conformal boundary of AdS$_5$ bulk manifold, and $g^{(0)}_{\mu\nu}$ with the leading term in the Fefferman-Graham  expansion of the 5d  bulk metric:
\be \label{res-3}
ds^2_{bulk} = L^2 {d\rho^2 \over 4\rho^2} + {1\over \rho}\left[g_{\mu\nu}^{(0)} + O(\rho) \right] dx^\mu dx^\nu \qquad \rho \to 0
\ee
where $\rho \to 0$ corresponds to the AdS boundary and $L$ is the AdS length.
This way, one can obtain 4d maximally symmetric metrics $g^{0}$ whose Ricci curvature $\bar{R}$ satisfies the relation:
\be \label{res-5}
\Lambda=\frac{1}{4} \left( \bar R-\frac{GN^2 \bar R^2}{48\pi} \right).
\ee
The Ricci curvature $\bar R$  of the  maximally symmetric background space-time can be positive, negative or vanishing. It is convenient to parametrize it in the various cases as follows:
{\be
\bar R= \left\{ \begin{array}{lll}
\displaystyle 12H^2,&\phantom{aa} &{\rm de~~ Sitter},\\ \\
\displaystyle 0 ,&\phantom{aa}&{\rm Minkowski} \\ \\
\displaystyle -12\chi^2 ,&\phantom{aa}& {\rm Anti~~ de~~ Sitter}
\end{array}\right.
\label{bc}\ee}
The parameter $N$, characterizing the number of degrees of freedom of the CFT,   is related to the bulk Planck scale $M$ and bulk AdS length $L$ by  $N^2 \propto (M L)^3 $.  The first term in equation (\ref{res-5})  is the contribution from the vacuum Einstein equation, and the second term is the CFT contribution.

To conclude, the {\it independent} parameters of the theory are the curvature of the background space\footnote{Even if  $\bar{R}$ is not a parameter in the action, we can trade  $\Lambda$  for $\bar{R}$ using (\ref{res-5}). Although the relation between $\Lambda$ and $\bar{R}$ is not one-to-one, by scanning over all values of $\Lambda$ we can obtain any value of $\bar{R}$.} $\bar R$, the four-dimensional Newton constant $G$, the two $R^2$ couplings $\a$ and $\b_\text{eff}$, where $\beta_\text{eff}$ is defined in (\ref{intro4.2})  and the number of colors\footnote{In fact, one should take the central charge as a parameter for the CFT.  We shall use $N$ as a proxy for the central charge} $N$ of the holographic CFT$_4$. It will be convenient to  express  quantities in terms of the following  ``reduced'' parameters:
\be\label{tildes}
\tilde{\alpha} = {\pi \alpha \over N^2}, \qquad \tilde{\beta}_\text{eff} = {\pi \beta_\text{eff} \over N^2}.
\ee

Our goal is to determine, as a function of the parameters of the model,  the spectrum of gravitational  fluctuations of the boundary metric  around any maximally symmetric  4d boundary metric $\bar{\zeta}_{\m \nu}$. We  use this information to determine the perturbative stability of the system.

The perturbed boundary metric is taken to be:
\be
g^{(0)}_{\mu\nu} = \bar{\zeta}_{\m \nu} + \delta \zeta^b_{\mu\nu}
\ee

In an appropriate gauge, the boundary perturbation  can be written as
\begin{equation} \label{res-6}
\d \z^b_{\omega\s}=\psi\bar{\z}_{\omega\s}+h_{\omega\s}^{(0)}
\end{equation}
where $\psi$ is a scalar degree of freedom, and $h_{\omega\s}^{(0)}$ is a tensor perturbation which is transverse and traceless with  respect to the boundary metric. The scalar is a pure boundary mode\footnote{ {The scalar mode couples to the trace of the stress-energy tensor of the CFT$_4$. Since this theory is conformally invariant, the two-point function of the trace vanishes. Therefore the non-trivial action for the scalar mode is generated by the boundary $R^2$ terms as well as the conformal anomaly of the CFT$_4$,  \protect{\cite{VilenkinStaro}}. If the theory is instead a QFT, extra contributions are expected for the dynamics of the scalar mode.}}, whereas the four-dimensional gravity tensor modes couple to tensor perturbations in the bulk.

The metric perturbations are coupled to the CFT via the  bulk dynamics: the boundary field $h_{\omega\s}^{(0)}(x)$ is the leading term in a near-boundary expansion of the perturbation of the {\it bulk} metric.
 \\

\noindent {\bf Spectral functions}\\
The spectral analysis around the holographic background is tightly connected to the holographic two-point function of the  boundary stress tensor.   When working at linear order in fluctuations  both in the bulk and on the boundary, all one needs is the structure of the effective action (\ref{intro2})  at quadratic order as a function of the boundary metric perturbation $\delta \zeta^b$:
\be \label{intro5}
S_\text{eff}^{(2)} =
\int d^4x  {1\over 2}\delta \zeta^b_{\mu\nu} O_\text{grav}^{\mu\nu,\rho\sigma} \delta \zeta^b_{\rho\sigma}   - {1\over 2} \int d^4x \int d^4y  \, \delta \zeta^b(x)_{\mu\nu}  \, \< T^{\mu\nu}  (x) T^{\rho\sigma}   (y) \>_{CFT}\,  \delta \zeta^b_{\rho\sigma}    (y)
\ee
These two terms correspond to the quadratic order approximation of each of the two terms in (\ref{intro2}):  $O_\text{grav} $ is the local  kinetic operator of the  quadratic term in the 4d gravity action  $S_\text{grav}$ in (\ref{res-0});   $\< T^{\mu\nu} T^{\rho\sigma}  \>_{CFT} $ is the holographic two-point function of the stress tensor, which is by definition:
\be \label{intro6}
\< T^{\mu\nu}(x) T^{\rho\sigma}(y) \>_{CFT} = -  {\delta  \over \delta \zeta^b_{\mu\nu}(x) } {\delta  \over \delta \zeta^b_{\rho\sigma}(y) } S^{on-shell}_{bulk}
\ee

The stress tensor two-point function contains both local and non-local contributions. The local contributions simply renormalize the coefficients of local terms which are already present in $O_\text{grav} $. The non-local contributions are genuine new effects of the CFT which one cannot find in a local gravity theory.

Equation (\ref{intro5})  shows that by computing the holographic two-point function we have access to
the full   propagator, which we denote by ${\cal F}^{-1}$,  of the boundary metric fluctuations: the inverse propagator  is
\be \label{intro7}
{\cal F}^{\mu\nu\rho\sigma} \equiv  O_\text{grav}^{\mu\nu,\rho\sigma} - \< T^{\mu\nu} T^{\rho\sigma} \>_{CFT}
\ee
and the spectrum  of the system are the solutions of the integrodifferential equation
\be \label{intro8}
{\cal F}^{\mu\nu\rho\sigma}  \delta \zeta^b_{\rho\sigma} = 0.
\ee

The linear equation (\ref{intro8})  can be recast into two separate {\it scalar} spectral equations for the scalar and tensor modes defined in (\ref{res-6}),  by going to the appropriate ``momentum space'' of the boundary coordinate. This is done by  decomposing  the modes in eigenfunctions of the d'Alembert operator $\nabla^{2}$ of the background boundary metric $\bar{\z}_{\omega\s}$: in the positive, zero and negative curvature case we take the  fluctuation to satisfy
\be \label{res-8}
\left(\nabla^2 - r {\bar{R}\over 12}\right)\delta\f(x) = \left\{ \begin{array}{ll} -H^2\left(\nu^2 -{9\over4}\right)  \delta \f(x) & \quad dS \\ & \\ -  k^2    \delta  \f & \quad \text{Minkowski}  \\ & \\
\chi^2\left(\nu^2 -{9\over4}\right)  \delta \f(x) & \quad AdS \\\end{array}  \right.
\ee
where $\delta\f(x)$ stands for either $\psi$ or $h^{(0)}_{\mu\nu}$, $r$ is the spin of the perturbation ($r=0$ for $\psi$ and $r=2$ for $h^{(0)}_{\omega\s}$), $H$ is the  Hubble parameter in the case of positive curvature boundary (de Sitter) and  $\chi$ is  the inverse AdS length in the case of negative curvature boundary, as in (\ref{bc}).
For flat space, this is the usual Fourier decomposition where $k^2 = k^\mu k_\mu$.  In both curved cases,  $\nu$ is a dimensionless eigenvalue measuring the invariant ``momentum'' in units of the background curvature.

The values of  $\nu^2$ (or $k^2$ in the flat case) are determined by the spectral equation (\ref{intro8}), which in momentum space becomes a transcendental equation for $\nu^2$   (or $k^2$) of the form:
\be \label{res-9}
{\cal F}(\nu) = 0
\ee
where the precise form of the function ${\cal F}$ depends both on the nature of the mode (scalar or tensor) and the background curvature and the parameters in the action.

Before proceeding further with the results, we discuss here two ingredients that affect the results.

(a) The coordinate system used on the maximally symmetric space (AdS, dS, flat) on which the CFT$_4$ is defined. It is a well-known fact that quantum field theory data, like correlation functions, do depend crucially on the coordinate system. In flat space, we use only Minkowski coordinates. In AdS, we use both Poincar\'e and global coordinates.
In dS, we examine global coordinates, Poincar\'e coordinates and static patch coordinates.

(b) The state on which the two-point function of the energy-momentum tensor is calculated. In flat space, we choose the (unique) Poincar\'e invariant vacuum.
In global AdS and Poincar\'e AdS, similarly, we choose the AdS invariant vacuum.
In Global coordinates dS as well as Poincar\'e coordinates dS we choose again the unique dS Invariant state that in the latter case is known as the Bunch-Davis vacuum.
In dS with static coordinates, we choose the dS-invariant vacuum corresponding to outgoing boundary conditions at the cosmological horizon.

Overall the correlator we compute is the Lorentzian retarded correlator. This is defined for real eigenvalues of the Laplacian on (AdS, dS, flat) space.
In the space-like case, this correlator is similar to the Euclidean correlator. Its analytic continuation to the complex plane is unique. In the time-like case, extra imaginary parts arise from the logarithmic branch cut of the correlator, but these do not affect the analytic continuation.

The expressions obtained can be found below: \\

\noindent{\bf Scalar mode}
In this case, the inverse propagator is a polynomial in $\nu^2$ (or $k^2$), because it results from a quadratic action which is local on the boundary. The expression of the inverse propagator is given by
\begin{itemize}
\item {\bf Minkowski}
\be \label{res-10}
{\cal F}_{scalar}(k) =  -{3\over 16 \pi G}\left( k^2 -  {4\over \alpha G } \right)
\ee
\item {\bf de Sitter and Anti-de Sitter }
\be \label{res-11}
{\cal F}_{scalar}(\nu) =  -{1\over 64 \pi G}\left[\a G \bar{R} - 12 + {GN^2\bar{R}\over 2\pi}\right]\left\{ {4\over G\a} - {N^2 \bar{R} \over 6\pi\a} - {\bar{R}\over 12}\left(\nu^2 -{9\over 4}\right) \right\}.
\ee
This is the "physical" scalar inverse propagator. For the details, see section \ref{bdy pert}.
\end{itemize}

\noindent{\bf Tensor modes} For tensor modes, the non-local contribution from the CFT stress-tensor correlator in (\ref{intro7}) gives rise to non-polynomial expressions for the inverse propagators:
\begin{itemize}
\item {\bf Minkowski}
\be \label{res-12}
{\cal F}_\text{tensor,Mink}(k) = {N^2 \over 64\pi^2} k^2 \left\{-{2\pi \over GN^2} + {k^2\over 2}\left[{1\over 2} - 2\g_E - \log\left(GN^2k^2\right)- {{\pi\b_\text{eff}\over N^2}}\right] \right\}
\ee
\item {\bf de Sitter}
 \bea \label{res-13}
 {\cal F}_\text{tensor,dS}(\n) = && {N^2 H^2 \over 64\pi^2} \left(\nu^2 - {9\over 4}\right) \left\{1 - {2\pi\over GN^2H^2} + {2\pi \alpha\over N^2} - {1\over 2}\left(\n^2-{1\over 4}\right)\bigg[ \right.  \nn \\
  && \left.\left. 2\log\left(GN^2H^2\right) -{1\over 2} +  2\mathcal{H}\left(\n-\frac{1}{2}\right)  + {\pi\beta_\text{eff}\over N^2}\right]\right\}.
\eea
where $\mathcal{H}$ is the harmonic number function defined in (\ref{hn}).
The expression (\ref{res-13})  with  $\a=0$  was already obtained in \cite{Chesler}\footnote{In \cite{Chesler}  $\b$ was fixed but the renormalization scale $\mu$ (called E in that paper) was allowed  to vary.}. In this work,  we rederive it in our setup and generalise it to negative and zero curvature and arbitrary values of the $\alpha$ parameter.
\item {\bf Anti-de Sitter}\\
In this case, there are two connected  boundaries, corresponding to two - a priori independent - copies of the CFT. Therefore, one has freedom in how to couple 4d gravity to the system.
Here, we discuss two concrete cases: \\
\textbf{a) Dynamical gravity on one side:} In this case only one of the two CFTs is coupled to dynamical gravity, and the metric on the second boundary is frozen.
\bea\label{res-13a}
{\cal F}_\text{tensor,AdS}^{-}(\n) =&& {N^2 \chi^2 \over 64\pi^2}\left(\n^2-\frac{9}{4} \right) \left\{ 1 + {2\pi\over N^2}\left({1\over G \chi^2} + \a \right) +\right.  \nonumber \\
   && \left. - {1\over 2} (\n^2- 1/4)\left[ {\pi\beta_\text{eff}\over N^2} +  \log\left(GN^2\chi^2\right) - {1\over 2} +  \nn  \right. \right. \\
&&\left. \left. +  \mathcal{H}\left(-{1\over 2}-\n \right) + \mathcal{H} \left(-{1\over 2} + \n \right) \right] \right\}.
\eea
\textbf{b) Symmetric boundary conditions:} In this case, there is effectively a single boundary (see \cite{AdS1} for a recent discussion),  and there is again a  single dynamical gravity theory coupled to a  single 4d  CFT on AdS. This leads to the following spectral density:
\bea \label{res-14a}
{\cal F}_\text{tensor,AdS}^{sym}(\n) =&& {N^2 \chi^2 \over 64\pi^2}\left(\n^2-\frac{9}{4} \right)\left\{   1 + {2\pi\over N^2}\left({1\over G \chi^2} + \a \right)  + \right.  \nonumber \\
 && \left.  - {1\over 2} (\n^2- 1/4)\left[ {\pi\beta_\text{eff}\over N^2} + \log\left(GN^2\chi^2\right) - {1\over 2} \right. \right. \nn \\
 && \left. \left. + \mathcal{H}\left(\n-\frac{1}{2}\right) + \mathcal{H}\left(-\n-\frac{1}{2}\right) - {\pi\over \cos\pi\n}\right] \right\}.
\eea
\end{itemize}

\noindent {\bf Stability}\\
Instabilities of the system are encoded in the properties of the zeros of ${\cal F}$. We perform a full analysis of all parameter space, which we summarize below.   As a byproduct, by setting $N=0$  we  obtain the pure gravity spectral functions and study the corresponding zeros, which give indications about the stability of quadratic gravity around any constant curvature background.

When discussing the gravity + CFT system, we  always compare the results with those of pure gravity theories with the appropriate renormalized parameters. This allows us to identify the new effects (if any) which arise specifically from the coupling to the CFT.
 For a CFT with parameter $N$, the comparison should be made by choosing the pure gravity parameters $\alpha$ and $\beta$  such that $\alpha = \tilde{\alpha}/\pi$ and $\beta = \tilde{\beta}_\text{eff}/\pi$, in terms of the quantities  defined in (\ref{tildes}): these are the quantities which are expected to be of order unity after renormalization of the local terms by  the CFT is taken into account.

Depending on  the curvature, there are different criteria for  instabilities. On any background,  instabilities can be of  two types:
\begin{itemize}
\item {\bf Tachyonic instabilities} correspond to modes which grow  exponentially in time and are related  to the {\it position} of the root $\nu$ in the complex plane. Specifically, a root $\nu$ of ${\cal F}(\nu)$ is {\it tachyon-stable} in the  following cases:
\be \label{res-14}
\left\{ \begin{array}{ll} \left| Re (\nu)\right| \leq {3\over 2}   & \quad dS \\ & \\ k^2 \leq 0  & \quad \text{Minkowski}  \\ & \\
Re(\nu) \neq 0 & \quad AdS \\\end{array}  \right\} \quad \Rightarrow \quad \text{tachyon-stable}
\ee
In all other cases, the mode is tachyonic.

 Since they were derived requiring a bounded late-time behaviour for the modes, one may worry that the bounds (\ref{res-14}) depend on the coordinates chosen on the slice, and in particular on the choice of the time coordinate. In space-times endowed with a  global time-like killing vector, there is a preferred choice of time coordinate and the bounds (\ref{res-14}) for Minkowski and AdS translate into the usual ones, i.e. respectively positive mass squared and validity of the BF bound (the latter can be obtained equivalently both in global AdS coordinates or in Poincar\'e coordinates, although global time and Poincar\'e time do not coincide).

  In dS, however, things may be more subtle. We have derived  the tachyon-stability criterion for   the three more widely used local  coordinate systems in de Sitter, namely global coordinates, Poincar\'e coordinates and static patch  coordinates. In all three cases the condition one obtains is the same bound (\ref{res-14}). This analysis relies on tracking the time-dependence of eigenfunctions of the D'Alembertian in different coordinate systems and can be found -in the case of scalar modes- in section  \ref{sec:scalar tachyons}.  In the case of tensor modes,  after decomposing  them further into irreducible tensors of  the fixed time-slice symmetry group, one finds the same conditions as for scalars, as can be seen from the analysis in Appendix \ref{dS criterion}.

\item {\bf Ghost instabilities} correspond to a  mode  with eigenvalue $\nu_0^2$ (or $k^2_0$) developing a ``wrong sign'' kinetic term, and are related to the value of the residue of ${\cal F}$ at the pole:
\be \label{res-15}
\left\{ \begin{array}{ll} Res{\cal F}^{-1}(\nu_0^2) < 0   & \quad dS \\ & \\ Res{\cal F}^{-1}(k^2_0) < 0 & \quad \text{Minkowski}  \\ & \\
Res{\cal F}^{-1}(\nu_0^2) > 0 & \quad AdS \\\end{array}  \right\} \quad \Rightarrow \quad  \text{ghost-stable}
\ee
The sign conventions are discussed in  section \ref{decoupling}.  In the case of ghosts, the stability criterion  only depends on the sign of the residue  and not on the choice of coordinates.

A heavy ghost can be tolerated if its mass is above the cut-off of the theory because in this case it cannot be described in the context of effective theory (and it may become healthy in the UV-completion).

In this work, we compare ghost masses with two cut-offs: the 4d Planck scale $G^{-1/2}$ (the ultimate cut-off in the semiclassical approach) and the  species cutoff scale,
$M_{\rm species}$
\be
M_{\rm species}\equiv (GN^2)^{-1/2}\;,
\label{spec}\ee
which can be argued to be the true cut-off of a gravity theory coupled to $N^2$ degrees of freedom \cite{dvali2}. Moreover, it seems that  the latter  is the natural scale in which to measure boundary  curvature $R$  in the present set-up: it  always appears in the combination
$$G N^2 R\sim {R\over M_P^2 N^2}\sim {R\over M_{\rm species}^2}\;.$$
\end{itemize}

An unstable mode can be a ghost, a tachyon, or both.  In what follows we summarize our results in the scalar and tensor sector and for zero, positive and negative background curvature.
One important point to which we have to pay attention is whether the unstable mode is within the limits of effective field theory, i.e. whether it is light in Planck units (in the case of pure gravity) or light in units of the species scale (\ref{sp-a})  (in the case of gravity coupled to the CFT). \\

\noindent {\bf Stability in the scalar sector}\\
For the scalar mode, it is straightforward to read off the conditions (\ref{res-14}-\ref{res-15}) from equations (\ref{res-10}-\ref{res-11}): this leads to the following conclusions:
\begin{itemize}
\item In Minkowski space, the scalar spectral function (\ref{res-10})  is the same in pure gravity and in the presence of the CFT and does not depend on $N$ as the conformal anomaly of the CFT is not relevant. The scalar mode is never a ghost, and it is  tachyonic if $\alpha > 0$. This agrees with previous analysis (e.g. \cite{Q7a}). The tachyonic mode is within the bounds of the theory if its mass is below the cut-off, which in terms of the ``reduced'' $\tilde\alpha$ parameter defined in (\ref{tildes}) requires  $\tilde{\alpha} \gg 1$ (the same condition as in pure gravity, since $\alpha = \tilde{\alpha}$).

\item In de Sitter space,  {scalar} tachyon-stability requires:
\begin{equation} \label{res-17}
    \frac{1}{\a}\left(1-\frac{GN^2 H^2}{2\pi} \right)\leq 0.
    \end{equation}
For consistency, $GN^2 H^2\ll 1$ and therefore the second factor is always positive in effective field theory.
    Therefore, tachyon stability implies $\a<0$.

Moreover, the scalar mode is a ghost if:
\be\label{res-19}
\left({\pi \alpha \over N^2} + {1\over 2} \right) {G N^2 H^2 \over 12 \pi} > 1.
\ee
{ie. when $\a\gg 1$}.
For pure gravity, (\ref{res-17}) with $N=0$ is the same condition ($\alpha<0$) as for Minkowski space. In pure  de Sitter gravity, the scalar can also be a ghost, if $\alpha$ is very large (at least of order $1/(GH^2) \gg 1$). This mode is light in Planck units if $|\alpha| \gg 1$.

In the presence of the CFT, the tachyon stability condition is modified by the second term proportional to $N^2$ in  (\ref{res-17}). However, note that this term is small if we insist the curvature is below the species cutoff, which requires $GN^2H^2 \ll 1$. If this is the case,  the tachyon-stability condition is not affected much by the CFT in the context of low-energy EFT\footnote{A notable case in  which this condition is violated is the Starobinsky realization of de Sitter (or more generally, inflation), in which the cosmological constant term is absent and the de Sitter curvature is fixed to $GN^2H^2 = 4\pi$ \cite{St, VilenkinStaro}. In this case, the tachyon stability condition is reversed to $\alpha >0$. We comment on this case in Appendix \ref{staro}.}.
The scalar mode is below the species cut-off if $|\tilde{\alpha}| \gg 1$ (the same condition as for pure gravity). Finally, both with and without the CFT   the  time-scale $\tau$ of the tachyonic instability is roughly the inverse tachyon mass, $\tau \sim \sqrt{G|\alpha|}$. In effective field theory ($GH^2 \ll 1 $) this is much faster than the de Sitter Hubble rate $H^{-1}$ (i.e. the tachyon instability is very strong) unless $|\alpha| \gg (GH^2)^{-1}$. Therefore, for $\tilde{\alpha}$ in the interval
\be
1 \ll \tilde{\alpha} \ll {1\over GH^2 N^2}
\label{strtacdS}
\ee
we have a strong instability (faster than one Hubble time) within effective field theory. This condition also applies to pure gravity, if we set $N=1$ and $\tilde{\alpha} = \alpha$.

\item The discussion is similar for Anti-de Sitter. Scalar tachyon-stability requires:
\be \label{res-18}
 {9\over 4} - {4\over \a G\chi^2}\left(1 + {GN^2\chi^2\over 2\pi}\right) \geq 0.
\ee
and the scalar mode is a ghost if
\be\label{res-19-ii}
-\left({\pi \alpha \over N^2} + {1\over 2} \right) {G N^2 \chi^2 \over 12 \pi} > 1.
\ee
Like before, to be in the effective field theory we must require that
$G N^2 \chi^2\ll 1$.
The condition for the scalar modes (whether a tachyon or a ghost) to be within the bounds of effective field theory is $|\tilde{\alpha}| \gg 1$.
\end{itemize}

\noindent {\bf Stability in the tensor sector}\\
Unlike the case of the scalar, exploring the roots of the tensor spectral function can only be done numerically, except in some corners where analytic approximations for the transcendental functions can be used (in particular the large eigenvalue limit $\nu\to \infty$). Below we give the broad features of the stability results  in the three cases (zero, positive and negative curvature). More details can be found in the main body of the paper.

In each case, we emphasize what happens for two  special parameter values:

(a)  $N=0$ which corresponds to pure gravity with higher curvature terms;

(b) $\alpha=\beta_\text{eff}=0$, which  corresponds to setting the (renormalized) local quadratic curvature terms to zero. This   gives a measure  of the truly non-local contributions from the CFT.

\begin{itemize}
\item {\bf Minkowski}\\

\begin{itemize}

\item In the special case of  pure gravity ($N=0$), for $\beta \neq 0$ (and independently of $\alpha$)  the quadratic Ricci tensor term always generates a ghost, whose mass is,
\cite{Stelle1978},
\be
m^2_{ghost} = {4\over \beta G}.
\ee
For $\beta<0$ this is also a tachyon. This mode is light compared to the cut-off $G^{-1/2}$  when $|\beta| \gg 1$.
Therefore, the gravity theory is a good and stable effective theory only if $\beta$ is positive and $\beta\lesssim 1$.

\item In the presence of the CFT, the spectral function is  (\ref{res-12}) and its  non-trivial roots   are the solutions of a transcendental  equation of the type $X \log X = a$, where $X$ is proportional to $k^2$ and $a$ is a real constant. The analysis can be done semi-analytically and it leads to the conclusion that for any value of $\tilde{\alpha}$ and $\tilde{\beta}_{\text{eff}}$ {\it Minkowski space always contains two tachyonic tensor modes}. The theory becomes eventually tachyon-stable only in the extreme limit $\tilde{\beta}_{\text{eff}} \to +\infty$.
In this limit,  one always finds a light  ghost  (light compared to the ``species'' scale $(G N^2)^{-1/2}$), as in the pure gravity case. All in all, the masses of the unstable tensor modes are above the species cut-off for $O(1)$ values of $\beta_{\text{eff}}$ (this includes  the special case $\a=\beta_{\text{eff}}=0$), while Minkowski space is unstable within EFT iff $|\tilde{\beta}_{\text{eff}}|\gg 1$ and independently of $\tilde{\alpha}$.

\end{itemize}

\item {\bf de Sitter}

\begin{itemize}
\item  In the special case of pure higher curvature gravity ($N=0$), there are always two tensor modes, one of which is the massless graviton, and the other is massive. The massive mode is  tachyonic if the following condition is violated:
\be \label{ds tach intro}
{2\over \beta} \left(\alpha - {1\over GH^2}\right) < 1.
\ee
Because $GH^2 \ll 1$, the condition (\ref{ds tach intro}) is violated if $\beta< 0$ for $\alpha$ and $\beta$ of order unity (this matches the Minkowski result).

 Whether or not (\ref{ds tach intro}) holds, either mode is necessarily a ghost. If $\beta - 2\alpha < 2 (GH^2)^{-1}$, the ghost is the massless spin-2 mode, otherwise it is the massive mode. For $O(1)$ values of $\alpha$ and $\beta$, the ghost is the massive mode, and its mass is of  order  ${\cal O}(M_p)$. One can have a light ghost only if $\alpha \gg  (GH^2)^{-1} \gg 1$ (in which case the ghost is the massless graviton)  or if $|\beta| \gg 1$ (in this case which one is the ghost depends on the sign of $\beta$). { All in all,   in pure gravity the theory is stable and ghost-free within EFT (i.e. below the cut-off $M_p$) if $\alpha$ and $\beta$ are both $O(1)$. }

\item We now turn to the case of gravity coupled to the CFT. In de Sitter,
 the presence of  tachyonic  tensor modes depends on the curvature, on $N$  and the parameters $\tilde{\alpha}$ and $\tilde{\beta}_{\text{eff}}$. The dS curvature $H$  always enters in the combination $GN^2H^2$, i.e. the natural scale to which the curvature is compared is the ``species'' scale (\ref{sp-a}).

For a de Sitter background, the presence or absence of tachyonic instabilities is illustrated in figure \ref{fig:dS}. For a fixed  value of $GN^2H^2$,  tachyon-stability corresponds to values of $\tilde{\beta}_{\text{eff}}$ larger  than a certain critical value, which is typically of order unity. For fixed  $\tilde{\beta}_{\text{eff}}$, there are two regimes, depending on the value of $\tilde{\alpha}$: for small  $\tilde{\alpha}$,  and into negative values, the theory is tachyon-stable for
$GN^2H^2$  larger than  a certain critical value (generically of order unity); for large and positive $\tilde{\alpha}$ there are also intermediate regions of stability: the theory goes from unstable at small $GN^2H^2$,  to stable as  $GN^2H^2$ increases, to again unstable, and finally to stable at large $GN^2H^2$. In the specific case   $\tilde{\a} = \tilde{\beta}_{\text{eff}} = 0$ there is a critical value for $GN^2H^2$ below  which de Sitter space is  tachyon-unstable, as it was also shown in \cite{Chesler}. The critical value corresponds to $GN^2H^2 \approx 0.32$.  For small curvatures, and for $\tilde{\alpha}$ of order unity, the tachyon pole is generically located around the cut-off scale, unless one takes $|\tilde{\beta}_{\text{eff}}| \gg1$.

For any values of the parameters, there are tensor ghosts (tachyonic or not). However, generically, these ghosts are heavy (in units of the ``species'' cut-off $(GN^2)^{-1/2}$) or they occur for curvatures of the order of the cut-off.

\item In the special case $\tilde{\a}=\tilde{\beta}_\text{eff}=0$, like in the generic case above,  for any curvature (including zero-curvature flat space-time), the mass of the ghost is always larger but comparable to the species scale.

\end{itemize}

\item{\bf Anti-de Sitter.}
\begin{itemize}
\item In the special case of  pure gravity ($N=0$), the situation is similar to the one in de Sitter. There are two tensor modes,   one of them  massless and the other  massive.  For generic $O(1)$ parameters $\alpha$ and $\beta$, the massive mode  is a tachyon for  $\beta < 0$ (up to small corrections).  One of the two tensor modes is always a ghost, and it is light only when $\alpha$ and/or $\beta$ are very large. Therefore, as in de Sitter, for $O(1)$ values of the parameters, the theory does not have instabilities within EFT. This is what happens  in top-down string theory \cite{zwi}. On the other hand, this analysis means that one has to be careful in taking $\alpha$ and $\beta$ too large. This is standard practice to obtain  a  qualitatively different  behaviour from Einstein AdS gravity.
  This is common in phenomenological holographic models and some examples with commentary are \cite{r2,r3}.

\item In the presence of the CFT,  as in de Sitter,  tensor modes can be tachyonic or not depending on the parameters $\tilde{\alpha}$, $\tilde{\beta}_{\text{eff}}$.
 The situation is represented in figure \ref{fig:AdS}.
  For fixed  AdS  curvature, there are tachyonic modes  for large and negative values of $\tilde{\b}_{\text{eff}}$ up to a certain critical value (which depends on $\tilde{\alpha}$ for large curvatures but is independent of $\tilde{\alpha}$ for small curvatures) above which the theory is tachyon-stable. The critical value is generically $\mathcal{O}(1)$. For a fixed $\tilde{\beta}_{\text{eff}}$ there are different possibilities: the theory may be  tachyon-stable ($\tilde{\beta}_\text{eff}$ large and positive, $\alpha \gtrsim 0$ ), or be tachyon-stable only above a certain curvature ($\tilde{\beta}_{\text{eff}}$ large and negative) or cross from tachyon-stability to instability to stability again for $\tilde{\beta}_{\text{eff}}\sim O(1)$ and $\alpha <0$. Unless  $|\tilde{\beta}_{\text{eff}}| \gg1$, the tachyonic modes are above the species cut-off.

\item The special case  $\tilde{a} = \tilde{\b}_{\text{eff}}=0$. Here,  AdS space-time is tachyon-stable for any curvature below the species cut-off.
\end{itemize}

\end{itemize}

Finally, we note that until now it was the scalar instability in de Sitter (or near de Sitter) that was employed as a mechanism for exiting inflation.
However, our results show that, depending on the parameters, the ``fastest'' instability may be in  the scalar or the tensor sector.
It should be stressed  though that if the fastest instability is the spin-2 one, this is a disaster for cosmology. The reason is that this instability generates large transverse
variations of the background metric destroying fast its homogeneity and therefore the main principle of cosmology.
Consequently, for cosmology, spin-2 instabilities must be avoided.

Up to specific details which may vary depending on the parameters, the general features of the spectra  discussed above can be  summarised as follows: \\

{\bf Pure gravity:}
\begin{itemize}
\item {\bf Minkowski:}
\begin{itemize}
\item scalar tachyon if $\alpha>0$
\item  if $\beta\neq 0$, two tensor modes: one massless graviton and one  massive ghost (tachyonic or not).
\end{itemize}
\item {\bf dS and AdS}
\begin{itemize}
\item scalar tachyon if $\alpha>0$ if $GN^2H^2\ll 1$.
\item  scalar (light) ghost if $\alpha \gg 1/(GH^2)$.
\item If $\beta\neq 0$,  two tensor modes, one massless  and one  massive. One of them is necessarily a ghost.
\end{itemize}
\end{itemize}
In all these cases, these ghosts/tachyons are below the cutoff  $M_p$ {\it only} if $|\alpha| \gg 1$ and/or  $|\beta| \gg 1$.\\

{\bf Gravity coupled to the CFT:}
\begin{itemize}
\item[-]
The bounds on the ghost/tachyon regions  vary, and there may be more massive tensor modes in the spectrum  (in particular in AdS).
\item[-] The cut-off is now lowered to the species scale, $M_p/\sqrt{N}$
\item[-] The presence of {\it light} ghosts/tachyons still requires  the effective coefficients of the $R^2$ and $R_{\mu\nu}R^{\mu\nu}$ to be large $|\tilde{\alpha}| \gg 1$ and/or  $|\tilde{\beta}_\text{eff}| \gg 1$.
\end{itemize}

\subsection{Discussion}

Our findings show that there are whole regions of parameter space where  the  holographic matter + gravity theory suffers from both scalar and tensor instabilities, for all signs  of the curvature. In particular, the unstable region contains the whole of flat space except eventually in the limit where we decouple the CFT.

Even though ghosts and tachyons seem ubiquitous, as we argued earlier, only when  the unstable modes are lighter than our EFT cut-off (\ref{sp-a})  do they signal an unequivocal instability. From our analysis, it emerges that, at small curvatures (compared to the cut-off),  if the renormalized coefficients of the local  quadratic curvature terms,   (\ref{tildes}),   are $O(1)$,  unstable modes generically  have masses above the UV cut-off $M_p/N$.  On the other hand, the presence of light ghosts or tachyons  requires very large values of the parameters  (\ref{tildes}).

It turns out that, for  large values of the higher curvature parameters (\ref{tildes}), one {\it also} finds light ghosts in pure quadratic-curvature gravity (in dS, AdS or Minkowski) without the CFT. Based on our analysis of parameter space, we can make the following statement: \\

\noindent
{\it Within the validity of EFT, for parameter values for which pure gravity shows no pathologies, neither does the gravity+CFT system.}\\

In other words, for  background curvatures  below the cut-off, light unstable modes   in the gravity + CFT  system are due  essentially to (effective) large {\it local} higher curvature terms which would result in the same instabilities in the pure gravity with the same  parameters. It should be remembered though, that in pure gravity the cutoff is taken to be the Planck scale while in the gravity+CFT system, the cutoff is taken to be the (renormalized) species cutoff in (\ref{spec}).

{ Note that if we insist instead on taking the EFT cut-off to be the Planck scale (rather than the species scale) even in the presence of the CFT, this conclusion changes, and we are led to the fact that coupling a healthy CFT to gravity {\it does} introduce instabilities within EFT. This  is another indication that the correct  cut-off is indeed $M_p/{\sqrt{N}}$. }

From the holographic standpoint, in the case of a CFT coupled to gravity, the scalar modes are the simplest since they do not propagate in the bulk, the only dynamical scalars are boundary degrees of freedom whose dynamics are determined by the $R^2$ terms and the conformal anomaly  \cite{anomalyinflation1}.
Therefore, any  scalar instabilities can be simply traced purely to a local boundary gravity action.

{  The unstable scalar is a pure-gauge mode in Einstein gravity, but it becomes dynamical thanks to the higher curvature terms and the conformal anomaly, and depending on the coefficients,  it may become tachyonic and/or ghost-like.
In this context, scalar  instabilities were studied in 4-dimensional higher-curvature gravity  around flat space in \cite{Q1}. Around de Sitter, scalar instabilities were investigated in \cite{VilenkinStaro} in the (original) Starobinsky model \cite{St}, and here we recover the results obtained in the linearized version of Vilenkin's analysis. It is worth mentioning that this model falls outside of the EFT description: indeed, in \cite{St,VilenkinStaro} the 4d cosmological constant is set to zero, which fixes the dS curvature to satisfy $GN^2H^2 = 4\pi$. The value of $H$  is above the species  cut-off $1/(\sqrt{G}N)$  (although for large $N$ it may still be sub-Planckian). In this case, the no-tachyon condition  in the scalar sector is $\alpha >0$, (see equation (\ref{res-17}). However for phenomenological reasons, it is rather desirable to have a  scalar tachyon in order to leave the de Sitter solution in the early universe, and one should choose $\alpha <0$.}

 Similar considerations can be made if we want to make a comparison  with  what goes nowadays under the name of the Starobinsky model for inflation\footnote{$\hat \alpha$ in equation (\ref{R+R2}) is related to our $\alpha$ as $\hat \alpha = {\alpha\over 384 \pi}$.} ,
\be \label{R+R2}
S = -\int\sqrt{-g}\left(  {1\over 16\pi G} R - \hat\alpha R^2 \right).
\ee
This may be thought of as a simplified version  of the   anomaly-driven realization of de Sitter in \cite{St,VilenkinStaro} in which one neglects the non-local contribution from the conformal anomaly $\sim R^2\log R$. For this model, one does not need a CFT, but pure higher curvature gravity is enough.  The model (\ref{R+R2}) does not admit de Sitter solutions since the absence of the logarithmic term  pushes this solution to infinite curvature. However,  it admits quasi-de Sitter slowly-rolling FRW solutions (in an appropriately defined Einstein frame)  for $\hat \alpha <0$,  i.e. precisely where one expects a scalar tachyonic instability (see equation  (\ref{res-17}) for large curvature): it is this tachyon that eventually pushes the solution  away from the near-de Sitter geometry, thus ending inflation. This is the choice made in phenomenological models of inflation,  where the parameter
$$\alpha=384\pi\hat \alpha \sim - 5.95\times 10^{11}$$
 to reproduce the amplitude of the primordial perturbation spectrum.  For $\hat \alpha<0$, the scalar mode is not a ghost in pure $R+R^2$ gravity, as can be seen by setting $N=0$ in (\ref{res-19}).
 More details about how our results compare to  these models can be found in Appendix \ref{staro}.

A general discussion of higher-derivative gravity including the tensor modes can be found in \cite{Stelle1978}, where it was pointed out that the $R_{\mu\nu}R^{\mu\nu}$ term gives rise to a tensor ghost around flat space. Here, we also extend this discussion to  de Sitter  and  AdS.

In the holographic context of gravity coupled to a large-$N$ CFT,  instabilities were already found in certain corners of parameter space by \cite{Chesler}, and we agree with their results. A general analysis of the gravitational spectrum around de Sitter space was also performed in \cite{anomalyinflation1}  for a specific value of the dS Hubble parameter (namely $GN^2H^2 = 4\pi$)  for which the (renormalized) cosmological term is zero (this case however, is outside of the range of EFT since $R\sim H^2 > M_{species}$).    As we mentioned above, here we find that   these tensor instabilities are either outside of the EFT validity, or they require large effective values of the higher curvature coefficients which would make the pure gravity theory pathological as well. A more detailed comparison with \cite{Chesler} and \cite{anomalyinflation1} appears in appendix \ref{compa}.

When there are light tensor tachyons or ghosts, it is  interesting to ask  which direction in solution space the instability leads to. The scalar instability contains a homogeneous mode which can be understood as an instability of the de Sitter solution towards a more general FRW. These are the instabilities of the type considered in  \cite{VilenkinStaro}. However, non-homogeneous scalar instabilities and tensor instabilities break FRW.

A related  question is whether  this analysis around maximally symmetric space-times persists in more realistic cosmological solutions such as FRW. The same holographic setup used here can in principle be applied to FRW boundary metrics, by generalizing the bulk solution  along the lines of \cite{tetradis1,tetradis2,tetradis3,KoyamaSoda}.

This paper is organized as follows.
The setup of our work is presented in section \ref{section:action}, where we start from a theory of gravity in $AdS_5$ and obtain the boundary action with higher curvature terms induced on a regularized boundary.

Metric perturbations are set up in section \ref{bulk pert} for the bulk perspective, and in section \ref{bdy pert} for the boundary theory. Section \ref{bdy pert} also studies the dynamics of the pure boundary scalar perturbation.
The five remaining degrees of freedom for metric perturbations are contained in a transverse-traceless tensor studied in section \ref{linearized eq}, where its equation of motion is obtained.
In section \ref{decoupling} we discuss  tensor instabilities in pure gravity with quadratic curvature terms.
Tensor instability in the general case of the CFT  coupled to gravity is studied in section \ref{sec:flat correlator} for flat space-time, in section \ref{dS poles} for positive curvatures and in section \ref{AdS poles} for negative curvatures.

The appendix contains some of the technical details of this paper. We briefly review here the different sections.
In appendix \ref{app:free scalars}, we provide an explicit example of effective field theory which develops instabilities (ghosts and tachyons) due to the IR expansion. We also find that the mass of these unstable modes is always above the EFT cut-off.
Appendix \ref{renorm} reviews the computation of the counterterms for the bulk renormalization procedure \cite{HSS}.
In appendix \ref{staro}, we relate our setup to Starobinsky's inflation \cite{St}.
The geometry of AdS-slicing coordinates is reviewed in appendix \ref{app:AdS-slicing} where we map them to the more usual \textit{global coordinates} of AdS. We also remark that AdS-slicing coordinates are \textit{global} in the sense that they cover the whole AdS manifold.
Appendix \ref{bulk schrodinger} relates the bulk radial equation of spin-2 perturbation to a Schrodinger problem, which allows us to study the normalizability of its solutions.
In appendix \ref{app:tacscale}, we compute the decay rate of Minkowski space-time in terms of the mass of a tachyonic pole.
Appendices \ref{dS criterion} and \ref{AdS criterion} derive the criteria (\ref{res-14}) for the spin-2 modes in dS and AdS respectively.
appendix \ref{app:legendre} studies the asymptotic behaviour of associated Legendre functions which enter into the solution for spin-2 perturbations in AdS-slicing coordinates of AdS.
Appendix \ref{quadratic action} computes the quadratic terms of the boundary action, which enter into the definition of the two-point functions
for metric perturbations.
Appendix \ref{gauge scalar} proves that one can discard an unphysical scalar mode appearing in the quadratic action, by showing that this mode is constrained.
Appendix \ref{compa} compares our results to previous papers which have used a similar setup. We find the values of our parameters which reduce our setup to their case.
Finally, appendix \ref{moresnapshots} provides supplementary material concerning the poles of the spin-2 propagators in curved space.

The arXiv webpage of this paper contains supplementary material, including 5 animated gifs showing the poles of the spin-2 propagator for different choices of parameters ($GN^2\bar{R}$, $\tilde{\a}$ and $\tbe$). These gifs and their associated parameters are presented in the ``animated\_gifs.pdf" ancillary file.

\section{The theory}
\label{section:action}

We use the following notation for the various metrics:
\begin{center}
\begin{tabular}{|c|c|c|}
\hline
symbol & name & relation/definition\\
\hline
$\mathcal{G}_{ab}$ & 5d bulk metric & \makecell{Fefferman-Graham \\coordinates (\ref{Graham-Fefferman})} \\
\hline
$ g_{\omega\sigma} $ & Fefferman-Graham slice metric & defined in (\ref{Graham-Fefferman}) \\
\hline
$g^{(0)}_{\omega\sigma} $ & 4d space-time metric & leading term of (\ref{hr10})\\
\hline
$\gamma_{\omega\sigma} $ & \makecell{induced metric on \\ the regulated boundary} & related to $g^{(0)}_{\omega\s}$ as (\ref{im})\\
\hline
$ \bar{\zeta}_{\omega\sigma} $ & \makecell{background 4d \\space-time metric} & \makecell{maximally symmmetric\\ background of $g^{(0)}_{\omega\s}$ (\ref{scalar0})} \\
\hline
$ \delta \zeta_{\omega\sigma} $ &  slice metric perturbation & defined in (\ref{b6a})\\
\hline
$ \delta \zeta^b_{\omega\sigma} $ & \makecell{$\delta \zeta_{\omega\s}$ evaluated \\ on the boundary} & \makecell{defined in (\ref{scalar0}),\\ gauge fixed in (\ref{scalar2})} \\ \hline
\end{tabular}
\end{center}

\subsection{Setup}
We consider a semi-classical theory of gravity in four dimensions,  described by a 4d metric $g^{(0)}_{\omega\sigma}$,   including quadratic curvature terms  and coupled to a 4d Conformal Field Theory (CFT). The total action is
\be
S = S_\text{grav} + S_\text{CFT}.
\label{s0}
\ee
The first term, $S_\text{grav}$, is the gravity action:
\be
S_\text{grav} = S_\text{EH} + S_\a + S_\b,
\label{s01}
\ee
which includes the Einstein-Hilbert plus the cosmological constant term\footnote{In our notation, the curvature tensors are understood to be those built from the metric $g^{(0)}_{\omega\sigma}$, unless otherwise specified explicitly.},
\be
 S_\text{EH} = -{1 \over 16\pi G} \int d^4x \sqrt{g^{(0)}}(R - 2\Lambda),
 \label{s02}
 \ee
 as well as  two quadratic curvature terms:
 \be
S_\a = { \alpha\over 384\pi}\int d^4x \sqrt{g^{(0)}}R^2,
\label{s03}
\ee
\be
S_\b =  {\beta \over 64\pi} \int d^4x \sqrt{g^{(0)}}\left(R^{\omega\s}R_{\omega\s}-{1\over 3}R^2\right).
\label{s1}
\ee
Here, $G$ is the Newton constant,  $\Lambda$  the cosmological constant,  $\alpha$ and $\beta$ are the dimensionless $R$-squared couplings.

The second term $S_\text{CFT}$ in  (\ref{s0}) is the quantum effective action of a  CFT in  a background metric $g^{(0)}_{\omega\s}$,
and it is a functional of the background  metric $g^{(0)}_{\omega\s}$.

The action (\ref{s0}) is meant to be the {\em renormalized} action, in which all the divergences have been renormalized. The parameters $G,\Lambda,\alpha,\beta$ are therefore to be interpreted as  finite, physical  parameters left after the renormalization procedure (which will be described in detail in  subsection \ref{sec:renorm-action} and reference \cite{GKNW}).

 Variation of the action with respect to the boundary metric yields the following Einstein equation:
\be
R_{\omega\s}-\frac{1}{2}R g^{(0)}_{\omega\s}+\Lambda g^{(0)}_{\omega\s} +8 \pi G \left( ^{(\a)}H_{\omega\s}+^{(\b)}H_{\omega\s} \right) =8\pi G   \langle T_{\omega\s} \rangle.
\label{EE}
\ee
where
\be
^{(\a)}H_{\omega\s}= \frac{\a}{96\pi}\left\{ \nabla_\omega \nabla_\sigma R-R R_{\omega\s}-\left(\Box R-\frac{1}{4}R^2 \right)g^{(0)}_ {\omega\s} \right\},
\label{Halpha}
\ee
\be
^{(\b)}H_{\omega\s}=\frac{\b}{32\pi}\left\{ \frac{1}{2} \left(R_{\kappa\lambda}R^{\kappa\lambda}-\frac{1}{3}R^2+\frac{1}{3} \Box R \right)g^{(0)}_{\omega\s}-2 R_{\omega\kappa\s\lambda}R^{\kappa\lambda}-\Box R_{\omega\s}  \right. \nn
\ee
\be
\qquad\qquad\qquad \left. +\frac{1}{3} \left( 2 R R_{\omega\s}+\nabla_\omega \nabla_\s R \right) \right\}.
\label{Hbeta}
\ee
In Eq. \eqref{EE}, the right-hand side is the renormalized CFT stress-energy tensor expectation value,
\be
\label{tcft} \langle T_{\omega\s} \rangle=\frac{2}{\sqrt{g_{(0)}}}\frac{\d S_{CFT} }{\d g^{(0)\omega\s} }.
\ee

Before we present the computation of the CFT stress tensor, we comment on the terms which are explicit on the left-hand side of (\ref{EE}). It is  important to remark that $ ^{(\b)}H_{\omega\s}$ is traceless. Furthermore, $^{(\a)}H_{\omega\s}$ is also traceless if the boundary has a constant curvature.

For convenience, in the following, we define the tensor
\begin{equation}
E_{\omega\s}\equiv -{16\pi G \over \sqrt{g^{(0)}}}{\delta S\over \delta g^{(0)\omega\s}}.
\end{equation}
We then write Einstein's equations as:
\be
E_{\omega\s}=0.
\end{equation}
 \par

From now on, we shall assume the CFT is a large-$N$ theory which has a holographic description in terms of a (semiclassical)  five-dimensional gravity dual.  We  shall  review how the renormalized stress tensor (\ref{tcft}) is computed in this context \cite{HSS}, and how the renormalized parameters of the effective gravity theory arise.
However, as we shall argue, our results do not depend on this assumption.
 Below we present the main results  and give more details in appendix \ref{renorm}.

\subsection{Constructing the renormalized action} \label{sec:renorm-action}

To arrive  at (\ref{s0})  we replace the CFT contribution with its dual description, namely an  Einstein-Hilbert theory on a  5-dimensional manifold $\mathcal{M}$ (the bulk, on which the metric will be denoted by ${\mathcal G}$) together with covariant boundary terms on the boundary,  $\pa\mathcal{M}$. However, this is divergent and to regulate the divergences we move the boundary $ \pa\mathcal{M}$ to the regulated boundary $\pa\mathcal{M}_{\e}$ which is inside $\mathcal{M}$ and this also defines the regulated bulk space $\mathcal{M}_{\e}$. $\gamma$ is the induced metric on the regulated boundary.

 The bare regularized gravity dual  action is:
\be
S_\text{reg} = S_\text{bulk} + S_\text{grav}^0.
\label{hr}
\ee
The first term is the usual Einstein-Hilbert action with a boundary $\partial \mathcal{M}_{\e}$,
\be
S_\text{bulk} = -M^3\left[\int_{\mathcal{M}_{\e} } d^5x \sqrt{\mathcal{G}}(R[\mathcal{G}] - 2\Lambda_{5}) -2 \int_{\partial\mathcal{ M}_{\e} } d^4x \sqrt{\gamma}K\right]. \label{hr0}
\ee
 where $M$ is the $5$-dimensional Planck mass,  $\gamma$ is the determinant of the induced metric on $\partial \mathcal{M}_{\e} $ and $K$ is the corresponding extrinsic curvature
 \footnote{Geometrical tensors follow the same conventions as Wald's book \textit{General Relativity}}.
The second term in (\ref{hr}) is a boundary term which depends only on intrinsic tensors on $\de {\mathcal M}_{\e}$,

\be
S_\text{grav}^0  = S_\text{EH}^0 + S_\a^0 + S_\b^0,  \label{hr0a}
\ee
where
\be
S_\text{EH}^0 = -{1 \over 16\pi G^0}\int d^4x \sqrt{\gamma}(R[\gamma] - 2\Lambda^0),
 \label{hr0b}
\ee
a $R^2$ term
\be
S_\a^0 = {\alpha^0 \over 384\pi}\int d^4x \sqrt{\gamma}\left(R[\g]\right)^2,
  \label{hr0c}
\ee
and an additional term proportional to the 4-dimensional Weyl anomaly\footnote{In our notation, the Latin letters will denote the bulk coordinates, and the Greek indices such as $\omega,\s$ denote $4$-dimensional slice coordinates.} \cite{duff}
\be
S_\b^0 = {\beta^0\over 64\pi} \int d^4x  \sqrt{\gamma}\left(R[\gamma]^{\kappa\lambda}R[\gamma]_{\kappa\lambda}-{1\over 3}\left(R[\g]\right)^2\right).
 \label{hr0d}
\ee
This action looks similar to (\ref{s1}), and depends on a set of bare parameters $\alpha^0$, $\beta^0$, $\Lambda^0$ and $G^0$.  Below we shall relate these bare parameters to the physical ones ($\alpha$, $\beta$, $\Lambda$ and $G$) in the renormalized action (\ref{s0}).

 We also define a length $L$ associated with the bulk cosmological constant,
\be
\Lambda_{5} = -{6\over L^2}. \label{hr5a}
\ee
We consider asymptotically AdS solutions, for which the ansatz for the full metric is written using Fefferman-Graham coordinates \cite{FG}, given by
\be
ds^2 = \mathcal{G}_{ab}dX^adX^b= L^2\frac{d\rho^2}{4\rho^2} + \frac{1}{\rho}g_{\omega\s}(x,\rho)dx^{\omega} dx^\s,
\label{Graham-Fefferman}
\ee
where $L$ and $x^\s$ have the dimension of a length and $\rho$ is dimensionless. This coordinate system is the one of an asymptotically AdS space with a conformal boundary located at $\rho \rightarrow 0$. We define the  regulated boundary $\pa{\mathcal M}_{\e}$ as the hypersurface $\rho=\epsilon$, on which the induced metric  is:
\be
\gamma_{\omega\s}(\epsilon, x) = {1\over \epsilon}g_{\omega\s}(\epsilon,x) .
\label{im}
\ee
The  metric $g_{\omega\s}$ is determined by solving the bulk Einstein equation order by order in $\rho$ as $\rho \to 0$, starting with  an arbitrary metric $g^{(0)}_{\omega\s}$ to lowest order \cite{HSS}:
\be
g_{\omega\s}(x,\rho) = g^{(0)}_{\omega\s} +  \rho g^{(2)}_{\omega\s} + \rho^{2}g^{(4)}_{\omega\s} + \hat{g}_{\omega\s}\rho^{2}\log{\rho} + \mathcal{O}(\rho^{3}).
\label{hr10}
\ee
The leading term in this expansion, $g^{(0)}_{\omega\s}$, is identified with the  metric of the dual field theory side.  The terms  $g^{(2)}_{\omega\s}$ and $\hat{g}_{\omega\s}$ are given by\footnote{Recall that all the geometrical tensors are built from the metric $g^{(0)}_{\omega\s}$ unless otherwise stated}:
 \be
g^{(2)}_{\omega\s} = -{L^2\over 2}\left(R_{\omega\s} - {R\over 6}g^{(0)}_{\omega\s}\right),
\label{fg161}
\ee
\be
\hat{g}_{\omega\s} = {L^4\over 16}\left\{ 2R_{\omega \kappa \s \lambda}R^{\kappa\lambda} - {1\over 3} \nabla_\omega\nabla_\s R + \nabla^2 R_{\omega\s} - {2\over 3}RR_{\omega\s} + \left({1\over 6}R^2 -,\right.\right.
\label{fg261}
\ee
$$
\left.\left.- {1\over 6}\nabla^2 R - {1\over 2}R_{\kappa\lambda}R^{\kappa\lambda}\right)g^{(0)}_{\omega\s} \right\}
$$
These expressions are found by solving the bulk Einstein equation in a near-boundary expansion  \cite{HSS}.  Note that, comparing Eqs. \eqref{Hbeta} and \eqref{fg261}, we can write:
\begin{equation} \label{trghat}
\tensor[^{(\beta)}]{H}{_\omega_\s}  = -{\beta\over 2\pi }\hat{g}_{\omega\s}.
\end{equation}
 Unlike $g^{(2)}_{\omega\s}$ and $\hat{g}_{\omega\s}$, $g^{(4)}_{\omega\s}$ is not fully determined from $g^{(0)}_{\omega\s}$, except for its trace, which is given by\footnote{As in \cite{HSS}, when matrix components are not written, it means that both matrix multiplication and trace operations are done using the metric $g^{(0)}$.} \cite{HSS}:
 \be
 \tr\left[ g^{(4)}\right] = {1\over 4}\tr\left[ \left( g^{(2)}\right)^2\right].
 \label{trg4}
 \ee

Divergences of $S_\text{bulk}$, which arise when we remove the regulator and take $\epsilon \to 0$,  are made explicit when  $S_\text{bulk}$  is written in terms of $g^{(0)}_{\omega\s}$.
The method to obtain these divergences is briefly reviewed in appendix \ref{renorm}, resulting in:

\begin{equation}
S_\text{bulk} =  \frac{M^3}{L} \int d^4 x\sqrt{g^{(0)}}\left\{-{6\over \epsilon^{2}}  + {1\over 8} \log\epsilon \left(R^{\kappa\l}R_{\kappa\l} - {1\over 3}R^2\right) \right\}
+ \mathcal{O}(\epsilon^0), \label{Sdiv1}
\end{equation}
The first term in curly brackets contains all the divergent terms of $S_\text{bulk}$. These can also be written covariantly in a series expansion involving curvature tensors of the induced metric on the boundary.
\begin{equation}
S_\text{bulk} = \frac{M^3}{L} \int d^4x \sqrt{\gamma}\left\{ - 6  - {L^2\over 2} R[\g] + \right. \nn
\ee
\be
 + \left. {L^4\over 8}\left( {1\over 2} + \log \epsilon\right)\left(R^{\kappa\lambda}[\g]R_{\kappa\lambda}[\g] - {1\over 3}\left(R[\g]\right)^2\right)\right\} + ... \label{hr30}
\ee
where $...$ indicates higher curvature invariants.
The explicit $\epsilon$ dependence in (\ref{hr30}) reflects the conformal anomaly.

The  quadratic curvature term in (\ref{Sdiv1}) can be shifted by a finite amount by redefining the cut-off.   This scheme dependence is made explicit by introducing  as an extra parameter, a  scale $\m$, and defining the divergent part of the action $S_\text{div}$ as follows:
 \be
 S_\text{div} \equiv \frac{M^3}{L} \int d^4 x\sqrt{g^{(0)}}\left\{-{6\over \epsilon^{2}}  + {1\over 4} \log(\sqrt{\epsilon} \m L) \left(R^{\omega\s}R_{\omega\s} - {1\over 3}R^2\right) \right\}.
  \label{ig1e}
 \ee

We now turn to the bare ``boundary''  gravitational action (\ref{hr0a}). It is also divergent in the limit $\epsilon \to 0$. This  can be made manifest by expressing it  in terms of curvature tensors of $g^{(0)}_{\omega\s}$ using the expansion (\ref{hr10}) and the expressions (\ref{fg161}-\ref{fg261}). The result for the Einstein-Hilbert part is
\be
 S_\text{EH}^0 = -{1\over 16\pi G^0} \int d^4x \sqrt{\gamma}(R[\gamma] - 2\Lambda^0),
\ee
\be
 =  -{1\over 16\pi G^0} \int d^4x \sqrt{g^{(0)}} \left\{-{2\Lambda_0 \over \epsilon^2 } + {1\over \epsilon}\left(1 + {\Lambda_0L^2\over 6}\right)R + \right.
 \nonumber
 \ee
 \be
\left.  + {L^2\over 4}\left(2 + {\Lambda_0L^2\over 4}\right)\left(R^{\omega\s}R_{\omega\s} - {1\over 3}R^2\right) + \mathcal{O}(\epsilon)\right\},
\label{ig1}
\ee
while $S_{\a}^{0}$ and $S_\b^0$ are finite. Note that  additional finite quadratic terms appear when we expand $S_{EH}^0$ in powers of $\epsilon$.

Therefore, the full boundary action  $S_\text{grav}^0$  written in terms of $g^{(0)}_{\omega\s}$, including   all divergent and finite terms,
has the form:
\be
S_\text{grav}^0 = -{1\over 16\pi G^0} \int d^4x \sqrt{g^{(0)}} \left\{-{2\Lambda_0 \over \epsilon^2 } + {1\over \epsilon}\left(1 + {\Lambda_0L^2\over 6}\right)R \right\} +
\nonumber
 \ee
 \be
 + \left[\beta^0 - {1\over 16\pi G^0}{L^2\over 4}\left(2 + {\Lambda_0L^2\over 4}\right)\right]\int d^4x \sqrt{g^{(0)}} \left(R^{\omega\s}R_{\omega\s} - {1\over 3}R^2\right)
 \nonumber
 \ee
 \be
 + \alpha^0\int d^4x \sqrt{g^{(0)}} R^2 + \mathcal{O}(\epsilon).
 \label{ig1bis}
\ee

The renormalisation procedure we adopt consists in taking the limit $\epsilon \to 0$  by choosing appropriately the bare parameters  ($\Lambda^0$, $G^0$, $\a^0$ and $\b^0$) as a function of the cut-off, such that the quantity
\be\label{ig1b}
S_\text{grav} \equiv S^0_\text{grav} + S_{div}
\ee
remains finite\footnote{This is different from  the standard holographic renormalization procedure \cite{HSS}, in which the bare parameters ($\Lambda^0$, $G^0$, $\a^0$ and $\b^0$) are independent on the cut-off and  a counterterm action (whose coefficients are completely fixed) is introduced to cancel all divergences coming from $S_\text{bulk}$ (\ref{hr30}). This  leaves only finite quadratic curvature terms in the renormalized action and no Einstein-Hilbert term.} in the limit $\epsilon\to 0$.

We write the resulting finite action in terms of new \textit{physical} parameters ($\Lambda$, $G$, $\alpha$ and $\beta$), each  corresponding  to the one  boundary term:

\be
\Lambda = {1\over \epsilon}\left(1 + {\Lambda^0L^2\over 6}\right)^{-1} \left[\Lambda^0 - {48 \pi G^0 M^3 \over  L} \right],
\label{ig7}
\ee
\be
G = {\epsilon G^0\over 1+{\Lambda^0L^2\over 6}},
\label{ig8}
\ee
\be
\alpha = \alpha^0,
\label{ig9}
\ee
\be
\beta = \beta^0 + 16\pi M^3 L^3 \log(\sqrt{\epsilon} L\m) - {2 L^2 \over G^0}\left(1 + {\Lambda^0 L^2\over 8}\right).
\label{ig10}
\ee
and we take $\epsilon \to 0$ together with appropriate limits of ($\Lambda_0$, $G_0$, $\alpha_0$ and $\beta_0$) so that the left hand sides are finite.

Finally, combining (\ref{hr}) and (\ref{ig1b}) we write the renormalized action as
\be
S = \underset{\epsilon \rightarrow 0}{\lim}\left[S_\text{bulk} - S_\text{div} \right] + S_\text{grav},
\label{ig11}
\ee
\be
 \equiv S_\text{CFT} + S_\text{grav},
 \label{ig11a}
\ee
i.e. equation (\ref{s0}). The bulk contribution inside the square brackets is interpreted as the renormalized effective action of the CFT.

\subsection{The induced stress tensor\label{aa}}
The renormalized stress tensor is defined by
 \be
 \left<T_{\omega\s}\right> = \underset{\epsilon\rightarrow 0}{\lim}  \left[{1\over \epsilon}{2\over \sqrt{\g}}{\delta S_{CFT}\over \delta \g^{\omega\s}} \right]  =  {2\over \sqrt{g^{(0)}}}{ \delta S_\text{CFT}\over \delta g^{(0)\omega\s}}.
 \label{hr42}
 \ee
It can be shown that this definition leads to
\be
{2\over \sqrt{\g}}{\delta S_{bulk}\over \delta \g^{\omega\s}} = 2 M^3 (K_{\omega\s} - K\gamma_{\omega\s}).
\label{hr38}
\ee
As shown in \cite{HSS} the divergent pieces of (\ref{hr38}) cancel the ones of the  $S_{\text{div}}$. We are then left with the renormalized stress tensor given by \footnote{For a notational simplification, there is no difference in subscript and superscript in Fefferman-Graham metric expansion. }
\be
\left<T_{\omega\s}\right>  = - {2 M^3\over  L} \left\{2\left[ 2\log(\m L) - 1\right] \hat{g} - 2g^{(4)} + \left( g^{(2)}\right)^2 - {1\over 4}g^{(0)}\tr\left[ \left( g^{(2)}\right)^2\right] +  \right. \nn
\ee
\be
\left. + {1\over 4}g^{(0)}\left( \tr\left[ g^{(2)}\right]\right)^2 -{1\over 2}g^{(2)}\tr\left[ g^{(2)}\right]\right\}_{\omega\s},
\label{stress}
\ee
 where $g^{(2)}_{\omega\s}$, $g^{(4)}_{\omega\s}$ and $\hat{g}_{\omega\s}$ are the terms of the Fefferman-Graham expansion (\ref{hr10}), the expressions for $g^{(2)}_{\omega\s}$ and $\hat{g}_{\omega\s}$ in terms of $g^{(0)}_{\omega\s}$ are given in Eqs. (\ref{fg161},\ref{fg261}). $g^{(4)}_{\omega\s}$ must be obtained from solving the bulk dynamics. The stress-tensor expectation value in (\ref{stress}) is to be inserted into the right-hand side of the Einstein equation (\ref{EE}).

Even if the stress-tensor is not fully constrained by the boundary data, its trace is known using (\ref{trg4}). It gives
 \be
g^{(0)\omega\s } \left<T_{\omega\s}\right> = {(M L)^3\over 4  }\left(R^{\omega\s}R_{\omega\s} - {1\over 3}R^2\right).
\label{hr43aa}
 \ee

In a generic  CFT with a 5d gravity dual, the parameter $M^3L^3$ is large, and  proportional to the central charge\footnote{Recall that in holographic  CFTs, the two    central charges $a$ and $c$ are equal, up to $1/N^2$ corrections} $a$:
\be
 (M L)^3 = {a \over 2\pi^2}.
 \label{hr43ab-ii}
 \ee

When the CFT is a large-$N$ gauge theory in 4d, then  $a\propto N^2$
For example, in ${\cal N}=4$ SYM we have, in the large-$N$ limit:
\be
a =  {N^2\over 4}.
 \label{hr43ab-iii}
\ee

In what follows  we  assume, for definiteness, the  ${\cal N}=4$ relation (\ref{hr43ab-iii}), and set:
 \be
 M^3 L^3 = {N^2\over 8 \pi^2}.
 \label{hr43ab}
 \ee
This will allow us to replace  $M^3 L^3$ with $N^2$ and write all equations which pertain to the field theory side purely in terms of 4d parameters.  Readers can keep in mind that, for any other CFT (even  for those that are not large-$N$) they can substitute
$$N^2 \to  4 a.$$

\subsection{Background solutions}
\label{slicings}

In this section, we  discuss the background (i.e. homogeneous) solutions of the equations of motion for the 5d theory (\ref{hr}).
 We take these solutions to be the  $AdS_5$ metric with three different maximally symmetric slicings,
\be
ds_5^2 = L^2du^2 + a^2(u) \bar{\zeta}_{\omega\s} dx^\omega dx^\s,
\label{b0}
\ee
where $a(u)$ is a dimensionless scale factor and  the slice metric $\bar{\zeta}_{\omega\s}$ is a $u$-independent  maximally symmetric 4d metric.   This results in three possible coordinate systems for AdS$_5$,  that correspond to the dual CFT on three distinct four-dimensional maximally symmetric metrics: AdS$_4$, dS$_4$ and M$_4$.

\begin{itemize}
\item $\bar \zeta$ being the Minkowski metric. In this case
\be
a(u) = e^{u},
\label{b0flat}
\ee
where $u >0$ and the AdS boundary is located at $u\rightarrow +\infty$.

\item $\bar \zeta$ being the de Sitter metric  with Hubble curvature $H$, in which case
\be
a(u) = LH \sinh u,
\label{b0dS}
\ee
where $u\in \mathbb{R}$. $u=0$ is a horizon. From now on, we take $u$ positive. Therefore, $u \rightarrow + \infty$ is the $AdS_5$ boundary. The curvature of dS is given by
\begin{equation}
\bar{R} =12 H^2.
\label{b0H}
\end{equation}

\item $\bar \zeta$ being the Anti-de Sitter metric  with radius $\chi^{-1}$, in which case
\be
a(u) = L\chi \cosh u.
\label{b0AdS}
\ee
In this case, there are two  asymptotic boundaries, located at $u=\pm \infty$. These two boundaries are connected. More details for the geometry of AdS-slicing coordinates are given in appendix \ref{app:AdS-slicing}.
In the field theory interpretation, these correspond to two independent copies of the CFT on AdS$_4$ that interact via their common AdS$_4$ boundary\footnote{A conformal rescaling of such a setup corresponds to an interface between two copies of the same CFT in flat space, see the extended discussion in \cite{AdS}.}.  It is a matter of choice whether only one or both are coupled to dynamical metric perturbations, as we shall discuss in section \ref{AdS prop}. Since there is no horizon at $u=0$, we shall observe  that both sides are reachable by the bulk metric perturbations. The boundary curvature is related to $\chi$ by:
\begin{equation}
\bar{R} =-12\chi^2.
\label{b0chi}
\end{equation}
\end{itemize}

The bulk metric (\ref{b0})  can also be written in Fefferman-Graham coordinates as
\be
ds_5^2 = L^2 {d\rho^2\over 4\rho^2} +{1\over \rho} f(\rho) ds_4^2.
\label{b1}
\ee
where $ds_4^2 = \bar{\zeta}_{\omega\s}dx^\omega dx^\s $ and the function $f$ for each different slicing is then given by
\be
\label{tabular slicings}
\begin{tabular}{|c|C|C|C|}
\hline
space-time & ds_4^2 & \rho & f(\rho)\\
\hline
M$_{4}$ & ds^2_\text{flat} & \rho = e^{-2u} & 1 \\
\hline
dS$_4$ & ds^2_{dS} & \rho = \left({2\over LH}\right)^2 e^{-2u} &  f_{dS}(\rho) = 1 - {(LH)^2\over 2}\rho + \left({LH\over 2}\right)^4\rho^2 \\
\hline
AdS$_4$ & ds^2_{AdS} & \rho = \left({2\over L\chi}\right)^2 e^{-2\text{sign}(u)u} & f_{AdS}(\rho) = 1 + {(L\chi)^2\over 2}\rho + \left({L\chi\over 2}\right)^4\rho^2 \\
\hline
\end{tabular}
\ee

In these coordinates, $\rho >0$ and the $AdS_5$ boundary is located at $\rho = 0$. In de Sitter slicing,  we are free to choose a sign of $u$ (here we took positive $u$) because of the horizon $u=0$ which separates the two sides of the bulk.
However, in $AdS$ slicing where there is no such horizon, the bulk $AdS_5$ needs two different Fefferman-Graham patches such that $\rho \rightarrow 0$ is the $AdS_5$ boundary $u\rightarrow \pm \infty$. Hence the $\text{sign}(u)$ in the expression for $\rho$ in AdS-slicing.\footnote{The global embedding of dS and AdS slices in AdS$_5$ is discussed in detail in \cite{CdL} for dS, and in \cite{AdS1} for AdS.}

The background solutions (\ref{b1}) are then related to the general Fefferman-Graham expansion (\ref{Graham-Fefferman}) by
\be
\left. g_{\omega\s}(x,\rho)\right|_\text{background} = f(\rho)\bar{\zeta}_{\omega\s},
\label{b2}
\ee
from which every term of the expansion (\ref{hr10}) are fixed. In particular,   we can read off  the corresponding boundary theory metric $g^{(0)}_{\omega\s}$ as the leading term as $\rho \to 0$:
\be \label{bc-iii}
g^{(0)}_{\omega\s} = \bar{\zeta}_{\omega\s}
\ee
and it  is either the  Minkowski metric or the de Sitter metric with Hubble scale $H$, or the anti-de Sitter metric with AdS length  $\chi^{-1}$. We denote   the background curvature  $R[\bar{\zeta}] \equiv \bar{R}$,
{\be
\bar R= \left\{ \begin{array}{lll}
\displaystyle 12H^2,&\phantom{aa} &{\rm de~~ Sitter},\\ \\
\displaystyle 0 ,&\phantom{aa}&{\rm Minkowski} \\ \\
\displaystyle -12\chi^2 ,&\phantom{aa}& {\rm Anti~~ de~~ Sitter}
\end{array}\right.
\label{bc-ii}\ee}
For a maximally symmetric background, the trace of the Einstein equation (\ref{EE}) gives
\be
\Lambda = {1\over 4}\left(\bar{R} - {GN^2\bar{R}^2\over 48\pi}\right),
\label{scalar7}
\ee
where $N$ was defined in (\ref{hr43aa}). Note that, for each value of the boundary parameters $\Lambda$ and $G$, there are either two  values of the curvature $\bar{R}$ satisfying equation (\ref{scalar7}), or there are none. On the other hand, by scanning all values of $\Lambda$, we can obtain any value of $\bar{R}$. Therefore, it is convenient to  trade $\Lambda$ for $\bar{R}$: in what follows we shall replace  $\Lambda$ in terms of  $\bar{R}$ using  (\ref{scalar7}) in all equations.
This leaves $GN^2\bar{R}$ as the only dimensionless background curvature parameter.

Equation (\ref{scalar7}) does not depend on $\alpha$ or $\beta$ since they multiply tensors that are traceless when evaluated on the background metric $\bar{\zeta}_{\omega\s}$ with constant curvature.

The maximally symmetric backgrounds were discussed in detail (for  holographic CFTs and holographic RG flows on de Sitter) in \cite{GKNW}. We now move to the perturbations around these background solutions. These are described by turning on perturbations in both the bulk and the boundary metric and solving the corresponding Einstein's equation and boundary conditions. This will be the subject of the next section.

\section{Bulk metric perturbations}
\label{bulk pert}

Equation (\ref{b2})  holds for the unperturbed, background metric. In this section, we study perturbations of the bulk metric, by adopting  the same gauge invariant decomposition of metric perturbations as in \cite{anomalyinflation1}.

In a perturbed geometry, the bulk metric reads
\be
ds_5^2 = (\mathcal{G}_{ab} + \delta \mathcal{G}_{ab})dX^a dX^b.
\label{b6}
\ee
Using (\ref{b0}), one can relate the slice component perturbations to actual perturbations of the slice metric $\delta\zeta_{\omega\s}$ defined as
\be
\delta \mathcal{G}_{\omega\s} = a^2(u)\delta \zeta_{\omega\s},
\label{b6a}
\ee
such that the full metric can now be written as
\be
ds_5^2 = (\mathcal{G}_{uu} + \delta \mathcal{G}_{uu})du^2 + 2(\mathcal{G}_{u\s} +\delta \mathcal{G}_{u \s}) dudX^\s + a^2(u)(\bar{\zeta}_{\omega\s} + \delta \zeta_{\omega\s})dx^\omega dx^\s.
\label{b6b}
\ee

Even if a $5\times5$ symmetric matrix contains 15 independent elements,
only 10 degrees of freedom are invariant under the gauge transformation
\be
\delta \mathcal{G}_{ab} \rightarrow \delta \mathcal{G}_{ab} + 2\nabla^{(\mathcal{G})}_{\left(a\right.}\xi_{\left.b\right)}.
\label{b7}
\ee
One can construct these 10 invariant quantities by decomposing the perturbation  $\mathcal{G}_{ab}$ into transverse and traceless elements for the slice covariant derivative $\hat{\nabla}$ built with $\zeta_{\omega\s}$ as follows \cite{anomalyinflation1}:
\be
\delta \mathcal{G}_{uu} = A,
\label{b8}
\ee
\be
\delta \mathcal{G}_{u\s} = B_\s + \hat{\nabla}_\s B,
\label{b9}
\ee
\be
\delta \zeta_{\omega\s} = h_{\omega\s} + 2\hat{\nabla}_{(\omega}\chi_{\s)} + \bar{\zeta}_{\omega\s}\psi + \hat{\nabla}_{(\omega}\partial_{\s)} \phi,
\label{b10}
\ee
where $B_\s$, $\chi_\s$ are transverse and $h_{\omega\s}$ is transverse-traceless:
\be
\hat{\nabla}^\s\chi_\s = 0 = \bar{\zeta}^{\omega\s}h_{\omega\s},
\label{b11}
\ee
\be
\hat{\nabla}^\omega h_{\omega\s} = 0.
\label{b12}
\ee
As it is well known (and rederived in \cite{anomalyinflation1} in the present context), the only propagating degree of freedom in the bulk of  $AdS$ with pure gravity is the \textit{tensor} (transverse-traceless) perturbation $h_{\omega\s}$  which contains 5 degrees of freedom. On top of this, there exists a  scalar mode which has purely boundary dynamics, and that will be discussed in the next section. Therefore, here we set to zero all components of the perturbation except for the tensor mode.

The next step is  to obtain the equation of motion for this tensor mode. The bulk Einstein equation is:
\be
R_{ab}[\mathcal{G}] = -{4\over L^2}\mathcal{G}_{ab}.
\label{b13}
\ee

When linearized with respect to $h_{\omega\s}$, the above  equation yields:
\be
(L^2 \nabla^{(\mathcal{G)}2}  + 2 )a^2(u)h_{\omega\s} = 0,
\label{b14}
\ee
where the differential operator into parenthesis is known as the Lichnerowicz operator for AdS. This operator can be decomposed into the $(u,x^\omega)$ slicing coordinates (\ref{b0}). Equation (\ref{b14}) then takes the following form
\be
\left\{\partial_u^2 + 4 {a'\over a} \partial_u + 2\left[1-\left({a'\over a}\right)^2\right]  + L^2a^{-2}\hat{\nabla}^2\right\} h_{\omega\s} = 0.
\label{b14bis}
\ee
This equation will be specialized to different slicings and solved in section \ref{linearized eq}.

Tensor perturbations $h_{\omega\s}$ can be expanded in a similar way as in (\ref{Graham-Fefferman}):
\be
h_{\omega\s} = h^{(0)}_{\omega\s} + \rho h^{(2)}_{\omega\s} + \rho^2 h^{(4)}_{\omega\s} + \rho^2 \log \rho \hat{h}_{\omega\s} + \mathcal{O}(\rho^3).
\label{b15}
\ee

We shall now  linearize the  boundary Einstein field equation (\ref{EE}) and obtain an equation which involves the various terms in the near-boundary expansion (\ref{b15}). To this end, we need to relate perturbations of the metric $g_{\omega\s}$ defined in (\ref{Graham-Fefferman}) to the slice perturbations $h_{\omega\s}$ defined in (\ref{b10}).
We introduce the following notation, for any tensor $A$ of the slice metric:
\be
(\delta_h A)[\bar{\zeta}] \equiv \underset{\varepsilon \rightarrow 0}{\lim} {A[\bar{\zeta} + \varepsilon h^{(0)}] - A[\bar{\zeta}]\over \varepsilon}.
\label{dS14a}
\ee
We identify term by term the expansion (\ref{hr10}) with the expansion of the bulk metric (\ref{b0}) close to the boundary $\rho\rightarrow 0$ where $\rho(u)$ is given in table \ref{tabular slicings}. The result for both AdS and dS is given by
\begin{subequations}
\label{dS13aaa}
\be
 \delta_h g^{(0)}_{\omega\s} = h^{(0)}_{\omega\s},
 \label{dS13ab}
\ee
\be
\delta_h g^{(2)}_{\omega\s}  =  h^{(2)}_{\omega\s} - {L^2R\over 24}h^{(0)}_{\omega\s},
\label{dS13ac}
\ee
\be
\delta_h g^{(4)}_{\omega\s} =  h^{(4)}_{\omega\s} -  {L^2R\over 24} h^{(2)}_{\omega\s} + \left({L^2R\over 48}\right)^2h^{(0)}_{\omega\s},
\label{dS13ad}
\ee
\be
\delta_h \hat{g}_{\omega\s} = \hat{h}_{\omega\s}.
\label{dS13ae}
\ee
\label{i11b}
\end{subequations}

The quantities  $h^{(2)}$ and $\hat{h}$ can be written in  terms of $h^{(0)}$ and  of boundary curvature tensors: as is summarized in appendix \ref{renorm}, $g^{(2)}_{\omega\s}$ and $\hat{g}_{\omega\s}$ are obtained in terms of $g^{(0)}$ by solving perturbatively the bulk Einstein equation for small $\rho$ \cite{HSS}. By varying the solution for $g^{(2)}$ and $\hat{g}$ given in (\ref{fg16}, \ref{fg26}) with respect to $h^{(0)}_{\omega\s}$, we obtain using (\ref{dS13aaa}):
\be
h^{(2)}_{\omega\s} = {L^2\over 4}\left(\nabla^2- {R\over 6}\right)h^{(0)}_{\omega\s},
\label{b16}
\ee
\be
\hat{h}_{\omega\s}= -{2\pi L^4\over\b}  \delta_h \tensor[^{(\b)}]{H}{_\omega_\s} = -{L^4\over 32}\left( \nabla^2 - \frac{R}{6}\right)\left( \nabla^2 - \frac{R}{3}\right)h^{(0)}_{\omega\s},
\label{b17}
\ee
where the laplacian operator $\nabla^2$ is constructed with the Fefferman-Graham metric $g^{(0)}_{\omega\s}$.
On the contrary, we need to solve (\ref{b14}) in the whole bulk (with appropriate conditions in the interior) to find $h^{(4)}_{\omega\s}$. We postpone this to section 5.

Using the equations above, all linearized quantities can be expressed purely in terms of  $h^{(0)}_{\omega\s}$ and $h^{(4)}_{\omega\s}$,  which for now are independent.

The variation of the holographic stress-tensor (\ref{stress}) in the presence of a perturbation $\delta h_{\mu\nu}$ is given by:
\be
\delta_h \left<T_{\omega\s}\right> = {N^2\over 2\pi^2L^4}\left[\delta_h g^{(4)}_{\omega\s} - \left({L^2R\over 24}\right)^2\delta_h g^{(0)}_{\omega\s} + (1-2\log (\m L))\delta_h \hat{g}_{\omega\s}\right],
\label{dSTh}
\ee
where $\delta_h g^{(4)}_{\omega\s}$, $\delta_h g^{(0)}_{\omega\s}$ and $\delta_h \hat{g}_{\omega\s}$ have to be written using equations (\ref{dS13ab}-\ref{dS13ae}) and  (\ref{b16}-\ref{b17}).

We now turn to the linearization of the left-hand side of Einstein's equation (\ref{EE}), in which the cosmological constant can be replaced by a function of the background curvature $\bar{R}$ using (\ref{scalar7}).  Note that the CFT stress-tensor also contributes to the value of $\Lambda$ through the trace of the background Einstein equation (\ref{scalar7}). By moving all the CFT contributions (i.e. those proportional to $N$) to the right-hand-side of the linearized Einstein equation, we find
\be
\left(-\nabla^2 + {R\over 6}\right)h^{(0)}_{\omega\s} + 8\pi G\delta_h (\tensor[^{(\alpha)}]{H}{_\omega_\s} + \tensor[^{(\beta)}]{H}{_\omega_\s})  =  8\pi G \delta_h \left<T_{\omega\s}\right>^T,
\label{dSTh1}
\ee
where
\be
 \left<T_{\omega\s}\right>^T \equiv  \left<T_{\omega\s}\right> - {1\over 4}g^{(0)}_{\omega\s}\left<T^\m_\m\right>.
\label{dSTh2}
\ee
The curvature squared terms $\tensor[^{(\a)}]{H}{_\omega_\s}$ and $\tensor[^{(\b)}]{H}{_\omega_\s}$ are then linearized with respect to the tensor perturbation. Then, equation (\ref{dSTh1}) is written as a sum of contributions from the Fefferman-Graham terms $h^{(0)}_{\omega\s}$, $h^{(4)}_{\omega\s}$ and $\hat{h}_{\omega\s}$:
\be
h^{(4)}_{\omega\s} + {L^4R\over 24}\left\{{3\pi \over GN^2R}- {1\over 4} - {\pi\alpha\over 4 N^2} \right\}\left(\nabla^2 -{R\over 6} \right)h^{(0)}_{\omega\s} + \left(1 - 2\log(\m L) + \beta {\pi\over N^2}\right)\hat{h}_{\omega\s} = 0,
\label{dS13b}
\ee
where $\hat{h}_{\omega\s}$ is to expressed in terms of $h^{(0)}_{\omega\s}$ using (\ref{b17}).

Equation (\ref{dS13b}) is a linear equation relating $h^{(4)}_{\omega\s}$ to $h^{(0)}_{\omega\s}$. As usual in holography however,  $h^{(4)}_{\omega\s}$ is determined by  $h^{(0)}_{\omega\s}$ by solving the bulk equation and imposing a regularity condition in the interior. This makes $h^{(4)}_{\omega\s}$ into a (non-local) linear functional of $h^{(0)}_{\omega\s}$.  Therefore, all in all, equation (\ref{dS13b}) takes the form of a dynamical equation for  $h^{(0)}_{\omega\s}$, of the form:
\be \label{bb21}
{\mathcal F}^{\mu\nu\omega\sigma}(\nabla^2, \bar{R})  h^{(0)}_{\omega\s} = 0.
\ee

 Determining the explicit form of the functional ${\mathcal F}$ will be the goal of section 5. Here  we conclude by the remark that equation  (\ref{bb21}) can also be obtained by varying the quadratic part of the action (\ref{s0}) evaluated on-shell: indeed, as  shown in appendix  \ref{quadratic action},  once it is evaluated on the solution of the linear  bulk equation,  the quadratic part of the action (\ref{s0}) is equal to the boundary expression:
\be
S^{(2)}[h^{(0)}] =  {N^2\over 2\pi^2}\int d^4 x\,  \sqrt{\bar{\zeta}}h^{(0)\omega\s }\left\{h^{(4)}_{\omega\s} + \left({{\pi\beta\over N^2}} + 1 - 2\log(\m L)\right)\hat{h}_{\omega\s} + \right.
\nonumber
\ee
\be
\left. + {RL^4\over 24}\left(\nabla^2 - {R\over 6} \right) \left({3\pi\over GN^2R}- {1\over 4} -{\pi\a\over 4N^2} \right) h^{(0)}_{\omega\s}  \right\}.
\label{res11}
\ee
Using (\ref{b17}) for $\hat{h}$ and the determination of $h^{(4)}$ in terms of $h^{(0)}$ from the bulk solution, this expression can be written again as a quadratic functional of $h^{(0)}$:
\be
S^{(2)}[h^{(0)}]  = \int d^4x \,\sqrt{g^{(0)}}\int d^4y\,\sqrt{g^{(0)}}  h^{(0)\mu\nu} (x){\mathcal F}_{\mu\nu\omega\sigma}(\nabla^2, \bar{R})[x,y]  h^{(0)\omega\s(y)}
\label{S2_1}
\ee
where ${\mathcal F}$ is the same functional which gives the  equation of motion (\ref{bb21}), as it is clear  by varying (\ref{S2_1}) with respect to $ h^{(0)\mu\nu}$.  The quantity ${\mathcal F}_{\mu\nu\omega\sigma}$  is the  inverse propagator of the induced boundary gravity tensor fluctuations $h^{(0)}_{\omega\s}$:

 \be
 \mathcal{F}_{\mu\nu\omega\sigma} \equiv {1\over \sqrt{g^{(0)}(x)}} {1\over \sqrt{g^{(0)}(y)}} {\delta^2 S^{(2)}\over \delta h^{(0)\mu\nu}(x)\delta h^{(0)\omega\s}(y)}
 \label{b20}
 \ee

Using   the definition (\ref{ig11a}) in the quadratic action,
\be
S^{(2)} = S^{(2)}_\text{grav} +  S^{(2)}_{CFT},
\ee
the right-hand side of equation (\ref{b20}) can be seen as the sum of two contributions, one from  $S_\text{grav}$ and one from $S_{CFT}$.

As $S_\text{grav}$ is local (it is  quadratic in the boundary curvature), the   first contribution is a {\it local} 4-derivative differential operator,
\be  \label{b22-i}
 {1\over \sqrt{g^{(0)}(x)}} {1\over \sqrt{g^{(0)}(y)}} {\delta^2 S^{(2)}_\text{grav}\over \delta h^{(0)\mu\nu}\delta h^{(0)\omega\s}} =  \delta(x,y) O_{\mu\nu\rho\sigma}(\nabla^2,\bar{R}) .
\ee
The part coming from the CFT is by definition the renormalized stress tensor correlator of the  CFT:
\be\label{b22}
 {1\over \sqrt{g^{(0)}(x)}} {1\over \sqrt{g^{(0)}(y)}} {\delta^2 S^{(2)}_{CFT}\over \delta h^{(0)\mu\nu}\delta h^{(0)\omega\s}} = - \< T_{\mu\nu}(x) T_{\omega\s}(y) \>_{CFT}.
\ee
Therefore, the full inverse graviton propagator (\ref{b20}) has the form :
\be \label{b23}
\mathcal{F}_{\mu\nu\omega\sigma} = \delta(x,y) O_{\mu\nu\rho\sigma}(\nabla^2,\bar{R})   - \< T_{\mu\nu}(x) T_{\omega\s}(y) \>_{CFT}.
\ee
The non-local part is fully contained in the term $h^{(4)}$ in equation (\ref{res11}), and to determine it one has to solve the bulk radial equations.

We make a final comment on the appearance  of the bulk AdS radius $L$ in equation (\ref{res11}). As $L$ is {\it not} a parameter of the 4d theory (only $M L$ is, see equation (\ref{hr43ab-ii}), this quantity should not enter the full spectral operator ${\cal F}_{\mu\nu\rho\sigma}$. This is indeed the case: as will become obvious  with the explicit computations in section \ref{linearized eq}, there is a similar logarithmic contribution  to (\ref{res11}) coming  from $h^{(4)}$, which will effectively replace $\log \mu L \to -2\log (2\mu \sqrt{G} N)$.  These terms come from the variation of Weyl anomaly in the CFT, which has the form of the term with coefficient $\beta$ in the gravitational action (see  equation (\ref{s1})).  This implies that effectively  $\beta$ and $\mu$ will always appear in the combination:
\be \label{b24}
\beta_\text{eff} = \beta - {N^2 \over \pi}\log (4\mu^2 G N^2).
\ee

\section{The boundary scalar perturbation}
\label{bdy pert}

In this section, we focus on scalar perturbations\footnote{Here by scalar we mean with respect to the slice isometry group.}.  In the special case of pure Einstein-Hilbert gravity, the scalar mode is not dynamical because it is constrained by the non-diagonal components of Einstein's equations. But this perturbation is rendered dynamical by the higher curvature terms in the 4d gravitational action.\footnote{If the QFT is not conformal, then the coupling of the QFT to gravity will also contribute to the scalar dynamics via the two-point function of the trace of the energy-momentum tensor.}

In the present setup,  where gravity lives on the boundary of  $AdS_5$, the scalar mode exists only on the boundary because only tensor perturbations are dynamical in the bulk (see e.g. \cite{anomalyinflation1}) as the trace of the energy-momentum tensor has trivial dynamics in a CFT.

\subsection{Gauge fixing}

To study the dynamics of this  boundary scalar mode, we define  boundary metric perturbations:
\be
ds_4^2 =  g^{(0)}_{\omega\s} dx^\omega dx^\s = (\bar{\zeta}_{\omega\s} + \delta \zeta_{\omega\s}^{b})dx^\omega dx^\s,
\label{scalar0}
\ee
where $\bar{\zeta}_{\omega\s}$ is the background metric (flat, dS or AdS) defined in \ref{slicings} and $\delta \zeta_{\omega\s}^{b}$ is a perturbation which, unlike the general perturbation in equation (\ref{b6b}),  depends only on the slice coordinates $x^\mu$.

The decomposition (\ref{b10}) still applies, and the boundary gauge transformations are:
\be
\delta \zeta_{\omega\s}^b \rightarrow \delta \zeta_{\omega\s}^b + 2\hat{\nabla}_{(\omega}\xi_{\s)}.
\label{scalar1}
\ee
One can do a gauge transformation to eliminate the transverse and longitudinal vector components, by choosing $\xi_b$ in (\ref{b7}) to be:
\be
\xi_\s = - \chi_\s - {1\over 2}\partial_\s \phi.
\label{scalar2}
\ee
Keeping only the scalar mode, one is left with:
\be
\delta \zeta_{\omega\s}^b = \psi\, \bar{\zeta}_{\omega\s}.
\label{scalar2a}
\ee
Equation (\ref{scalar2a})  is the definition of the scalar perturbation, and we study its dynamics in the following subsections.

\subsection{Scalar equation of motion}
\label{scalar}
 The classical equations of motion for $\psi$ are obtained by linearizing the Einstein equation (\ref{EE}), which we rewrite here for convenience:
\be
0= E_{\omega\s}[g^{(0)}] \equiv  - {16\pi G\over \sqrt{g^{(0)}}}{\delta S[g^{(0)}] \over \delta g^{(0)\omega\s}}
 \label{def E}
 \ee
 \be
= R_{\omega\s} - {1\over 2}Rg^{(0)}_{\omega\s} + \Lambda g^{(0)}_{\omega\s} + 8\pi G(\tensor[^{(\alpha)}]{H}{_\omega_\s} + \tensor[^{(\beta)}]{H}{_\omega_\s}) - 8\pi G \left<T_{\omega\s}\right>,
\label{scalar4}
\ee
where quadratic curvature terms $\tensor[^{(\alpha)}]{H}{_\omega_\s}$, $\tensor[^{(\beta)}]{H}{_\omega_\s}$ are defined in (\ref{Halpha}) (\ref{Hbeta}) and the CFT stress tensor is given in (\ref{stress}).
The equation of motion for $\psi$ is obtained by linearizing $E_{\omega\s}[g^{(0)}]$ around $E_{\omega\s}[\bar{\zeta}]$. Linear and quadratic curvature terms are linearized using the definitions (\ref{scalar0}-\ref{scalar2a}). However, to obtain the CFT stress tensor, one needs to solve the bulk equations.  Nevertheless, its trace (\ref{hr43aa}) and its divergence (which is zero) are fully constrained by the boundary geometrical tensors. Hence, one can take a shortcut and perturb the trace of Einstein's equation.
It will then be convenient to define the trace of the generalized Einstein tensor (\ref{def E}) as
\be
E[g^{(0)}] \equiv g^{(0)\omega\s} E_{\omega\s}[g^{(0)}].
\label{scalar17a}
\ee
Then, the full, non-linear, traced Einstein equation is given by
\be
 0  = E[g^{(0)}] = -R + 4\Lambda - {\alpha G\over 4} \Box R - {GN^2\over 4\pi}\left(R^{\omega\s}R_{\omega\s} - {1\over 3}R^2\right),
\label{psi6}
\ee
where $\Box$ is the Laplacian operator $\nabla^2$ applied to a scalar quantity.
Equation (\ref{psi6}) only contains scalar geometric quantities of the boundary. When evaluated on the background metric $\bar{\z}$, equation (\ref{psi6}) reduces to (\ref{scalar7}).

The linearization of geometrical quantities which appear in (\ref{psi6}) for an arbitrary perturbation (\ref{scalar0}) around $\bar{\z}$ are given by
\be
\delta R = -\left(3\Box + \bar{R} \right)\psi,\qquad \d \left(R^{\omega\s}R_{\omega\s} \right)=\frac{\bar{R}}{2}\d R.
\label{scalar9}
\ee
We observe that these linearized scalar quantities depend on $\psi$ only, due to $h^{(0)}$ being traceless.
This leads to  the linearized version of equation (\ref{psi6}):
\be
\left[1 + {\a G\over 4}\Box - {GN^2 \bar{R}\over 24\pi}\right](3\Box + \bar{R})\psi = 0.
\label{scalar16}
\ee
Equation (\ref{scalar16}) is  a linear {\it local} equation for $\psi(x)$. However, this equation is misleading, as only one of the two modes appearing in this equation is propagating.
 The correct computation of the scalar propagator comes from  varying the action as in (\ref{def E}). In appendix   \ref{gauge scalar}, we carefully derive the  propagator of the single scalar propagating mode.
\be
\displaystyle{\mathcal{F}^{-1} = -{64\pi G}\left[\a G \bar{R} -12 + {GN^2 \bar{R}\over 2\pi}\right]^{-1}\left\{{1\over \Box+ { 4\over G\a} - {N^2 \bar{R}\over 6\pi \a}} \right\}.}
\label{2pt Psi b-ii}
\ee
Both the position of the pole and the sign of the residue depend on the parameters of the model: there are regions in parameter space where the scalar can be a ghost or a tachyon, \cite{St,VilenkinStaro}.
In our scheme, there is no (scheme-dependent) $\square R$ contribution to the conformal anomaly, and therefore the kinetic term for this mode originates in the $R^2$ term of the gravity action.

There is also a second mode that is not propagating\footnote{This was discussed in \cite{Q1} for flat space with quadratic curvature terms only.}, as we show in Appendix  \ref{gauge scalar}.
 It should be however stressed that the mode, although non-propagating, can affect the dynamics as a non-propagating mode, if boundary conditions are non-trivial, \cite{AM,TWn}.
In more general situations such a mode may become propagating. We expect this to happen if the quantum theory we couple gravity to is a QFT rather than a CFT. In
 such a case, the two-point function of the trace of the energy-momentum tensor contributes non-trivially to scalar modes, and a novel analysis needs to be done in the scalar sector.
Another such example can be found in \cite{MotGW}.

We now focus on the propagating scalar mode. We first  discuss  under which condition the scalar is ghostlike.

 In any background, the scalar mode  is a ghost if the residue of the pole has the ``wrong'' sign, the ``right'' sign being that of the pole of the massless spin-2 pole in pure gravity, which  in our conventions is:
\be
{\cal F}_{massless spin2}^{-1} = 32\pi G{1\over \nabla^2 - {\bar{R}\over 6}}.
\ee

Therefore, the scalar mode is a ghost if the residue of (\ref{2pt Psi b-ii})  is positive, i.e:
\be
\left({\pi\a\over N^2} + {1\over 2}\right){GN^2 \bar{R}\over 12\pi} > 1 \quad  \Rightarrow \quad \text{scalar mode is a ghost}.
\label{2pt Psi c}
\ee
This condition is valid both for positive and negative $\bar{R}$. In sections \ref{dS poles} and \ref{AdS poles}, we shall see that this inequality also appears in the context of the tensor sector, but not exactly for the same reasons.

It is useful to compare the  equation of motion (\ref{scalar16}) to other discussions in the literature. If either $\bar{R} = 0$ or $N = 0$, the equation of motion (\ref{scalar16}) agrees with the $R+R^2$ modified gravity analysis of \cite{Stelle1978}.
The first case, $N =0$, corresponds to pure gravity with no CFT; the second case, $\bar{R}=0$, corresponds to flat space, in which the boundary scalar mode decouples even in the presence of the CFT.%

Neglecting the unphysical solution in (\ref{scalar16}), we are left with a single scalar mode satisfying  a massive Klein-Gordon equation:
\be
\Box\psi = {4\over G\a}\left[{GN^2 \bar{R}\over 24\pi} - 1\right]\psi.
\label{nd15}
\ee

\subsection{Scalar tachyonic instabilities} \label{sec:scalar tachyons}

We define a mode to be {\it tachyonic} if the associated wavefunction in coordinate space grows exponentially at late times. To carry out the analysis we have to specify the background, upon which the form of the mode solutions of equation (\ref{nd15}) depends.

\subsubsection*{Minkowski}

We label modes by the eigenvalue of the D'Alembertian operator,
\be
\Box \psi =  - k^2 \psi.
\label{psi9-i}
\ee
 Then, equation (\ref{nd15})  translates, for $\bar{R}=0$, to:
\be
k^2 = {4\over G\alpha}. 
\label{psi10-i}
\ee
 The theory is tachyon-stable if the invariant four-momentum $k^2$  is timelike or null, $k^2 \leq 0$. In any other case, the solution to (\ref{psi9-i}) will contain solutions which are real exponentials in time, and will generically diverge as $t\to +\infty$.

From equation (\ref{psi10-i}) we conclude that:
\be
\alpha >0 \quad \Rightarrow \quad \text{scalar mode is tachyonic.}
\ee
Note that for  $\a = 0$, the scalar mode is decoupled (at quadratic order).

\subsubsection*{anti-de Sitter}

It is useful to parametrize the eigenvalues in terms of a complex `` total momentum eigenvalue''  $\nu^2$  as follows:
\be
\Box \psi = -{\bar{R} \over 12}\left(\n^2-{9\over 4}\right)\psi,
\label{psi9-ii}
\ee

 Equation (\ref{nd15}) then translates into
\be
\n^2-{9\over 4} = {48 \over G\alpha \bar{R}}\left(1 - {GN^2 \bar{R}\over 24\pi}  \right),
\label{psi10}
\ee

In AdS, a free massive scalar  $\phi$ satisfying $\Box \phi = m^2 \phi$  is tachyonic if it violates the BF bound,
 \cite{BF-1982}, which in 4 space-time dimensions means
\be
m^2 \chi^2 < - {9\over 4},
\ee
where  $\chi$ is the AdS length.  Comparing  with equation (\ref{psi9-ii}) with $\bar{R}=-12\chi^2$,   violation of the BF bound is equivalent to:
\be
\n^2 < 0.
\label{psi15}
\ee
Using (\ref{psi10}) we then conclude:
\be
 {9\over 4} - {4\over \a G\chi^2}\left(1 + {GN^2\chi^2\over 2\pi}\right) < 0 \quad \Rightarrow \quad \text{scalar mode is tachyonic.}
 \label{psi16}
\ee
\begin{itemize}
\item For $\alpha = 0$ the scalar mode decouples as in the other cases.
\item For pure gravity, in the absence of the CFT ($N^2=0$), the tachyonic condition (\ref{psi16}) becomes
\be
{16\over  \alpha   G \chi^2}~~ >~~ 9.
\label{psi17}
\ee
 In particular, this cannot be satisfied  for $\alpha<0$ (this is the opposite compared to the de Sitter case, as we shall see below).
\end{itemize}

\subsubsection*{de Sitter.}

Since de Sitter has no global time-like killing vector,  there may not be a universal (coordinate-independent) definition of what a tachyon is. We use the practical criterion that, {\it in a given coordinate system,} a tachyon is a mode whose amplitude diverges exponentially at late times. We establish this criterion in the three most widely used cases, i.e.  Poincar\'e coordinates (covering the expanding (or cosmological)  patch),  global coordinates,  and  static coordinates (covering the static  patch).

We use the same parametrization (\ref{psi9-ii}) of the D'Alembertian eigenvalues, where now $\bar{R} = 12 H^2$.

\begin{itemize}
\item[]{\bf Cosmological patch (Poincar\'e) coordinates.}

In cosmological (Poincar\'e)  coordinates, the de Sitter metric is given by
\be
ds^2_{dS} = {1\over (H\tau)^2}\eta_{\omega\s}dx^\omega dx^\s = -dt^2 + e^{2Ht}\eta_{\omega\s}dx^\omega dx^\s,   \qquad \bar{R} = 12H^2,
\label{psi1}
\ee
where $\tau = {e^{-H t}\over H}$ is the conformal time.
$\tau\to 0$ or $t\to +\infty$ is the future boundary of dS, ${\cal I}^+$. $\tau\to +\infty$ or $t\to -\infty$ is a past horizon that touches the past boundary of dS, ${\cal I}^{-}$ at one point.

The solutions of Eq. \eqref{psi9-ii}  are given in terms of Bessel functions (see for example \cite{Chesler}):
\be
\psi = (H\tau)^{3\over 2} J_{\pm\n}(\tau p){\sim} e^{Ht(\pm \n-3/2)}, \quad  , t\to +\infty \;  (\tau \to 0)
\label{psi9a}
\ee
where $p\equiv \sqrt{\delta_{ij}p^ip^j}$ is the norm of the 3-dimensional Fourier momentum in the spatial directions. At the  far past horizon ($\tau \to +\infty$) one should impose infalling  boundary conditions for the bulk wave-function, i.e. a Hankel function. This  correspond both  to the retarded correlator in the  holographic calculation,  and to  picking the Bunch-Davis vacuum.

Both solutions  (\ref{psi9a})  are bounded as $\tau\to 0$   if :
\be
|\text{Re}(\n)|  \leq 3/2.
\label{psi2}
\ee
Defining a tachyon as a mode which grows exponentially in time,  equation (\ref{psi10}) then translates into  the statement\footnote{In an expanding background such a mode is still ok to have around as long as the growth rate is much smaller than the Hubble time, i.e.  $|Re(\nu)|- 3/2| \ll 1$. In this case one can say that de Sitter space is unstable but long-lived.}:
\be
{1\over \a}\left(1- {GN^2H^2\over 2\pi}\right) > 0 \quad \Rightarrow \quad \text{scalar mode is tachyonic.}
\label{psi_tachyonic}
\ee
By inserting  equation  (\ref{psi10}) into the exponent of (\ref{psi9a}), we obtain the ``decay rate'' $\Gamma$ of de Sitter due to the tachyon instability:
\be
\G = H\left[ \sqrt{{9\over 4} + {4\pi \over GN^2H^2 \tilde{\a}}\left(1 - {GN^2H^2\over 2\pi}\right)} -{3\over 2} \right],
\label{Gamma-scalar}
\ee
which is real and positive if we are in the tachyonic regime (\ref{psi_tachyonic}).

We now  discuss a few special cases.
\begin{itemize}
\item As in flat space, if $\alpha=0$, the scalar mode is non-propagating.
\item The special case $GN^2H^2 = 4\pi$ corresponds to a vanishing 4d cosmological constant  $\Lambda= 0$ (by equation (\ref{scalar7})).  This is the case  studied in \cite{anomalyinflation1}. We find, in agreement with that work, that scalar tachyonic  instabilities occur for $\alpha<0$. This also includes the homogeneous scalar mode from the original Starobinsky model \cite{St,VilenkinStaro}.
\item In the absence of the  CFT, i.e. for  $N^2= 0$,   de Sitter is  tachyon-unstable in the scalar sector for $\alpha> 0$. This is the opposite sign compared to the previous paragraph case.
There is no contradiction here,  since (\ref{scalar7}) shows that de Sitter is not a solution when both $\Lambda$ {\it and} $N^2$ vanish.
\end{itemize}

\item[]{\bf Global de Sitter coordinates}

In global coordinates, the de Sitter metric reads:
\be \label{globalC}
ds^2_{dS} = H^2 \left( -d T^2 +  \cosh^2 T d\Omega^2 \right)
\ee
where $-\infty < T < +\infty$ and $d\Omega^2$ is the metric on the unit 3-sphere. As their name suggests, these coordinates cover the entirety of dS space.
$T\to -\infty$ is the past dS boundary, ${\cal I}^-$ and $T\to+\infty$ is the future dS boundary, ${\cal I}^+$.

At late times $T\to \infty$,  this metric  looks like the metric  (\ref{psi1}) with $t \to  H T$,  except for the fact that Euclidean $3$-space is replaced by a sphere. Therefore,  the solutions of the D'Alembert equation at late times will again take the form (\ref{psi9a}), with $p^2$ now  appropriately quantized. Consequently, the condition that the solution diverges at late time is the same in global coordinates as in cosmological coordinates, namely (\ref{psi2}), leading again to (\ref{psi_tachyonic}).

The difference with respect to the cosmological patch is that now one is free to chose any solution at the past boundary, and this defines different quantum states in de Sitter. Our calculation translates into selecting a dS-invariant state.

\item[]{\bf Static Patch coordinates}

The static patch of de Sitter is described by the metric:
\be
ds^2  = - (1 - H^2 r^2) d\hat{t}^2 + {dr^2  \over 1-H r^2} + r^2 d\Omega_2^2
\ee
where $0 < r < H^{-1}$ and $d\Omega^2$ is the metric on the unit 2-sphere.
This metric contains only a single point from each of the ${\cal I}^+, {\cal I}^-$.

This metric has a (cosmological) horizon at $r_h = H^{-1}$, where one should impose infalling\footnote{This way, the holographic correlator is fixed to be the retarded one} or normalizable boundary conditions for the wave-function, in addition to regularity at the origin $r=0$.

The calculation of the spectrum of the de Sitter D'Alembertian with these boundary condition results in obtaining the static patch quasi-normal modes, with time dependence $e^{-i\omega \hat{t}}$. The spectrum of quasi-normal frequencies for a scalar field of mass $m^2 = -H^2(\nu^2 - 3/2)$  can be found for example in \cite{lopez}, and reads:
\be
\omega_{n,l} = -i H \left(l + 2n + {3\over 2} \pm \nu \right),
\ee
where $l$ and $n$ are non-negative integers. Stability requires all the quasi-normal frequencies to lie in the lower complex plane (so  that the time dependence $e^{-i\omega \hat{t}}$ is exponentially damped at late times). The most stringent requirement occurs for $l=0, n=0$, and it translates into $|Re(\nu)| < 3/2$, i.e. the same condition  (\ref{psi2}) we found in cosmological and global coordinates.

This establishes the validity of the the condition (\ref{psi_tachyonic}) about  the tachyonic (in)stability of scalar modes in the static patch of de Sitter as well.

\end{itemize}

\section{The  spin-two spectral equations}
\label{linearized eq}

We now move to  tensor perturbations $h^{(0)}_{\omega\s}$ defined in (\ref{b20}). The linearized Einstein equation (\ref{dS13b}) contains  $h^{(4)}$, which can only be specified by solving the perturbation equations in the bulk.  This section is devoted to expressing $h^{(4)}$ in terms of the boundary perturbation $h^{(0)}$ by solving the bulk tensor equation (\ref{b14}). This has to be done separately for each slice geometry  (flat, positive and negative curvature). We treat each case in a separate subsection.

\subsection{Flat-slicing} \label{sec:flat prop}
The bulk equation of motion for the tensor mode (\ref{b14}) simplifies significantly in flat slicing coordinates.
First, one can write the bulk metric as a conformally flat space by defining the usual Poincar\'e coordinate $Z$ as
\be
Z \equiv e^{-u} = \sqrt{\rho}.
\label{flat1}
\ee
The perturbed bulk metric (\ref{b6b}) in which we only keep the propagating tensor $h_{\omega\s}$ is then written as
\be
ds_5^2 = {1\over Z^2}[L^2 dZ^2 + (\eta_{\omega\s} + h_{\omega\s}) dx^\omega dx^\s].
\label{flat1a}
\ee
The bulk equation of motion (\ref{b14}) describes the dynamics of a massless graviton in AdS. In flat slicing coordinates (\ref{flat1a}), we can insert $a = e^{u}$ into (\ref{b14bis}), which boils down to the massless scalar equation
\be
\Box_5 h_{\omega\s} = 0,
\label{flat2}
\ee
where $\Box_5$ is the $AdS$ scalar Laplacian in Poincar\'e coordinates (\ref{b0flat}), given by
\be
L^2\Box_5 = Z^2(\partial_Z^2 + L^2\eta^{\k\l}\partial_\k\partial_\l) - 3Z\partial_Z.
\label{flat3}
\ee

Now the strategy is to search for separable solutions, which we write as
\be
h_{\omega\s}(Z,x) = F(Z,k) \tilde{h}^{(0)}_{\omega\s}(x,k),
\label{flat4}
\ee
where $\tilde{h}^{(0)}$ solves the eigenvalue equation parametrized by $k^2$ as
\be
\partial^\s\partial_\s \tilde h^{(0)}_{\omega\k} = -k^2 \tilde h^{(0)}_{\omega\k} \equiv (m_2)^2 \tilde h^{(0)}_{\omega\k},
\label{flat5}
\ee
and $(m_2)^2$ can be a complex number in general.

The second equation is an ordinary differential  equation for $F(Z,k)$,
\be \label{flat5a}
\left(Z^2\de_Z^2 - 3 Z \de_Z -  k^2 L^2 Z^2\right) F(Z,k) = 0.
\ee
Note that, ultimately,  we want to write an equation for the boundary tensor perturbation of the form (\ref{bb21}) in Fourier space,  with $\nabla^2$ replaced by $-k^2$. The solution will be the physical mass$^2$ of a propagating 4d mode. As the solutions of this equation  may be complex, we must allow for complex values of $k^2$ beyond the usual choices of timelike ($k^2<0$) and spacelike ($k^2>0$) momentum one obtains for real wavenumbers.
With this caveat, it is now convenient to write (\ref{flat5a}) as:
\be
\left[y^2{d^2\over dy^2} - y^2 - 3 y{d\over dy}\right]F(y) = 0, \qquad  y\equiv L k Z.
\label{flat6a}
\ee
where  we define $k$, for any complex  $k^2$ outside of the negative real axis, as  the complex root of $k^2$ with {\it positive} real part\footnote{This prescription is enough to identify tachyonic modes, for which $Re(k) \neq 0$.  Instead, real negative $k^2$ corresponds to non-tachyonic propagating particles, and as usual, their propagator needs a further  prescription. We  use the analytic continuation of the results to purely imaginary values of $k$, which corresponds to taking the retarded stress tensor 2-point function .} and  with a slight abuse of notation, we have replaced $F(Z, k)$ by $F(y)$.

We now solve the equations for $\tilde{h}^{(0)}$ (\ref{flat5}) and for $F$ (\ref{flat6a}). First, the solution of (\ref{flat5}) are the Fourier modes :
\be
\tilde h^{(0)}_{\omega\k}(k^\s) = e^{\pm ik^\s x_\s} = e^{\pm i (-\omega t + \bm{k}.\bm{x})},\;   \bm{k}\in \mathbb{R}^3, \; \omega \equiv \sqrt{-k^2 + \bm{k}^2}.
\label{flat6}
\ee

For a non-negative eigenvalue $k^2$, modes with $|\bm{k}| < |k|$  will necessarily feature an imaginary part in $\omega$ and one of the two solutions in (\ref{flat6}) will diverge with time. This is the usual tachyon instability for flat space, which occurs for massive Klein-Gordon equations with negative mass square, and more generally it persists also for a complex mass.
Therefore, the condition that a mode characterised by $k^2$ is non-tachyonic is:
\be
\text{Re}(k) = 0.
\label{flat7a}
\ee
where, as above, we have defined $k$ as the complex root of $k^2$ with positive real part\footnote{This is simply a convention since the equation has a symmetry in $k\to -k$.}.

Equation (\ref{flat6a})  is solved by modified Bessel functions,
\be
F(y) = y^2 (\lambda_1 K_2(y) +  \lambda_2 I_2(y)).
\label{flat7}
\ee

We must impose that the solution (\ref{flat7}) is  regular at the horizon $Z \rightarrow +\infty$. This   requires $\lambda_2 = 0$ because in this limit $I_2(y) \sim exp[k L Z]$ and   by definition  $\text{Re}(k)>0$. The remaining solution $K_2(y)$ is a vanishing exponential at $Z\to +\infty$.

We fix the remaining parameter $\lambda_1$ by choosing  the normalization at the AdS$_5$ boundary $Z=0$ so that:
\be
F(Z=0) = 1.
\label{flat11}
\ee
This way, the solution (\ref{flat4}) for the bulk tensor perturbation $h_{\omega\s}(x,Z)$ coincides at $Z=0$ with the boundary tensor mode $h^{(0)}_{\omega\s}$ defined in the FG expansion  (\ref{b15}). For this reason, the leading term $\tilde h^{(0)}_{\omega\s}$ of $h_{\omega\s}$ in (\ref{flat4}) is identified as the Fourier mode of the leading term in the Fefferman-Graham expansion (\ref{b15}) which was defined as $h^{(0)}_{\omega\s}$.
We drop the tilde from now on.

For small $y$, the Bessel function $K_2$ behaves as:
\be
K_2(2y) \underset{y\rightarrow 0}{=} {1\over 2}\left\{y^{-2} - 1 + {3\over 4}y^2 - y^2\left[ \gamma_E + \log(y)\right] \right\}+ \mathcal{O}(y^4),
\label{flat10}
\ee
where $\g_E$ is the Euler-Mascheroni constant.
Then, equation (\ref{flat11})  fixes  $\lambda_1 = 1/2$ in (\ref{flat7}).

Having completely fixed $F(Z,k)$, we can read-off   $h^{(4)}_{\omega\s}$ and $\hat{h}_{\omega\s}$ from its near-boundary expansion,  (\ref{flat10}) and (\ref{flat4}) and compare with the corresponding  terms in equation (\ref{b15}), recalling that $Z=\sqrt{\rho}$ . We find:
\begin{subequations}
\label{n21a}
\be
h^{(2)}_{\omega\k} = - \left({kL\over 2}\right)^2 h^{(0)}_{\omega\k},
\label{n21}
\ee
\be
h^{(4)}_{\omega\k} =\left({kL\over 2}\right)^4\left[ {3\over 4} - \gamma_E - \log\left({kL\over 2}\right)\right] h^{(0)}_{\omega\k},
 \label{n26}
\ee
\be
\hat{h}_{\omega\k} = -{1\over 2}\left({kL\over 2}\right)^4 h^{(0)}_{\omega\k}.
\label{n26a}
\ee
\end{subequations}
The terms $h^{(2)}_{\omega\k}$ and $\hat{h}_{\omega\k}$ agree with the perturbative solutions of the bulk Einstein equation (\ref{hr20}) that are given in appendix \ref{renorm} by (\ref{fg16}) and (\ref{fg26}).  To perform this comparison and check that they agree, it is enough to linearize $g^{(2)}_{\omega\k}$ (\ref{fg16}) and $\hat{g}_{\omega\k}$ (\ref{fg26}) with respect to the transverse traceless perturbation $h^{(0)}_{\omega\s}$.

The linearization of the stress tensor (\ref{stress}) around a flat background for the tensor perturbation is given by\footnote{The term $h^{(2)}$ does not contribute in (\ref{n26b}) because it always appears in the CFT stress tensor (\ref{stress}) multiplied by $g^{(2)}[\bar{\zeta}]$, which vanishes on a flat background.}:
\be
\delta_h \left<T_{\omega\k}\right> = {N^2\over 2\pi^2L^4}\left[h^{(4)}_{\omega\k} + \left(1- 2\log \mu L\right)\hat{h}_{\omega\k}\right].
\label{n26b}
\ee

We can use the bulk solutions (\ref{n21a}) into (\ref{n26b}), to obtain the perturbed stress-tensor in terms of $h^{0}$ alone:
\be
\delta_h \left<T_{\omega\k}\right> = {N^2\over 2\pi^2}  \left( {k\over 2}\right)^4\left[ {1\over 4} - \gamma_E - \log\left({k\over 2\m}\right)  \right] h^{(0)}_{\omega\k}.
\label{n27}
\ee

As a final step, inserting the expressions  (\ref{n21a})  in   (\ref{dS13b}) we obtain  the  linearized Einstein equation specialized to a flat background, in the form of an equation for $h^{(0)}$ alone:

\be
{N^2\over 64\pi^2} k^2 Q_\text{flat}(k)h^{(0)}_{\omega\s} = 0,
\label{n28ab}
\ee
where
\be
Q_\text{flat}(k) \equiv \left\{-{2\pi \over GN^2} + k^2\left[{1\over 4} - \g_E - \log\left({k\over 2\m}\right)- {1\over 2}{{\pi\beta\over N^2}}\right]\right\}.
\label{n28b}
\ee
From (\ref{n28b}), as anticipated in Section \ref{bulk pert}, we  can observe that the contributions from the renormalization scale $\m$ coming from the CFT and the quadratic curvature term proportional to $\b$ combine into the parameter $\beta_\text{eff}$ given in equation (\ref{b24}).
For convenience, we also define
\be
\tilde{\b}_\text{eff} \equiv {\pi\b_\text{eff}\over N^2}.
\label{n28d}
\ee
The spectral equation (\ref{n28ab}) is then written as
\be
Q_\text{flat}(k) \equiv \left\{-{2\pi \over GN^2} + {k^2\over 2}\left[{1\over 2} - 2\g_E - \log\left({GN^2k^2}\right)- \tilde{\b}_\text{eff}\right]\right\}.
\label{n28e}
\ee

The quantity multiplying $h^{(0)}$ in (\ref{n28ab}) is the inverse propagator (\ref{b20}) for a flat space-time. Its expression is given by\footnote{For the overall coefficient in this expression, see  Appendix \ref{quadratic action}.}:

\be
\mathcal{F}_\text{flat}(k) =  {N^2\over 64\pi^2} k^2 Q_\text{flat}(k).
\label{n28a}
\ee
Non-trivial solutions ($h^{(0)}_{\omega\s}\neq 0$) to the equation of motion (\ref{n28ab})  correspond to the propagating momentum modes $k$ of the boundary perturbation $h^{(0)}$ and are found by solving the  spectral equation
\be
k^2 Q_\text{flat}(k) = 0.
\label{n29}
\ee
Solutions of this equation are the poles of the propagator $\mathcal{F}_\text{flat}^{-1}$.
An obvious solution to that equation is the massless mode $k^2=0$ which is  present in pure Einstein-Hilbert gravity. ``Exotic'' Modes with $k^2\neq 0$ satisfy:
\be
1 = {GN^2 k^2 \over 4\pi}\left({1\over 2} - 2\gamma_E - \log\left(GN^2k^2\right) - \tilde{\b}_\text{eff} \right).
\label{n31}
\ee
The only solutions which are non-tachyonic are those for which $k^2<0$.
The absence of tachyon-instabilities of flat space is then equivalent to the absence of solutions $k$ to (\ref{n31}) with a non-zero real part. We study the existence of such unstable solutions in section \ref{sec:flat correlator}.

\subsection{dS-slicing}
\label{dS prop}
We now consider the CFT on de Sitter and we turn to equation (\ref{b14bis}) applied to dS slicing coordinates (\ref{b0dS}). As a result, we obtain
\be
\left\{\partial_u^2 + 4\coth{u}\partial_u + \frac{H^{-2}\hat{\nabla}^2 - 2}{\sinh^2u}\right\}h_{\omega\s} = 0.
\label{dS3}
\ee
The operator inside curly brackets is similar to the expression of the Laplace operator of $AdS_5$ acting on scalars, in which case the numerator of the last term would be replaced by the 4-dimensional slice scalar Laplacian.

Similarly to the flat slicing case (\ref{flat4}), we search for separable solutions of the form:
\be
h_{\omega\s}(x ,u) = F(u,\n)\tilde{h}^{(0)}_{\omega\s}(x,\n).
\label{dS4}
\ee
This results in two equations: the first one is an eigenvalue problem on the slice, which we write as:
\be
 (\hat{\nabla}^2 - 2H^2)\tilde{h}^{(0)}_{\omega\s} =  -H^2\left(\n^2 - {9\over4}\right) \tilde{h}^{(0)}_{\omega\s} \equiv (m_2)^2\tilde{h}^{(0)}_{\omega\s}.
 \label{dS5}
\ee
The second equation is an ODE in the radial direction:
 \be
 \left\{{d^2\over du^2} + 4 \coth u{d\over du} - {\n^2 - {9\over 4} \over \sinh^2 u}\right\}F(u,\n) = 0.
 \label{dS7}
 \ee

The information about tachyonic instabilities is contained in the value of $\n$.
As shown in Appendix \ref{dS criterion}, and as it is pointed out in \cite{anomalyinflation1,Chesler}, modes with
\be
|\text{Re}(\n)| >3/2
\label{dS6}
\ee
are tachyonic because (\ref{dS5}) contains a solution which diverges with time (see appendix \ref{dS criterion} for the details).

In pure 4d gravity, the only propagating mode would be the transverse-traceless graviton, which is a zero eigenvalue for the  Lichnerowicz operator of de Sitter (the left-hand side of equation (\ref{dS5})
and corresponds to  $\n = \pm 3/2$.

Turning on the CFT matter content and the quadratic curvature terms will allow for modes with different values of $\nu$. These will be determined by solving the  boundary spectral equation, which we derive below.

As in the flat case, to obtain the boundary spectral equation we have to solve the radial equation (\ref{dS7}).  The most general solution of (\ref{dS7}) is a linear combination of two hypergeometric functions given by
\be
F(u,\n) = C_+ \tanh u^{\n-\frac{3}{2}}\tensor[_2]{F}{_1}\left[\frac{1}{2}\left(\nu-\frac{3}{2}\right),\frac{1}{2}\left(\nu-\frac{1}{2}\right);1+\nu;\tanh^2u\right] +\nonumber
\ee
\be
 + C_- \tanh u^{-\nu-\frac{3}{2}}\tensor[_2]{F}{_1}\left[\frac{1}{2}\left(-\nu-\frac{3}{2}\right),\frac{1}{2}\left(-\nu-\frac{1}{2}\right);1-\nu;\tanh^2u\right],
\label{dS8}
\ee
where $C_\pm$ are integration constants.
This solution may have a   singularity at the horizon $u = 0$, depending on the real part of $\n$. As we show in appendix \ref{bulk schrodinger},
requiring the solution (\ref{dS8}) to be normalizable at $u=0$ gives the following constraints :
\begin{itemize}
\item If $\text{Re}(\n) > 0$, we need to set $C_-=0$ for normalizability at $u=0$.
\item If $\text{Re}(\n) < 0$, we need to set $C_+=0$ for normalizability at $u=0$.
\item If $\text{Re}(\n) = 0$, both solutions oscillate at the horizon $u=0$.
\end{itemize}
Since the problem (\ref{dS7}) is symmetric in $\n\leftrightarrow -\n$, we can choose $\text{Re}(\n) \geq 0$ without loss of generality. In this case, the most general regular solution at $u=0$ is the one with $C_- = 0$. The case where $\n$ is imaginary lies in the stable region (\ref{dS6}) and needs a further prescription (e.g. infalling boundary conditions). We shall define the spectral function by analytic continuation to purely imaginary $\nu$.

We fix  the normalization of $F$ by imposing  that in the UV, at $u\rightarrow +\infty$:
\be
F(u,\nu) \underset{u \rightarrow +\infty}{\longrightarrow} 1.
\label{dS10}
\ee
This condition ensures that
$\tilde{h}^{(0)}_{\omega\s}$ defined in  (\ref{dS4}) identifies with the leading term $h^{(0)}_{\omega\s}$ of the Feffeman-Graham expansion (\ref{b15}). From now on, we drop the tilde on $h^{(0)}$ although it only represents a single mode $\n$ (\ref{dS5}).

Using Gauss' hypergeometric theorem
\be
\tensor[_2]{F}{_1}\left[a,b;c;1\right] = \frac{\Gamma(c)\Gamma(c-a-b)}{\Gamma(c-a)\Gamma(c-b)},
\label{dS11}
\ee
valid for $\text{Re}(c)>\text{Re}(a+b)$, the boundary condition (\ref{dS10}) fixes the value of the integration constant to
\be
C_+ = \frac{\Gamma\left(\frac{5}{2}+\nu\right) \sqrt{\pi}2^{-\nu-\frac{3}{2}}}{\Gamma(1+\nu)}.
\label{dS12}
\ee

The near-boundary expansion of $F(u,\n)$ can be obtained using a hypergeometric transformation (page 49 of \cite{Magnus}). It allows us to transform $F(u,\n)$ into a power series of $e^{-u}$ instead of $\tanh^2u$ (hypergeometric functions are defined as a power series of their last argument). The first few terms of the result are given by
\be
F(u,\nu) = 1 - e^{-2u}\left(\n^2 - {9\over 4}\right) - e^{-4u}\left(\n^2 - {9\over 4}\right)\bigg\{1 + \nonumber
\ee
\be
+ \left(\n^2 - {1\over 4}\right)\left[-u - {3\over 4} + \mathcal{H}\left(\n-{1\over 2}\right)\right]\bigg\} + \mathcal{O}(e^{-6u}),
\label{dS13}
\ee
where $\mathcal{H}$ is the harmonic number function defined in terms of the Euler Gamma function, $\G$ as
\be
\mathcal{H}(z) = {\G'(z+1)\over \G (z+1)} + \g_E.
\label{hn}
\ee
The terms in (\ref{dS13})  are enough to read all of the Fefferman-Graham expansion (\ref{b15}) using the relation between positive $u$ and $\rho$ in tabular \ref{tabular slicings}:
\begin{subequations}
\label{dS13aa}
\be
h^{(2)}_{\omega\s} = -h^{(0)}_{\omega\s}\left({LH\over 2}\right)^2\left(\nu^2-{9\over 4}\right)
\label{i60d1}
\ee
\be
h^{(4)}_{\omega\s} = - h^{(0)}_{\omega\s} \left({LH\over 2}\right)^4 \left(\nu^2-{9\over 4}\right)\left\{ 1 + \left(\nu^2-{1\over 4}\right)\left[ \log\left({LH\over 2}\right)- {3\over 4}
+ \mathcal{H} \left( \n - {1\over 2}  \right)\right]\right\}
\label{i611}
\ee
\be
\hat{h}_{\omega\s} =  - {h^{(0)}_{\omega\s}\over 2} \left({LH\over 2}\right)^4 \left(\nu^2-{9\over 4}\right) \left(\nu^2-{1\over 4}\right).
\label{i61a1}
\ee
\end{subequations}

As we already discussed for the flat slicing, $h^{(2)}$ and $\hat{h}$ can be found using an independent method discussed in appendix \ref{renorm}. This method consists in solving perturbatively the bulk Einstein equation at small values of the Fefferman-Graham coordinate $\rho$ for an arbitrary boundary metric $g^{(0)}$. The solution for $g^{(2)}$ and $\hat{g}$, given in (\ref{fg16}) and (\ref{fg26}), can be linearized with respect to $h^{(0)}$ to obtain the same result as (\ref{i60d1}) and (\ref{i61a1}) using the formulae (\ref{dS13aaa}). However,   this alternative method does not determine $h^{(4)}$ in terms of $h^{(0)}$, but only its trace and divergence.

Inserting \eqref{i611}-\eqref{i61a1} into the linearized Einstein equation for the tensor mode (\ref{dS13b}), we find the equation of motion of the boundary spin-2 perturbation in momentum space given by
\be
{N^2 H^4 \over 64\pi^2}\left(\n^2-{9\over 4}\right)Q_{dS}(\n) h^{(0)}_{\omega\s} = 0,
\label{dS14}
\ee
where
\be
Q_\text{dS}(\n) \equiv 1 - {2\pi\over GN^2H^2} + 2\tilde{\a} - {1\over 2}\left(\n^2-{1\over 4}\right)\left[\log\left(GN^2H^2\right)-{1\over 2} + 2\mathcal{H}(\n-1/2)  + \tilde{\b}_\text{eff}\right],
\nonumber
\ee
\be
\text{Re}(\n)>0,
\label{dS15}
\ee
where, as in flat space,  we have   combined the contributions from $\m$ and $\b$ into a single parameter  $\tilde{\b}_\text{eff}$ defined via  equations (\ref{b24}) and (\ref{n28d}). We have also defined the parameter $\tilde{\a}$ as
\be
\tilde{\a} \equiv {\pi\a\over N^2}.
\label{dS15a}
\ee

From now on, we shall always refer to the new quadratic curvature coefficients $\tilde{\a}$ and $\tilde{\b}_\text{eff}$, except in subsection \ref{decoupling} where we  set $N=0$ in which case these new quantities become ill-defined.

The inverse spin-2 propagator defined in (\ref{b20}) is then given by
\be
\mathcal{F}_\text{dS}(\n) = {N^2H^4\over 64\pi^2}\left(\n^2-{9\over 4}\right)Q_{dS}(\n), \quad \text{Re}(\n)>0.
\label{dS17}
\ee
The overall coefficient of (\ref{dS17}) is obtained in appendix \ref{quadratic action}.
The expression for $Q_\text{dS}$ given in (\ref{dS15}) is only valid for positive real parts of $\n$ because we chose $C_- = 0$ for normalizability of the bulk solution (\ref{dS8}) at $u=0$.
By symmetry of the bulk equation (\ref{dS7}) in $\n\leftrightarrow -\n$, and in $\n \leftrightarrow \n^*$, the propagator $\mathcal{F}_\text{dS}(\n)$ must also obey the same symmetries. The combination of these two symmetries implies that both the real and imaginary axes of $\n$ are axes of symmetry for $\mathcal{F}_\text{dS}$. As a  consequence, the inverse propagator for $\text{Re}(\n)<0 $ is obtained by replacing $\n \rightarrow -\n$ in (\ref{dS15}).

Each value of $\n\in\mathbb{C}$ solving equation (\ref{dS14}) is a pole of the  2-point function $\mathcal{F}_\text{dS}$ and corresponds to a propagating mode.
The positions and residues of these poles depend on the Hubble rate $H$ of the boundary metric $\bar{\zeta}_{\omega\s}$ (\ref{b0}), the quadratic curvature coefficient $\tilde{\a}$ (\ref{dS15a}, \ref{s03}), the scheme-dependent quadratic curvature coefficient $\tilde{\b}_\text{eff}$ (\ref{n28d}, \ref{s1}) and the colour number $N^2$.

The existence of tachyonic modes $\left|\text{Re}(\n)\right|>3/2$ will be studied in section \ref{dS poles}.

\subsection{AdS-slicing}
\label{AdS prop}
Deriving the analogous spin-2 spectral equation (\ref{dS14}) for AdS slicing follows the same steps as in the previous subsection, with the difference that now there are two UV boundaries, located at $u\rightarrow \pm \infty$, corresponding to two  CFTs, \cite{AdS1}.

The equation of motion for bulk metric perturbations (\ref{b14}) in AdS slicing (\ref{b0AdS}) is
\be
\left\{\partial_u^2 + 4\tanh{u} \partial_u + {\chi^{-2}\hat{\nabla}^2 + 2 \over \cosh^2{u}}\right\}h_{\omega\s} = 0,
\label{AdS1}
\ee
which is  the analogue of equation (\ref{dS3}).
As it was done for dS slicing, we search for separable solutions of the form:
\be
h_{\omega\s} = F(u,\n)\tilde{h}^{(0)}_{\omega\s}(x,\n).
\label{AdS2}
\ee
We then separate equation (\ref{AdS1}) into an eigenvalue problem on the slice,
\be
(\hat{\nabla}^2 + 2\chi^2)\tilde{h}^{(0)}_{\omega\s} = \chi^2\left(\n^2-{9\over 4}\right)\tilde{h}^{(0)}_{\omega\s} \equiv (m_2)^2\tilde{h}^{(0)}_{\omega\s},
\label{AdS3}
\ee
and a  radial equation,
\be
\left\{{d^2 \over du^2} + 4\tanh{u} {d\over du} + {\n^2 - {9\over 4} \over \cosh^2{u}}\right\}F(u,\nu) = 0.
\label{AdS5}
\ee

Before solving the radial equation, we first comment on the role the eigenvalues $\n$ play in the tachyonic instability.
Note that the massless graviton is associated with the eigenvalue $\n = \pm 3/2$. Unlike  de Sitter, where  the eigenvalue for a massless spin-2 graviton separates between tachyonic and non-tachyonic modes,  in AdS, some negative masses are non-tachyonic because they are allowed by the BF bound \cite{BF-1982}. Thus, in AdS, the massless graviton does not saturate the stability bound.

For general complex $\nu$,  we study the stability of metric perturbations in the Poincar\'e patch of AdS in appendix \ref{AdS criterion}. To obtain a condition on the value of $\n$ in the complex plane, we study the existence of normalizable tachyonic modes of $AdS_4$ for an arbitrary $\n$. As a result, such normalizable tachyonic modes exist if and only if
\be
\n^2<0.
\label{AdS4}
\ee
When $\n^2$ is real, this statement reduces to the usual BF bound. Furthermore, we find that any complex $\n$ with a non-zero real part is not tachyonic.

The most general solution of equation  (\ref{AdS5}) is given by associated Legendre functions
\be
F(u,\n) = (\cosh u)^{-2}\left(\lambda_1 P_{\nu -1/2}^2(\tanh{u}) + \lambda_2 Q_{\nu-1/2}^2(\tanh{u})\right).
\label{AdS6}
\ee

Since AdS-slicing coordinates (\ref{b0AdS}) do not contain a horizon at $u=0$, tensor perturbations $h(u,\n)$ can propagate in the whole bulk, between the two UV boundaries at $u\rightarrow \pm\infty$. As a consequence, we are left with a choice of boundary conditions that we did not have for dS-slicing (in which case we imposed normalizability at the horizon).
In AdS-slicing coordinates, different linear combinations  of the two independent bulk solutions (\ref{AdS6}) correspond to different combinations of sources coupled to the CFT  on each  boundary. In our  case, the boundary source is $h^{(0)}_{\omega\s}$, the boundary metric perturbation. Therefore, generically,  this setup corresponds to  a  bimetric theory.

One possible choice is that only  the boundary metric at $u\rightarrow -\infty$ is chosen to be dynamical. Then,  one should impose that on the  other boundary, at $u\rightarrow +\infty$, the source term of boundary metric perturbation vanishes. Another possibility to have a single dynamical metric is to identify the two boundaries, which corresponds to imposing a $\mathbb{Z}_2$ symmetry $u \to -u$  on the solution. We consider each of these  cases in the following two subsections.

The discussion above is relevant for a holographic CFT.
For a generic CFT on AdS$_4$, the two-point function of the energy-momentum tensor depends on boundary conditions.
For the simplest boundary conditions, Neumann or Dirichlet, the two-point function of the energy-momentum tensor can be calculated by mapping it to flat space by a conformal transformation and then using the method of images.
We shall not pursue this further in the present paper.

 \subsubsection{Dynamical gravity on one side}
\label{AdS one sided}

We choose to turn off leading (i.e. source-like) metric perturbations on the boundary at $u \to +\infty$. This corresponds to the boundary conditions:
\be
\label{as1a}
F(u,\n) \underset{u \rightarrow \pm \infty}{\longrightarrow}
\left\{
\begin{array}{ll}
2 \lambda_2 &= 0,\\ & \\
{4\over \pi}\lambda_1 \cos(\pi\n) - 2\lambda_2 \sin(\pi\n) &= 1,
\end{array}
\right.
\ee
such that $\tilde{h}^{(0)}$ (\ref{AdS2}) coincides with the leading term $h^{(0)}$ of the Fefferman-Graham expansion (\ref{b15}) at the $u\rightarrow -\infty$ side of the $AdS_5$ boundary. These conditions fix the coefficients in (\ref{AdS6}) as
\be
\lambda_2 = 0,
\label{as3}
\ee
\be
\lambda_1 = {\pi\over 4\cos(\pi\n)},
\label{as4}
\ee
valid for $\nu \neq n + {1\over2 }$, $n\in \mathbb{Z}$. In the case where $\nu = n + {1\over2 }$, we still have the requirement that  $\lambda_2=0$ but $\lambda_1$ is unconstrained.

The case of $\n = n +1/2$ is special because it corresponds to the spectrum of normalizable modes in $AdS_5$, which can be seen from the asymptotic behaviour of Legendre functions in (\ref{i55}).
Since $\l_1$ diverges in the limit $\n\rightarrow n + 1/2$, this discrete series of modes correspond to poles of the stress-tensor propagator or zeros of the propagator for tensor metric perturbations. Therefore, they cannot be the solution of the spin-2 spectral equation. As a consequence, we can ignore them for the rest of the paper.

The unique solution (\ref{AdS6}) which satisfies the boundary conditions (\ref{as1a}) can then be expanded near the dynamical boundary $u\rightarrow -\infty$. This expansion is given by
\be
F(u,\n) =  1+ \left(\nu^2-{9\over 4}\right)\bigg\{ e^{2u} - e^{4u}\bigg[1+ \left(\nu^2-{1\over 4}\right)\Big(u - {3\over 4}
\nonumber
\ee
\be
+ {1 \over 2} \mathcal{H}\left(\n -{1\over 2}\right)+ {1\over 2}\mathcal{H}\left(-\n -{1\over 2}\right)\Big)\bigg]\bigg\} + \mathcal{O}(e^{6u}),
\label{as5}
\ee
where $\mathcal{H}$ is again the harmonic-number function defined in (\ref{hn}).
To read-off the Fefferman-Graham terms of the spin-2 perturbation from (\ref{as5}), we need to replace $u$ by the Fefferman-Graham coordinate given in  \ref{tabular slicings} by
\be
e^{2u} = \left({L\chi\over 2}\right)^2\rho.
\label{as6}
\ee
Then, each term of the Fefferman-Graham expansion (\ref{b15}) can be identified from (\ref{as5}) as
\begin{subequations}
\label{as7}
\be
h^{(2)}_{\omega\s} = h^{(0)}_{\omega\s}\left({L\chi\over 2}\right)^2\left(\nu^2-{9\over 4}\right)
\label{as7a}
\ee
\be
h^{(4)}_{\omega\s} = -h^{(0)}_{\omega\s} \left({L\chi\over 2}\right)^4 \left(\nu^2-{9\over 4}\right)\left\{ 1 + \left(\nu^2-{1\over 4}\right)\left[\log\left({L\chi\over 2}\right)\right.\right.
\nonumber
\ee
\be
\left. \left. - {3\over 4}
+ {1\over 2} \mathcal{H}\left(-{1\over 2}-\n \right) + {1\over 2}\mathcal{H} \left(-{1\over 2}-\n \right)\right]\right\}
\label{as7b}
\ee
\be
\hat{h}_{\omega\s} =  -{h^{(0)}_{\omega\s} \over 2} \left({L\chi\over 2}\right)^4 \left(\nu^2-{9\over 4}\right) \left(\nu^2-{1\over 4}\right).
\label{as7c}
\ee
\end{subequations}
The relation between the expansion of $h_{\omega\s}(\rho,x^\a)$ and $\delta_h g_{\omega\s}(\rho,x^\a)$ is given by (\ref{dS13aaa}). If we compare with the de Sitter slicing case (\ref{dS13aa}), the analytic continuation $H^2\rightarrow -\chi^2$ does not hold for $h^{(4)}_{\omega\s}$ given in AdS by (\ref{as7b}) and in dS by (\ref{i611}). The only difference between the two resides in the combination of harmonic functions $\mathcal{H}$. For AdS (\ref{as7b}), the combination is symmetric in $\n\leftrightarrow -\n$, which is not the case in de Sitter because the singularity at the horizon forced us to pick a sign for $\text{Re}(\n)$ and break the $\mathbb{Z}_2$ symmetry in $\n$ for the bulk solution (\ref{dS8}).

Inserting the relations (\ref{as7}) into the linearized Einstein equation for the tensor perturbation (\ref{dS13b}), we obtain the  spectral equation $h^{(0)}$:
\be
{N^2 \chi^4\over 64\pi^2}\left(\n^2-{9\over 4}\right)Q_\text{(-)}(\n)h^{(0)}_{\omega\s} = 0,
\label{as10}
\ee
where
\be
Q_\text{(-)}(\n) = 1 + {2}\left({\pi\over GN^2 \chi^2} + \tilde{\a} \right)  - {1\over 2} (\n^2- 1/4)\left[ \tilde{\b}_\text{eff}\right.  \nonumber
\ee
\be
 \left. + \log\left(GN^2\chi^2\right) - {1\over 2} +  \mathcal{H}\left(-{1\over 2}-\n \right) + \mathcal{H} \left(-{1\over 2} + \n \right) \right].
 \label{as111}
\ee
We have used $\tilde{\a}$ defined in (\ref{dS15a}) and $\tilde{\b}_\text{eff}$ defined in (\ref{n28d}).
The inverse propagator for tensor perturbations
is then  given by\footnote{The overall coefficient is determined in Appendix \ref{quadratic action} }:
\be
\mathcal{F}_\text{(-)} = {N^2\chi^4\over 64\pi^2}\left(\n^2-{9\over 4}\right)Q_\text{(-)}(\n).
\label{as12}
\ee

\subsubsection{Symmetric boundary conditions}
\label{AdS z2 symmetric}

As an alternative way to couple AdS boundary gravity to the holographic sector, here we impose that the bulk tensor perturbation  $h_{\a\b}$ has equal sources $h^{(0)}$ on both boundaries $u\rightarrow \pm \infty$.
This is implemented by the boundary condition:
\be
F(u,\n) \underset{u \rightarrow \pm \infty}{\longrightarrow}
\left\{
\begin{array}{ll}
2 \lambda_2 &= 1,\\ & \\
{4\over \pi}\lambda_1 \cos(\pi\n) - 2\lambda_2 \sin(\pi\n) &= 1.
\end{array}
\right.
\label{sym1}
\ee
As for the previous boundary conditions, the case where $\n = n+1/2$ where $n$ is an integer leaves $\lambda_1$ unconstrained. However, we need to distinguish between the two following cases :
\begin{itemize}
\item If $n$ is odd, $\l_2 = 1/2$ solves (\ref{sym1}) and $\l_1$ is unconstrained. This constant can therefore be set to an arbitrary value while (\ref{sym1}) still holds.
\item If $n$ is even, there is no solution for (\ref{sym1}). Such modes are then forbidden in the symmetric case.
\end{itemize}
If $\n$ is not a half-integer then the integration constants are given by
\be
\lambda_2 = {1\over 2},
\label{sym2}
\ee
\be
\lambda_1 = {\pi\over 4}\left(\tan(\pi\n) + {1\over \cos(\pi\n)}\right).
\label{sym3}
\ee
One can observe from (\ref{sym3}) that the limit $\n\rightarrow n +1/2$ can possibly make $\l_1(\n)$ diverge. As already discussed in the asymmetric case \ref{AdS one sided}, these modes are the discrete spectrum of normalizable modes in $AdS_5$. We again discuss the two cases :
\begin{itemize}
\item If $\n \rightarrow n + 1/2$ with $n$ even, then $\l_1(\n)$ diverges. This limit corresponds to a pole of the stress-tensor correlator. Therefore, half-integer $\n$ with even $n$, which are forbidden as discussed below (\ref{sym1}), cannot be a solution to the spectral equation.
\item If $ \n \rightarrow n + 1/2$ with $n$ odd, then $\l_1(\n)\rightarrow 0$. As a reminder, $\l_1(\n = n+1/2)$ can be set to an arbitrary value when $n$ is odd. We can therefore extend $\l_1$ by continuity to half integers with odd $n$. Then, the solution (\ref{sym3}) for $\l_1(\n)$ is continuous at $\n = n+1/2$, $n$ odd, and we can include these half integers into our analysis while working with (\ref{sym3}).
\end{itemize}
The solution for $F(u,\n)$ given by (\ref{sym2}) and (\ref{sym3}) is then  symmetric under $u\leftrightarrow -u$.  Its behaviour near both sides of the boundary $u\rightarrow \pm\infty$ is obtained in appendix \ref{app:legendre}, and the result is given by equation (\ref{sym9}). We observe that the terms of the Fefferman-Graham expansion are all identical to (\ref{as7}) except  $h^{(4)}_{\omega\s}$
which now reads:
\bea\label{sym4}
h^{(4)}_{\omega\s} && = -h^{(0)}_{\omega\s}\left(\n^2-{9\over 4}\right)\left[1 + {1\over 2}\left(\nu^2-{1\over 4}\right)\left(2\log\left({L\chi\over 2}\right) + \right. \right. \nonumber \\  && \left. \left.- {3\over 2 } + \mathcal{H}(\n-1/2) + \mathcal{H}(-\n-1/2) - {\pi\over \cos\pi\n}\right) \right].
\eea

Inserting the bulk data (\ref{sym4}) into the equation of motion (\ref{dS13b})  we obtain our final result for the spectral equation in AdS with symmetric boundary conditions:
\be
{N^2 \chi^4\over 64\pi^2}\left(\n^2-{9\over 4}\right)Q_\text{sym}(\n)h^{(0)}_{\omega\s} = 0,
\label{sym5}
\ee
where
\be
Q_\text{sym}(\n) =  1 + {2}\left({\pi\over G N^2\chi^2} + \tilde{\a} \right)  - {1\over 2} (\n^2- 1/4)\left[ \tilde{\b}_\text{eff} + \right.  \nonumber
\ee
\be
 \left. + \log\left(GN^2\chi^2\right) - {1\over 2}
 + \mathcal{H}(\n-1/2) + \mathcal{H}(-\n-1/2) - {\pi\over \cos\pi\n}\right].
 \label{sym6}
\ee
This expression  is similar to the one obtained  in the asymmetric case (\ref{as111}), except for the last term ${\pi/\cos(\pi\n)}$. This term becomes negligible if $\n$ is far from the real axis, as it decreases exponentially with the imaginary part of $\n$.

The inverse propagator for an AdS space-time with symmetric sources is then:
\be
\mathcal{F}_\text{sym} = {N^2\chi^4\over 64\pi^2}\left(\n^2-{9\over 4}\right)Q_\text{sym}(\n).
\label{sym7}
\ee

\subsection{Identifying ghosts from poles in the propagator}
\label{subsec:residues}
Ghost instabilities are determined from the residue of the poles of the propagator. Whether a mode is a ghost is determined by the sign of the residue of the pole in $\nu^2$: if it has the same sign as for the massless graviton in Einstein GR theory (on the same background), then the mode is healthy, otherwise, it is a ghost.

It is convenient to identify the residue from the  derivative of $\mathcal{F}(\n)$ with respect to the real part of $\n$: indeed using the holomorphic property of $\mathcal{F}$ on the  complex half-plane with positive real part, we have:
\be
\mathcal{F}'(a+ i b) =\left. {\partial \mathcal{F}\over\partial a}\right|_{a+ib}.
\label{dSpole1}
\ee
By symmetry of $\mathcal{F}(\n)$ under $\n \leftrightarrow -\n$, and  by subtracting the Taylor expansion of $\mathcal{F}$ close to a pole $\n_0$ with the expansion close to $-\n_0$, one finds:
\be
{1\over \mathcal{F}(\n)} \underset{\n \rightarrow \n_0}{=} {\n_0\over \mathcal{F}'(\n_0)}{1\over \n^2 - \n_0^2} + \mathcal{O}(1).
\label{dSpole2}
\ee
Therefore, the residue of the pole in $\n^2$ can be obtained as:
\be
\text{Res}[\mathcal{F}^{-1}](\n_0^2) \equiv {\n_0\over \mathcal{F}'(\n_0)}.
\label{dSpole3}
\ee
%

In pure gravity, the sign of the residue of the massless spin-2 pole in de Sitter is negative in our conventions. Therefore, ghosts are defined to be poles with a residue which is not real and negative. It can be real and positive, or even complex. If $\mathcal{F}'(\n_0) = 0$, then $\n_0$ is a higher order pole. In AdS, however, the massless spin-2 pole of Einstein-Hilbert gravity has a positive residue.

In flat space, $\n$ has to be replaced by $k$ in equation (\ref{dSpole3}). The residue of a pole $k_0$ in the $k^2$ plane is related to $\mathcal{F}_\text{flat}'(k_0)$ as
\be
\text{Res}[\mathcal{F}^{-1}_\text{flat}](k_0^2) \equiv {k_0\over \mathcal{F}'(k_0)}.
\label{flat_residue}
\ee
In Einstein-Hilbert gravity, our conventions lead to a negative residue of the massless spin-2 pole. Positive and complex residues (\ref{flat_residue}) will then be associated with ghost-like poles.

\section{Tensor instabilities in pure gravity}
\label{decoupling}
Before we discuss the instabilities arising from tensor modes in the gravity coupled to the holographic CFT, we pause here to give a brief overview of the tensor instabilities in pure gravity with higher curvature terms, described by the action (\ref{s01}).

The spectral functions in  pure gravity can be obtained simply by taking $N= 0$ in the expressions obtained in the previous section\footnote{Even if the $N\rightarrow 0$ limit cannot be treated in holography, taking $N=0$ in our setup is a quick way  to decouple the bulk gravity theory from the boundary, and retrieve the results one would have obtained in a 4d modified gravity theory with Einstein-Hilbert plus quadratic curvature terms $\a$ and $\b$ given by the action (\ref{s01}).}, namely equations (\ref{n28e}-\ref{n28a}) for flat space, (\ref{dS15}-\ref{dS17}) for de Sitter and (\ref{as111}-\ref{as12}) or (\ref{sym6}-\ref{sym7})  for Anti-de Sitter.

\begin{itemize}
\item[]{\bf Minkowski}\\
 In the flat case, setting the $N=0$ in  $\mathcal{F}_\text{flat}$ (\ref{n28a}) leads to:
\be
\mathcal{F}_\text{flat} \underset{N= 0}{=} -{k^2\over 64\pi}\left\{{2\over G} + {\b\over 2}k^2\right\}.
\label{N0}
\ee
The propagator is then a sum of two simple poles given by
\be
\mathcal{F}_\text{flat}^{-1} = -32\pi G\left\{{1\over k^2} - {1\over k^2 + {4\over \b G} } \right\}.
\label{N0a}
\ee
Therefore, the poles of the 2-point functions are the massless solutions $k^2 = 0$ and an additional massive solution,
\be
k^2 = - {4\over \b G}.
\label{N01}
\ee
If $\b$ is negative, the 4-momentum corresponding to this solution is space-like, which implies a tachyonic instability. Thus,  when the CFT is removed, flat space  is then tachyon-unstable for strictly negative $\b$ and tachyon-stable for positive $\b$. As $\beta \to 0$ the massive mode decouples  and one recovers Einstein gravity.
These results agree with \cite{Q7a} concerning the stability of flat space with a quadratic curvature action. Such perturbations around Minkowski space were first derived in \cite{Stelle1978}.

Note that, in our conventions, the  residue of the massless mode is negative. As we have seen above,  the massive pole could be tachyonic or not depending on the sign of $\b$. However, it always corresponds to a ghost because its residue is positive for any $\beta$. This pole is the usual ghost of quadratic gravity theories \cite{Stelle1978, Q1}.

Note that, when the mass of the ghost is above the cut-off (which for pure gravity is the 4d Planck scale  $G^{-1/2}$) our results are not trustworthy in the context of a low-energy effective field theory of gravity. This is the case for $|\beta|\lesssim 1$.

In conclusion, in pure gravity,  flat space-time is tachyon-stable when  $\b >0 $ and ghost-unstable for any $\b$, with the caveat that  for $|\beta|$ small or of order unity the mass of the ghost is above the cutoff for the analysis to be trusted.

\item[]{\bf (anti-)de Sitter}\\
First, we remark that (as  shown in Appendix \ref{dS criterion}), the criterion for tachyonic stability in de Sitter  in the tensor sector can be reduced to the one in the scalar sector (it applies to the spatially transverse traceless tensors). Therefore, one can repeat the analysis we performed for the scalar mode in section \ref{sec:scalar tachyons}, where we found that the condition for tachyonic stability is the same in global, cosmological and static dS coordinates, namely $|Re(\nu)|< 3/2$. 

For de Sitter, the pure gravity spectral function is obtained by setting $N= 0$ in (\ref{dS17}), which gives:
\be
\mathcal{F}_{dS}(\n) \underset{N= 0}{=}  - {H^4\over 64\pi}\left(\n^2-{9\over 4}\right)\left\{{2\over GH^2}- 2\a + {\b\over 2}\left(\n^2 - {1\over 4}\right)\right\}.
\label{N02}
\ee
One simply needs to replace $H^2\rightarrow -\chi^2$ to obtain the result for AdS, so we treat positive and negative curvature together.

The propagator can then be written as a sum of two poles,
\be
\mathcal{F}_{dS}^{-1}(\n) \underset{N= 0}{=} - {64\pi \over H^4}\left[{2\over GH^2} - 2\a + \b \right]^{-1}\left\{{1\over \n^2 - {9\over 4}} - {1\over \n^2 - {1\over 4} + {4\over \b}\left({1\over GH^2} - \a\right) }\right\}.
\label{N02a}
\ee
The first pole is the massless graviton, which is the only propagating mode that remains for $\beta=0$. If $\b \neq 0$,  the second pole is located at
\be
\n^2 = {1\over 4} + {4\over \b}\left(\a - {1\over GH^2} \right).
\label{N03}
\ee
This equation shows that the $\b \rightarrow +\infty$ limit (while keeping $\a$ and $GH^2$ fixed) always makes a solution converge to $\n = \pm {1\over 2}$. In the opposite limit $\beta \to 0$ this massive mode disappears and only the massless graviton remains.

Note that if $\b = 0$, the factor in curly braces in (\ref{N02}) vanishes for:
\be
\a = {1\over GH^2}.
\label{N04}
\ee
Therefore, in the $\b=0$ case, and for any value of $\a$, there exists a special value of the curvature  scale $H$ such that the   tensor mode has a vanishing quadratic kinetic term  and therefore it is strongly coupled\footnote{In such cases, there is a possibility of a Vainshtein-like mechanism operating. We do not know whether this has been investigated in this context.}.
  In other words, the theory is strongly coupled for  $\b = 0$ and $\alpha$ and $H$ related by (\ref{N04}).

From (\ref{N02}) we observe that  zeros of ${\cal F}$ correspond to real $\nu^2$.  The de Sitter tachyon-stability condition (\ref{dS6}) becomes  $\n^2 \leq 9/4$, while the anti-de Sitter condition (\ref{AdS4}) becomes $\nu^2>0$. In pure gravity, these conditions translate to
\be
{2\over \b}\left(\a - {1\over GH^2}\right) < 1 \quad \Rightarrow \quad \text{dS tachyon stable},
\label{N05}
\ee
and anti-de Sitter is tachyon-stable if
\be
{4\over \b}\left(\a + {1\over G\chi^2}\right) > -{1\over 4}  \quad \Rightarrow \quad \text{AdS tachyon stable}.
\label{N06}
\ee
\end{itemize}

We now turn to ghost instabilities, starting   with de Sitter.  When the prefactor in the square brackets of (\ref{N02a}) is positive, then the massless pole located at $\n^2 = 9/4$ has the same sign as the massless pole in pure gravity ($\alpha=\beta=0$) and therefore it is not a ghost,  whereas the massive  pole represents a ghost.
On the contrary, if the prefactor in square brackets is negative, then the  massless pole is  ghost-like  and the massive pole becomes ghost-free. All in all, the  higher derivative pure gravity theory always has a ghost, be it the massless graviton or the massive mode.

For AdS  the conclusions are opposite: in our conventions, the massless graviton in pure Einstein gravity has positive residue in AdS, as can be seen by setting $\alpha=\beta=0$ and replacing  $H^2\rightarrow -\chi^2$ in (\ref{N02a}).

\subsection{When are tensor ghosts  light?}

The discussion above holds if we take the spectral functions at face value.  However, these conclusions can be trusted only when the poles lie within the validity of effective field theory, i.e. when the masses of the unstable modes are below the cut-off, which in pure gravity can be taken to be the Planck scale $M_p = (8\pi G)^{-1/2}$. For the same reasons,  all the  expressions above make sense in effective field theory if the curvature is sub-Planckian, i.e. if $GH^2 \ll 1$.

We shall verify when the unstable mode mass  is sub-Planckian  in the various cases.
\begin{itemize}
\item {\bf Flat space.} By equation (\ref{N01}),  the modulus of the  massive pole in Planck units is roughly $G |m^2| = 4/|\beta|$. Therefore, we conclude that:
 \be
|\beta| \gg  1 \quad \Rightarrow  \quad \text{flat space gravity has a light tensor ghost.}
\ee
If in addition $\beta <0$, this is also a light tachyon.

\item {\bf de Sitter.}  In this case, we have to distinguish two situations, depending on the sign of the prefactor in (\ref{N02a}):
\begin{enumerate}
\item If $\beta - 2\alpha < - 2/(GH^2)$, then the massless mode is a ghost, and it is by definition below the cut-off.
Since $GH^2 \ll 1 $, this requires either $\beta$ very large and negative, or $\alpha$ very  large and positive.
\item $\beta - 2\alpha > - 2/(GH^2)$  then the massive mode is a ghost. This is the most ``natural'' situation, as it does not require extreme values of $\alpha$ and $\beta$.  By equation (\ref{N03}) the modulus of its mass squared in Planck units is\footnote{Recall that the mass is $H^2(\nu^2 - 9/4)$.}:
\be
G |m^2_{ghost}| = \left|-2GH^2 + {4\over \beta}\left(\alpha GH^2 - 1\right)\right|.
\ee
The ghost is sub-Planckian when the second term on the left-hand side is smaller than unity (since the first term $GH^2$ is always small in effective theory).
For $\alpha$ not too large, this is the case if $|\beta| \gg 1$:
\be
\alpha \sim O(1), \; |\beta| \gg 1   \Rightarrow  \quad \text{de Sitter  gravity has a light tensor ghost.}
\ee
\end{enumerate}
All in all, we observe that for large $|\beta|$, there is {\it always} a light ghost in de Sitter (it may be massive, massless, or tachyonic).

\item  {\bf anti-de Sitter.}
For AdS, the situation is the same as for de Sitter (with $H^2 \to -\chi^2$), except that the role of points 1 and 2 above are exchanged:
\begin{enumerate}
\item If $\beta - 2\alpha >  2/(G\chi^2)$ then the massless pole is a ghost. Note that this is the generic situation for  $O(1)$ values of $\alpha$ and $\beta$, since the right-hand side of that inequality is a small number.
\item If instead  $\beta - 2\alpha <  2/(G\chi^2)$, then the ghost is the massive mode. Its mass in Planck units is, in modulus:
\be
G |m^2_{ghost}| = \left|2G\chi^2 - {4\over \beta}\left(\alpha G\chi^2 + 1\right)\right|.
\ee
For the ghost to be sub-Planckian this again requires $\beta \gg 1$, but now this must be accompanied by a fine-tuning $\alpha \simeq \beta/2 \gg 1$ to ensure the ghost is the massive mode.
\end{enumerate}
We conclude that {\it generically}, AdS higher-curvature gravity has a light tensor ghost, unless  $\beta - 2\alpha <  2/(G\chi^2) \ll 1 $.
\end{itemize}

\section{Poles of the Minkowski spin-two propagator and stability}
\label{sec:flat correlator}
In this section, we  analyse the flat-space spectral function found in section \ref{sec:flat prop} and determine for which values of the parameters flat space-time are unstable under tensor perturbations.

All the information about the tachyonic instability is contained in the location of the poles of the propagator, i.e. the zeros of the spectral function (\ref{n28a}). As we have explained in section \ref{sec:flat prop}, in our conventions,  a zero of (\ref{n28a})  at a value $k$ with a non-zero real part corresponds to a tachyonic mode\footnote{Recall that, in terms of momenta of Fourier modes, $k$ represents  the square root of $k^2 = -(k^0)^2 + \bm{k}^2$ with positive real part.}.

Ghost instabilities  are determined by computing the residue of these poles. In particular, a pole is not a ghost if its residue is negative, as is the case for the massless pole of the propagator in pure gravity (\ref{N0a}).

The spectral equation for non-trivial modes on Minkowski is given  in equation (\ref{n31}), which we rewrite  here for convenience:
\be
1 = {GN^2 k^2 \over 4\pi}\left[ {1\over 2} - 2\gamma_E - \log\left(GN^2k^2\right) - \tilde{\b}_\text{eff} \right].
\label{p31}
\ee
As in our conventions we are taking Re$(k)>0$, we can  use the identity
\be
\log(k^2) = 2 \log k.
\label{n42}
\ee
Equation (\ref{n31}) can  then be written in  the  simpler form:
\be
X \log X = -a,
\label{n43}
\ee
where $X$ and $a$ are defined as:
\be
X \equiv {GN^2k^2} \exp\left\{{-{1\over 2} + 2\gamma_E + \tilde{\b}_\text{eff} }\right\}, \quad a \equiv {4\pi} e^{-{1\over 2} + 2\gamma_E + \tilde{\b}_\text{eff}}.
\label{n44}
\ee

The new variable $X$ contains all information about tachyonic instabilities, the same way that $k$ did. The stability condition (\ref{flat7a}) translates into
\be
X <0.
\label{n45a}
\ee
Our problem is now simply to solve (\ref{n43}) for $X$, which is a $W(-a)$ Lambert's function, containing two branches. As a result, one or at most two solutions exist for a complex $X$. To determine this, we write
\be
X = x e^{i\theta} \label{n46}
\ee
and inject this expression into (\ref{n43}).
Then (\ref{n43}) is equivalent to the two real equations
\be
x\theta=a\sin\theta\sp \theta\cot\theta=-\log x
\label{n46b}\ee

The stability condition (\ref{n45a}) translates into  $\theta = \pm \pi$.
If a solution for any other value for $\theta$ exists, it corresponds to a tachyonic mode.
\begin{itemize}
\item First, we  study the existence of purely real tachyonic solutions $k^2>0$, for which $\theta = 0$.
This is the case which was  considered in \cite{anomalyinflation1}. If $0\leq a<e^{-1}$,  equation (\ref{n43}) has two solutions (one with a larger mass than the other), which merge at $a = e^{-1}$. If on the other hand $a>e^{-1}$,  there is no solution  with $\theta = 0$.  Using this property, we can write a  condition on $\tilde{\b}_\text{eff}$ such that flat space has two  tachyonic modes with $k^2>0$:
\be
\tilde{\b}_\text{eff} \leq \tilde{\b}_\text{eff}^\text{merge} \equiv -\log\left(4 \pi e^{{1\over 2} + 2\g_E}\right) \quad \Rightarrow \quad \text{Two tachyonic modes.}
\label{n47}
\ee
 When (\ref{n47}) is an equality, we observe a double pole located at $X=e^{-1}$. In $k^2$ space, this double pole is located at
\be
GN^2k^2 =  4\pi .
\label{n48}
\ee
\item We  now consider the general case $\theta \neq 0$. The imaginary and real parts of equation (\ref{n43}) give (\ref{n46b}), that can be rewritten as
\end{itemize}
\begin{subequations}
\label{n50}
\be
\log x = -\theta \cot \theta,
\label{n50a}
\ee
\be
e^{-\theta \cot \theta} = a {\sin \theta \over \theta},
\label{n50b}
\ee
\end{subequations}
The number of solutions of (\ref{n50b}) depends on the values of $a$.
Since $\theta \cot \theta < 1$ and ${\sin\theta\over \theta}<1$ for $\theta\neq 0$, we can  conclude that  solutions of (\ref{n50b}) with $\theta \neq 0$ exist only in the range:
\be
a > e^{-1}.
\label{n51}
\ee
These solutions are tachyonic as long as they do not move to the negative real axis $\theta = \pm \pi$.  This occurs as $\tilde{\b}_\text{eff}\rightarrow +\infty$: in this limit, the instability approaches the imaginary axis, corresponding to  $k^2<0$. This was already noted in \cite{anomalyinflation1}, and can be shown  as follows:
The two extremal values  $\theta = \pm \pi$ can be reached only if we take $a\rightarrow +\infty$. Indeed, using the fact that the left-hand side of (\ref{n50b}) is bounded:
\be
\left| e^{-\theta \cot \theta} \right| \leq 1,
\label{n61}
\ee
then $a\sin{\t}/\t$ must also be bounded as $a\rightarrow +\infty$. Therefore,
\be
\theta \underset{a \rightarrow \infty}{\rightarrow} \pm \pi.
\label{n62}
\ee

The limit $a\rightarrow +\infty$ is reached by taking  $\tilde{\b}_\text{eff}\rightarrow +\infty$. From the definitions (\ref{b24}-\ref{n28d}), one way of obtaining this limit is by setting  $N=0$ which corresponds to decoupling the CFT as we have discussed in subsection \ref{decoupling}.
The pure gravity case for flat space was also studied in \cite{Q7a}, and the analysis we have presented here agrees with the conclusion of that work: in pure gravity, flat space is tachyon-stable for positive $\b$ and tachyon-unstable for negative $\b$.

To summarize, there are 2 real tachyonic poles when $0\leq a < e^{-1}$. They merge at $a=e^{-1}$ and they become complex at $a>e^{-1}$.

The complex poles of the spin-2 propagator can be found numerically, and  are shown in figure \ref{fig:flat}  for some illustrative values of $\tilde{\b}_\text{eff}$.  Each snapshot corresponds to a different $\tilde{\b}_\text{eff}$. Poles correspond to the intersections of the blue lines (zeros of the real part of the inverse propagator) and of the orange lines (zeros of the imaginary part). The massless pole is not shown in this Figure, it is always located at $k=0$ for any value of the parameters $\tilde{\a}$ and $\tilde{\b}_\text{eff}$.

In figure \ref{fig:flat}, we start at large and negative values of $\tilde{\b}_\text{eff}$ in the upper-left panel\footnote{Solutions of (\ref{n43}) are always a pair of complex conjugates due to the symmetry under $\theta\rightarrow -\theta$, which is explicit in equations (\ref{n50a},\ref{n50b}). This, and the fact that we chose the square root with $\text{Re}(k)>0$, are the reasons why we only display the top-right quarter of the complex plane in Figure \ref{fig:flat}.}. Of the  two tachyonic solutions, only the lighter one  close to the origin  is visible in the upper-left panel, while  the heavier one is far away along the real axis. As $\tilde{\b}_\text{eff}$ is increased, the heavier solution comes closer to the lighter solution as shown in snapshot (b). Then, they merge in snapshot (c) where $a=e^{-1}$, i.e where $\tilde{\b}_\text{eff}$ is chosen such that (\ref{n47}) is an equality. The complex instability continues to travel along the fixed orange curve and then becomes closer to the imaginary axis as $\tilde{\b}_\text{eff}$ is increased.

\begin{figure}[ht]
\centering
\includegraphics[width=0.9\textwidth]{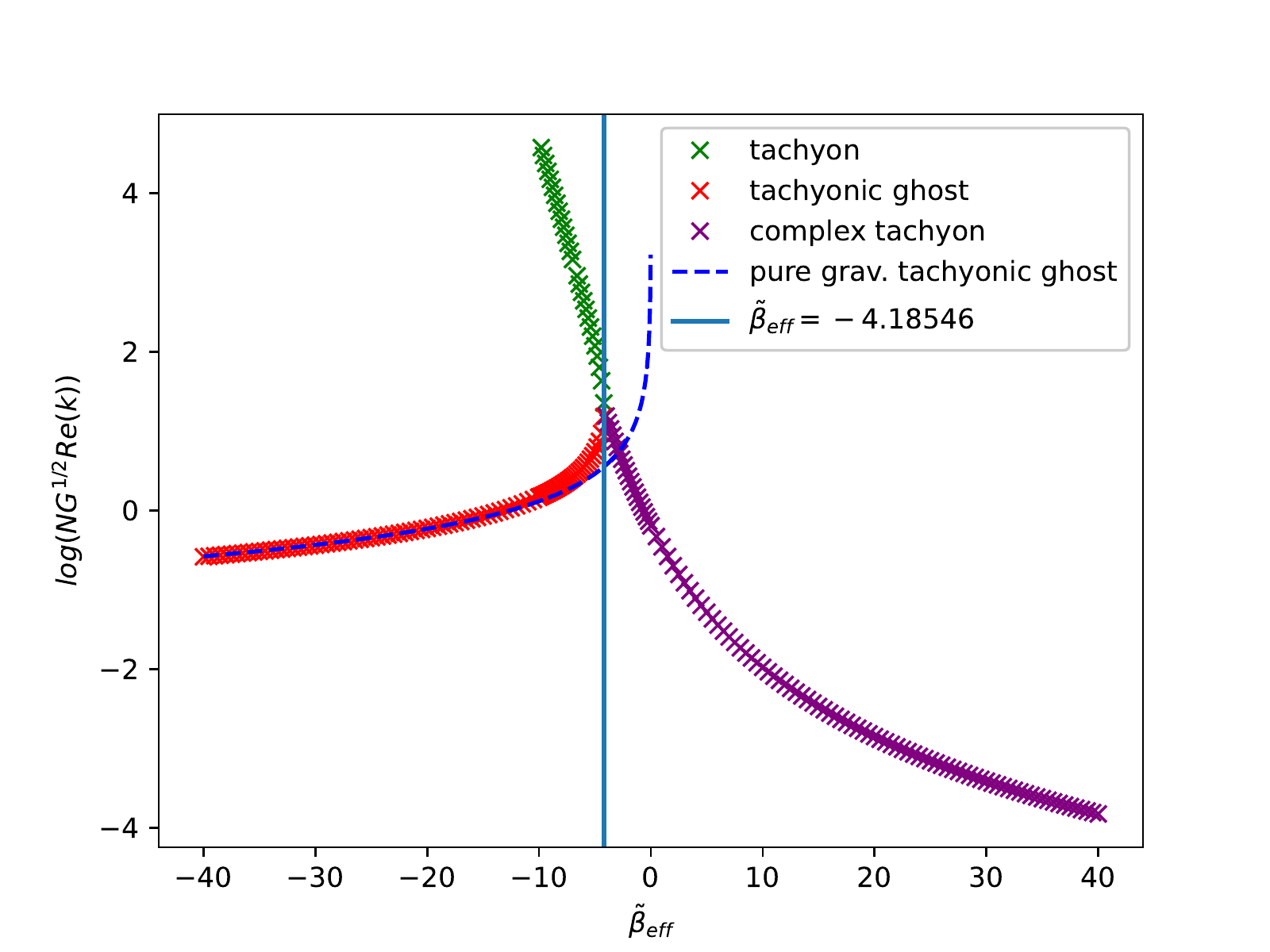}
\caption{\it \it In this plot, we show the real part of the two tachyonic spin-2 poles in flat space (which also gives the inverse time scale of the Minkowski space spin-2 tachyonic instability, see app. \ref{app:tacscale}) in units of the species cutoff, as a function of $\tilde{\b}_\text{eff}$.
 Red and green markers correspond respectively to the ghost-like (lighter) tachyon and a non-ghostly (heavier) tachyon.
Purple markers correspond to two complex conjugate ghost-like tachyonic poles.
The blue vertical line is the value of $\tilde{\b}_\text{eff}$ which saturates (\protect\ref{n47}), where these two poles merge and move to the complex plane as two complex conjugate poles as $\tilde{\b}_\text{eff}$ is further increased. Large  (positive and negative) values of $\tbe$ correspond to a long-lived tachyon.
 For comparison,
the black curve corresponds to the case of pure gravity with $\beta =\tilde{\beta}_\text{eff}/\pi$, where the tachyonic pole is given by equation (\protect\ref{N01}).
For this curve, the vertical axis is $\log(G^{1\over 2}Re(k))$ while the horizontal axis is $\pi\beta$.
}
\label{flatRek}
\end{figure}

\begin{figure}[ht]
\centering
\includegraphics[width=0.9\textwidth]{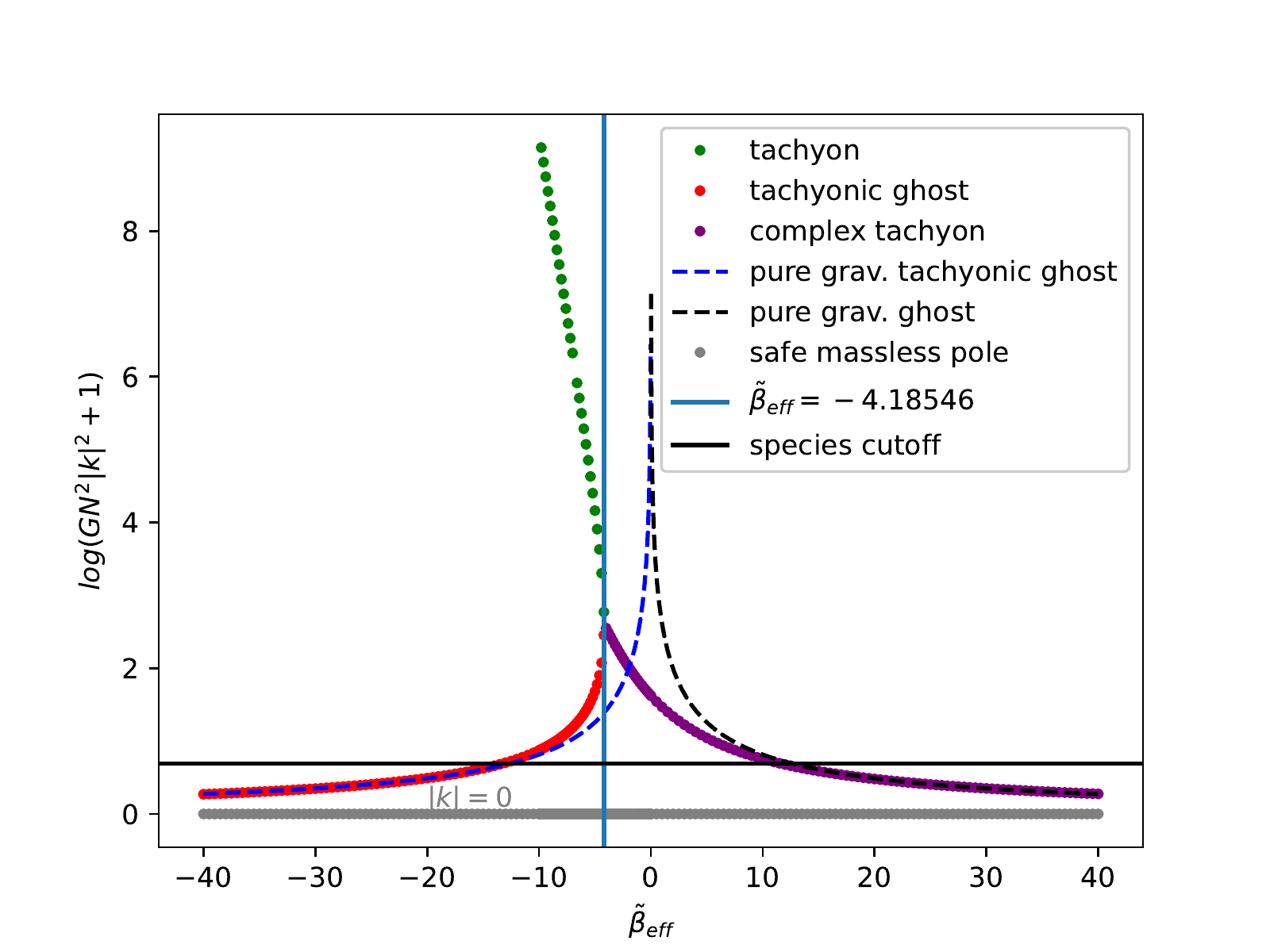}
\caption{\it \it
In this plot, we show the modulus of the mass of the spin-2 tachyonic poles in flat space, defined in (\protect\ref{flat5}), in units of the species scale (\protect\ref{sp-a}), as a function of $\tilde{\b}_\text{eff}$. Red and green markers correspond respectively to the light ghost-like tachyon and the heavy non-ghostly tachyon. The tachyon exists for large and negative $\tbe$ but its mass is too large to appear in the window.
Purple markers correspond to two complex conjugate ghost-like tachyonic poles.
The black curves correspond to the massive ghost in pure gravity with $\beta = \tbe /\pi$: the solid line is the ghost-like tachyon ($\beta<0$), and the dashed line is the non-tachyonic ghost ($\beta >0$)
The blue vertical line is the value of $\tilde{\b}_\text{eff}$ which saturates (\protect\ref{n47}), where these two poles merge and move to the complex plane as two complex conjugate poles as $\tilde{\b}_\text{eff}$ is further increased.
The black horizontal line marks the species cut-off (or the Planck scale cut-off for pure gravity).
For this curve, the vertical axis is $\log(G^{1\over 2}Re(k))$ while the horizontal axis is $\pi\beta$.
There are unstable modes  lighter than  the species cutoff only for  large values of $|\tbe|$.}
\label{flatModk}
\end{figure}

We now turn to the analysis of ghosts. In the $a<e^{-1}$ regime, there are two tachyonic modes, the heavier one is a ghost and the lighter is not. The mass of the ghost is always comparable to $GN^2$ in this regime.

When $\tilde{\b}_\text{eff}$ is increased, after the merging at $a=e^{-1}$ has occurred in snapshot (c) of Figure \ref{fig:flat}, the tachyonic complex ghost pole  moves on the complex plane and approaches towards the imaginary axis as $\tilde{\b}_\text{eff}$ becomes large and positive. The ghost becomes lighter and lighter in units of $GN^2$.

In the large-$\tilde{\b}_\text{eff}$ regime, the ghost sticks to the imaginary axis where its residue becomes real and  positive. More precisely, the imaginary part of the residue becomes smaller and smaller compared to the real part.

 In the limit $\tilde{\b}_\text{eff}\rightarrow +\infty$, the mass squared of the spin-2 mode $(m_2)^2$ defined in (\ref{flat5}) becomes negative. This limit can be taken in (\ref{n31}) to obtain the value
\be
{GN^2(m_2)^2} \underset{|\tilde{\b}_\text{eff}|\rightarrow +\infty }{\sim} {4\pi\over \tilde{\b}_\text{eff}}.
\label{n63}
\ee
This equation agrees with what is seen in Figure \ref{fig:flat}.

In Figures \ref{flatRek} and \ref{flatModk} we show the behaviour, respectively, of the real part and the complex modulus of the spin-2  poles.
As it is shown in appendix \ref{app:tacscale}, the real part of $k$ corresponds to the typical inverse time scale of the tachyonic instability. From these figures, one can follow the trajectory of the poles as a function of $\tbe$.  As one  can observe, for $O(1)$ values of $\tbe$, the poles are above the (species) cut-off, therefore they are outside of the regime of our EFT analysis. It is only for  $\tbe$ very large and positive or very large and negative that at least one pole becomes lighter than the cut-off scale. However, this is the same regime in which even pure gravity (black curves in figures \ref{flatRek} and \ref{flatModk}) has a light instability (although in the case of pure gravity, large positive $\beta$ corresponds to a ghost which is not also a tachyon, unlike in the presence of the CFT). In any regime where pure gravity does not have light unstable modes, adding the CFT does not make the effective field theory unstable.

The behaviour of the complex solutions for $X$ right after the merging can be described analytically by performing an expansion for small $\theta$ in (\ref{n50}):
\begin{subequations}
\label{n51a}
\be
\log x = -1 + {\theta^2\over 3} + \mathcal{O}(\theta^3),
\label{n51aa}
\ee
\be
x \left[\log x \left(1 - {\theta^2\over 2}\right) - \theta^2 + \mathcal{O}(\theta^3)\right] = -a.
\label{n51ab}
\ee
\end{subequations}
We can therefore eliminate $x$ to find a solution for $\t$ given by
\be
\theta^2 \approx 2(ae - 1).
\label{n53}
\ee
The two complex branches of $X$ then start at $a = e^{-1}$.
The solution for $X=xe^{i\t}$ is then
\be
x = e^{-1 + {\theta^2\over 3} + \mathcal{O}(\theta^3)},
\label{n54}
\ee
and
\be
\theta \approx \pm \sqrt{2(ae-1)}.
\label{n55}
\ee

\section{Poles of the dS spin-two propagator and stability}
\label{dS poles}

We now consider the positive curvature case and study the stability under tensor perturbations of 4d gravity plus a holographic CFT on de Sitter space-time.  Both tachyonic and ghost instabilities will be determined numerically, but we also provide analytical insight into these results.

Tachyonic instability is identified by studying the location of zeros of the de Sitter tensor inverse propagator (\ref{dS17}) in the complex $\nu$ domain: such instability is characterized by the condition $\text{Re}(\n)>3/2$ (\ref{dS6}).

Whether or not the mode is a ghost is determined by the sign of the residue of the pole, as explained in subsection \ref{subsec:residues}

\subsection{Numerial results for two typical sets of parameters}
\label{dS analysis}

Before performing   a full analysis in the parameter space spanned by ($GN^2H^2$, $\tilde{\a}$, $\tilde{\b}_\text{eff}$), in this subsection we present,  as an illustrative example,  the results for  two distinct sets of parameters which give different results but are typical cases of the more general behaviour of the system.
For each of these two sets of parameters, we fix  ($GN^2H^2$, $\tilde{\a}$) and solve numerically the equation (\ref{dS14}) for several values of $\tilde{\b}_\text{eff}$.

The first example, shown  in Figure \ref{dS_H0.01_alpha0} is an example of the small curvature regime.
For instance, we observe that in this figure, the theory is tachyon-unstable from the snapshot (a) to snapshot (e) because one (or two) solutions are tachyonic ($\text{Re}(\n)>3/2$). The two tachyons merge in snapshot (c) to form a double pole, where $\mathcal{F}_\text{dS}'(\n)$ vanishes\footnote{Theories with a double pole have been recently discussed in \cite{Tseytlin}.}.

After the merging, the double pole separates into two complex conjugate solutions. We only display positive imaginary parts in this Figure. In snapshot (f) the tachyon with complex $\n$ enters the stability region $|\text{Re}(\n)|<3/2$ because its real part decreases as $\tilde{\b}_\text{eff}$ is increased. Therefore, the theory is tachyon-stable from snapshot (f) to snapshot (i) and will continue to be stable for even larger values of $\tilde{\b}_\text{eff}$.
As $\tbe$ is increased, the pole which became tachyon-stable in (f) goes to the imaginary axis and then forms a double pole at the intersection of the real and imaginary axes.

When a tachyon is present, it is important to determine the time scale of the instability, which is fixed by the value of $\nu$ in the same way for both scalars and tensors (see appendix \ref{dS criterion} for details).   For a tachyonic mode with  ${Re}(\n)>3/2$, the solution of (\ref{psi9a}) which dominates at large $t$ behaves as\footnote{For $\text{Re}(\n)<-3/2$, there would be a sign flip $\n\rightarrow -\n$ in (\ref{ts1}).}:
\be
\tilde{\theta}_{ij} \underset{Ht \rightarrow +\infty}{\propto} e^{-Ht(3/2 - \n)}.
\label{ts1}
\ee
The characteristic rate $\G$ of the exponential divergence in (\ref{ts1}) is therefore given by
\be
\G = H( |\text{Re}(\n)| - 3/2 ).
\label{ts2}
\ee

We now turn to the analysis of ghost instabilities.
The sign of the residue of each pole is obtained by computing numerically $\mathcal{F}'(\n)$ and applying the formula (\ref{dSpole3}). The resulting sign is encoded in the colour of each dot in Figure \ref{dS_H0.01_alpha0}. Green dots correspond to negative residues, indicating a ghost-free pole. Red dots correspond to ghosts with positive residue, and purple dots correspond to complex residues.
One can observe that from Figure \ref{dS_H0.01_alpha0}, a ghost (either positive or complex residue) is always present, for any value of $\tilde{\b}_\text{eff}$. For generic values of $\tilde{\b}_\text{eff}$, the mass of the ghost defined by (\ref{dS5}) is  large compared to the Hubble rate $H$. For large values of $\tilde{\b}_\text{eff}$ (both positive and negative), the ghost pole approaches $\n = 1/2$, matching the $N=0$ case (\ref{N02a}) in the limit $\beta \to \infty$.

\begin{figure}[ht]
\centering
\includegraphics[width=0.9\textwidth]{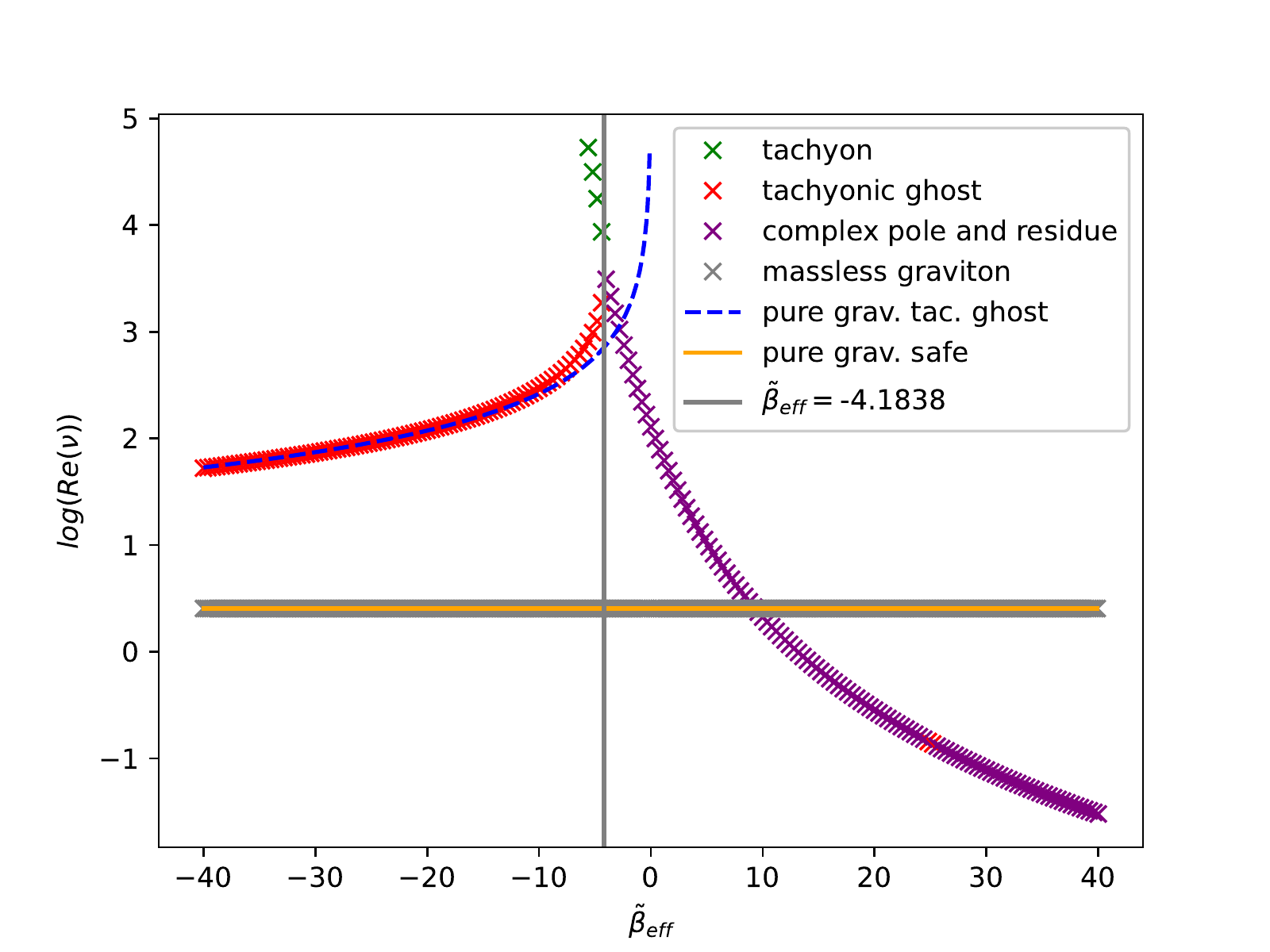}
\caption{\it
In this plot, we show the real part of the spin-2 poles in de Sitter with the same parameters as in Figure \ref{dS_H0.01_alpha0}. The real part also gives the inverse time scale, or \textit{strength}, (\ref{ts2}) of the dS tachyonic instability, in units of the species cutoff, as a function of $\tilde{\b}_\text{eff}$.
 Red and green markers correspond respectively to the ghost-like (lighter) tachyon and a non-ghostly (heavier) tachyon.
Purple markers correspond to two complex conjugate ghost-like poles. For comparison,
the blue dashed line and the yellow line correspond to the case of pure gravity with $\beta =\tbe/\pi$ (\protect\ref{N02}).
For pure gravity curves, the horizontal axis is $\pi\beta$.
The grey vertical line is the value of $\tilde{\b}_\text{eff}$ which saturates (\protect\ref{Ln1}), (corresponding to snapshot (c) of Figure \protect\ref{dS_H0.01_alpha0}), where two poles merge and move to the complex plane as two complex conjugate poles as $\tilde{\b}_\text{eff}$ is further increased.
Each pole above the green line at $\text{Re}(\n) = 3/2$, is tachyonic. The complex pole crosses this green line around ${\tilde{\b}_\text{eff}} \approx 9.3$, it then becomes non-tachyonic for larger values of $\tbe$.}
\label{omegadS_alpha0_H0.01}
\end{figure}

The divergence rate $\G$ is computed numerically as a function of $\tilde{\b}_\text{eff}$, and the results are shown in Figure \ref{omegadS_alpha0_H0.01} for the set of parameters we have used  in Figure \ref{dS_H0.01_alpha0}. Figure \ref{omegadS_alpha0_H0.01} shows in green the tachyonic pole, in red the tachyonic ghost, in purple the complex pole and in grey the massless pole which is neither a ghost nor a tachyon. The two poles of pure gravity, are also shown for comparison. The massive pole (\ref{N03}) is a blue dashed curve, while the massless pole is an orange line.
As $\tilde{\b}_\text{eff}$ is increased, the tachyonic rate $\G$ (\ref{ts2}) decreases, down to the point where the spin-2 mode becomes tachyon-stable around ${\tilde{\b}_\text{eff}}\sim 9.3$.
The pure gravity massive pole for large and negative $\b$ coincides with the ghost pole for large and negative $\tbe$ if we set $\tbe = \pi\b$, i.e $N=1$ in the definition of $\tbe$ (\ref{n28d}). Large and positive $\tbe$ also agree with the pure gravity result, even if this is not visible from this figure. One would need to look at much higher values of $\tbe$ (a few thousand) to see that the complex pole goes to the real axis, as it is shown in snapshots (g,h,i) of Figure \ref{dS_H0.01_alpha0}. In this case, both the massive pole of pure gravity and the ghost asymptote at $\n = 1/2$, as expected by a naive $\tbe \rightarrow +\infty$ limit of (\ref{dS17}).
For very large and negative $\tbe$ the massless pole becomes a ghost. According to (\ref{N02a}), one would need to have $\tbe \leq -{2\pi\over GN^2H^2} \approx - 628.3$. In the presence of the CFT, however, we find numerically that the massless pole is a ghost for $\tbe \lesssim -624.2$. This critical value of $\tbe$ corresponds to the merging of the ghost with the massless pole which were both present in snapshot (a) of Figure \ref{dS_H0.01_alpha0}.

The results of Figure \ref{omegadS_alpha0_H0.01} are compatible with the paper \cite{Chesler} because they have $\a = 0$ and study the $GN^2H^2<<1$ regime where they also find a complex pole, which is also present in flat space (see Figure \ref{flatModk}). In this paper, the authors have found an approximate value of $\tbe$ at which the complex pole crosses the massless line displayed in orange. Larger values of $\tbe$ then correspond to an absence of tachyonic instabilities. However, the pole is still complex up to $\tbe \approx 5026.43$ where it becomes a \textit{real} ghost.

\begin{figure}[ht]
\centering
\includegraphics[width=0.9\textwidth]{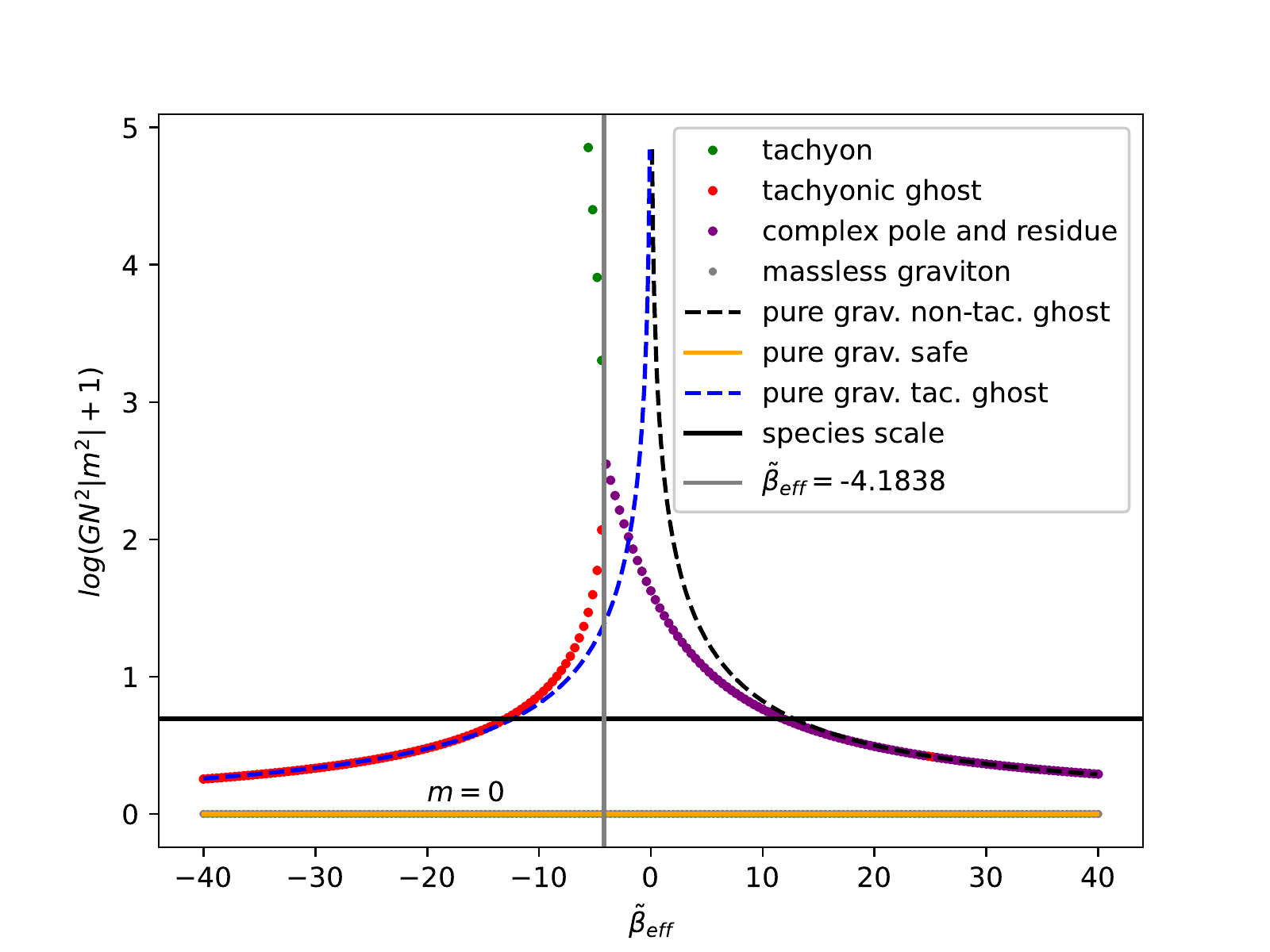}
\caption{\it \it
In this plot, we show, for the same parameters as in Figure \ref{dS_H0.01_alpha0}, the complex modulus of the mass of the spin-2 tachyonic poles in de Sitter, defined in (\protect\ref{dS5}), in units of the species scale (\protect\ref{sp-a}), as a function of $\tilde{\b}_\text{eff}$. Red and green markers correspond respectively to the light ghost-like tachyon and the heavy non-ghostly tachyon.
Purple markers correspond to two complex conjugate ghost-like poles.
The species scale is shown by a horizontal solid black line.
For comparison, different curves show the poles in pure gravity with $\beta = \tbe /\pi$: the dashed blue and black curves are respectively tachyonic and non-tachyonic ghosts. The orange curve is safe.
For pure gravity curves, the vertical axis is $\log(G^{1\over 2}|m|^2 + 1)$ while the horizontal axis is $\pi\beta$.
The grey vertical line is the value of $\tilde{\b}_\text{eff}$ which saturates (\protect\ref{Ln1}), (corresponding to snapshot (c) of Figure \protect\ref{dS_H0.01_alpha0}), where these two poles merge and move to the complex plane as two complex conjugate poles as $\tilde{\b}_\text{eff}$ is further increased.
The black horizontal line marks the species cut-off (or the Planck scale cut-off for pure gravity).
There are unstable modes  lighter than  the species cutoff only for  large values of $|\tbe|$.}
\label{mod_dS_alpha0_H0.01}
\end{figure}

Figure \ref{mod_dS_alpha0_H0.01} shows the modulus of the mass squared of the tensor modes in de Sitter, plotted in units of the species scale (\ref{sp-a}). This figure is a numerical evaluation for the same parameters as the ones chosen in Figures \ref{dS_H0.01_alpha0} and \ref{omegadS_alpha0_H0.01}. As in Figure \ref{omegadS_alpha0_H0.01}, the green, red and blue curves are respectively the tachyonic, the ghost and the massive mode of pure gravity. The modulus of the ghost mass agrees with the pure gravity massive mode for large values of $|\tbe|$. For generic values, both the ghost and the tachyon lie above the species cutoff.
The complex pole, which appears for $\tbe\gtrsim -4.1838$,  goes beyond the species scale for large and positive values of $\tbe$.

Our second example  corresponds to $GN^2H^2$ of order unity (rather than $GN^2H^2\ll 1$ as was the case in figure \ref{dS_H0.01_alpha0}). Specifically, we take  $GN^2H^2 = \pi/4$ and $\tilde{\a} = 10$.
The spin-2 poles of de Sitter in this case are  shown in Figure \ref{dS_Hpiover4_alpha10}. For  large and negative values of $\tilde{\b}_\text{eff}$ we find a different behaviour  than in Figure \ref{dS_H0.01_alpha0}:  in the present case there is only one tachyon, and it is not a ghost. However, the massless pole at $\n=3/2$ is now a ghost. As $\tilde{\b}_\text{eff}$ is increased, the tachyon becomes lighter and lighter from snapshot (a) to (c). It stays on the positive real axis  (another difference from Figure \ref{dS_H0.01_alpha0}).

Snapshots (a-f) correspond to a tachyon-unstable theory because the heaviest solution has a real part larger than 3/2. The tachyon merges with the massless graviton in snapshot (g) and becomes a ghost when $\tilde{\b}_\text{eff}$ is increased. This ghost then moves towards $\n=1/2$ for large and positive values of $\tilde{\b}_\text{eff}$ in snapshot (i). This matches  the decoupling limit $N=0$, as can be seen from (\ref{N02a}).  For large and negative $\tilde{\b}_\text{eff}$, the massless pole $\n^2 = 9/4$ is a ghost whereas the $\n^2=1/4$ is not. For large and positive $\tilde{\b}_\text{eff}$, the respective signs of their residues are switched. This is what is observed by comparing snapshots (a) and (i), where the red and green poles are interchanged.

\begin{figure}[ht]
\centering
\includegraphics[width=0.9\textwidth]{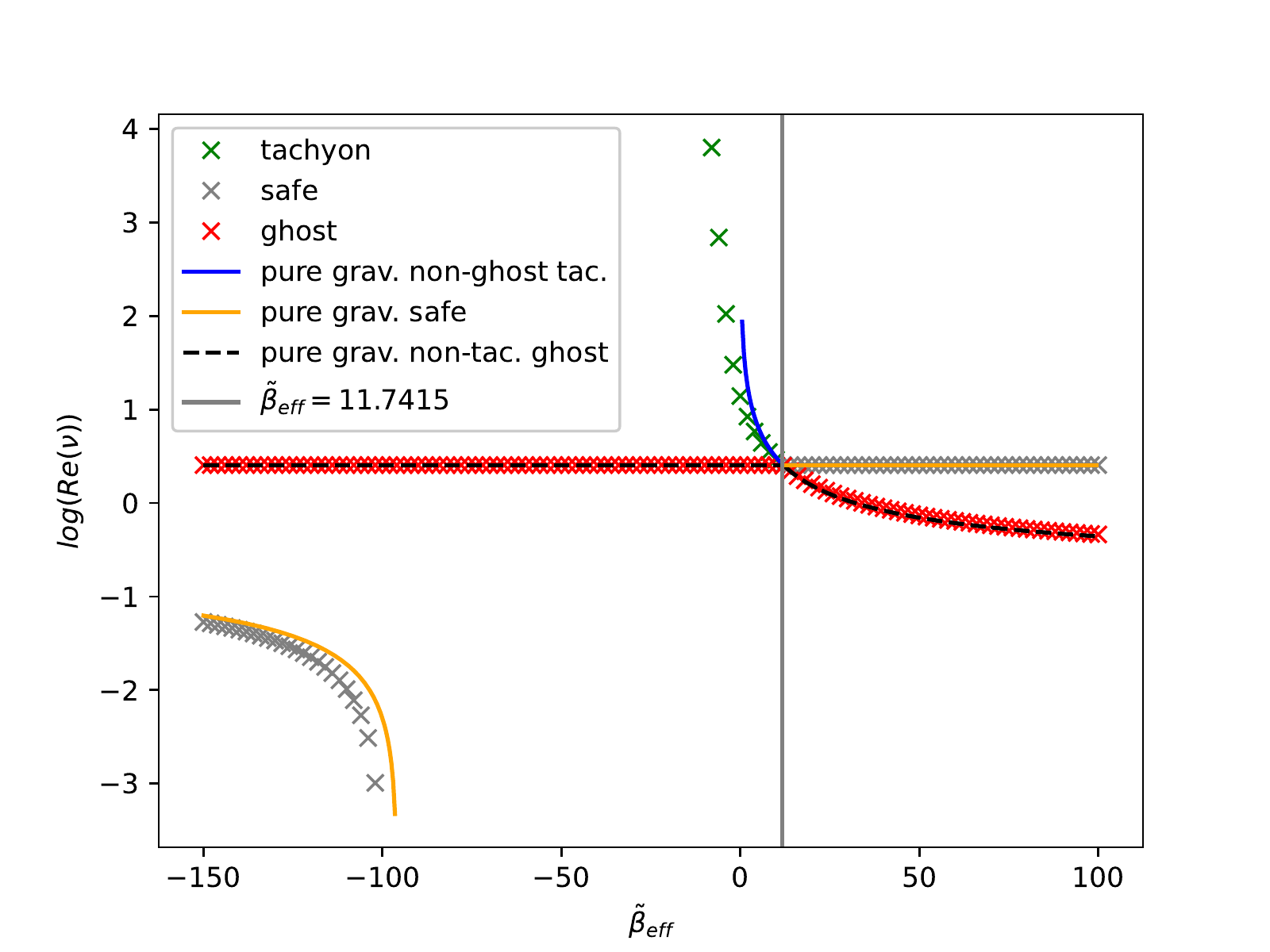}
\caption{\it
In this plot, we show the real part of the spin-2 poles in de Sitter, with the same parameters as in Figure \protect\ref{dS_Hpiover4_alpha10}.
 The real part also gives the inverse time scale (or \textit{strength}) of the dS tachyonic instability (\ref{ts2}), in units of the species cutoff, as a function of $\tilde{\b}_\text{eff}$.
 Grey markers correspond to safe poles (one for large and negative $\tbe$ disappearing around $\sim -100$, the other is massless for $\tbe>11.7415$), while red and green markers correspond respectively to the non-ghostly tachyonic pole and the light ghost pole.
For comparison, different curves show the case of pure gravity with $\beta =\tbe/\pi$ (\protect\ref{N03}): the solid blue line is a non-ghostly tachyon, the solid yellow line is non-ghostly-non-tachyonic, and the black dashed line is a non-tachyonic ghost. There is a gap at $-90\lesssim \tbe<0$ for pure gravity because the massive pole (\protect\ref{N03}) is purely imaginary in this interval. In the presence of the CFT, the safe pole disappears into negative (i.e not allowed) values of $\text{Re}(\n)$.
For pure gravity curves, the horizontal axis is $\pi\beta$.
 The vertical grey line is the value of $\tilde{\b}_\text{eff}$ corresponding to the transition from tachyonic to non-tachyonic (\protect\ref{Ln4}), where the tachyon merges with the massless ghost at $\n=3/2$. This value is also displayed in snapshot (g) of Figure \protect\ref{dS_Hpiover4_alpha10}.
}
\label{omegadS_alpha10_Hpiover4}
\end{figure}

The characteristic rate of the tachyonic instability is plotted in Figure \ref{omegadS_alpha10_Hpiover4}  as a function of $\tilde{\b}_\text{eff}$. This figure is obtained with the same parameters as  Figure \ref{dS_Hpiover4_alpha10}. Compared with  \ref{omegadS_alpha0_H0.01},  in Figure \ref{omegadS_alpha10_Hpiover4} there is no merging between two unstable massive poles (which is denoted by a grey vertical line in Figure \ref{omegadS_alpha0_H0.01}). Here instead, we have a single tachyon moving along the real axis as $\tbe$ is increased, it merges with the massless pole at $\n=3/2$ to form a safe massless pole and a ghost. The grey vertical line in Figure \ref{omegadS_alpha10_Hpiover4} marks the value of $\tilde{\b}_\text{eff}$ for above which the  theory becomes tachyon-stable.  This corresponds to the value  in snapshot (g) of Figure \ref{dS_Hpiover4_alpha10}.

\begin{figure}[ht]
\centering
\includegraphics[width=0.9\textwidth]{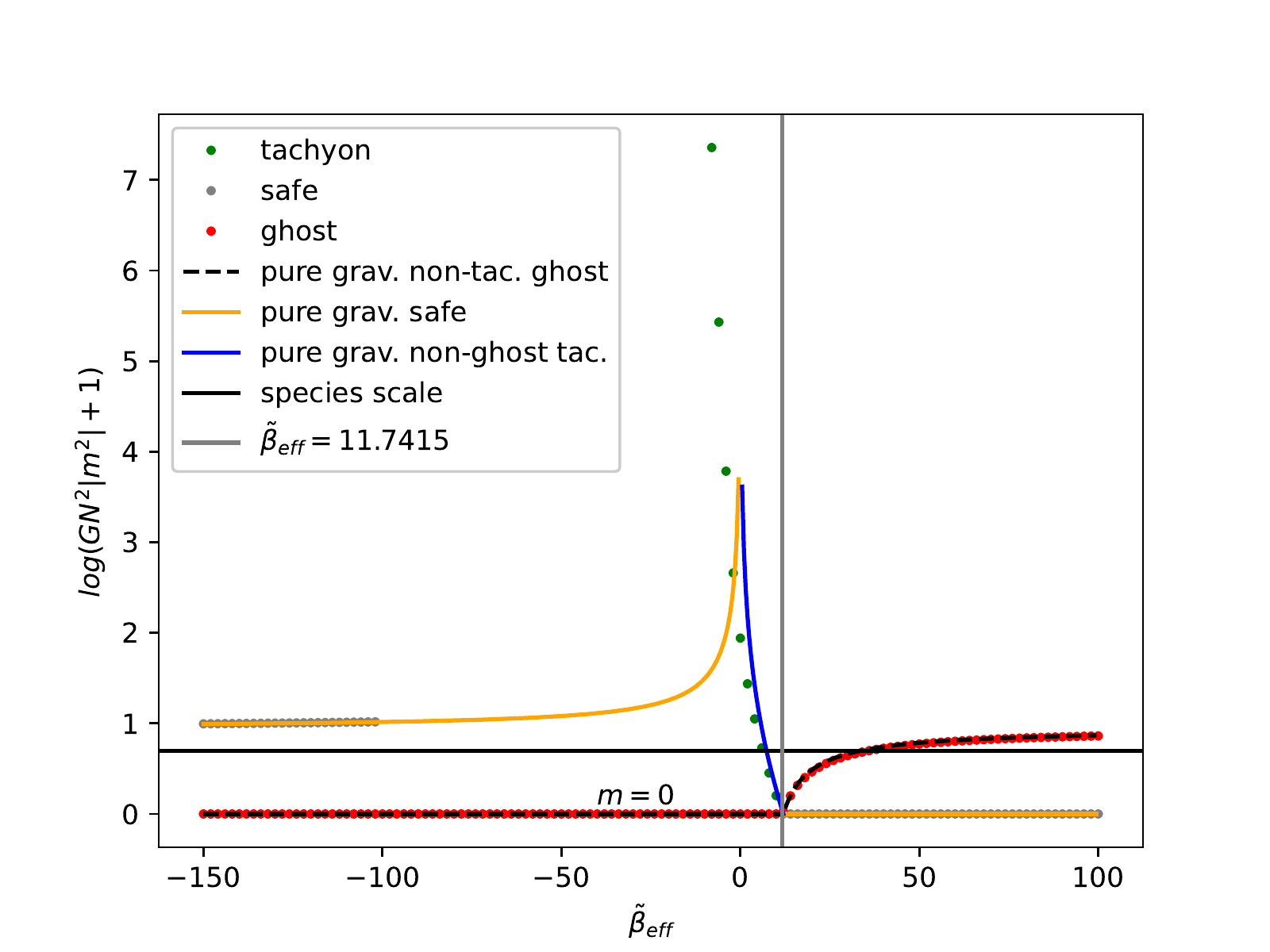}
\caption{\it
In this plot, obtained with the same parameters as in Figures \protect\ref{dS_Hpiover4_alpha10} and \protect\ref{omegadS_alpha10_Hpiover4}, we show the complex modulus of the mass of the spin-2 tachyonic poles in de Sitter, defined in (\protect\ref{dS5}), in units of the species scale (\protect\ref{sp-a}), as a function of $\tilde{\b}_\text{eff}$. Grey markers correspond to safe poles (one for large and negative $\tbe$ disappearing around $\sim -100$, the other is massless for $\tbe>11.7415$), while red and green markers correspond respectively to the non-ghostly tachyonic pole and the light ghost pole.
The species scale is shown by a horizontal solid black line.
For comparison, different curves correspond to the pure gravity modes with $\beta = \tbe /\pi$: the blue curve is tachyonic, the black dashed curve is a non-tachyonic ghost, and the orange curve is safe.
For pure gravity curves, the vertical axis is $\log(G^{1\over 2}|m|^2 + 1)$ while the horizontal axis is $\pi\beta$.
The vertical grey line is the value of $\tilde{\b}_\text{eff}$ corresponding to snapshot (g) of Figure \protect\ref{dS_Hpiover4_alpha10}), where the tachyonic pole becomes massless, and therefore stops being tachyonic.}
\label{mod_dS_alpha10_Hpiover4}
\end{figure}

Figure \ref{mod_dS_alpha10_Hpiover4} shows the modulus of the mass squared of the tensor modes in de Sitter, plotted in units of the species scale (\ref{sp-a}). This figure is a numerical evaluation for the same parameters as the one chosen in Figures \ref{dS_Hpiover4_alpha10} and \ref{omegadS_alpha10_Hpiover4}. As in Figure \ref{omegadS_alpha10_Hpiover4}, the green, red and blue curves are respectively the tachyonic, the ghost and the massive mode of pure gravity. The modulus of the ghost mass agrees with the pure gravity massive mode for large values of $|\tbe|$. For generic values, both the ghost and the tachyon lie above the species cutoff.
One can observe in this Figure that the massless pole is a ghost for $\tbe \lesssim 12$, whether the CFT is present or not. More precisely, the massless pole is a ghost for $\pi \b <12$ in pure gravity (\ref{N02}), whereas the actual value in the presence of the CFT is $\tbe \lesssim 11.7415$ as one can observe in snapshot (g) of Figure \ref{dS_Hpiover4_alpha10}.

Qualitatively, the cases displayed in Figures \ref{dS_H0.01_alpha0} and \ref{dS_Hpiover4_alpha10} (which, we remind the reader, correspond  to small $GN^2H^2$ and $O(1)$ $GN^2H^2$ respectively)  have a different behaviour as a function of $\tilde{\b}_\text{eff}$: in the first case,  a complex tachyonic ghost becomes non-tachyonic through the complex plane when $\tilde{\b}_\text{eff}$ is increased; in the second case,  a real tachyonic pole  becomes tachyon-stable on the real axis as $\tilde{\b}_\text{eff}$ is increased.
We chose to present only two different cases because they are paradigmatic of what happens in the whole parameter space. We span more values of $GN^2H^2$ and $\a$ in appendix \ref{moresnapshots}. As a result, any point choice of $(GN^2H^2,\tilde{\a})$ space should be similar to one of these two cases discussed above.
The ArXiv webpage of this paper contains ancillary files, including animated gifs. Each snapshot of these gifs corresponds to a different value of $\tbe$ for fixed  $(GN^2H^2,\tilde{\a})$.

In the next subsection, we  present an analytic approximation which explains these two different behaviours.

Our findings indicate that once we stay below the species cutoff,  qualitatively the behaviour
 is similar to the case without the CFT if
 we rescale the parameters of the effective gravity theory with a factor of $N^2$.
 We also find  that a richer set of phenomena can happen above the cutoff, but we cannot trust our description.
In previous works, \cite{anomalyinflation1}, the analysis was done
 for situations that were in the general area of the cutoff or above.

\subsection{Analytic results for tensor tachyonic modes in dS at large $|\n|$}
In this subsection, we provide  approximate analytical results for the location of the tachyonic poles in the tensor propagator (\ref{dS17}) on de Sitter. These analytics provide a better understanding of the qualitative picture presented in the previous subsection

We  focus on the ``non-trivial'' poles, i.e. away from the massless graviton pole $\nu=9/4$. Therefore, we  look for the zeros of $Q_\text{dS}(\n)$ defined in (\ref{dS15}). There is no simple analytic expression to this function for an arbitrary location in the complex plane. However, it can be approximated by a logarithm when   $|\n|$ is large. This occurs in particular for small curvature,  as it was the case in Figure \ref{dS_H0.01_alpha0}:  indeed, with $GN^2H^2 \ll 1 $, and  finite $\tilde{\a}$ and $\tilde{\b}_\text{eff}$, solving $Q_\text{dS}(\nu) = 0$ requires cancelling the large value of  ${2\pi \over GN^2H^2}$ against a large value of $|\n|$, as it is argued in \cite{Chesler}.

With these considerations,  in the rest of this subsection, we develop an analytic approximation for the poles in the limit of large $|\nu|$. In this limit, we can use the Stirling formula

\be
\mathcal{H}(z) \underset{|z|\rightarrow \infty}{=} \log{z} + \gamma_E + \mathcal{O}(z^{-1}),
\label{sH1}
\ee
where $\g_E$ is the Euler-Mascheroni constant.
The large $\n$ expansion of (\ref{dS15}) is then given by
\be
Q_\text{dS} =  1 -  {2\pi \over GN^2H^2} + 2\tilde{\a} - {\n^2\over 2} \left[{\tilde{\b}_\text{eff}} - {1\over 2} +\log\left(GN^2H^2\right)+ 2\log \n - 2\g_E + \mathcal{O}(|\n|^{-1})\right].
\label{sH3}
\ee
We have $2\log(\n) = \log(\n^2)$, since we have  chosen $\text{Re}(\n)>0$
in (\ref{dS8}), insofar as the branch-cut of the log function in (\ref{sH1}) is on the negative real axis. The equation of motion (\ref{dS14}) then takes a similar form to the one of flat space (\ref{n43}),
\be
X\log X = -a,
\label{sH4}
\ee
where now $X$ and $a$ depend on the curvature and are given by:
\be
X \equiv {\n^2GN^2 H^2}\exp \left\{{\tilde{\b}_\text{eff}} - {1\over 2} + 2\g_E \right\},
\label{sH5}
\ee
\be
a \equiv 2GN^2H^2\left[2\left({\pi\over GN^2H^2}-\tilde{\a}\right)-1\right]\exp\left\{{\tilde{\b}_\text{eff}} - {1\over 2} + 2\g_E\right\}.
\label{sH6}
\ee
We can observe that there are similar definitions for $X$ and $a$ in flat space-time (\ref{n44}). However, $a$ can now be negative using specific combinations of $\tilde{\a}$ and the curvature.

Equations (\ref{sH4},\ref{sH5},\ref{sH6})  hold for large $|\n|$ and any curvature. In particular, they can be used to understand  the flat space limit: indeed,  by comparing the  eigenvalue equations for de Sitter (\ref{dS5}) to the one for Minkowski (\ref{flat5}), we observe that the flat limit can be taken by defining
\be
\n^2H^2 \underset{H^2\rightarrow 0}{\rightarrow} k^2,
\label{sH7}
\ee
where the limit is taken by sending $|\nu|\to \infty$ so that $k^2$ is kept finite. In this limit, $\tilde{\a}$ becomes negligible and we find:
\be
\mathcal{F}_\text{dS}  \overset{k^2\rightarrow \n^2H^2}{\underset{H\rightarrow 0 }{\rightarrow}} \mathcal{F}_\text{flat}.
\ee
which coincides with the result we have obtained by the direct flat space calculation, equation (\ref{n43}). We have therefore shown  that the dS propagator matches  continuously onto the flat space propagator when we take the curvature to zero.

 The de Sitter tachyon-stability condition (\ref{dS6}) also becomes the flat space condition  (\ref{flat7a}) in this limit. Indeed, taking  the flat space limit of  de Sitter condition $|\text{Re}(\n)| < 3/2$ we obtain
\be
\text{Re}(k) \sim H|\text{Re}(\n)| < {3H\over 2} \rightarrow 0,
\label{sH8}
\ee
i.e. the flat space tachyon-stability condition.

We now turn to arbitrary curvatures, but still, search for solutions satisfying $|\n| \gg 1$. This allows us to use equation (\ref{sH4}) to better understand the results we have found in the two typical examples in the last subsection.
We keep  $GN^2H^2$ and $\tilde{\a}$ finite, so $X$ (\ref{sH5}) and $a$ (\ref{sH6}) differ from their flat space analogs (\ref{n44}). In particular, $a$ can be negative for finite curvature, unlike  in flat space where $a>0$.

When looking for solutions of  (\ref{sH4}), one can   distinguish three cases:
\begin{itemize}
\item If $a<0$, the unique solution $X$ of (\ref{sH4}) is real.
\item If $0\leq a \leq e^{-1}$, there are two real solutions, one degenerate solution if $a = e^{-1}$.
\item If $e^{-1} < a$, there are two complex conjugate solutions.
\end{itemize}

The second and third items describe Figure \ref{dS_H0.01_alpha0}. Snapshots (a) and (b) correspond to $0\leq a \leq e^{-1}$.  As $\tilde{\b}_\text{eff}$ increases, $a$ increases up to $e^{-1}$ where the two solutions merge in snapshot (c). Using the definition of $a$ in (\ref{sH6}), one can  obtain approximately the critical value at the merging:
\be
\tilde{\b}_\text{eff}^\text{merge} \equiv -{1\over 2} - 2\g_E - \log\left(2GN^2 H^2\right) - \log\left[2 \left({\pi\over GN^2H^2} - \tilde{\a}\right) -1 \right].
\label{Ln1}
\ee
Applying this formula to the parameters of Figure \ref{dS_H0.01_alpha0}, we find $\tilde{\b}_\text{eff}^\text{merge} = -4.18386$, which corresponds to snapshot (c).
After the merging, the complex solution travels in the complex plane up to crossing the green stability line. The value of $\tilde{\b}_\text{eff}$ chosen to plot snapshot (h) corresponds to equation (\ref{sH13}) obtained in the small $H$ approximation.

We now consider the case  $a<0$, which corresponds to a  single real tachyonic solution. From equation (\ref{sH6}),   $a<0$ is equivalent to
  \be
{\pi\over  GN^2H^2 } < \tilde{\a} + {1\over 2}.
 \label{Ln2}
 \ee
It is intriguing that the inequality (\ref{Ln2}) turns out to be the same as the condition for the scalar mode to be a ghost (\ref{2pt Psi c}).

Equation (\ref{Ln2}) holds in every snapshot of Figure \ref{dS_Hpiover4_alpha10}, in which $\tilde{\a}$ and $GN^2H^2$ are fixed. Therefore in all these snapshots, we have $a<0$.  Even if we are not in the large-$|\n|$ regime, it is a remarkable fact the analysis above still gives an accurate qualitative description of the results: we have a single real tachyon which moves along the real axis towards the massless pole.

If the transition from tachyonic to non-tachyonic does indeed  happen on the real axis, then it must be at  $\n=3/2$. If this is the case,  it is  sufficient to evaluate $Q_\text{dS}(\n)$  at $\n=3/2$ to obtain a stability condition for the other parameters as follows.  Using $\mathcal{H}(1) = 1$ in (\ref{dS15}), we obtain
 \be
Q_\text{dS}(3/2) = -{1\over 2} - \log\left(GN^2H^2\right) - 2\left[{\pi\over GN^2H^2} + {\tilde{\b}_\text{eff}\over 2} - \tilde{\a} \right].
\label{Ln3}
 \ee
The transition between stability and instability corresponds to $Q_\text{dS}(3/2) = 0$, in which case we obtain
 \be
 \tilde{\b}_\text{eff}^\text{massless} \equiv -{1\over 2} - \log\left(GN^2H^2\right) + 2\left(\tilde{\a} - {\pi\over GN^2H^2}\right).
 \label{Ln4}
 \ee
 If the parameters of the theory satisfy (\ref{Ln4}), then $\n = 3/2$ is a double pole of the tensor two-point function. But if $\tilde{\b}_\text{eff} > \tilde{\b}_\text{eff}^\text{massless}$, then the theory is tachyon-stable.
Evaluating (\ref{Ln4}) for the parameters taken in Figure \ref{dS_Hpiover4_alpha10} gives the value chosen to plot snapshot (c). It is clear from this snapshot that the solution which was unstable in snapshot (b) crosses the stability line. Therefore, the assumption made for (\ref{Ln4}) that the transition would happen on the real axis is verified numerically for this particular set of parameters.

In the $a<0$ case, we were able to derive an exact formula for tachyonic stability as a function of $\tilde{\b}_\text{eff}$ in (\ref{Ln4}). This was obtained assuming that the tachyonic solution would cross the point $\n=3/2$. However, in the case of $a>0$, the tachyonic solution is complex and can cross the stability line with a generic imaginary part, as was shown in Figure \ref{dS_H0.01_alpha0}.

If we make the further assumption that $a \gg 1$, it is possible to perform an additional approximation to find $X$.
As argued also in \cite{Chesler} in the case  $\a=0$, a solution of equation (\ref{sH4}) for large and positive $a$ can be found using the ansatz
\be
X \underset{a\rightarrow +\infty}{\sim} = -{a\over \log(-a)}.
\label{sH9}
\ee
Injecting this ansatz in the original equation (\ref{sH4}), we then find
\be
X\log X = -a \left[1 - {\log(\log(-a))\over \log(-a)}\right] \underset{a\rightarrow 0}{\sim} -a,
\label{sH10}
\ee
which is a ``slow" convergence as ${\log(\log(-a))\over \log(-a)}\rightarrow 0$.
We then have an imaginary part in the solution $X$ (\ref{sH9}) since $\log(-a) = \pm i\pi + \log a$. The complex square root of the slowly converging solution (\ref{sH9}) is then
\be
\sqrt{X} =  i \sqrt{a\over \log a}\left[\pm 1 - {i\pi\over 2\log a}\right] + \mathcal{O}\left({1\over \log^{3/2} a}\right).
\label{sH11}
\ee
The branch with a positive real part has been taken because the real part of $\n$ is assumed to be positive in the bulk radial solution (\ref{dS13}).
Then, replacing $a$ and $X$ by their definitions (\ref{sH5},\ref{sH6}), one gets the approximate solution for $\n$
\be
\n \approx  {2\left( {\pi\over GN^2H^2} - \tilde{\a} - {1\over 2}\right)^{1/2}\over \left\{{\tilde{\b}_\text{eff}} - {1\over 2} + 2\g_E  + \log\left[4GN^2H^2\left( {\pi\over GN^2H^2} - \tilde{\a} - {1\over 2}\right)\right]\right\}^{1\over 2}} \times
\nonumber
\ee
\be
\times\left[ \pm i  + {\pi/2\over {\tilde{\b}_\text{eff}} - {1\over 2} + 2\g_E  + \log\left[4GN^2H^2\left( {\pi\over GN^2H^2} - \tilde{\a} - {1\over 2}\right)\right]}\right].
\label{sH12}
\ee
For $\a =0$,  the solution (\ref{sH12}) reduces to the results of \cite{Chesler} derived for small curvature.
 The stability condition (\ref{dS6}) is then
 \be
 \tilde{\b}_\text{eff} \geq \tilde{\b}_\text{eff}^\text{c} \equiv {1\over 2} - 2\g_E  - \log\left[4GN^2H^2\left( {\pi\over GN^2H^2} - \tilde{\a} - {1\over 2}\right)\right] + \left[{4\pi^2\over 9} \left( {\pi\over GN^2H^2} - \tilde{\a} - {1\over 2}\right)\right]^{1/3}.
 \label{sH13}
 \ee
 Applying this result to the case where $\a = 0$ and $GN^2H^2 = 0.01$, we obtain that stability is reached for ${\tilde{\b}_\text{eff}} \geq 7.94268$ which is chosen for snapshot (f) in Figure \ref{dS_H0.01_alpha0}.

The approximation used to arrive at (\ref{sH12}) cannot hold for large  negative values of ${\tilde{\b}_\text{eff}}$: in this case there are  two real tachyonic solutions,  which cannot be described by (\ref{sH12}). This is due to the fact that the large $a$ approximation, equivalent to (\ref{sH9}), cannot not hold for  large  negative values  ${\tilde{\b}_\text{eff}}$  because $a$ is proportional to $e^{\tilde{\b}_\text{eff}}$.

\subsection{Tachyonic and ghost-like instabilities for dS in parameter space}
\label{subsection ghost dS}

After having addressed the qualitative features of the spectrum in the previous sections, we now present a full numerical scan of parameter space,  identify the stability and instability regions (concerning ghosts and tachyons) and determine the characteristic scale of the instability. We do this first for tachyonic instabilities, then we move on to investigate ghost instabilities.

\begin{figure}[ht]
\centering
\includegraphics[width= \textwidth]{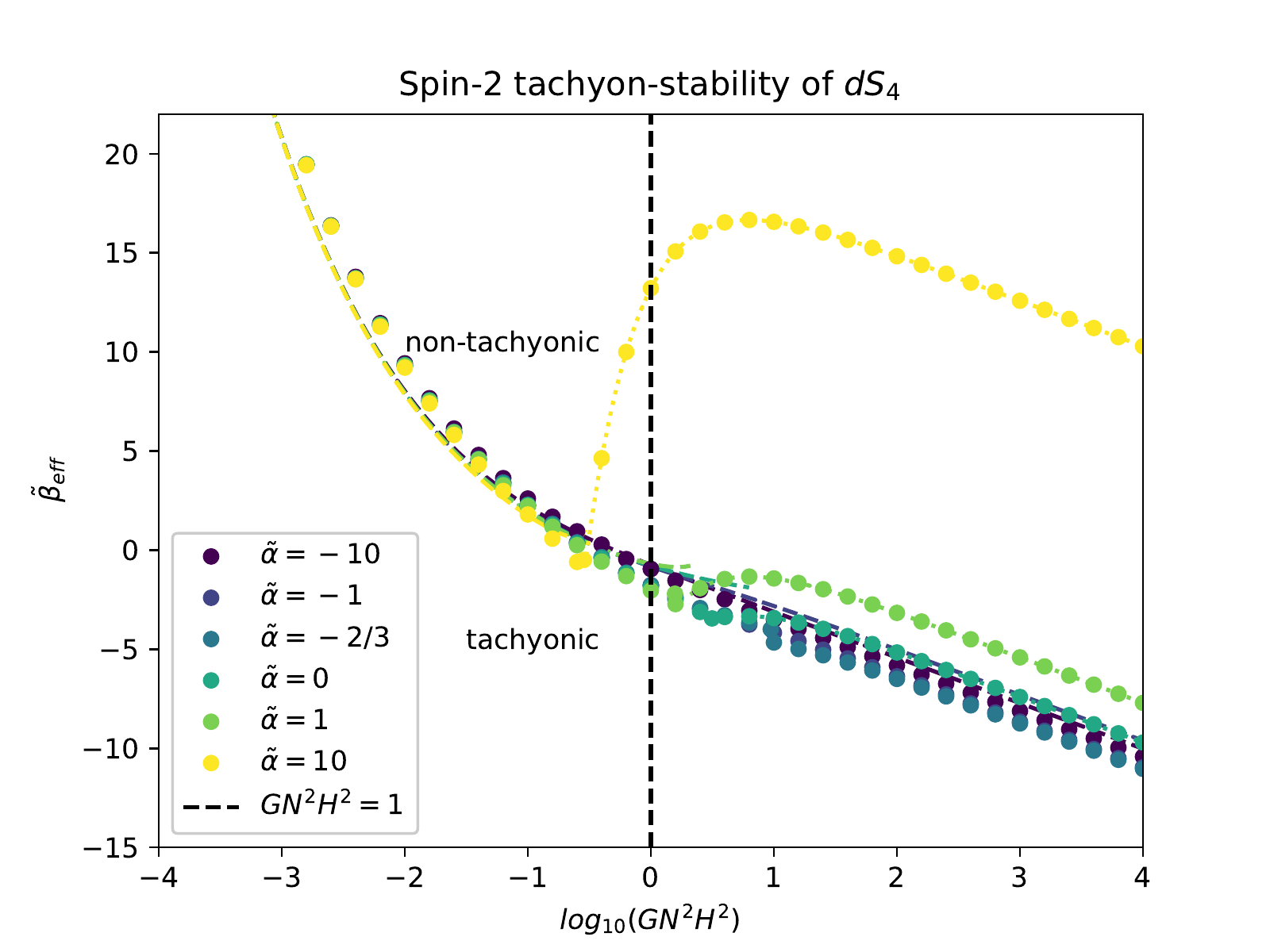}
\caption{\it Spin-2 tachyonic instability of de Sitter space depending on $\tilde{\b}_\text{eff}$ and $GN^2H^2$, for different values of $\tilde{\a}$.
  Dotted lines with large dots, are the boundaries between the stable and unstable regions and have been computed numerically.
    Above each curve, we are in a non-tachyonic regime ($\text{Re}(\n) \leq 3/2$), while the region below is tachyonic ($\text{Re}(\n) > 3/2$).
The dotted lines with small dots are given by the exact formula (\protect\ref{Ln4}) assuming that there is only one tachyon and it is located on the real axis.
There is one such line for each positive $\tilde{\a}$ but they are always very close to the exact boundary. The dashed lines are given by the large $|\n|$ and large $a$ approximation (\protect\ref{sH13}).}
\label{fig:dS}
\end{figure}

Figure \ref{fig:dS} shows the distinction between tachyon-stable and tachyon-unstable regions, as a function of the parameters $(GN^2H^2,\tilde{\a}, \tilde{\b}_\text{eff})$.
In this figure, the critical value for $\tilde{\b}_\text{eff}$ is given as a function of the curvature for several values of $\tilde{\a}$. Each curve corresponds to a different $\tilde{\a}$.
For a given $\tilde{\a}$, the region below the curve is unstable because it corresponds to lower values of $\tilde{\b}_\text{eff}$, which are tachyonic. The region above the curve is stable because it corresponds to higher values of $\tilde{\b}_\text{eff}$, for which the tachyon has entered the stability region $|\text{Re}(\n)|<3/2$ exactly at the critical value.

On each curve of figure \ref{fig:dS}, there is a regime (which roughly corresponds to small curvatures, and corresponds to the left part of the figure) in which  the critical value of $\tilde{\b}_\text{eff}$ decreases with increasing curvature, regardless of the value of $\tilde{\alpha}$. For larger curvatures, one may observe a different regime:  for large enough $\tilde{\alpha}$ we observe that $\tilde{\b}_\text{eff}$  starts increasing with the curvature  to then decrease again. This behaviour sets in  approximatively at  $\tilde{\a}\approx 0$. The larger $\tilde{\a}$ is, the higher the increase in the critical value of $\tilde{\b}_\text{eff}$.

 From the large-$|\n|$ approximation, we expect that the small curvature regime (left part of the figure) contains a complex tachyon, whereas the eventual \textit{bump} on the right part of the figure should contain a single tachyonic pole on the real axis. The boundary between these two regions should correspond to the value of the curvature when $a=0$, i.e. where (\ref{Ln2}) is an equality.

We  have checked how well the analytic large-$|\nu|$ approximation matches the numerical results: in the region where  $a>0$ (left part of  Figure \ref{fig:dS}),  the analytic approximation  (\ref{sH13}) is represented by dashed lines.  On the right, where $a<0$, the approximation (\ref{Ln4}) is represented by dotted lines.
The two analytic regimes are separated by a critical value of the curvature given by the value that saturates (\ref{Ln2}).
For curvatures above this value, there is a single tachyonic pole located on the real axis.
This critical curvature exists only for $\tilde{\a} >-1/2$. For $\tilde{\a}<-1/2$, the large-$|\n|$ approximation (\ref{sH13}) extends to all curvatures and leads to a monotonic behaviour of $\tilde{\b}_\text{eff}^\text{critical}$ as a function of the curvature.

The analytical approximations do not exactly match the numerical results, especially the dashed lines when curvatures are not small. However, we can observe in Figure \ref{fig:dS} that large curvatures are very well described by the exact formula (\ref{Ln4}), where (\ref{Ln2}) holds. In the large-$|\n|$ regime, a single tachyonic pole is located on the real axis, and we have assumed that it would stay on the real axis even for $\n=3/2$ while entering the stability zone. This hypothesis seems to be confirmed by the numerics because dotted lines (approximation) coincide with the large circles (numerics).

\begin{figure}[ht]
\centering
\includegraphics[width= \textwidth]{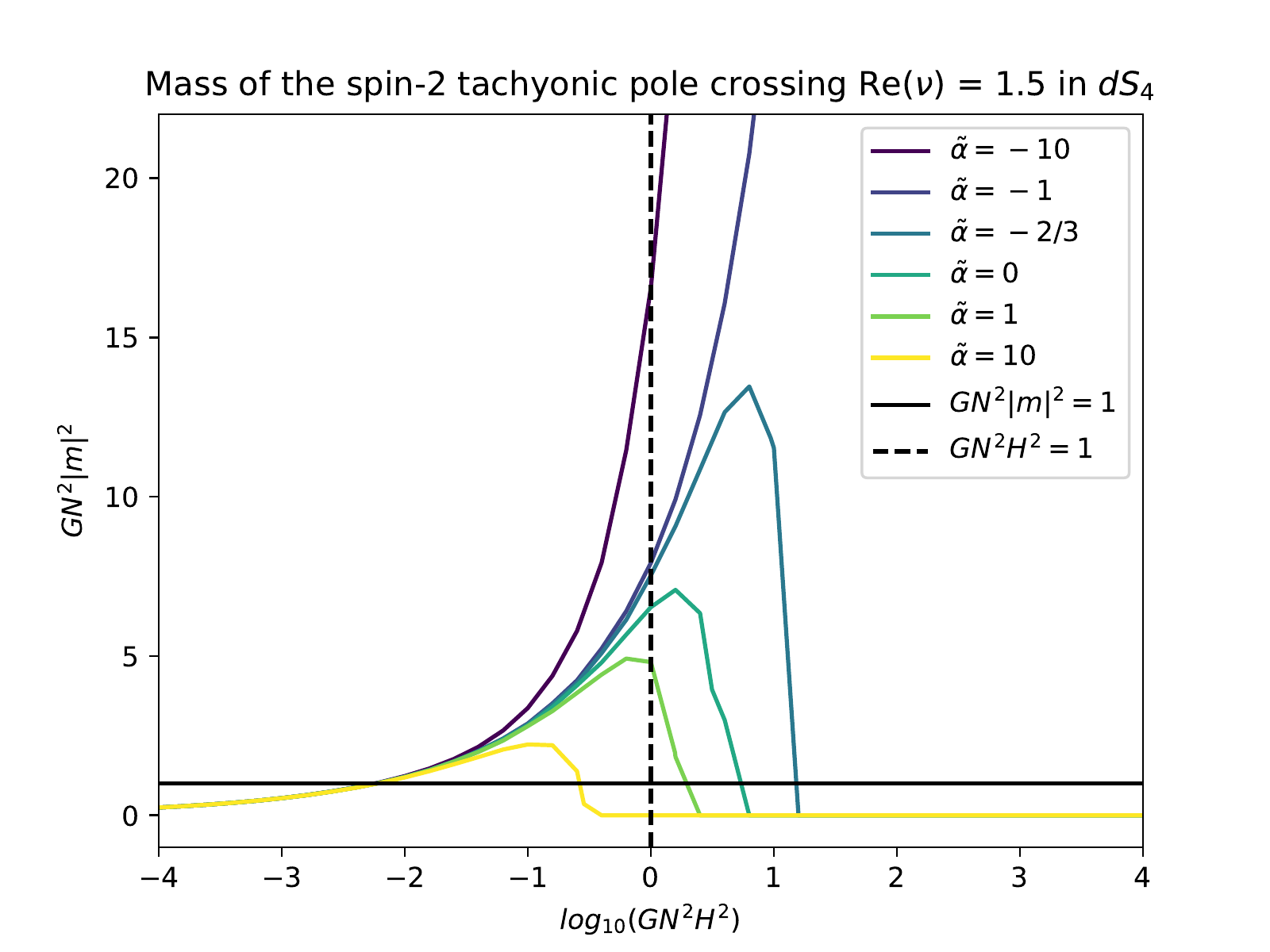}
\caption{\it Mass of the spin-2 tachyonic instability of de Sitter space when it crosses the $\text{Re}(\n) =  3/2$ line, for a given set of parameters $(\tilde{\a},GN^2H^2)$ while varying $\tbe$. The mass is plotted in units of the species scale (\ref{sp-a}).
 Each coloured curve is a different choice of $\tilde{\a}$.}
\label{dS_tacmass}
\end{figure}

\begin{figure}[ht]
\centering
\includegraphics[width= \textwidth]{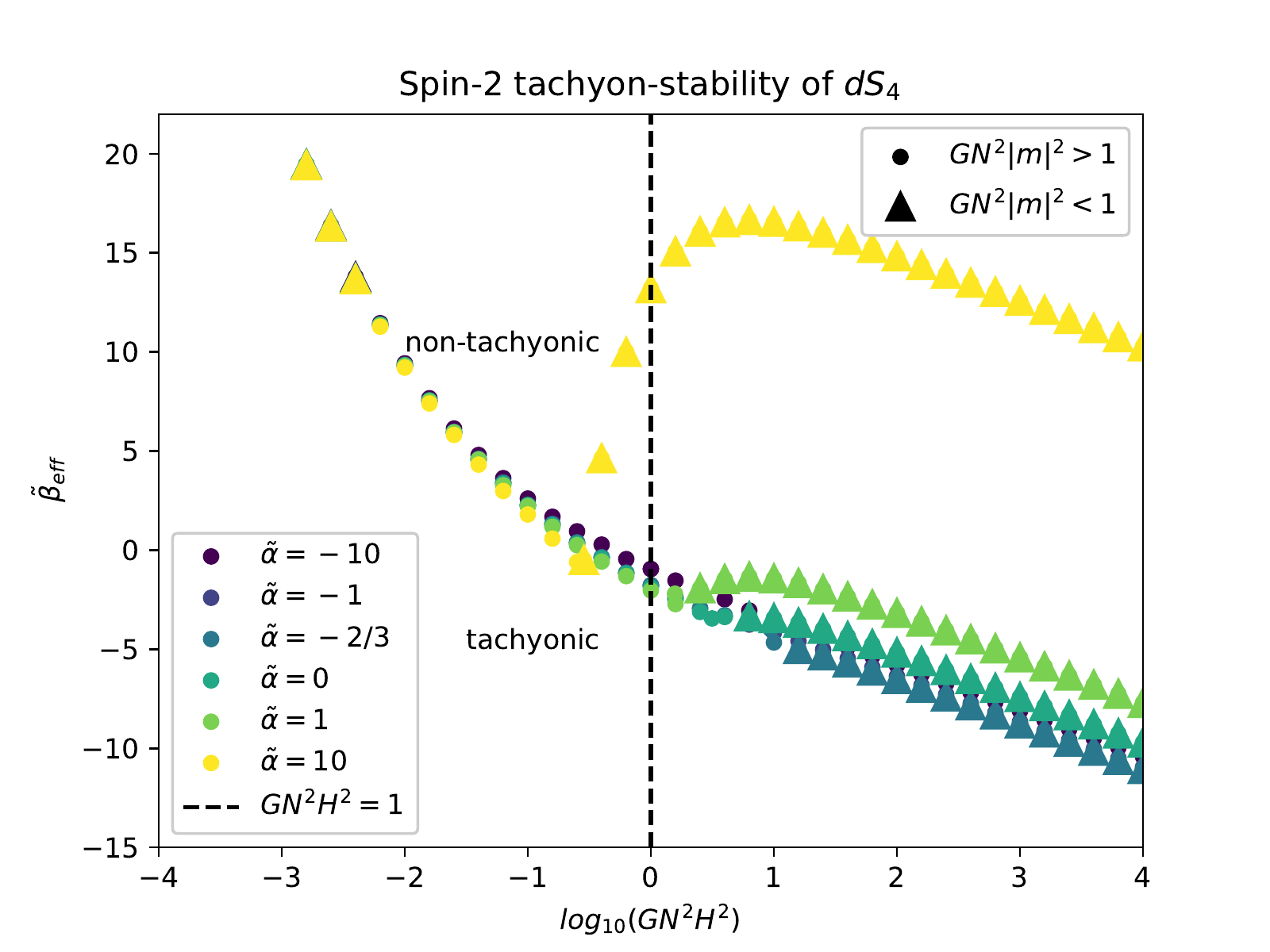}
\caption{\it Spin-2 tachyonic instability of de Sitter space depending on $\tilde{\b}_\text{eff}$ and the curvature $H$, for different values of $\tilde{\a}$.
 Markers are placed on the boundary $\text{Re}(\n) = 3/2$ between the tachyonic and non-tachyonic regions which have been computed numerically.
  Each colour corresponds to a given value of $\tilde{\a}$. For each coloured curve, we are in a tachyonic regime ($\text{Re}(\n) >3/2$) below the markers, while the region above is non-tachyonic ($\text{Re}(\n) \leq 3/2$).}
\label{fig:dS2}
\end{figure}

Figure \ref{dS_tacmass} shows the mass of the spin-2 tachyonic instability of de Sitter for the value of $\tbe$ at which it stops being tachyonic ($\text{Re}(\n) = 3/2$). The mass is plotted as a function of the curvature for different values of $\tilde{\a}$. The remaining parameter is then $\tbe$, which is then fixed by the $\text{Re}(\n)=3/2$ requirement. At this transition between tachyonic and non-tachyonic, we measure the mass numerically and report it on the figure. We observe in this figure that the mass is below the cutoff for small curvatures. The mass starts to move above the cutoff at curvatures around $GN^2H^2 \approx 10^{-2.22}$ for $\tilde{\a}=10$ and $GN^2H^2\approx 10^{-2.25}$ for $\tilde{\a} = -10$. The value of $\tilde{\a}$ does not play an important role in the regime of such small curvatures because we are close to the flat space case in which the spin-2 pole locations do not depend on $\tilde{\a}$.
The tachyonic pole eventually goes back beyond the species cutoff for large curvatures if $\tilde{\a}$ is not too negative. For example, for $\tilde{\a} = -2/3$, the mass goes beyond the cutoff at $GN^2H^2 \approx 10^{1.2}$ which is itself above the species cutoff at $GN^2H^2 = 1$.
It is then possible to identify the points of Figure \ref{fig:dS} which are above the species scale. This additional information is shown in Figure \ref{fig:dS2}, which is similar to Figure \ref{fig:dS}, except that triangles correspond to poles with mass below the species cutoff whereas large dots have a mass larger than the cutoff.

\begin{figure}[ht]
\centering
\begin{subfigure}{0.8\textwidth}
\includegraphics[width= \textwidth]{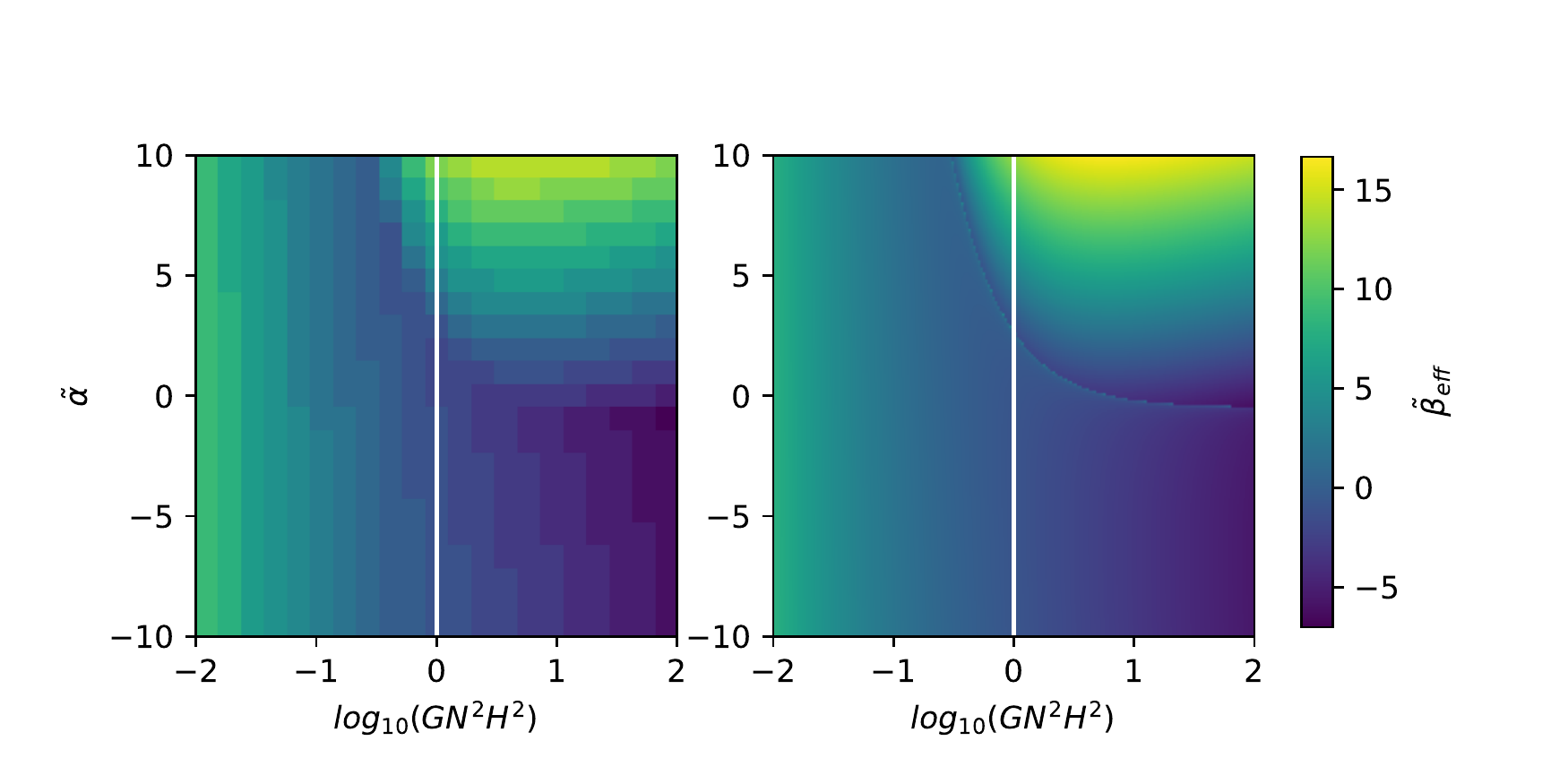}
\end{subfigure}

\begin{subfigure}{0.8\textwidth}
\includegraphics[width=\textwidth]{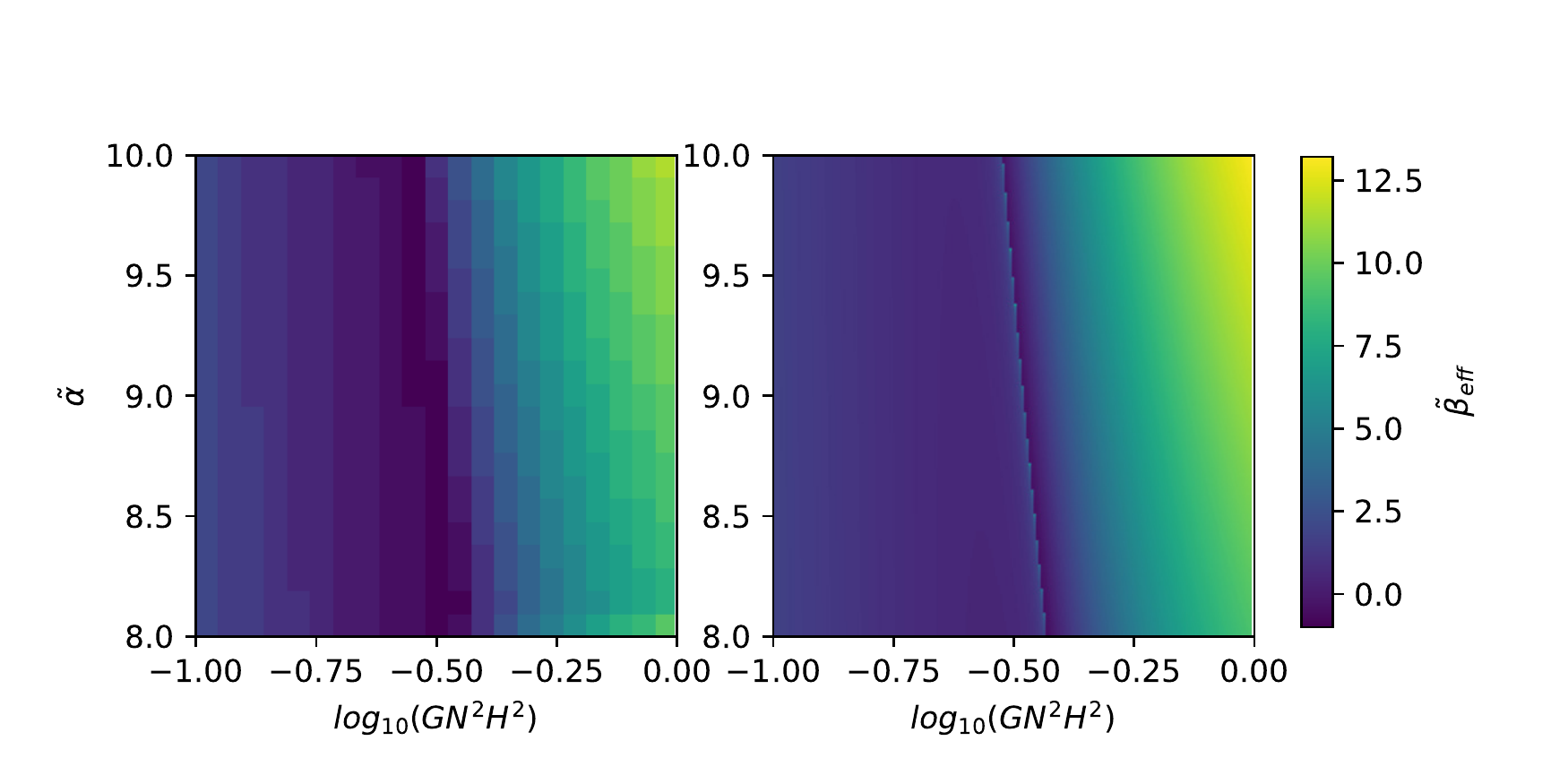}
\end{subfigure}
\caption{\it Values of $\tilde{\b}_\text{eff}$ such that de Sitter space goes from tachyonic-unstable to tachyonic-stable for the spin-2 mode, plotted in the ($\tilde{\a}$, $GN^2H^2$) plane. For each pixel, values of $\tilde{\b}_\text{eff}$ lower than the given colour corresponds to instability, and higher values correspond to stability.
The white vertical line on the top row panels is the separation between curvatures above and below the species cutoff (\protect\ref{sp-a}). The right panels are obtained using large-$|\n|$ analytical approximations, while the left panels are numerical results. As it was done for Figure \protect\ref{fig:dS}, approximations are split into two regimes: if the inequality (\protect\ref{Ln2}) holds, then (\protect\ref{Ln4}) is used. Otherwise, (\protect\ref{sH13}) is used. The bottom panels are zoomed on a smaller parameter space, where the analytical approximation is supposed to break down around $a\approx 0$ (\protect\ref{sH6}), which corresponds to the contour of the area on the top right of each panel.}
\label{fig:2Dplot_dS}
\end{figure}

Figure \ref{fig:2Dplot_dS} is a different representation of the critical value of $\tilde{\beta}_\text{eff}$ which separates the tachyon-stable from the tachyon-unstable regime: in this figure, the colour code corresponds to the critical value of $\tilde{\b}_\text{eff}$ which separates between tachyon-stable and tachyon-unstable in the ($\tilde{\a}$, $GN^2H^2$) parameter space. It also compares the value of $\tilde{\b}_\text{eff}$ obtained numerically with the analytical approximations (\ref{Ln4}) and (\ref{sH13}).
Each row of this figure gives a different window for ($\tilde{\a}$, $GN^2H^2$). The top row gives a more extensive view while the bottom row is a zoom on
 a space where the analytics are supposed to break down.
 The right panels correspond to the analytical approximations (\ref{sH13} - \ref{Ln4}) with a larger number of pixels than the numerics given on the left panels.

The apparent discontinuity in the right panels comes from a junction between approximation (\ref{Ln4}) for real-axis tachyon and (\ref{sH13}) for complex tachyon. Around this junction, the large-$|\n|$ approximation is not valid anymore. The discontinuity which is visible on both panels on the right is an artefact of the large-$|\n|$ approximation and is absent from the numerics in the middle panels. Instead of a discontinuity, one can observe a valley of values for $\tilde{\b}_\text{eff}$ which are lower than expected by the analytics. This behaviour could also be observed in Figure \ref{fig:dS} at the minima of $\tilde{\b}_\text{eff}$.

Figure \ref{fig:dS}  separates tachyonic from non-tachyonic  regions, but it does not contain any information on the location of the poles, which encodes the characteristic scale of the tensor tachyonic instability.
In what follows, we investigate this scale numerically.

\begin{figure}[ht]
\centering
\begin{subfigure}{0.45\textwidth}
\includegraphics[width=\textwidth]{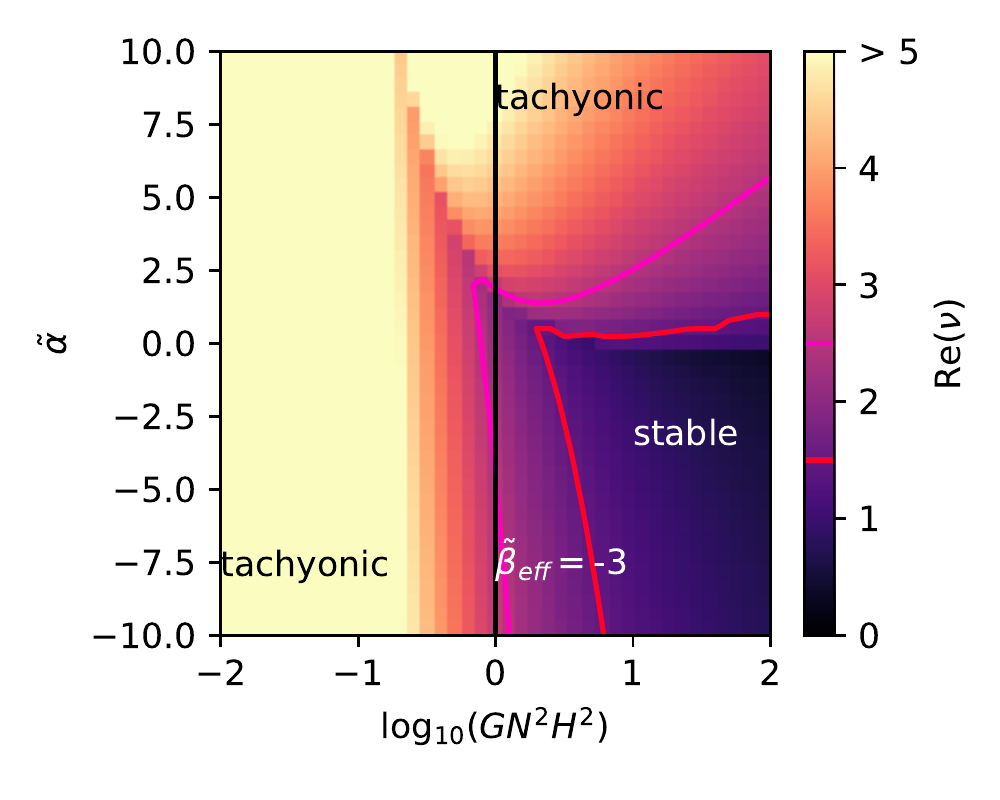}
\caption{\it }
\end{subfigure}
\begin{subfigure}{0.45\textwidth}
\includegraphics[width=\textwidth]{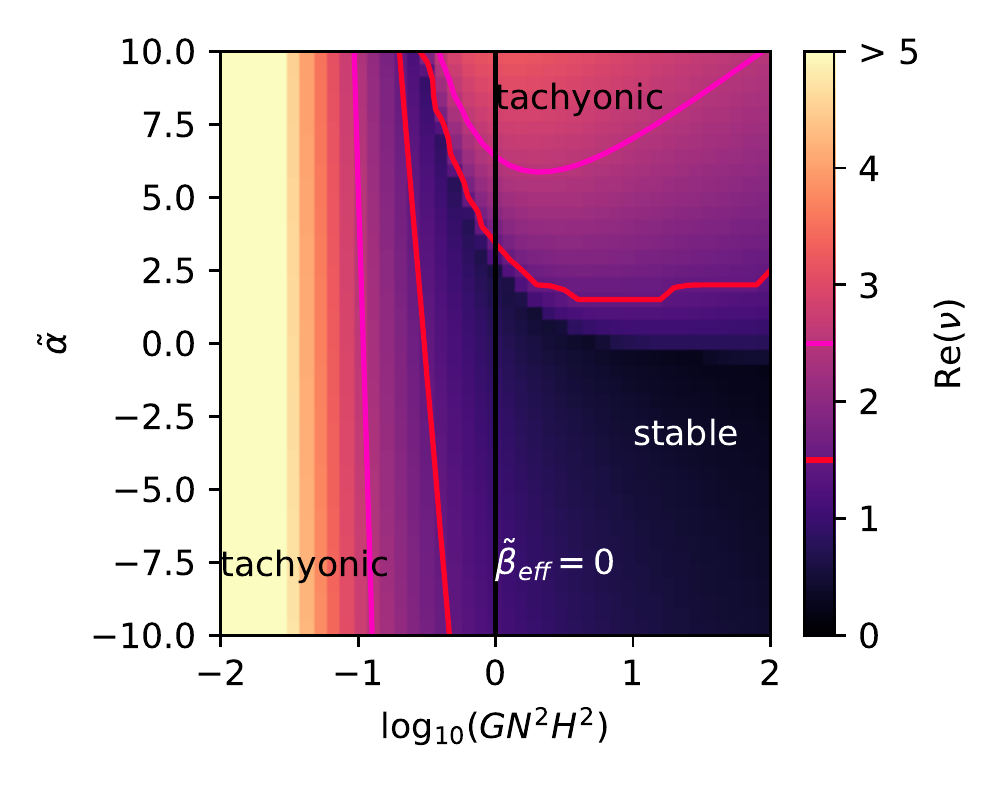}
\caption{\it }
\end{subfigure}
\begin{subfigure}{0.45\textwidth}
\includegraphics[width=\textwidth]{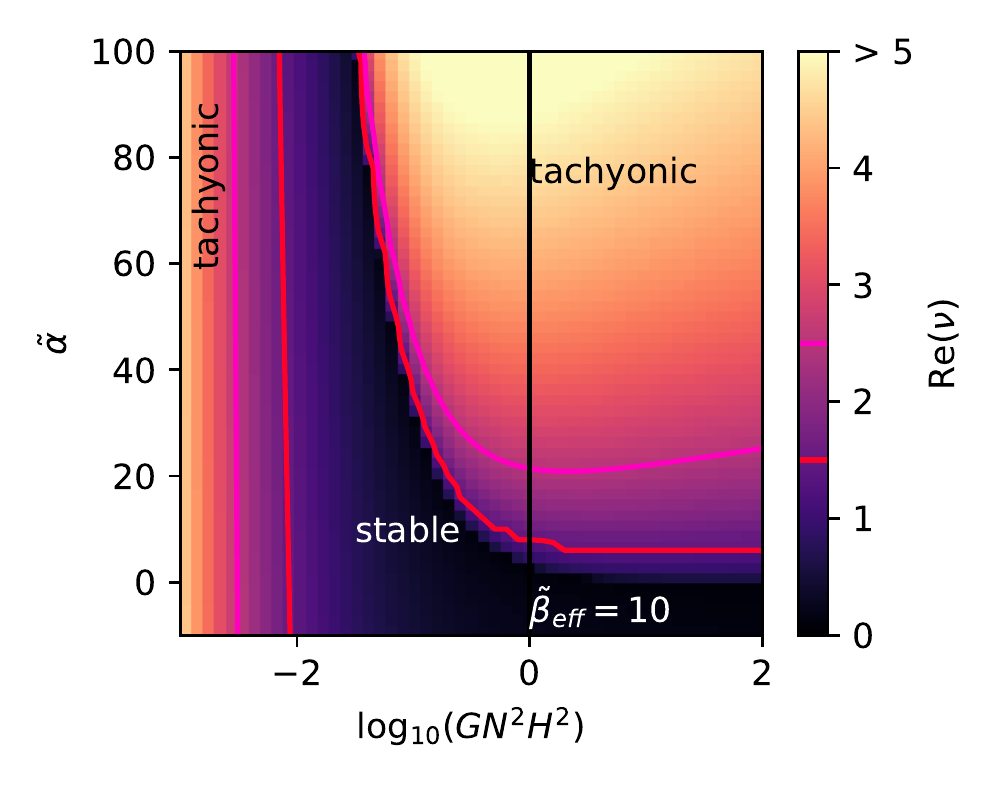}
\caption{\it }
\end{subfigure}
\begin{subfigure}{0.45\textwidth}
\includegraphics[width=\textwidth]{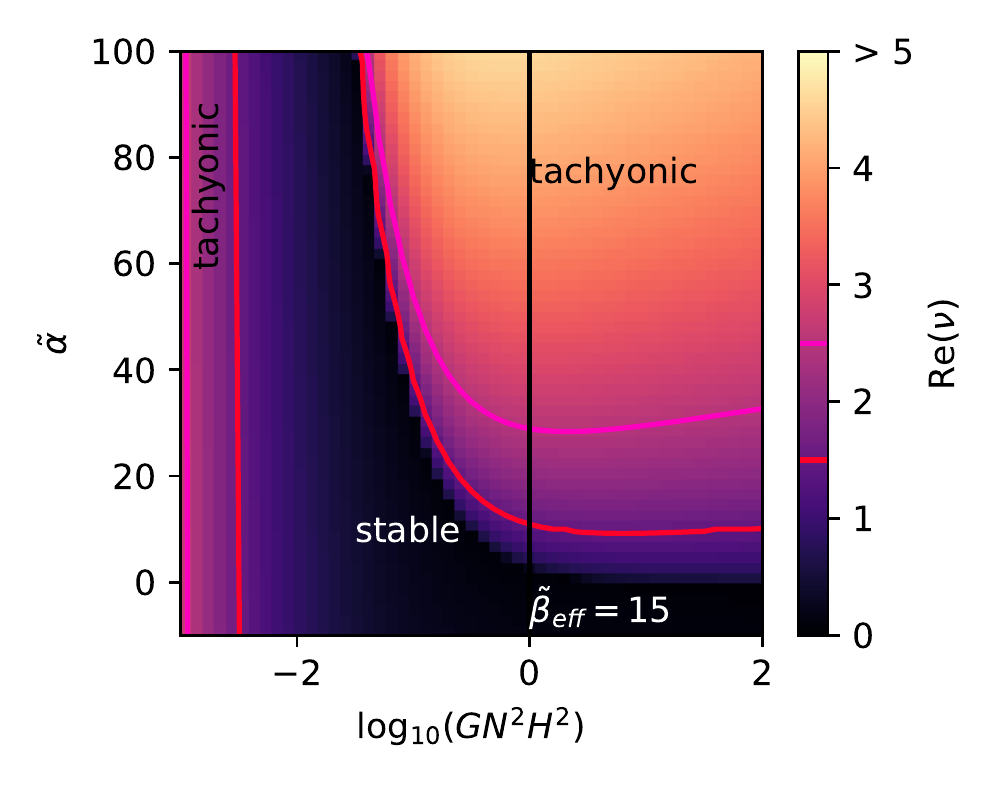}
\caption{\it }
\end{subfigure}
\caption{\it Regions and inverse time-scale of the tachyonic instability (defined in (\protect\ref{ts2}) for the spin-2 sector in the de Sitter case. The colour code of each of the figures above gives the value of the real part of $\n$ at the zero of the inverse propagator (\protect\ref{dS17}) for a given value of $\tilde{\b}_\text{eff}$. The black vertical line separates curvatures which are below the species cutoff defined in (\protect\ref{sp-a}) from curvatures which are above. The unstable region (\protect\ref{dS6}) is delimited by the red lines where $\text{Re}(\n) = 3/2$. The pink lines are placed at $\text{Re}(\n) = 5/2$, where the inverse time scale of the tachyonic instability (\protect\ref{ts2}) has the value $\G = H$. As $\tilde{\b}_\text{eff}$ increases, the stability region becomes larger.}
\label{dS_nu}
\end{figure}

The instabilities of the tensor sector are studied in Figure \ref{dS_nu}. The colour coding corresponds to the real part of $\nu$ for the tachyonic modes\footnote{It should be remembered that $\n$ controls the mode mass in units of the dS  Hubble scale.}. This is the quantity that controls  the divergence rate of the mode, via equation  (\ref{ts2}).
The red line $\text{Re}(\n) =3/2$  separates  tachyon-unstable from tachyon-stable regions. The four different sub-figures of Figure \ref{dS_nu} show, using a colour code, the size of $\text{Re}(\n)$ as a function of two of the parameters ($\tilde{\a}$, $GN^2H^2$), for fixed values of $\tilde{\b}_\text{eff}$. The four sub-figures correspond to different values of $\tilde{\b}_\text{eff}$.
We observe that there are two tachyonic regions, one for low enough values of $\log (GN^2H^2)$, the other for large values of both $\tilde{\a}$ and  $\log (GN^2H^2)$. As $\tilde{\b}_\text{eff}$ increases, these two regions move out in parameter space. In this figure, the word ``stable'' refers exclusively to the absence of tachyonic instabilities: we  remind the reader that there are always ghost-like spin-2 poles at all points in parameter space. We shall come back to these modes at the end of this section.

\begin{figure}[ht]
\centering
\begin{subfigure}{0.45\textwidth}
\includegraphics[width=\textwidth]{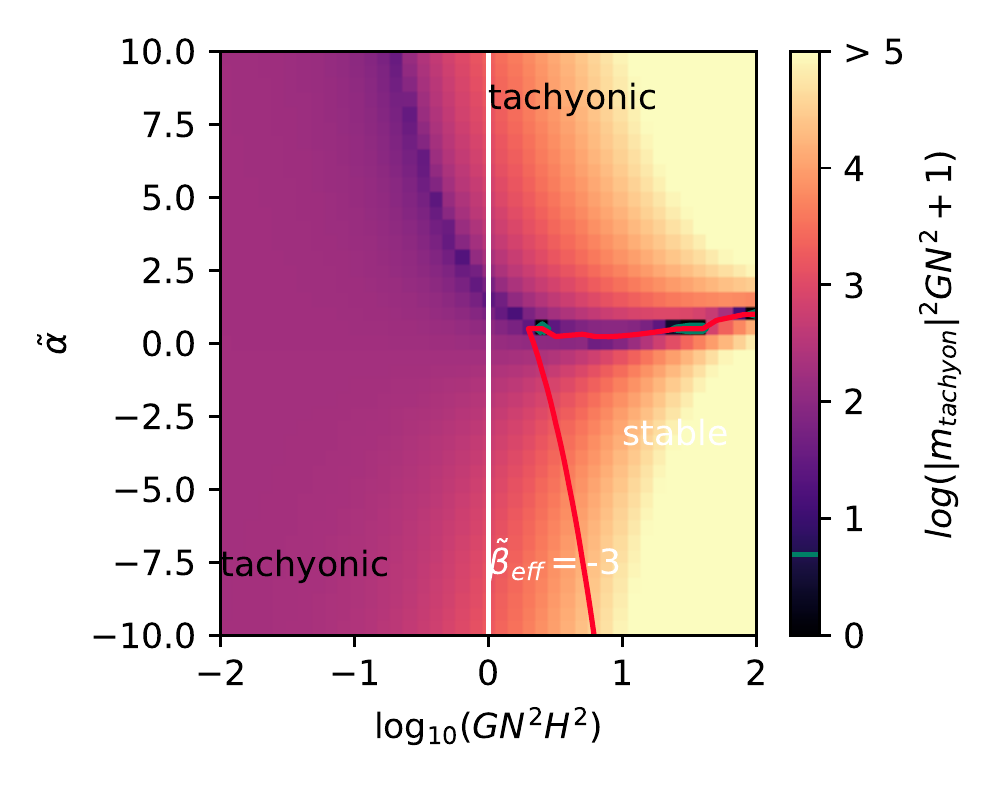}
\caption{\it }
\end{subfigure}
\begin{subfigure}{0.45\textwidth}
\includegraphics[width=\textwidth]{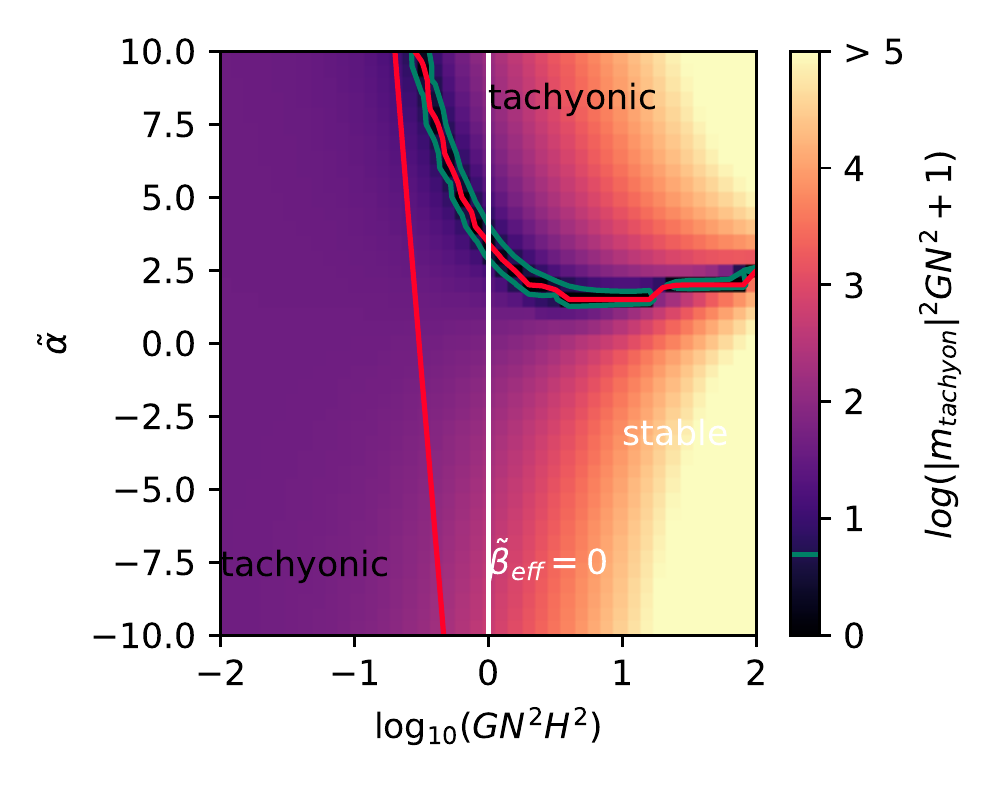}
\caption{\it }
\end{subfigure}
\begin{subfigure}{0.45\textwidth}
\includegraphics[width=\textwidth]{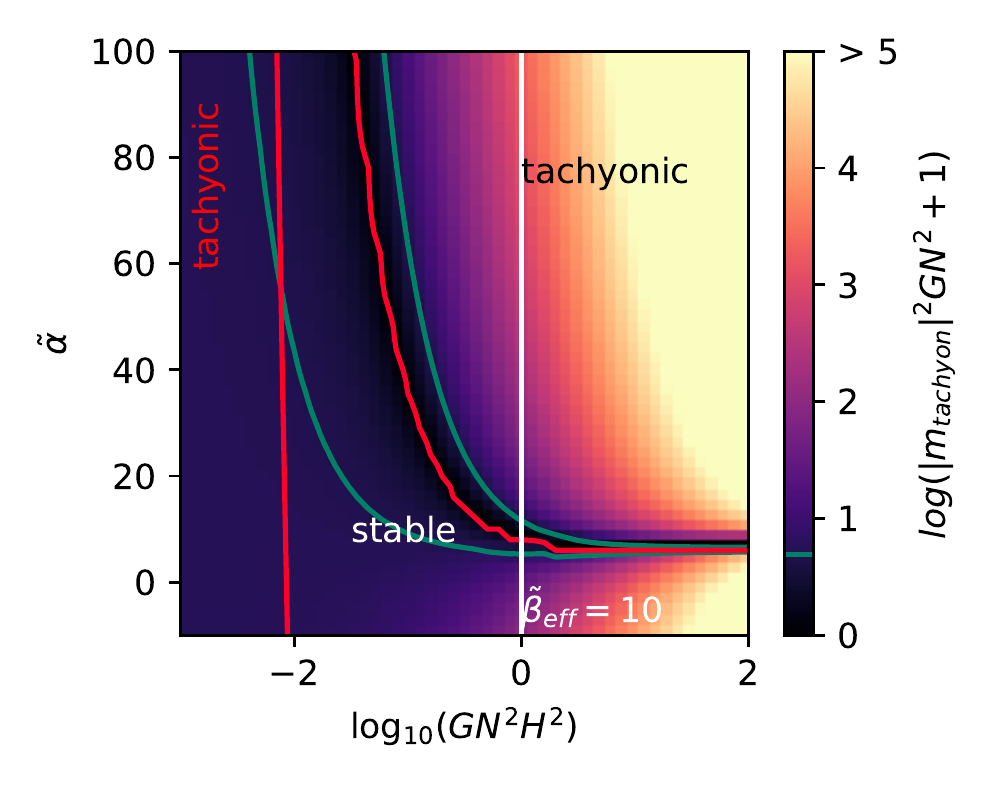}
\caption{\it }
\end{subfigure}
\begin{subfigure}{0.45\textwidth}
\includegraphics[width=\textwidth]{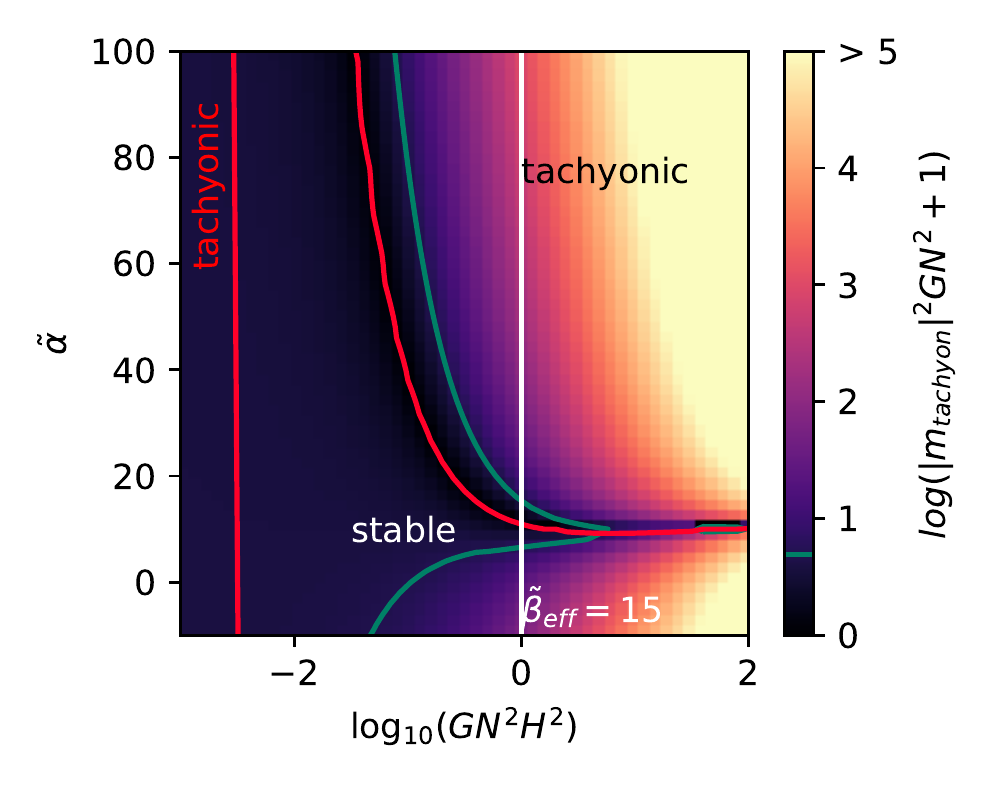}
\caption{\it }
\end{subfigure}
\caption{\it $|$Mass$|^2$ (\protect\ref{dS5}) of the de Sitter spin-2 lightest tachyonic pole in units of the species scale (\protect\ref{sp-a}), plotted in the $(\tilde{\a}, GN^2H^2)$ plane. The red line separates the tachyonic region from the non-tachyonic region obtained from Figure \protect\ref{dS_nu}. The green curve corresponds to the species scale $GN^2|m|^2 =1$. The vertical white line separates curvatures that are below the species scale (on the left) from curvatures which are above (on the right). In the two first panels, the mass of the tachyonic pole is always larger than the species scale. The two bottom panels show that a larger $\tilde{\b}_\text{eff}$ allows for some tachyonic poles under the species scale but only for small curvatures or large $\tilde{\a}$. In the last panel, the small curvature region is entirely below the species cutoff.}
\label{dS_tachyonmass}
\end{figure}

Figure \ref{dS_tachyonmass} shows the mass squared (\ref{dS5}) of the lightest spin-2 tachyonic pole in units of the species scale (\ref{sp-a}). The red curves obtained from the previous Figure \ref{dS_nu} delimit the tachyonic regions in the plane $(\tilde{\a},GN^2H^2)$. Whereas the green curve corresponds to the species scale $GN^2|m_2|^2=1$. The darker regions which are delimited by the green curve are then below the species cutoff. Each panel of Figure \ref{dS_tachyonmass} corresponds to a different value of $\tbe$. Negative values, such as $\tbe = -3$ plotted in panel (a) contain a large tachyonic region, but the tachyon is always above the cutoff. When $\tbe$ is increased, the non-tachyonic region becomes larger. The small areas in panel (b) where the pole is below the cutoff are included in the non-tachyonic regions. Therefore, the tachyon is always above the cutoff in panel (b) too. For larger values of $\tbe$, such as $\tbe = 10$ in panel (c), we finally observe some overlap between the tachyonic and light (below the cutoff) regions. It means that around $\tbe \approx 10$ and above, de Sitter can contain a tachyon which lies below the effective cutoff of the species scale. Increasing $\tbe$ even more, such as in panel (d), the light tachyonic regions increase in size, the small curvature tachyon is then always below the cutoff, whereas the larger curvatures necessitate a large $\tilde{\a}$ to get a light tachyon.

It seems from Figure \ref{dS_tachyonmass}, that around $\tbe \approx 10$ and above, the small curvature region goes below the cutoff. This can be understood from inserting equation (\ref{sH9}) into $|m_2|^2 \approx GN^2H^2|\n|^2 \leq 1$, which would correspond to the relation between $\tbe$, $\tilde{\a}$ and the curvature such that the complex pole (with $\text{Re}(\n) = 3/2$) is below the cutoff. For small curvatures ($GN^2H^2<< 1$), this relation is
\be
\tbe \geq \tbe^\text{species} \equiv \sqrt{16 \pi^2 -1} - \log(4\pi) + {1\over 2}- 2\g_E \approx 9.34.
\label{sH14}
\ee
If the inequality (\ref{sH14}) holds, the small curvature region lies below the species cutoff. This value agrees with panel (d) of Figure \ref{dS_tachyonmass}, where we observe that small curvatures are below the cutoff. However, in panel (c), $\tbe = 10 >9.34$ so small curvatures should lie below the cutoff. However, we observe that this is not the case. The value obtained in (\ref{sH14}) does not only rely on a small curvature approximation but also on the validity of the ansatz (\ref{sH9}) for large $a$, which converges rather slowly with first corrections given in (\ref{sH10}). In particular, for $\tbe = \tbe^\text{species} \approx 9.34$ and $GN^2H^2=10^{-3}$, the $log(log(-a))/log(-a)$ correction term in (\ref{sH10}) is approximatively equal to $0.2$. This error propagates into the value obtained for $\tbe^\text{species}$, which could explain why the left of the panel (b) disagrees with (\ref{sH14}). Moreover, the correction term in (\ref{sH10}), does not go to zero when $GN^2H^2\rightarrow 0$ but converges to a finite value around $0.2$ for $\tbe = \tbe^\text{species}$. The only way to make this error vanish is the $\tbe \rightarrow +\infty$ limit.

\begin{figure}[ht]
\centering
\begin{subfigure}{0.45\textwidth}
\includegraphics[width=\textwidth]{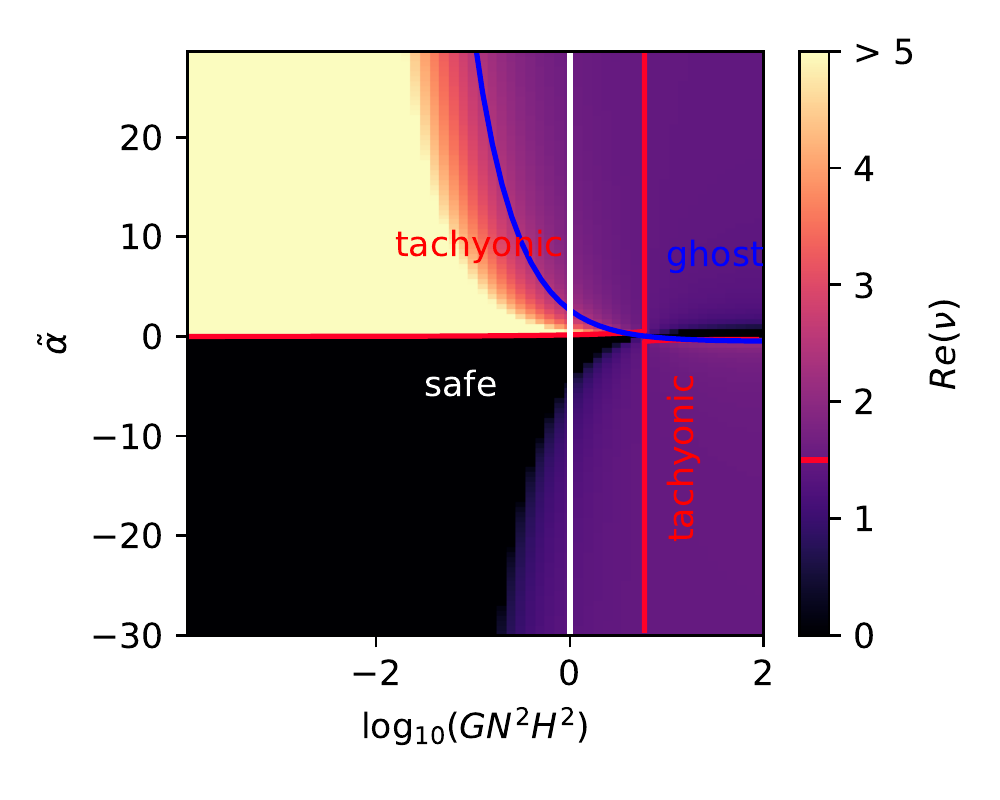}
\caption{\it }
\end{subfigure}
\begin{subfigure}{0.45\textwidth}
\includegraphics[width=\textwidth]{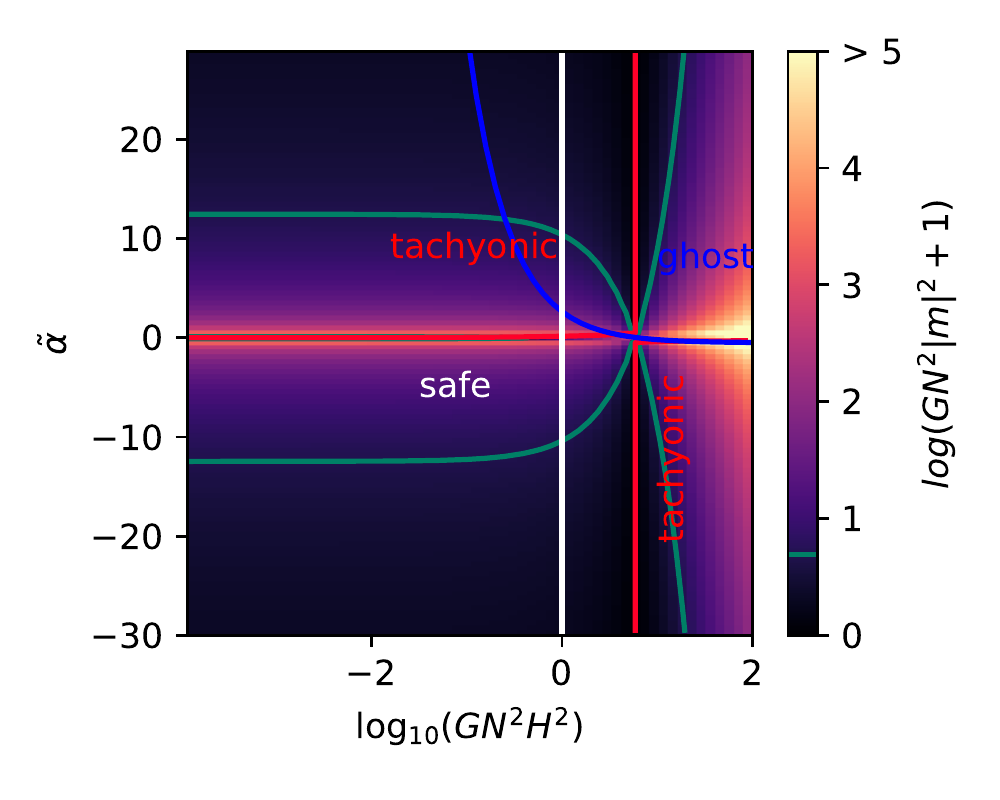}
\caption{\it }
\end{subfigure}
\caption{\it Regions of instabilities for the spin-0 sector in the de Sitter case. The colour code on the left panel indicates the real part of the solution $\n$ of the scalar sector plotted in the ($\tilde{\a}$, $GN^2H^2$) plane. Tachyonic regions (\protect\ref{psi_tachyonic}) are delimited by red lines. They correspond to regions where $\text{Re}(\n)>3/2$. The blue line delimits the region given by (\protect\ref{2pt Psi c}) where the scalar solution becomes a ghost. A white vertical line separates curvatures which are above (on the right) and below (on the left) the species scale (\protect\ref{sp-a}).
The green curve corresponds to the species scale $GN^2|m|^2 =1$
On the right panel, we compare the mass of the scalar solution with the cutoff of the theory, given by the species scale. For most of the ghost regions, the ghost is very massive compared to the cutoff. For $\a=0$, the kinetic term of the scalar mode vanishes. The tachyon is also heavier than the species cutoff, except in the top left region for small curvatures and large $\tilde{\a}$.}
\label{dS_scalar}
\end{figure}

Figure  \ref{dS_scalar} shows the analysis of the instabilities in the scalar sector. Plotting this figure does not require a numerical approach, since the scalar propagator (\ref{2pt Psi b-ii}) is directly written as a pole for the Laplacian operator $\Box$. Formulae (\ref{2pt Psi c}) and (\ref{psi10}) are used to plot the criterion for tachyonic and ghost-like instabilities respectively. In this case (unlike for the tensor) we can display  both ghost-like and tachyonic instabilities on the same figure because there is only one  pole in the scalar case (\ref{2pt Psi b-ii}).  On the left subfigure, we plot the real part of $\n$ as a function of the different parameters (in the scalar sector, the equation of motion does not depend on $\tilde{\b}_\text{eff}$). Tachyonic regions are delimited by red lines and ghost-like regions by  blue lines. On the right subfigure, we plot the effective mass of the scalar mode in units of the species scale $(G N^2)^{-1/2}$, see equation (\ref{sp-a}).

From figure \ref{dS_scalar} we see that the scalar ghost is below the species scale (\ref{sp-a}) for large enough values of $\tilde{\a}$.
For reasonable values of $GN^2H^2$ (below or comparable to the species scale), this ghost is also a tachyon. Therefore, we focus on tachyonic stability for the scalar mode in the following.

\begin{figure}[ht]
\centering
\begin{subfigure}{.4\textwidth}
\includegraphics[width= \textwidth]{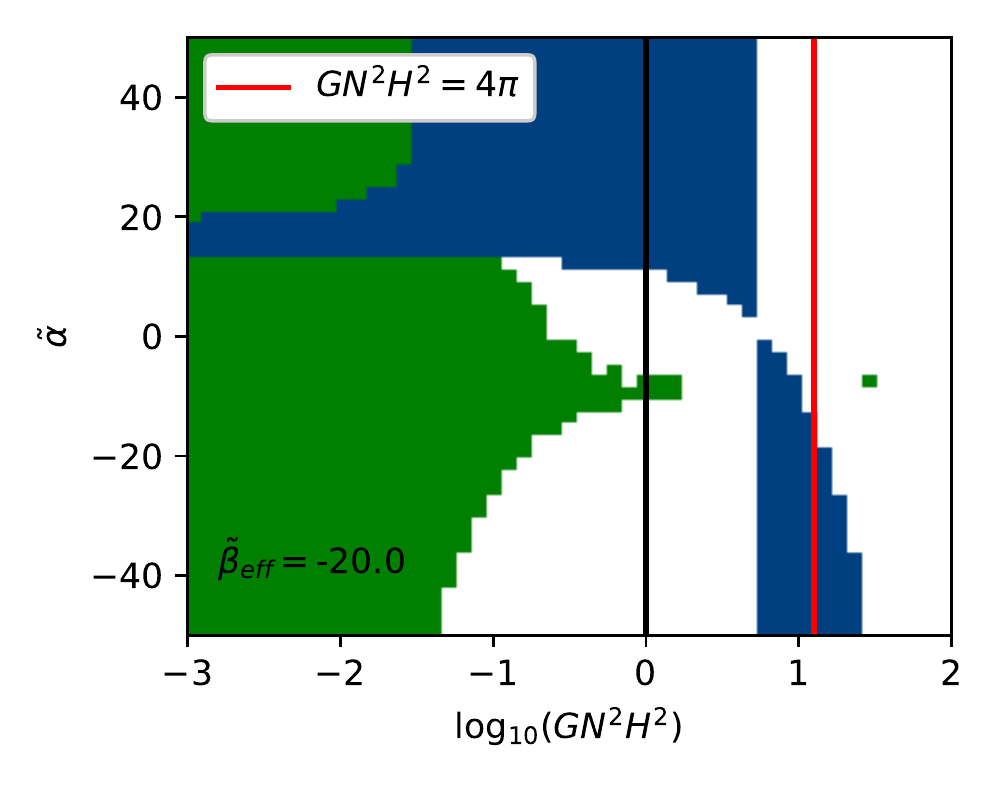}
\caption{\it $ \tbe = -20$}
\end{subfigure}
\begin{subfigure}{.4\textwidth}
\includegraphics[width=\textwidth]{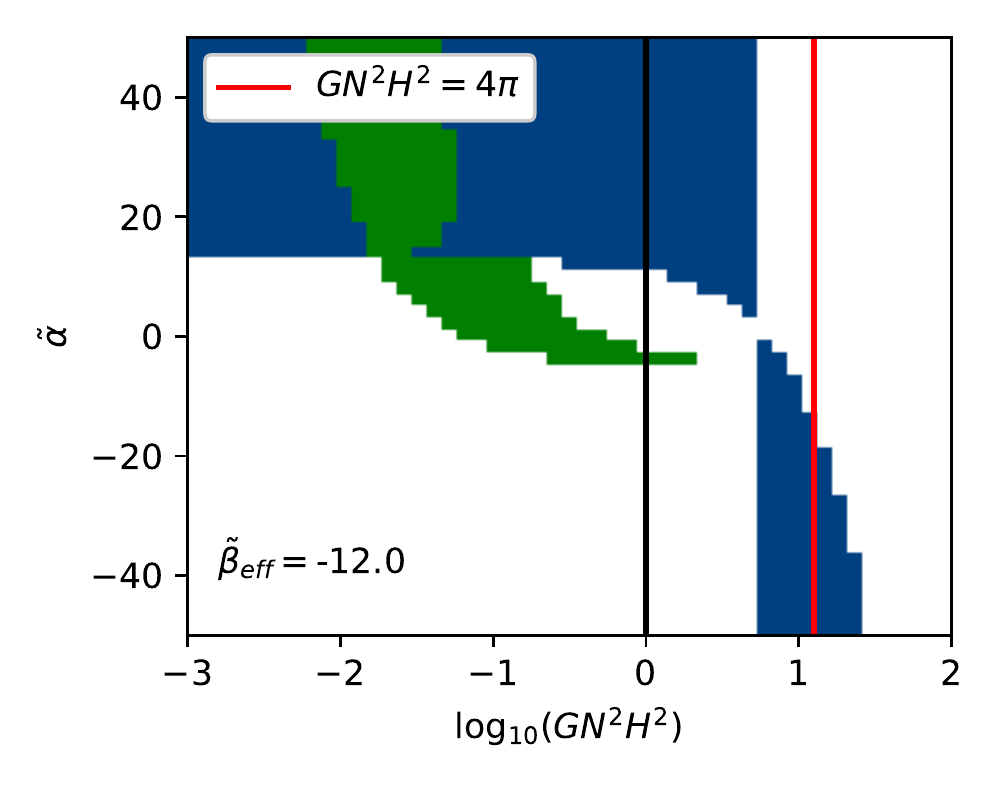}
\caption{\it $\tbe = -12$}
\end{subfigure}
\begin{subfigure}{.4\textwidth}
\includegraphics[width= \textwidth]{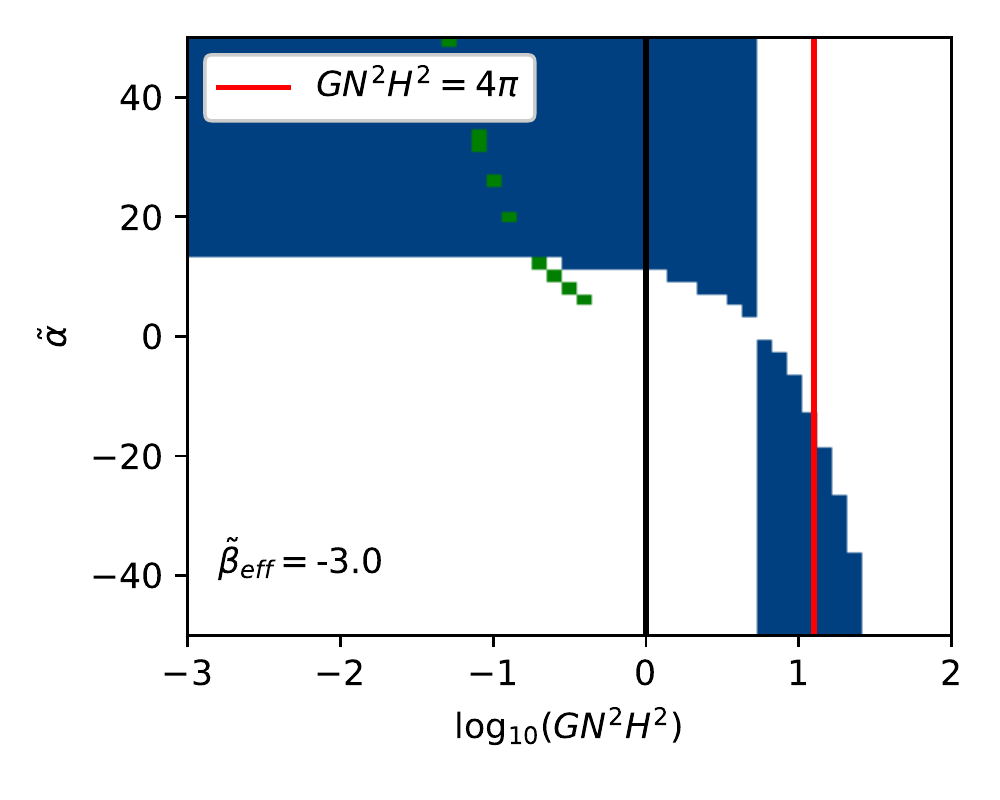}
\caption{\it $\tbe= -3$}
\end{subfigure}
\begin{subfigure}{.4\textwidth}
\includegraphics[width=\textwidth]{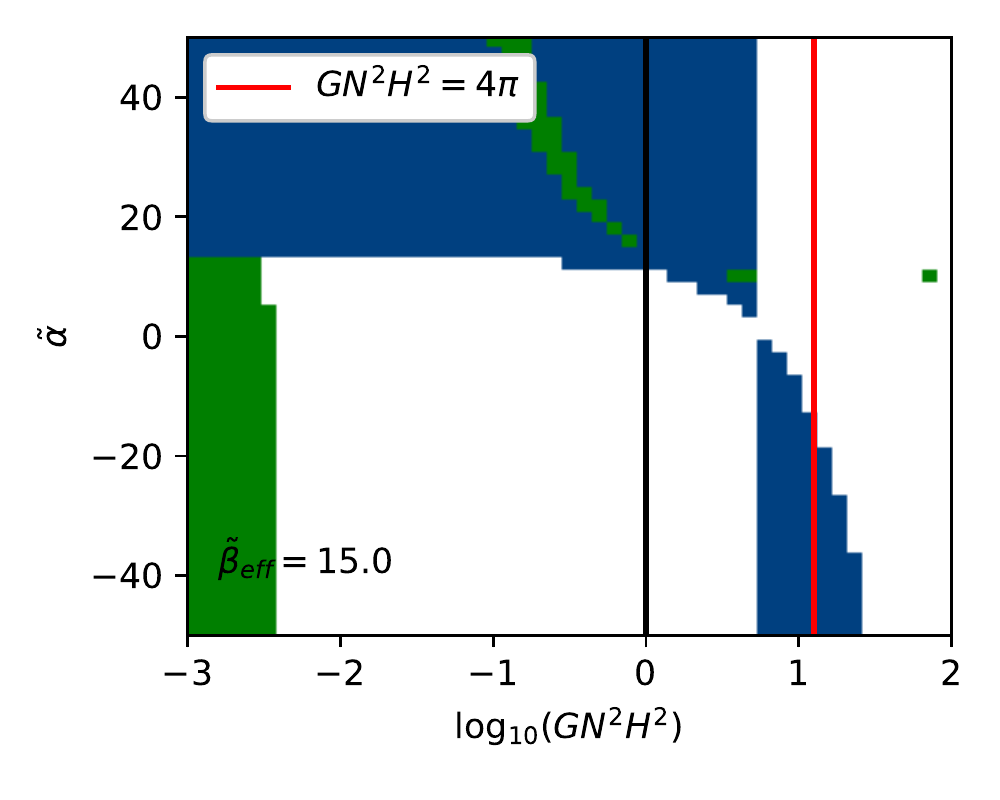}
\caption{\it $\tbe = 15$}
\end{subfigure}
\caption{\it Regions of parameter space in de Sitter, for several values of $\tbe$, showing whether the strength of the tensor tachyonic instability is larger or smaller than the strength of the scalar tachyonic instability. In green regions, the tensor tachyon instability dominates. In blue regions, the scalar tachyon instability dominates. In white regions, there are no tachyonic instabilities (if there is one, its mass is above the cutoff). In this figure, we only consider tachyonic poles which are below the species cutoff. The vertical black line separates curvatures which are below (left) and above (right) the species cutoff. The red line in the plots corresponds to the value of the curvature chosen in \protect\cite{VilenkinStaro}, which can be obtained by setting the renormalized cosmological constant $\Lambda$ to zero, as it was done in (\protect\ref{scalar7}). At this fixed curvature, the scalar instability dominates over the tensorial instability for negative $\tilde{\a}$. The last panel shows the region which is tachyonic for large values of $\tbe$ and small curvatures (see (d) of Figure \ref{dS_tachyonmass}). Increasing $\tbe$ will make this region disappear from the selected window because it will move to even smaller curvatures.}
\label{fig:scalarvstensor}
\end{figure}

In Figure \ref{fig:scalarvstensor} we compare the ``strength'' of the tensor tachyonic instability with that of the scalar tachyonic instability. By strength, we mean the inverse time scale associated with the instability, defined by  $\G$ in (\ref{ts2}). The decay rate $\G$ of the scalar sector is given in (\ref{Gamma-scalar}), while it is computed numerically for the tensor sector in Figure \ref{dS_nu}. For  fixed $\tilde{\b}_\text{eff}$,   the regions in the $(\a,GN^2H^2)$ plane where the  tensor  instability is stronger than the  scalar one are  coloured in green;  in blue regions, the scalar tachyon instability is stronger;  in white regions, there are no tachyonic instabilities.

It is interesting to compare our  results with those obtained by Vilenkin in \cite{VilenkinStaro} for the (original) Starobinsky model. In  this work, the renormalized cosmological constant was chosen to vanish, i.e.  $\Lambda = 0$ in our equation (\ref{scalar7}). This choice corresponds to the red vertical line in the four subfigures of Figure \ref{fig:scalarvstensor}. In \cite{VilenkinStaro},  the value of $\tilde{\b}_\text{eff}$ was irrelevant as this work concerned only the scalar mode. Vilenkin  found that the scalar mode was unstable for large negative $\tilde{\a}$. According to our results in Figure \ref{fig:scalarvstensor}, in his case, the scalar instability was indeed the strongest instability for small values of $\tilde{\b}_\text{eff}$. However, our results extend also to other regions. We observe that for small values of $\tilde{\b}_\text{eff}$ and large and negative values of $\tilde{\a}$, for sufficiently small $GN^2H^2$, the tensor tachyonic instability dominates over the scalar one.
There are other regions in which the spin-2 instability is the strongest. Moreover, there are smaller regions (in white) which are tachyon-stable. These regions grow in size as $\tilde{\b}_\text{eff}$ becomes large and positive, and shrink as $\tilde{\b}_\text{eff}$ becomes large and negative.

\begin{figure}[ht]
\centering
\begin{subfigure}{0.45\textwidth}
\includegraphics[width=\textwidth]{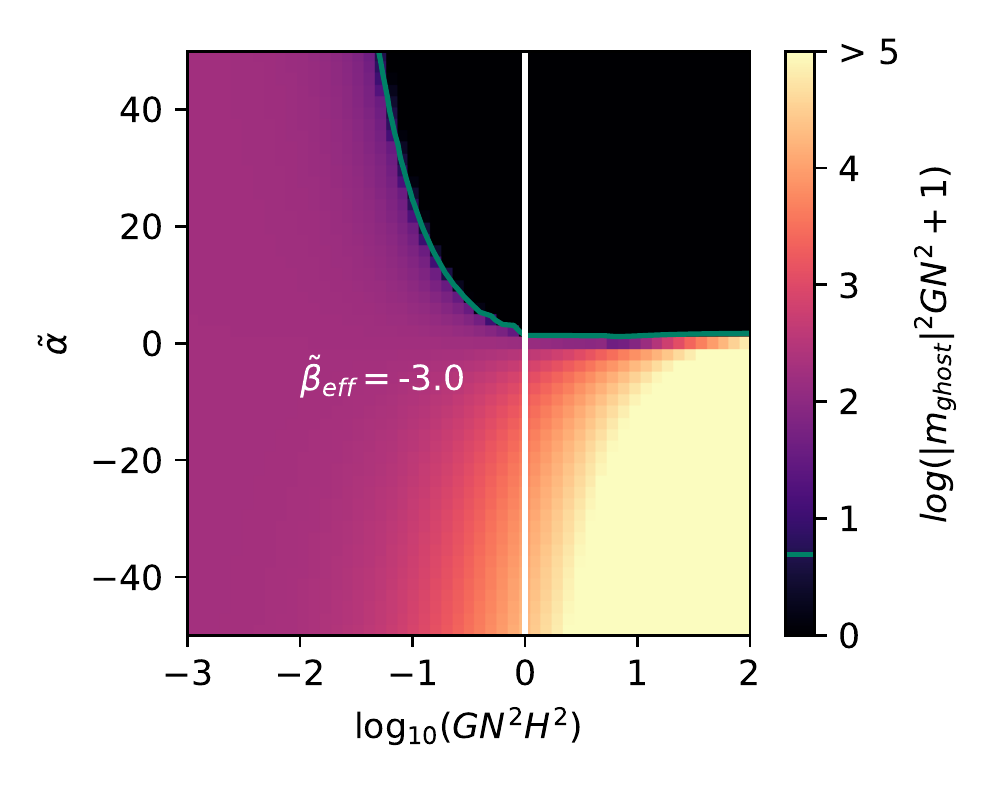}
\caption{\it }
\end{subfigure}
\begin{subfigure}{0.45\textwidth}
\includegraphics[width=\textwidth]{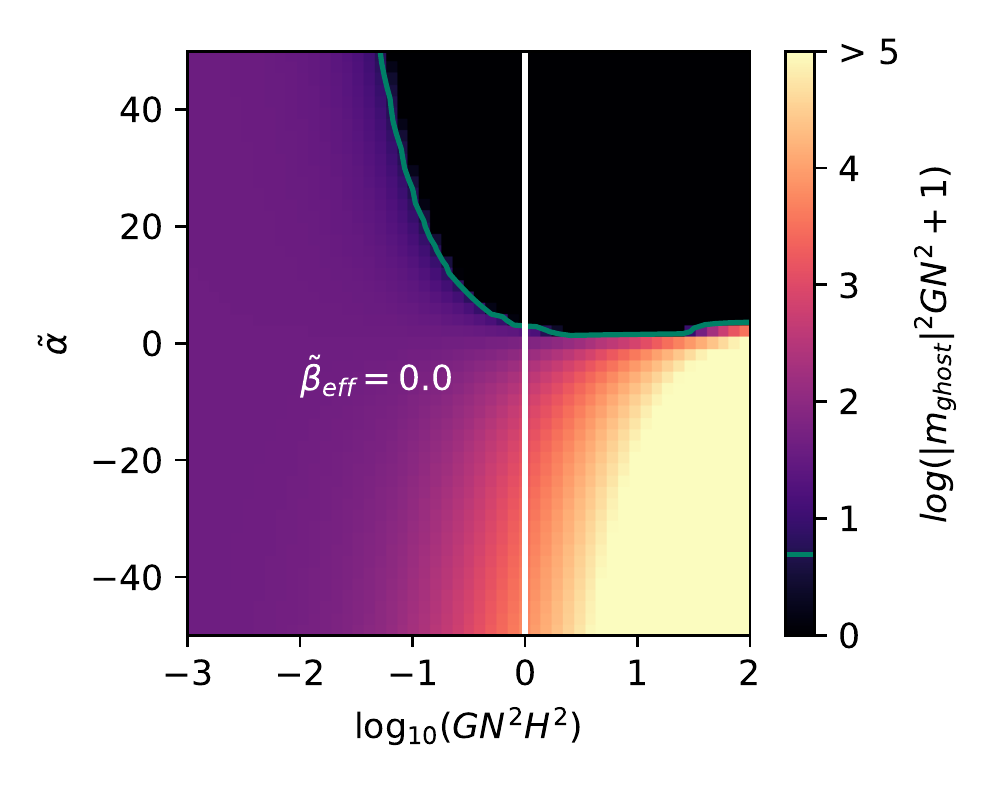}
\caption{\it }
\end{subfigure}
\begin{subfigure}{0.45\textwidth}
\includegraphics[width=\textwidth]{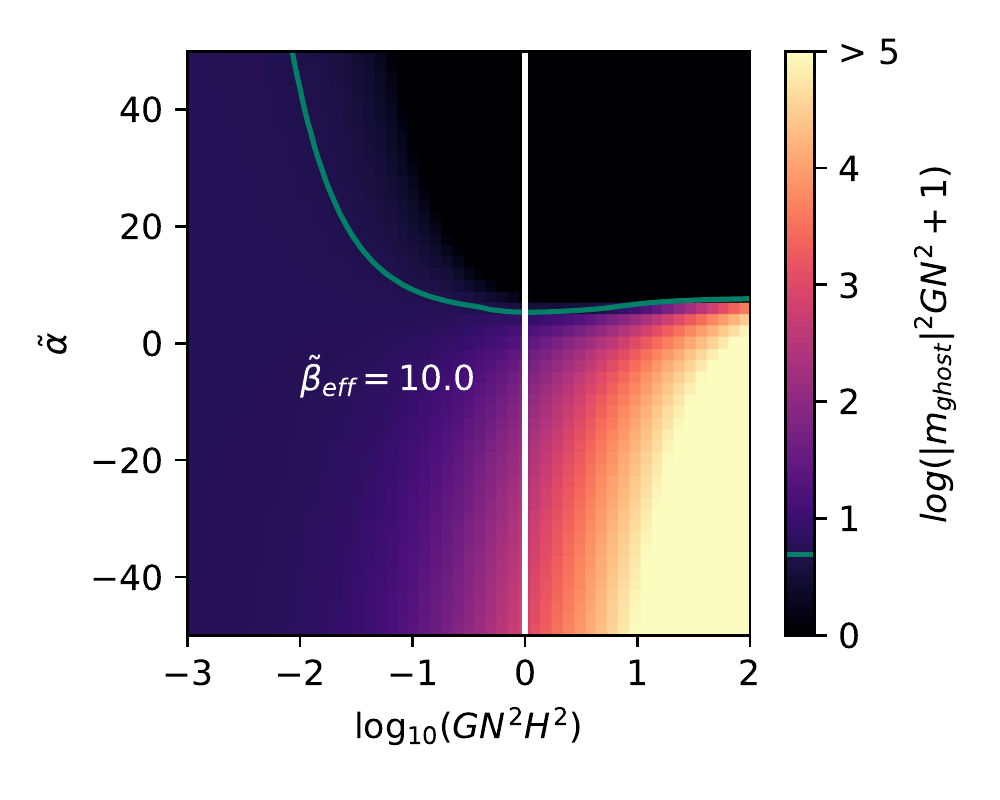}
\caption{\it }
\end{subfigure}
\begin{subfigure}{0.45\textwidth}
\includegraphics[width=\textwidth]{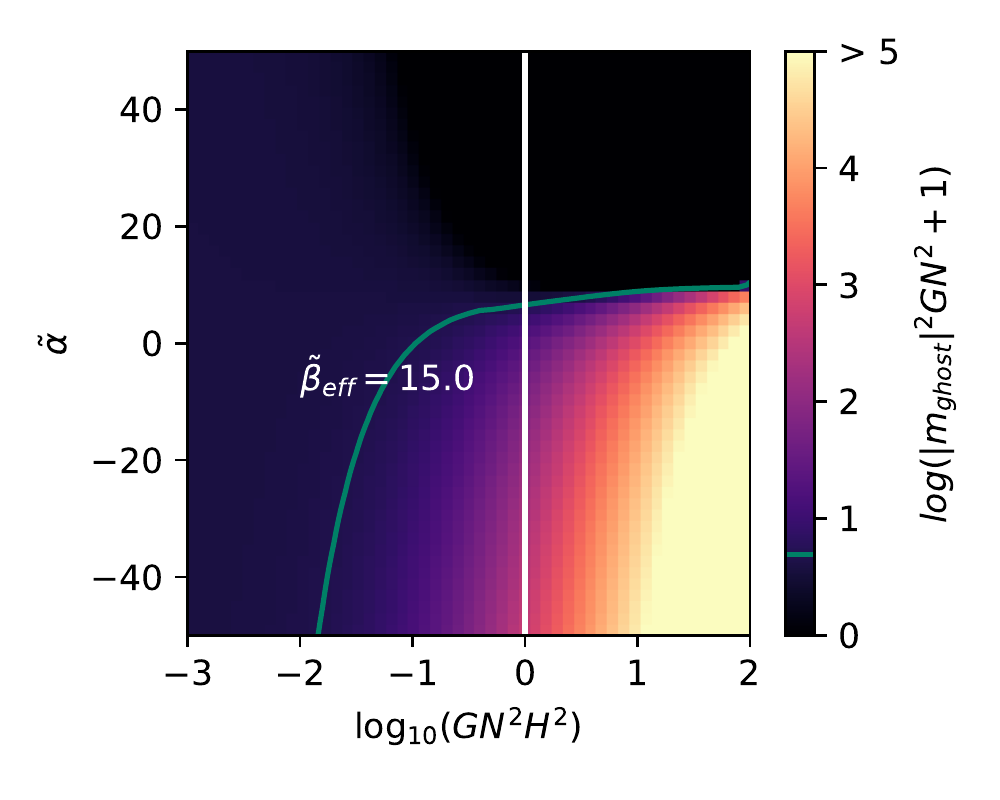}
\caption{\it }
\end{subfigure}
\caption{\it This figure indicates the modulus of the mass squared defined in (\protect\ref{dS5}), for the ghost pole of the spin-2 propagator (\protect\ref{dS17}) in units of the species cutoff defined in (\protect\ref{sp-a}), in the de Sitter case. The parameter space is spanned by the dimensionless curvature on the horizontal axis and the $R^2$ parameter $\tilde{\a}$ on the vertical axis. The vertical white line separates curvatures that are above and below the species cutoff. The green line is the boundary of the region where the ghost mass is below the cutoff, which corresponds to darker areas.}
\label{dS_ghost}
\end{figure}

As we have mentioned, in the regions denoted ``stable'' in Figure \ref{dS_nu}, all tensor modes are non-tachyonic. However, even in those regions, as we have seen in subsection \ref{dS analysis}, there is always one ghost pole\footnote{Except for some fine-tuned values of the parameters for which two poles merge to form a double pole.}. Its residue can be positive or  complex. This was already seen in the two examples given by Figures \ref{dS_H0.01_alpha0} and \ref{dS_Hpiover4_alpha10}.

One important question  is whether  the ghost pole is above or below the UV cutoff  scale (\ref{sp-a}): if the ghost mass is below the cutoff, then it must be regarded as a true instability of the low energy effective theory.  The results of this analysis are shown in Figure \ref{dS_ghost},  in which the colour code represents the mass squared for the spin-2 ghost in units of the species cutoff (\ref{sp-a}). The green line separates the regions where the ghost mass is below the cutoff (darker colours) from the regions where it is above (lighter colours).
In this figure, we observe that the ghost becomes lighter and lighter as $\tbe$ becomes large and positive, as well as when the parameter $GN^2H^2$ decreases.
In conclusion, de Sitter space-time is generally ghost-unstable for large and positive $\tbe$, whereas generic values of $\tbe$ are ghost-unstable only in the top-right corner of Figure \ref{dS_ghost}, corresponding to large curvatures and/or large $\tilde{\a}$.

\section{Poles of the AdS spin-two propagator and stability}
\label{AdS poles}
We now turn to the negative curvature case, and repeat the same analysis as in the previous for AdS.  First, we study two paradigmatic regions of parameters. Then, we provide analytical approximations to understand these two examples. Finally, we study numerically the stability of the system of gravity plus holographic CFT in AdS.

A new feature we find in this case is the presence of an infinite tower of stable solutions which only exists in AdS-slicing. These solutions are found near the poles of the stress-tensor two-point function, which in AdS appear in an infinite discrete set.

\subsection{Results for two typical sets of parameters}
\label{AdS_examples}

In this subsection, we focus on two examples (with small and large curvature, respectively),  solve the spectral equation (\ref{as10})  for tensor modes numerically and follow the evolution of the solutions as we change the parameters. In the following, when it is not specified,  it will be understood that we are using the single-boundary condition (\ref{as1a}), which leads to the inverse propagator $\mathcal{F}_{(-)}$ given in (\ref{as12}).

Recall that in the negative curvature case, a tachyon corresponds to a pole on the imaginary axis, see section \ref{AdS prop} and in particular equation (\ref{AdS4}).

The results of our first example are shown in Figure \ref{AdS_H0.01_alpha0}.  In this case, we choose a  small value of $G N^2 \chi^2$ (i.e. the AdS curvature in species-scale units) and $\tilde{\a} =0$. As we can observe from figure \ref{AdS_H0.01_alpha0},  large and negative values of $\tilde{\b}_\text{eff}$ always display two tachyons lying on the imaginary axis, with opposite signs for their residues: snapshot (b) shows that the lightest tachyon is also a ghost whereas the heavy tachyon is not. These two tachyon-unstable solutions merge for a value of $\tilde{\b}_\text{eff}$ close to $-4.18$,  snapshot (c).

For larger values of $\tilde{\b}_\text{eff}$, there are no imaginary solutions and the theory is tachyon-stable.  Following snapshots from (d) up to (g), a complex solution moves closer and closer to the real axis when $\tilde{\b}_\text{eff}$ is increased. This solution merges with the lightest stable pole (close to $\n=3/2$) and forms a double pole in snapshot (h). If we continue to increase $\tilde{\b}_\text{eff}$, two single poles appear. First, a pole stays close to the massless $\n=3/2$, with a negative residue. A second pole moves towards $\n=1/2$ with a positive residue.
The cloud of stable poles denoted by green points on the real axis is an infinite series of poles lying close to every half-integer: these half-integers are not poles, but zeros of the propagator, corresponding to poles of the stress-tensor two-point function, for which the inverse propagator $\mathcal{F}_{(-)}$ defined in (\ref{as12}), diverges. These can be traced back to poles in the harmonic number $\mathcal{H}$ appearing in $Q_\text{(-)}$ given in (\ref{as111}).

The mass of every pole in the snapshots of Figure \ref{AdS_H0.01_alpha0} (except the infinite series of safe poles on the real axis) is plotted in Figure \ref{AdS_mass_alpha0_chiem2}. By plotting the masses in units of the species scale, this figure allows us to see directly which pole is above or below the cutoff given by $GN^2|m|^2 = 1$. As a result, for generic values of $\tbe$, the tachyon, the ghost and the complex ghost are all above the cutoff. Whereas large values of $|\tbe|$ have a ghost-like pole lying below the cutoff, as is also the case for the pure gravity massive pole shown in dashed lines. This figure shows that the two tachyons (the ghost shown in red and the ghost-free in green) merge to form a pair of complex conjugate poles (non-tachyonic). This merging happens at masses that are much above the species cutoff.

\begin{figure}[ht]
\centering
\includegraphics[width=0.9\textwidth]{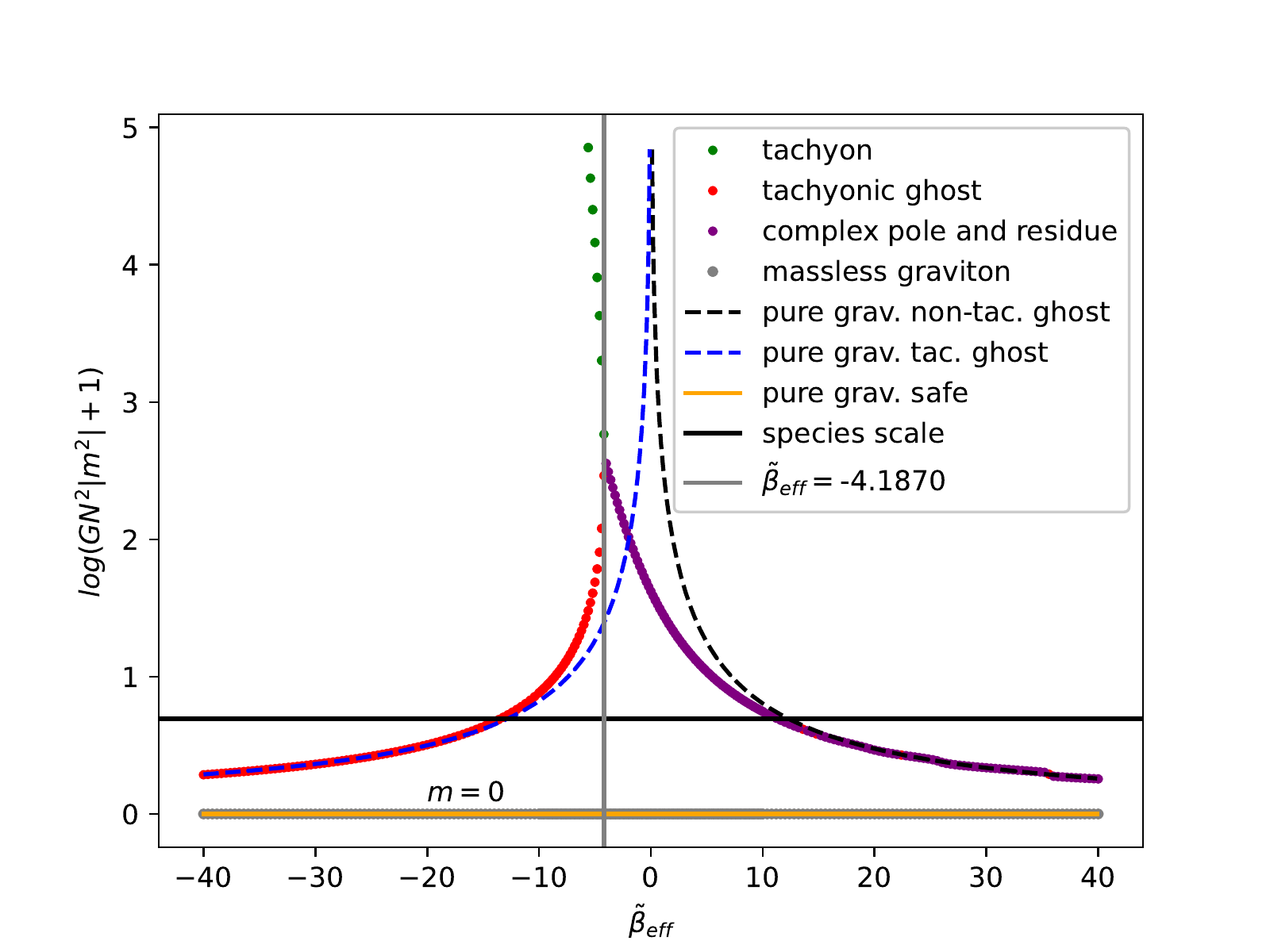}
\caption{\it
In this plot, obtained with the same parameters as in Figure \protect\ref{AdS_H0.01_alpha0}, we show the modulus of the mass of the spin-2 tachyonic poles in AdS, defined in (\protect\ref{AdS3}), in units of the species scale (\protect\ref{sp-a}), as a function of $\tilde{\b}_\text{eff}$. Red and green markers correspond respectively to the light ghost-like tachyon and the heavy non-ghostly tachyon.
Purple markers correspond to the complex conjugate pair of poles, which also have a complex residue.
The species scale is shown by a horizontal solid black line.
Black and blue curves correspond to the pure gravity modes with $\beta = \tbe /\pi$: the blue line is tachyonic and the black line is non-tachyonic, both are non-ghostly.
The vertical grey line is the value of $\tilde{\b}_\text{eff}$ at which the two tachyonic poles merge and move off the imaginary axis as $\tbe$ is further increased. This merging is displayed in snapshot (c) of Figure \ref{AdS_H0.01_alpha0}.}
\label{AdS_mass_alpha0_chiem2}
\end{figure}

Our second example is displayed in Figure \ref{AdS_chipiover4_am10}. In this case, we take the AdS curvature to be of the order of the species scale, specifically $G N^2\chi^2 = \pi/4$. We observe a very different approach to stability as  $\tilde{\b}_\text{eff}$ is increased: we have a single tachyon on the imaginary axis, which we observe entering the region shown in snapshot (d), and moves from large imaginary values to small imaginary values until it reaches the origin $\n=0$ in snapshot (g). As we further increase  $\tilde{\b}_\text{eff}$, the mode becomes stable because the pole moves off the imaginary axes. The tachyon-safe solution converges to $\n = 1/2$ when $\tilde{\b}_\text{eff}$ is increased to high positive values.

It is important to remark that a pole with negative residue (corresponding to a ghost) is present for all values of $\tilde{\b}_\text{eff}$. For large and negative $\tilde{\b}_\text{eff}$, it is close to $\n=1/2$,  whereas for large and positive values of $\tilde{\b}_\text{eff}$, it approaches $\n=3/2$. This can be understood using equation (\ref{N02a}) which is valid for asymptotically large values of $\tilde{\b}_\text{eff}$ since it was derived for $N=0$. One can observe that two poles are present in this formula. For large and negative $\tilde{\b}_\text{eff}$, the massless $\n=3/2$ poles are healthy residue and the $\n=1/2$ is a ghost. This is verified in snapshot (a) where the lightest pole is a ghost, and the first pole after $\n=3/2$ is safe. For large and positive $\tilde{\b}_\text{eff}$ the sign of their residue switch. This is observed in the last snapshot where the $\n=1/2$ pole is safe whereas the massless $\n=3/2$ is a ghost.\\

\begin{figure}[ht]
\centering
\includegraphics[width=0.9\textwidth]{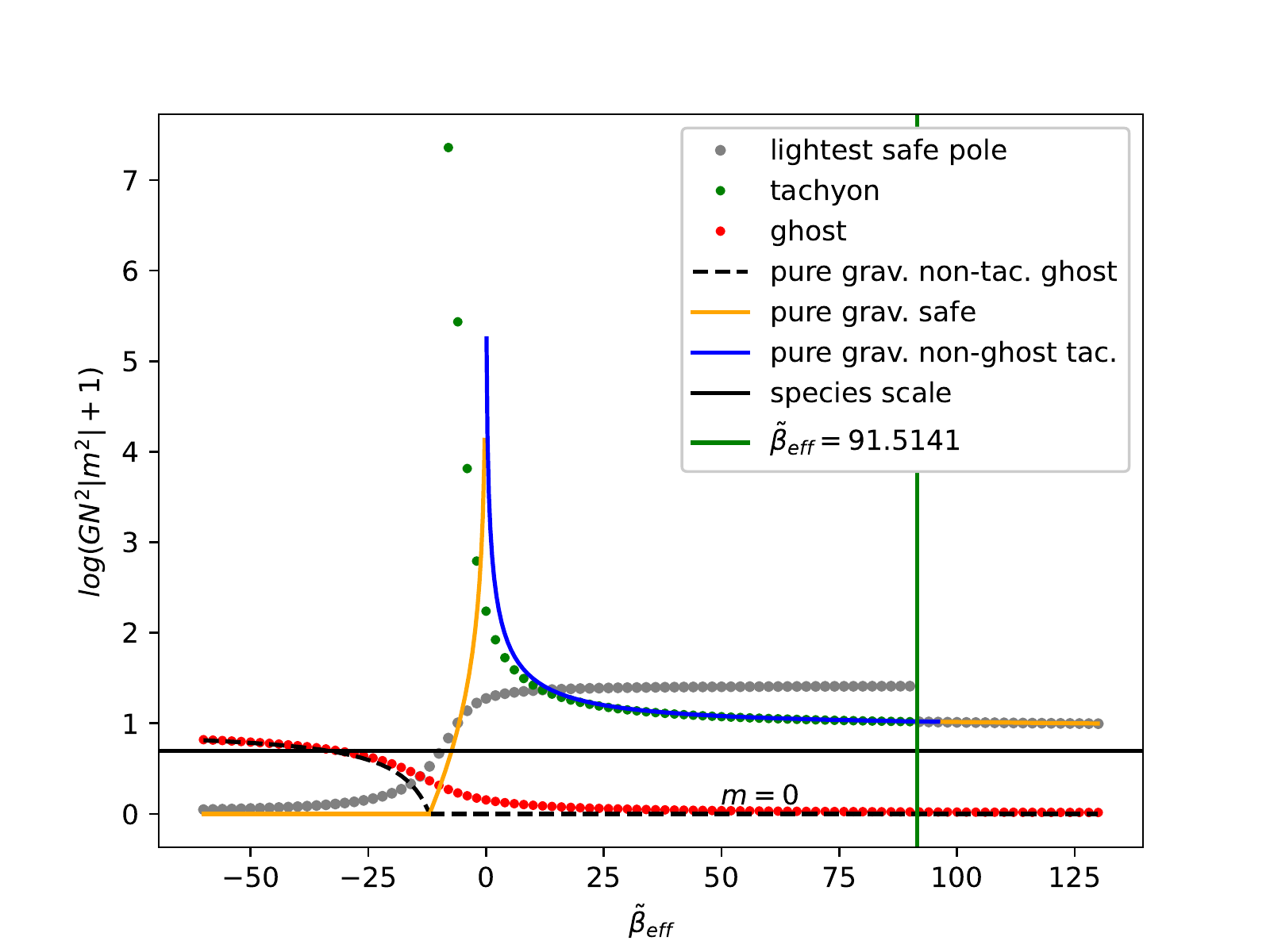}
\caption{\it
In this plot, obtained with the same parameters as in Figure \protect\ref{AdS_chipiover4_am10}, we show the modulus of the mass of the spin-2 tachyonic poles in AdS, defined in (\protect\ref{AdS3}), in units of the species scale (\protect\ref{sp-a}), as a function of $\tilde{\b}_\text{eff}$. Grey markers show the mass of the lightest safe mode. Red and green markers correspond respectively to the light non-tachyonic ghost and the heavy non-ghostly tachyon.
The species scale is shown by a horizontal solid black line.
The black, blue and orange curves correspond to the pure gravity modes with $\beta = \tbe /\pi$: the blue line is tachyonic, the black dashed line is the non-tachyonic ghost, and the orange line is safe.
For pure gravity curves, the vertical axis is $\log(G^{1\over 2}|m|^2 + 1)$ while the horizontal axis is $\pi\beta$.
The vertical green line is the value of $\tilde{\b}_\text{eff}$ at which the tachyonic pole forms a double zero at $\n=0$ and becomes real (non-ghostly) as $\tbe$ is further increased. This double zero is displayed in snapshot (g) of Figure \protect\ref{AdS_chipiover4_am10}, where the transition between tachyonic and non-tachyonic happens. For larger values of $\tbe$, the pole becomes safe. In pure gravity, the tachyonic pole becomes stable at $\pi\b = 96$, which is slightly different from the value with the CFT given by the vertical green line, where the tachyonic pole stops being tachyonic, crosses the origin at $\n=0$ and becomes the lightest safe pole. Larger values of $\tbe$ coincide with pure gravity.
}
\label{AdS_mass_am10_chipiover4}
\end{figure}

Figure \ref{AdS_mass_am10_chipiover4} shows the mass of the poles that are found in Figure \ref{AdS_chipiover4_am10}. The mass of the poles which are obtained in the presence of the CFT is shown in coloured dots, while the poles which were already present in pure gravity are shown using coloured curves.
In this plot, we observe that all poles are above the species cutoff except a safe (non-tachyonic, non-ghost) pole for large and negative values of $\tbe$ (as in pure gravity) which is massless in the $\tbe \rightarrow -\infty$ limit. This pole becomes massive when $\tbe$ is increased, while the ghost (in red) moves below the species cutoff. For large and positive values of $\tbe$, only the ghost is below the cutoff. Its mass goes to zero in the $\tbe\rightarrow +\infty$ limit.
This analysis holds in the pure gravity case. Indeed, the only regime where the CFT plays a role is for generic values of $\tbe$, where most of the poles have a mass higher than the species cutoff and must therefore be discarded from the EFT analysis.

Qualitatively, the two cases displayed in Figures \ref{AdS_H0.01_alpha0} and \ref{AdS_chipiover4_am10} represent quite different behaviours when $\tbe$ is varied. We have chosen to discuss only these two cases because they turn out to be paradigmatic of what happens in the whole parameter space. More cases are shown in appendix \ref{moresnapshots}, where each point in $(GN^2\chi^2,\a)$ space corresponds to a set of snapshots. Each example turns out to have the same behaviour as either one of the two cases already discussed.
The ArXiv webpage of this paper contains ancillary files, including animated gifs. Each snapshot of these gifs corresponds to a different value of $\tbe$ for fixed  $(GN^2\chi^2,\tilde{\a})$.

In the following subsection, we present an analytical argument which explains why these two cases are typical of what happens more generally, and how we can distinguish between these two types of behaviour.

\subsection{Analytic results for tensor tachyonic modes in AdS in the large-$|\n|$ regime}
\label{sec:AdS-analytics}

Tachyonic modes in AdS correspond to purely imaginary $\n$. In this section, we provide an analytical approximation which allows us to better understand the two examples given in the previous subsection. This approximation is the limit for large $|\nu|$, in which case the pole mass is much larger than the AdS curvature scale (but it may still lie below the species cut-off).

Interestingly, as was the case in dS, we shall observe that the approximation for large $|\n|$ turns out to be still valid for poles with values of $|\n|$ which may even be close to 1. We shall observe that the two cases studied in the previous subsection are paradigmatic: the whole parameter space may be separated into two regions, in which the behaviour of the poles is similar to the one shown in  Figures \ref{AdS_H0.01_alpha0}   and \ref{AdS_chipiover4_am10}, respectively. \\

In the large-$|\n|$ regime, we can use Stirling's approximation (\ref{sH1}) to replace the harmonic number $\mathcal{H}$ with a simpler log function.
The validity of the large-$|\n|$ approximation will be checked afterwards, by comparing the analytical predictions with numerical evaluations of the inverse propagator.

Using Stirling formula (\ref{sH1}), equation (\ref{as10}) becomes
\bea
Q_{(-)}(\n) && =  1 + 2\left({\pi \over GN^2\chi^2} +\tilde{\a}\right) -  {\n^2\over 2} \left[{\tilde{\b}_\text{eff}} - {1\over 2} + \log\left(GN^2\chi^2\right) +  \right. \nonumber \\
&& \left. + \log (\n) + \log(-\n)- 2\g_E + \mathcal{O}(|\n|^{-1}) \right].
\label{fAdS1}
\eea
One can already see the difference with the de Sitter case (\ref{sH3}): the log is split in a sum which is symmetric in $\n\leftrightarrow -\n$.
If we write $\n =|\n|e^{i \text{arg}(\n)}$, then
\bea
Q_\text{(-)}(\n) && =  1 + 2\left({\pi \over GN^2\chi^2} +\tilde{\a}\right) - {\n^2\over 2} \left[{\tilde{\b}_\text{eff}} - {1\over 2} + \log\left(|\n|^2GN^2\chi^2\right) +  \right. \nonumber \\   &&  \left.  +2i\text{arg}(\n) -  i\pi \text{sign}(\text{arg}(\n))- 2\g_E + \mathcal{O}(|\n|^{-1})\right].
\label{fAdS2}
\eea
 We now apply (\ref{fAdS2}) it to tachyonic modes,
\be
\n = ix, \quad \text{$x$ real.}
\ee
The complex phases cancel in (\ref{fAdS2}). Then, the equation of motion (\ref{as10}) can be written as
\be
X\log X = - a,
\label{sc1}
\ee
where
\be
X \equiv {x^2 GN^2\chi^2}\exp \left\{{\tilde{\b}_\text{eff}} - {1\over 2} + 2\g_E \right\},
\label{sc2}
\ee
\be
a \equiv 2GN^2\chi^2 \left[2\left({\pi\over GN^2\chi^2} + \tilde{\a}\right) + 1\right]\exp\left\{{\tilde{\b}_\text{eff}} - {1\over 2} + 2\g_E\right\}.
\label{sc3}
\ee
This is similar to the corresponding equations we found in the de Sitter case, (\ref{sH4}), (\ref{sH5}) and (\ref{sH6}), up to a few sign flips.

The large $x$ regime described by equations (\ref{sc1}), (\ref{sc2}) and (\ref{sc3}) is valid both for the asymmetric condition in the bulk (\ref{as1a}), and for the symmetric boundary condition (\ref{sym1}). To see why this regime is independent of boundary conditions, we observe that the difference between $Q_{(-)}$ (\ref{as111}) for single-boundary condition (\ref{as1a}) and $Q_\text{sym}$ for the symmetric case (\ref{sym1}) vanishes exponentially with $x$.

Similarly to the large $|\n|$ regime for de Sitter space-time, to discuss equation (\ref{sc1}) we  distinguish three cases :
\begin{itemize}
\item If $0<a<e^{-1}$ there are two tachyonic solutions ($x$ real).
\item If $a>e^{-1}$, no real solution for $x$, the theory is then tachyonic-stable.
Therefore, large $x$ solutions are always stable when $a>e^{-1}$. This is equivalent to
\be
\tilde{\b}_\text{eff} > \tilde{\b}_\text{eff}^\text{merge} \equiv -{1\over 2} - 2\g_E - \log \left(2GN^2 \chi^2\right) - \log\left[2\left({\pi\over GN^2\chi^2} + \tilde{\a}\right) + 1 \right],
 \label{sc4}
\ee
which is similar to (\ref{Ln1})  in de Sitter. However, the physics of the poles is different:  in de Sitter, $\beta_\text{merge}$ does not correspond to a  transition from instability to stability, but rather to the merging of two real solutions which then move off the real axis. In AdS on the other hand,  (\ref{sc4}) indicates the critical value at which solutions leave the imaginary axis, and therefore it represents a  stability condition, valid for large $x$ and $a>0$.
This condition is valid in the example of Figure \ref{AdS_H0.01_alpha0}, where the transition between tachyonic and non-tachyonic behaviour occurs around the value of $\tilde{\b}_\text{eff}$ chosen for the snapshot (c) for which $\tbe = \tbe^\text{merge}$ (\ref{sc4}).

\item If $a<0$, there is a single tachyonic solution $x$, whose value decreases when $a$ is increased. In terms of the parameters, the condition $a<0$ is equivalent to
\be
{\pi\over GN^2\chi^2} < -\tilde{\a} - {1 \over 2}.
\label{sc5}
\ee
The condition (\ref{sc5}) for having only one solution $\n = ix$ is analogous to equation (\ref{Ln2}) in de Sitter, which in that case was the condition for having a single $\n$ on the real axis. However, unlike in dS,  in the AdS case,  this equation also gives information about the number of tachyons. If  (\ref{sc5})   is verified, we have one tachyon, and if it is not, then we have either none or two tachyons depending on the value of $\tilde{\b}_\text{eff}$ through inequality (\ref{sc4}).

Interestingly the condition   (\ref{sc5}), like the analogous inequality for de Sitter (\ref{Ln2}), turns out to be the same as the condition (\ref{2pt Psi c}) for the scalar to be a ghost\footnote{We do not know whether there is a deep reason for this.}.

If we select the parameters such that (\ref{sc5}) is verified, and if the tachyonic pole stays on the imaginary axis even for small values of $x$ where the approximation above breaks down, then this pole should cross the origin $\n^2 = 0$ (and becomes stable) as $\tilde{\b}_\text{eff}$ is increased. This corresponds to the usual BF bound, which is respected for positive $\n^2$. Increasing $\tilde{\b}_\text{eff}$ increases $a$, such that the unique solution for $x$ decreases. At some point, $x$ eventually crosses the origin at $x=0$ and the pole becomes non-tachyonic. Therefore, the tachyon-stability condition for negative $a$ corresponds to
\be
 \tilde{\b}_\text{eff} \geq \tilde{\b}_\text{eff}^\text{BF}\equiv {1\over 2} - \log\left(GN^2\chi^2\right) - 2\mathcal{H}(-1/2) - 8\left[1+ 2\left({\pi\over GN^2\chi^2} + \tilde{\a}\right)\right],
\label{sc6}
\ee
where $\tilde{\b}_\text{eff}^\text{BF}$ corresponds to the value for which we have $Q_{(-)}(0) = 0$.
The stability condition (\ref{sc6}) is different from (\ref{sc4}) because it applies to the $a<0$ case. In the large-$x$ approximation,  the tachyon stays on the imaginary axis as we vary $a$. If this statement continues to hold for small $x$ down to $x=0$ where this solution becomes non-tachyonic, then the formula (\ref{sc6}) would give an exact stability condition. This is indeed what happens in the example of Figure \ref{AdS_chipiover4_am10}: the formula (\ref{sc6}) describes exactly the transition and gives an accurate condition for the onset of the tachyonic instability, as we chose $\tilde{\b}_\text{eff} = \tbe^\text{BF}$ in snapshot (g) where the theory is at the transition from tachyon-unstable to tachyon-stable.
 \end{itemize}

 Large-$|\n|$ solutions exist as long as a term in $Q_\text{(-)}(\n)$ (\ref{as111}) (or $Q_\text{sym}(\n)$) (\ref{sym6}) is large and positive, which is the case for example when $\tilde{\b}_\text{eff}$ is large and negative or when $GN^2\chi^2$ is small.
In the small curvature regime (or in the large and negative $\tilde{\b}_\text{eff}$ regime), the cancellation in the spectral equation can be done using the $\n$ dependent terms $\mathcal{H}(-1/2\pm \n)$. This includes two types of solutions. First, large-$|\n|$ type of solutions were studied in this subsection. Second, $\n$ can be close to a pole of the harmonic number $\mathcal{H}$. All these poles are located on the real axis, for each half-integer. This second type of solution, which was not present in dS, will be studied in the next subsection. Before that, we first comment on the flat space limit of the AdS spin-2 propagator.

 \paragraph*{Flat space limit of the AdS spin-2 propagator}

In the limit of vanishing curvatures $GN^2\chi^2\rightarrow 0$, the curvature-dependent term of the propagator (\ref{as111} or \ref{sym6} depending on IR conditions) diverges, as it was the case in de Sitter. Indeed, the term ${\pi\over GN^2\chi^2}$ in $Q_\text{(-)}$ (\ref{as111}), must be cancelled by the harmonic numbers $\mathcal{H}(-1/2\pm\n)$. This can be done by taking $|\n|$ large as it was done for de Sitter \cite{Chesler}. However, in AdS, the bulk normalizable modes are present in the inverse-propagator in the form of poles of the harmonic number $\mathcal{H}(-1/2-\n)$.

In the flat space limiting procedure, we exclude real-valued $\n$ because they would go to non-tachyonic poles in flat space. To see why we first ask that the flat space limit should be taken such that the eigenvalues of both Laplacians (the AdS$_4$ Laplacian and the Minkowski Laplacian) match. This requires
\be
 \n^2 \chi^2 \underset{\chi\rightarrow 0}{\sim} -k^2.
\label{fAdS3}
\ee
We then directly observe that real-valued $\n$ corresponds to $k^2<0$, which was excluded from the flat space propagator in (\ref{n28a}). We can therefore ignore real-valued $\n$ and therefore avoid the poles of the harmonic number located on the real axis.

Inserting the large-$|\n|$ limit into the asymmetric 2-point function of AdS (\ref{as10}) or the symmetric one (\ref{sym5}) leads in both cases to
\be
\mathcal{F}_{(-)} \underset{\chi\rightarrow 0}{\rightarrow} {N^2 k^2\over 64\pi^2}\left\{{2\pi \over GN^2k^2} + \right.\nn
\ee
\be
\left. {k^2\over 2} \left[{\tilde{\b}_\text{eff}} - {1\over 2} + \log\left(|\n|^2GN^2\chi^2\right) +2i\text{arg}(\n) - i\pi \text{sign}(\text{arg}(\n))- 2\g_E\right]\right\}.
\label{fAdS4}
\ee
Comparing this with the flat space propagator (\ref{n28a}), we find that
\be
\mathcal{F}_{(-)}(\n) \overset{k^2\rightarrow -\n^2\chi^2}{\underset{\chi\rightarrow 0 }{\rightarrow}} \mathcal{F}_\text{flat}.
\label{fAdS5}
\ee
The terms involving the complex phase $\text{Arg}(\n)$ in (\ref{fAdS4}) coincide with the expression of the Minkowski propagator in this limit only if we require
\be
\chi \n \sim \text{sign}(\text{Im}(\n)) i k.
\label{fAdS6}
\ee
If $\n$ is imaginary, we recover the purely tachyonic modes where $k$ is real, and therefore $k^2>0$.
On the other hand, when $\n$ is real, this constraint is ill-defined because $\n^2>0$ corresponds to $k^2<0$. In Minkowski space, we have defined the propagator away from the real axis which contains all the healthy propagating modes.

\subsection{Infinite series of stable solutions}
\label{AdS massive series}
As one can observe in Figure \ref{AdS_chipiover4_am10}, there is an infinite set of massive solutions on the real axis. It is also remarkable that these poles are not displayed in every snapshot, and this is due to the lack of numerical precision when these poles are too close to a half-integer. There is a pole for every blue circle near each half-integer. In Figure \ref{AdS_H0.01_alpha0}, these poles can also be seen (albeit less clearly) for some regions on the real axis. These poles are present for every half-integer but most of them are too close to a pole at half-integers to be resolved by the numerics.

Numerically, one finds that the solutions on the real axis are near each half-integer $\n = n+1/2$, where  $\mathcal{H}(-\n-1/2)$ has poles. To understand this feature is then instructive to expand the harmonic-number function close to its poles:
\be
\mathcal{H}(\epsilon-n) = -{1\over \epsilon} + \mathcal{H}(n-1) + \mathcal{O}(\epsilon).
\label{poleH}
\ee
Using this expression  in equation (\ref{as10}), the result for $n>1$ is:
\be
\mathcal{F}_{(-)}(n+1/2+\epsilon) = {N^2 \chi^4\over 64\pi^2}\left[a_1 \epsilon^{-1} + a_0 +\mathcal{O}(\epsilon)\right],
\label{poleHa}
\ee
where
\be
a_1 = -{1\over 2}(n-1)n(n+1)(n+2),
\ee
\be
a_0 = a_1 \left\{ {2n + 1\over (n-1)(n+2)} - 2{n(n+1)}\left[2\left({\pi\over GN^2\chi^2} + \tilde{\a}\right) + {1\over 2} - n \right] + \right. \nonumber
\ee
\be
\left.-  {1\over 2} + {\tilde{\b}_\text{eff}} + \log\left(GN^2\chi^2\right) + 2\mathcal{H}(n)\right\}.
\label{poleH0}
\ee
In the vicinity of half integer $\n = n+1/2 + \epsilon$, with $\epsilon <<1$, solutions to the spectral equation  (\ref{as10}) are then given approximately by choosing $\epsilon$ small but finite and approximately given by
\be
\epsilon \simeq -{a_1\over a_0}.
\label{poleH1}
\ee
A zero of $\mathcal{F}_{(-)}(\n)$ is  then  found at
\be
\n_0 \equiv n +{1\over 2} + \epsilon.
\label{poleH1a}
\ee

Neglecting $\mathcal{O}(\epsilon)$ terms in (\ref{poleHa}), it is then possible to conclude from (\ref{poleH1}) that there is always a solution near a half-integer value of $\n$ as long as $a_1/a_0 \ll 1$. If $a_1/a_0$ turns out to be of order $1$ or larger, then (\ref{poleH1}) cannot be trusted. Since we have a formula both for $a_1$ and $a_0$ in (\ref{poleH0}), we can check the value for $\epsilon$ for every $n$. The value of $\epsilon(n)$ is plotted by the green curve in the left panels of Figure \ref{AdS_tower}. The green curves are continuous (not a discrete set of values) because $n$ is replaced by $\text{Re}(\n) - 1/2$ in this plot.
It turns out that most of the $\epsilon(n)$ are small. There are some values of $n$ however, for which $\epsilon(n)$ is large. These values of $n$ which correspond to a large $\epsilon$ are centred around a particular value for which $\epsilon^{-1} \sim 0$, where we observe a sign flip of $\epsilon$. This region of the real axis is displayed in Figure \ref{AdS_tower}.

\begin{figure}[ht]
\centering
\begin{subfigure}{.35\textwidth}
\includegraphics[width=\textwidth]{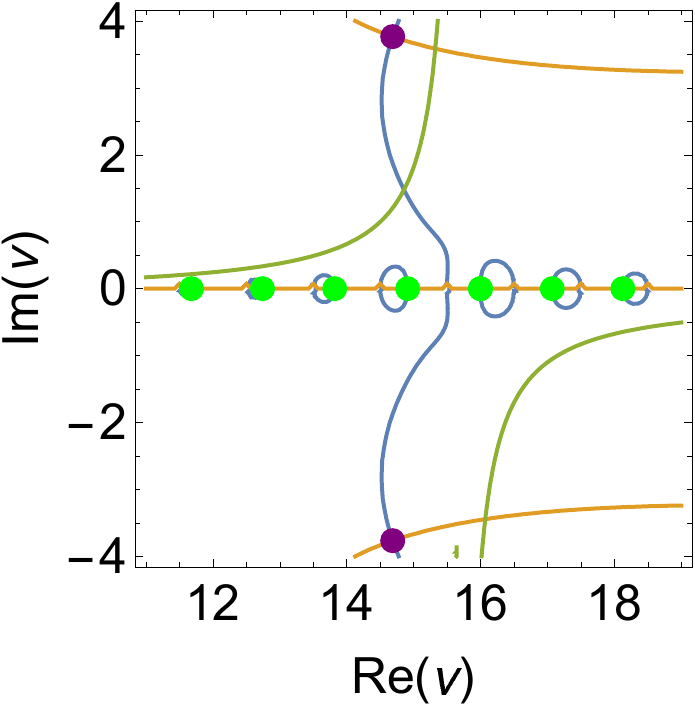}
\caption{\it ${\tilde{\b}_\text{eff}} = 3.4453$}
\end{subfigure}
\begin{subfigure}{.35\textwidth}
\includegraphics[width=\textwidth]{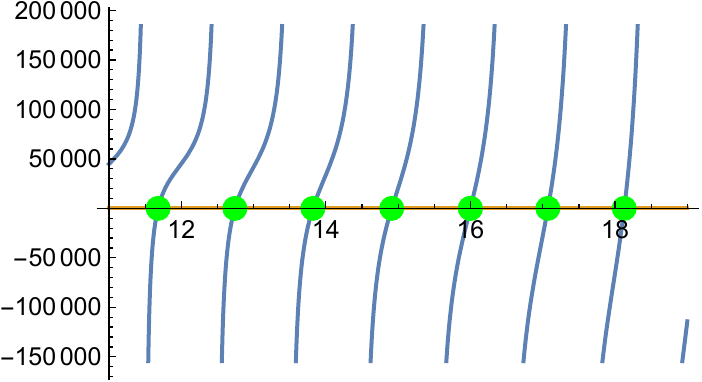}
\caption{\it ${\tilde{\b}_\text{eff}} = 3.4453$}
\end{subfigure}
\begin{subfigure}{.35\textwidth}
\includegraphics[width=\textwidth]{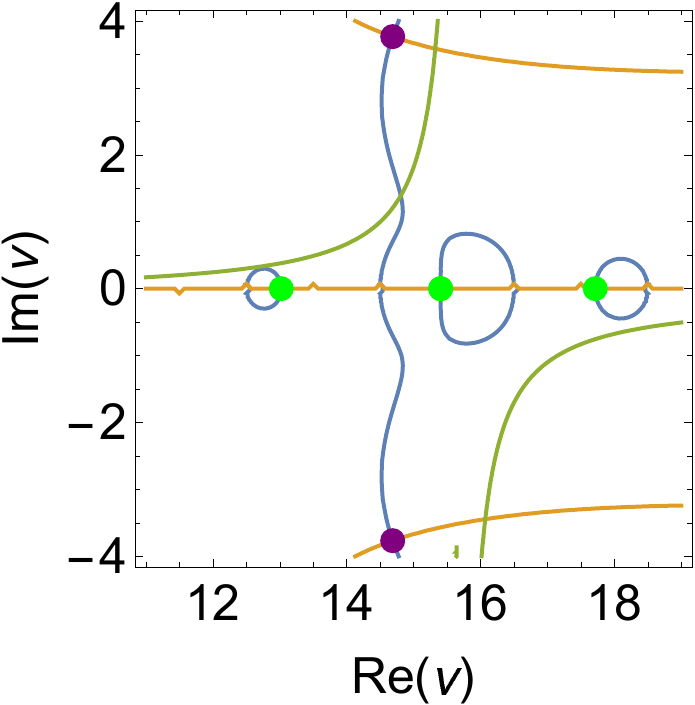}
\caption{\it ${\tilde{\b}_\text{eff}} = 3.3150$}
\end{subfigure}
\begin{subfigure}{.35\textwidth}
\includegraphics[width=\textwidth]{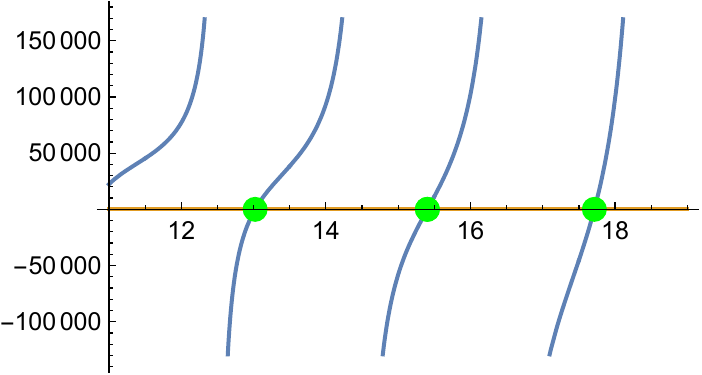}
\caption{\it ${\tilde{\b}_\text{eff}} = 3.3150$}
\end{subfigure}
\caption{\it Zeros of the AdS spin-2 inverse correlators $\mathcal{F}_{(-)}$ and $\mathcal{F}_\text{sym}$, for a small region in the complex plane of $\n$ (\protect\ref{AdS3}), where the $\epsilon$-approximation (\protect\ref{poleH}) is expected to break down. Parameters are chosen as $GN^2\chi^2 = 0.01$ and $\tilde{\a} = 0$. The left-hand side snapshots show zeros of the real part (blue) and zeros of the imaginary part (yellow) of $\mathcal{F}_{(-)}$ (top plot) and $\mathcal{F}_\text{sym}$ (bottom plot). The green curve shows the value of $\epsilon(n)$ given by (\protect\ref{poleH1}), it should give an estimate for the size of the blue circles. The approximation breaks down exactly at $n = 15$ since we have chosen $\tilde{\b}_\text{eff}$ to make the circle of the radius diverge exactly there. This is done using (\protect\ref{poleH1}) by solving $\epsilon^{-1}= 0$ for $n=15$. The right-hand side shows the real part of $\mathcal{F}_{(-)}$ which has a positive slope everywhere it crosses the real axis. Each intersection with the real axis is then a ghost-free pole of the tensor propagator.}
\label{AdS_tower}
\end{figure}

In Figure \ref{AdS_tower}, we observe a small part of the real axis centred to the point where $\epsilon^{-1} \sim 0$. The green lines in this figure show the expected value of $\epsilon(n)$ (\ref{poleH1}) which can be compared to the size of the blue circles. The radius of these circles roughly corresponds to the distance between a half-integer and the closest pole of the propagator. The actual pole of the propagator found numerically, is the intersection between the blue circles and the real axis, where green dots are placed. The green line predicts well the size of the blue circles everywhere, except where it is above 1, as one could have expected.
The place where $\epsilon$ is supposed to diverge according to (\ref{poleH1}) corresponds to the place in Figure \ref{AdS_tower} where the unique open blue curve crosses the real axis. The exact place where it crosses lies exactly at a half-integer, and this half-integer corresponds to a value of $n$ for which $\epsilon(n)$ is maximum.

It turns out that the breakdown of the small $\epsilon$ expansion, roughly at $\epsilon^{-1}\sim 0$, corresponds to the middle of the cloud of solutions in Figure \ref{AdS_H0.01_alpha0} where the open blue curve crosses the real axis and $\epsilon$ changes sign. This region of large values for $\epsilon$ is the same as in Figure \ref{AdS_H0.01_alpha0} where the numerics can find these zeros. Large values of $\epsilon$ make these blue circles large enough to be resolved by the numerics.

This analysis shows that there is always a solution near a half-integer $\n = n +1/2$. The small $\epsilon$ expansion is valid for every $n$, except in an interval where these solutions are not close enough ($\epsilon \sim 1$) to a half-integer. In this region where the approximation cannot be trusted, the numerics in Figure \ref{AdS_tower} confirm the existence of such solutions even if $\epsilon$ is not small.

We shall now investigate whether solutions corresponding to (\ref{poleH1a}) are ghost-like. For this purpose, we need to expand the inverse propagator $\mathcal{F}_{(-)}$ (\ref{as12}) close to a solution of the form (\ref{poleH1a}), such that
\be
 \n = n + 1/2 - {a_1\over a_0} + \varepsilon,
\label{poleH1b}
\ee
where $\varepsilon$ is a book-keeping parameter defined in order to expand $\mathcal{F}_{(-)}$ close to a given zero.
When $\varepsilon = 0$, we sit exactly at the zero of the inverse propagator found perturbatively in (\ref{poleHa}).
The expansion of $\mathcal{F}_{(-)}$ near the zero at $\varepsilon=0$ then reads
\be
\mathcal{F}_{(-)}\left(n+{1\over 2} - {a_1\over a_0}+\varepsilon\right) =  {N^2 \chi^4\over 64\pi^2}\left[-{a_0^2\over a_1}\varepsilon + \mathcal{O}(\varepsilon^2)\right].
\label{poleH2}
\ee
As a reminder of what was done in the dS case (\ref{dSpole2}), the residue of the pole of $\mathcal{F}_{(-)}^{-1}$ near $\n_0$ in the $\n^2$ plane is given by
\be
\text{res}[\mathcal{F}^{-1}_{(-)}](\n_0^2) = {\n_0\over \mathcal{F}_{(-)}(\n_0)}{1\over \n^2-\n_0^2}.
\label{poleH3}
\ee
Therefore, applying this formula to (\ref{poleH2}), we find that the residue of a pole lying close to a half-integer is given by
\be
\text{Res}[\mathcal{F}_{(-)}^{-1}\left(\left[n+1/2 - a_1/a_0\right]^2\right)] = - {64\pi^2\over N^2\chi^4}{a_1\over a_0^2}\left(n+ {1\over 2}- {a_1\over a_0}\right).
\label{poleH4}
\ee
Since $a_1<0$ the residue (\ref{poleH4}) is positive for the whole tower of massive particles close to half integers. They have the same sign as the massless graviton in AdS with pure gravity. The argument that the residue is positive near the real axis is verified numerically in Figure \ref{AdS_tower}. This Figure shows some poles of the tensor propagators $\mathcal{F}_{(-)}$ and $\mathcal{F}_\text{sym}$ on the real axis near the region where the $\epsilon$ expansion breaks down. This figure also confirms that $\epsilon$ changes sign where the open blue curve (not the circles) crosses the real axis. This crossing happens at the position of the half-integer for which we have $\epsilon^{-1}\sim 0$.

For the symmetric boundary condition, the propagator $\mathcal{F}_\text{sym}$ (\ref{sym5}) can also be expanded in $\epsilon$ as in (\ref{poleH1}) but with an additional term on the right-hand-side coming from the ${\pi\over \cos{\pi\n}}$ piece in (\ref{sym5}).
Using
\be
{\pi\over \cos{\pi\n}} =  {(-1)^n \over \epsilon} + \mathcal{O}(\epsilon),
\label{poleH2a}
\ee
we obtain new expressions for $a_0$ and $a_1$ defined in (\ref{poleHa}) for asymmetric boundary conditions. For symmetric boundary conditions, we define
\be
\mathcal{F}_\text{sym}(n+1/2+\epsilon) = {N^2 \chi^4\over 64\pi^2}\left[a_1^\text{sym} \epsilon^{-1} + a_0^\text{sym} +\mathcal{O}(\epsilon)\right].
\label{poleH2b}
\ee
In the case of $n$ odd,
\be
a_1^\text{sym} = 0,
\label{poleH3a}
\ee
which does not allow for a solution near odd half integers.
However, if $n$ is even, then
\be
a_1^\text{sym} = 2a_1,
\label{poleH5}
\ee
which allows for a solution near an even half-integer, but where $\epsilon$ is approximatively twice as big as in the asymmetric case (\ref{poleH1}).

As a consequence, we do not find a linear solution near $n+{1\over 2}$ if $n$ is odd for the symmetric boundary condition.
This perturbative result is confirmed numerically in the bottom part of Figure \ref{AdS_tower}, where only even integers present a blue circle, which is twice the size of the same circles in the asymmetric case (top panels of Figure \ref{AdS_tower}).

\subsection{Tachyons and ghosts in parameter space for the AdS case}

\begin{figure}[h!]
\centering
\includegraphics[width=\textwidth]{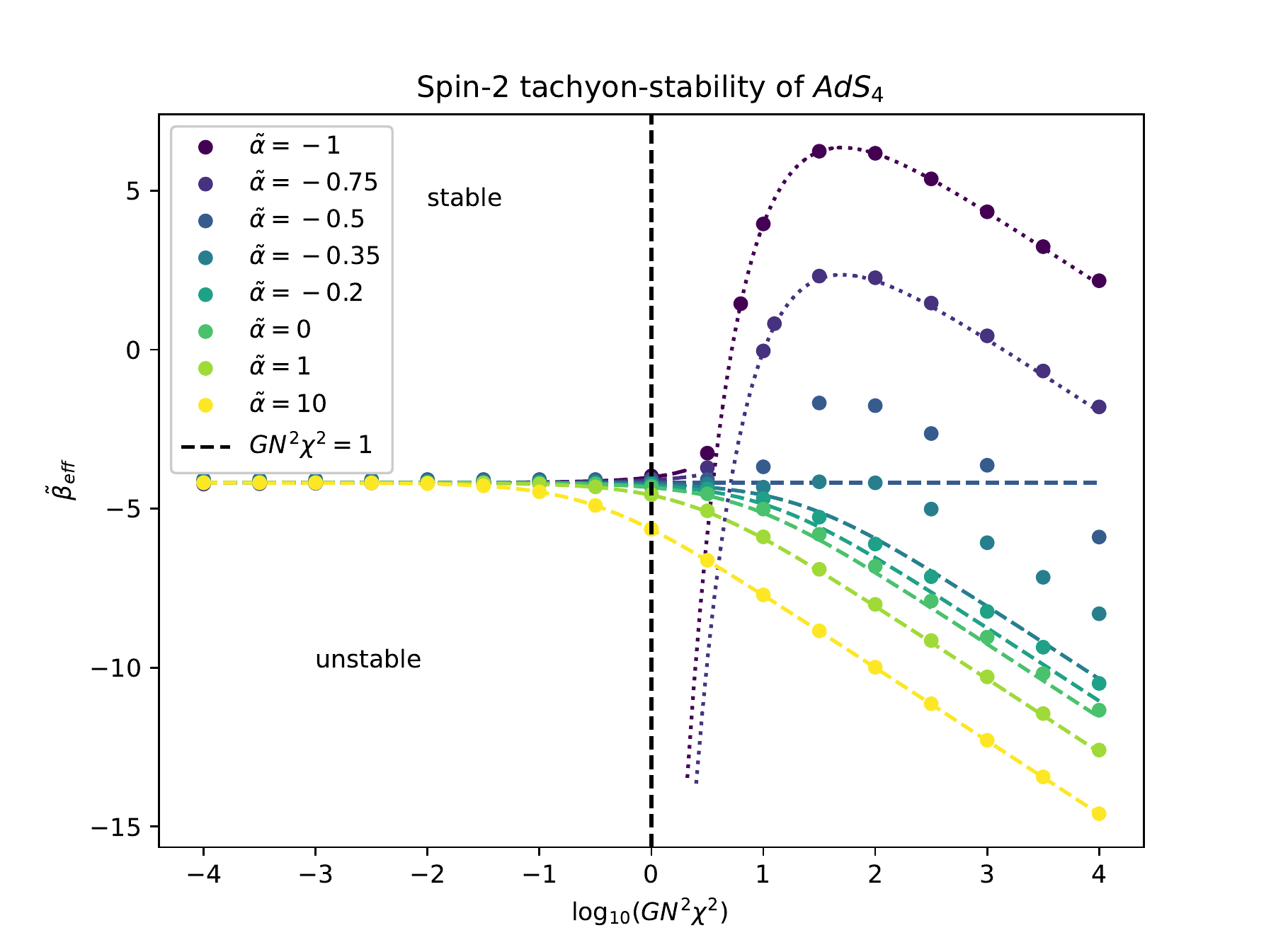}
\caption{\it Tachyonic regions of the spin-2 AdS perturbations. Tachyonic regions correspond to the existence of a solution $\n$ to the spectral equation (\protect\ref{as10}) such that $\text{Re}(\n)=0$ (\protect\ref{AdS4}). Several values of $\tilde{\a}$ are taken and stability is plotted in the space of ($\tilde{\b}_\text{eff}$, $GN^2\chi^2$). Dashed lines are the analytical predictions obtained from the large $|\n|$ and large $a$ approximation (\protect\ref{sc4}). Lines made of small squares are obtained assuming that the tachyonic pole stays on the imaginary axis and becomes stable at the origin (\protect\ref{sc6}). Large dots are numerical results. }
\label{fig:AdS}
\end{figure}

We first discuss the tensor sector.
The regions of parameter space where tachyonic tensor modes occur in  AdS can be read-off from   Figure \ref{fig:AdS}, which was obtained by solving the spectral equation numerically. This figure is the negative-curvature analogue of Figure \ref{fig:dS}.

Figure \ref{fig:AdS} shows the value of $\tbe$ at which the tachyonic pole becomes non-tachyonic, for a given set of parameters ($GN^2\chi^2$,$ \tilde{\a}$).
When $\tbe$ is above the curve shown in this figure, the theory is tachyon-free. As we have seen in the two typical examples in Figures \ref{AdS_H0.01_alpha0},\ref{AdS_chipiover4_am10}, the would-be tachyonic pole leaves the imaginary axis at a particular value of $\tbe$ and never goes back to the imaginary axis as $\tbe$ goes to $+\infty$. Therefore, the critical value of $\tbe$ shown in Figure \ref{fig:AdS} is the border in parameter space between tachyonic and non-tachyonic theories. The dashed coloured lines correspond to the large-$|\nu|$ analytical approximation obtained in section \ref{sec:AdS-analytics} for the case $a>0$ (\ref{sc4}), whereas the dotted lines correspond to the case $a<0$ (\ref{sc6}).

\begin{figure}[ht]
\centering
\includegraphics[width= \textwidth]{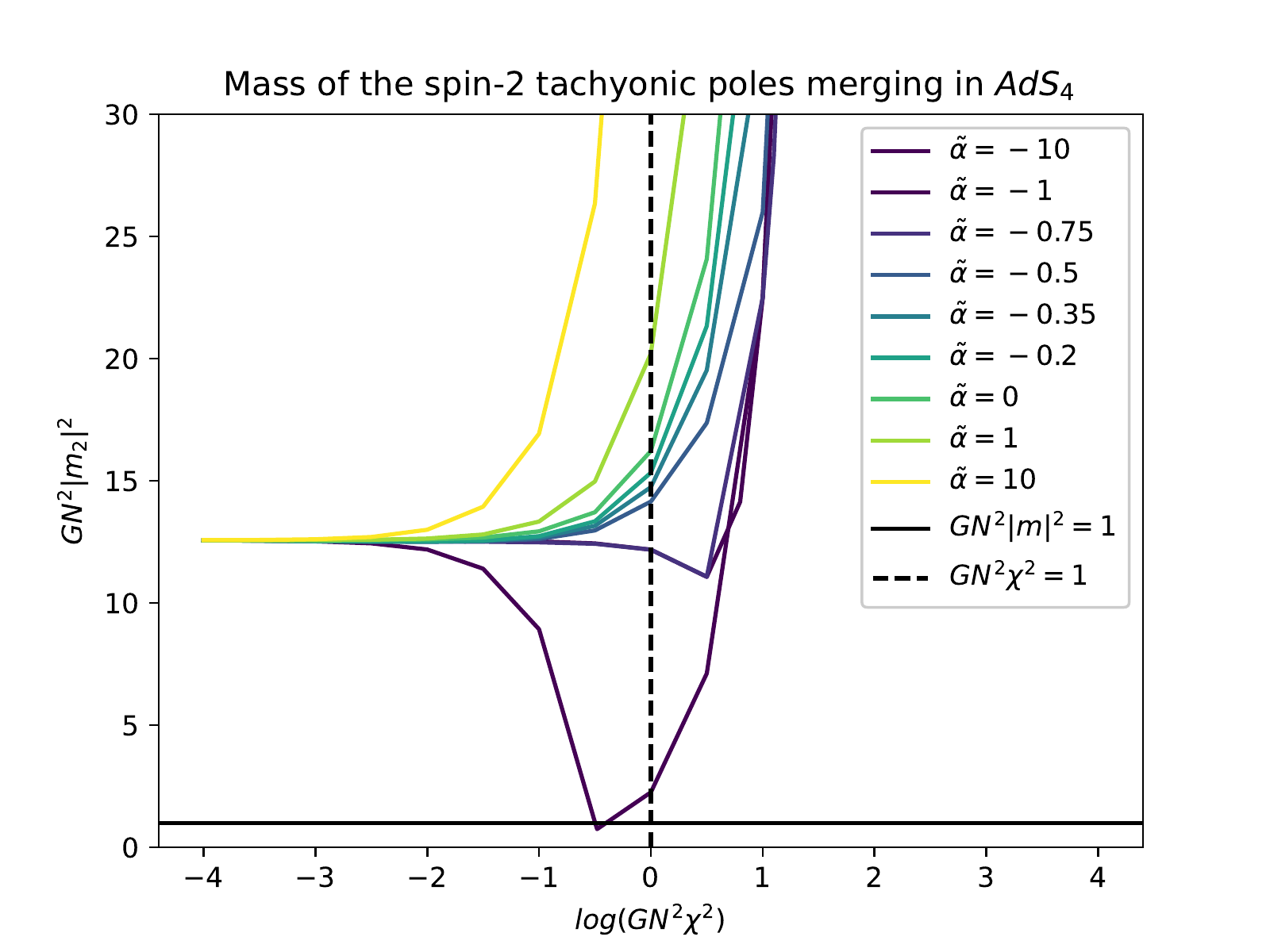}
\caption{\it Mass of the spin-2 tachyonic pole of AdS space-time for the value of $\tbe$ given by Figure \ref{fig:AdS}. This value of $\tbe$ corresponds to the merging of two tachyons on the imaginary axis $\text{Re}(\n) = 0$, creating a complex pole, which is non-tachyonic (Re$(\n)\neq 0$). The mass is plotted in units of the species scale (\ref{sp-a}).
 Each colored curve is a different choice of $\tilde{\a}$.}
\label{AdS_tacmass}
\end{figure}

For large $\tilde{\alpha}$ the interpolating curves are monotonic in the curvature, and as   $\tilde{\alpha}$ decreases they start displaying a maximum.
From the large-$|\n|$ approximation we expect there to be a critical curvature, given by equation (\ref{sc5}), above which $a$ is negative. This is where we decide to start the dotted lines. In the $a<0$ case, the large-$|\n|$ approximation suggests that there is only one single tachyon on the imaginary axis. We then make the further hypothesis that the transition from tachyon-instability to tachyon-stability occurs at the origin $\n=0$, where the large-$|\n|$ approximation cannot be valid. However, this hypothesis is verified numerically since the dotted lines agree perfectly with the numerics. An example of such transition was already shown in snapshot (g) of Figure \ref{AdS_chipiover4_am10}.

Figure \ref{AdS_tacmass} shows the mass of the spin-2 tachyonic pole of AdS corresponding to the circles of Figure \ref{fig:AdS}, at the value of $\tbe$ corresponding to the transition between tachyonic and non-tachyonic regimes. Therefore, the mass plotted in this figure corresponds to a tachyonic pole, lying on the imaginary axis, which is about to merge with another tachyon and leave the imaginary axis for larger values of $\tbe$. According to this figure, the transition between tachyonic and non-tachyonic regimes appears to happen always above the species scale, except for large and negative values of $\tilde{\a}$, for which the masses are below the species scale in a small interval of curvatures.

\begin{figure}[ht]
\centering
\begin{subfigure}{0.45\textwidth}
\includegraphics[width=\textwidth]{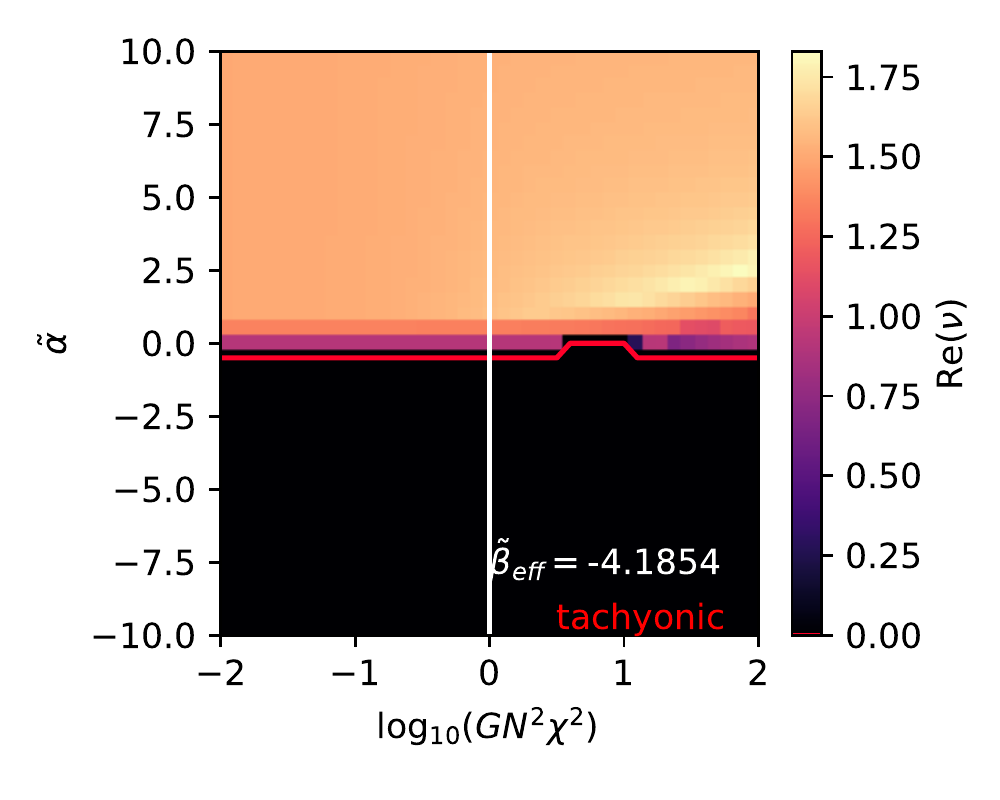}
\caption{\it }
\end{subfigure}
\begin{subfigure}{0.45\textwidth}
\includegraphics[width=\textwidth]{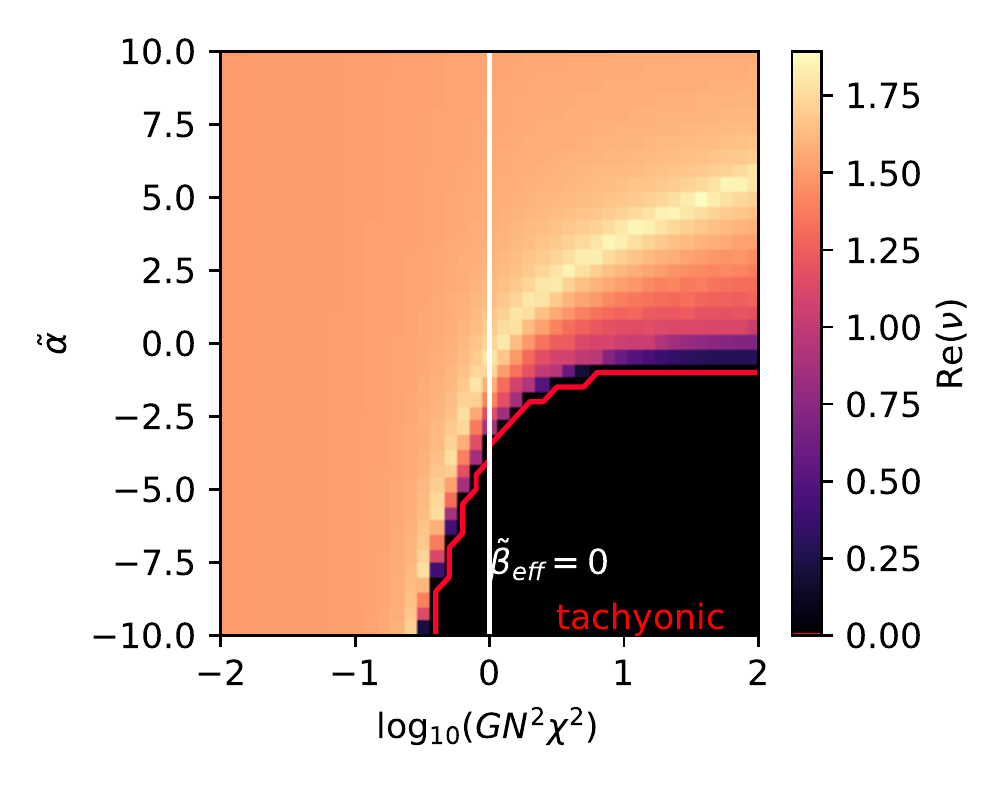}
\caption{\it }
\end{subfigure}
\begin{subfigure}{0.45\textwidth}
\includegraphics[width=\textwidth]{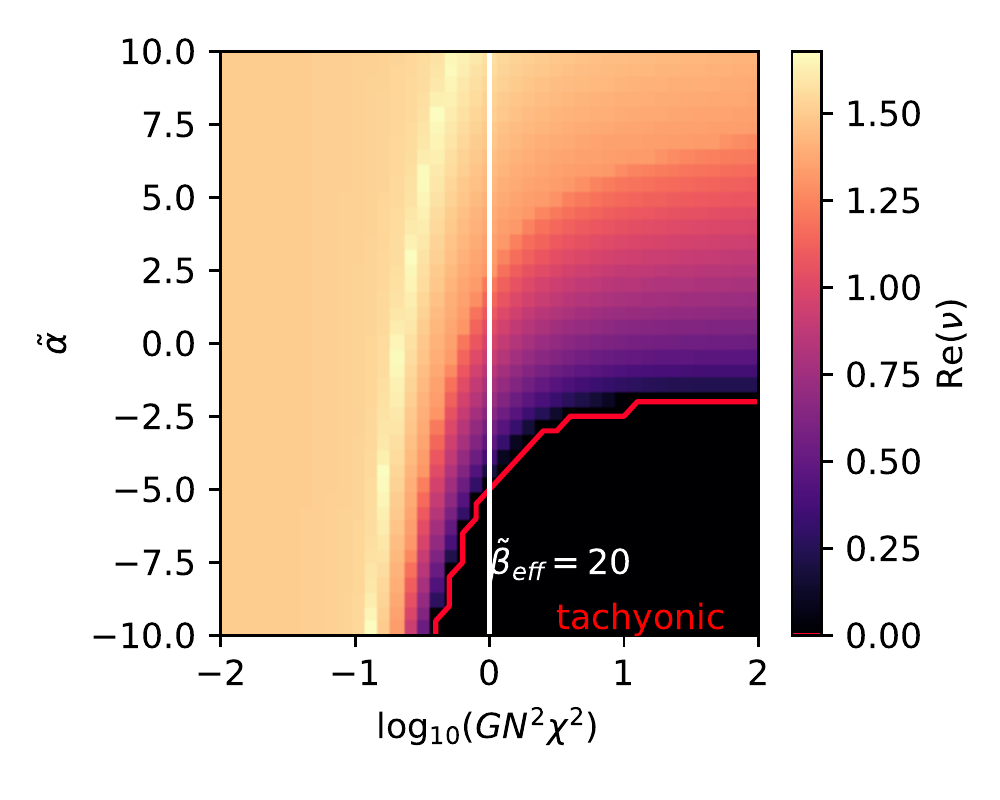}
\caption{\it }
\end{subfigure}
\begin{subfigure}{0.45\textwidth}
\includegraphics[width=\textwidth]{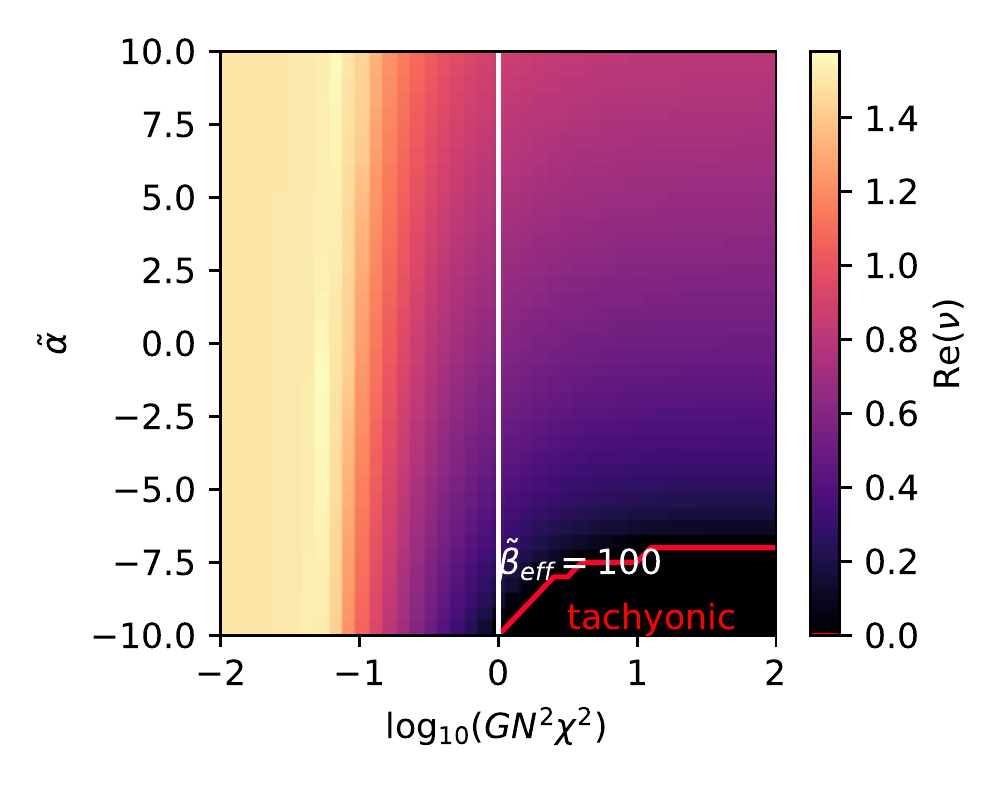}
\caption{\it }
\end{subfigure}
\caption{\it Regions of tachyon-stability of the spin-2 mode in the AdS case. The colour code of each of the subfigures above gives the minimum value of the real part of $\n$ among all the spin-2 poles, at a fixed value of $\tilde{\b}_\text{eff}$.
Different panels correspond to different values of $\tilde{\b}_\text{eff}$.
The vertical white line separates curvatures that are above and below the species cutoff. The red line separates tachyonic regions with $\text{Re}(\n) = 0$ as shown in appendix \protect\ref{AdS criterion}, from the tachyon-safe regions with $\text{Re}(\n) \neq 0$. The first panel takes the value corresponding to the zero-curvature limit of (\protect\ref{sc4}). Increasing $\tilde{\b}_\text{eff}$ moves the tachyonic regions to lower values of $\tilde{\a}$ and larger values of $GN^2\chi^2$.}
\label{AdS_nu}
\end{figure}

\begin{figure}[ht]
\centering
\begin{subfigure}{0.45\textwidth}
\includegraphics[width=\textwidth]{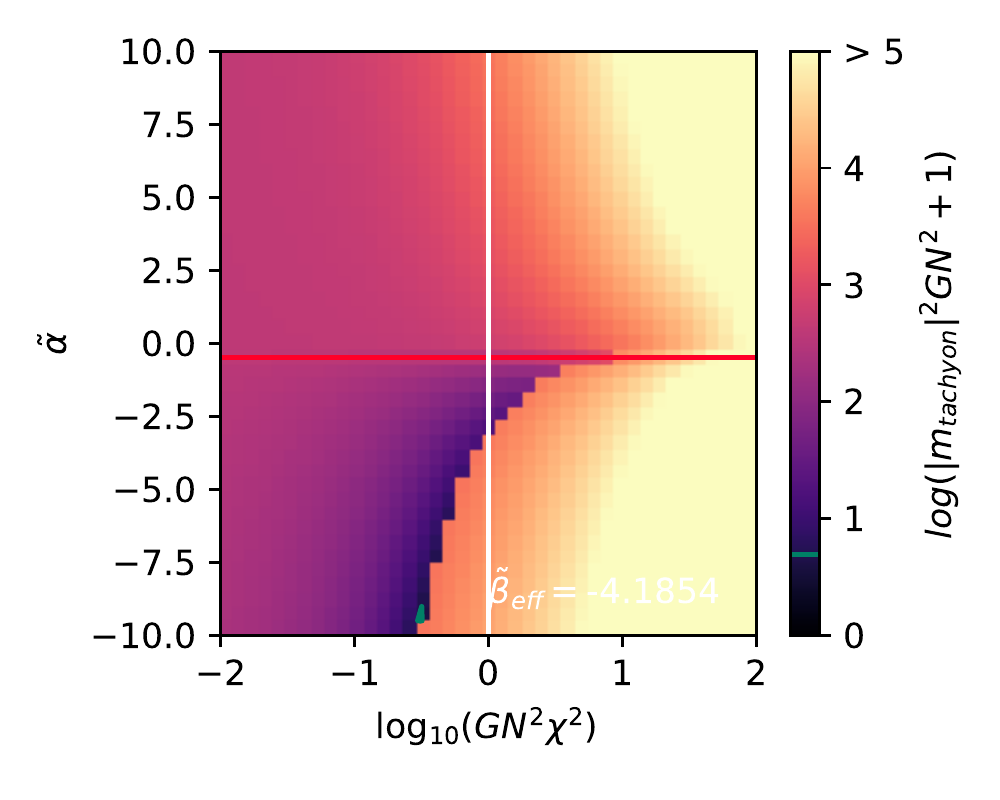}
\caption{\it }
\end{subfigure}
\begin{subfigure}{0.45\textwidth}
\includegraphics[width=\textwidth]{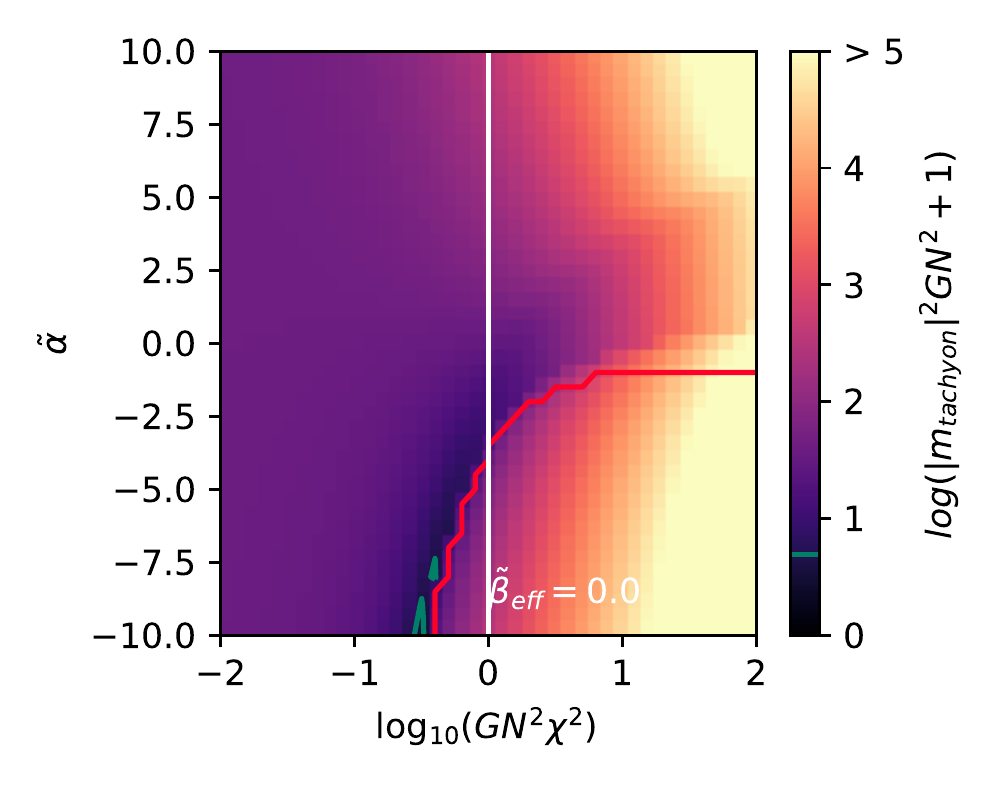}
\caption{\it }
\end{subfigure}
\begin{subfigure}{0.45\textwidth}
\includegraphics[width=\textwidth]{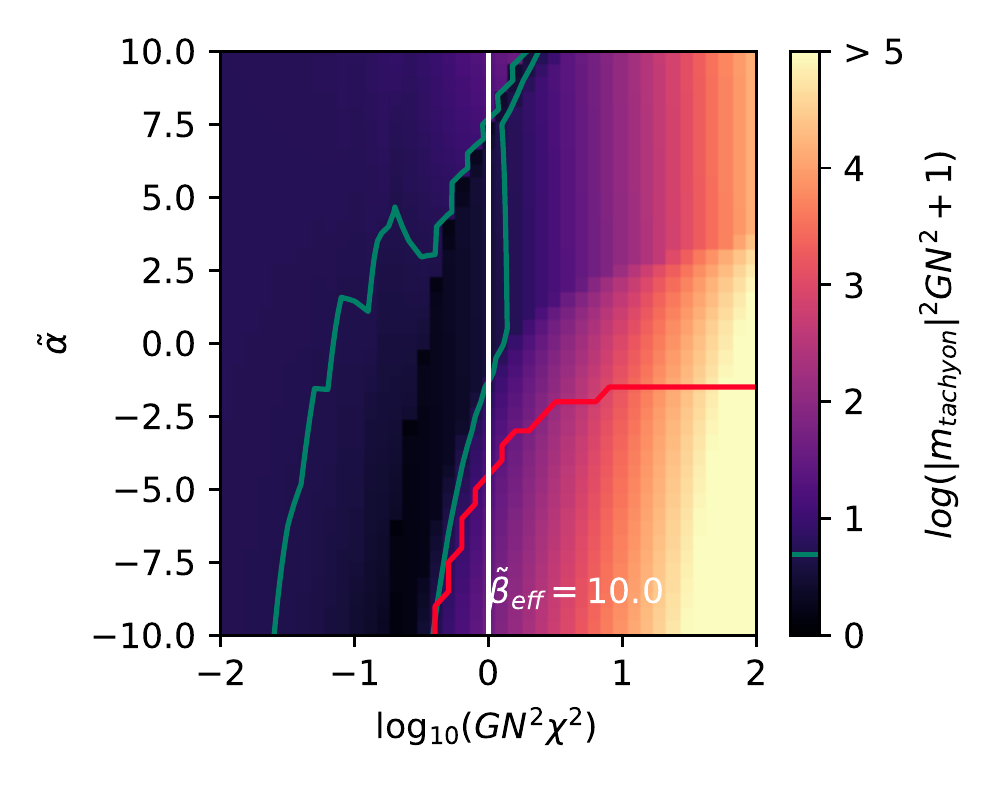}
\caption{\it }
\end{subfigure}
\begin{subfigure}{0.45\textwidth}
\includegraphics[width=\textwidth]{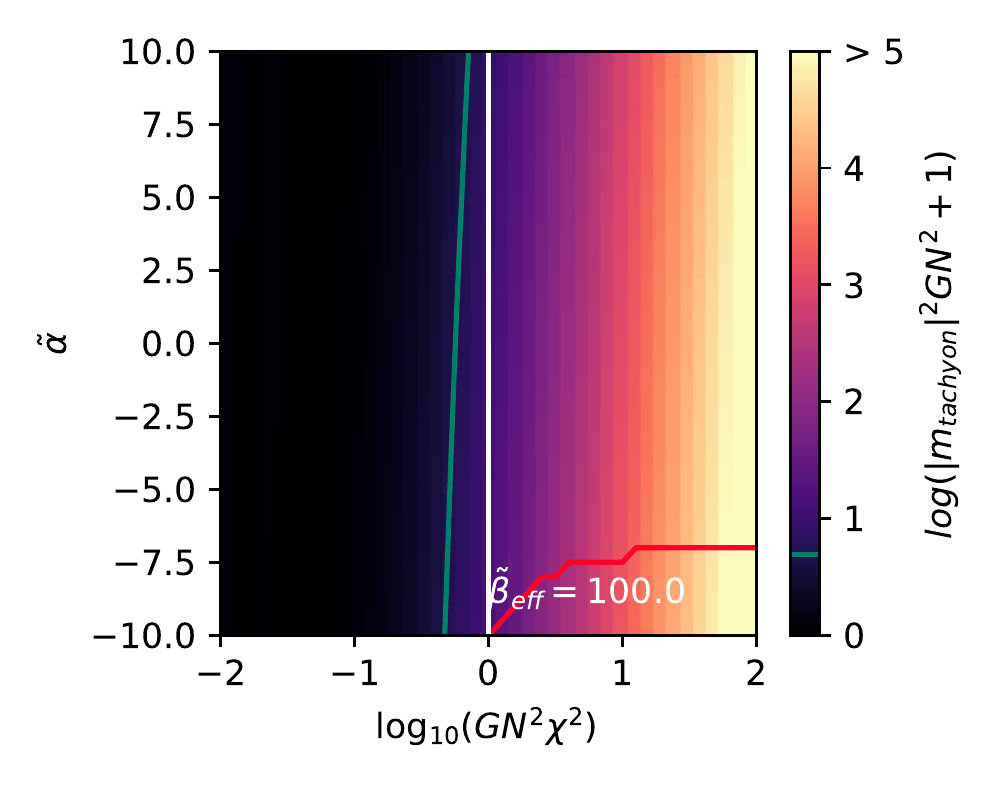}
\caption{\it }
\end{subfigure}
\caption{\it The \textit{would-be} tachyonic spin-2 mode in AdS. The colour code of each of the subfigures above represents the mass (\protect\ref{AdS3}) of this pole in units of the species scale (\protect\ref{sp-a}), at a fixed value of $\tilde{\b}_\text{eff}$. The different panels correspond to different values of $\tilde{\b}_\text{eff}$. The first panel takes the value for $\tbe$ corresponding to the zero-curvature limit of (\protect\ref{sc4}), for which asymptotically small curvatures are critical between tachyon-stable and tachyon-unstable. The tachyonic region on the bottom right corner of each panel is delimited by the red line. The cut-off at $GN^2|m|^2 = 1$ corresponds to the green line, which separates masses which are above the cutoff (bright colours) from masses which are below the cutoff (darker colours). The first panel presents a discontinuity in the bottom half (tachyonic region) because the bottom-right corner has only 1 heavy tachyonic pole (as in Fig. \ref{AdS_mass_am10_chipiover4}) while the bottom-left corner has a second tachyonic pole, which is lighter and ghost-like (as in Fig. \ref{AdS_mass_alpha0_chiem2}).}
\label{AdS_tachyonmass}
\end{figure}

Figure \ref{AdS_nu} shows the occurrence of tachyon-instability in the tensor sector for a few fixed values of $\tilde{\b}_\text{eff}$, with additional information shown about the value of the real part of the pole  which is closest to the imaginary axis (recall that a tachyon corresponds to a purely imaginary $\nu$.)
In the first panel (a)  the value of  $\tilde{\b}_\text{eff}$ corresponds to the critical value separating tachyon-stability and instability in the zero-curvature limit of equation (\ref{sc4}).
As  $\tilde{\b}_\text{eff}$ increases above this value (panels (b), (c) and (d))  the small curvature region becomes tachyon-stable as expected from equation  (\ref{sc4}), and the size of the tachyon-stable region increases.

The region marked ``stable''  in Figure \ref{AdS_nu} are such only concerning tachyonic instabilities: even in these regions there is always one ghost-like tensor mode. The mass of the ghost (in units of the species scale) is represented by the colour code in Figure \ref{AdS_ghost}. Lighter colours correspond to heavier ghosts. The green lines indicate the boundary of the region beyond which the ghost is heavier than the species cut-off, and therefore is outside of the regime of validity of effective field theory.

\begin{figure}[ht]
\centering
\begin{subfigure}{0.45\textwidth}
\includegraphics[width=\textwidth]{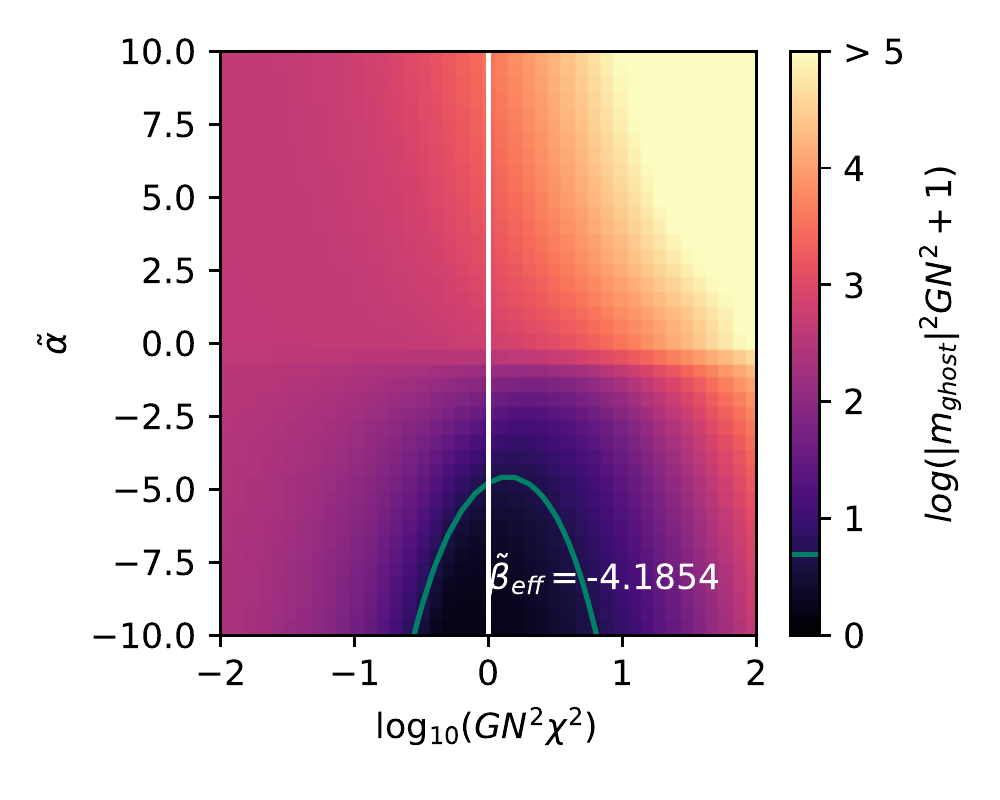}
\caption{\it }
\end{subfigure}
\begin{subfigure}{0.45\textwidth}
\includegraphics[width=\textwidth]{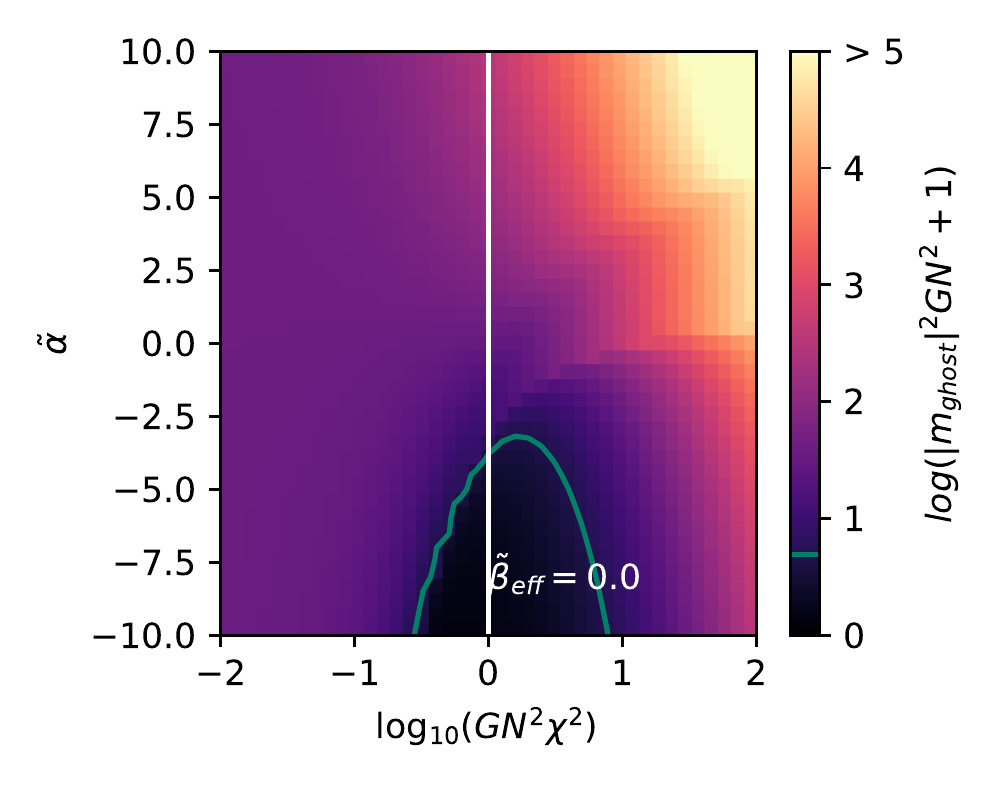}
\caption{\it }
\end{subfigure}
\begin{subfigure}{0.45\textwidth}
\includegraphics[width=\textwidth]{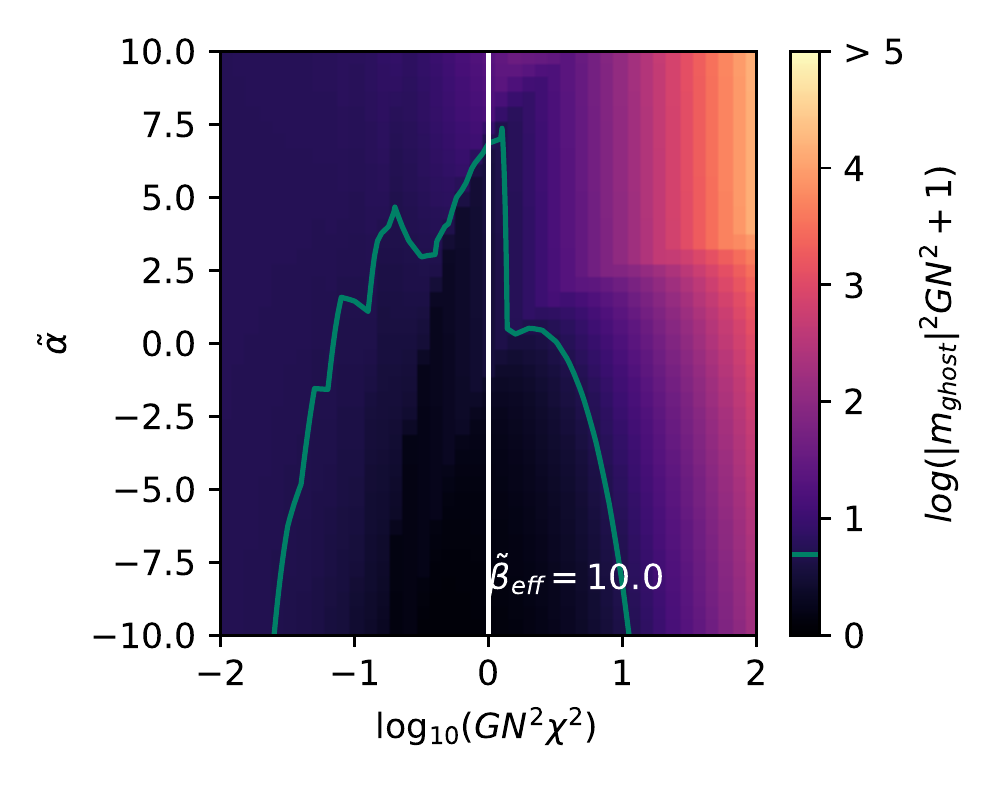}
\caption{\it }
\end{subfigure}
\begin{subfigure}{0.45\textwidth}
\includegraphics[width=\textwidth]{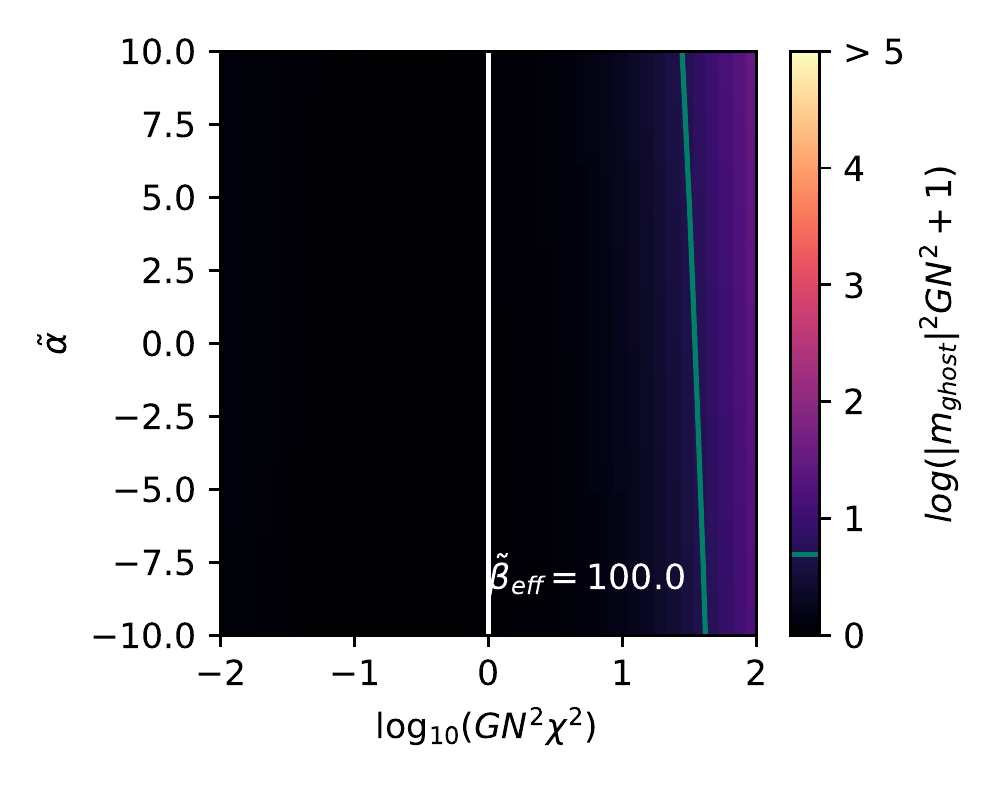}
\caption{\it }
\end{subfigure}
\caption{\it This figure displays the modulus of the mass of the ghost tensor pole in AdS, in units of the species scale (\protect\ref{sp-a}). The colour coding shows the mass of the ghost in units of the species scale in a log scale. Each panel of this figure corresponds to one particular value of $\tilde{\b}_\text{eff}$. The green line shows where the mass is equal to the species cutoff  $GN^2|m|^2 = 1$. Darker colours are below the cutoff.}
\label{AdS_ghost}
\end{figure}

For small values of $\tilde{\b}_\text{eff}$, the ghost mass is always above the species scale except in a small region for negative values of $\tilde{\a}$ (panels (a) and (b)).  As $\tilde{\b}_\text{eff}$ is increased to large and positive values, the ghost becomes lighter and lighter. Ghost masses that are below the species scale appear for small curvatures in the last two panels of Figure \ref{AdS_ghost}.  Increasing $\tilde{\b}_\text{eff}$ even more than $100$ will not change the result since the ghost stabilizes at $\n=3/2$.

We now turn to the scalar sector which has a single excited mode given by the solution of equation (\ref{psi10}). Therefore, the graphical representation of ghost-like and tachyon-like instabilities can be given in a single figure. Figure \ref{AdS_scalar} shows the mass of the scalar solution (\ref{psi10}) in units of the $AdS_4$ scale $\chi^{-1}$ (left panel) and in units of the species scale (right panel). In the scalar sector, only $\tilde{\alpha}$ and the curvature are relevant parameters since $\tilde{\b}_\text{eff}$ does not enter the scalar spectral equation.
The red lines separate tachyon-unstable from tachyon-stable regions and correspond to the points where equation (\ref{psi16}) is saturated. The blue lines separate the regions in which the scalar mode is a ghost from those in which the scalar is healthy, as prescribed by equation (\ref{2pt Psi c}).

\begin{figure}[ht]
\centering
\begin{subfigure}{0.45\textwidth}
\includegraphics[width=\textwidth]{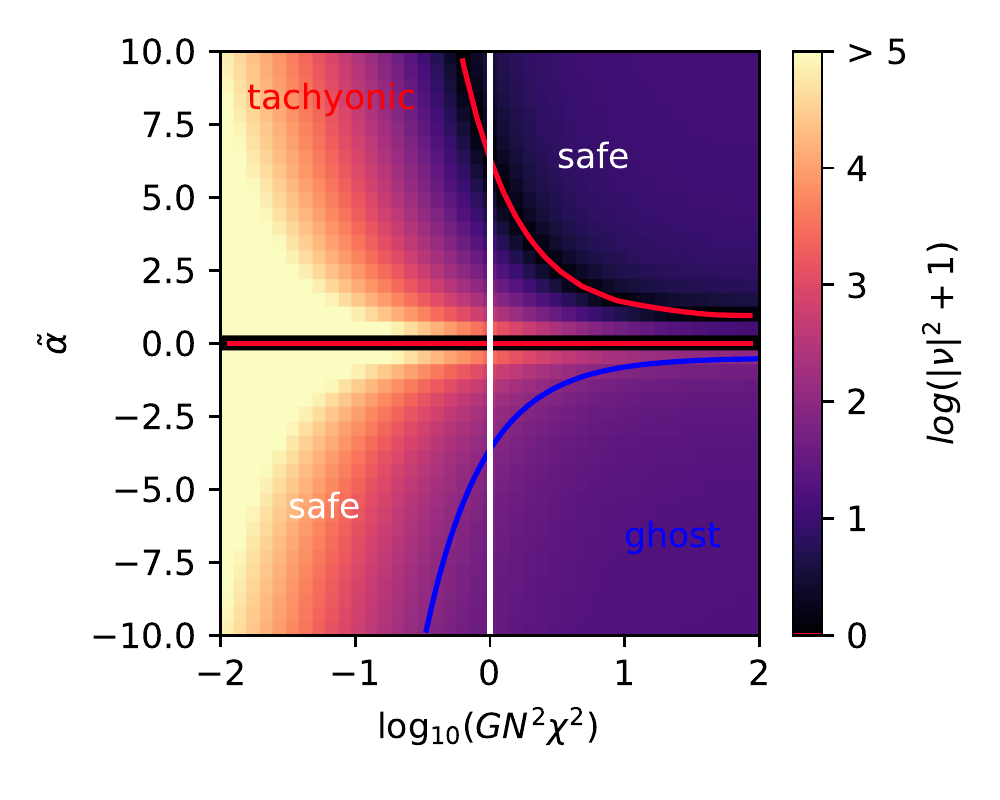}
\caption{\it }
\end{subfigure}
\begin{subfigure}{0.45\textwidth}
\includegraphics[width=\textwidth]{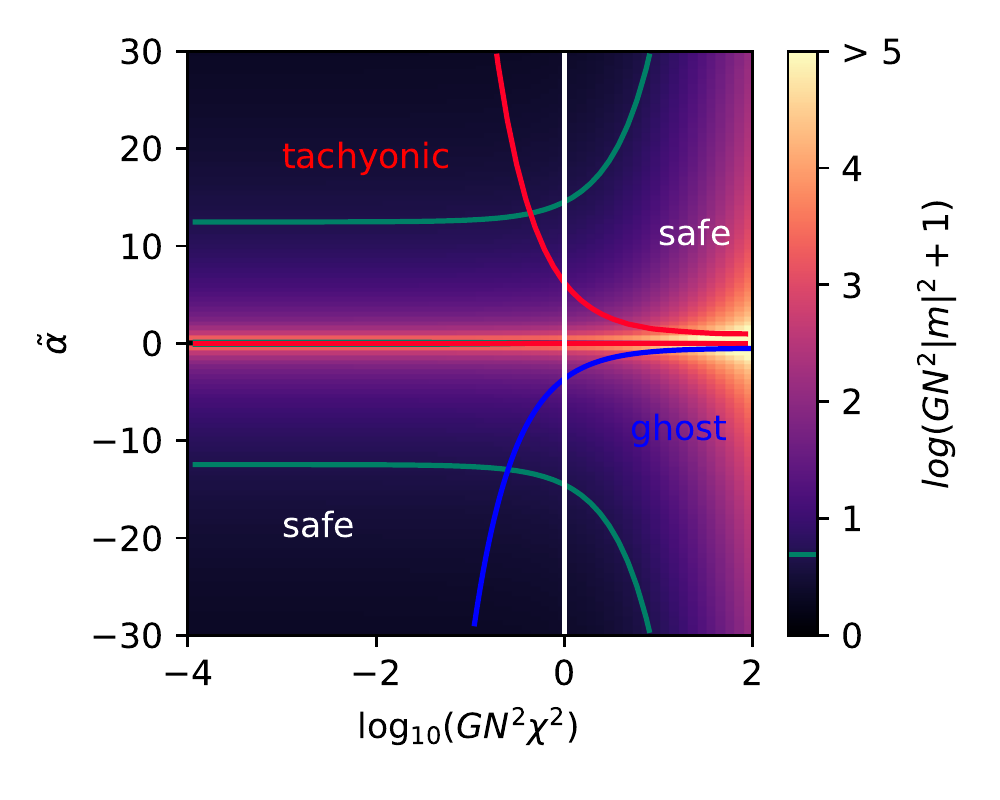}
\caption{\it }
\end{subfigure}
\caption{\it Regions of instabilities for the spin-0 sector in the anti-de Sitter case. The colour code on the left panel indicates $|\n|^2$. It is plotted as a function of $\tilde{\a}$ and $GN^2\chi^2$. Tachyonic regions, where $\text{Re}(\n) = 0$, are delimited by the red lines and the ghost region is delimited by the blue line using the inequality (\protect\ref{2pt Psi c}).
The vertical white line separates the curvatures which are above (on the right) and below (on the left) the species cutoff (\protect\ref{sp-a}).
The green curve corresponds to the species scale $GN^2|m|^2 =1$
On the right panel, we compare the mass of the scalar solution with the species scale (\protect\ref{sp-a}).
These diagrams do not depend on $\tilde{\b}_\text{eff}$.
On the left of the vertical white dashed line, the AdS scale is below the species cutoff.}
\label{AdS_scalar}
\end{figure}

\section*{Acknowledgements}
\addcontentsline{toc}{section}{Acknowledgements}

We would like to thank D. Anninos, M. Kleban, D. Mateos,  M. Montero, V. Niarchos, A. Porfyriadis, C. Rosen and I. Valenzuela.
E. Kiritsis, F. Nitti and V. Nourry are supported in part by CNRS grant IEA 199430.

\newpage
\appendix
\renewcommand{\theequation}{\thesection.\arabic{equation}}
\addcontentsline{toc}{section}{Appendix\label{app}}
\section*{Appendices}

\section{Ghosts and tachyons in Effective Field Theory
} \label{app:free scalars}

It is well known that, when starting from a healthy UV theory, ghosts and/or tachyons can appear in effective field theories as an artefact of integrating out some degrees of freedom and performing the low-energy expansion. In these cases,  the mass of the unstable mode is always of the order of, or above the cut-off (the mass of the states which were integrated out).

We give an example of this phenomenon in a simple model based on free scalar fields.

\subsection{A simple  model}

Consider two massive scalars coupled to each other.
\be
L={1\over 2}\left[(\pa\f_1)^2-m_1^2\f_1^2\right]+{1\over 2}\left[(\pa\f_2)^2-m_2^2\f_2^2\right]+g\f_1\f_2
\label{eft1}\ee

This action can be diagonalized by an orthogonal transformation
\be
\f_1=\cos\theta~\f_++\sin\theta~\f_-\sp \f_2=-\sin\theta~\f_++\cos\theta~\f_-
\label{eft2}\ee
with
\be
\tan(2\theta)={2g\over m_1^2-m_2^2}
\label{eft3}\ee
and the action becomes
\be
L={1\over 2}\left[(\pa\f_+)^2-m_+^2\f_+^2\right]+{1\over 2}\left[(\pa\f_-)^2-m_-^2\f_-^2\right]\sp 2m^2_{\pm}=m_1^2+m_2^2\pm\sqrt{(m_1^2-m_2^2)^2+4g^2}
\label{eft4}\ee
When
\be
g\leq m_1m_2
\label{eft5}\ee
the theory contains two non-interacting scalars with positive kinetic terms and with $m_{\pm}^2\geq 0$.
Of course if $m_2>m_1$ $\f_2$ is unstable to decay (convert) to $\f_1$, but $\f_{\pm}$ are stable.
This is a typical example that will cause oscillations like in the case of neutrinos.
$\f_{\pm}$ are the eigenstates of the Hamiltonian.
So we have in terms of one-particle states, for example
\be
|\f_1(p=0,t)\rangle =\cos\theta e^{im_+ t}|\f_+\rangle+\sin\theta e^{im_- t}|\f_-\rangle
\label{eft5b}
\ee

\subsection{EFT}

We now assume $m_2\gg m_1$ and we integrate out\footnote{We do this by solving the classical equation of motion, but since the theory is Gaussian this is the same as performing the path integral over $\f_2$.} $\f_2$ .
The two equations of motion are
\be
(\square+m_1^2)\f_1=g\f_2\sp (\square+m_2^2)\f_2=g\f_1
\label{eft6}\ee
We solve for $\f_2$
\be
\f_2=g(\square +m_2^2)^{-1}\f_1
\label{eft7}\ee
and substitute in the equation for $\f_1$
\be
(\square+m_1^2)\f_1=g^2(\square +m_2^2)^{-1}\f_1
\label{eft8}\ee
which is obtained from the effective action
\be
L_\text{eff}=-{1\over 2}\f_1\left[(\square +m_1^2)-g^2(\square +m_2^2)^{-1}\right]\f_1
\label{eft9}\ee

$L_\text{eff}$ is completely equivalent to $L$ and the effective propagator is
\be
D^{-1}_{L'}={\square+m_2^2\over  (\square +m_1^2)(\square +m_2^2)-g^2}={R_+\over \square +m_+^2}+{R_-\over \square +m_-^2}=\langle \f_1\f_1\rangle
\label{eft10}\ee
with
\be
R_{\pm}={\pm(m_1^2-m_2^2)+\sqrt{(m_1^2-m_2^2)^2+4g^2}\over 2\sqrt{(m_1^2-m_2^2)^2+4g^2}}
\label{eft11}
\ee
Both residues are positive.

Similarly
\be
{R_-\over \square +m_+^2}+{R_+\over \square +m_-^2}=\langle \f_2\f_2\rangle
\label{eft12}
\ee

\subsection{The IR expansion}

We shall now evaluate the EFT by taking a low-energy approximation to our integrating-out procedure.
We expand (\ref{eft7}) in the IR
\be
\f_2={g\over m_2^2}\left(1-{\square\over m_2^2}+{\square^2\over m_2^4}+\cdots\right)\f_1
\label{eft13}
\ee
Substituting in (\ref{eft6}) we obtain
\be
(\square+m_1^2)\f_1={g^2\over m_2^2}\left(1-{\square\over m_2^2}+{\square^2\over m_2^4}+\cdots\right)\f_1\ar
\ee
\be
\ar \left[m_1^2-{g^2\over m^2_2}+\left(1+{g^2\over m_2^4}\right)\square-{g^2\over m_2^6}\square^2+{\cal O}(\square^3)\right]\f_1=0
\ee
Stopping our expansion at that order, we can write the relevant action as
\be
L_{IR}=-{1\over 2}\f_1\left[m_1^2-{g^2\over m^2_2}+\left(1+{g^2\over m_2^4}\right)\square-{g^2\over m_2^6}\square^2\right]\f_1
\ee
with propagator
\be
D^{-1}_{IR}={1\over m_1^2-{g^2\over m^2_2}+\left(1+{g^2\over m_2^4}\right)\square-{g^2\over m_2^6}\square^2}={\tilde R_-\over \square+\tilde m_-^2}-{\tilde R_+\over
\square+\tilde m_+^2}
\ee
with
\be
\tilde m^2_{\pm}=-{m_2^2\over 2g^2}\left[m_2^4+g^2\pm\sqrt{(m_2^4+g^2)^2+4g^2(m_1^2m_2^2-g^2)}\right]
\ee
\be
\tilde R_{\pm}= {m_2^4\over \sqrt{(m_2^4+g^2)^2+4g^2(m_1^2m_2^2-g^2)}}
\ee
Using $m_2 {\gg } m_1$ and (\ref{eft5}) we can simplify the expressions above as
\be
\sqrt{(m_2^4+g^2)^2+4g^2(m_1^2m_2^2-g^2)}\simeq m_2^4+g^2+2g^2{m_1^2\over m_2^2}
\ee
\be
\tilde m_-^2\simeq m_1^2\sp \tilde m_+^2\simeq -{m_2^6\over g^2}\sp \tilde R_{\pm}\simeq \mp 1
\ee
The pole associated $\tilde m_-$ has the correct positive residue and the correct position corresponding to the slightly corrected light state $\f_1$,
The extra pole at $\tilde m_+$ is ghost-like and tachyonic with a mass scale
\be
|m_+^2|={m_2^6\over g^2}\geq {m_2^4\over m_1^2}\geq m_2^2
\ee
that is above the cutoff of the theory that is the mass of the heavier state.

This simple example shows that integrating out degrees of freedom in an IR expansion of the EFT may generically create ghosts and tachyons, even if the underlying UV theory is perfectly healthy.
However, these ghosts/tachyons always have masses above the cutoff scale.

In conclusion, in the context of EFT, only unstable modes whose masses are parametrically smaller than the UV cut-off can be considered as giving rise to true instabilities of the theory. Conversely,   one cannot reach any conclusion about the stability of the theory based on the presence of ghosts or tachyons whose mass is at or above the cut-off.

\section{Renormalized action}
\label{renorm}
In this appendix we briefly review the results of \cite{HSS} in $d=4$ for the computation of divergent terms of the bulk action (\ref{hr0}) evaluated on-shell. These divergences are then cancelled by the bare gravity action defined in (\ref{hr0a}).

The Gibbons-Hawking term of $S_\text{bulk}$ contains the extrinsic curvature which is defined by
\be
K = G^{ab}\nabla_a n_b, \label{hr1}
\ee
where $n^a$ is the unit vector normal to the boundary which points to the exterior.
 The induced metric and normal vector $n^a$ on the $\rho = \epsilon$ boundary are given by
\be
\gamma_{\omega\s}(\epsilon, x) = {1\over \epsilon}g_{\omega\s}(\epsilon,x) \label{hr4a}
\ee
\be
n^a =\left. { \partial^a\rho \over \sqrt{G_{cd}\partial^c\rho\partial^d\rho}}\right|_{\rho = \epsilon} = {2\epsilon\over L}\delta^{\rho a} \label{hr4}
\ee
The bulk action (\ref{hr0}) evaluated on-shell can then be written  in terms of $g_{\alpha\beta}(\rho,x)$ as
\be
S_\text{bulk} = {1\over 16\pi G_NL}\int d^dx\left[\int_\epsilon^{+\infty}d\rho{d\over \rho^{{d\over 2}+1}}\sqrt{g}+ {1\over \rho^{d\over 2}}\left(-2d \sqrt{g} + 4\rho\partial_\rho \sqrt{g}\right)_{\rho=\epsilon}\right]
\label{hr8}
\ee

This action can be written as a power series of $\epsilon$ by inserting the expansion of the metric (\ref{hr10}) into (\ref{hr8}).
Furthermore, the first few terms of (\ref{hr10}) are obtained in terms of $g^{(0)}_{\omega\s}$ by solving perturbatively the bulk Einstein field equation \cite{HSS}
\be
L^2 R_{ab}[G] + dG_{ab} = 0.
\label{hr20}
\ee
The linear term is then given by
 \be
g^{(2)}_{\omega\s} = -{L^2\over d-2}\left(R_{\omega\s} - {R\over 2(d-1)}g^{(0)}_{\omega\s}\right),
\label{fg16}
\ee
where, in our notation $R_{\omega\s} \equiv R_{\omega\s}[g^{(0)}]$.
However, only the trace and the divergence of $g^{(4)}$ are constrained by the near-boundary reconstruction of the bulk. We shall obtain $g^{(4)}$ and its perturbation starting from the $AdS_5$ bulk in section \ref{linearized eq}.
The log-term $\hat{g}$ is given by
\be
\hat{g}_{\omega\s} = {L^4\over 16}\left\{ 2R_{\omega \kappa \s \l}R^{\kappa\l} - {1\over 3} \nabla_\omega\nabla_\s R + \nabla^2 R_{\omega\s} - {2\over 3}RR_{\omega\s} + ({1\over 6}R^2 - {1\over 6}\nabla^2 R - {1\over 2}R_{\kappa\l}R^{\k\l})g^{(0)}_{\omega\s} \right\},
\label{fg26}
\ee
which is traceless.
Inserting (\ref{hr10}) into the bulk action (\ref{hr8}) gives a power series in $\epsilon$, given in $d=4$ by
\be
S_\text{bulk} = {1\over 16\pi G_NL}\int d^4 x\sqrt{g^{(0)}}\left(\epsilon^{- 2}a_{(0)} + \epsilon^{-1}a_{(2)} - \log\epsilon a_{(4)}\right) + \mathcal{O}(\epsilon^0),
 \label{hr9}
\ee
where
\be
a_{(0)} = 2(1-d) = -6,
 \label{hr16}
\ee
\be
a_{(2)} = {(4-d)(d-1)\over d-2} \tr g^{(2)}  = 0,
\label{hr17}
\ee
\be
a_{(4)} = {1\over 2}((\tr (g^{(2)}))^2 - \tr((g^{(2)})^2))  = -{L^4\over 8}\left(R^{\omega\s}R_{\omega\s}- {1\over 3}R^2\right).
\label{hr18}
\ee
The divergent piece of the bulk action
\be
S_\text{bulk} = S_\text{div} + \mathcal{O}(\epsilon^0) \label{hr20b}
\ee
is then given by
\be
S_\text{div} =  {1\over 16\pi G_N L} \int d^4 x\sqrt{g_{(0)}}\left\{-6\epsilon^{-2}  + {1\over 8} \log\epsilon \left(R^{\omega\s}R_{\omega\s} - {1\over 3}R^2\right) \right\}.
\label{hr20c}
\ee
Inverting perturbatively series for $\sqrt{g}$ and $R_{\omega\s}[g]$ in powers of $\epsilon$ allows one to express $\sqrt{g_{(0)}}$ and $R_{\omega\s} = R_{\omega\s}[g_{(0)}]$ covariantly in a power series of curvature tensors of the induced metric $\gamma_{\omega\s}$. These useful formulae are given by
\be
\sqrt{g_{(0)}} = \epsilon^{2}\sqrt{\gamma}\left[1- {\epsilon\over 2} \tr{g_{(2)}} + {\epsilon^2\over 8}(\tr{g_{(2)}^2} + (\tr{g_{(2)}})^2) + \mathcal{O}(\epsilon^3)\right],
\label{hr27}
\ee
\be
R = {1\over \epsilon}\left\{R[\gamma]-{L^2\over 2}\left(R^{\omega\s}[\gamma]R_{\omega\s}[\gamma]-{1\over 6}R[\gamma]^2\right) + \mathcal{O}(R[\gamma]^3)\right\}.
\label{hr28}
\ee
Using these expansions into (\ref{hr20c}) allows us to obtain the covariant counterterms written in the main text (\ref{hr30}).

 \section{Comparison with the Starobinsky model\label{staro}}

In this appendix, we relate our analysis to the Starobinksy $R+R^2$ model of inflation which is one of the most favoured single-field inflationary models by CMB observations \cite{Planck:2018inflation}. This model is obtained from the original Starobinsky model of anomaly-driven inflation without a cosmological constant \cite{St,VilenkinStaro}, by neglecting the non-local anomaly terms and keeping only the local $R^2$ term. This can be justified when the coefficient $\alpha$ of the $R^2$ term dominates. In our setup, this amounts to ignoring the CFT contribution (setting $N=0$) as well as setting $\beta = 0$, and keeping only the $\alpha R^2$ pure gravity term.

Dropping the non-local terms pushes the de Sitter solution to infinite curvature: in equation (\ref{scalar7})  with $\Lambda=0$, the de Sitter solution is the non-trivial one with $\bar{R} = 48\pi/GN^2$, and in the limit $N\to 0$ the curvature diverges. However, by writing the model as a scalar-tensor theory and performing a Weyl transformation to the Einstein frame, one obtains a single-field inflationary model with a quasi-de Sitter solution with a finite Hubble parameter.

The action for the simplified $R^2$ Starobinsky model is
\be
S = -\int d^dx \sqrt{-g} \left\{{R\over 16\pi G} -\hat\alpha  R^2\right\}.
\label{staro1}
\ee
Identifying this action to the $R^2$ action of our model (\ref{s03}) gives the relation between $\hat \alpha$ and $\alpha$:
\be
\hat \alpha =  {\alpha\over 384\pi}
\label{staro2}
\ee

The favoured observational value is
\be
\alpha \approx -5.95 \times 10^{11}\;,
\label{staro3}
\ee
 obtained from the amplitude of the power spectrum of primordial curvature fluctuations \cite{Faulkner:2006}.
As mentioned above, the model (\ref{staro1}) corresponds to the $H\to\infty$ limit of our analysis. We can still compare our results with the full model (including the CFT)  \cite{VilenkinStaro, St} with the same value of  $\alpha$ as the one favoured by data.  In this case, the curvature is fixed to
 \be \label{staro3-i}
GN^2H^2=4 \pi.
\ee
 The large  $R^2$ term (\ref{s03})  makes the scalaron $\psi$ light and tachyonic as we see below.   Notice that this model, due to (\ref{staro3-i}), falls outside of the regime of effective field theory.

\paragraph*{Scalar sector in Starobinsky inflation}

We first discuss the scalar mode (scalaron), which is the one that, in the pure $R+R^2$ model, can be identified with the inflaton and in the presence of the CFT makes de Sitter unstable.

Indeed, by inserting (\ref{staro3-i}) into the condition (\ref{psi_tachyonic}), we conclude that $\alpha<0$ is in the tachyonic regime. The characteristic decay rate $\G$ of the  scalaron instability can be read off by substituting (\ref{staro3-i}) into  (\ref{Gamma-scalar}):
\be
\G = H\left[-{3\over 2} + \sqrt{{9\over 4} - {1 \over  \tilde{\a}}} \right],
\label{staro4}
\ee
where $\tilde{\a} = {\pi\a\over N^2}$, which agrees with the value found in \cite{VilenkinStaro} close to the de Sitter solution. For a long-lived de Sitter, we need  $|\tilde{\alpha}|\sim |\alpha|/N \gg 1$, which also implies  $|\alpha |\gg 1$. This model then matches qualitatively the features of the pure $R +R^2$ model, with an unstable de Sitter replaced by a slowly-rolling FRW space-time.

\paragraph*{Tensor sector in Starobinsky inflation}

As explained above, our more general setup can retrieve the $R+R^2$ model (\ref{staro1}) by setting $N = \b = 0$. In this case,
the only propagating tensor mode is the massless graviton $\n = 3/2$ as one can see from equation (\ref{N02}) applied to $\b = 0$.
Therefore, there is no ghost or tachyonic spin-2 mode in the Starobinsky model.

We now  turn on $\b \neq 0$ while keeping $|\tilde{\a}|\gg 1$. Now the tensor sector acquires an additional propagating mode.
In such a regime, the $R^2$ term of the action dominates over the CFT. Therefore, the spin-2 propagator with the CFT (\ref{dS17}) can be approximated by the pure (modified) gravity propagator (\ref{N02}).
This propagator contains the usual massless pole and a massive one.

The massless pole is a ghost if
\be
{1\over 2\pi} - 2\tilde{\a} + \tilde{\b} <0, \quad \Rightarrow \quad \text{massless ghost}
\label{staro5}
\ee
otherwise, the massive pole is a ghost.
The second case holds for large and negative $\tilde{\a}$ and generic values of $\tilde{\b}$.
In addition to ghost-like instabilities, the massive pole is a tachyon if
\be
{2\over \tilde{\b}}\left(\tilde{\a} - {1\over 4}\right) < 1. \quad \Rightarrow \quad \text{tachyonic}
\label{staro6}
\ee
Thus, large and negative $\tilde{\a}$ are associated with tachyonic spin-2 perturbations for positive and generic values of $\tilde{\b}$.

However, the massive mode lies below the species scale when $|\tilde{\b}|\gg|\tilde{\a}|\gg 1$. If $\b$ is positive, the spin-2 pole is tachyonic and ghost-like. If $\b$ is negative, the spin-2 pole is only ghost-like.

If we decide to take the CFT contribution into account, one must refer to Figure \ref{fig:scalarvstensor} instead of equation (\ref{staro6}). This figure shows which sector (scalar or tensor) represents the strongest tachyonic instability. The vertical red line is for $\L = 0$ as is the case in Starobinsky's model. This figure shows that negative $\tilde{\a}$  are associated with scalar tachyonic instability, which is convenient for an inflationary scenario. However, small curvatures can be associated with tensor tachyonic instabilities dominating the usual scalaron.

\section{AdS slicing coordinates}
\label{app:AdS-slicing}

In this appendix, we describe the AdS$_{d+1}$ metric in AdS$_{d}$ slice coordinates.

Lorentzian $AdS_{d+1}$ is the hyperboloid
\be
\eta_{AB}X^AX^B = -L^2. \label{i0a}
\ee
where $A,B=-1,...,d+1$ and $\eta_{AB} = \text{diag}(-1,-1,1,...,1)$.
global $AdS_{d+1}$ coordinates are obtained by choosing
\begin{subequations}
\be
X^{-1} = L\cos{t}\cosh{\rho},
\ee
\be
X^0 = -L\sin{t}\cosh{\rho},
\ee
\be
X^\mu = L\Omega^\mu\sinh{\rho},
\ee
\label{i0b}
\end{subequations}
where $\m = 1,...,d$ and
\be
\delta_{\m\n}\Omega^\m\Omega^\n = 1\label{i0c}
\ee
AdS slicing is obtained by choosing $u$ as a radial coordinate crossing Lorentzian $AdS_d$ slices. Global coordinates can be chosen to describe the $d-$dimensional slice. The $AdS$ slicing coordinates are then given by
\begin{subequations}
\be
X^{-1} = L\cosh{u}\cos{\tau}\cosh{r},
\ee
\be
X^0 = -L\cosh{u}\sin{\tau}\cosh{r},
\ee
\be
X^i = Ln^i\cosh{u}\sinh{r},
\ee
\be
X^d  =L\sinh{u},
\ee
\label{i0d}
\end{subequations}
where $i,j=1,...,d-1$ and
\be
\delta^{ij}n^in^j=1\label{i0e}.
\ee
Using this coordinate system, we can reach the infinity of the embedding space $X^A$ either by taking $u\rightarrow \pm \infty$ or $r\rightarrow +\infty$. Therefore, the boundary of the hyperboloid has two pieces (both infinities for $u$) which are connected by the common boundary $r\rightarrow +\infty$ of the slice $AdS_4$.

To obtain a map between these two coordinate systems, we first rewrite $(X^\mu)^2$ using both (\ref{i0b}) and (\ref{i0d}). It gives the relation
\be
\sinh^2\rho = \cosh^2u\sinh^2r + \sinh^2u.\label{i0f}
\ee
This can be rewritten as
\be
\sinh \rho = \left(\cosh^2u\cosh^2r - 1\right)^{1\over 2}.\label{i0g}
\ee
For all $u$ and $r$ we have $\cosh u\cosh r >1 $ (global AdS is ill-defined when $\rho=0$), so we can use the formula $\sqrt{x^2-1} = \sinh (\text{Arccosh}x)$. Therefore we obtain
\begin{subequations}
\be
t = \tau \label{i0h}
\ee
\be
\cosh{\rho} = \cosh{u}\cosh{r}.\label{i0i}
\ee
\be
\Omega^d = \sinh{u}(\cosh^2u\cosh^2r-1)^{-{1\over 2}}\label{i0j}
\ee
\be
\Omega^i = \cosh u \sinh r (\cosh^2u\cosh^2r-1)^{-{1\over 2}}n^i\label{i0k}
\ee
\end{subequations}
The inverse transformation can be obtained using
\be
\cosh{r} = {\cosh{\rho}\over \cosh{u}} \label{i0l}
\ee
and the expression for $X^d$ in the two sets of coordinates which gives
\be
\sinh u = \Omega^d \sinh{\rho}, \label{i0l1}
\ee
\be
\cosh{u} = \sqrt{1+(\Omega^d\sinh{\rho})^2} \label{i0l2},
\ee
so that we can replace each $u$ and $r$ into every equation in (\ref{i0d}). The transformation from AdS slicing to global AdS is then
\begin{subequations}
\be
\tau = t  \label{i0l3}
\ee
\be
u = \text{Arcsinh}\left(\Omega^d \sinh{\rho}\right),\label{i0l4}
\ee
\be
\cosh{r} = \cosh{\rho}\left(1+(\Omega^d \sinh{\rho})^2\right)^{-{1\over 2}}\label{i0l5}
\ee
\be
n^i = \Omega^i\left(1-(\Omega^d)^2\right)^{-{1\over 2}}\label{i0l6}
\ee
\end{subequations}
One can easily check that $\delta_{ij}n^in^j = 1$ is still true using (\ref{i0c}).
Indeed,
\be
\delta_{ij}n^in^j = {\delta_{ij}\Omega^i\Omega^j\over 1- (\Omega^d)^2} =  {\delta_{\m\n}\Omega^\m\Omega^\n - (\Omega^d)^2 \over 1- (\Omega^d)^2}  = 1. \label{i0l7}
\ee
We can also choose a parametrisation of the $(d-1)$-sphere $\Omega^\m$ and specifically pick $\Omega^d$ to be the polar axis (so that it is parametrized by only one angle $\theta \in [0,\pi]$):
\be
\begin{array}{ccl}
\Omega^d &=& \cos\theta,\\
\Omega^1 &=& \sin\theta\cos\f_1,\\
 & \vdots & \\
\Omega^{i} &= & \sin\theta\sin\f_1...\sin\f_{i-1}\cos\f_i, \\
& \vdots & \\
\Omega^{d-1} & = & \sin\theta\sin\f_1...\sin(\f_{d-2}).
 \label{i0l8}
\end{array}
\ee
where $\f_i \in [0,2\pi[$.
Now the change of coordinates from AdS slicing to global AdS is written as
\begin{subequations}
\be
\tau = t  \label{i1a}
\ee
\be
u = \text{Arcsinh}\left(\cos\theta \sinh{\rho}\right),\label{i1b}
\ee
\be
\cosh{r} = \cosh{\rho}\left(1+(\cos\theta \sinh{\rho})^2\right)^{-{1\over 2}}\label{i1c}
\ee
\be
n^i = {\Omega^i \over \sin\theta}\label{i1d}
\ee
\end{subequations}

The induced metric in AdS slicing is given by
\be
ds^2 = L^2\left\{du^2 + (\cosh u)^2 ds^2_d \right\} = G_{ab}dx^adx^b,\label{i1}
\ee
where $ds^2_d$ is the metric of unit $AdS_d$.

\section{Schrodinger problem in the bulk}
\label{bulk schrodinger}

{In this appendix, we write the bulk radial equation for the spin-2 perturbation (\ref{b14bis}) as a Schrodinger equation for each slicing (flat, de Sitter and anti-de Sitter). This procedure gives a physical interpretation for bulk solutions with different values of the slice momentum $\n$ (in dS and AdS) and $k$ (in Minkowski) which is identified to the energy of this Schrodinger problem.

Moreover, the Schrodinger problem provides a norm which we can use to check the normalizability of the solutions. In particular, we want to check the normalizability of solutions near the horizon $u = 0$ in de Sitter slicing coordinates.}

The easiest way to write the bulk equation for the spin-2 modes (\ref{b14bis}) as a Schrodinger equation is to write the background metric (\ref{b0})
\be
ds^2_{d+1} = L^2 du^2 + a^2\bar{\zeta}_{\omega\s}dx^\omega dx^\s
\label{b0bis}
\ee
into conformal coordinates
\be
= a^2\left[\ell^2 dr^2 + \bar{\zeta}_{\omega\s}dx^\omega dx^\s \right],
\label{schro1}
\ee
where $\ell$ is the radius of the slice metric $\bar{\zeta}_{\omega\s}$ which we write here in order to keep $r$ dimensionless like $u$.
To find such a coordinate $r$, we need to solve
\be
\ell^2 dr^2 = L^2 a^{-2}du^2,
\label{schro1a}
\ee
where the conformal factors we consider are $a = L\chi \cosh{u}$ (\ref{b0AdS}) in AdS slicing, $a = LH \sinh{u}$ (\ref{b0flat}) in dS slicing and $a=e^{-u}$ (\ref{b0dS}) in flat slicing.
We just need to find the appropriate conformal coordinate $r$, compute $a(r)$ and transform the bulk equation of motion (\ref{b14}) into a Schrodinger equation in this new coordinate.
\subsection*{AdS slicing}
For AdS slicing (\ref{b0AdS}), $\ell = \chi^{-1}$, we find
\be
\tan\left({r\over 2}\right) = \tanh\left({u\over 2}\right).
\label{schro2}
\ee
The conformal boundary of AdS$_5$ located at $u\rightarrow \pm \infty$ corresponds to $r=\pm \pi$.
From (\ref{schro2}), we obtain
\be
\left\{
\begin{tabular}{C C C}
\sinh{u} & = & \tan{r}\\
\cosh{u} & = & {1\over \cos{r}}.
\end{tabular}
\right.
\label{schro2a}
\ee
Using (\ref{schro2a}), it is then possible to write equation (\ref{AdS5}) in terms of the conformal coordinate $r$. The equation is
\be
\left\{\partial_r^2 + (d-1)\tan{r}\partial_r + \n^2 - \left({d-1\over 2}\right)^2\right\}F = 0.
\label{schAdS3}
\ee
Defining the rescaled field $\Psi$ as
\be
\Psi = a^{d-1\over 2}F,
\label{schAdS4}
\ee
the equation becomes a Schrodinger problem
\be
\left\{-{d^2\over dr^2} + V(r)\right\}\Psi = E\Psi,
\label{schAdS5}
\ee
where
\be
V(r) \equiv \left({d-1\over 2}\right)\left[1+ \left({d+1\over 2}\right)\tan^2{r}\right],
\label{schAdS6}
\ee
\be
E \equiv -\n^2 + \left({d-1\over 2}\right)^2
\label{schAdS7}
\ee

\subsection*{dS slicing}
For dS slicing (\ref{b0dS}), $\ell = H^{-1}$ and the conformal coordinate $r$ is a solution of (\ref{schro1a}), which for positive $u$ is given by
\be
e^r = \tanh\left({u\over 2}\right).
\label{schro3}
\ee
The limit $u\rightarrow +\infty$ which goes to the conformal boundary of AdS$_5$ then corresponds to taking $r\rightarrow +\infty$. The horizon at $u=0$ then corresponds to $r = -\infty$.

The following steps are identical to the ones done in AdS-slicing. Instead of (\ref{schro2a}), we have the following relations
\be
\left\{
\begin{tabular}{C C C}
\sinh{u} & = & {1\over \sinh{r}}\\
\cosh{u} & = & - \coth{r}.
\end{tabular}
\right.
\label{schro3a}
\ee
The bulk radial equation of motion for spin-2 perturbations (\ref{dS7}) is then written in terms of the conformal coordinate $r$ as
\be
\left\{\partial_r^2 - (d-1)\coth{r}\partial_r - \n^2 + \left({d-1\over 2}\right)^2\right\}F = 0.
\ee
Using the same redefinition as in (\ref{schAdS4}) with $a = LH\sinh{u}$, we find the Schrodinger equation (\ref{schAdS6},\ref{schAdS7}) satisfied by $\Psi = a^{d-1\over 2}F$. The potential and energy are respectively given by
\be
V(r) \equiv \left({d-1\over 2}\right)\left[- 1 + \left({d+1\over 2}\right)\coth^2{r}\right],
\label{schdS6}
\ee
\be
E \equiv \n^2 - \left({d-1\over 2}\right)^2.
\label{schdS7}
\ee
The scalar product associated with this Schrodinger problem is defined as
\be
\left<f,g\right> = \int_{\mathbb{R}_-^*} dr f^*(r)g(r).
\label{schdS8}
\ee
The two linearly independent solutions obtained in (\ref{dS8}) should be normalizable according to the norm associated with the scalar product (\ref{schdS8}). The asymptotic behaviour of the Schrodinger field near $u=0$ for $d=4$ is given by
\be
 \Psi(u) = (\sinh u)^{3\over 2} F(u)  \underset{u\rightarrow 0}{\sim} u^{\pm\n}.
 \label{schdS9}
 \ee
  The normalizable condition near the horizon at $u=0$ for the Schrodinger field $\Psi$ for $d=4$ is then
\be
|\Psi|^2 = \left<\Psi(r,\n),\Psi(r,\n)\right> = \int_{-\pi\over 2}^{\pi\over 2}dr |\Psi(r)|^2 \sim \int_0 {du\over u} u^{\pm\n} < \infty.
\label{schdS10}
\ee
This integral converges on the horizon $u=0$ if $\text{Re}(\n)>0$ for the ``$C_-$''
solution in (\ref{dS8}), which has a negative sign exponent in (\ref{schdS10}).
 Conversely, it converges if $\text{Re}(\n)<0$ for the ``$C_+$'' solution.
In conclusion, the sign of $\text{Re}(\n)$ determines which solution $C_\pm$ is
 normalizable. In section \ref{dS prop}, we decide to take a positive real part
 of $\n$ and therefore need to choose $C_- =0$ which is not normalizable because
 the norm (\ref{schdS10}) diverges near the horizon $u=0$.

\section{Flat space tachyonic time scale}
\label{app:tacscale}

In this appendix, we study the time dependence of a tachyonic perturbation using the formalism of
Green functions. We show that a tachyonic pole of the Minkowski propagator is associated with
 a runaway in the retarded Green function of the perturbation.

Both the scalar (\ref{psi9-i}) and the tensor (\ref{flat5}) perturbations are decomposed into eigenmodes
 of the Minkowski Laplacian operator $\partial^2$. Then, a single mode $\f$ perturbation
 associated with the eigenvalue $k^2$ is a solution of the Klein-Gordon equation
\be
(\partial^2  -m^2)\f(x) = 0,
\label{green1}
\ee
where $x$ stands for the 4-dimensional coordinate vector $x \equiv (t,\bm{x})$.
Equation (\ref{green1}) can be separated into 3 different cases. First, we study the case $m^2>0$, which corresponds to the usual Klein-Gordon equation for a positive mass squared. Second, we study the $m^2<0$ case. Finally, we study the case where $m^2$ is complex but away from the real axis. We shall observe that the retarded Green function contains a runaway in the two last cases, and obtain the characteristic time of this runaway.

The spectral equation for (\ref{green1}) is obtained by performing a Fourier transform over the four space-time coordinates:
\be
\f(t,\bm{x}) = \int d^4x e^{-ik.x}\tilde{f}(\omega,\bm{k}) = \int dt d^3\bm{x} e^{-i\bm{k}.\bm{x} + i\omega t} \tilde{\f}(\omega,\bm{k}) .
\label{green2}
\ee
The spectral equation is then
\be
(\omega^2 - \bm{k}^2 -m^2)\tilde{\f} = 0
\label{green3}
\ee
The most general solution of (\ref{green3}) is given by
\be
\tilde{\f} = \a(\bm{k})\delta(\omega - E_k) + \b(\bm{k})\delta(\omega+E_k),
\label{green4}
\ee
where we have defined $E_k$ as one of the square roots of
\be
E_k^2 \equiv \bm{k}^2 + m^2.
\label{green5}
\ee
We can choose arbitrarily one of the two square roots. Taking one or the other would simply exchange $\a(\bm{k})$ and $\b(\bm{k})$ in the solution (\ref{green4}).
We  specify which square root is chosen for each following subsection ($m^2$ positive, negative or complex).

The Green function $G$ associated to equation (\ref{green1}) is defined as
\be
(\partial^2 - m^2)G(x) = \delta(x)
\label{green6}
\ee
The most general solution for $G$ is then given by
\be
G(x) = \f(x)  + G_p(x),
\label{green7}
\ee
where $\delta(x)$ is a Dirac distribution centered at $x = 0$, $\f$ is the homogeneous solution (\ref{green4}) and $G_p$ is a particular solution. This particular solution can be obtained via the inverse 4-dimensional Fourier transform of (\ref{green6}). The result is
\be
G_p(x) = \int {d^3\bm{k}\over (2\pi)^3} e^{-i\bm{k}.\bm{x}}\int {d\omega\over 2\pi} e^{i\omega t}{1\over \omega^2 - E_k^2}.
\label{green8}
\ee
\subsection*{Positive mass squared}
In that case, $E_k$ is real, and we choose it to be the positive square root of (\ref{green5}).
The integral over $\omega$ is evaluated using the residue theorem, and the contour can be chosen arbitrarily (one can choose either the retarded, advanced or the Feynman prescription). For example, the retarded prescription circles the two poles $E_k = \pm \omega$ for positive $t$. The result is then
\be
G^\text{R}_p(t>0, \bm{x}) = -\int {d^3\bm{k}\over (2\pi)^3} e^{-i\bm{k}.\bm{x}} {\sin(E_k t)\over E_k}.
\label{green9}
\ee
This is the usual retarded Green function of the Klein-Gordon operator. Its time dependence appears only in a sine function and therefore does not contain a runaway. Another prescription would have given another combination of complex exponentials with positive and negative signs. Therefore, all the different prescriptions are safe.
For example, the Feynman prescription would have given
\be
G^F_p(x) = i \int {d^3\bm{k}\over (2\pi)^3} e^{-i\bm{k}.\bm{x}} {e^{iE_k|t|}\over 2E_k}.
\label{green9a}
\ee

\subsection*{Negative mass squared $m^2 = - \m^2$}
In that case, it is necessary to separate the values of $|\bm{k}|$ into two distinct regimes.
\begin{itemize}
\item When $|\bm{k}|\geq  \m$, then $E_k$ is real and we choose the positive square root of (\ref{green5}) as we did in the last subsection.
\item When $|\bm{k}| <\m$, then $E_k$ is purely imaginary, and we choose the positive imaginary square root of (\ref{green5}).
\end{itemize}
We can then write the integral (\ref{green8}) as a sum of two integrals :
\be
G_p(x) =  \left[\int_{|\bm{k}|<\m} {d^3\bm{k}\over (2\pi)^3} + \int_{|\bm{k}|\geq \m} {d^3\bm{k}\over (2\pi)^3} \right] e^{-i\bm{k}.\bm{x}}\int {d\omega\over 2\pi} e^{i\omega t}{1\over \omega^2 - E_k^2}
\label{green10}
\ee
\be
\equiv G_1(x) + G_2(x)
\ee
The first term of $G_p(x)$, for which $E_k$ is imaginary, is called $G_1$. The second term, for which $E_k$ is real, is called $G_2$.
The $G_2$ integral has poles on the real axis of $\omega$. The prescription for the contour can be chosen similarly as in the previous case with positive mass squared. In particular, the retarded prescription gives (\ref{green8}) with a UV cutoff at $|\bm{k}| = \m$. $G_1$, however, has poles on the imaginary axis of $\omega$. Therefore, the path along the real axis does not encounter any poles. As a consequence, the contour prescription is fixed.
The residue theorem then gives
\be
G_1(x) = i \int_{|\bm{k}|<\m} {d^3\bm{k}\over (2\pi)^3} e^{-i\bm{k}.\bm{x}} {e^{i E_k |t|} \over 2 E_k}.
\label{green11}
\ee
In this case, we already observe the absence of a runaway, because $iE_k<0$, so the integrand of (\ref{green11}) decreases exponentially with time. However, $G_1$ breaks the causality of $G_p$ because it is not zero for negative times, and this cannot be cancelled by $G_2$.

Following the same idea as in \cite{acausality}, we remark that an acausality can be traded with a runaway by changing the prescription of the Green function. To retrieve causality and build a retarded propagator, we now add the homogeneous solution $\f$ (\ref{green4}) to $G_p$.
First, we choose the Feynman prescription for $G_2$ (\ref{green9a}), such that the integrand of $G_2$ coincides with the one of (\ref{green11}) at $|\bm{k}| = k$, where $E_k = 0$. The most general Green function (\ref{green7}) is then given by
\be
G(x) =  \left[\int_{|\bm{k}|<\m} {d^3\bm{k}\over (2\pi)^3} + \int_{|\bm{k}|\geq \m} {d^3\bm{k}\over (2\pi)^3} \right]  e^{-i\bm{k}.\bm{x}} \left[ i {e^{i E_k |t|} \over 2 E_k}  + \a(\bm{k}) e^{iE_k t} +  \b(\bm{k}) e^{-iE_k t}  \right]
\nonumber
\ee
\be
 = \int_{\mathbb{R}^3} {d^3\bm{k}\over (2\pi)^3} e^{-i\bm{k}.\bm{x}} \left[ i {e^{i E_k |t|} \over 2 E_k}  + \a(\bm{k}) e^{iE_k t} +  \b(\bm{k}) e^{- i E_k t}  \right]
 \label{green12}
\ee
The retarded Green function corresponding to $G^\text{R}(t<0) = 0$ is obtained by setting
\begin{subequations}
\label{green13a}
\be
\a(\bm{k}) = 0,
\label{green13}
\ee
\be
\b(\bm{k}) = - {i \over 2 E_k}.
\label{green14}
\ee
\end{subequations}
The result for positive $t$ is given by
\be
G^\text{R}(t>0,\bm{x}) = - \int_{\mathbb{R}^3} {d^3\bm{k}\over (2\pi)^3} e^{-i\bm{k}.\bm{x}} {\sin(E_k t)\over E_k}.
\label{green15}
\ee
Again, this integral can be written as a sum of two integrals as $G^\text{R} = G_1^\text{R} + G_2^\text{R}$, where $G_2^\text{R}$ is the same as (\ref{green9}) with a UV cutoff at $|\bm{k}| = k$, and $G_1^\text{R}$ is different because $E_k = i\sqrt{\m^2 - \bm{k}^2}$ is purely imaginary. More explicitly,
\be
G_1^\text{R}(t>0,\bm{x}) = - \int_{|\bm{k}|<\m} {d^3\bm{k}\over (2\pi)^3} e^{-i\bm{k}.\bm{x}} {\sin(E_k t)\over E_k}.
\label{green16}
\ee
Since the integrand only depends on the modulus of $\bm{k}$, we define $k = |\bm{k}|$ and $r \equiv |\bm{x}|$ as new variables for our integral which is now written as
\be
G_1^\text{R}(t>0,r) = - {1\over 4\pi^2} \int_0^\m dkk^2 {\sin sr\over sr} {\sinh(t\sqrt{\m^2 - k^2})\over \sqrt{k^2 - s^2}}.
\label{green17}
\ee
For $r=0$, we have
\be
G_1^\text{R}(t>0,r=0) = - {1\over 4\pi^2} \int_0^\m dk k^2 {\sinh(t\sqrt{\m^2 - k^2})\over \sqrt{\m^2 - k^2}}
\nonumber
\ee
\be
= - {1\over 4\pi^2} \int_0^\m dE_k \sqrt{E_k^2 - k^2}\sinh(E_k t)
\label{green18}
\ee
For small times, it is easy to expand the sinh in (\ref{green18}) and integrate over $E_k$. The result is
\be
G_1^\text{R}(t\sim 0^+,r=0) = -{\m^2\over 16\pi} t + \mathcal{O}(t^2)
\label{green19}
\ee
The large-time asymptotic behaviour of (\ref{green18}) is obtained by remarking that the integral is proportional to the Struve function denoted $L_1(\m t)$. As a result,
\be
G_1^\text{R}(t>0, r=0) = -{1\over 8\pi}{\m\over t}L_1(\m t).
\label{green20}
\ee
This function behaves as the modified Bessel $K_1$ for large arguments. Therefore, we have
\be
G_1^\text{R}(t>0, r=0) \underset{\m t\rightarrow +\infty}{\propto} {e^{\m t} \over (\m t)^{3/2}}
\label{green21}
\ee
This diverges exponentially with time, where $\m$ is the inverse time scale.

\subsection*{Complex $m^2$}
In this subsection, we assume $m^2$ is complex, and we write its real and imaginary parts as
\be
m^2  = a + ib
\label{green22}
\ee
First, we relate $E_k$ (\ref{green5}) to the real and imaginary parts of $m^2$.
We  define the real and imaginary parts of $E_k$ as $A$ and $B$ respectively. We find that
\be
A = \pm {1\over 2}\left(\tilde{a} + \sqrt{\tilde{a}^2 + b^2}\right),
\label{green23}
\ee
\be
B = {b\over 2 A},
\label{green24}
\ee
where
\be
\tilde{a} \equiv a + \bm{k}^2.
\label{green25}
\ee
As in the previous cases, we choose arbitrarily one of the two square roots for $E_k$. We  pick the sign in (\ref{green23}) such that $B>0$. This is always possible since the case $b = 0$ was already studied in the previous two subsections. Therefore, the sign to pick in (\ref{green23}) should be the same sign as $b$ such that we have $B>0$.

The two poles $\pm E_k$ of $G_p$ (\ref{green8}) are now located on the complex plane, away from the real axis. Therefore, the contour prescription is fixed, as in the negative mass squared case. We then obtain the particular solution
\be
G_p(x) = i \int_{\mathbb{R}^3} {d^3\bm{k}\over (2\pi)^3} e^{-i\bm{k}.\bm{x}} {e^{i E_k |t|} \over 2 E_k}.
\label{green26}
\ee
This does not contain a runaway since $\text{Im}(E_k)>0$. However, the retarded Green function is obtained the same way as in (\ref{green12}), by adding the homogenous solution to $G_p$. By fixing $\a$ and $\b$ to the same values as in (\ref{green13a}), we obtain the retarded Green function
\be
G^\text{R}(t>0,\bm{x}) = - \int_{\mathbb{R}^3} {d^3\bm{k}\over (2\pi)^3} e^{-i\bm{k}.\bm{x}} {\sin(E_k t)\over E_k},
\label{green27}
\ee
and $G^\text{R}(t\leq 0) = 0$. This retarded Green function now contains a runaway, coming from the imaginary part of $E_k$. We would need to perform the integral over $\bm{k}$ to obtain the exact time dependence of this runaway. However, we can observe directly that the strongest runaway for large times will come from the largest imaginary part of $E_k$, $B$, which was chosen to be positive. In order to maximize $B$ (\ref{green24}) for a fixed $a$ and $b$, the only possibility is to minimize $\tilde{a}$, and therefore take $\bm{k}^2 = 0$. Therefore, the strongest runaway comes from the homogeneous mode $\bm{k} = \bm{0}$.  In this case, $E_k^2 = m^2$. $\sin(E_kt)$ diverges exponentially with time, and the inverse time scale is given by
\be
B = {|b|\over a + \sqrt{a^2 + b^2}}.
\label{green28}
\ee
Figure \ref{flatRek} shows the inverse time scale of the runaway corresponding to the worst possible mode $\bm{k}=\bm{0}$, as a function of the parameters of the theory in flat space. The inverse time scale corresponds to the real part of $k$. Indeed, for $\bm{k}^2 = 0$, we have
\be
E_k = \pm m.
\label{green29}
\ee
If $k$ is the square root of $k^2$ with the positive real part, then we have to take the $+$ branch of (\ref{green29}) and it follows that
\be
B = \text{Im}(m).
\label{green30}
\ee
In the main text, we are searching for poles of the propagator for the Laplacian eigenvalue defined by (\ref{flat5}), where $k^2 \equiv -m^2$. Therefore, the strength of a tachyonic pole where $k^2>0$ is then given by
\be
B = \text{Re}(k).
\ee

\section{Tachyonic tensor eigenmodes in de Sitter}
\label{dS criterion}

In this appendix, we determine which tensor perturbations are tachyonic.
 Since these perturbations are decomposed into eigenmodes of the
 covariant de Sitter Laplacian, we study the tachyon nature  of the  spin-two perturbation for a
 single eigenvalue $\n\in \mathbb{C}$ defined in (\ref{dS5}).
 We then obtain a criterion on the value of $\n$.

The perturbed expanding Poincar\'e coordinates of $d$-dimensional de Sitter are given by
\be
ds^2_{dS} =g^{(0)}_{\omega\s}dx^\omega dx^\s = \left[(H\tau)^{-2}\eta_{\omega\s} + \delta\zeta^b_{\omega\s}\right]dx^\a dx^\b.
\label{lic1}
\ee
The equation of motion for metric perturbations follows from the linearization of the Einstein equation (\ref{EE}). One can substitute the cosmological constant using the background equation (\ref{scalar7}).
The Einstein equation is then a sum of linear curvature terms, quadratic curvature terms and matter content in the stress tensor.
The Lichnerowicz operator $L[\delta \zeta^b]_{\omega\s}$ is defined by the variation of the linear curvature terms with respect to $\delta\zeta^b_{\omega\s}$ by
\be
L[\delta\zeta^b]_{\omega\s} = \delta \left(R_{\omega\s} - {1\over 2}R g^{(0)}_{\omega\s} - H^2{d(d-1)\over 2} g^{(0)}_{\omega\s} \right).
\label{lic2}
\ee
If we restrict to the transverse traceless perturbations $h^{(0)}_{\omega\s}$ (\ref{scalar2a}), then it takes the very simple form
\be
L[h^{(0)}]_{\omega\s} = \left(H^2 - {1\over 2}\nabla^2\right)h_{\omega\s},
\label{lic3}
\ee
where we recognize the last term of (\ref{dS3}).
The eigenvalue problem (\ref{dS5}) for $h_{(0)\a\b}$
is then just written in terms of the Lichnerowicz operator as
\be
-2H^{-2}L[h^{(0)}]_{\omega\s} = -\left(\n^2 - {(d-1)^2\over4}\right) h^{(0)}_{\omega\s}.
\label{lic5}
\ee
In order to write this equation as a set of scalar equations acting on each component of $h^{(0)}_{\omega\s}$, we define the new metric perturbation $\g_{\omega\s} = (H\tau)^2h^{(0)}_{\omega\s}$ such that the metric (\ref{lic1}) is now written
\be
ds^2_{dS} = (H\tau)^{-2}(\eta_{\omega\s} + \g_{\omega\s})dx^\omega dx^\s.
\label{lic6}
\ee
We now define a differential operator acting on the new metric perturbation $\g_{\a\b}$ as
\be
D[\g]_{\omega\s} \equiv -2\tau^2L\left[{\g\over\tau^2}\right]_{\omega\s}
\label{lic6a}
\ee
Equation (\ref{lic5}) is then written as
\be
D[\g]_{\omega\s} =  -\left(\n^2 - {(d-1)^2\over4}\right) \g_{\omega\s}.
\label{lic7}
\ee

We now try to write an explicit formula for $D[\g]$ in terms of $\g$. A direct computation using Poincar\'e coordinates (\ref{lic1}) gives
\be
D[\g]_{\omega\s} = H^{-2} \Box \g_{\omega\s} - 2d\g_{0(\omega}\delta^0_{\s)} + 4\tau \partial_{(\omega}\g_{\s)0} + 2\eta_{\omega\s}\g_{00},
\label{lic8}
\ee
where $\Box$ is the de Sitter Laplacian acting on scalars.
Derivatives with respect to $\tau$ are now labelled with the index $0$. The expression (\ref{lic8}) shows us that different components of $\g_{00}$ are coupled to each other in equation (\ref{lic7}). The strategy is now to further decompose (\ref{lic7}) into 3 equations; a scalar equation for $\g_{00}$, a vector equation for $\g_{0i}$ and a tensor equation for $\g_{ij}$ \footnote{Index $i$ and $j$ will refer to the spatial coordinates of the $d$-dimensional metric.}. This splitting is done by using
\begin{subequations}
\label{lic9}
\be
D[\g]_{00} = \left\{H^{-2}\Box - 2(d+1) + 4\tau\partial_0 \right\} \g_{00},
\label{lic9scalar}
\ee
\be
D[\g]_{0i} = \left\{ H^{-2}\Box - d + 2\tau\partial_0 \right\} \g_{0i} + 2\tau\partial_i \g_{00},
\label{lic9vector}
\ee
\be
D[\g]_{ij} = H^{-2}\Box \g_{ij} + 4\tau\partial_{(i }\g_{j)0} + 2\delta_{ij}\g_{00}.
\label{lic9tensor}
\ee
\end{subequations}
These three equations are still coupled because $\g_{00}$ contributes to $D[\g]_{0i}$, and $\g_{0\a}$ contributes to $D[\g]_{ij}$. Furthermore, these components are also related to each other through the transverse-traceless property of $h_{\omega\s}$. Indeed, the tracelessness condition (\ref{b11}) gives
\be
\eta^{\omega\s}\g_{\omega\s} = 0,
\label{lic10}
\ee
and transversality conditions (\ref{b12}) give
\be
\g_{0\omega} + {\tau\over d}\partial^\s\g_{\s\omega} = 0.
\label{lic11}
\ee
After taking these constraints into account, $\g_{\omega\s}$ only propagates 5 degrees of freedom.
We now solve equations (\ref{lic9}). To do that, we first give the scalar Laplacian of the de Sitter background (\ref{lic1}) as
\be
H^{-2}\Box = \tau^2(-\partial_0^2 + \partial_i^2 )+ \tau(d-2)\partial_0.
\label{lic12}
\ee
The first step is to diagonalize the $(d-1)$-dimensional euclidean Laplacian $\delta^{ij}\partial_i\partial_j = \partial_i^2$ using a Fourier transform \footnote{The convention of the Fourier transform is chosen to be $\g_{\omega\s} = \int {dk^i\over (2\pi)^{d-1}}e^{-ik^iy^i}\tilde{\g}_{\omega\s}$}.
In Fourier space, the scalar de Sitter Laplacian  acting on a Fourier mode $\tilde{\g}_{\omega\s}$ is given by \footnote{a slight abuse of notation allows us to write the scalar Laplacian in momentum space the same way as we write it in real space.}
\be
H^{-2}\Box= -\tau^2\left(\partial_0^2 + \bm{k}^2 \right) +  (d-2)\tau\partial_0,
\label{lic13b}
\ee
where the momentum squared is defined as
\be
\bm{k}^2 \equiv \delta_{ij}k^ik^j.
\label{lic13d}
\ee
\paragraph*{Scalar equation}\ \\
The only equation in (\ref{lic9}) which involves only one component of the perturbation $\g_{\a\b}$ is the scalar equation (\ref{lic9scalar}). The solution to the eigen-problem (\ref{lic7}) is then any linear combination of the solutions $\tilde{\g}_{00}^\pm$ written as
\be
\tilde{\g}_{00}^\pm = (k\tau)^{3+d\over 2}J_{\pm\n}(k\tau) \underset{\tau\rightarrow 0}{\sim} (k\tau)^{{3+d\over 2} \pm \n},
\label{lic14}
\ee
where $\lambda_\pm$ are integration constants, they do not depend on $\tau$. We also defined $k=\sqrt{\bm{k}^2}$. Since we are only interested in the eventual divergence of $\tilde{\g}_{00}$ for large time ($H \tau = e^{-2Ht} \rightarrow 0$), then (\ref{lic14}) shows that this scalar quantity is unstable if
\be
|\text{Re}(\n)| > {3+d\over 2}.
\label{lic14a}
\ee
The exact solution (\ref{lic14}) for $\tilde{\g}_{00}$ can then be used to find an exact solution for $\tilde{\g}_{0i}$ (\ref{lic9vector}).

\paragraph*{Vector equation}\ \\
Instead of solving (\ref{lic9vector}), which is an inhomogeneous equation for $\g_{0i}$ coupled to $\g_{00}$, we can use the transversality constraint (\ref{lic11}). We split $\g_{0i}$ into a transverse and a longitudinal component. In Fourier space (of the 3-dimensional spatial coordinates)
\be
\tilde{\g}_{0i} = b_i + k_i b,
\label{lic14b}
\ee
such that
\be
k_i b_i = 0,
\label{lic14c}
\ee
then transversality (\ref{lic11}) fixes the longitudinal part in terms of $\g_{00}$ which we already solved. It gives
\be
ik^2 b = \left(-\partial_0 + {d\over \tau}\right)\tilde{\g}_{00}.
\label{lic14d}
\ee
We already know the solution (\ref{lic14}) for $\tilde{\g}_{00}$. Therefore, $b$ is given by a linear combination of the two independent solutions $b^\pm$ given by
\be
i k b^\pm = (k\tau)^{1+d\over 2} \left[{d-3\over 2} - (k\tau) {d\over d(k\tau)}\right]J_{\pm \n}(k\tau),
\label{lic14e}
\ee
which for large time evaluates to
\be
i k b^\pm \underset{\tau\rightarrow 0}{\sim} \left({d-3\over 2} - \n \right)(k\tau)^{{1+d\over 2} \pm \n}.
\label{lic14f}
\ee
The transverse part of (\ref{lic9vector}) in Fourier space is obtained by applying the transverse projection operator given by
\be
b_i = \left(\delta_{ij} - {k_ik_j\over k^2}\right)\tilde{\g}_{0j}.
\label{lic15a}
\ee
This projection is applied to the "$0i$" component of equation (\ref{lic7}) to obtain
\be
\left\{H^{-2}\Box + \n^2 - {(d-1)^2\over 4}  - d + 2\tau\partial_0\right\}\tilde{b}_i = 0.
\label{lic15b}
\ee
Again, this equation is solved using Bessel functions as any linear combination of the two independent solutions $b_i^\pm$ given by
\be
b_i^\pm = (k\tau)^{1+d\over 2} J_{\pm\n}(k\tau).
\label{lic16}
\ee
To conclude on the vector perturbation, both longitudinal and transverse parts of $\g_{0i}$ are unstable if
\be
|\text{Re}(\n)| > {d+1\over 2}.
\label{lic17}
\ee
By comparing with the result from the scalar component, the scalar instability criterion (\ref{lic14a}) is stronger than the vector result (\ref{lic17}). A scalar instability implies a vector instability.

\paragraph*{Tensor equation}\ \\
As we did for the vector perturbation, we decompose $\tilde{\g}_{ij}$ into
\be
\tilde{\g}_{ij} = \theta_{ij} + k_ik_j \f + 2k_{(i}\mathcal{V}_{j)} + \delta_{ij}\Psi,
\label{lic18}
\ee
such that
\be
\delta^{ij}\t_{ij} = k_i\mathcal{V}_i = 0
\label{lic19}
\ee
and
\be
k_i\t_{ij} = 0.
\label{lic20}
\ee
Every quantity defined in (\ref{lic18}) can be expressed as a projection of $\tilde{\g}_{ij}$, where the projector is a function of $k_i$, as it was the case for the vector transverse projection in (\ref{lic15a}).
We now search for a solution for each quantity in (\ref{lic18}).

First, the tracelessness constraint (\ref{lic10}) written in terms of the decomposition (\ref{lic18}) is
\be
\g_{00} = (d-1)\Psi + k^2\f.
\label{lic20a}
\ee
The spatial components of the transversality constraint (\ref{lic12}) are given by
\be
k_i\left[\left({d\over \tau} - \partial_0\right)b - i\Psi - i k^2 \f \right] + \left(\partial_0- {d\over \tau}\right)b_i - \partial_j^2\mathcal{V}_i = 0.
\label{lic20b}
\ee
The longitudinal part of equation (\ref{lic20b}) is
\be
\left[\left({d\over \tau} - \partial_0\right)b -i(\Psi +k^2 \f) \right] = 0,
\label{lic20c}
\ee
and the transverse part is obtained using the same projection as in (\ref{lic15a}):
\be
-i k^2 \tilde{\mathcal{V}}_i + \left({d\over \tau} - \partial_0\right)b_i = 0.
\label{lic20d}
\ee
The three constraints in momentum space (\ref{lic20a},\ref{lic20c},\ref{lic20d}) are solved algebraicaly for $\f$, $\Psi$ and $\mathcal{V}_i$ in terms of the known solutions $\g_{00}$, $b$ and $b_i$. The result is
\begin{subequations}
\label{lic20e}
\be
\Psi  = {1\over d-2}\left[\tilde{\g}_{00} + i\left({d\over \tau}- \partial_0\right)\tilde{b}\right],
\label{lic20ePsi}
\ee
\be
k^2\tilde{\f}  = - {1\over d-2}\left[\tilde{\g}_{00} + i (d-1)\left({d\over \tau} - \partial_0\right)\tilde{b}\right],
\label{lic20efi}
\ee
\be
-k^2\tilde{\mathcal{V}}_i  = \left(\partial_0- {d\over \tau}\right)\tilde{b}_i.
\label{lic20eV}
\ee
\end{subequations}
From the solutions for $\tilde{\g}_{00}$ (\ref{lic14}), for $b$ (\ref{lic14f}) and $b_i$ (\ref{lic16}), one can observe that each of these three perturbations scales for large time as $\sim \tau^{{d-1\over 2} \pm\n}$. We now solve the transverse-traceless part of the eigenproblem (\ref{lic7}) which is simply given by
\be
\left\{\Box  + \n^2 -\left({d-1\over 2}\right)^2 \right\}\theta_{ij} = 0
\label{lic24}
\ee
Therefore, the solution to the spatial tensor part of (\ref{lic7}) (which is also a Bessel equation) is
\be
\tilde{\t}_{ij} = \lambda_{ij}^{\pm}(k^i)(k\tau)^{d-1\over 2} J_{\pm}(k\tau).
\label{lic25}
\ee
We therefore find that the instability conditions on $\t_{ij}$, $\f$, $\Psi$ and $\mathcal{V}_i$ all read
\be
|\text{Re}(\n)| > {d-1\over 2},
\label{lic26}
\ee
which is weaker than the vector criterion (\ref{lic17}).
Therefore, the existence of an instability relies only on the tensor component $\g_{ij}$. The criterion (\ref{lic26}) was also derived in \cite{Chesler} but using a different decomposition which made all 5 degrees of freedom equally unstable.

\section{Tachyonic tensor eigenmodes in anti de Sitter}
\label{AdS criterion}

In this appendix, we derive a criterion on the eigenvalue $\n$ which determines if a given mode is tachyonic or not.
All the steps done in the previous appendix \ref{dS criterion} can be adapted to AdS in Poincar\'e coordinates by changing the metric (\ref{lic6}) to
\be
ds^2_{AdS} = (\chi z)^{-2}(dz^2 -dt^2 + d\bm{x}^2) +  h_{\a\b}dx^\a dx^\b,
\label{cr1}
\ee
\be
 = \left[(\chi z)^{-2}\eta_{\omega\s} + h_{\omega\s}\right]dx^\omega dx^\s.
 \label{cr1a}
\ee
where $\bm{x}$ is a (d-2)-dimensional vector, and we define $\eta_{\omega\s} = \text{diag}(+1,-1,+1,...,+1)$.
Greek indices $\omega,\s$ are for the full set of $d$-dimensional coordinates. We also use the index "0" for the time coordinate $t$, and roman letters such as $i,j$ for $(t,\bm{x})$.
As we did for de Sitter slicing, the eigenproblem (\ref{AdS3}) is studied using the rescaled perturbation
\be
\g_{\omega\s} = (\chi z)^2 h_{\omega\s}.
\label{cr2}
\ee
Similarly to de Sitter (\ref{lic7}), we define a differential operator from the left-hand side of the eigenproblem (\ref{AdS3}). It is written as
\be
\bar{D}[\g]_{\omega\s}\equiv z^2(\chi^{-2}\nabla^2 + 2)(\g_{\omega\s}z^{-2}),
\label{cr3}
\ee
And the eigen-problem (\ref{AdS3}) is then written as
\be
\bar{D}[\g]_{\omega\s} = \left[\n^2 - \left({d-1\over 2}\right)^2\right]\g_{\omega\s}.
\label{cr3a}
\ee
Doing the same computation which resulted in (\ref{lic8}), we obtain
\be
\bar{D}[\g]_{\omega\s} = \Box \g_{\omega\s} + 2d\delta^z_{(\omega}\g_{\s)z} - 4z\partial_{(\omega}\g_{\s)z} + 2\eta_{\omega\s}\g_{zz}.
\label{cr4}
\ee
We are now going to solve this equation in momentum space as we did for de Sitter because it allows us to solve ordinary differential equations.
The scalar AdS Laplacian appearing in (\ref{cr4}) in Poincar\'e coordinates is given by
\be
\chi^{-2}\Box = z^2(\partial_z^2 - \partial_t^2 + \partial_{\bm{x}}^2) - (d-2)z\partial_z
\label{cr5}
\ee
In de Sitter, we diagonalized the $(d-1)$-dimensional Laplacian $\partial_i^2$ using Fourier modes. However,
 in AdS, the Laplacian we want to diagonalize is the $(d-1)$-dimensional Minkowski space Laplacian as
\be
(-\partial_0^2 + \partial_{\bm{x}}^2)\g_{\omega\s} = \left[-(k^0)^2 + \bm{k}^2\right]\g_{\omega\s} \equiv  -k^2 \g_{\omega\s}.
\label{cr5aa}
\ee
 which can contain complex eigenvalues $k^2$ if the Laplacian is not self-adjoint. In particular, if we assume self-adjointness of the spatial part $\partial_{\bm{x}}^2$ but allow for solutions which diverge with time, the imaginary part of $k^2$ will be contained in $(k^0)^2$. A frequency squared $(k^0)^2$ has two square roots $k^0$. Its imaginary part is then responsible for an exponentially growing solution of (\ref{cr5aa}) corresponding to one of the two square roots $k^0$.
Using (\ref{cr5aa}), the scalar Laplacian of AdS in Poincar\'e coordinates is given by
\be
\chi^{-2}\Box = z^2(\partial_z^2- k^2) - (d-2) z\partial_z.
\label{cr5a}
\ee
We then decompose $\g_{\omega\s}$ into as we did for de Sitter. We had seen that it was enough to solve the eigenproblem (\ref{AdS3}) for 3 quantities $\g_{00}$, $b_i$ and $\t_{ij}$ defined in (\ref{lic14b}) for the transverse vector and (\ref{lic18}) for the tensorial decomposition. All the other quantities (namely $b$, $\f$, $\Psi$ and $\mathcal{V}$) are constrained by the transverse-traceless properties of $h_{\omega\s}$ (\ref{b11},\ref{b12}), so they cannot represent new instabilities. In Poincar\'e AdS (\ref{cr1a}), the constraints read
\be
\eta^{\omega\s}\g_{\omega\s} = 0,
\label{cr6traceless}
\ee
\be
{\tau\over d}\partial^\s\g_{\omega\s} = \g_{0\omega}.
\label{cr6transverse}
\ee
It is therefore enough to study three equations obtained from (\ref{cr3a}). They are given by:
\begin{itemize}
\begin{subequations}
\label{cr6d}
\item the $zz$ component of (\ref{cr3a})
\be
\left\{\chi^{-2}\Box - \n^2 + \left({d-1\over 2}\right)^2 + 2(d+1) - 4z\partial_z\right\}\g_{zz} = 0,
\label{cr6a}
\ee
\item the transverse part of the $0i$ component of (\ref{cr3a})
\be
\left\{\chi^{-2}\Box -\n^2 +  \left({d-1\over 2}\right)^2 + d - 2z \partial_z\right\} b_i = 0,
\label{cr6b}
\ee
\item and the transverse-traceless part of the $ij$ component of (\ref{cr3a})
\be
\left\{\chi^{-2}\Box - \n^2 + \left({d-1\over 2}\right)^2\right\} \t_{ij} = 0.
\label{cr6c}
\ee
\end{subequations}
\end{itemize}
Since all these equations are solved by (modified) Bessel functions, we can bring them to the same form.
We define the spin $s=$ 0,1 or 2 perturbation $\mathcal{Z}_s$ being either equal to $\tilde{\g}_{00}$ for $s=0$, $b_i$ for $s=1$ or $\t_{ij}$ for $s=2$. A rescaled function $\psi_s$ is also defined as
\be
\psi_s(w,k^i) = e^{-n_s w}\mathcal{Z}_s(w,k^i),
\label{cr10}
\ee
with the radial coordinate $w \equiv \log z$.
A direct computation using (\ref{cr5a}) into equations (\ref{cr6d}) shows that each $\psi_s$ satisfies the same Schrodinger equation
\be
\left[-{d^2\over dw^2} +  k^2 e^{2w}\right]\psi_s = - \n^2 \psi_s
\label{cr11}
\ee
 if
\be
n_s = {d+3\over 2} - s.
\label{cr12}
\ee
The most general solution of (\ref{cr11}) is
\be
\psi_s = \lambda_s^\pm I_{\pm\n}(ke^w),
\label{cr7}
\ee
where $k$ is the square root of $k^2$ with positive real part.
We now study its solutions depending on the sign of $\text{Re}(k)$.
\begin{itemize}
\item If $\text{Re}(k) = 0$, this corresponds to timelike $k^2<0$ modes which possess real frequencies $k^0$. These modes do not diverge with time and are timelike. They are therefore stable. The solution (\ref{cr7}) is then evaluated at an imaginary argument, which gives Bessel functions (not modified)
\be
\psi_s = \lambda_s^\pm J_{\pm\n}(|k|e^w).
\label{cr7a}
\ee
The timelike solution which is regular at $z\rightarrow 0$ is given by one of these two Bessel functions depending on the sign of the real part of $\n$.
In conclusion, timelike solutions allow any eigenvalue $\n\in \mathbb{C}$.

\item
We now turn to complex $k^2$, with $|\text{Arg}(k^2)|<\pi$ which are necessarily unstable. For example, $k^2>0$ corresponds to a tachyonic mode. Since $k^2$ contains two square roots, we take the $k$ with a positive real part.
The linear combination of (\ref{cr7}) solutions which is regular at the horizon $z\rightarrow +\infty$ is the Bessel $K_\n$:
\be
\psi_s = \lambda_s K_{\n}(ke^w) \underset{z\rightarrow +\infty}{\rightarrow} \sqrt{ \pi\over 2 k z} \exp\left\{-ke^w\right\}.
\label{cr8}
\ee
\end{itemize}
As one could already observe from (\ref{cr1}), the radial coordinate of AdS is not time but $z$. Furthermore, the eigenproblem (\ref{AdS3}) defines a momentum $\n$ which is dual to $z$ in AdS. In dS, the value of $\n$ was fixing the characteristic time of the instability. In AdS, however, the boundary does not correspond to infinite time.
In order to address stability in AdS, we turn to a viewpoint similar to the BF bound analysis \cite{BF-1982}. The stability condition is following:
{For a given $\n\in \mathbb{C}$, AdS is unstable if there exist a regular solution $\psi_s$ with $\text{Re}(k)\neq 0$.}

To address the stability of a given $\n$, we then need to look if there is a $\psi_s$ solution (\ref{cr8}) which is regular at $w\rightarrow -\infty$, since (\ref{cr8}) is already exponentially decreasing at $w\rightarrow +\infty$.
\begin{itemize}
\item If $Re(\n)> 0$, the leading behavior is then
\be
K_\n(ke^w) \underset{w \rightarrow- \infty}{\sim} {\pi\over 2\sin(\pi\n)} {2^{\n} k^{-\n} e^{- w\n} \over \G(1-\n)} \rightarrow \infty.
\label{cr15}
\ee
\item If $Re(\n) < 0$
\be
K_\n(ke^w) \underset{w \rightarrow- \infty}{\sim} {\pi\over 2\sin(\pi\n)} {2^{-\n} k^{\n} e^{w\n} \over \G(1+\n)} \rightarrow \infty.
\label{cr15a}
\ee
\item If $\text{Re}(\n) = 0$, or $\n = i\m$ for $\m\in \mathbb{R}$, the solution is a combination of plane waves:
\be
\psi_s(ke^w)  \underset{w \rightarrow- \infty}{\rightarrow}  {1\over 2i\m} \left[\left({k\over 2}\right)^{-i\m} \G(1+i\m) e^{-i\m w} - \left({k\over 2}\right)^{i\m} \G(1-i\m) e^{i\m w}  \right].
\label{cr16}
\ee
\end{itemize}

We therefore find plane-wave normalizable $\psi_s(w)$ for imaginary $\n$ and spacelike ($k^2>0$) modes. We conclude that $\n = i\m$ allows for a regular solution with non-zero $\text{Re}(k)$. These modes are unstable eigenvalues of the operator $\bar{D}[\g]$ defined for AdS slicing in (\ref{cr3}).

To conclude, the only possibility for a spacelike $k^2>0$ mode to be regular both at the AdS boundary $w \rightarrow -\infty$ and the horizon $w\rightarrow +\infty$ is to have $\text{Re} (\n) = 0$. It agrees with the usual BF bound $\n^2<0$ \cite{BF-1982} in the limit where $\n^2$ is real. Indeed, the usual BF bound analysis was done for real mass squared, which is the eigenvalue of the scalar Laplacian. By looking at the equation for $\t_{ij}$ (\ref{cr6c}), we can observe that the mass squared is identified to $\n^2-{(d-1)^2\over 4}$.

\section{Asymptotic behaviour of Legendre functions}
\label{app:legendre}

This appendix is devoted to the asymptotic behaviour of associated Legendre functions of argument $x \in ]-1,1[$, which enter into the solution of tensor modes in the bulk with AdS$_3$ slicing. This appendix allows us to relate boundary conditions (see \ref{AdS one sided} and \ref{AdS z2 symmetric}) to the value of constants of integrations $\l_1,\l_2$ in (\ref{AdS6}).

In the AdS-slicing case (\ref{AdS prop}), the bulk equation of motion for tensor perturbations (\ref{AdS1}) can be written as a Legendre function for the holographic coordinate $u$ (\ref{AdS5}). Its solutions are given by the linear combination
\be
F(u,\n) = (\cosh u)^{-2}\left(\lambda_1 P_{\nu -1/2}^2(\tanh{u}) + \lambda_2 Q_{\nu-1/2}^2(\tanh{u})\right).
\label{legendre1}
\ee
In the single-boundary case \ref{AdS one sided}, it is easy to fix the integration constants of (\ref{legendre1}) because Legendre functions are defined as hypergeometric series in $\tanh{u}$ and $\lambda_1 (\cosh u)^{-2} P_{\nu -1/2}^2(\tanh{u})$ is the only solution which vanishes at $u\rightarrow +\infty$. Therefore, imposing $F(u,\n)\rightarrow 0$ at $u \rightarrow +\infty$ sets $\lambda_2 = 0$ and one can obtain the expansion of the remaining solution using the usual expression for associated Legendre functions in terms of hypergeometric series :
\be
P_\n^\m(x) = {1\over \Gamma(1-\m)}\left({1+x\over 1-x}\right)^{\m\over 2} \tensor[_2]{F}{_1}\left(-\n, \n + 1; 1-\m;{1-x\over 2}\right), x\in[-1,1].\label{v20}
\ee
The formula (\ref{v20}) is applied to (\ref{legendre1}) for
\be
x = \tanh{u}.
\label{k4}
\ee
The formula (\ref{v20}) is convenient for an expansion close to $x\rightarrow 1$ using the definition of the hypergeometric series (from now we use the short notation $F = \tensor[_2]{F}{_1}$)
\be
F(a,b;c;x) = \sum_{n=0}^{+\infty}{(a)_n(b)_n\over (c)_n}{x^n\over n!}, \label{l0}
\ee
where
\be
(a)_n \equiv a(a+1)...(a+n-1).
\label{l1}
\ee
Using (\ref{v20}) into (\ref{v20}) gives
\be
(1-x^2)P_{\n-1/2}^2(x) = 2\left({1-x\over 2}\right)^2  \left(\n^2-{9\over 4}\right) \left(\n^2-{1\over 4}\right) + \mathcal{O}((1-x)^3).
\label{k1a}
\ee
Therefore, the only contribution to $F(x=1)$ comes from $Q^2_{\n-1/2}$, which is fixed to zero by the boundary condition (\ref{as1a}).
However, for other boundary conditions, we may need to expand (\ref{legendre1}) close to $x=-1$.
The hypergeometric transformation adapted to the limit $\m\rightarrow 2$. This transformation is given on page 49 of \cite{Magnus}, or (15.3.12) of
\href{http://www.math.ubc.ca/~cbm/aands/page_560.htm}{Abramowitz and Stegun - Hypergeometric functions}. We reproduce it here for consistency :
\be
F(a,b;a+b-\m;y) = {\Gamma(\m)\Gamma(a+b-\m)\over \Gamma(a)\Gamma(b)}\sum^{m-1}_{n=0}{(a-\m)_n(b-\m)_n\over n!(1-\m)_n}(1-y)^{n-m} \nonumber
\ee
\be
-{(-1)^\m\Gamma(a+b-\m)\over \Gamma(a-\m)\Gamma(b-\m)}\sum^{+\infty}_{n=0}{(a)_n(b)_n\over n!(n+\m)!}(1-y)^n\left[\log(1-y) \right.\nonumber
\ee
\be
\left. -\psi(n+1)-\psi(n+\m+1)+\psi(a+n)+\psi(b+n)\right], \label{k2}
\ee
valid for $|1-y|<1$ and $|\text{Arg}(1-y)|<\pi$. This is the case because we take
\be
y = {1-x\over 2},\label{k3}
\ee
Therefore, the small parameter of our expansion at $u \rightarrow -\infty$ is going to be
\be
1-y = {1+x\over 2} \in ]0,1[.
\label{k5}
\ee
Applying (\ref{k2}) to
\be
a = {1\over 2} - \n, \label{k6a}
\ee
\be
 b = \n + {1\over 2} \label{k6b},
\ee
we obtain
\be
(1-x^2)P^2_{\n-1/2}(x) = - 4{\cos(\pi\n)\over \pi}\bigg\{1 + \left(\n^2-{9\over 4}\right)\left({1+x\over 2}\right) - \nn
\ee
\be
{1\over 2}\left({1+x\over 2}\right)^2 \left(\n^2-{9\over 4}\right) \left(\n^2-{1\over 4}\right) \nonumber
\ee
\be
 \times \left[\log\left({1+x\over 2}\right) - {3\over 2} + \mathcal{H}\left(\n-{1\over 2}\right) + \mathcal{H}\left(-\n-{1\over 2}\right) \right]\bigg\} +\mathcal{O}\left((1+x)^3\right) \label{k7}
\ee
We now compute the expansions of $Q_{\n-1/2}^\m(x)$ near the two boundaries.
The first one can be obtained using the formula from page 170 of \cite{Magnus}
\be
P^\m_\n(-x) = \cos(\pi(\n+\m))P^\m_\n(x) - {2\over \pi}\sin(\pi(\n+\m))Q^\m_\n(x) \label{k8},
\ee
valid for $0<x<1$.
Isolate $Q^\m_\n(x) $ and take $\m=2$:
\be
Q^2_{\n-1/2}(x) = -{\pi\over 2\cos(\pi\n)}\left[\sin(\pi\n)P^2_{\n-1/2}(x) - P^2_{\n-1/2}(-x)\right]. \label{k9}
\ee
The expansion in powers of $(1-x)$ for the second term is simply given by (\ref{k7}) evaluated at $-x$.
The first term is (\ref{k1a}).
The result is
\be
(1-x^2)Q^2_{\n-1/2}(x) = - 2\left\{ 1 + \left(\n^2-{9\over 4}\right)\left({1-x\over 2}\right)- {1\over 2}\left({1-x\over 2}\right)^2 \left(\n^2-{9\over 4}\right) \left(\n^2-{1\over 4}\right)\right. \nonumber
\ee
\be
\times \left.\left[\pi\tan(\pi\n) + \log\left({1-x\over 2}\right) - {3\over 2} + \mathcal{H}\left(\n-{1\over 2}\right) + \mathcal{H}\left(-\n-{1\over 2}\right) \right]\right\}   + \mathcal{O}\left((1-x)^3\right). \label{k10}
\ee

The last expansion we need is $Q^2_{\n-1/2}(x)$ close to $x = -1$. This can be obtained using another formula from page 170 of \cite{Magnus}:
\be
Q^\m_\n(-x) = -\cos(\pi(\n+\m))Q^\m_\n(x) - {\pi\over 2} \sin(\pi(\n+\m))P^\m_\n(x) \label{k11}
\ee
valid for $0<x<1$.
Therefore, for $-1<x<0$, we can use
\be
Q^\m_\n(x) = -\cos(\pi(\n+\m))Q^\m_\n(-x) - {\pi\over 2} \sin(\pi(\n+\m))P^\m_\n(-x) \label{k12}
\ee
where the first term expansion is given by (\ref{k10}) evaluated at $-x$ and the second term is given by (\ref{k1a}) evaluated at $-x$ too.
The result is given by
\be
(1-x^2)Q^2_{\n-1/2}(x) = 2\sin(\pi\n)\left\{1 + \left(\n^2-{9\over 4}\right)\left({1+x\over 2}\right)- \right. \nn
\ee
\be
{1\over 2}\left({1+x\over 2}\right)^2 \left(\n^2-{9\over 4}\right) \left(\n^2-{1\over 4}\right) \bigg[\pi\tan(\pi\n) - \pi\cot(\pi\n) + \nn
\ee
\be
\left.\left. \log\left({1+ x\over 2}\right) - {3\over 2} + \mathcal{H}\left(\n-{1\over 2}\right) + \mathcal{H}\left(-\n-{1\over 2}\right) \right]\right\}  + \mathcal{O}\left((1+x)^3\right). \label{k13}
\ee

Finally, the two different expansions for $F(x)$ are given by
\be
F(x) \underset{x\rightarrow 1}{=} 2\lambda_1 \left({1-x\over 2}\right)^2  \left(\n^2-{9\over 4}\right) \left(\n^2-{1\over 4}\right)  \nonumber
\ee
\be
+ 2\lambda_2 \left\{ 1 + \left(\n^2-{9\over 4}\right)\left({1-x\over 2}\right)- {1\over 2}\left({1-x\over 2}\right)^2 \left(\n^2-{9\over 4}\right) \left(\n^2-{1\over 4}\right)\right. \nonumber
\ee
\be
\times \left.\left[\pi\tan(\pi\n) + \log\left({1-x\over 2}\right) - {3\over 2} + \mathcal{H}\left(\n-{1\over 2}\right) + \mathcal{H}\left(-\n-{1\over 2}\right) \right]\right\}   + \mathcal{O}\left((1-x)^3\right) \label{k13a}
\ee
\be
F(x) \underset{x\rightarrow -1}{=} \left( {4\lambda_1\cos\pi\n\over \pi}- 2\lambda_2\sin\pi\n \right) \left\{ 1 + \left(\n^2-{9\over 4}\right)\left({1+x\over 2}\right)- \right. \nonumber
\ee
\be
{1\over 2}\left({1+x\over 2}\right)^2 \left(\n^2-{9\over 4}\right) \left(\n^2-{1\over 4}\right) \left.\left[ \log\left({1+x\over 2}\right) - {3\over 2} + \mathcal{H}\left(\n-{1\over 2}\right) + \mathcal{H}\left(-\n-{1\over 2}\right) \right]\right\}  \nonumber
\ee
\be
+ {\pi\lambda_2\over \cos\pi\n}\left({1+x\over 2}\right)^2 \left(\n^2-{9\over 4}\right) \left(\n^2-{1\over 4}\right) + \mathcal{O}\left((1-x)^3\right) \label{k13b}
\ee
The limits of $F(x)$ on both direction are
\be
F(u) \longrightarrow
\left\{
\begin{array}{lcr}
2\lambda_2& \text{when} &u\longrightarrow +\infty \\
{4\lambda_1\over \pi}\cos(\pi\nu) - 2\lambda_2\sin(\pi\nu) & \text{when} & u\longrightarrow -\infty
\end{array}
\right.
\label{i55}
\ee
In order to read out the vacuum expectation value term $h^{(4)}$ from the Fefferman-Graham expansion (\ref{b15}), we first take the coordinate transformation (\ref{k4}) to obtain
\be
{1\pm x\over 2} = e^{\pm 2u}(1-e^{\pm 2u}) + \mathcal{O}(e^{\pm 4u}) .\label{k13d}
\ee
And the Fefferman-Graham coordinate $\rho^{\pm}$ is related to $u$ by
\be
e^{\mp 2u} = \left({L\chi\over 2}\right)^2\rho^\pm. \label{k13e}
\ee
In terms of $u$, the expansions of $F$ near each boundary is then given by
\be
F(u) \underset{u\rightarrow +\infty}{=} 2\lambda_1e^{-4u}\left(\n^2-{1\over 4}\right)\left(\n^2-{9\over 4}\right) +
\nonumber
\ee
\be
2\lambda_2\left\{1 + \left(\n^2 - {9\over 4}\right)e^{-2u} - e^{-4u}\left(\n^2-{9\over 4}\right)\bigg[ 1 + \right.
\nonumber
\ee
\be
\left.\left.{1\over 2}\left(\n^2-{1\over 4}\right)\left(\pi\tan(\pi\n) -2u - {3\over 2} + \mathcal{H}\left(\n-{1\over 2}\right) + \mathcal{H}\left(-\n-{1\over 2}\right)\right)\right]\right\}
\label{Fu exp plus}
\ee
\be
F(u)  \underset{u\rightarrow -\infty}{=} \left({4\lambda_1\over \pi}\cos (\pi\n) -2\lambda_2 \sin(\pi\n) \right)\left\{ 1 + \left(\n^2-{9\over 4}\right)e^{2u} - \right.
\nonumber
\ee
\be
\left. e^{4u} \left(\n^2-{9\over 4}\right)\left[1 + {1\over 2}\left(\n^2-{1\over 4}\right)\left(2u - {3\over 2} + \mathcal{H}\left(\n-{1\over 2}\right) + \mathcal{H}\left(-\n-{1\over 2}\right)\right)\right]\right\} +
\nonumber
\ee
\be
\lambda_2 {\pi\cos(\pi\n)} e^{4u}\left(\n^2-{1\over 4}\right)\left(\n^2-{9\over 4}\right)
\label{Fu exp minus}
\ee

Replacing in (\ref{Fu exp plus}, \ref{Fu exp minus}) the integration constants $\lambda_1$ and $\lambda_2$ by boundary conditions (\ref{as1a}) gives the result (\ref{as5}).
In the case of symmetric boundary conditions (\ref{sym1}), the bulk radial solution is given by
\be
F_\text{sym}(u) \underset{u\rightarrow \pm\infty}{=} 1 + e^{\mp 2u}\left(\n^2-{9\over 4}\right)- e^{\mp 4u}\left(\n^2-{9\over 4}\right)\bigg[ 1+
\nonumber
\ee
\be
{1\over 2}\left(\n^2-{1\over 4}\right)\left(\mp 2u -{\pi\over \cos(\pi\n)}-{3\over 2} +  \mathcal{H}\left(\n-{1\over 2}\right) + \mathcal{H}\left(-\n-{1\over 2}\right)\right)\bigg].
\label{sym9}
\ee
From this solution, one can then identify equations (\ref{AdS2}) with (\ref{b15}) to obtain all the terms in (\ref{as7}).

\section{Quadratic action for tensor modes}
\label{quadratic action}

In this appendix, we obtain the quadratic terms of the boundary action (\ref{s0}) for tensor perturbations.
The scalar case is subtle and we discuss it separately in the next appendix.

%
We start from the action (\ref{ig11a}):
\be
S[g^{(0)}] = S_\text{CFT}[g^{(0)}] - {1\over 16\pi G}\int\sqrt{g^{(0)}}\left\{R - 2\Lambda + {\a G\over 24}R^2 + 4\b G \left(R^{\omega\s}R_{\omega\s} - {1\over 3}R^2\right)\right\}
\label{res1}
\ee
Doing an expansion in the metric with respect to a spin-2 TT perturbation $h^{(0)}_{\omega\s}$ as
\be
g^{(0)}_{\omega\s} = \bar{\zeta}_{\omega\s} + \epsilon h^{(0)}_{\omega\s},
\label{res2}
\ee
where $\epsilon$ is a book-keeping parameter.
A Taylor expansion of the action (\ref{res1}) in $\epsilon$ is written as
\be
S[g_{(0)}] = S[\bar{\zeta}] + \epsilon S^{(1)}[\bar{\zeta}] + {\epsilon^2\over 2}S^{(2)}[\bar{\zeta}] + \mathcal{O}(\epsilon^3).
\label{res3}
\ee
The linear order $S^{(1)}$ is already given by the generalized Einstein tensor computed in (\ref{scalar4}). This has to be evaluated on the background metric $\bar{\zeta}_{\omega\s}$, and the general metric perturbation is specialized to the TT mode $h^{(0)}$, so $\delta g^{(0)\omega\s} = h^{(0)\omega\s}$. We recall here the definition (\ref{def E}) of the generalized Einstein tensor:
\be
S^{(1)}[g_{(0)}] = -{1\over 16\pi G}\int \sqrt{g_{(0)}}h^{(0)\omega\s}E_{\omega\s}[g^{(0)}].
\label{res4}
\ee
Taking the second derivative of the action (\ref{res1}) is equivalent to taking the first derivative of $E_{\omega\s}[g^{(0)}]$ with respect to $\epsilon$.
We give some useful linearization formulae in the following.
First, we rewrite the generalized Einstein tensor where we replace $\Lambda$ by its expression (\ref{scalar7}) which comes from the trace of the background Einstein equation (\ref{scalar4}). The result is
\be
E_{\omega\s} = \tilde{G}_{\omega\s} + 8\pi G(\tensor[^{(\alpha)}]{H}{_\omega_\s} + \tensor[^{(\beta)}]{H}{_\omega_\s} - \left<T_{\omega\s}\right>^T),
\label{res5}
\ee
where
\be
\tilde{G}_{\omega\s} \equiv R_{\omega\s} - {1\over 2}Rg^{(0)}_{\omega\s} + {R \over 4} g^{(0)}_{\omega\s},
\label{res6}
\ee
and
\be
 \left<T_{\omega\s }\right>^T \equiv  \left<T_{\omega\s}\right> - {1\over 4}\bar{\zeta}_{\omega\s} \left<T^\kappa_\kappa\right>.
 \label{res7}
\ee
The linearization of each term in (\ref{res5}) is given by
\be
\delta_h \tilde{G}_{\omega\s} = \left(-{\nabla^2\over2} + {R\over 12} \right) h^{(0)}_{\omega\s},
\label{res8}
\ee
\be
\delta_h \tensor[^{(\a)}]{H}{_\omega_\s}={\a\over 8\pi} {R\over 12}\left({\nabla^2\over2} - {R\over 12} \right) h^{(0)}_{\omega\s},
\label{res9}
\ee
\be
\delta_h \tensor[^{(\b)}]{H}{_\omega_\s}  = \frac{\beta}{32\pi}\left( \frac{1}{2}\Box^2 h_{\omega\s}-\frac{R}{4}\Box h_{\omega\s}+\frac{R^2}{36}h_{\omega\s} \right)= -{\b\over 2\pi} {\hat{h}_{\omega\s}\over L^4},
\label{res10}
\ee
where $\hat{h}_{\a\b}$ is defined in (\ref{b15}).
And the linearization of the CFT stress tensor depends on the background choice (flat, dS or AdS). Since $\delta_h \left<T_{\omega\s}\right>$ is only given for a single Laplacian mode, we first need to specify the background geometry and then decompose the action (\ref{res1}) into a complete basis of Laplacian eigenvalues.

Therefore, the quadratic part of the action is
\be
S^{(2)}[\bar{\zeta}] =  -{1\over 16\pi G}\int d^4x \sqrt{\bar{\zeta}}h^{(0)\omega\s }\left\{ \left(-{\nabla^2\over2} + {R\over 12} \right) \left(1 - {\a GR\over 12}\right) h^{(0)}_{\omega\s} - 4G\b {\hat{h}_{\omega\s}\over L^4} - 8\pi G  \left<T_{\omega\s}\right>^T \right\}.
\label{g21}
\ee
Replacing the stress tensor by its expression in terms of the expansion $h^{(n)}$, we obtain
\be
S^{(2)}[\bar{\zeta}] =  {N^2\over 2\pi^2}\int d^4 x \sqrt{\bar{\zeta}}h^{(0)\omega\s }\left\{h^{(4)}_{\omega\s} + \left({\tilde{\b}_\text{eff}} + 1 - 2\log(\m L)\right)\hat{h}_{\omega\s} + \right.
\nonumber
\ee
\be
\left. {RL^4\over 24}\left(\nabla^2 - {R\over 6} \right) \left({3\pi\over GN^2R} -{\pi\a\over 4N^2} - {1\over 4}\right) h^{(0)}_{\omega\s}  \right\}.
\label{res12}
\ee

\paragraph*{Flat space quadratic action}

The quadratic action given in (\ref{g21}) is applied to zero curvature. It gives
\be
S^{(2)}[\eta] = -{1\over 16\pi G}\int d^4x h^{(0)\omega\s}\left\{-{\nabla^2\over2} h^{(0)}_{\omega\s} - {4 G \b\over L^4}\hat{h}_{\omega\s} - 8\pi G \delta_h \left<T_{\omega\s}\right>^T\right\}
\label{flatres1}
\ee
Since the last term is only known in momentum space (\ref{n27}), we first apply a usual Fourier transform to the metric perturbation
\be
h^{(0)}_{\omega\s} = \int {d^4k\over (2\pi)^4} e^{-ik.x}\tilde{h}^{(0)}_{\omega\s},
\label{flatres2}
\ee
in order to write the quadratic action (\ref{flatres1}) as
\be
S^{(2)}[\eta] =  -{1\over 16\pi G}\int {d^4k\over (2\pi)^4} \tilde{h}^{(0)\omega\s }(-k) \left\{{k^2\over2} \tilde{h}^{(0)}_{\omega\s}(k) - {4 G \b\over L^4}\hat{\tilde{h}}_{\omega\s}(k) - 8\pi G \delta_h \left<\tilde{T}_{\omega\s}\right>^T(k)\right\}
\label{flatres3}
\ee
Using (\ref{n27}) and the fact that the trace anomaly is zero on a flat background, we obtain the result
\be
S^{(2)}[\eta]  =  \int {d^4k\over (2\pi)^4} \tilde{h}^{(0)\omega\s }(-k)\tilde{h}^{(0)}_{\omega\s}(k)  \mathcal{F}_\text{flat}(k),
\label{flatres4}
\ee
where
\be
\mathcal{F}_\text{flat}(k) \equiv {N^2\over 64\pi^2} k^2 \left\{-{2\pi \over GN^2} + {k^2\over 2}\left[{1\over 2} - 2\g_E - \log\left(GN^2k^2\right)- {\tilde{\b}_\text{eff}}\right]\right\}
\label{flatres5}
\ee
If we add a source to the action (\ref{flatres4}) as
\be
S^{(2)}[J_{\omega\s}] \equiv \int {d^4k\over (2\pi)^4} \tilde{h}^{(0)\omega\s}(-k) \left\{\tilde{h}^{(0)}_{\omega\s}(k)\mathcal{F}_\text{flat}(k) - \tilde{J}_{\omega\s}(k) \right\},
\label{flatres6}
\ee
then the classical solution which cancels the functional derivative of the quadratic action (\ref{flatres6}) with respect to $h^{(0)\omega\s }$ can be written as
\be
h^{(0)}_{\omega\s}(x) = \int d^4z J_{\omega\s}(z)G(x-z),
\label{flatres7}
\ee
where
\be
G(x) \equiv \int {d^4k\over (2\pi)^4}{e^{-ik.x} \over \mathcal{F}_\text{flat}(k)}.
\label{flatres8}
\ee

\paragraph*{Curved space-time quadratic action}

The spin-2 perturbation is decomposed into a basis of tensor (transverse-traceless) eigenmodes $\t_{\omega\s}(x)$ as

\be
h^{(0)}_{\omega\s}(x) = \int d\n \t_{\omega\s}(\n, x) \tilde{h}(\n),
\label{adsres1a}
\ee
where
\be
(\nabla^2 - {\bar{R}\over 6})\t_{\omega\s} = - {\bar{R}\over 12}\left(\n^2-{9\over 4}\right)\t_{\omega\s},
\label{adsres2}
\ee
where eigenvectors $\t_{\omega\s}$ of different eigenvalues are orthogonal, and normalized such that
\be
\int d^4x \sqrt{\bar{\zeta}}\t^{\omega\s}(\n,x)\t_{\omega\s}(\m,x) = \delta \left( \m-\n \right).
\label{adsres3}
\ee
This orthogonality relation allows us to write
\be
\int d^4x \sqrt{\z} \t^{\omega\s}(\n,x)h^{(0)}_{\omega\s}(x) = \int d \m \delta\left( \m-\n \right) \tilde{h}(\m) = \tilde{h}(\n).
\label{adsres4}
\ee
which tells that $\tilde{h}$ is the scalar product between the eigenvector and $h_{(0)\a\b}$.
In momentum space (\ref{adsres1a}), $h^{(4)}$ can be written explicetly in terms of $h^{(0)}$ if we specialize to a specific boundary geometry.
\begin{itemize}
\item For AdS with single-sided boundary conditions, we use (\ref{as7b}) to replace $h^{(4)}$ and write the action (\ref{res12}) only in terms of $h^{(0)}$. Using the orthogonality (\ref{adsres3}), we obtain
\be
S^{(2)}[\bar{\zeta}] =\int d\n \tilde{h}(\n)^2 \mathcal{F}_\text{(-)}(\n),
\label{adsres5}
\ee
where
\be
\mathcal{F}_\text{(-)} \equiv  {N^2 \chi^4\over 64\pi^2}\left(\n^2-{9\over 4}\right)Q_\text{(-)}(\n)
\label{adsres6}
\ee
and
\be
Q_\text{(-)}(\n) = 1 + {2\pi\over N^2}\left({1\over G \chi^2} + \a \right)  - {1\over 2} (\n^2- 1/4)\left[ \tilde{\b}_\text{eff}\right.  \nonumber
\ee
\be
 \left. + \log\left(GN^2\chi^2\right) - {1\over 2} +  \mathcal{H}\left(-{1\over 2}-\n \right) + \mathcal{H} \left(-{1\over 2} + \n \right) \right].
 \label{adsres7}
\ee
\item
For de Sitter, we replace $h^{(4)}$ using (\ref{i611}) into (\ref{res12}). The result is
\be
\mathcal{F}_\text{dS}(\n) = {N^2H^4\over 64\pi^2}\left(\n^2-{9\over 4}\right)Q_{dS}(\n), \quad \text{Re}(\n)>0.
\label{adsres8}
\ee
where
\be
Q_\text{dS}(\n) \equiv 1 - {2\pi\over GN^2H^2} + 2\tilde{\a} - {1\over 2}\left(\n^2-{1\over 4}\right)\left[\log\left(GN^2H^2\right)-{1\over 2} + 2\mathcal{H}(\n-1/2)  + \tilde{\b}_\text{eff}\right],
\nonumber
\ee
\be
\text{Re}(\n)>0,
\label{adsres9}
\ee
\end{itemize}

\section{Dynamics of scalar modes}
\label{gauge scalar}

As we have shown in section \ref{bdy pert},  the trace of the linearized Einstein equation results in (\ref{scalar16}), which is of fourth order in derivatives and therefore propagates two scalar modes. However, that equation is not the whole story because $\psi$ also gets constrained from off-diagonal components of Einstein equation (\ref{EE}).
To obtain the full set of equations, we need to know how $g^{(4)}_{\omega\s}$ depends on $\psi$ to linear order.

Since $\psi$ does not propagate in the bulk, one way to proceed is  to perform a diffeomorphism in the bulk as shown in \cite{Imbimbo,HSS}. This is will give   the full set of equations of motion for $\psi$.
In order to obtain constraints on the variation of $g_{\omega\s}^{(4)}$ under a conformal transformation $\psi$, we  solve order by order in the Fefferman-Graham expansion, a particular class of bulk diffeomorphisms which evaluate at $\psi$ on the boundary $\rho = \epsilon$. In \cite{Imbimbo}, such a diffeomorphism is defined as
\be
\left\{
\begin{array}{l}
\tilde{\rho} = \rho e^{-\psi(x)} \approx \rho(1-\psi(x)),\\
\tilde{x}^\s = x^\s + a^\s.
\end{array}
\right.
\label{nd1}
\ee
 Under $\rho \rightarrow \tilde{\rho}$ and $x^\s \rightarrow \tilde{x}^\s$, the slice metric transforms as
\be
g_{\omega\s}(x,\rho) \rightarrow g_{\omega\s}(x,\rho) + \psi_{\omega\s}(x,\rho),
\label{nd1a}
\ee
where $\psi_{\omega\s}(x,\rho)$ is given by
\be
\psi_{\omega\s} = \psi(1-\rho\partial_\rho)g_{\omega\s} + 2\nabla^g_{(\omega}a_{\s)}.
\label{nd2}
\ee
The covariant derivative $\nabla^g$ is the one compatible with $g_{\omega\s}(x,\rho)$. The bulk diffeomorphism $\psi_{\omega\s}(x,\rho)$ admits the Fefferman-Graham expansion
\be
\psi_{\omega\s}(x,\rho) = \psi g^{(0)}_{\omega\s} + \rho \psi^{(2)}_{\omega\s} + \rho^2 \psi^{(4)}_{\omega\s} + \rho^2\log\rho\hat{\psi}_{\omega\s} + \mathcal{O}(\rho^3),
\label{nd2a}
\ee
where we have already used the leading order $\mathcal{O}(\rho^0)$ of (\ref{nd2}), which gives $\psi g^{(0)}_{\omega\s}$.
One can already observe that $\psi$ in (\ref{nd1}) is indeed the same scalar variation as in (\ref{scalar2a}) because requiring that (\ref{nd1}) preserves the tensorial structure of the Fefferman-Graham metric (such that cross-terms $dx^\omega d\rho$ vanish), one can relate the near-boundary expansion of $a^\s$ with the one of $g_{\omega\s}$ \cite{Imbimbo}. The two first terms are enough in our case, they read
\be
a^\s(x,\rho) = \rho{L^2\over 4}\partial^\s \psi(x) - \rho^2 {L^2\over 8}g^{(2)\omega\s}\partial_\omega \psi(x) + \mathcal{O}(\rho^3),
\label{nd3}
\ee
where indices are now raised and lowered using the boundary metric $g^{(0)}_{\omega\s}$. Inserting (\ref{nd3}) into (\ref{nd2}) then leads to the result
\be
\psi_{\omega\s} = \psi g^{(0)}_{\omega\s}
\label{nd4}
\ee
\be
\psi^{(2)}_{\omega\s} = {L^2\over 2} \nabla_\omega\partial_\s \psi
\label{nd5}
\ee
\be
\psi^{(4)}_{\omega\s} = - \psi (g^{(4)}_{\omega\s} + \hat{g}_{\omega\s}) + {L^2\over 4}\left[\partial^\kappa \psi \nabla_\kappa g^{(2)}_{\omega\s}-\nabla_{(\omega}\left\{ g^{(2)}_{\s)\k}\partial^\k\psi \right\} +2 g^{(2)}_{\k(\omega}\nabla_{\s)} \partial^\kappa \psi \right]
\label{nd6}
\ee
\be
\hat{\psi}_{\omega\s} = -\psi \hat{g}_{\omega\s}
\label{nd7}
\ee
One can check the above formulae for $g^{(0)}_{\omega\s}$, $g^{(2)}_{\omega\s}$ and $\hat{g}_{\omega\s}$ starting with the solutions (\ref{fg16}) and (\ref{fg26}) and linearize them with $\psi$. However, the variation (\ref{nd6}) can only be obtained from the bulk diffeomorphism (\ref{nd3}).
Substituting all of these variation formulae into the CFT stress-tensor we obtain that it transforms under (\ref{nd1}) as
\be
\left<T_{\omega\s}\right>[(1+\psi)g^{(0)}] - \left<T_{\omega\s}\right>[g^{(0)}]  = - \psi \left<T_{\omega\s }\right> - {N^2\over 4\pi^2 L^4}\left\{ 2\psi \hat{g}_{\omega\s} - {L^2\over 2} \left[\partial^\kappa \psi \nabla_\kappa g^{(2)}_{\omega\s}-\nabla_{(\omega}\left\{ g^{(2)}_{\s)\k}\partial^\k\psi \right\}  \right. \right.
\nonumber
\ee
\be
\left.\left.+2 g^{(2)}_{\k(\omega}\nabla_{\s)} \partial^\kappa \psi\right] + 2(g^{(2)}\psi^{(2)})_{\omega\s} - {1\over 2}g^{(0)}_{\omega\s}\tr\left[ g^{(2)} \psi^{(2)}\right] +{1\over 2}g^{(0)}_{\omega\s}\tr\left[ g^{(2)} \right]\tr\left[ \psi^{(2)}\right]\right.  \nn
\ee
\be
\qquad\qquad\qquad \left.  - {1\over 2}\left( \psi^{(2)}_{\omega\s}\tr\left[ g^{(2)}\right] +g^{(2)}_{\omega\s}\tr \left[ \psi^{(2)}\right] \right) \right\}. \label{nd8}
\ee

We now restrict to a variation around (A)dS background with constant $\bar{R}$. The background values of $g^{(2)}_{\omega\s}$ and $g^{(4)}_{\omega\s}$ are then given by table (\ref{tabular slicings}). The stress-tensor variation (\ref{nd8}) then evaluates to the much simpler expression given by
\be
\left<T_{\omega\s}\right>[(1+\psi)\bar{\zeta}] - \left<T_{\omega\s}\right>[\bar{\zeta}]  = -\psi \left<T_{\omega\s}\right> - {N^2\bar{R}\over 192\pi^2}(\nabla_\omega\partial_\s - \bar{\zeta}_{\omega\s}\Box)\psi,
\label{nd9}
\ee
where now, all geometrical quantities such as $\bar{R}$ and $\nabla_\omega$ are built with the background metric $\bar{\zeta}_{\omega\s}$.
The first term corresponds to the classical scaling law for the stress tensor under a Weyl transformation. However, the second term of (\ref{nd9}) brings a correction coming from the conformal anomaly of the quantum vacuum expectation value $\left<T_{\omega\s}\right>$.
We now have all the ingredients to linearize the Einstein equation (\ref{EE}) under a conformal transformation $\psi$. One can use the conformal variation of the Ricci tensor
\be
R_{\omega\s}[(1+\psi)\bar{\zeta}] - R_{\omega\s}[\bar{\zeta}] = -{1\over 2}\left(\bar{\zeta}_{\omega\s}\Box + 2\nabla_\omega\partial_\s\right)\psi.
\label{nd10}
\ee
Using (\ref{nd9}) and linearizing all other terms with (\ref{nd10}), we obtain the following equation
\be
\left\{ \bar{\zeta}_{\omega\s} \Box - \nabla_\omega\partial_\s + \Lambda  \bar{\zeta}_{\omega\s} + {G\alpha\over 4}\left[\left({\bar{R}\over 4} + \Box\right) \bar{\zeta}_{\omega\s} - \nabla_\omega\partial_\s \right]\Box \right.
\nonumber
\ee
\be
\left. - {GN^2 \bar{R}\over 24\pi} \left[{\bar{R}\over 8}  \bar{\zeta}_{\omega\s} + ( \bar{\zeta}_{\omega\s}\Box - \nabla_\omega\partial_\s)\right]\right\}\psi = 0.
\label{nd11}
\ee
Inserting the value of $\Lambda$ (\ref{scalar7}) allows us to write (\ref{nd11}) in a simple, factorized form given by
\be
\left\{\left(\Box + {\bar{R}\over 4}\right)\bar{\zeta}_{\omega\s} - \nabla_\omega\partial_\s \right\}\left[1+{G\a\over 4}\Box - {GN^2\bar{R}\over 24\pi}\right]\psi = 0.
\label{nd11a}
\ee
 Taking the trace of (\ref{nd11a}) gives back equation (\ref{scalar16}). The factor in curly braces is responsible for the solution we are now trying to discard.

The squared brackets in (\ref{nd11a}) are absorbed into an auxiliary field $\Psi$ defined by
\be
\Psi \equiv \left[1 + {G\alpha\over 4}\Box  - {GN^2\bar{R}\over 24\pi}\right]\psi.
\label{nd14}
\ee

As we  show in the next subsection, $\Psi$ is non-propagating and it can be consistently set to zero in the absence of sources. This does not occur in the case of conformal matter considered here. As a consequence, the equation of motion for the physical $\psi$ is
\be
\left[1 + {G\alpha\over 4}\Box  - {GN^2\bar{R}\over 24\pi}\right]\psi=0
\label{psi eq}
\ee

\subsection{Constrained  scalar mode}
This subsection is devoted to showing that the only propagating  mode for $\psi$ is the solution to the equation $\Psi = 0$ as it is defined in (\ref{nd14}).
First, we observe that imposing flat space ($R=0$) in (\ref{nd11a}) reduces to the case studied in \cite{Q1}. More precisely, our equation (\ref{nd11a}) for $R=0$ is their equation (8) for $K_2 = {\a\over 192\pi}$.
We are going to follow their arguments and generalize them to space-time with non-zero curvature.

\paragraph*{In flat space,}
equation (\ref{nd11a}) reduces to
\be
(\Box \eta_{\omega\s} - \partial_\omega \partial_\s)\Psi = 0,
\label{w1}
\ee
where
\be
\Psi  = \psi + {G\a\over 4}\Box \psi.
\label{w2}
\ee

If there is an additional matter content with stress-tensor $\t_{\omega\s}$, then
(\relax\ref{w1})   becomes:
\be
(\Box \eta_{\omega\s} - \partial_\omega \partial_\s)\Psi =8\pi G  \delta_\psi \t_{\omega\s},
\ee
where $\delta_\psi$ means the linearization operator with respect to the conformal scalar $\psi$, as in the left-hand-side of (\ref{nd9}). The scalar $\Psi$ does not propagate because it is constrained by the ``$00$'' and ``$0i$'' components of Einstein's equation as we now show.  The ``$00$" component of this equation is given by
\be \label{constraint}
-\delta^{ij}\pa_i\pa_j \Psi = 8\pi G \delta_\Psi \t_{00}.
\ee
 Thus, $\Psi$ solves a static equation, similar to an electric potential in classical E\&M. If we demand that $\Psi$ vanishes  as $|\vec{x}| \to \infty$, then $\Psi$ is completely fixed by equation (\ref{constraint}).

 To see even more explicitly that $\Psi$ is non-propagating, it is enough to look at the solutions in the absence of sources:  from the $i\neq j$ and $0i$ components of (\ref{w1}) we obtain:
\be
\de_i\de_j \Psi = 0, \qquad \de_0 \de_i \Psi = 0
\ee

which implies that $\Psi$ is the sum of separate linear functions of the spatial coordinates  with time-independent  coefficients, plus an arbitrary function of time only. Demanding that $\Psi$ vanishes  at spatial infinity forces us to set  all the coefficients to zero, showing that $\Psi=0$ is the only physical solution.

\paragraph*{In de Sitter,}

$\Psi$ defined in (\ref{nd14}) is solution to the equation (\ref{nd11a}) rewritten as
\be
\left\{\left(\Box + {\bar{R}\over 4}\right)\zeta_{\omega\s} - \nabla_\omega\partial_\s \right\}\Psi = 0.
\label{w12}
\ee

 As in the flat space-time case, we show that $\Psi$ solves a \textit{static} equation and can be consistently set to zero.
 If there was an additional matter content with a stress-tensor $\t_{\m\n}$, the linearized Einstein equation for $\psi$ would read
 \be \label{fullEinstein}
 \left\{\left(\Box + {\bar{R}\over 4}\right)\zeta_{\omega\s} - \nabla_\omega\partial_\s \right\}\Psi =8\pi G \delta_\psi \t_{\omega\s}.
 \ee
 The ``00" component of this equation written in Poincar\'e coordinates is given by
 \be \label{fullEinstein00}
  \left\{ - \pa_i^2 -3\left(\tau^{-1}\pa_0 + \tau^{-2}\right) \right\}\Psi =8\pi G \delta_\psi \t_{00}.
 \ee
 We see that the second order time derivative is absent  and we are left with a non-propagating  equation.
We now  set the source to zero on the right hand side. In this case,  the two terms of the previous equation must   separately vanish, as one can see using the  $\omega = 0, \s = i $ components of (\ref{w12}):

\be
(\partial_0 + \tau^{-1})\partial_i \Psi = 0,
\label{w15}
\ee
The most general solution is
\be
\Psi = f(\tau) + \tau^{-1}A(x_i).
\label{w16}
\ee
Inserting the above expression  in (\ref{fullEinstein00})  (with the right hand side set to zero)
 gives the two equations
\be
-\nabla^2 A = 0,  \qquad (\partial_0 + \tau^{-1}) f=0.
\label{w17}
\ee
The equation  for $f$  is solved by $f = {cnst/\tau}$, so that  we can include this as a constant term in $A$ and just write $\Psi$ as
\be
\Psi = {A(x_i)\over \tau}, \qquad \nabla^2 A =0.
\label{w18}
\ee
Finally, inserting this result in the $\omega = i, \s = j$ component of (\ref{w12}),
 we obtain
\be
\partial_i \partial_j A = 0,
\label{w19}
\ee
which is  solved by $A = a + b^ix_i$. Therefore, the most general solution of (\ref{w12})  is
\be
\Psi = \tau^{-1}(a + b^ix_i).
\label{w20}
\ee

Demanding that field configurations vanish at spatial infinity, we arrive at the same conclusion as in flat space, that $\Psi$ is constraint to vanish.

By a similar computation as in dS, where $\tau^2$ is replaced by $-z^2$, the conclusion that $\Psi$ is constrained to vanish for AdS.

\section{Comparison with previous results for a dS boundary
\label{compa}}

In this appendix, we map our parameters $(\tilde{\a},{\tilde{\b}_\text{eff}},GN^2R)$ and compare some of our results  to previous papers which have used a similar setup.

The first study of de Sitter stability with a non-perturbative CFT obtained holographically was done in \cite{anomalyinflation1}.
 This paper corresponds to the particular case where the (renormalized) cosmological constant is zero.
 So the only contribution to the background curvature comes from the CFT.
We  also compare our results to the more recent paper \cite{Chesler}, which also studies the stability of de Sitter with a holographic CFT. In their study, the $R^2$ coefficient $\a$ is set to zero.

In our paper, stability conditions rely on the poles of the propagator of metric perturbations and on the residue of these poles. The variable of these propagators is the eigenvalue $\n^2$ of the laplacian operator. Thus, we must compare our definition of $\n$ (and $k$ for flat space) to the ones of previous papers.
First, the definition of \cite{Chesler} for $\n$ is the same as ours (\ref{dS5}).
In addition, the radial part of the bulk equation of motion (\ref{dS7}) coincides with eq. (49) of \cite{Chesler}.
Since one must choose the sign of $\text{Re}(\n)$ (see discussion below (\ref{dS8})), their solution corresponds to taking $\text{Re}(\n) <0$, which is equivalent to $C_+ = 0$ in our case. As discussed already in the paper, negative real parts of $\n$ are obtained by taking $\n \rightarrow -\n$ in (\ref{dS15}).

To compare with the second paper \cite{anomalyinflation1}, we relate $\n$ to their eigenvalue labeled by $p$ as
\be
H^{-2}\nabla^2 h^{(0)}_{\omega\sigma} = (2- p(p+3))h^{(0)}_{\omega\sigma}.
\label{comp1}
\ee
Comparing this equation with (\ref{dS5}) leads to
\be
p = \pm \n - {3\over 2}.
\label{comp2}
\ee
Similarly with the choice of $\text{Re}(\n)$ positive or negative, the normalizability of the bulk solution depends on the choice of $p$ in (\ref{comp2}). In \cite{anomalyinflation1}, they choose $\text{Re}(p)>- 3/2$. Therefore, one needs to use the replacement $p = -\n - 3/2$ to retrieve the results of \cite{Chesler} which has negative real parts for $\n$ and $p = \n - 3/2$ to compare with our results for which $\n$ has positive real parts.

We now compare the spin-2 equation of motions in these different papers.
In \cite{Chesler}, the equation of motion is given by
\be
\left(\n^2 - {9\over 4}\right)Q_\text{C}(\n) = 0,
\label{comp2a}
\ee
where
\be
Q_C(\nu) \equiv 256 + \frac{8GN^2H^2}{\pi - GN^2H^2}\left\{ 16 + (1-4\nu^2)\left[3 - 4\mathcal{H}\left(-\frac{1}{2}-\nu\right) + 4\log\left({2E\over H}\right)\right]\right\},
\label{comp3}
\ee
where $E$ must be related to our parameter $\b$.
On the other hand, the inverse propagator $F(p)$ in eq. (3.82) of \cite{anomalyinflation1} is given by
\be
F_H(p,\beta) = \Psi(p) + \frac{4\pi R^2}{GN^2}(p^2 + 3p + 6) + 2\beta_H p(p+1)(p+2)(p+3) - 4\a_H p(p+3),
\label{comp4}
\ee
\be
\Psi(p) = p(p+1)(p+2)(p+3)[\psi(p/2+5/2)+\psi(p/2+2)-\psi(2) - \psi(1)]\nonumber\ee \be
+ p^4 + 2p^3 - 5p^2 - 10p - 6\;,
\label{comp5}
\ee
where $\a_H$ and $\b_H$ must be related to the parameters $\a$ and $\b$ from our setup (\ref{s03}), (\ref{s1}).
Using (\ref{comp2}), we obtain an algebraic relation between $Q_C(\n)$ and $F(p)$ given by
\be
F_H(p) = {\pi - GN^2H^2\over 64GN^2H^2}Q_C(\nu),
\label{comp6}
\ee
if the parameters of \cite{anomalyinflation1} and \cite{Chesler} are related by
\be
\a_H = 0,
\label{comp7}
\ee
\be
GN^2H^2 = 4\pi,
\label{comp8}
\ee
\be
\log\left({H\over E}\right) = \b_H + {3\over 4}.
\label{comp9}
\ee
First,
(\ref{comp7}) is due to the absence of a $R^2$ term in the boundary action of  \cite{Chesler}. This additional term is proportional to $\a_H $ in \cite{anomalyinflation1}.
Second, (\ref{comp8}) is explained by the absence of a cosmological constant in the Einstein-Hilbert action of \cite{anomalyinflation1}, which fixes the dimensionless curvature $GN^2H^2$ to $4\pi$ since the CFT is the only contribution to the background curvature. This is equivalent to setting $\L = 0$ in (\ref{scalar7}), leading to (\ref{comp8}).
Before, we compare our results with these two previous papers. In our case, the curvature is not fixed (as in \cite{Chesler}), and the coefficient of the $R^2$ term is also arbitrary (as in \cite{anomalyinflation1}).

We now compare our inverse propagator $\mathcal{F}_\text{dS}(\n)$ (\ref{dS17}) with $F_H(p)$ (\ref{comp6}), which obey the algebraic relation
\be
-2 \mathcal{F}_\text{dS}(\n)  = F_H(p),
\label{comp10}
\ee
valid under the following conditions
\be
\tilde{\a} = \a_H,
\label{comp11}
\ee
\be
GN^2H^2 = 4\pi,
\label{comp12}
\ee
\be
\tilde{\b}_\text{eff} = {1\over 2} + 2\b_H - \log(16\pi).
\label{comp13}
\ee

{Our results agree with \cite{anomalyinflation1} if we set the curvature as in (\ref{comp12}). Their analysis is similar to the one done in the discussion of Figure \ref{dS_Hpiover4_alpha10}, which corresponds to the special case where the massless spin-2 pole is a ghost.}
{Our results also agree with $\cite{Chesler}$ for $\a = 0$. For instance, their final result is the obtention of (\ref{sH13}) for $\a = 0$, which tells the value of $\b$ from which the spin-2 sector is non-tachyonic. However, they do not discuss the presence of ghosts and whether their tachyon is above or below the species cutoff. }

\section{Snapshots for the spin-2 propagator}
\label{moresnapshots}

In this appendix, we provide snapshots for the poles of the tensor propagator in the $\n\in \mathbb{C}$ plane,
for a wide range of parameters $(\tilde{\a},\tbe, GN^2R)$.

\subsection{Minkowski}

 \begin{figure}[h!]
\centering
\begin{subfigure}{.3\textwidth}
\includegraphics[width=\textwidth]{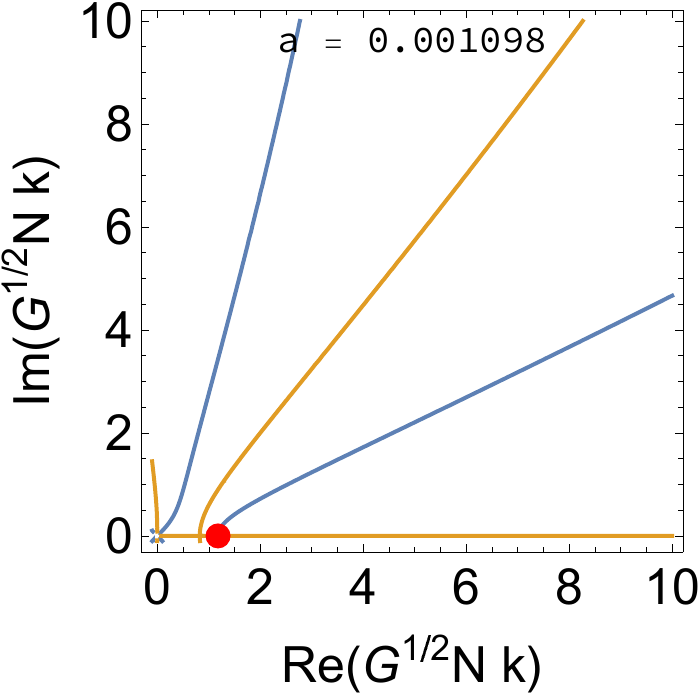}
\caption{\it ${\tilde{\b}_\text{eff}} = -10$}
\end{subfigure}
\begin{subfigure}{.3\textwidth}
\includegraphics[width=\textwidth]{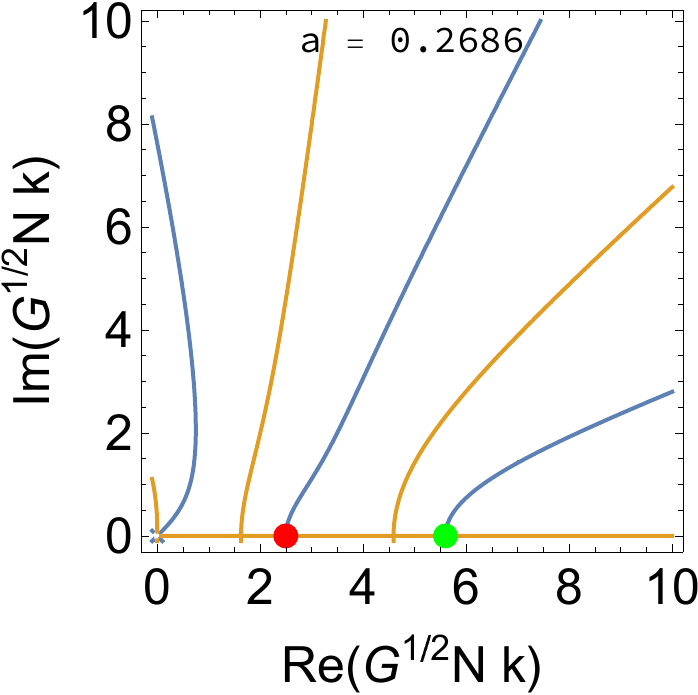}
\caption{\it ${\tilde{\b}_\text{eff}} = -4.5$}
\end{subfigure}
\\
\begin{subfigure}{.3\textwidth}
\includegraphics[width=\textwidth]{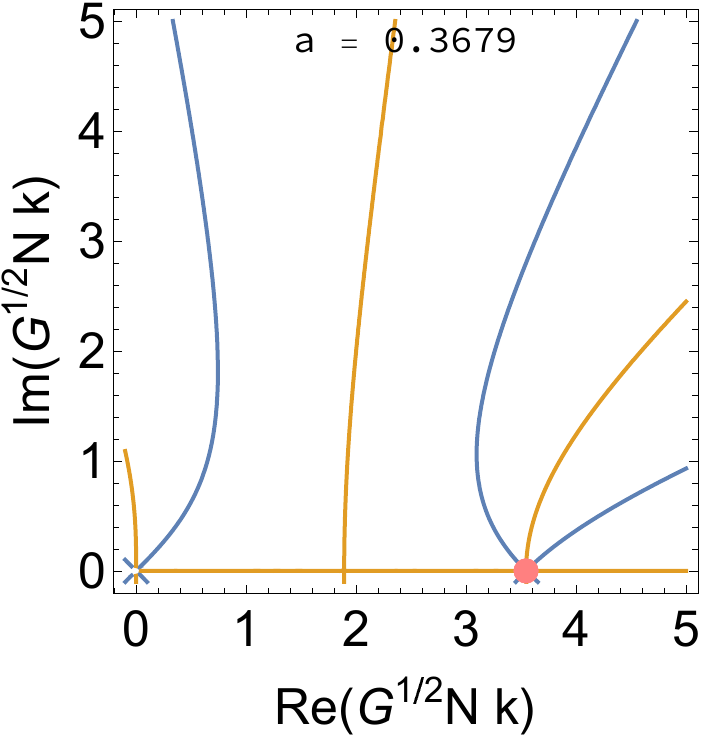}
\caption{\it ${\tilde{\b}_\text{eff}} = -4.18546$}
\end{subfigure}
\begin{subfigure}{.3\textwidth}
\includegraphics[width=\textwidth]{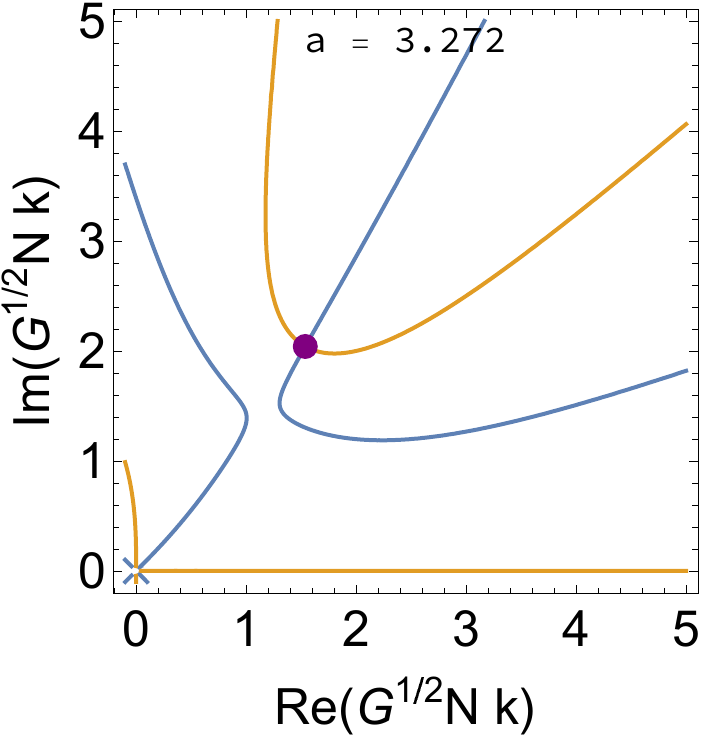}
\caption{\it ${\tilde{\b}_\text{eff}} = -2$}
\end{subfigure}
\begin{subfigure}{.3\textwidth}
\includegraphics[width=\textwidth]{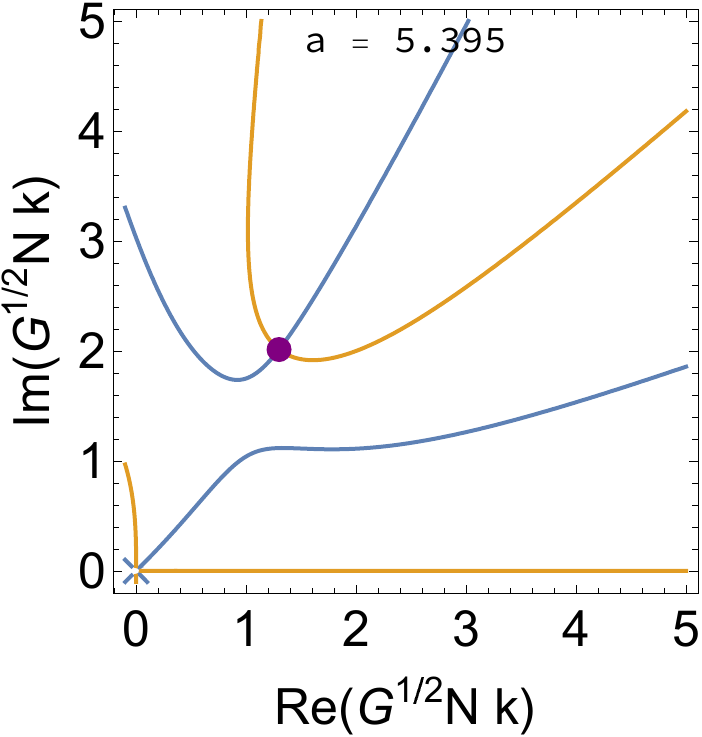}
\caption{\it ${\tilde{\b}_\text{eff}} = -1.5$}
\end{subfigure}
\begin{subfigure}{.3\textwidth}
\includegraphics[width=\textwidth]{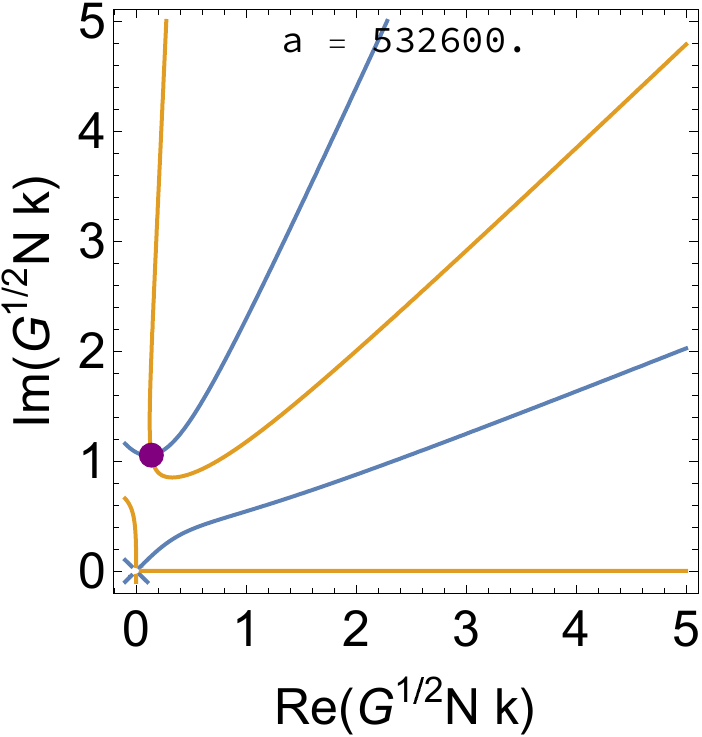}
\caption{\it ${\tilde{\b}_\text{eff}} = 10$}
\end{subfigure}
\begin{subfigure}{.3\textwidth}
\includegraphics[width=\textwidth]{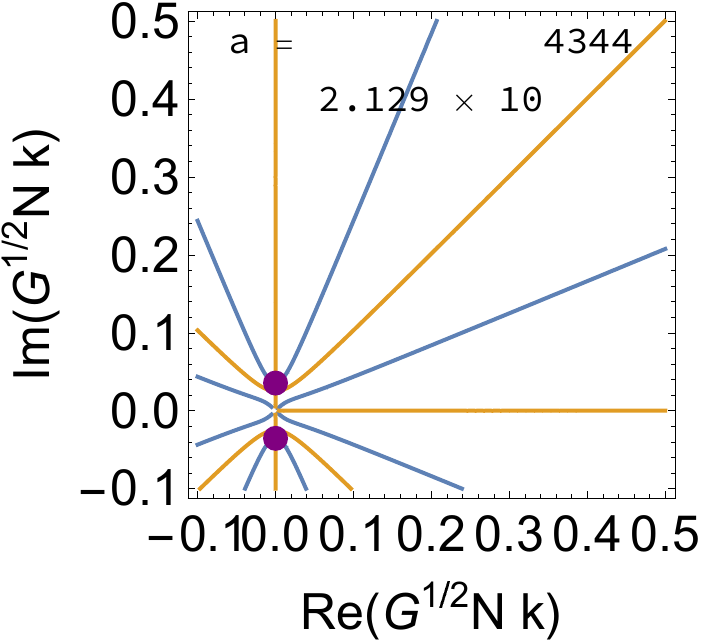}
\caption{\it ${\tilde{\b}_\text{eff}} = 10000$}
\end{subfigure}
\caption{\it These snapshots show the poles of the spin-2 propagator in Minkowski space-time, for selected values of the parameter $\tilde{\b}_\text{eff}$. The poles correspond to zeros of the real (blue curve) and imaginary (orange curve) parts of $\mathcal{F}_\text{flat}(k)$ (\protect\ref{n28a}).%
  and are  denoted by coloured dots. A green dot indicates a negative residue (ghost-free), while a red dot corresponds to a positive residue (ghost). A  purple dot is for complex residue (also a ghost).  As $\tilde{\b}_\text{eff}$ is increased going from upper left to lower right, two tachyons located on the real axis for negative merge in snapshot (c) to form a second order pole. The merging happens at $\tilde{\b}_\text{eff} = \tilde{\b}_\text{eff}^\text{merge}$ (\protect\ref{n47}).
   For $\tilde{\b}_\text{eff}> \tilde{\b}_\text{eff}^\text{merge}$ there is a complex conjugate pair, which moves to the origin for large and positive $\tilde{\b}_\text{eff}$.
}
\label{fig:flat}
\end{figure}

\newpage

\subsection{de Sitter}

In the main text, we have given two examples of parameters $(\tilde{\a},GN^2H^2)$ for which we find numerically the poles of the tensor propagator in de Sitter space-time in Figure \ref{dS_H0.01_alpha0} (small curvature, zero $\tilde{\a}$) and in Figure \ref{dS_Hpiover4_alpha10} (generic curvature, negative $\tilde{\a}$). These two examples give different behaviours when $\tbe$ is varied.
One could question if these two examples are paradigmatic or if another choice of $(\tilde{\a},GN^2H^2)$ would lead to different results.
In the following table, we give the links to snapshots for 9 different regimes.

\begin{tabular}{|c|c|c|c|}
\hline
\diagbox{$\tilde{\a}$}{$GN^2H^2$} & $<<1$ & $\sim \pi$ & $>>1$ \\ \hline
$-\tilde{\a} >>1$ &\textcolor{green}{Fig. \ref{dS_alpha-1000_H0.01}}& \textcolor{green}{Fig. \ref{dS_alpha-1000_H2pi}} &  \textcolor{green}{Fig. \ref{dS_alpha-1000_H1000}}\\ \hline
$|\tilde{\a}| \leq 1$  & \textcolor{green}{Fig. \ref{dS_H0.01_alpha0}} & \colorbox{blue!30}{\textcolor{green}{Fig. \ref{dS_alpha-1_H2pi}}, \color{red}{Fig. \ref{dS_H4pi_alpha0}} }& \colorbox{blue!30}{\textcolor{green}{Fig.} \ref{dS_alpha-1_H1000}, \textcolor{red}{Fig. \ref{dS_alpha0_H1000}}}\\ \hline
$\tilde{\a} >>1$ & \colorbox{blue!30}{\textcolor{green}{Fig.} \ref{dS_alpha100_H0.001}, \textcolor{red}{Fig.} \ref{dS_alpha1000_H0.01}} & \textcolor{red}{Fig. \ref{dS_alpha1000_H2pi}} & \textcolor{red}{Fig. \ref{dS_alpha1000_H1000}}  \\ \hline
\end{tabular}
\\

$a$ (\ref{sH6}) is either \textcolor{green}{positive} if ``Fig" is written in green, or \textcolor{red}{negative} if it is written in red. \colorbox{blue!30}{Blue boxes} correspond to regimes where the sign of $a$ must be determined by the inequality (\ref{Ln2}) :
\be
a<0\iff {\pi\over GN^2H^2}<\tilde{\a} + {1\over 2}
\label{sup1}
\ee

The conclusion is that a given point in the $(\tilde{\a},GN^2H^2)$ plane corresponds either to the behaviour of
\begin{itemize}
\item \textbf{Type A}: Figure \ref{dS_H0.01_alpha0}, where two tachyons merge on the real axis and form a pair of complex conjugate poles when $\tbe$ is increased.
\item \textbf{Type B}: Figure \ref{dS_Hpiover4_alpha10}, where all the poles are real valued $\n^2$ for any value of $\tbe$ (no complex pole).
\end{itemize}

In general, for a given point in the $(\tilde{\a},GN^2H^2)$ plane, a positive $a$ is type A and a negative $a$ is type B as predicted by the large-$|\n|$ approximation (\ref{sH4}). However, this is not necessarily true for large $GN^2H^2$ and $a$ close to zero, where the large-$|\n|$ analysis breaks down.

\begin{figure}[ht]
\centering
\begin{subfigure}{.3\textwidth}
\includegraphics[width=\textwidth]{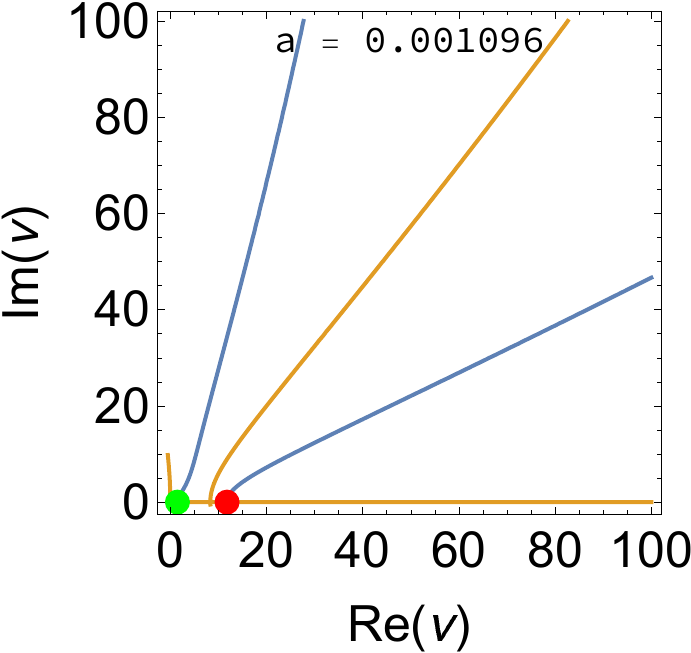}
\caption{\it ${\tilde{\b}_\text{eff}} = -10$}
\end{subfigure}
\begin{subfigure}{.3\textwidth}
\includegraphics[width=\textwidth]{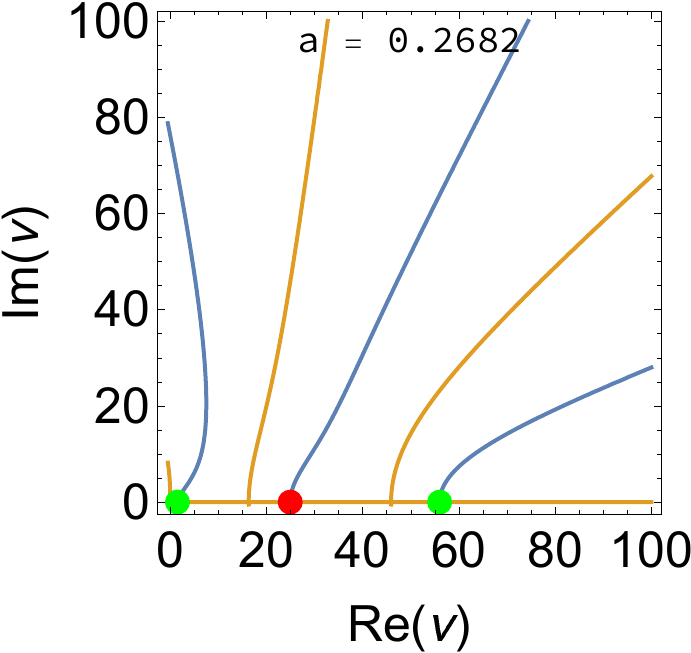}
\caption{\it ${\tilde{\b}_\text{eff}} = -4.5$}
\end{subfigure}
\begin{subfigure}{.3\textwidth}
\includegraphics[width=\textwidth]{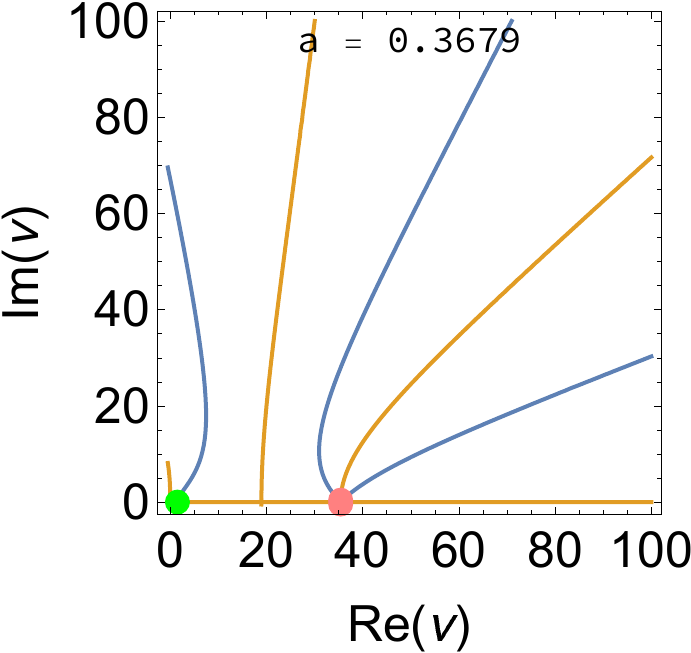}
\caption{\it ${\tilde{\b}_\text{eff}} = -4.18386$}
\end{subfigure}
\begin{subfigure}{.3\textwidth}
\includegraphics[width=\textwidth]{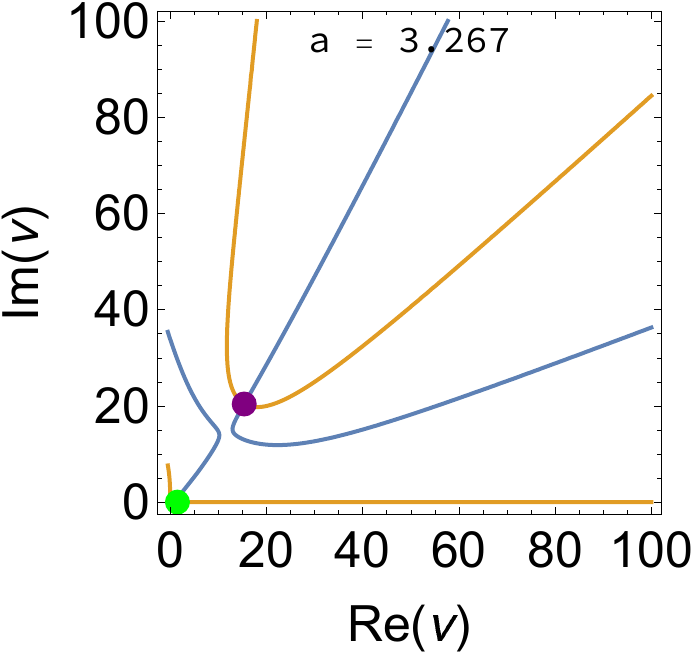}
\caption{\it ${\tilde{\b}_\text{eff}} = -2$}
\end{subfigure}
\begin{subfigure}{.3\textwidth}
\includegraphics[width=\textwidth]{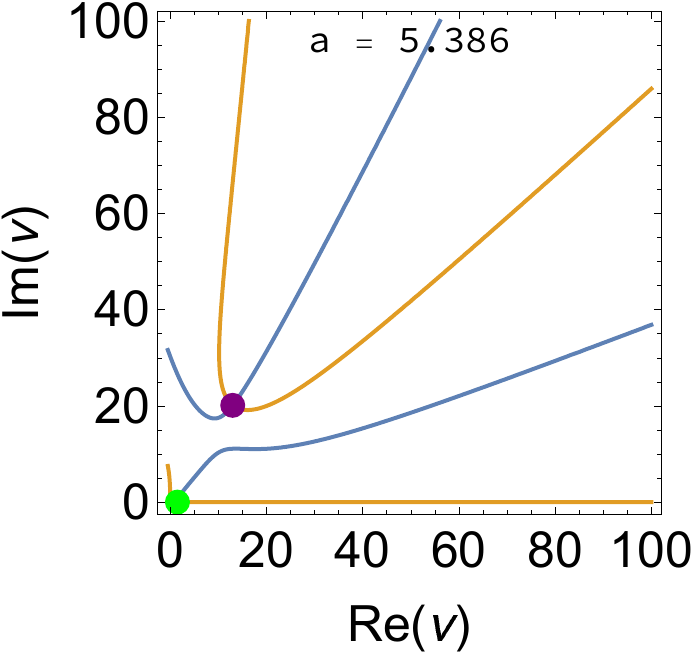}
\caption{\it ${\tilde{\b}_\text{eff}} = -1.5$}
\end{subfigure}
\begin{subfigure}{.3\textwidth}
\includegraphics[width=\textwidth]{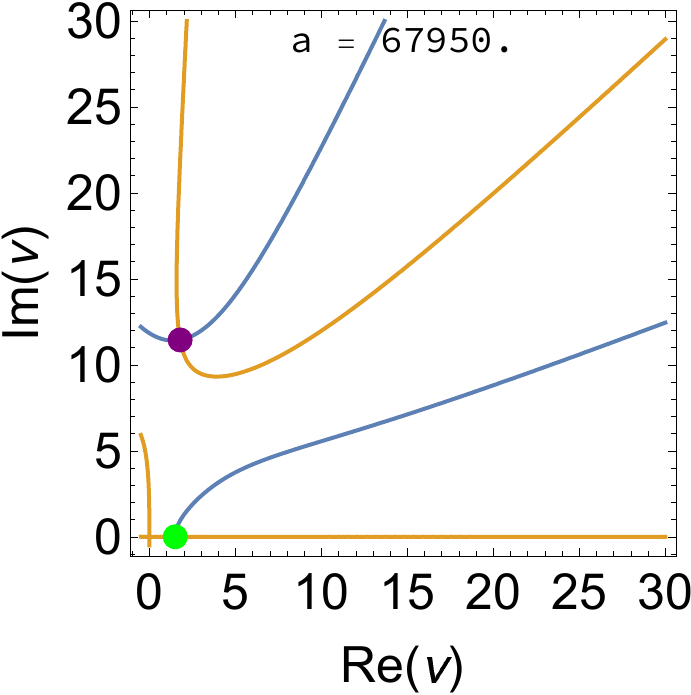}
\caption{\it ${\tilde{\b}_\text{eff}} = 7.94268$}
\end{subfigure}
\begin{subfigure}{.3\textwidth}
\includegraphics[width=\textwidth]{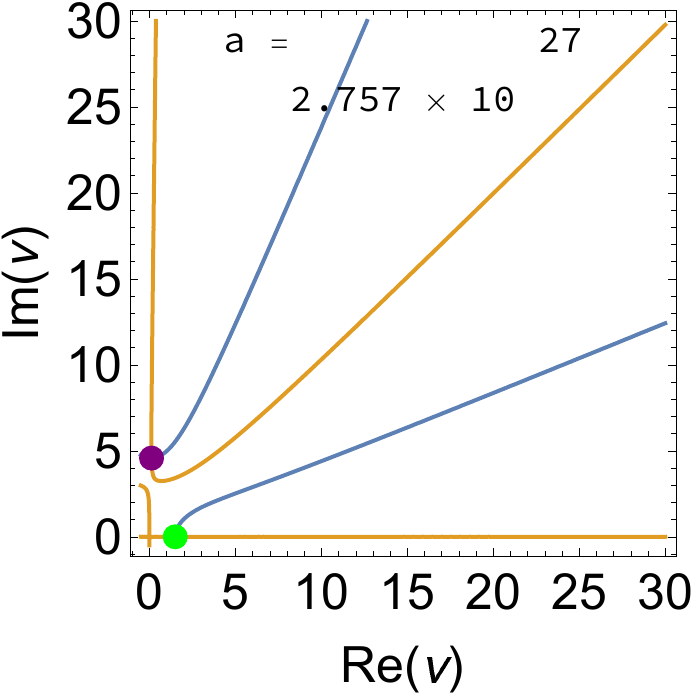}
\caption{\it ${\tilde{\b}_\text{eff}} = 60$}
\end{subfigure}
\begin{subfigure}{.3\textwidth}
\includegraphics[width=\textwidth]{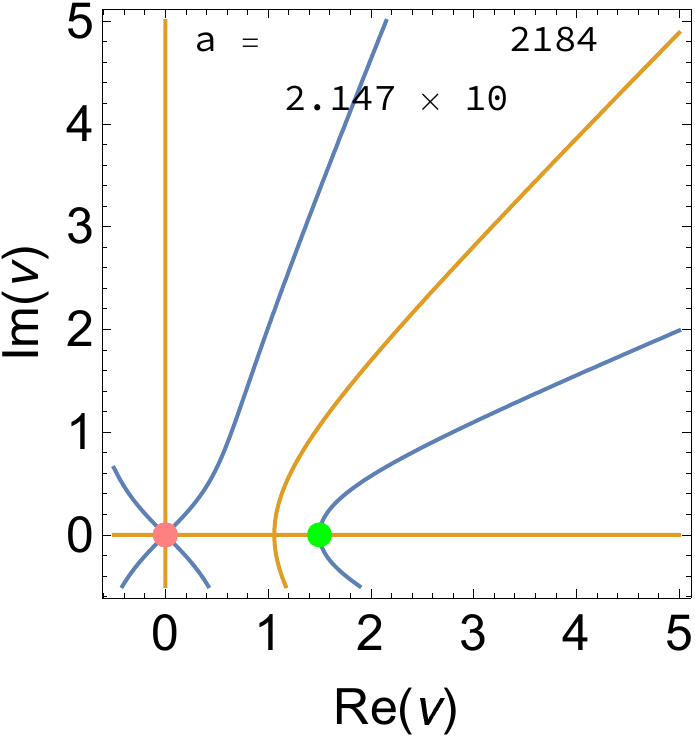}
\caption{\it ${\tilde{\b}_\text{eff}} = 5026.43 $}
\end{subfigure}
\begin{subfigure}{.3\textwidth}
\includegraphics[width=\textwidth]{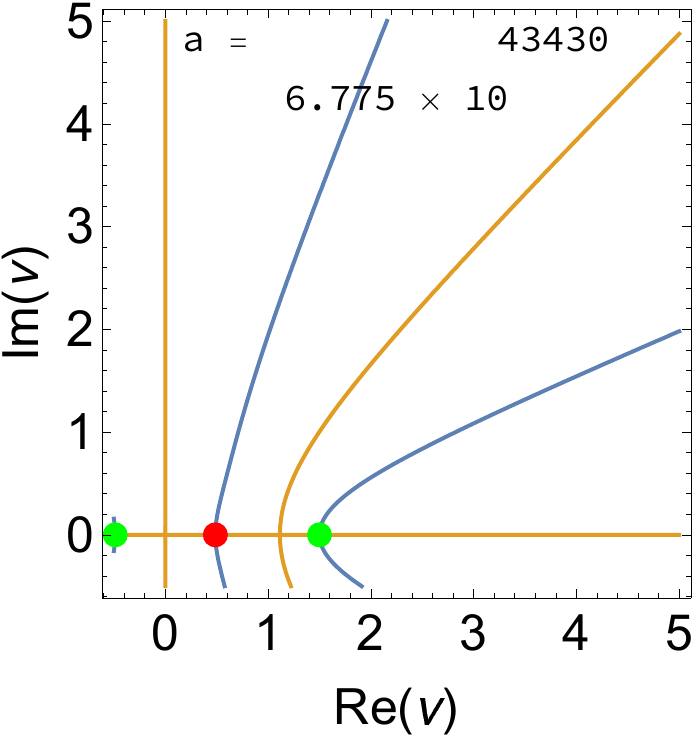}
\caption{\it ${\tilde{\b}_\text{eff}} = 100000$}
\end{subfigure}
\caption{\it Zeros of the real (blue curve) and imaginary (orange curve) parts of the de Sitter tensor propagator $\mathcal{F}_\text{dS}(\n)$ (\protect\ref{dS14}) for different values of $\tilde{\b}_\text{eff}$, and $\a = 0$, $GN^2H^2 = 0.01$. Solutions are therefore given by the intersection of blue and orange lines. Small values of $\tilde{\b}_\text{eff}$, have two tachyons on the real axis. They merge in snapshot (c) where $\tilde{\b}_\text{eff}$ is given by (\protect\ref{Ln1}). This merging forms a double zero because the first derivative of $\mathcal{F}_{dS}$ vanishes. The instability crosses the green stability line $\text{Re}(\n) = 3/2$ at another critical value of $\tilde{\b}_\text{eff}$ (\protect\ref{sH13}).}
\label{dS_H0.01_alpha0}
\end{figure}

\begin{figure}[ht]
\centering
\begin{subfigure}{.3\textwidth}
\includegraphics[width=\textwidth]{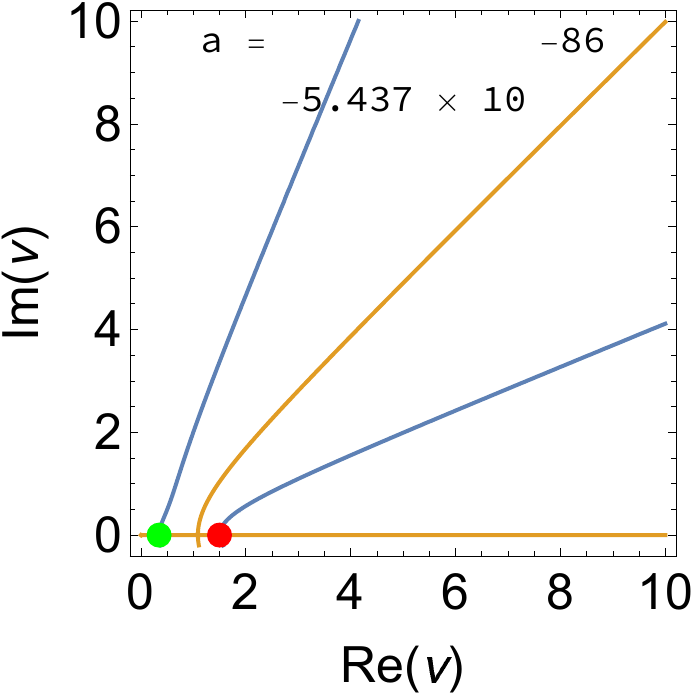}
\caption{\it \it${\tilde{\b}_\text{eff}} = -200$}
\end{subfigure}
\begin{subfigure}{.3\textwidth}
\includegraphics[width=\textwidth]{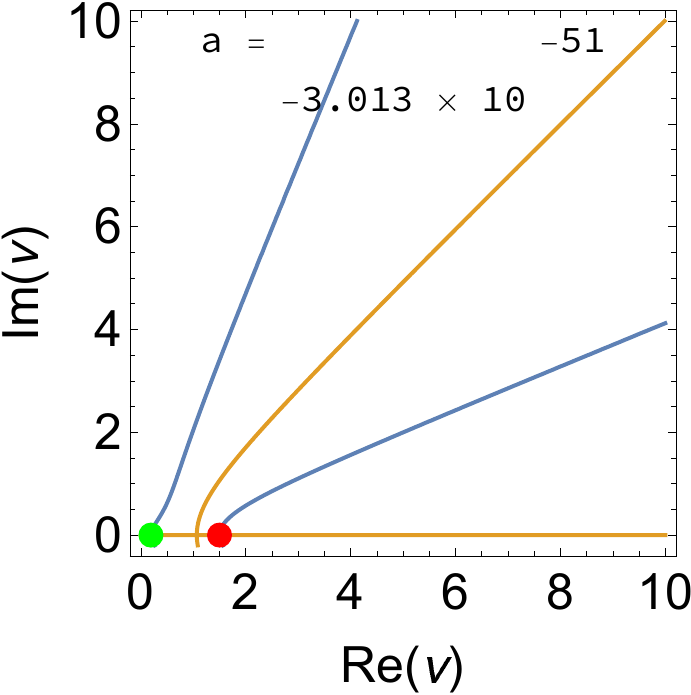}
\caption{\it \it${\tilde{\b}_\text{eff}} = -120$}
\end{subfigure}
\begin{subfigure}{.3\textwidth}
\includegraphics[width=\textwidth]{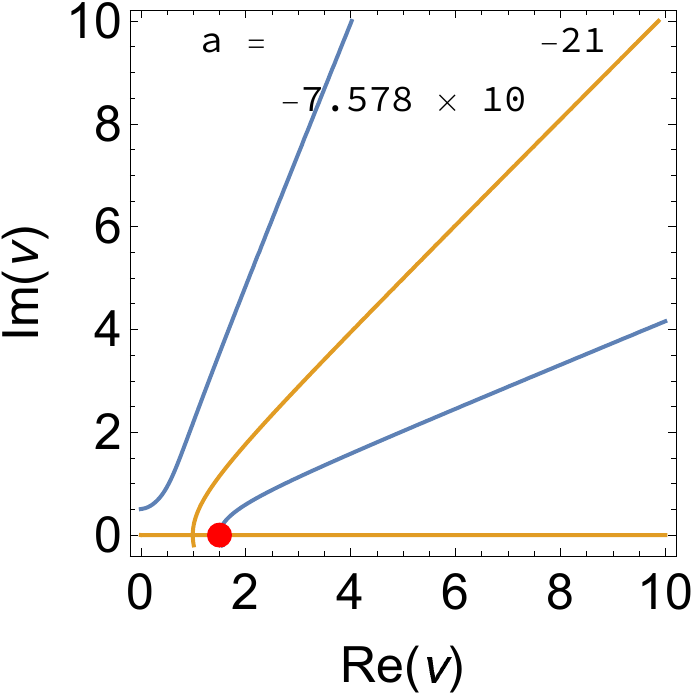}
\caption{\it \it${\tilde{\b}_\text{eff}} = -50$}
\end{subfigure}
\begin{subfigure}{.3\textwidth}
\includegraphics[width=\textwidth]{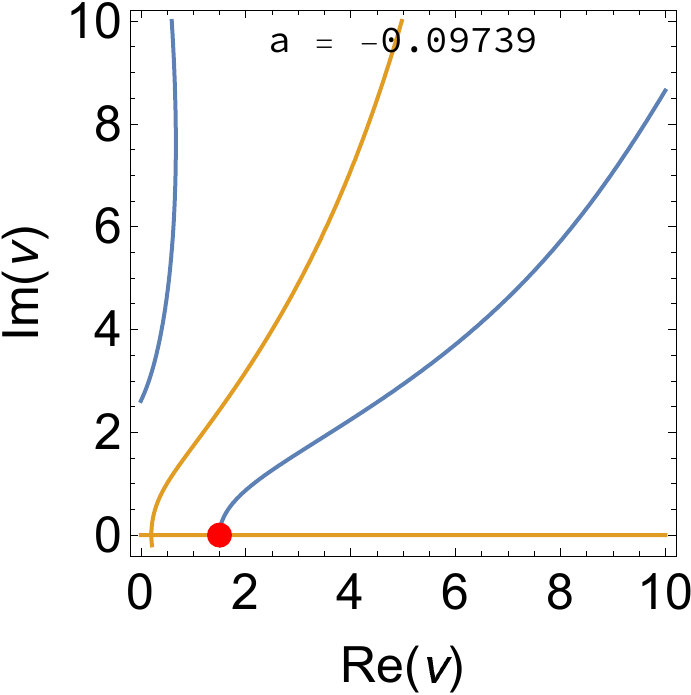}
\caption{\it \it${\tilde{\b}_\text{eff}} = -6$}
\end{subfigure}
\begin{subfigure}{.3\textwidth}
\includegraphics[width=\textwidth]{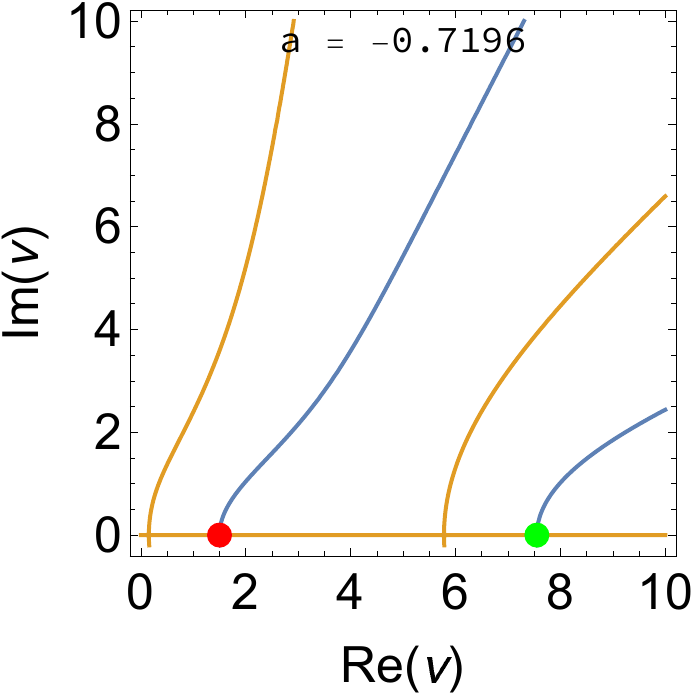}
\caption{\it \it${\tilde{\b}_\text{eff}} = -4$}
\end{subfigure}
\begin{subfigure}{.3\textwidth}
\includegraphics[width=\textwidth]{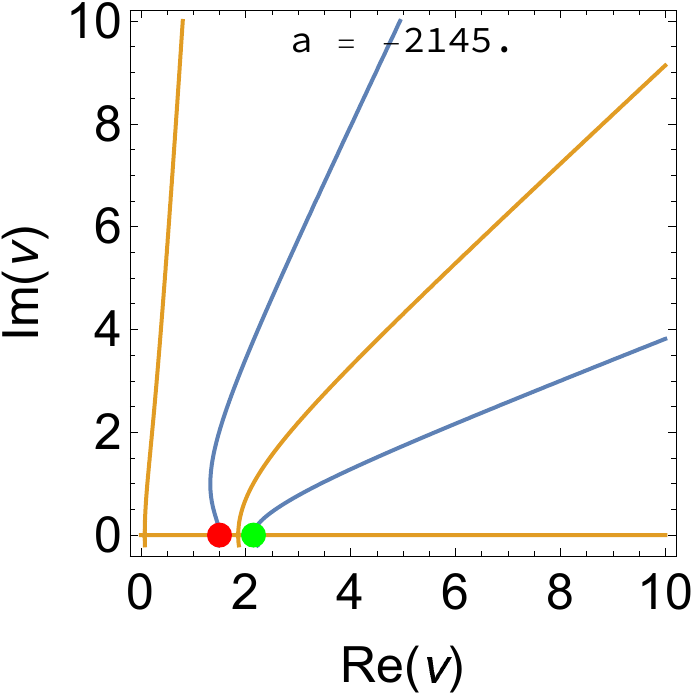}
\caption{\it \it${\tilde{\b}_\text{eff}} = 4$}
\end{subfigure}
\begin{subfigure}{.3\textwidth}
\includegraphics[width=\textwidth]{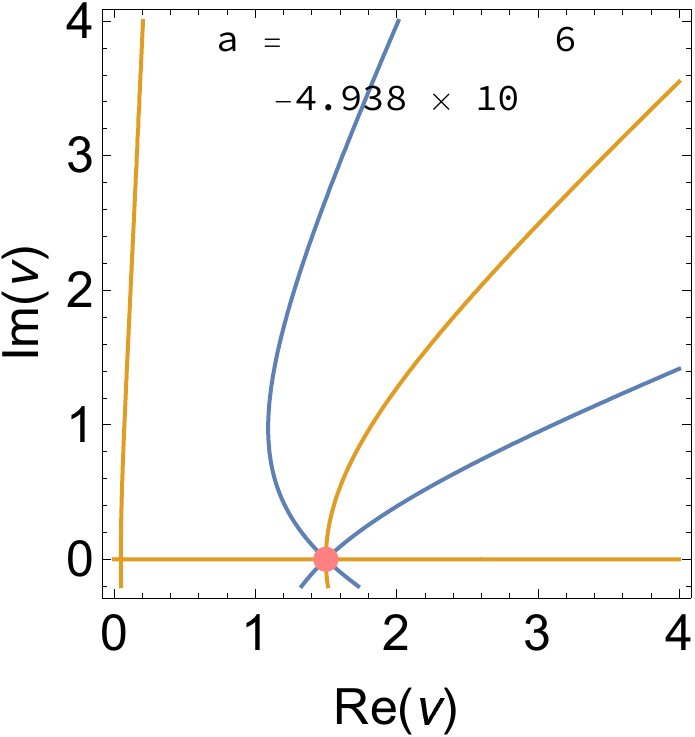}
\caption{\it \it${\tilde{\b}_\text{eff}} = 11.7416$}
\end{subfigure}
\begin{subfigure}{.3\textwidth}
\includegraphics[width=\textwidth]{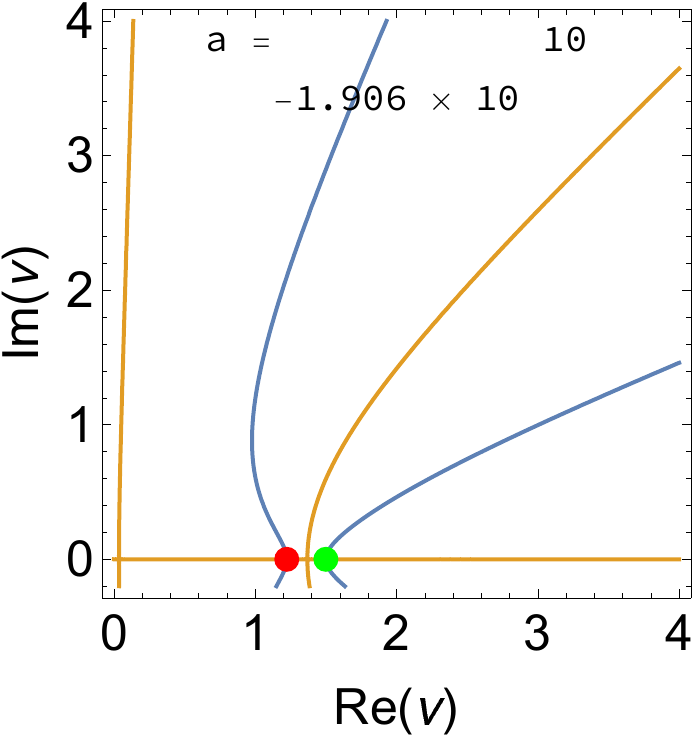}
\caption{\it \it${\tilde{\b}_\text{eff}} = 20$}
\end{subfigure}
\begin{subfigure}{.3\textwidth}
\includegraphics[width=\textwidth]{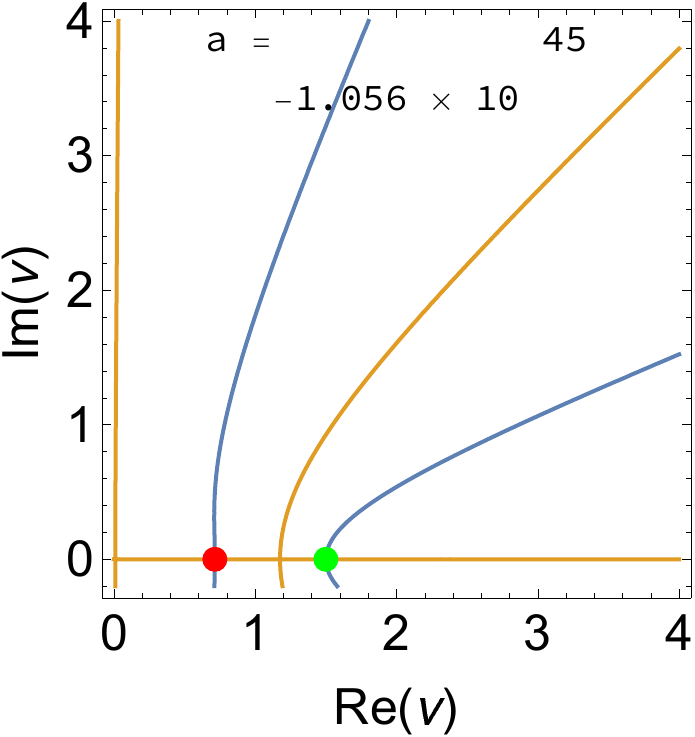}
\caption{\it \it${\tilde{\b}_\text{eff}} = 100$}
\end{subfigure}

\caption{\it Zeros of the real (blue curve) and imaginary (orange curve) parts of the inverse spin-2 propagator of de Sitter $\mathcal{F}_\text{dS}^{-1}(\n)$ (\protect\ref{dS14}) for different values of $\tilde{\b}_\text{eff}$, with fixed $\tilde{\a} = 10$ and $GN^2H^2 = \pi/4$.  Solutions are therefore given by the intersection of blue and orange lines. The tachyon, which was outside the window, arrives from snapshot (e) and merges at the critical value (\protect\ref{Ln4}) in snapshot (g). Increasing $\tilde{\b}_\text{eff}$ to large and positive values makes the ghost pole converge at $\n = 1/2$.}
\label{dS_Hpiover4_alpha10}
\end{figure}

\begin{figure}[ht]
\centering
\begin{subfigure}{.3\textwidth}
\includegraphics[width=\textwidth]{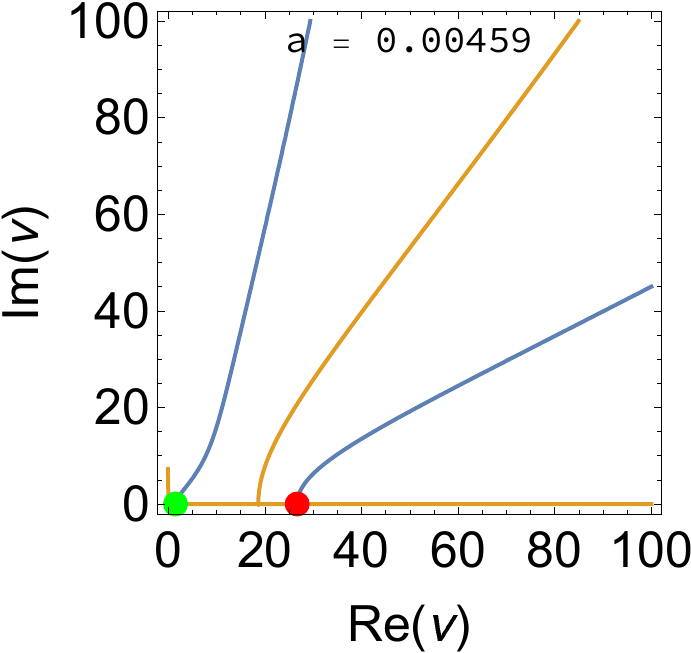}
\caption{\it ${\tilde{\b}_\text{eff}} = -10$}
\end{subfigure}
\begin{subfigure}{.3\textwidth}
\includegraphics[width=\textwidth]{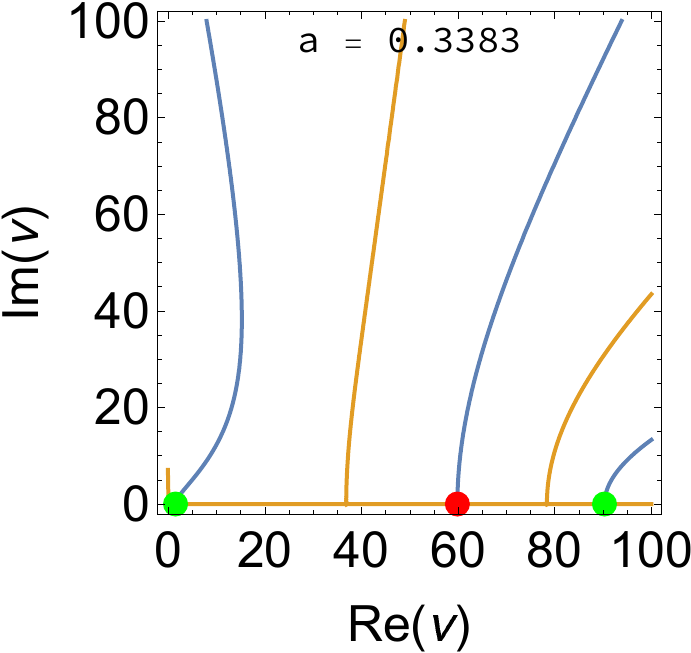}
\caption{\it ${\tilde{\b}_\text{eff}} = -5.7$}
\end{subfigure}
\begin{subfigure}{.3\textwidth}
\includegraphics[width=\textwidth]{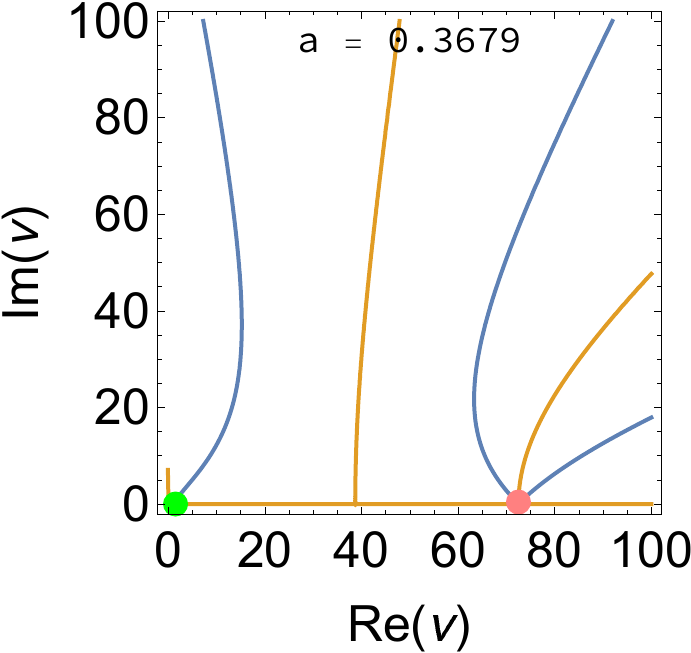}
\caption{\it ${\tilde{\b}_\text{eff}} = -5.61613$}
\end{subfigure}
\begin{subfigure}{.3\textwidth}
\includegraphics[width=\textwidth]{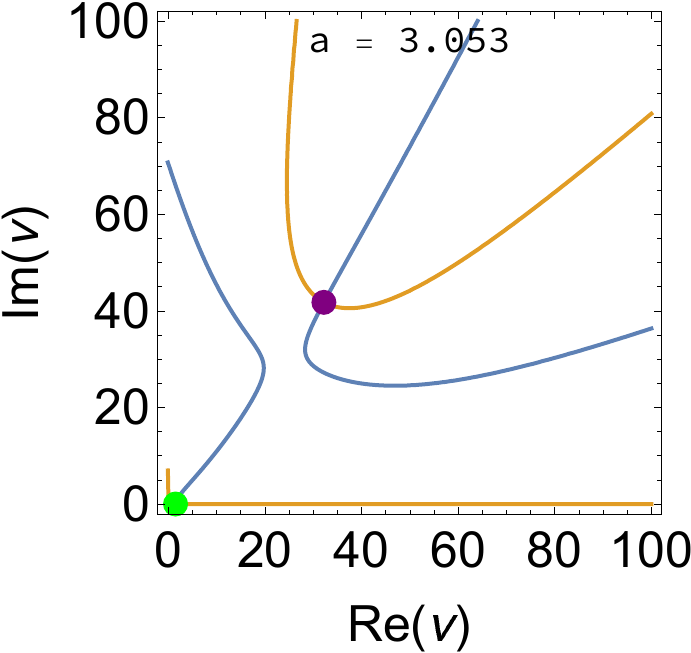}
\caption{\it ${\tilde{\b}_\text{eff}} = -3.5$}
\end{subfigure}
\begin{subfigure}{.3\textwidth}
\includegraphics[width=\textwidth]{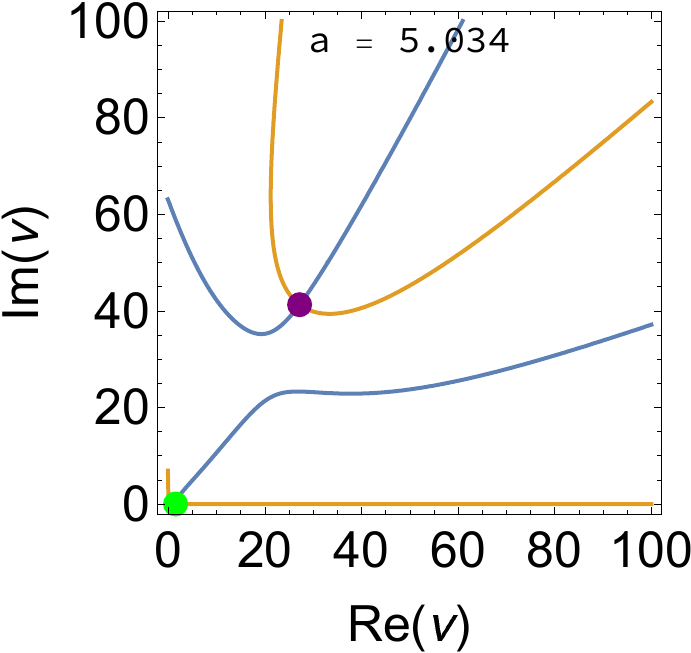}
\caption{\it ${\tilde{\b}_\text{eff}} = -3$}
\end{subfigure}
\begin{subfigure}{.3\textwidth}
\includegraphics[width=\textwidth]{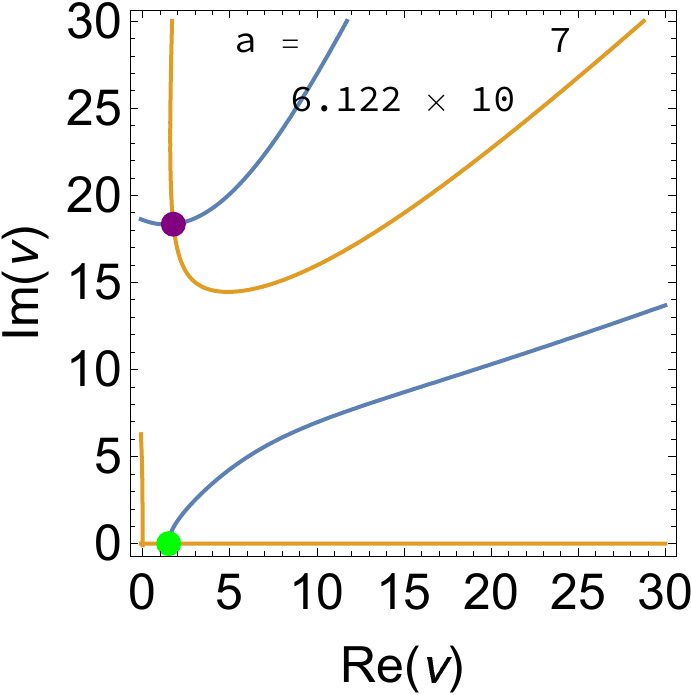}
\caption{\it ${\tilde{\b}_\text{eff}} = 13.3138$}
\end{subfigure}
\begin{subfigure}{.3\textwidth}
\includegraphics[width=\textwidth]{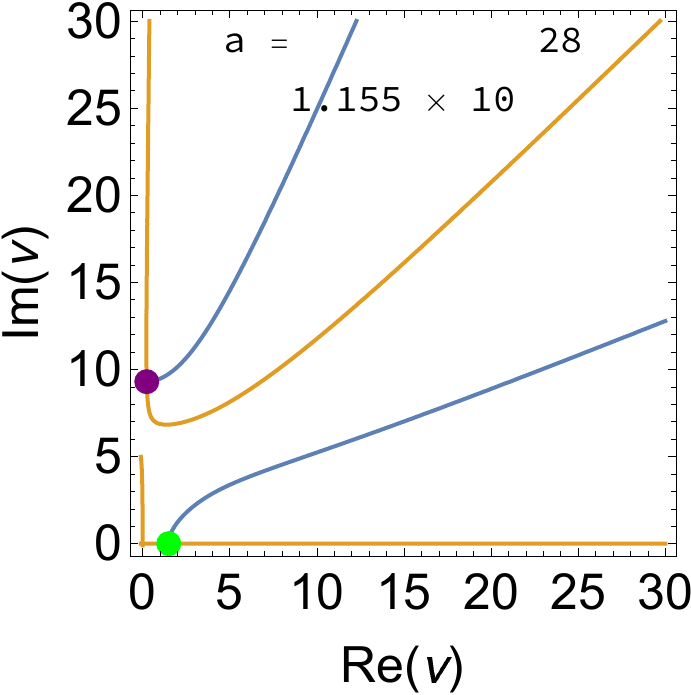}
\caption{\it ${\tilde{\b}_\text{eff}} = 60$}
\end{subfigure}
\begin{subfigure}{.3\textwidth}
\includegraphics[width=\textwidth]{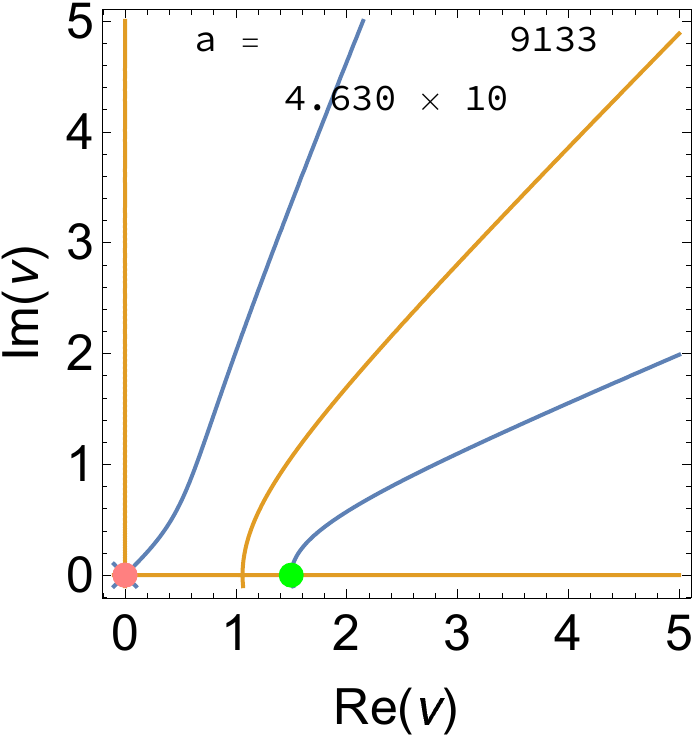}
\caption{\it ${\tilde{\b}_\text{eff}} = 21026.4$}
\end{subfigure}
\begin{subfigure}{.3\textwidth}
\includegraphics[width=\textwidth]{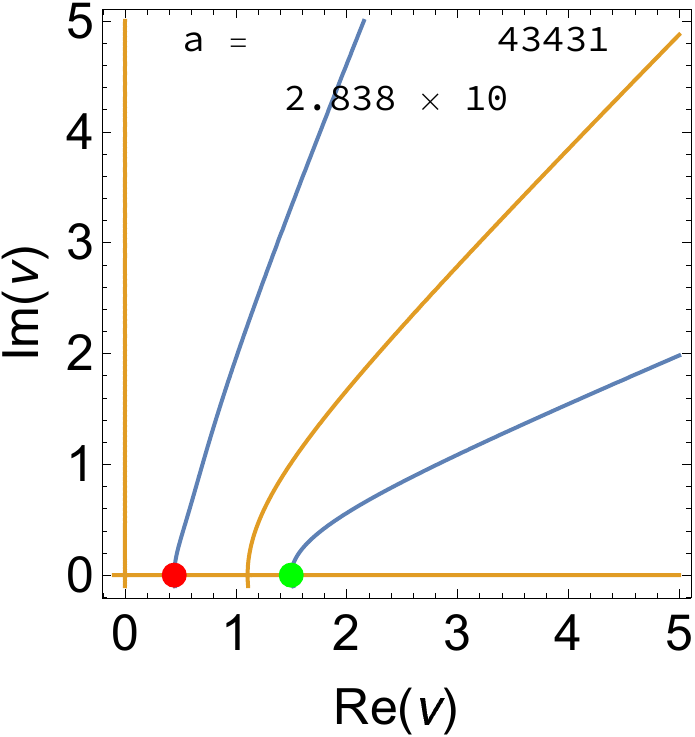}
\caption{\it ${\tilde{\b}_\text{eff}} = 100000$}
\end{subfigure}
\caption{\it de Sitter, $\tilde{\a} = -1000$, $GN^2H^2 = 0.01$.}
\label{dS_alpha-1000_H0.01}
\end{figure}

\begin{figure}[ht]
\centering
\begin{subfigure}{.3\textwidth}
\includegraphics[width=\textwidth]{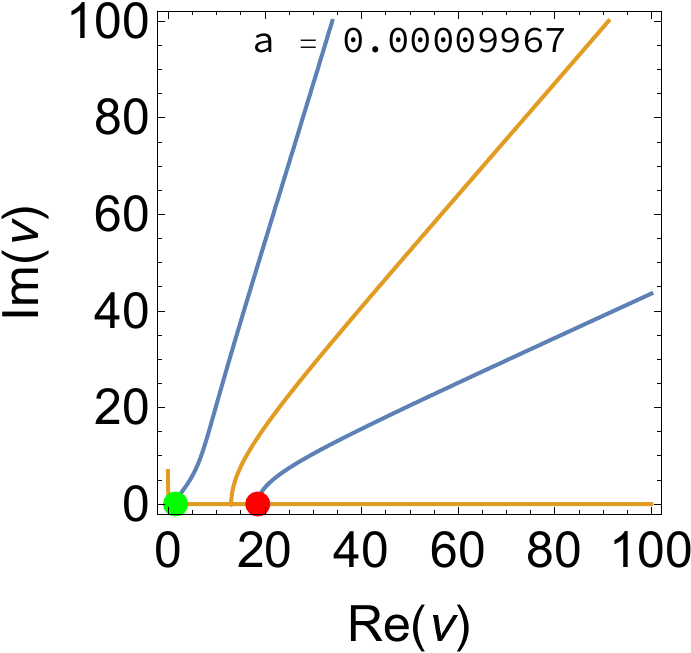}
\caption{\it ${\tilde{\b}_\text{eff}} = -20$}
\end{subfigure}
\begin{subfigure}{.3\textwidth}
\includegraphics[width=\textwidth]{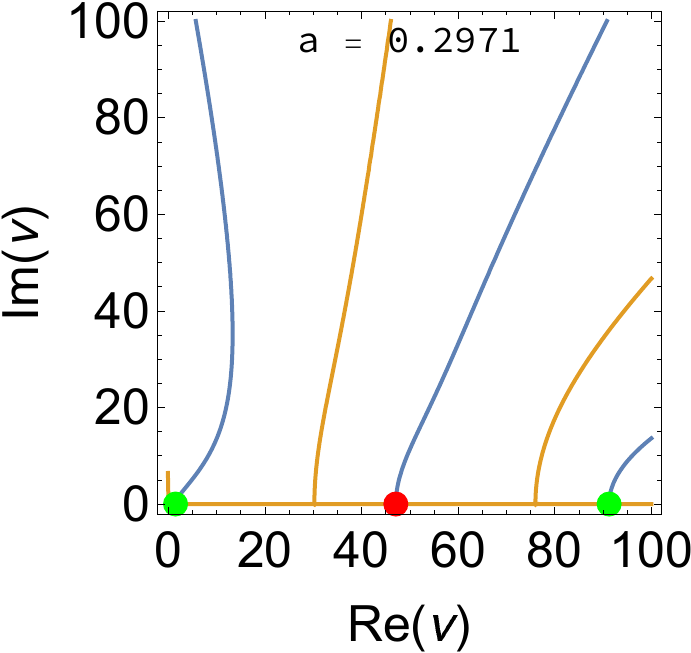}
\caption{\it ${\tilde{\b}_\text{eff}} = -12$}
\end{subfigure}
\begin{subfigure}{.3\textwidth}
\includegraphics[width=\textwidth]{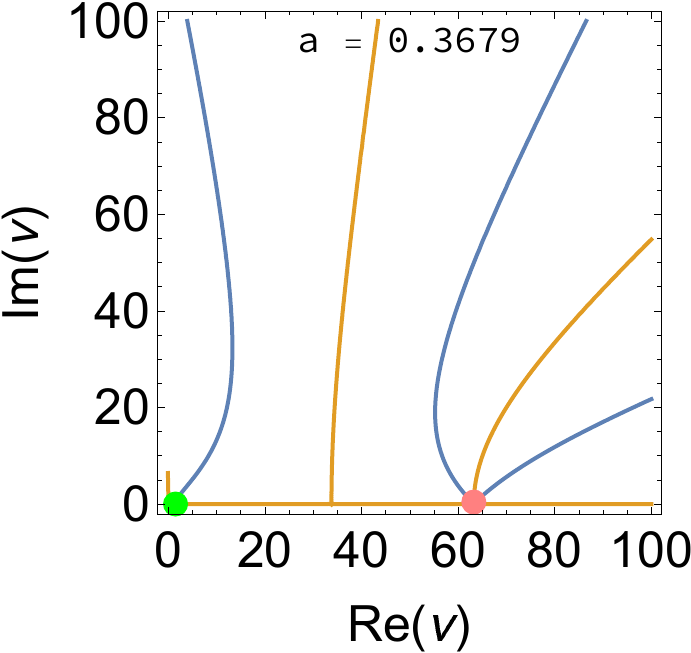}
\caption{\it ${\tilde{\b}_\text{eff}} = -11.7864$}
\end{subfigure}
\begin{subfigure}{.3\textwidth}
\includegraphics[width=\textwidth]{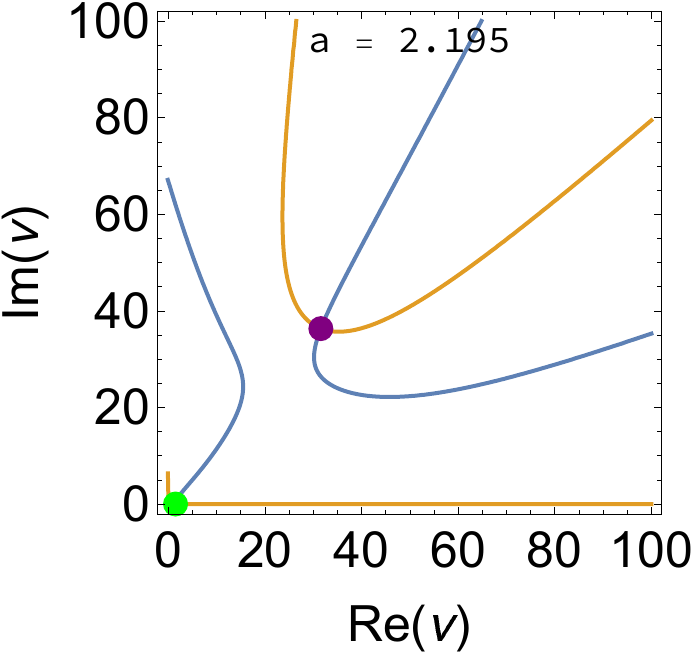}
\caption{\it ${\tilde{\b}_\text{eff}} = -10$}
\end{subfigure}
\begin{subfigure}{.3\textwidth}
\includegraphics[width=\textwidth]{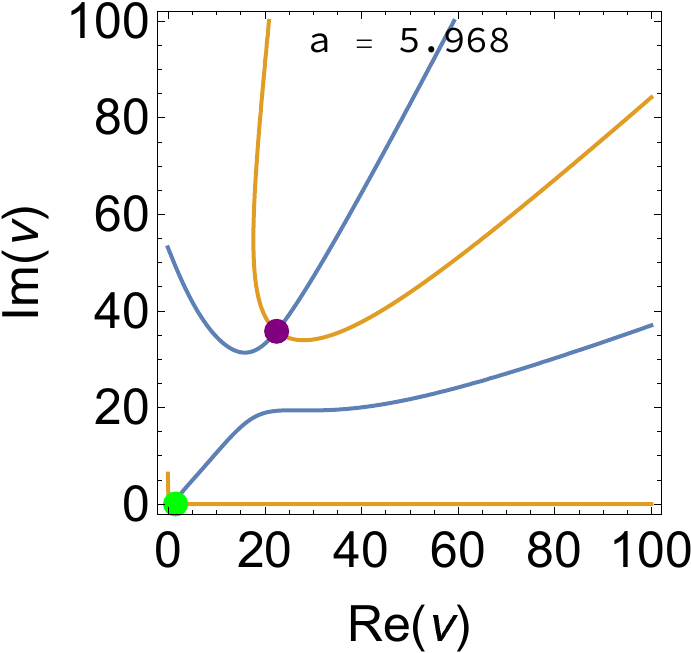}
\caption{\it ${\tilde{\b}_\text{eff}} = -9$}
\end{subfigure}
\begin{subfigure}{.3\textwidth}
\includegraphics[width=\textwidth]{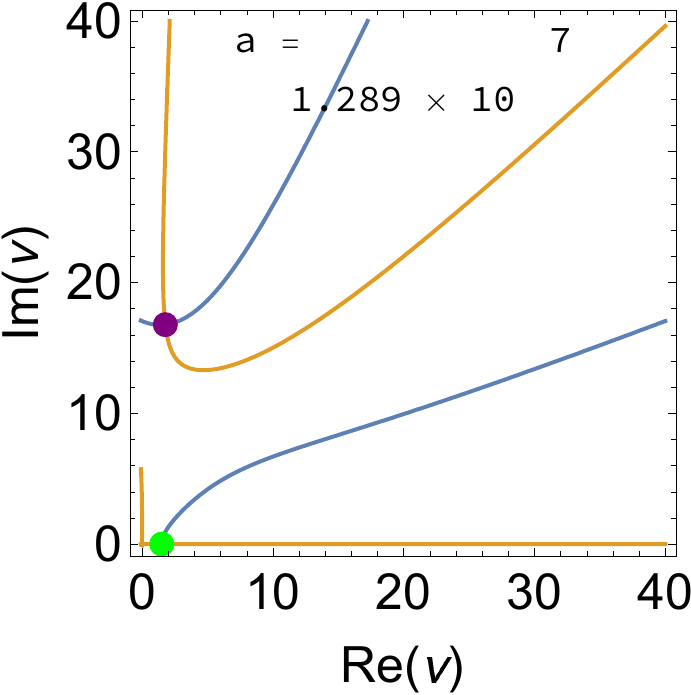}
\caption{\it ${\tilde{\b}_\text{eff}} = 5.58551$}
\end{subfigure}
\begin{subfigure}{.3\textwidth}
\includegraphics[width=\textwidth]{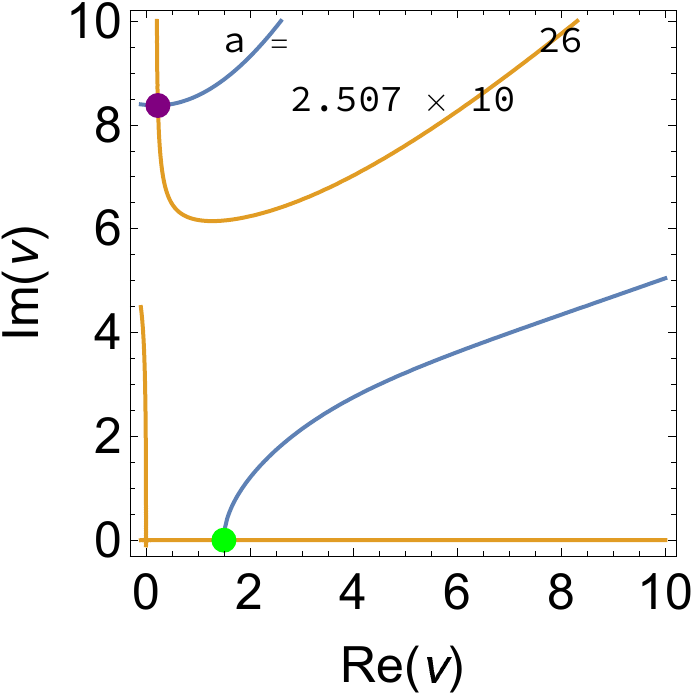}
\caption{\it ${\tilde{\b}_\text{eff}} = 50$}
\end{subfigure}
\begin{subfigure}{.3\textwidth}
\includegraphics[width=\textwidth]{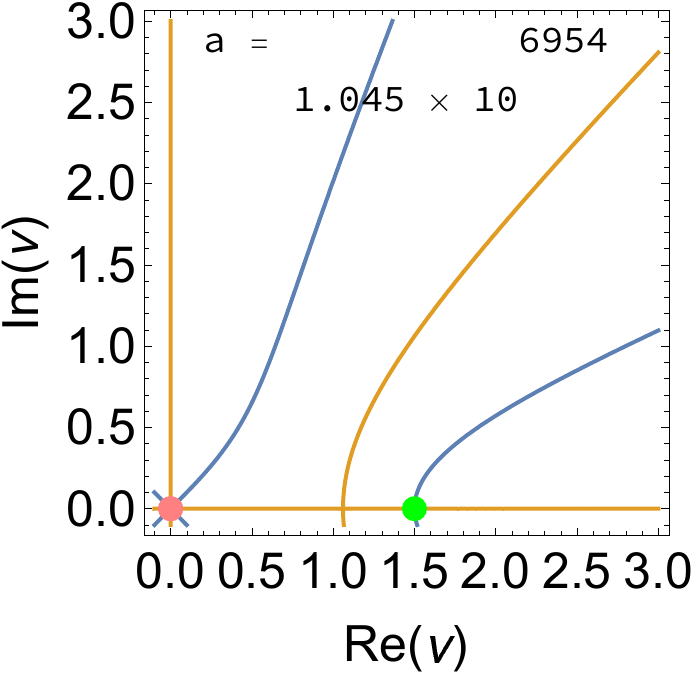}
\caption{\it ${\tilde{\b}_\text{eff}} = 16001.4$}
\end{subfigure}
\begin{subfigure}{.3\textwidth}
\includegraphics[width=\textwidth]{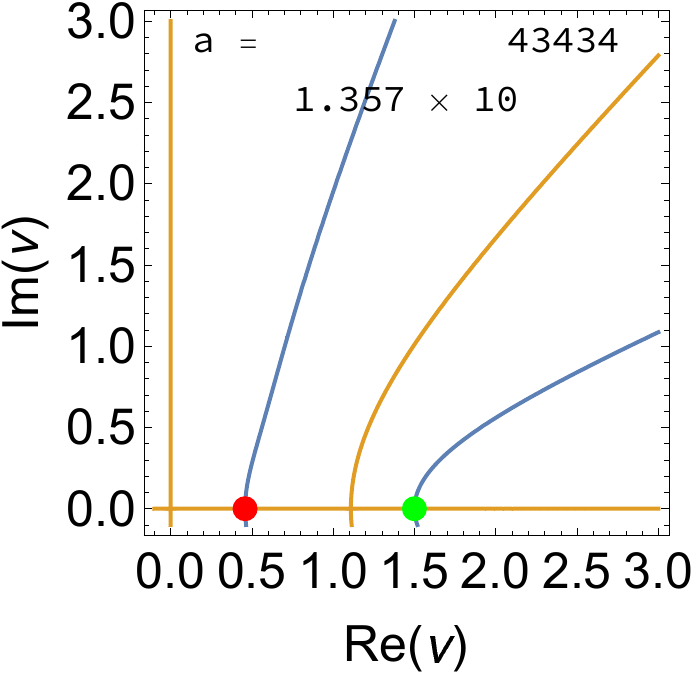}
\caption{\it ${\tilde{\b}_\text{eff}} = 100000$}
\end{subfigure}
\caption{\it dS, $\tilde{\a} = -1000$, $GN^2H^2 = 2 \pi$.}
\label{dS_alpha-1000_H2pi}
\end{figure}

\begin{figure}[ht]
\centering
\begin{subfigure}{.3\textwidth}
\includegraphics[width=\textwidth]{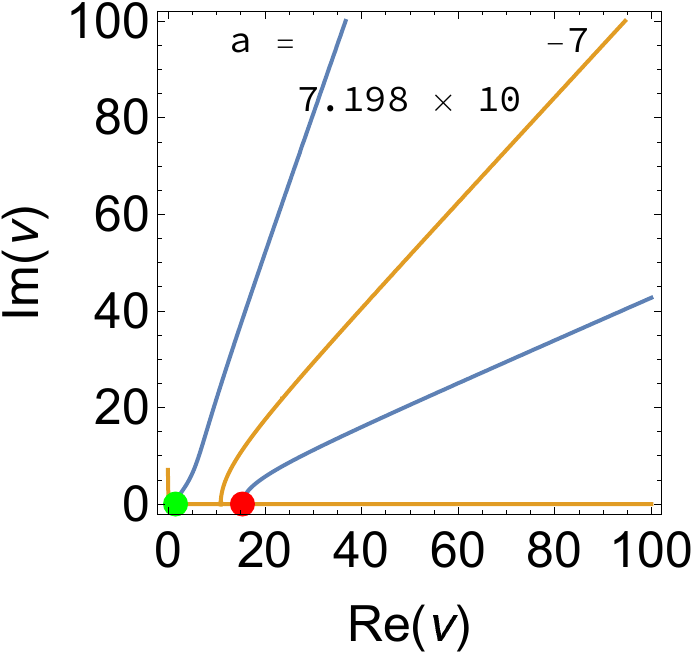}
\caption{\it ${\tilde{\b}_\text{eff}} = -30$}
\end{subfigure}
\begin{subfigure}{.3\textwidth}
\includegraphics[width=\textwidth]{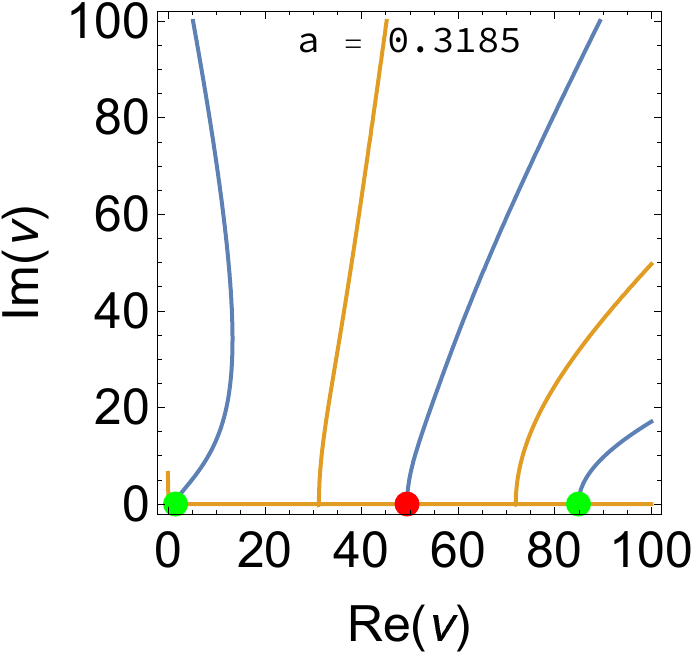}
\caption{\it ${\tilde{\b}_\text{eff}} = -17$}
\end{subfigure}
\begin{subfigure}{.3\textwidth}
\includegraphics[width=\textwidth]{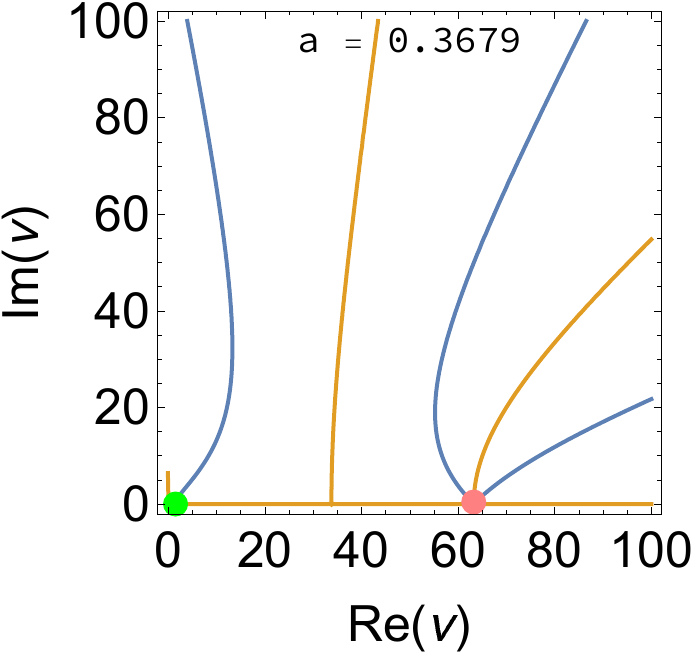}
\caption{\it ${\tilde{\b}_\text{eff}} = -16.8557$}
\end{subfigure}
\begin{subfigure}{.3\textwidth}
\includegraphics[width=\textwidth]{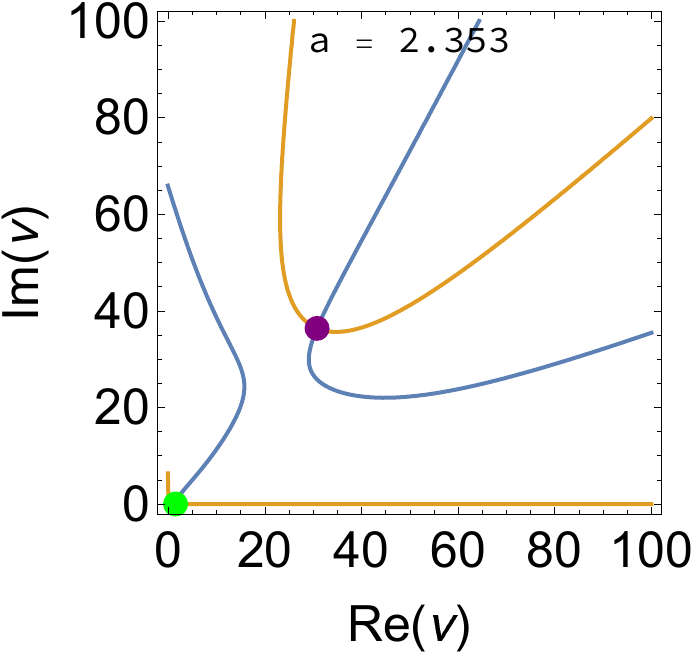}
\caption{\it ${\tilde{\b}_\text{eff}} = -15$}
\end{subfigure}
\begin{subfigure}{.3\textwidth}
\includegraphics[width=\textwidth]{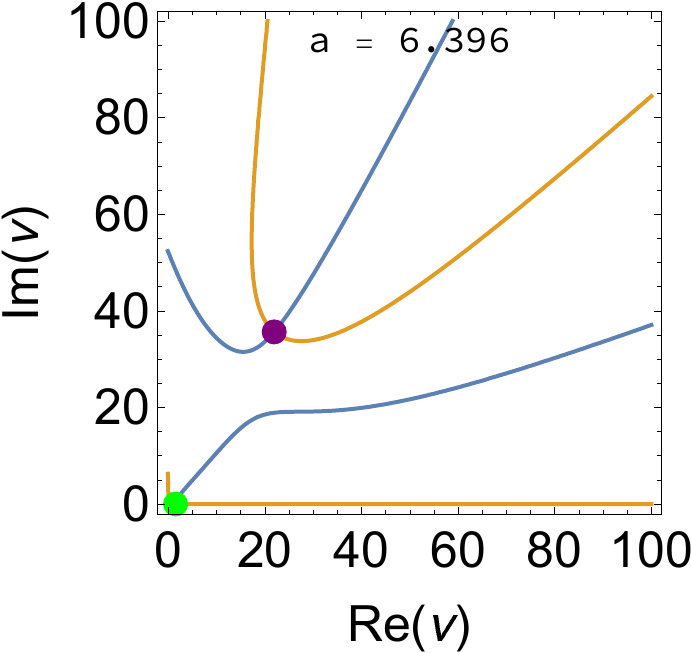}
\caption{\it ${\tilde{\b}_\text{eff}} = -14$}
\end{subfigure}
\begin{subfigure}{.3\textwidth}
\includegraphics[width=\textwidth]{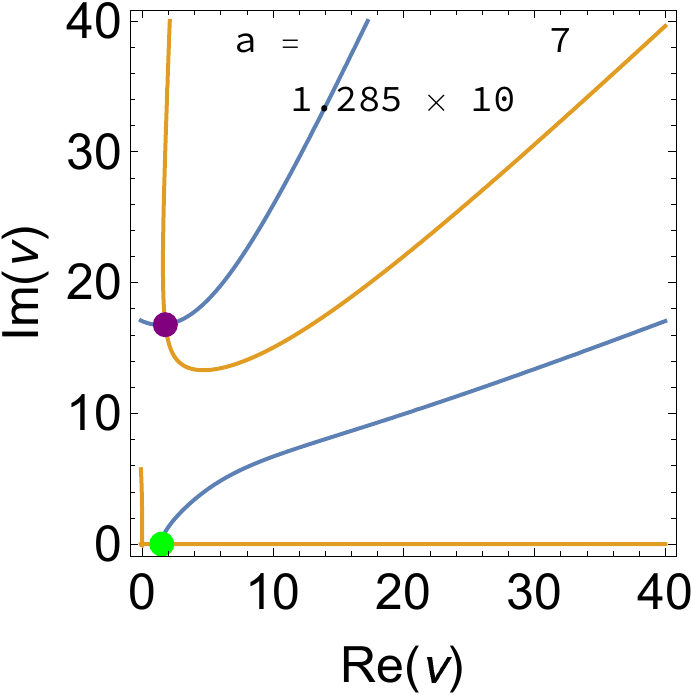}
\caption{\it ${\tilde{\b}_\text{eff}} = 0.513416$}
\end{subfigure}
\begin{subfigure}{.3\textwidth}
\includegraphics[width=\textwidth]{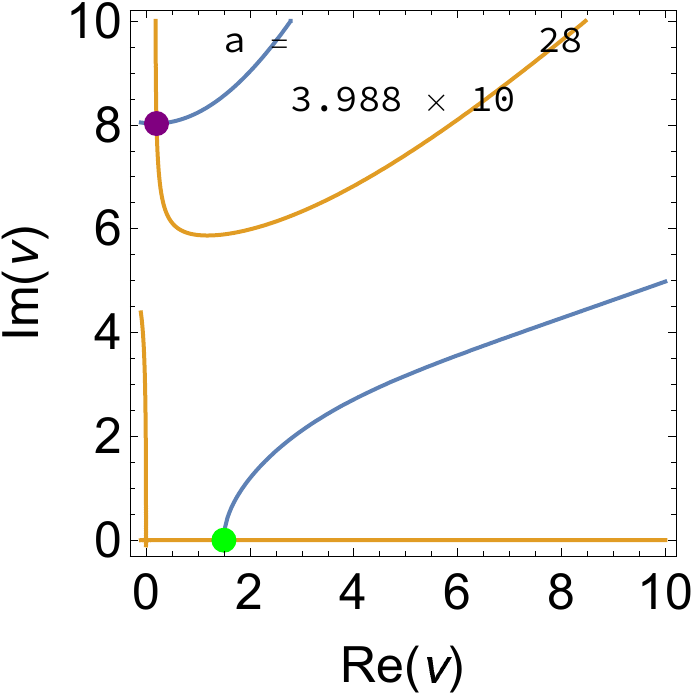}
\caption{\it ${\tilde{\b}_\text{eff}} = 50$}
\end{subfigure}
\begin{subfigure}{.3\textwidth}
\includegraphics[width=\textwidth]{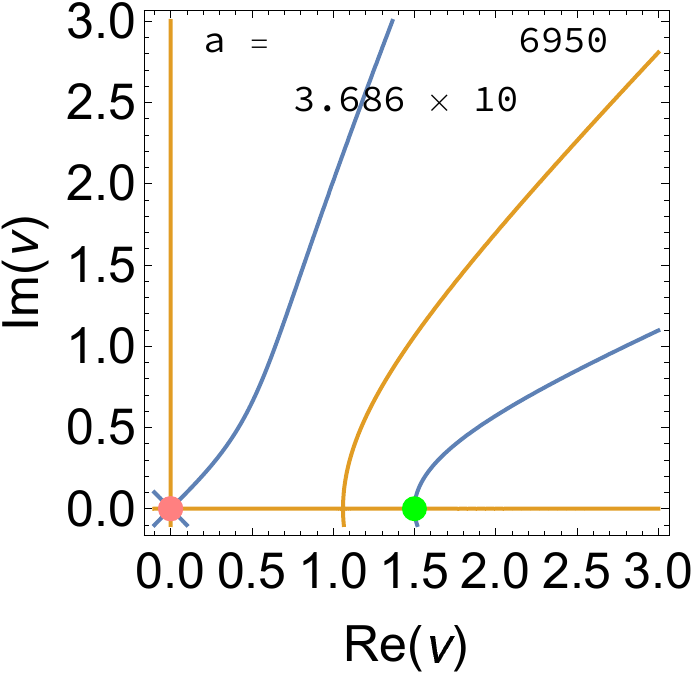}
\caption{\it ${\tilde{\b}_\text{eff}} = 15988.4$}
\end{subfigure}
\begin{subfigure}{.3\textwidth}
\includegraphics[width=\textwidth]{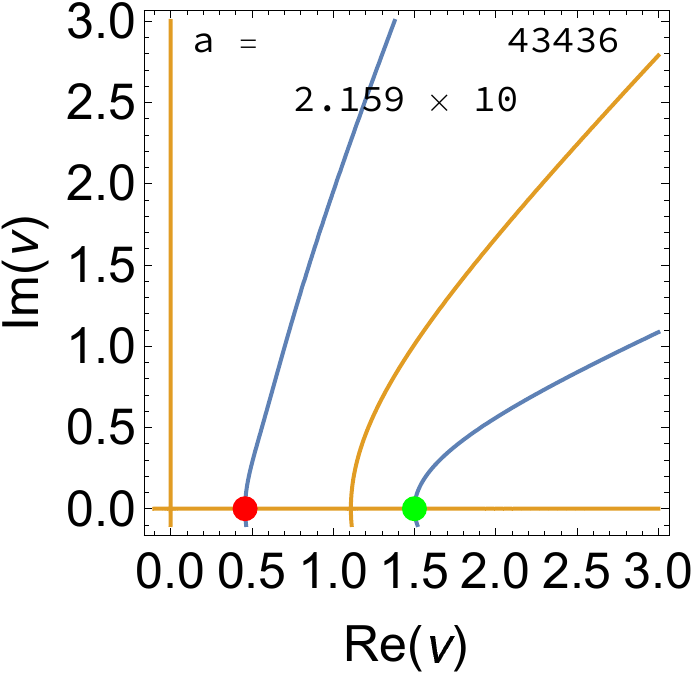}
\caption{\it ${\tilde{\b}_\text{eff}} = 100000$}
\end{subfigure}
\caption{\it dS, $\tilde{\a} = -1000$, $GN^2H^2 = 1000$.}
\label{dS_alpha-1000_H1000}
\end{figure}


\begin{figure}[ht]
\centering
\begin{subfigure}{.3\textwidth}
\includegraphics[width=\textwidth]{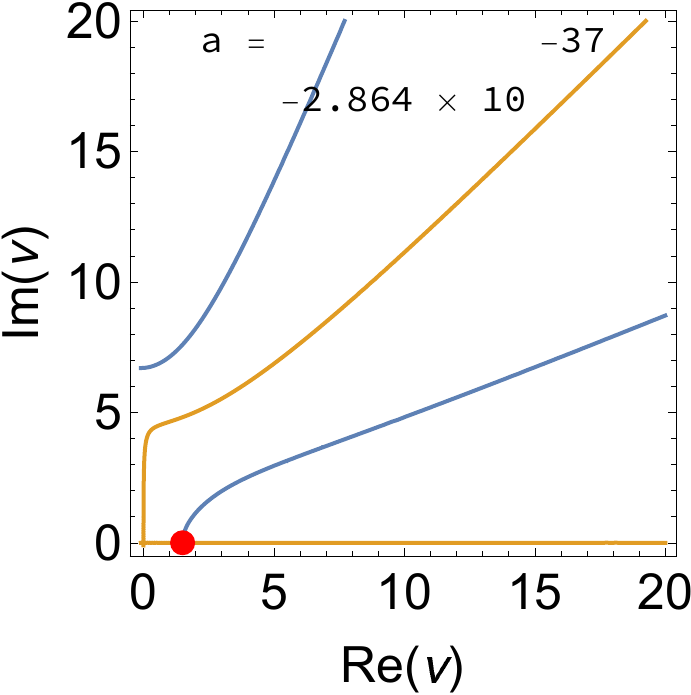}
\caption{\it ${\tilde{\b}_\text{eff}} = -100$}
\end{subfigure}
\begin{subfigure}{.3\textwidth}
\includegraphics[width=\textwidth]{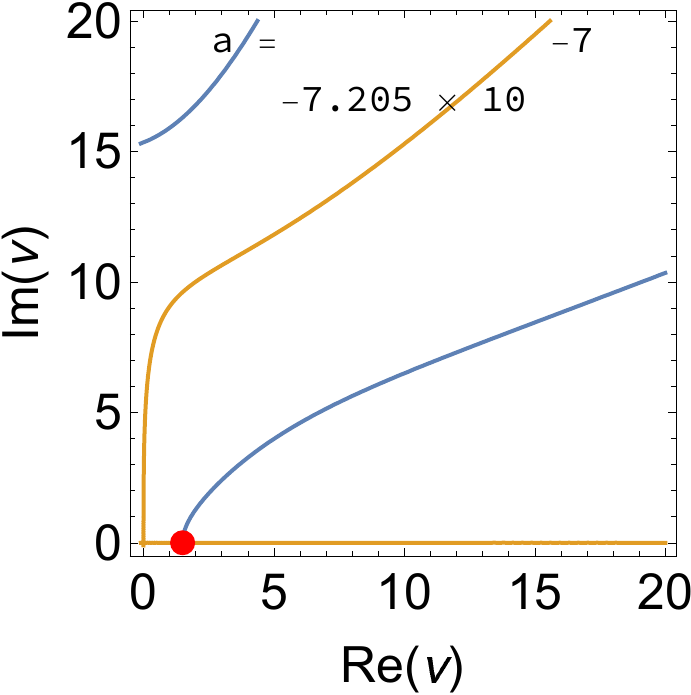}
\caption{\it ${\tilde{\b}_\text{eff}} = -30$}
\end{subfigure}
\begin{subfigure}{.3\textwidth}
\includegraphics[width=\textwidth]{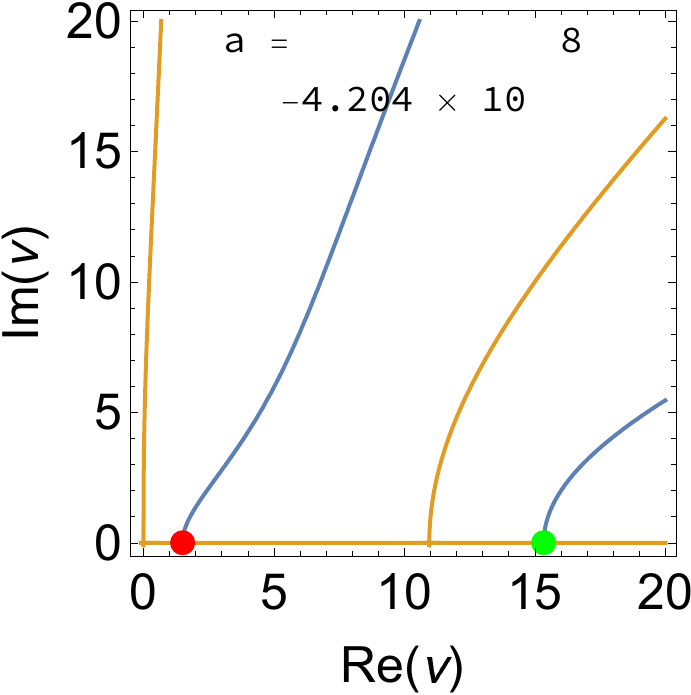}
\caption{\it ${\tilde{\b}_\text{eff}} = 4$}
\end{subfigure}
\begin{subfigure}{.3\textwidth}
\includegraphics[width=\textwidth]{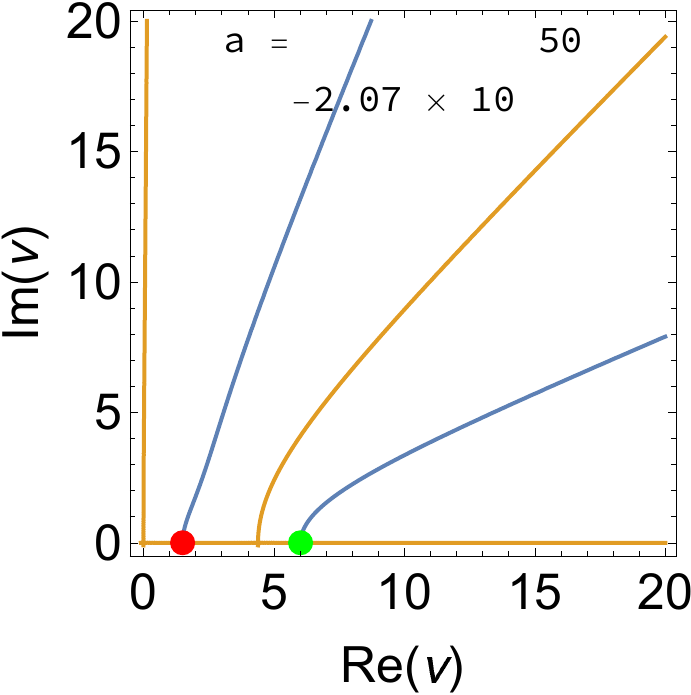}
\caption{\it ${\tilde{\b}_\text{eff}} = 100$}
\end{subfigure}
\begin{subfigure}{.3\textwidth}
\includegraphics[width=\textwidth]{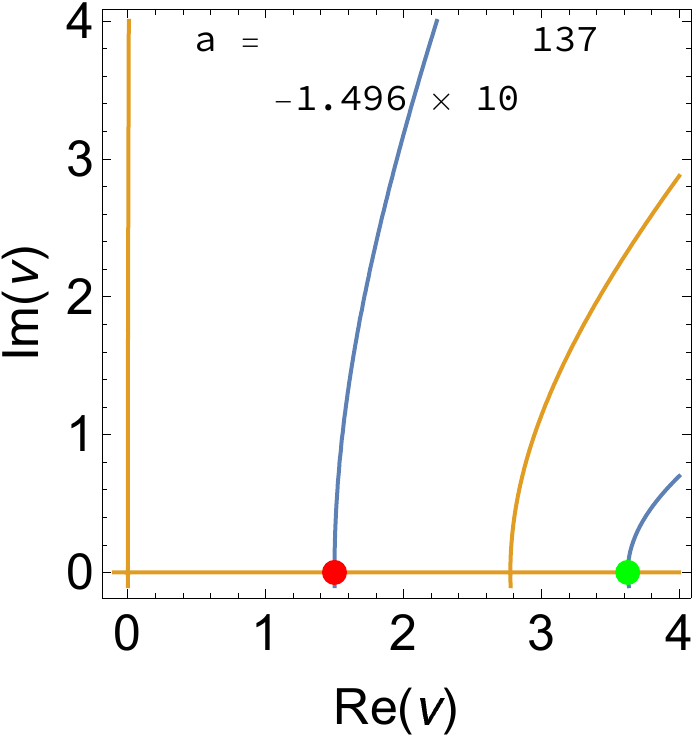}
\caption{\it ${\tilde{\b}_\text{eff}} = 300$}
\end{subfigure}
\begin{subfigure}{.3\textwidth}
\includegraphics[width=\textwidth]{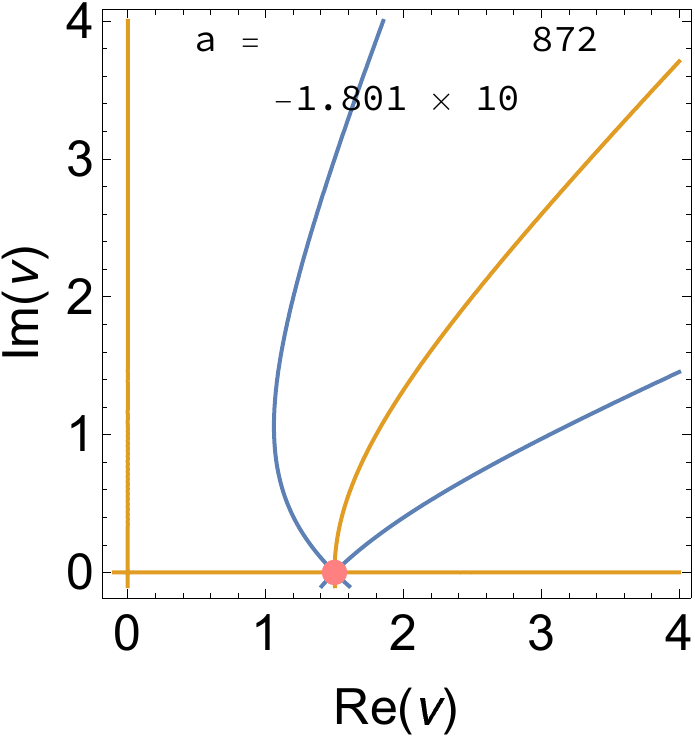}
\caption{\it ${\tilde{\b}_\text{eff}} = 1992.59$}
\end{subfigure}
\begin{subfigure}{.3\textwidth}
\includegraphics[width=\textwidth]{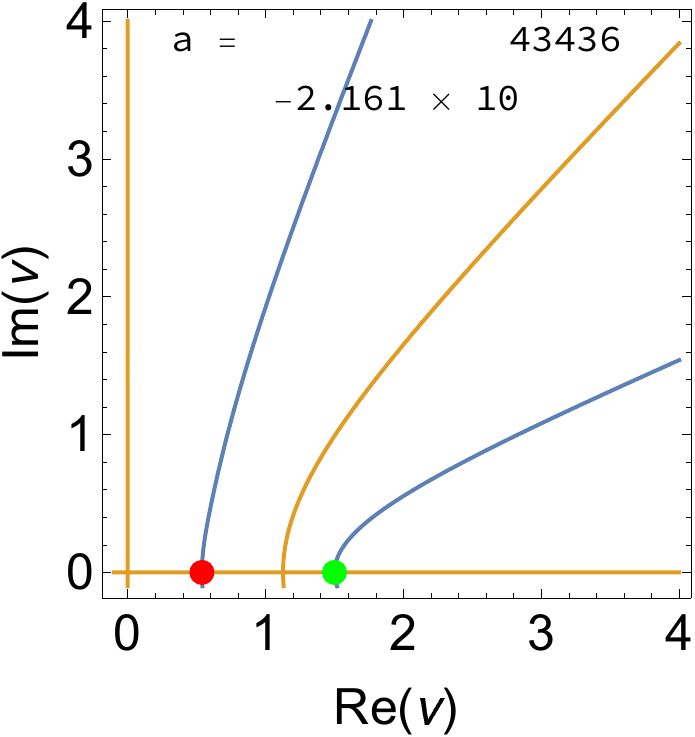}
\caption{\it ${\tilde{\b}_\text{eff}} = 100000$}
\end{subfigure}
\caption{\it dS, $\tilde{\a} = 1000$, $GN^2H^2 = 1000$.}
\label{dS_alpha1000_H1000}
\end{figure}


\begin{figure}[ht]
\centering
\begin{subfigure}{.3\textwidth}
\includegraphics[width=\textwidth]{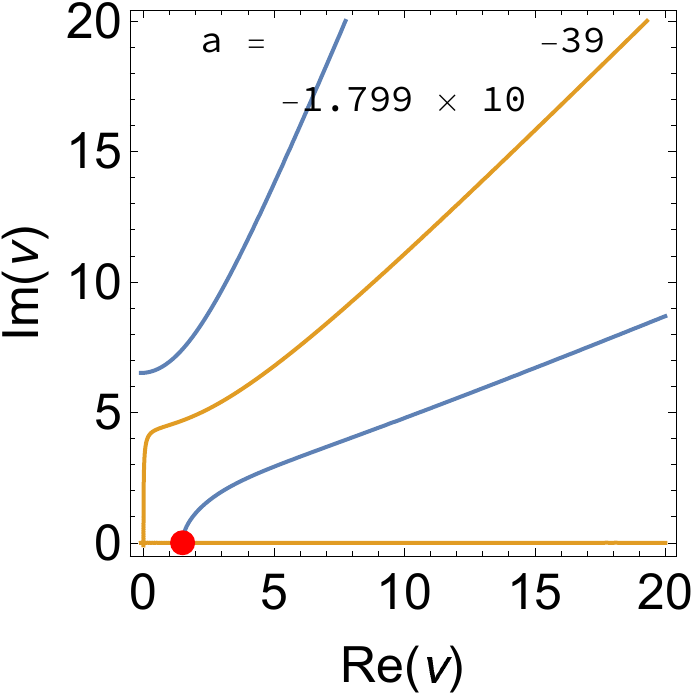}
\caption{\it ${\tilde{\b}_\text{eff}} = -100$}
\end{subfigure}
\begin{subfigure}{.3\textwidth}
\includegraphics[width=\textwidth]{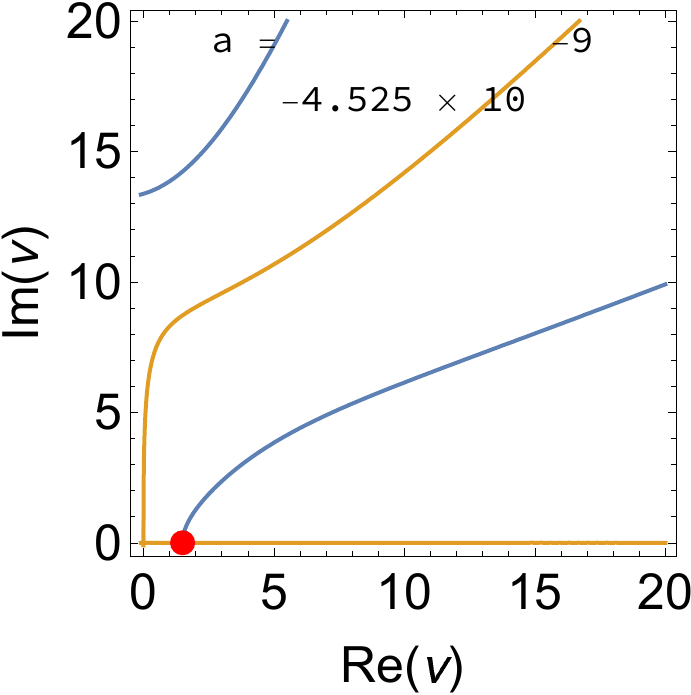}
\caption{\it ${\tilde{\b}_\text{eff}} = -30$}
\end{subfigure}
\begin{subfigure}{.3\textwidth}
\includegraphics[width=\textwidth]{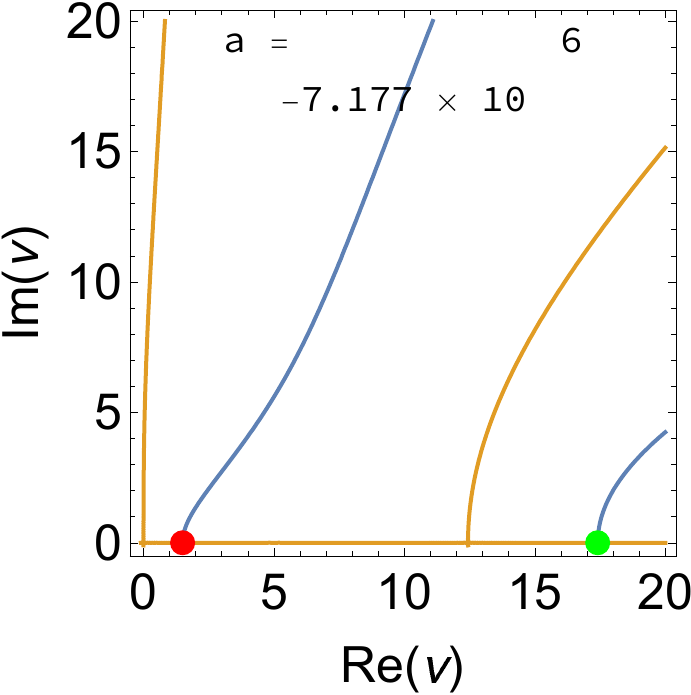}
\caption{\it ${\tilde{\b}_\text{eff}} = 5$}
\end{subfigure}
\begin{subfigure}{.3\textwidth}
\includegraphics[width=\textwidth]{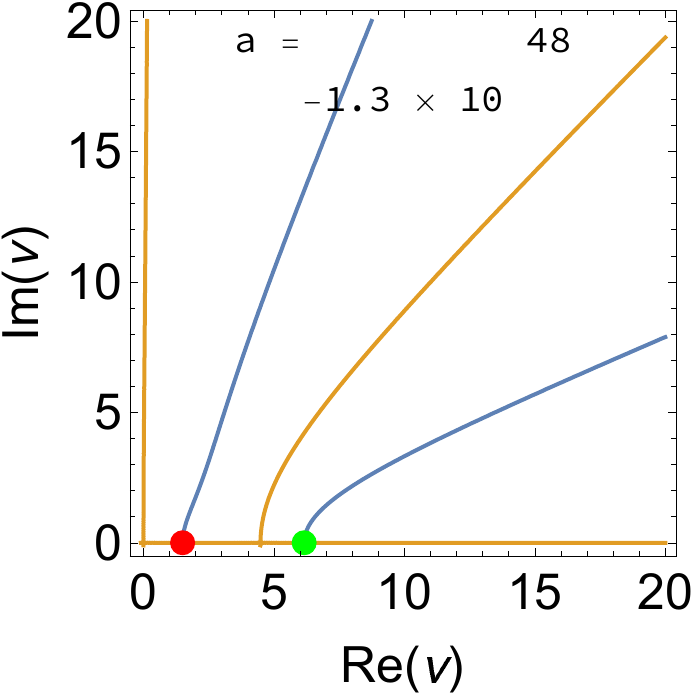}
\caption{\it ${\tilde{\b}_\text{eff}} = 100$}
\end{subfigure}
\begin{subfigure}{.3\textwidth}
\includegraphics[width=\textwidth]{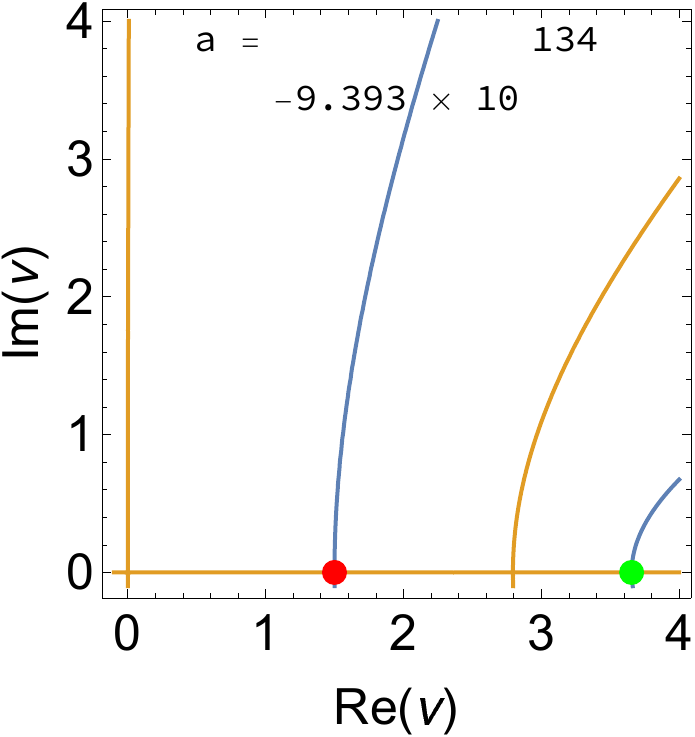}
\caption{\it ${\tilde{\b}_\text{eff}} = 300$}
\end{subfigure}
\begin{subfigure}{.3\textwidth}
\includegraphics[width=\textwidth]{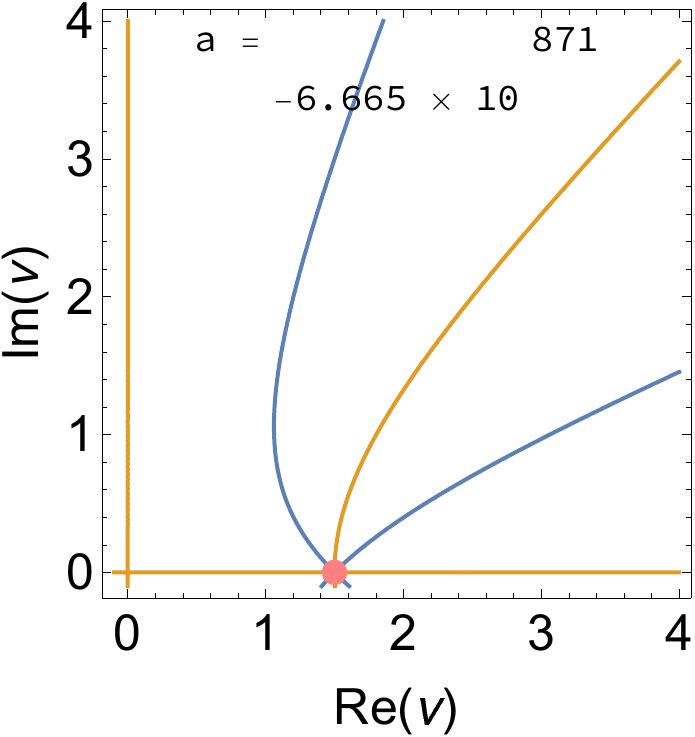}
\caption{\it ${\tilde{\b}_\text{eff}} = 1996.66$}
\end{subfigure}
\begin{subfigure}{.3\textwidth}
\includegraphics[width=\textwidth]{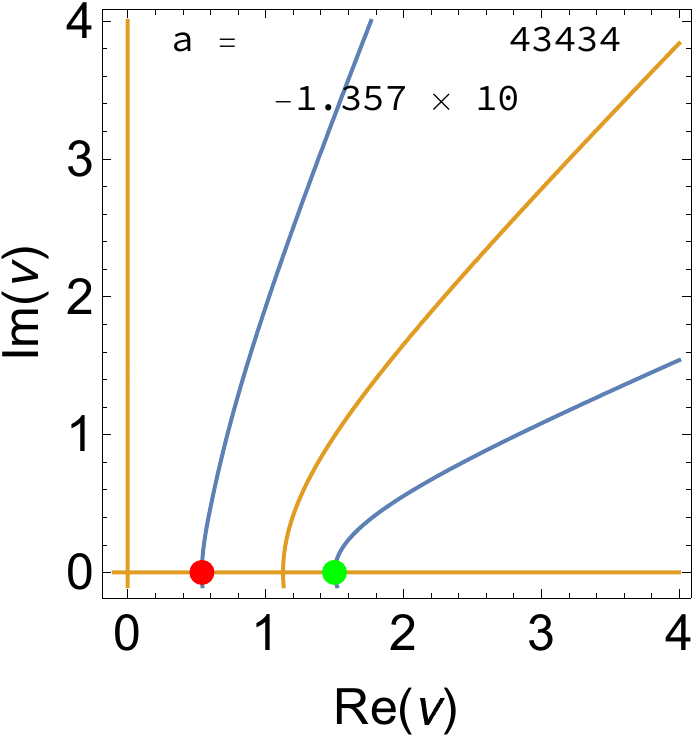}
\caption{\it ${\tilde{\b}_\text{eff}} = 100000$}
\end{subfigure}
\caption{\it dS, $\tilde{\a} = 1000$, $GN^2H^2 = 2 \pi $.}
\label{dS_alpha1000_H2pi}
\end{figure}


\begin{figure}[ht]
\centering
\begin{subfigure}{.3\textwidth}
\includegraphics[width=\textwidth]{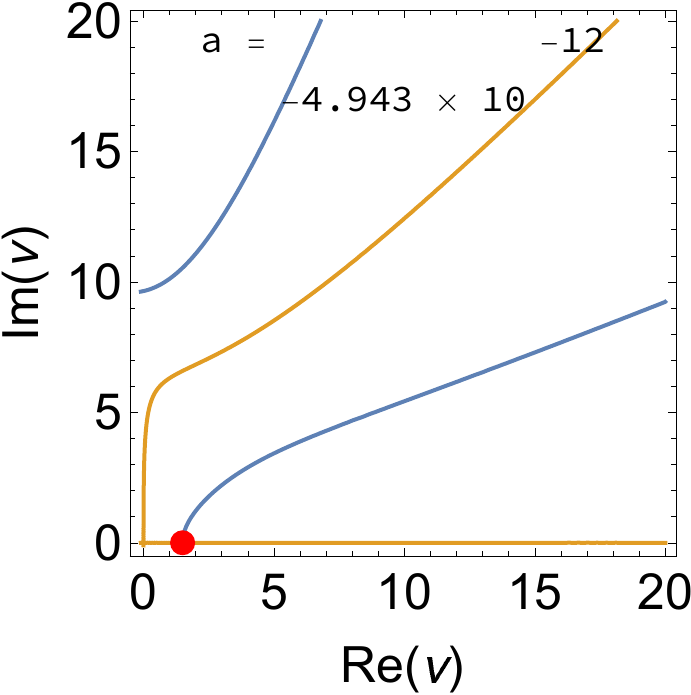}
\caption{\it ${\tilde{\b}_\text{eff}} = -30$}
\end{subfigure}
\begin{subfigure}{.3\textwidth}
\includegraphics[width=\textwidth]{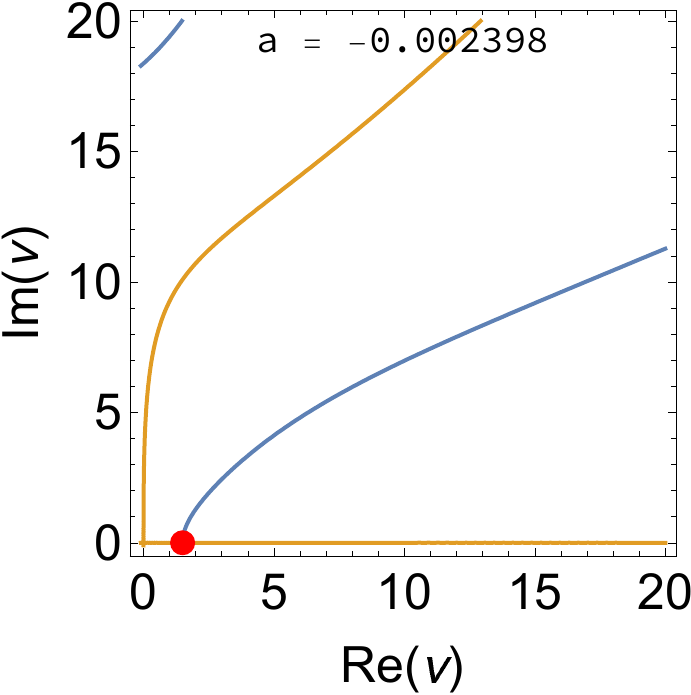}
\caption{\it ${\tilde{\b}_\text{eff}} = -10$}
\end{subfigure}
\begin{subfigure}{.3\textwidth}
\includegraphics[width=\textwidth]{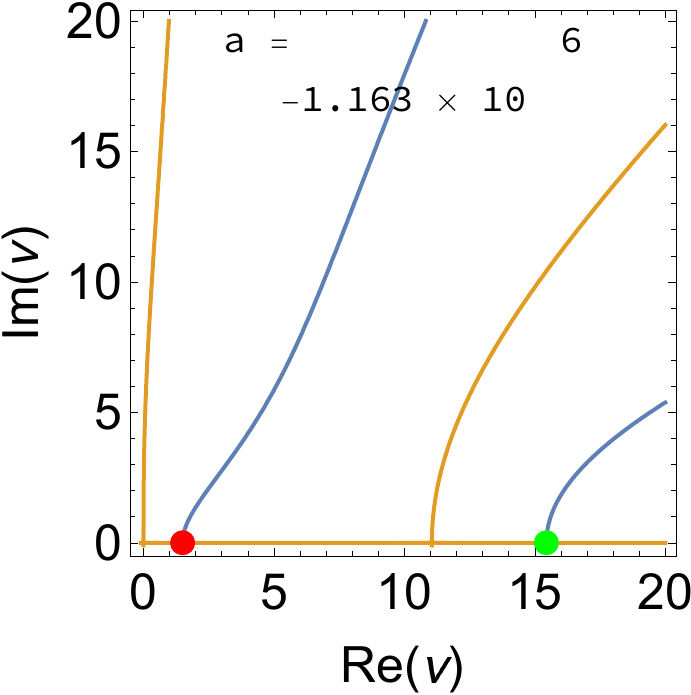}
\caption{\it ${\tilde{\b}_\text{eff}} = 5$}
\end{subfigure}
\begin{subfigure}{.3\textwidth}
\includegraphics[width=\textwidth]{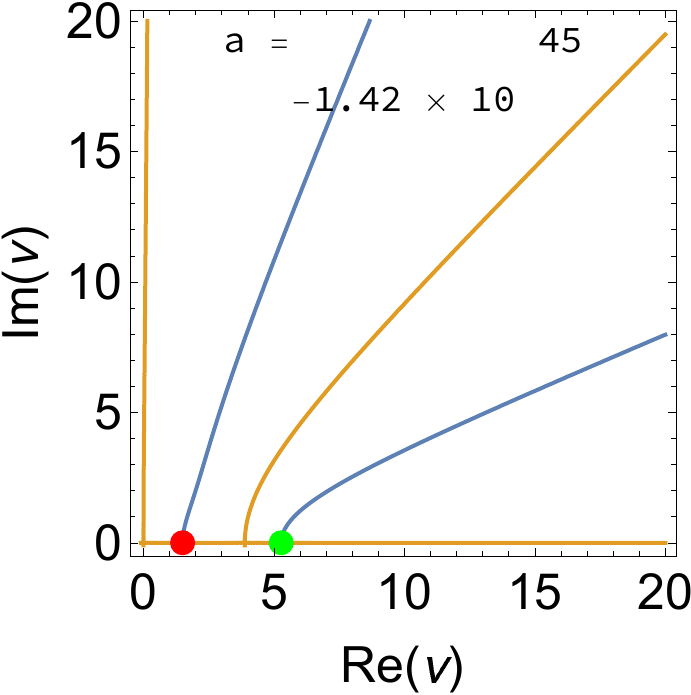}
\caption{\it ${\tilde{\b}_\text{eff}} = 100$}
\end{subfigure}
\begin{subfigure}{.3\textwidth}
\includegraphics[width=\textwidth]{Figures/dS_alpha1000_H2pi/beta300}
\caption{\it ${\tilde{\b}_\text{eff}} = 300$}
\end{subfigure}
\begin{subfigure}{.3\textwidth}
\includegraphics[width=\textwidth]{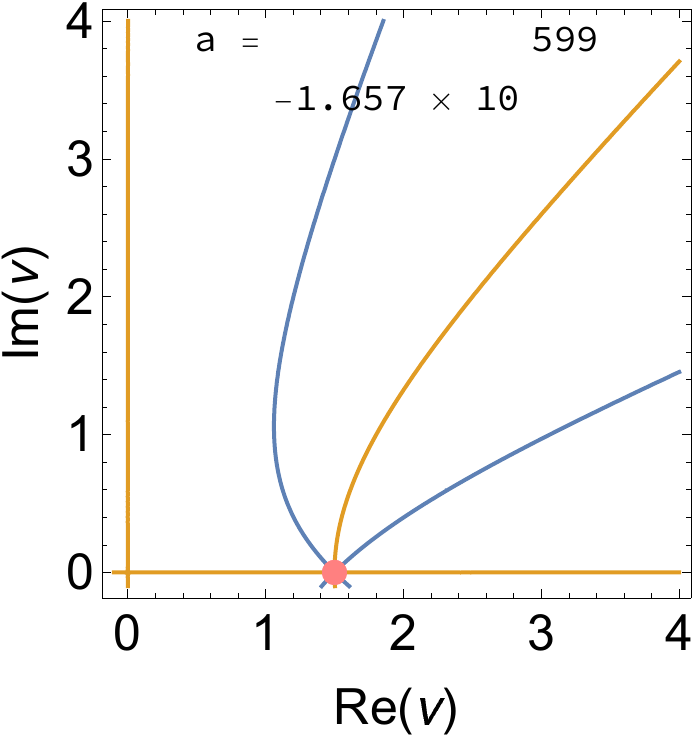}
\caption{\it ${\tilde{\b}_\text{eff}} = 1375.79$}
\end{subfigure}
\begin{subfigure}{.3\textwidth}
\includegraphics[width=\textwidth]{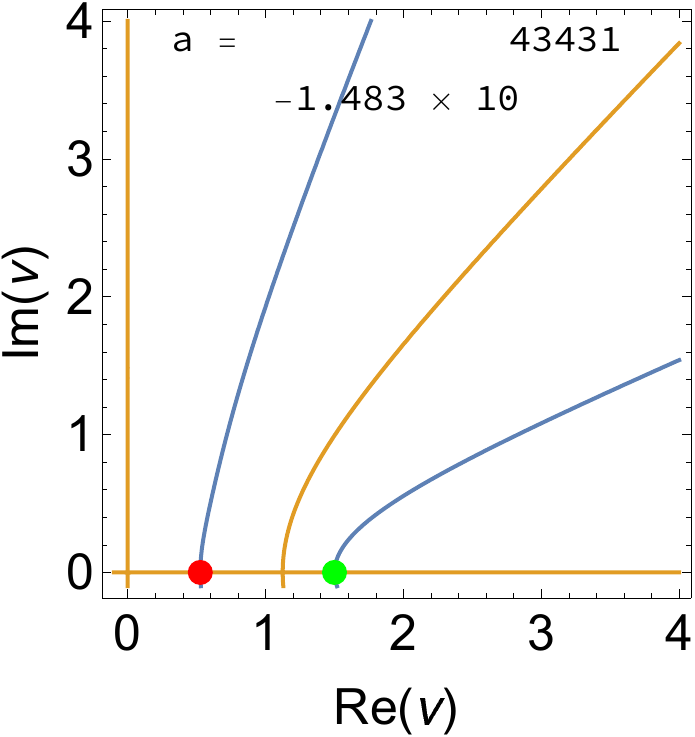}
\caption{\it ${\tilde{\b}_\text{eff}} = 100000$}
\end{subfigure}
\caption{\it dS, $\tilde{\a} = 1000$, $GN^2H^2 = 0.01 $.}
\label{dS_alpha1000_H0.01}
\end{figure}


\begin{figure}[ht]
\centering
\begin{subfigure}{.3\textwidth}
\includegraphics[width=\textwidth]{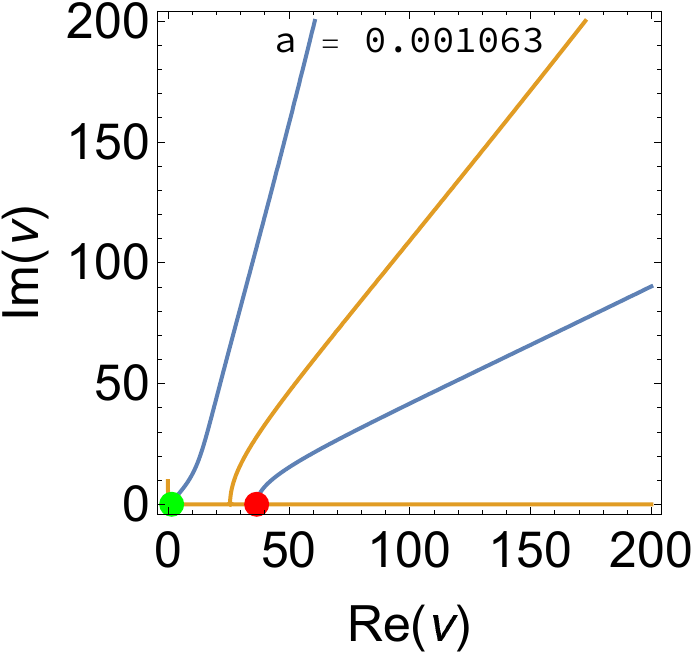}
\caption{\it ${\tilde{\b}_\text{eff}} = -10$}
\end{subfigure}
\begin{subfigure}{.3\textwidth}
\includegraphics[width=\textwidth]{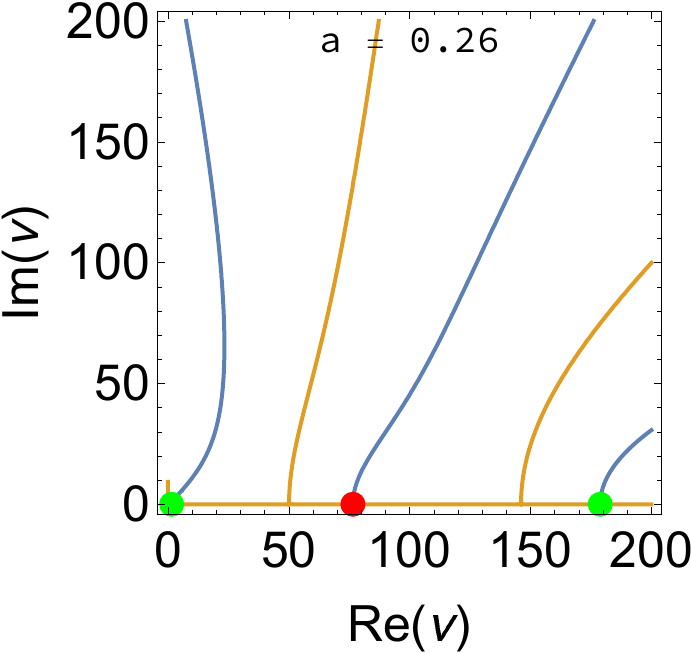}
\caption{\it ${\tilde{\b}_\text{eff}} = -4.5$}
\end{subfigure}
\begin{subfigure}{.3\textwidth}
\includegraphics[width=\textwidth]{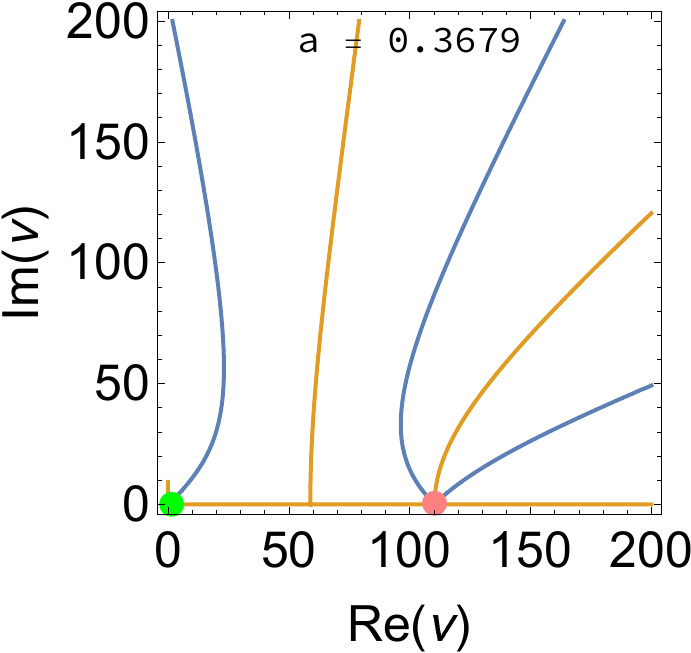}
\caption{\it ${\tilde{\b}_\text{eff}} = -4.15294$}
\end{subfigure}
\begin{subfigure}{.3\textwidth}
\includegraphics[width=\textwidth]{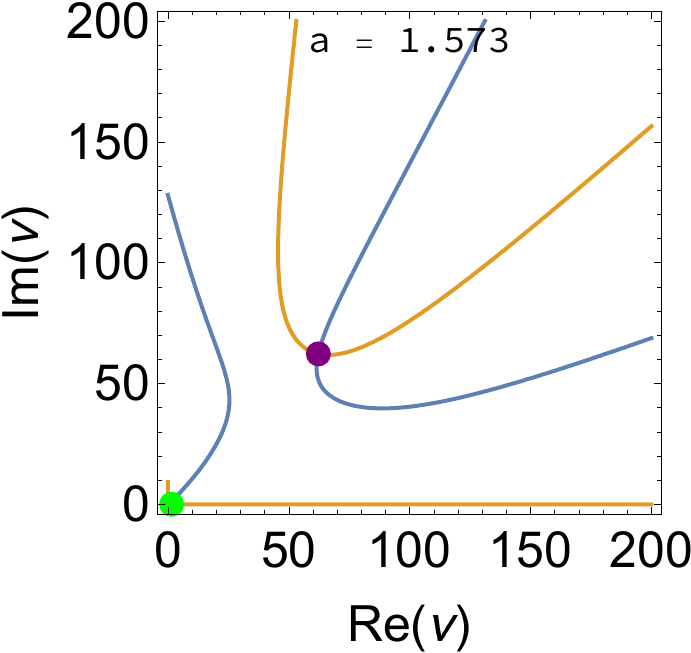}
\caption{\it ${\tilde{\b}_\text{eff}} = -2.7$}
\end{subfigure}
\begin{subfigure}{.3\textwidth}
\includegraphics[width=\textwidth]{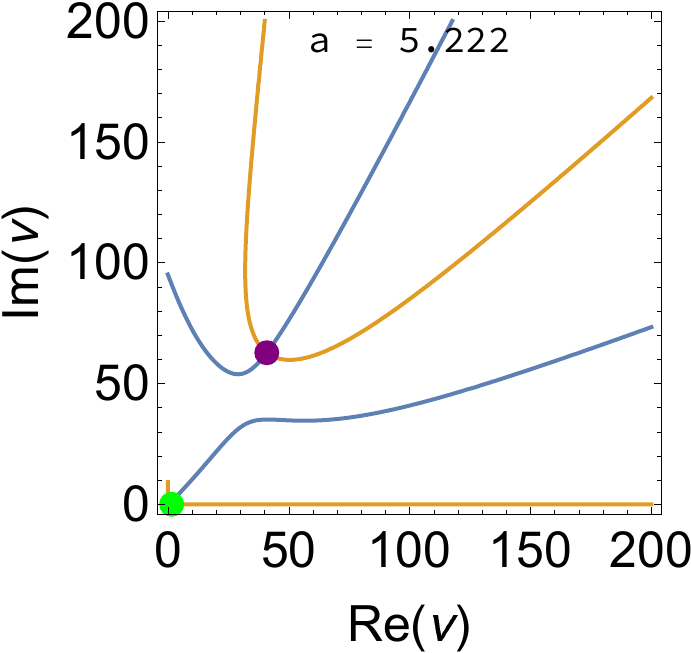}
\caption{\it ${\tilde{\b}_\text{eff}} = -1.5$}
\end{subfigure}
\begin{subfigure}{.3\textwidth}
\includegraphics[width=\textwidth]{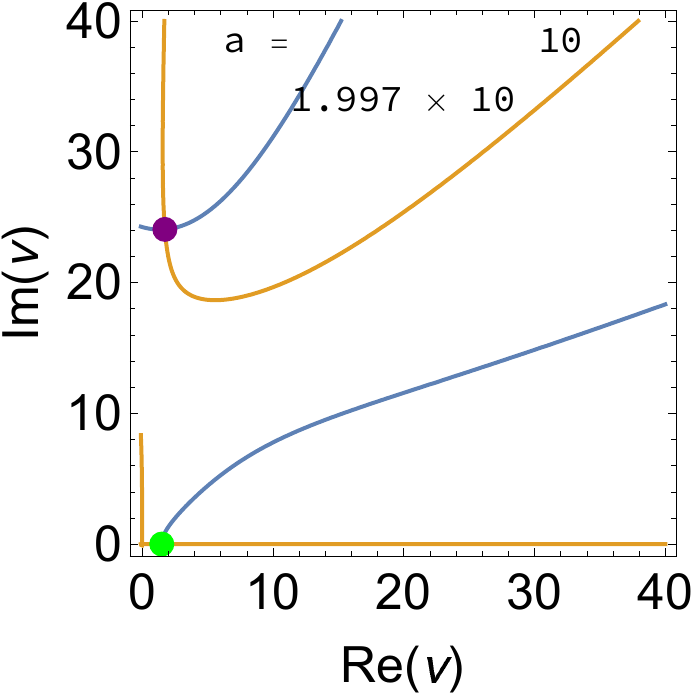}
\caption{\it ${\tilde{\b}_\text{eff}} = 20.5646$}
\end{subfigure}
\begin{subfigure}{.3\textwidth}
\includegraphics[width=\textwidth]{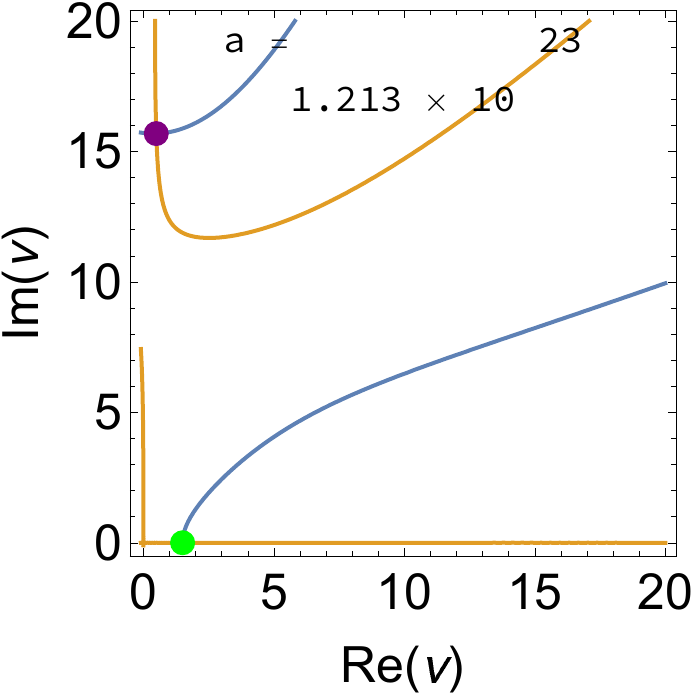}
\caption{\it ${\tilde{\b}_\text{eff}} = 50$}
\end{subfigure}
\begin{subfigure}{.3\textwidth}
\includegraphics[width=\textwidth]{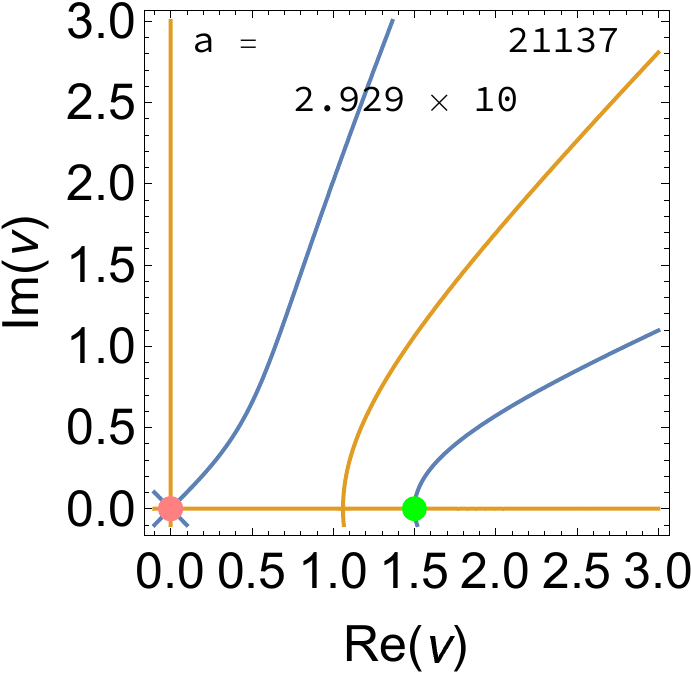}
\caption{\it ${\tilde{\b}_\text{eff}} = 48667.7$}
\end{subfigure}
\begin{subfigure}{.3\textwidth}
\includegraphics[width=\textwidth]{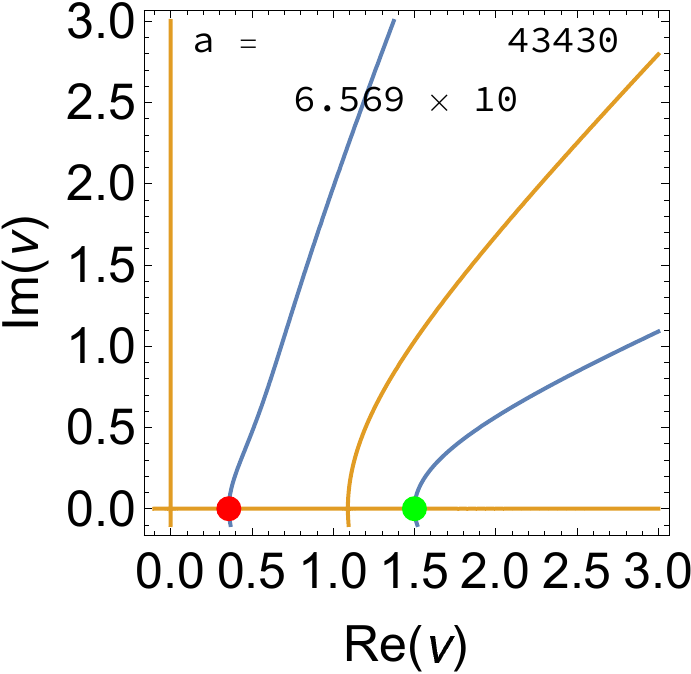}
\caption{\it ${\tilde{\b}_\text{eff}} = 100000$}
\end{subfigure}
\caption{\it dS, $\tilde{\a} = 100$, $GN^2H^2 = 0.001 $.}
\label{dS_alpha100_H0.001}
\end{figure}


\begin{figure}[ht]
\centering
\begin{subfigure}{.3\textwidth}
\includegraphics[width=\textwidth]{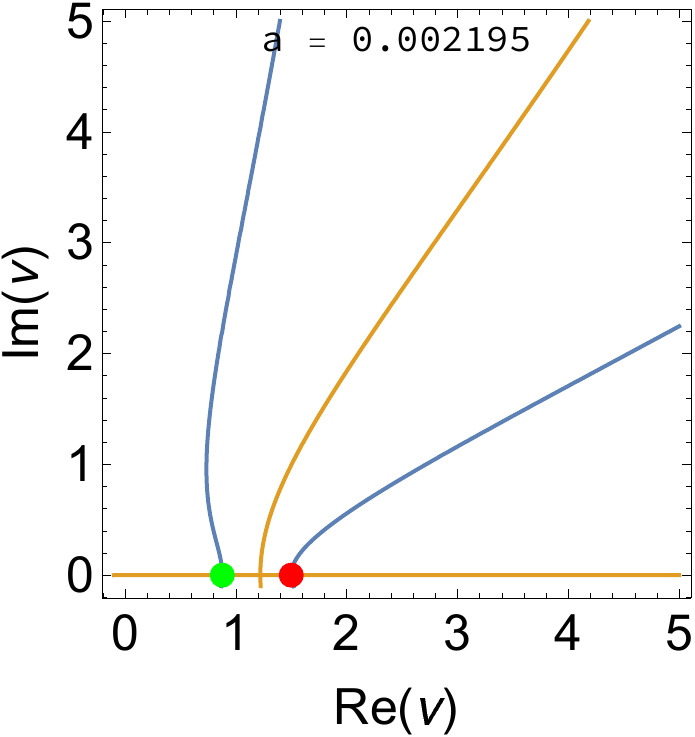}
\caption{\it ${\tilde{\b}_\text{eff}} = -10$}
\end{subfigure}
\begin{subfigure}{.3\textwidth}
\includegraphics[width=\textwidth]{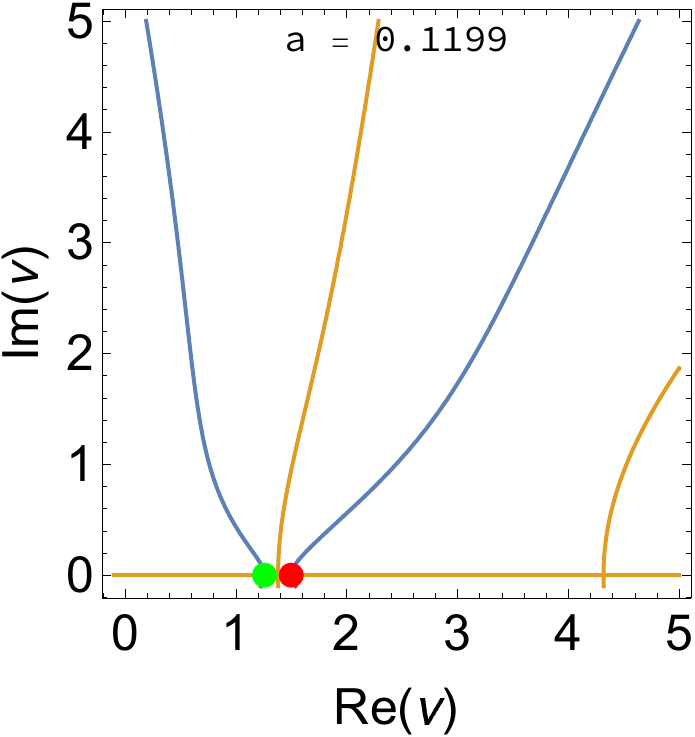}
\caption{\it ${\tilde{\b}_\text{eff}} = -6$}
\end{subfigure}
\begin{subfigure}{.3\textwidth}
\includegraphics[width=\textwidth]{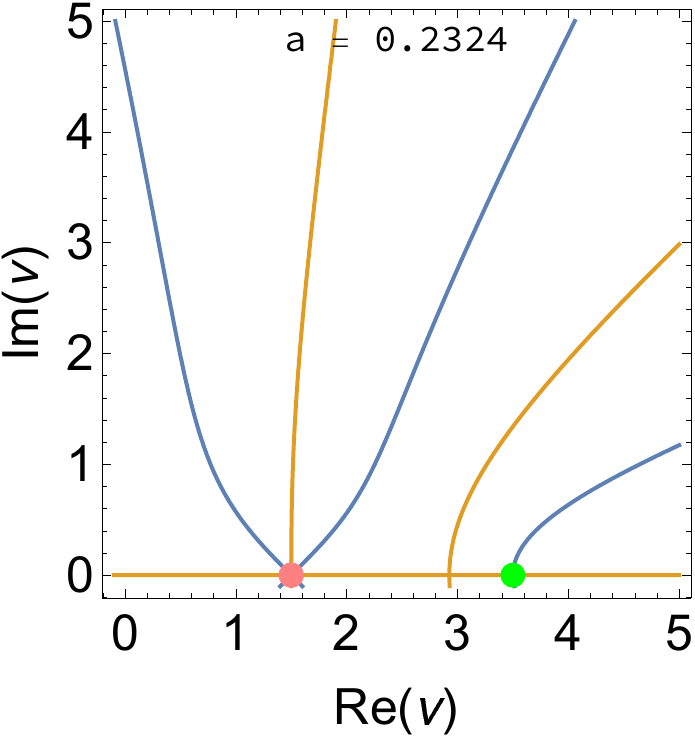}
\caption{\it ${\tilde{\b}_\text{eff}} = -5.33788$}
\end{subfigure}
\begin{subfigure}{.3\textwidth}
\includegraphics[width=\textwidth]{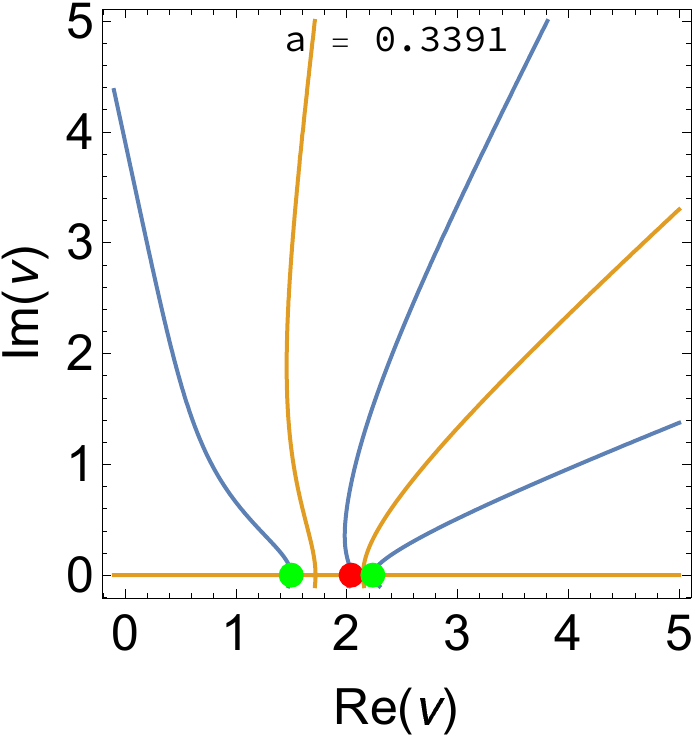}
\caption{\it ${\tilde{\b}_\text{eff}} = -4.96$}
\end{subfigure}
\begin{subfigure}{.3\textwidth}
\includegraphics[width=\textwidth]{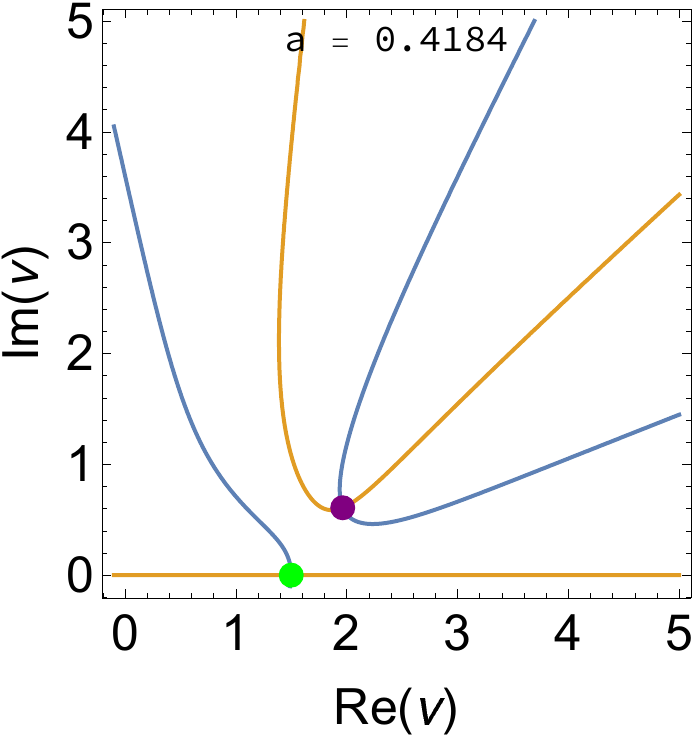}
\caption{\it ${\tilde{\b}_\text{eff}} = -4.75$}
\end{subfigure}
\begin{subfigure}{.3\textwidth}
\includegraphics[width=\textwidth]{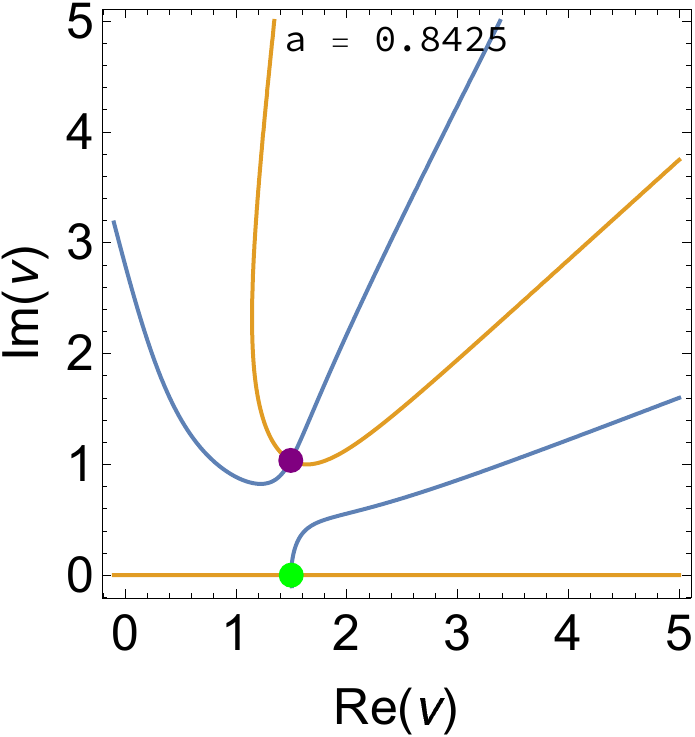}
\caption{\it ${\tilde{\b}_\text{eff}} = -4.05$}
\end{subfigure}
\begin{subfigure}{.3\textwidth}
\includegraphics[width=\textwidth]{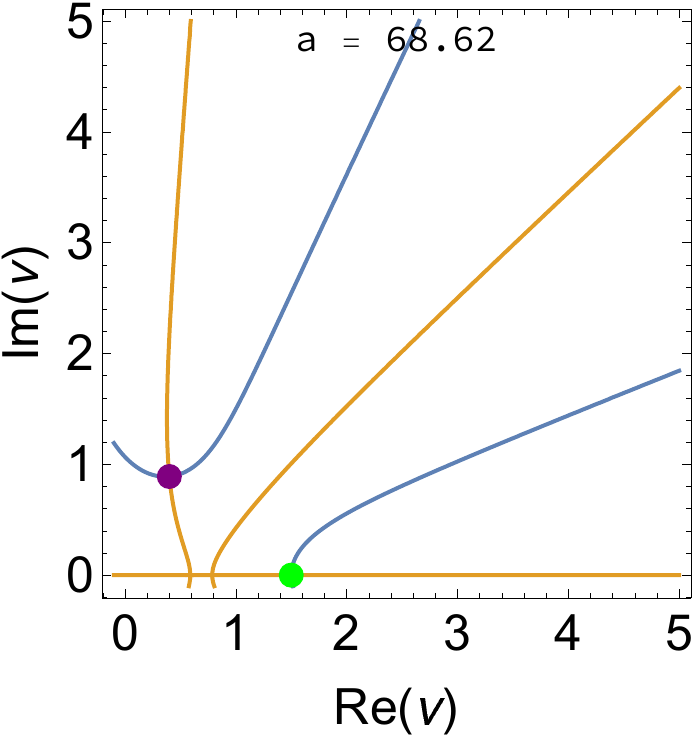}
\caption{\it ${\tilde{\b}_\text{eff}} = 0.35$}
\end{subfigure}
\begin{subfigure}{.3\textwidth}
\includegraphics[width=\textwidth]{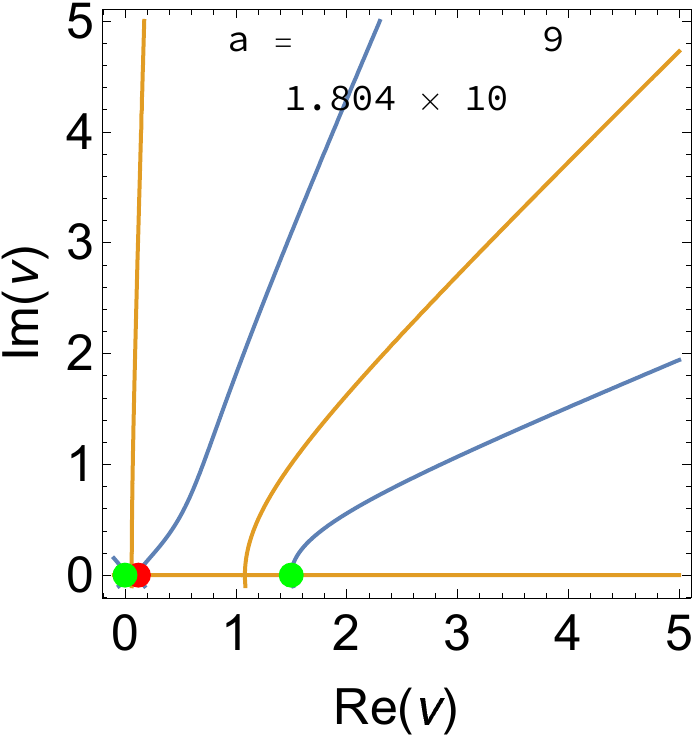}
\caption{\it ${\tilde{\b}_\text{eff}} = 17.4347$}
\end{subfigure}
\begin{subfigure}{.3\textwidth}
\includegraphics[width=\textwidth]{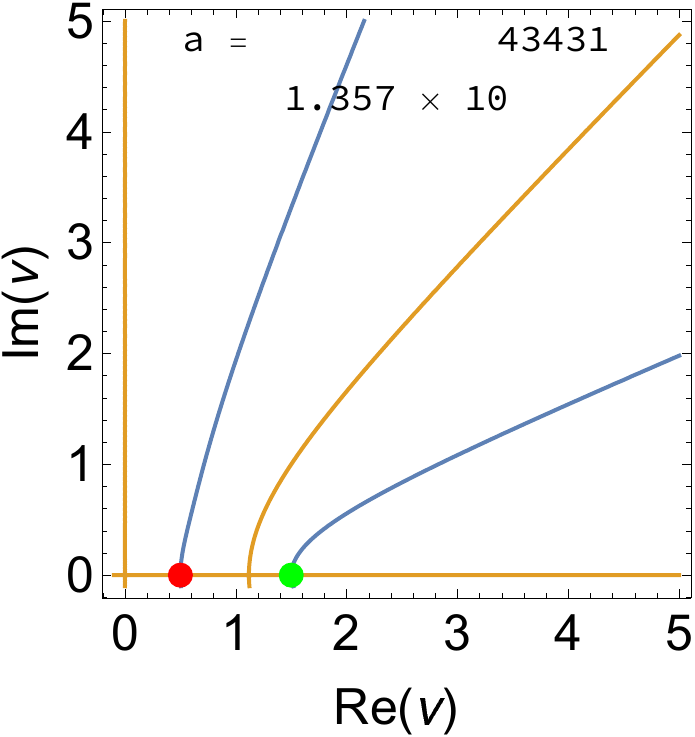}
\caption{\it ${\tilde{\b}_\text{eff}} = 100000$}
\end{subfigure}
\caption{\it dS, $\tilde{\a} = -1$, $GN^2H^2 = 2\pi $.}
\label{dS_alpha-1_H2pi}
\end{figure}

\begin{figure}[ht]
\centering
\begin{subfigure}{.3\textwidth}
\includegraphics[width=\textwidth]{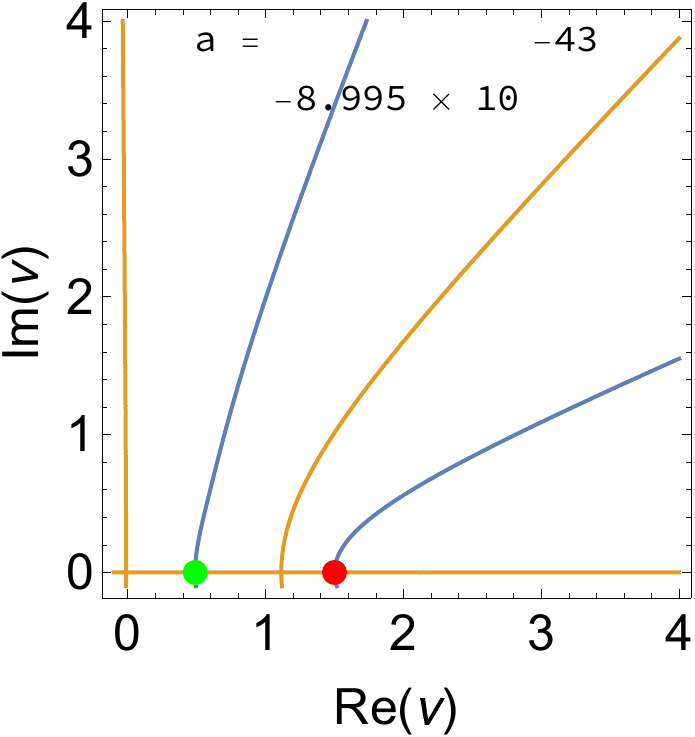}
\caption{\it \it${\tilde{\b}_\text{eff}} = -100$}
\end{subfigure}
\begin{subfigure}{.3\textwidth}
\includegraphics[width=\textwidth]{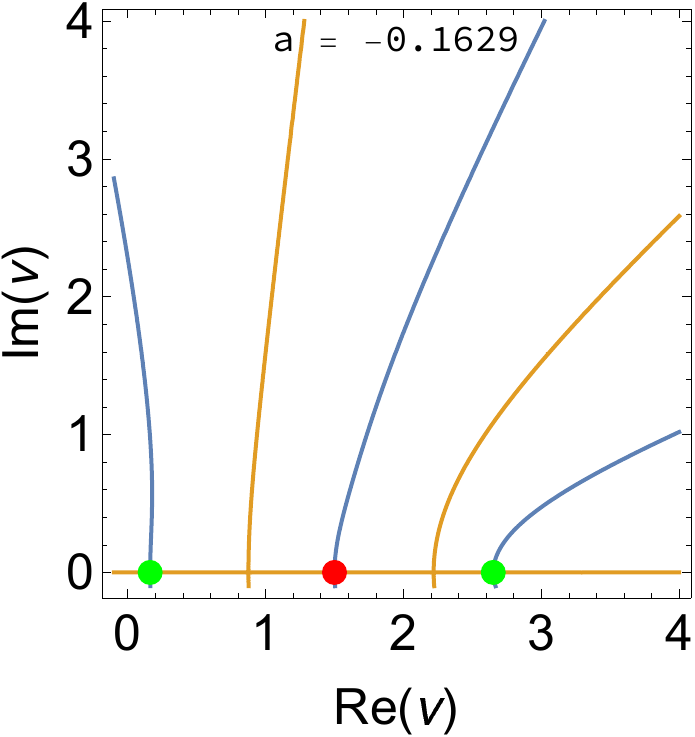}
\caption{\it \it${\tilde{\b}_\text{eff}} = -5$}
\end{subfigure}
\begin{subfigure}{.3\textwidth}
\includegraphics[width=\textwidth]{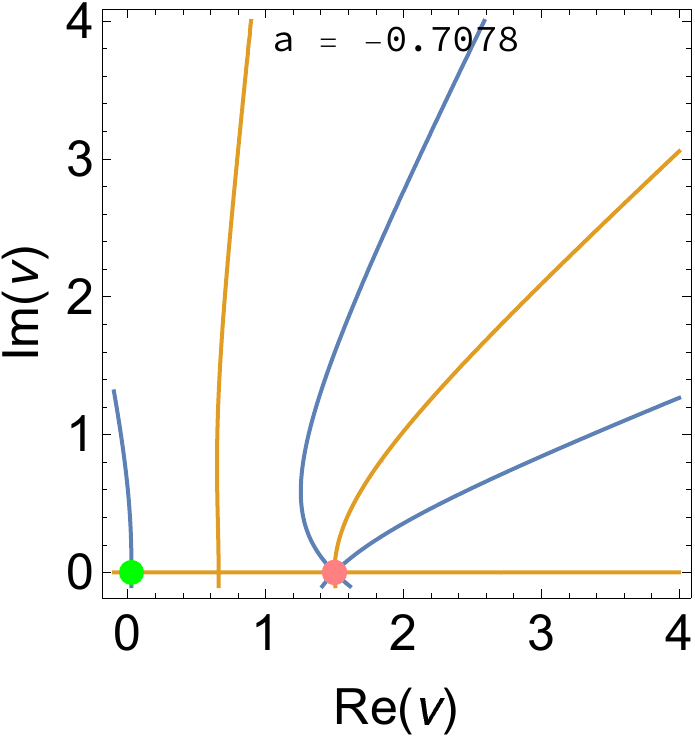}
\caption{\it \it${\tilde{\b}_\text{eff}} = -3.53102$}
\end{subfigure}
\begin{subfigure}{.3\textwidth}
\includegraphics[width=\textwidth]{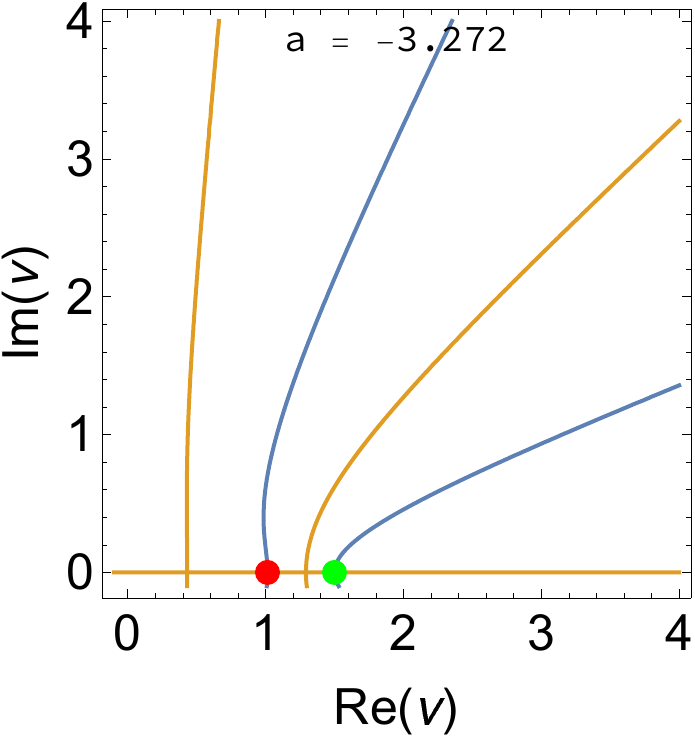}
\caption{\it \it${\tilde{\b}_\text{eff}} = -2$}
\end{subfigure}
\begin{subfigure}{.3\textwidth}
\includegraphics[width=\textwidth]{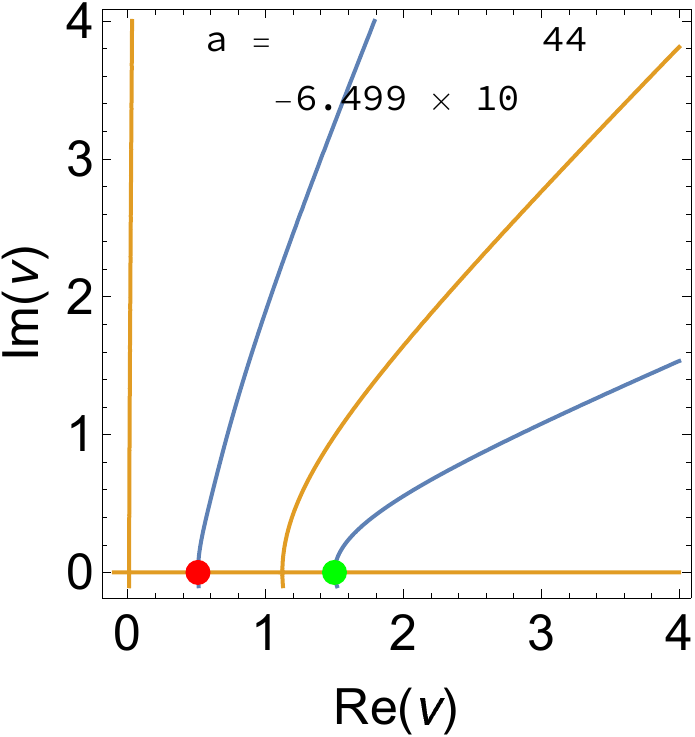}
\caption{\it \it${\tilde{\b}_\text{eff}} = 100$}
\end{subfigure}
\caption{\it $\a = 0$, $GN^2H^2 = 4\pi$.}
\label{dS_H4pi_alpha0}
\end{figure}


\begin{figure}[ht]
\centering
\begin{subfigure}{.3\textwidth}
\includegraphics[width=\textwidth]{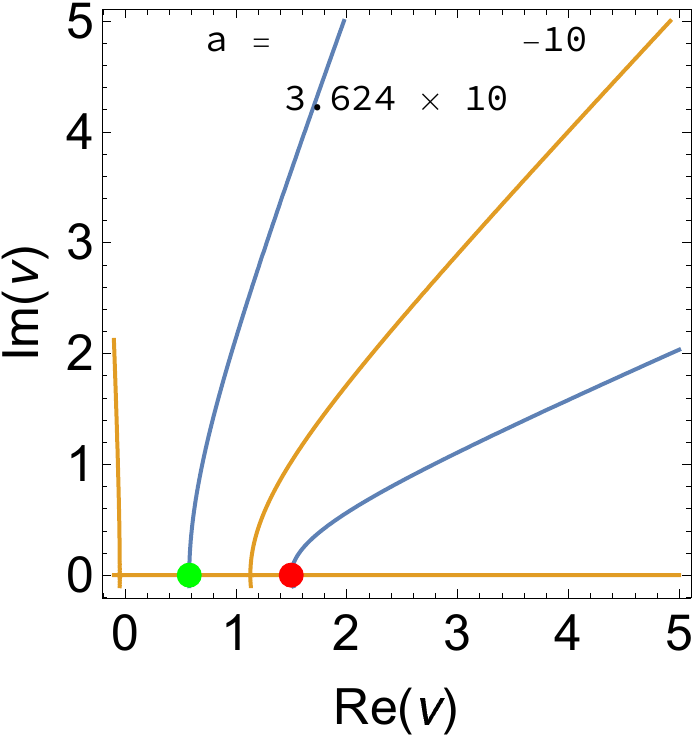}
\caption{\it ${\tilde{\b}_\text{eff}} = -30$}
\end{subfigure}
\begin{subfigure}{.3\textwidth}
\includegraphics[width=\textwidth]{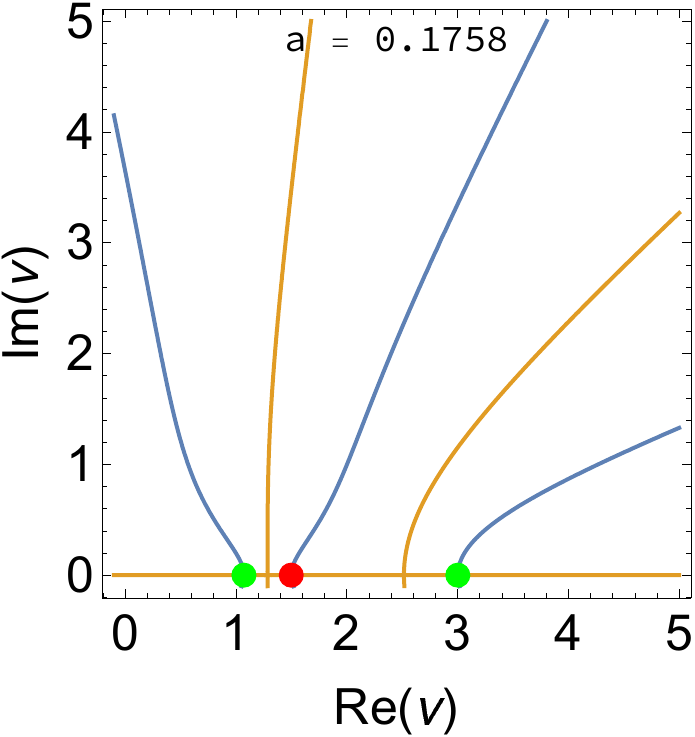}
\caption{\it ${\tilde{\b}_\text{eff}} = -10$}
\end{subfigure}
\begin{subfigure}{.3\textwidth}
\includegraphics[width=\textwidth]{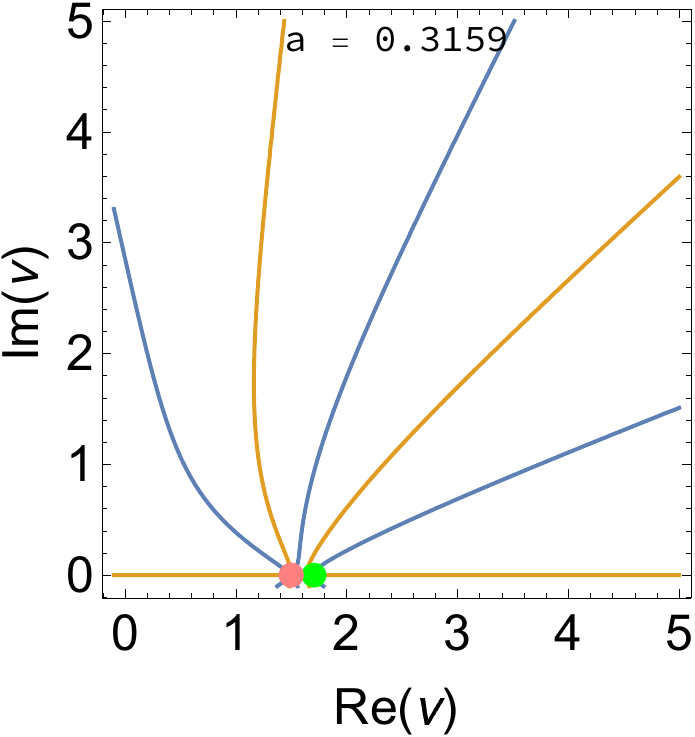}
\caption{\it ${\tilde{\b}_\text{eff}} = -9.41404$}
\end{subfigure}
\begin{subfigure}{.3\textwidth}
\includegraphics[width=\textwidth]{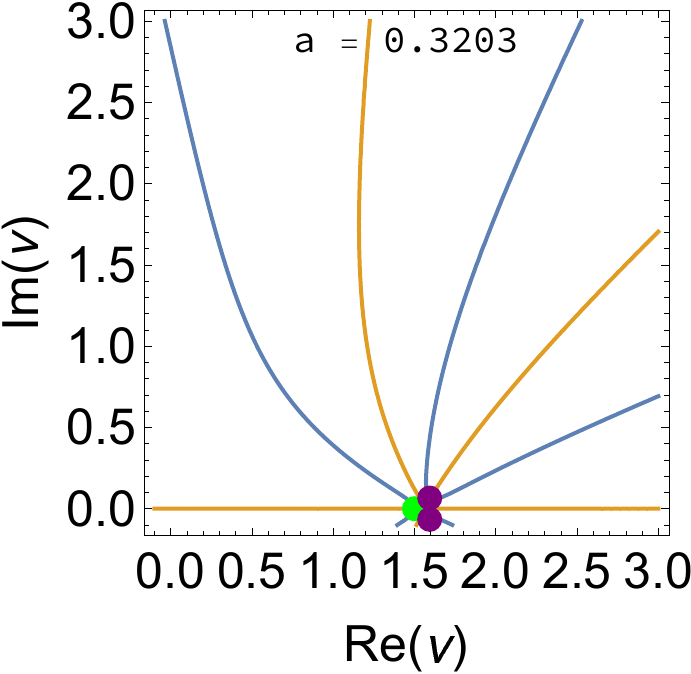}
\caption{\it ${\tilde{\b}_\text{eff}} = -9.4$}
\end{subfigure}
\begin{subfigure}{.3\textwidth}
\includegraphics[width=\textwidth]{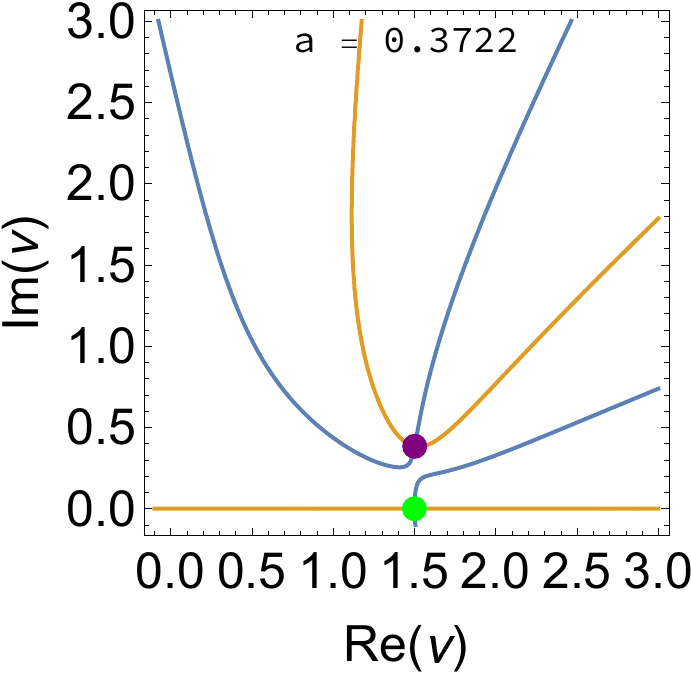}
\caption{\it ${\tilde{\b}_\text{eff}} = -9.25$}
\end{subfigure}
\begin{subfigure}{.3\textwidth}
\includegraphics[width=\textwidth]{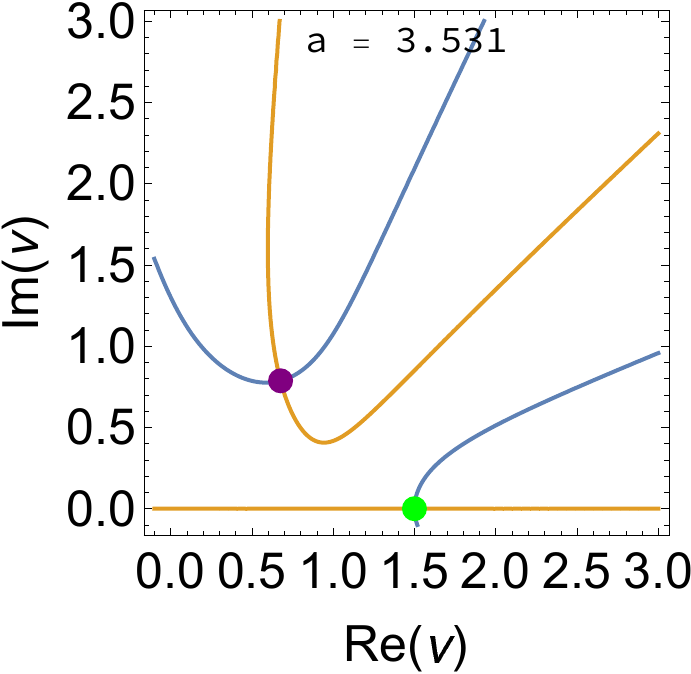}
\caption{\it ${\tilde{\b}_\text{eff}} = -7$}
\end{subfigure}
\begin{subfigure}{.3\textwidth}
\includegraphics[width=\textwidth]{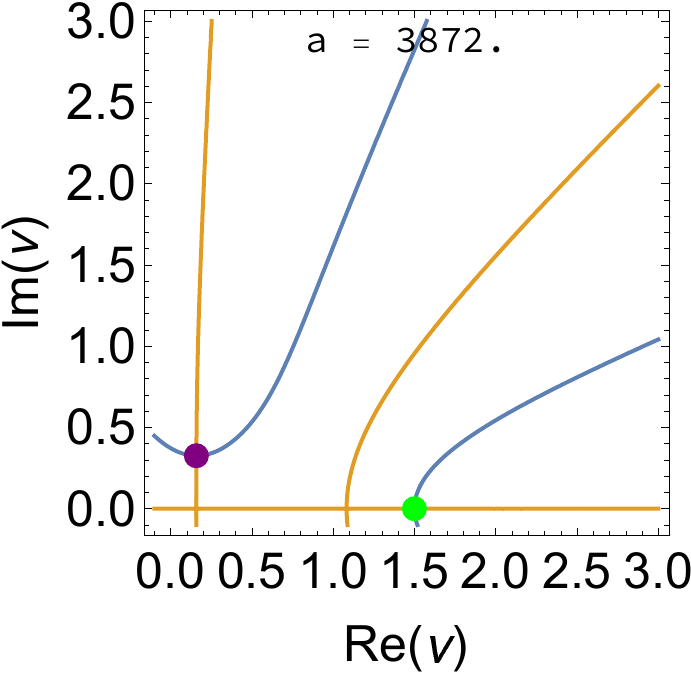}
\caption{\it ${\tilde{\b}_\text{eff}} = 0$}
\end{subfigure}
\begin{subfigure}{.3\textwidth}
\includegraphics[width=\textwidth]{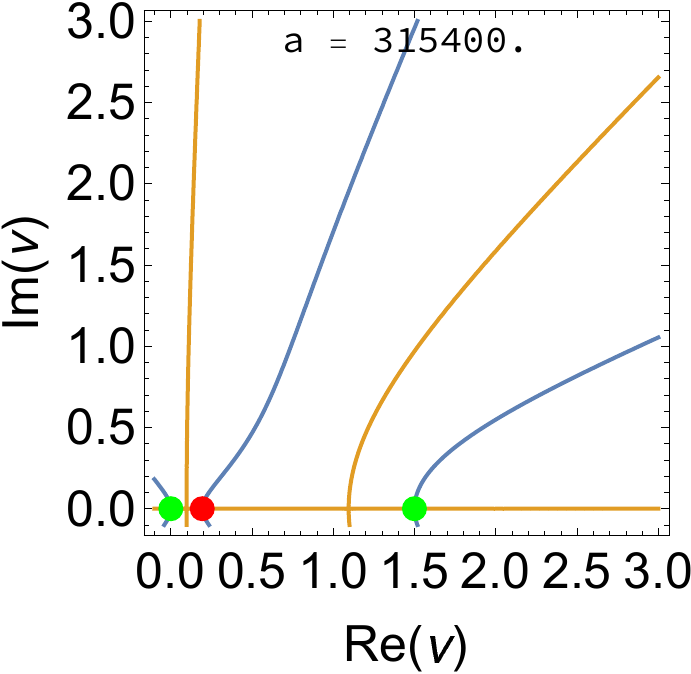}
\caption{\it ${\tilde{\b}_\text{eff}} = 4.4$}
\end{subfigure}
\begin{subfigure}{.3\textwidth}
\includegraphics[width=\textwidth]{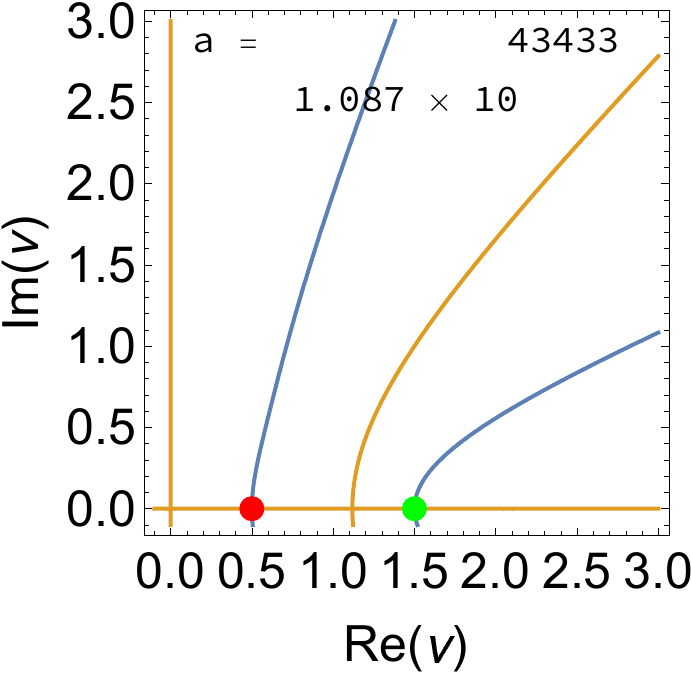}
\caption{\it ${\tilde{\b}_\text{eff}} = 100000$}
\end{subfigure}
\caption{\it dS, $\tilde{\a} = -1$, $GN^2H^2 = 1000 $.}
\label{dS_alpha-1_H1000}
\end{figure}


\begin{figure}[ht]
\centering
\begin{subfigure}{.3\textwidth}
\includegraphics[width=\textwidth]{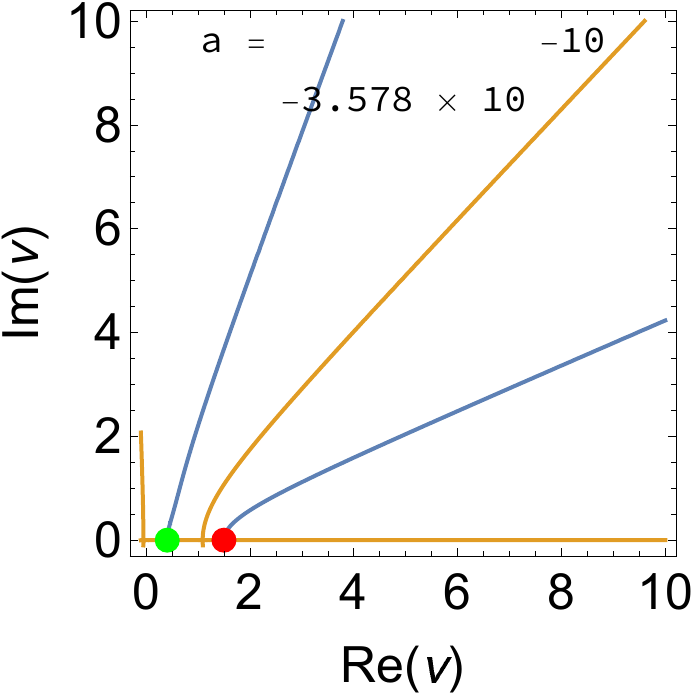}
\caption{\it ${\tilde{\b}_\text{eff}} = -30$}
\end{subfigure}
\begin{subfigure}{.3\textwidth}
\includegraphics[width=\textwidth]{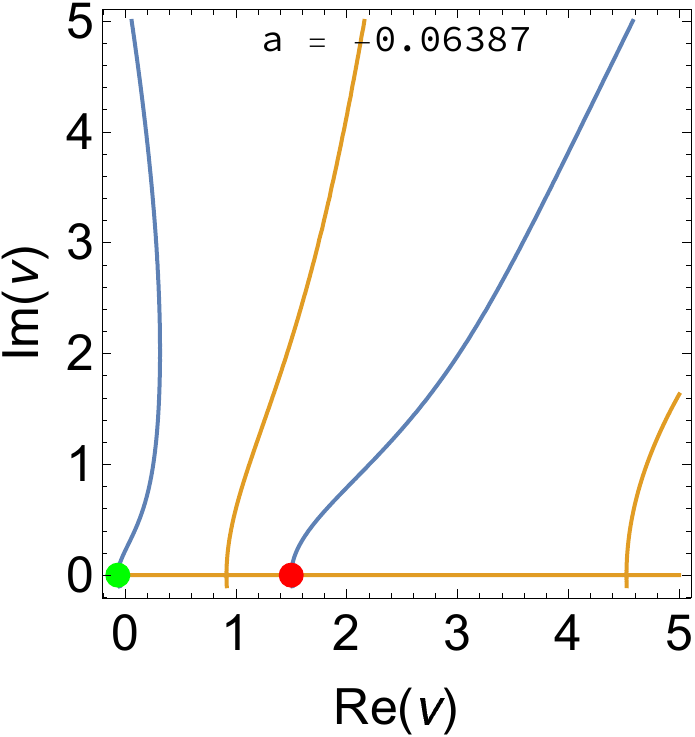}
\caption{\it ${\tilde{\b}_\text{eff}} = -11$}
\end{subfigure}
\begin{subfigure}{.3\textwidth}
\includegraphics[width=\textwidth]{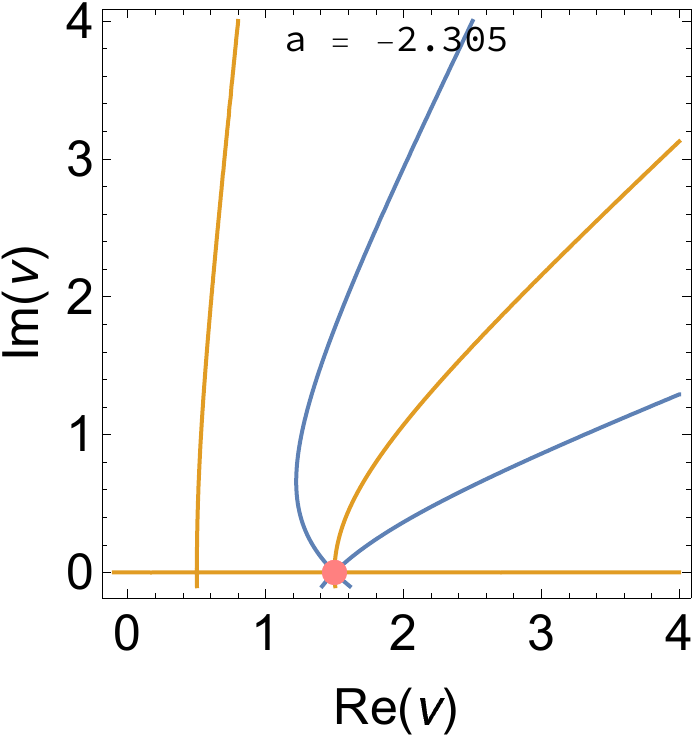}
\caption{\it ${\tilde{\b}_\text{eff}} = -7.41404$}
\end{subfigure}
\begin{subfigure}{.3\textwidth}
\includegraphics[width=\textwidth]{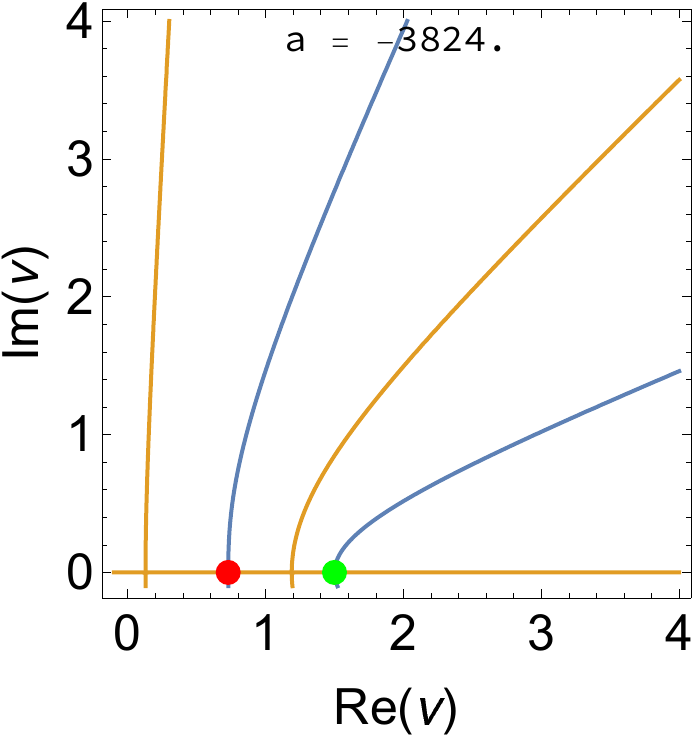}
\caption{\it ${\tilde{\b}_\text{eff}} = 0$}
\end{subfigure}
\begin{subfigure}{.3\textwidth}
\includegraphics[width=\textwidth]{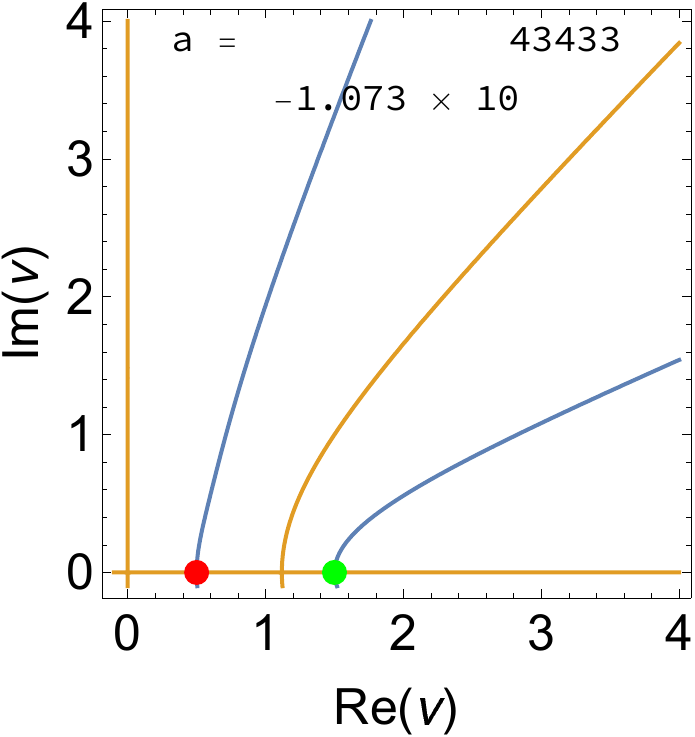}
\caption{\it ${\tilde{\b}_\text{eff}} = 100000$}
\end{subfigure}
\caption{\it dS, $\tilde{\a} = 0$, $GN^2H^2 = 1000 $.}
\label{dS_alpha0_H1000}
\end{figure}

\FloatBarrier
 \subsection{Anti-de Sitter}

 As we did for dS in the previous subsection, we provide snapshots for some examples of parameters $(\tilde{\a},GN^2H^2)$ while varying $\tbe$. They are summarized in the following table.
 \\

 \begin{tabular}{|c|c|c|c|}
 \hline
 \diagbox{$\tilde{\a}$}{$GN^2\chi^2$} & $<<1$ & $\sim \pi$ & $>>1$ \\ \hline
 $-\tilde{\a} >>1$ & \colorbox{blue!30}{\textcolor{green}{Fig.} \ref{AdS_alpha-100_chi0.001}, \textcolor{red}{Fig.} \ref{AdS_alpha-1000_chi0.01}} & \textcolor{red}{Fig. \ref{AdS_alpha-1000_chi2pi} }& \textcolor{red}{ Fig. \ref{AdS_alpha-1000_chi1000}}\\ \hline
 $|\tilde{\a}| \leq 2 $ & \textcolor{green}{Fig. \ref{AdS_H0.01_alpha0}} & \colorbox{blue!30}{ \textcolor{green}{Fig.} \ref{AdS_alpha0_chi2pi} , \textcolor{red}{Fig.} \ref{AdS_chi2pi_am2}} &\colorbox{blue!30}{\textcolor{green}{Fig.} \ref{AdS_alpha0_chi1000}, \textcolor{red}{Fig.} \ref{AdS_alpha-1_chi1000}}\\ \hline
 $\tilde{\a} >>1$ & \textcolor{green}{ Fig. \ref{AdS_alpha1000_chi0.01}} & \textcolor{green}{Fig. \ref{AdS_alpha1000_chi2pi} }& \textcolor{green}{Fig. \ref{AdS_alpha-1000_chi1000} } \\ \hline
 \end{tabular}
\\

$a$ (\ref{sc3}) is either \textcolor{green}{positive} if ``Fig" is written in green, or \textcolor{red}{negative} if it is written in red. \colorbox{blue!30}{Blue boxes} correspond to regimes where the sign of $a$ must be determined by the inequality (\ref{sc5}) :
 \be
 a<0\iff {\pi\over GN^2\chi^2}<- \left(\tilde{\a} + {1\over 2} \right)
 \ee

The conclusion is that a given point in the $(\tilde{\a},GN^2\chi^2)$ plane corresponds either to the behaviour of
\begin{itemize}
\item \textbf{Type A}: Figure \ref{AdS_H0.01_alpha0}, where two tachyons merge on the imaginary axis and form a pair of complex conjugate poles when $\tbe$ is increased.
\item \textbf{Type B}: Figure \ref{AdS_chipiover4_am10}, where all the poles are real valued $\n^2$ for any value of $\tbe$ (no complex pole).
\end{itemize}

\begin{figure}[h!]
\centering
\begin{subfigure}{.3\textwidth}
\includegraphics[width=\textwidth]{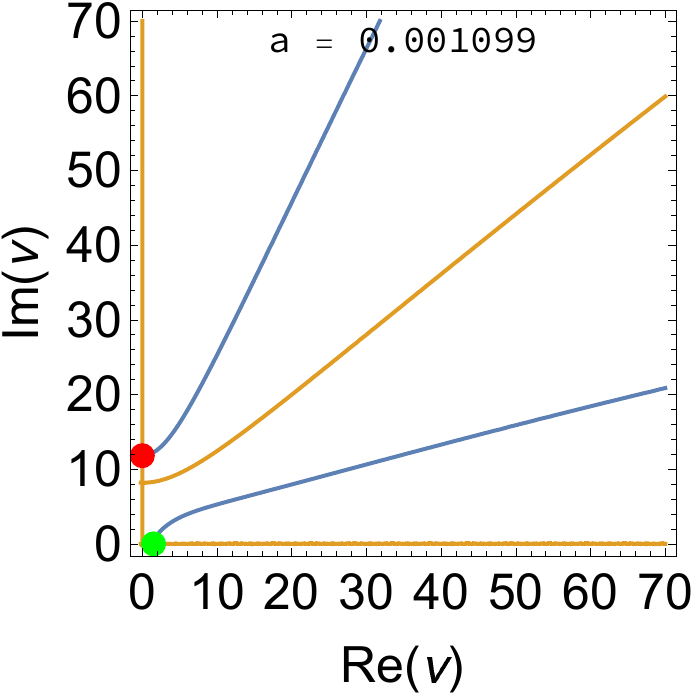}
\caption{\it ${\tilde{\b}_\text{eff}} = -10$}
\end{subfigure}
\begin{subfigure}{.3\textwidth}
\includegraphics[width=\textwidth]{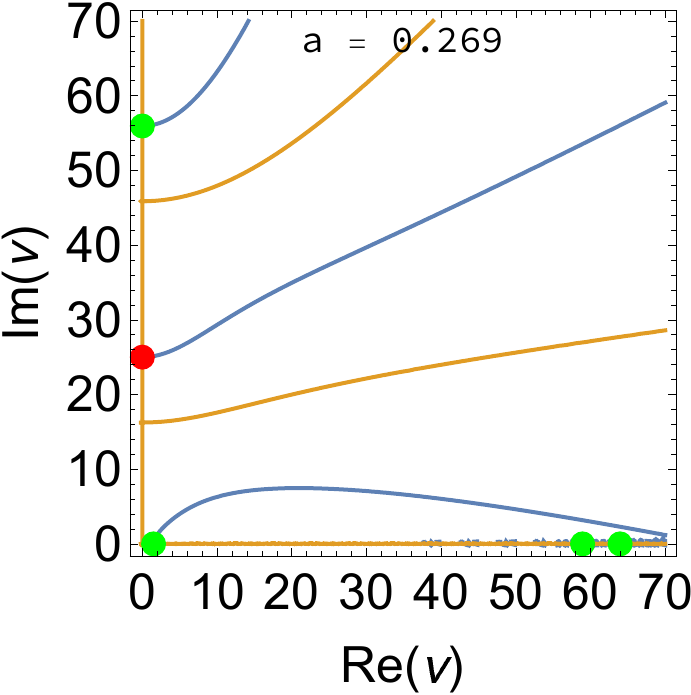}
\caption{\it ${\tilde{\b}_\text{eff}} = -4.5$}
\end{subfigure}
\begin{subfigure}{.3\textwidth}
\includegraphics[width=\textwidth]{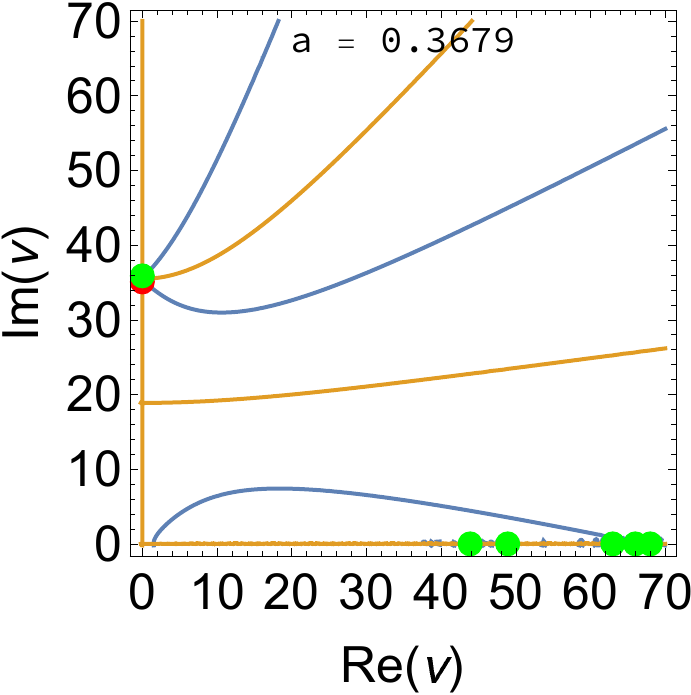}
\caption{\it ${\tilde{\b}_\text{eff}} = -4.18705$}
\end{subfigure}
\begin{subfigure}{.3\textwidth}
\includegraphics[width=\textwidth]{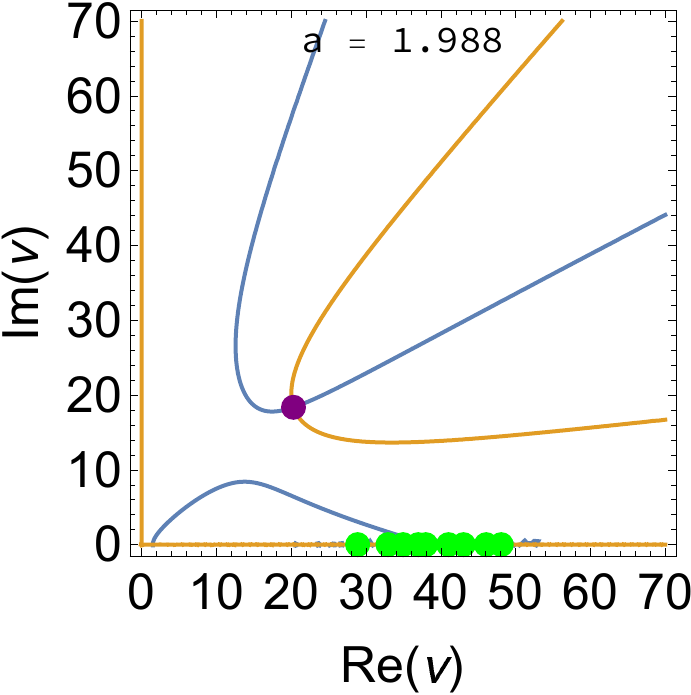}
\caption{\it ${\tilde{\b}_\text{eff}} = -2.5$}
\end{subfigure}
\begin{subfigure}{.3\textwidth}
\includegraphics[width=\textwidth]{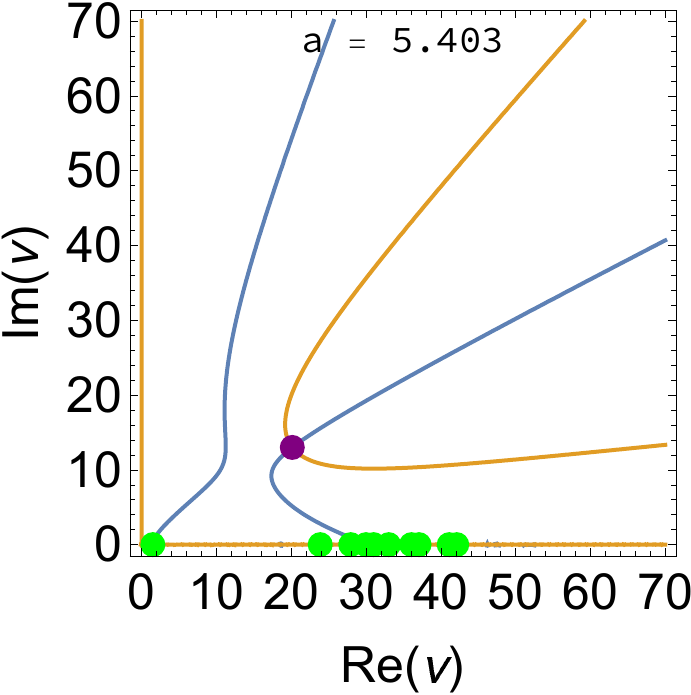}
\caption{\it ${\tilde{\b}_\text{eff}} = -1.5$}
\end{subfigure}
\begin{subfigure}{.3\textwidth}
\includegraphics[width=\textwidth]{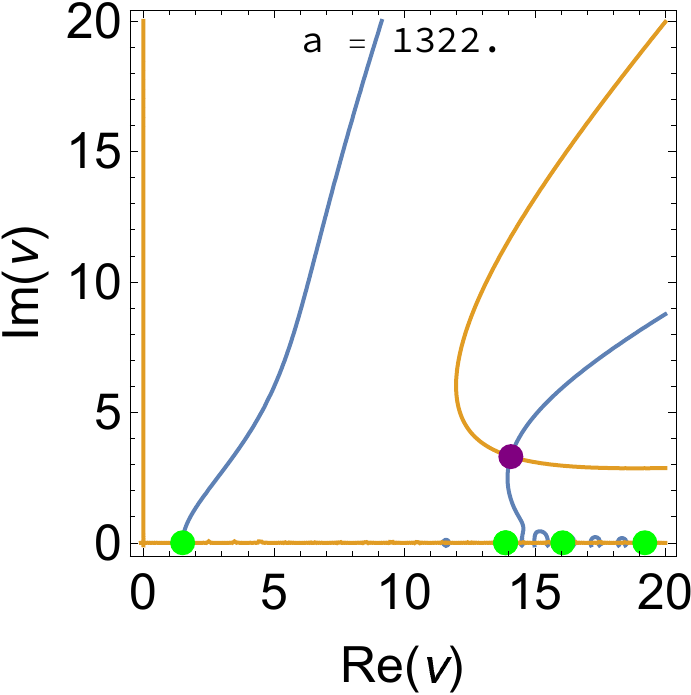}
\caption{\it ${\tilde{\b}_\text{eff}} = 4$}
\end{subfigure}
\begin{subfigure}{.3\textwidth}
\includegraphics[width=\textwidth]{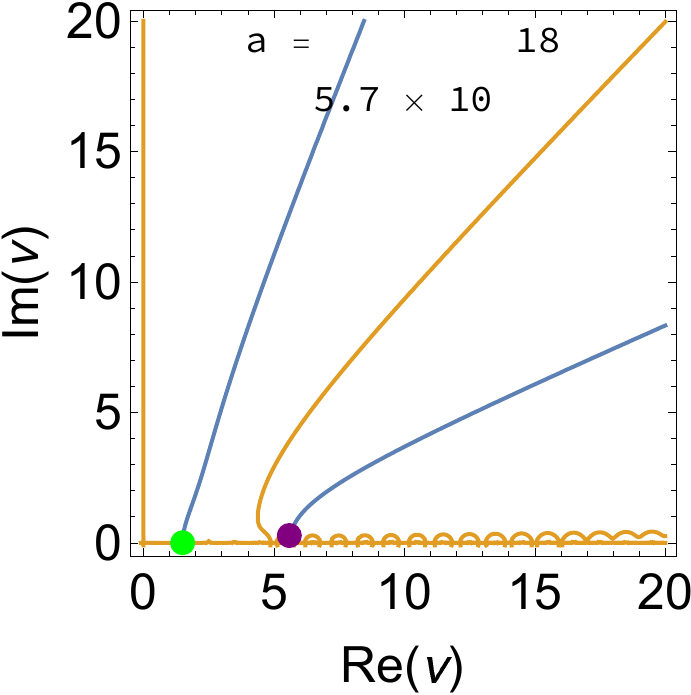}
\caption{\it ${\tilde{\b}_\text{eff}} = 40$}
\end{subfigure}
\begin{subfigure}{.3\textwidth}
\includegraphics[width=\textwidth]{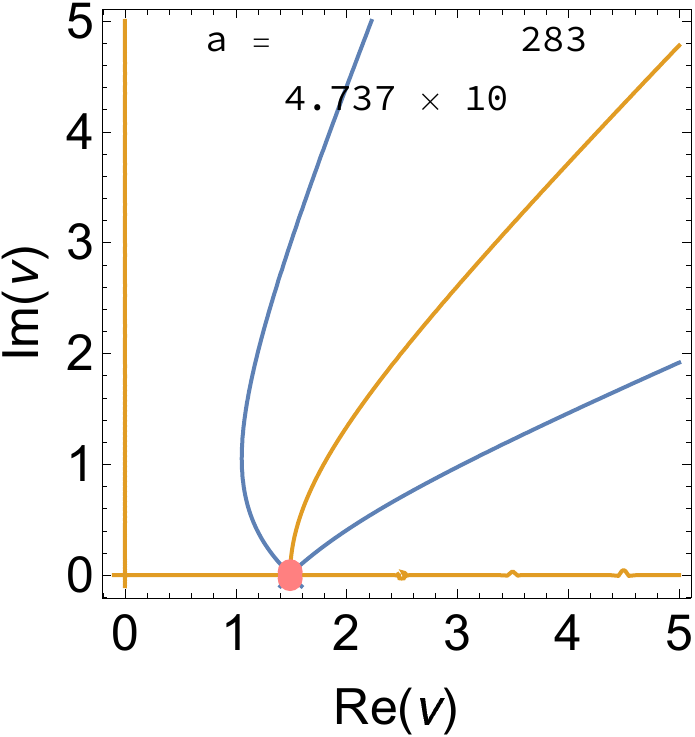}
\caption{\it ${\tilde{\b}_\text{eff}} = 650$}
\end{subfigure}
\begin{subfigure}{.3\textwidth}
\includegraphics[width=\textwidth]{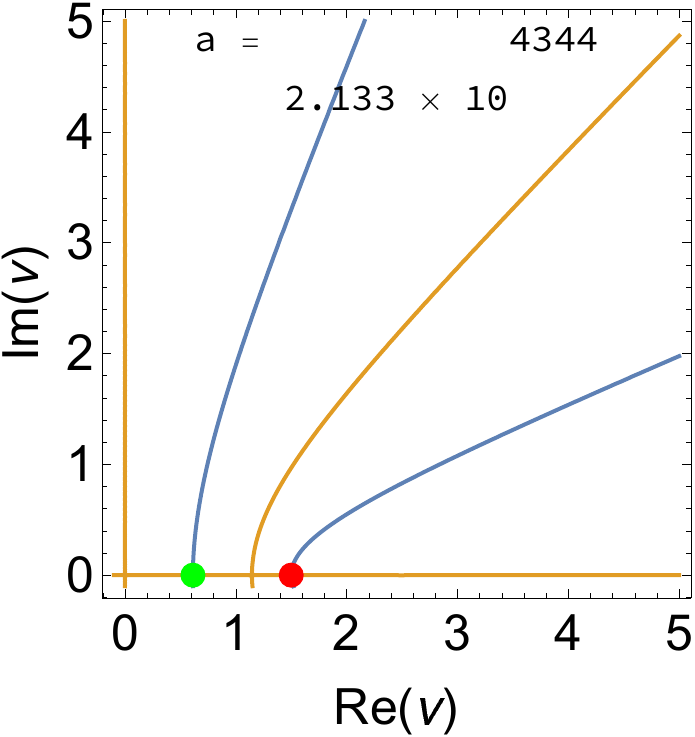}
\caption{\it ${\tilde{\b}_\text{eff}} = 10000$}
\end{subfigure}
\caption{\it Zeros of the real (blue curve) and imaginary (orange curve) parts of the AdS spin-2 inverse correlator $\mathcal{F}_\text{(-)}^{-1}(\n)$ (\protect\ref{as10}) for $\a = 0$, $GN^2\chi^2 = 0.01$. Each panel is obtained for a different value of $\tilde{\b}_\text{eff}$. The colour of a pole represents the sign of its residue. Green is for positive, red for negative and purple for complex residue. Two tachyons on the imaginary axis merge around the value of $\tilde{\b}_\text{eff}$ given in equation (\protect\ref{sc4}), shown in snapshot (c). This merging is a second-order pole because $\mathcal{F}_\text{dS}'$ vanishes.
The two complex conjugate poles move to the complex plane as $\tilde{\b}_\text{eff}$ is increased.
Two poles in (g) merge in snapshot (h), where they form a massless second-order pole.
Higher values of $\tilde{\b}_\text{eff}$ (i) have a massless ghost and a $\n=1/2$ stable pole.}
\label{AdS_H0.01_alpha0}
\end{figure}

 \begin{figure}[h!]
\centering
\begin{subfigure}{.3\textwidth}
\includegraphics[width=\textwidth]{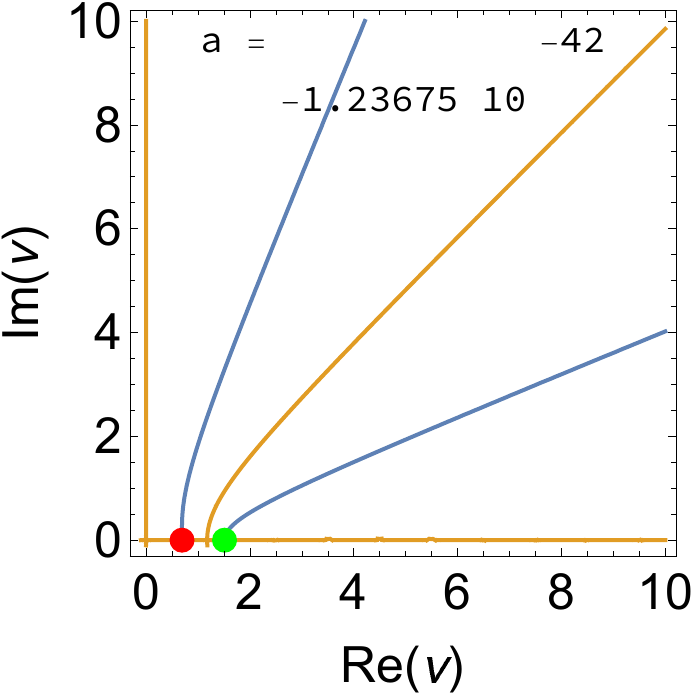}
\caption{\it ${\tilde{\b}_\text{eff}} = -100$}
\end{subfigure}
\begin{subfigure}{.3\textwidth}
\includegraphics[width=\textwidth]{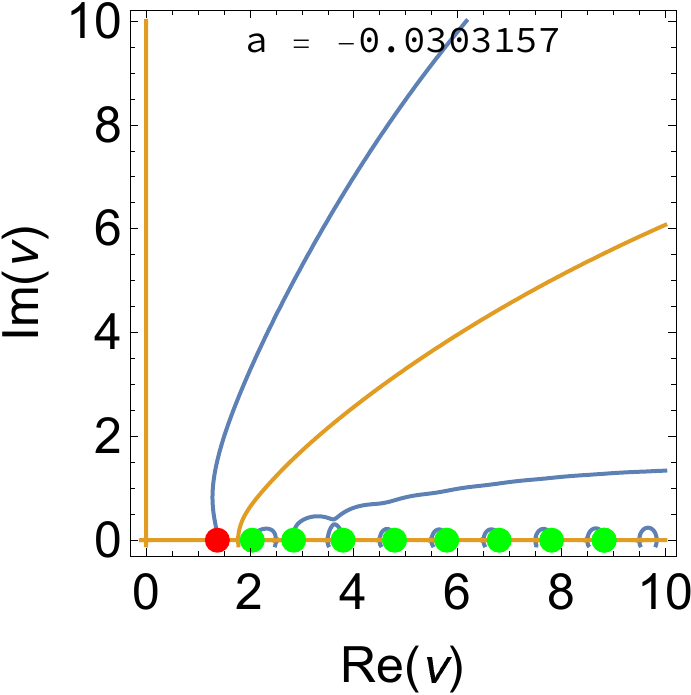}
\caption{\it ${\tilde{\b}_\text{eff}} = -7$}
\end{subfigure}
\begin{subfigure}{.3\textwidth}
\includegraphics[width=\textwidth]{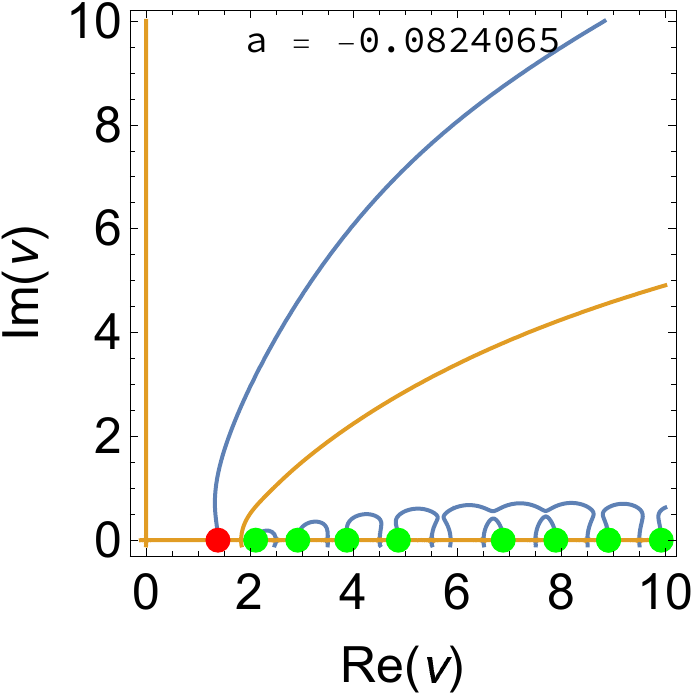}
\caption{\it ${\tilde{\b}_\text{eff}} = -6$}
\end{subfigure}
\begin{subfigure}{.3\textwidth}
\includegraphics[width=\textwidth]{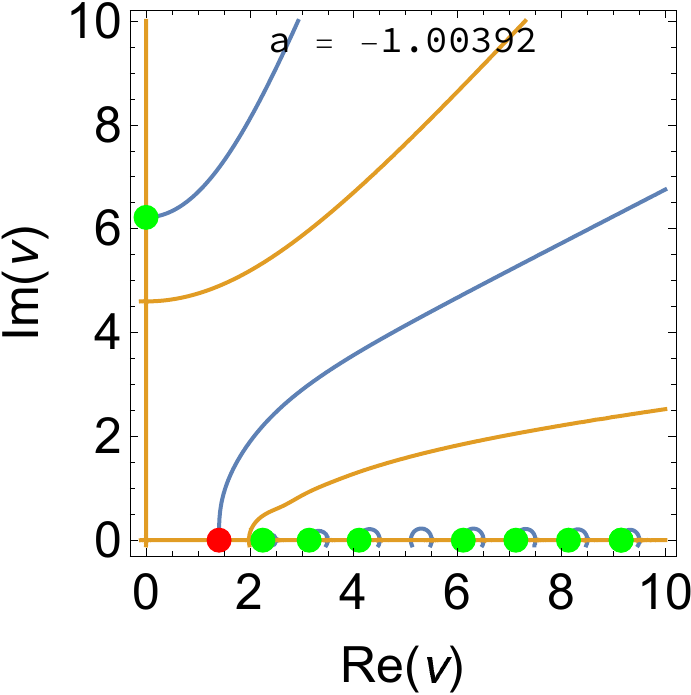}
\caption{\it ${\tilde{\b}_\text{eff}} = -3.5$}
\end{subfigure}
\begin{subfigure}{.3\textwidth}
\includegraphics[width=\textwidth]{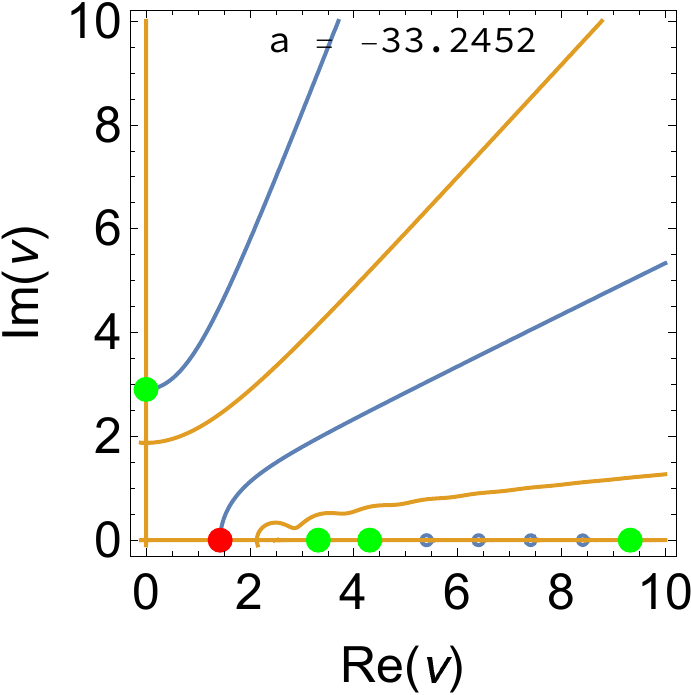}
\caption{\it ${\tilde{\b}_\text{eff}} = 0$}
\end{subfigure}
\begin{subfigure}{.3\textwidth}
\includegraphics[width=\textwidth]{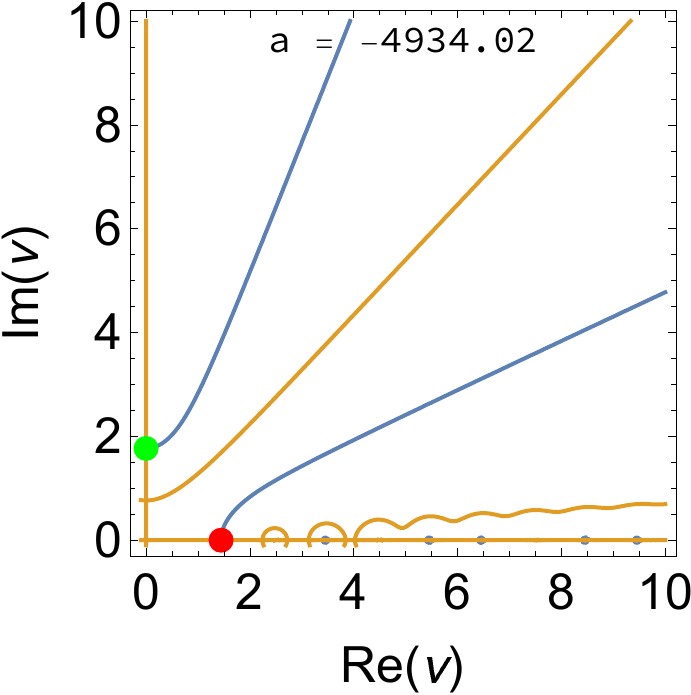}
\caption{\it ${\tilde{\b}_\text{eff}} = 5$}
\end{subfigure}
\begin{subfigure}{.3\textwidth}
\includegraphics[width=\textwidth]{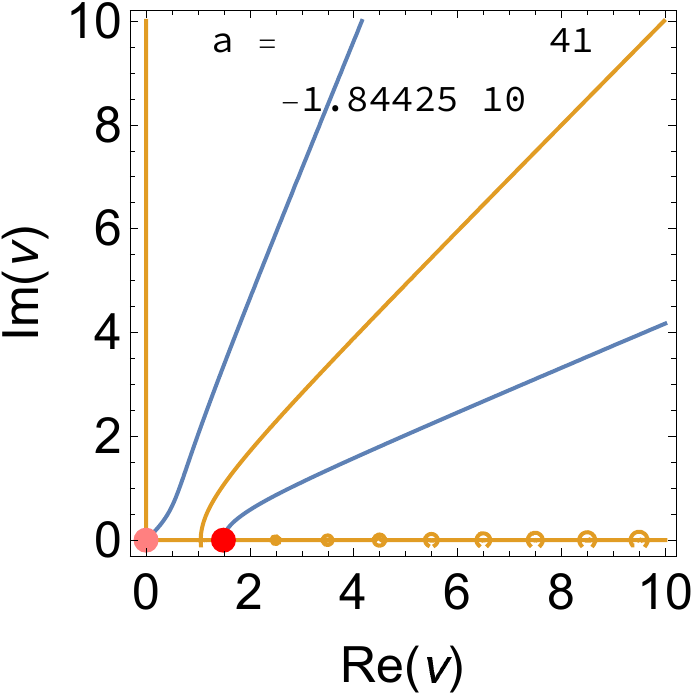}
\caption{\it ${\tilde{\b}_\text{eff}}={\pi\b_\text{BF}\over N^2} = 91.5142$}
\end{subfigure}
\begin{subfigure}{.3\textwidth}
\includegraphics[width=\textwidth]{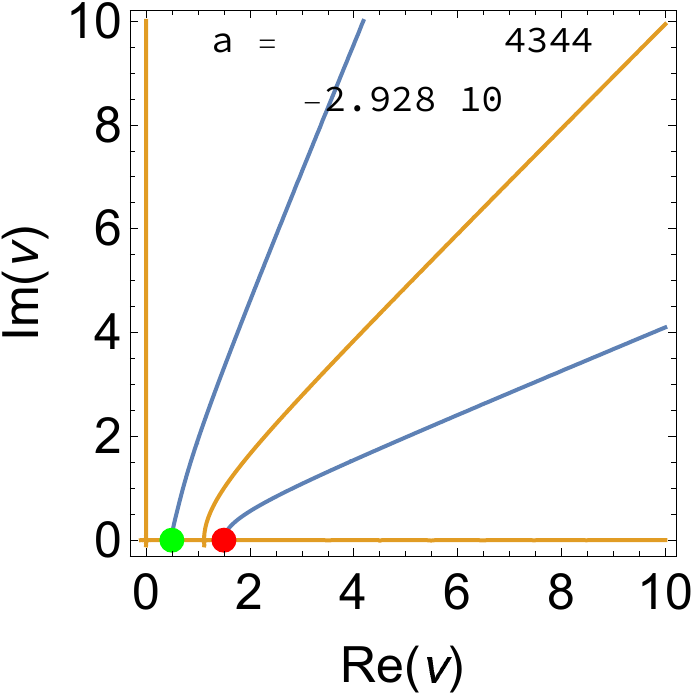}
\caption{\it ${\tilde{\b}_\text{eff}} = 10000$}
\end{subfigure}
\caption{\it Zeros of the real (blue curve) and imaginary (orange curve) parts of the AdS spin-2 inverse correlator $\mathcal{F}_\text{(-)}^{-1}(\n)$ (\protect\ref{dS14}) for different values of $\tilde{\b}_\text{eff}$, with fixed $\a = -10$ and $GN^2\chi^2 = \pi/4$.
The colour of a pole represents the sign of its residue. Green is for positive, red for negative and purple for complex residue
From negative values of $\tilde{\b}_\text{eff}$, up to some critical value in snapshot (g), there is one tachyon which cannot be seen in snapshot (a) because it is outside the window and moves towards the origin. Snapshot (g) shows the transition from tachyonic instability to tachyonic stability, where the origin is a double pole.}
\label{AdS_chipiover4_am10}
\end{figure}

 \begin{figure}[ht]
 \centering
 \begin{subfigure}{.3\textwidth}
 \includegraphics[width=\textwidth]{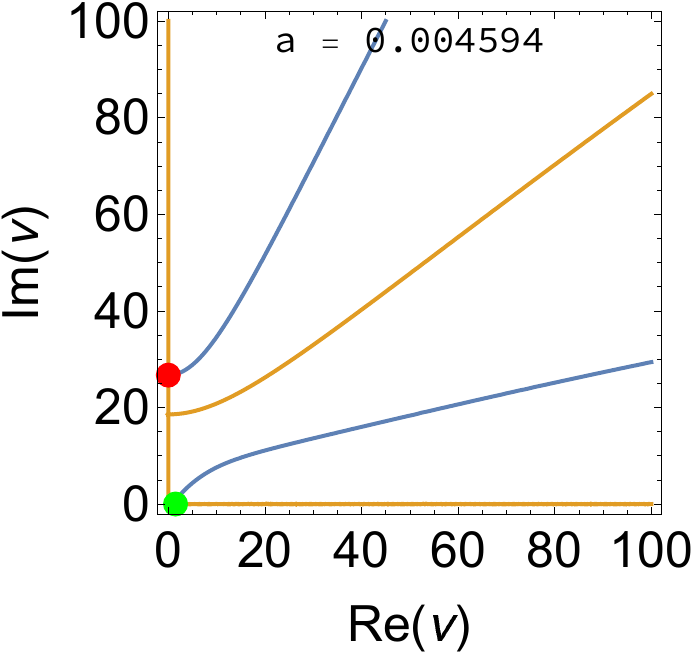}
 \caption{\it ${\tilde{\b}_\text{eff}} = -10$}
 \end{subfigure}
 \begin{subfigure}{.3\textwidth}
 \includegraphics[width=\textwidth]{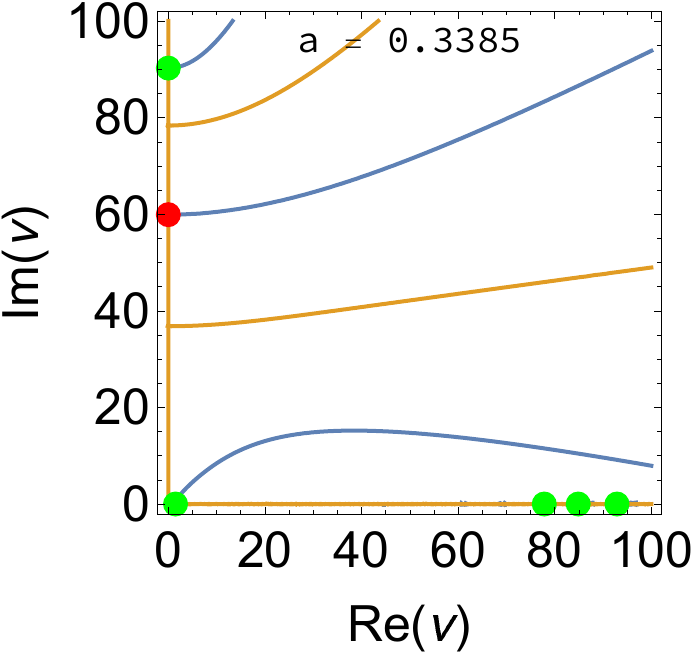}
 \caption{\it ${\tilde{\b}_\text{eff}} = -5.7$}
 \end{subfigure}
 \begin{subfigure}{.3\textwidth}
 \includegraphics[width=\textwidth]{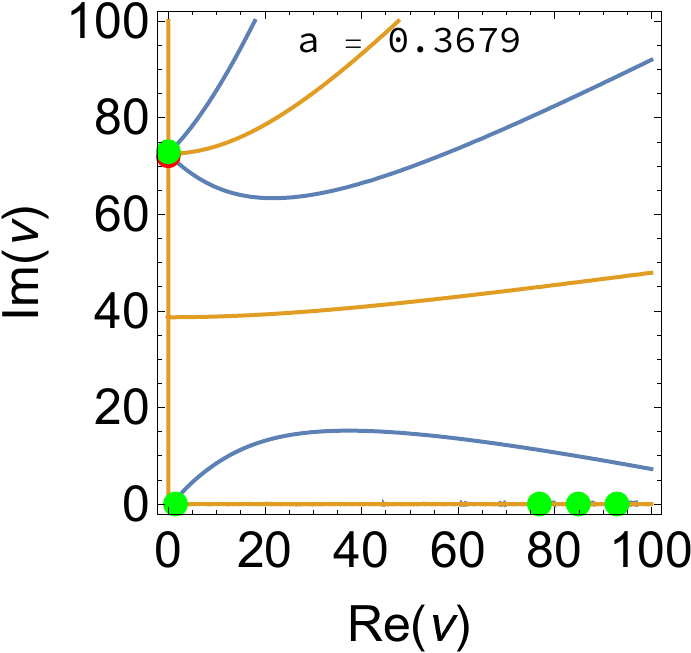}
 \caption{\it ${\tilde{\b}_\text{eff}} = -5.61689$}
 \end{subfigure}
 \begin{subfigure}{.3\textwidth}
 \includegraphics[width=\textwidth]{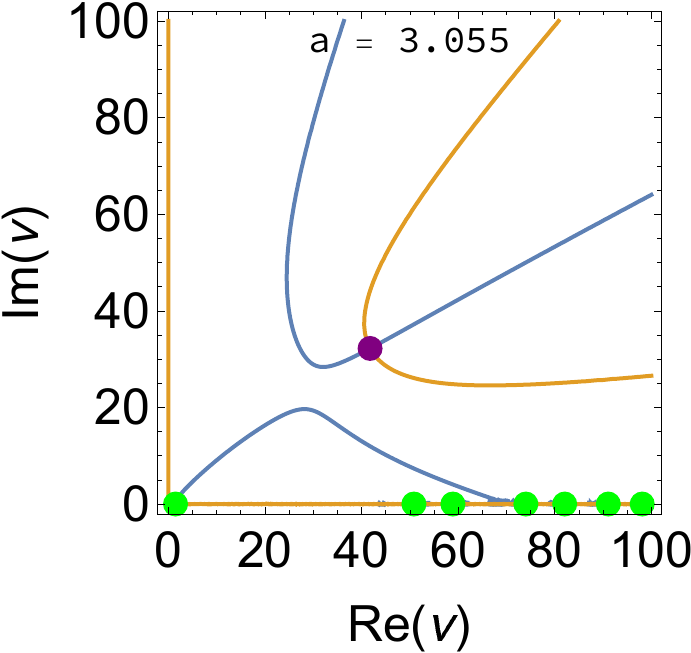}
 \caption{\it ${\tilde{\b}_\text{eff}} = -3.5$}
 \end{subfigure}
 \begin{subfigure}{.3\textwidth}
 \includegraphics[width=\textwidth]{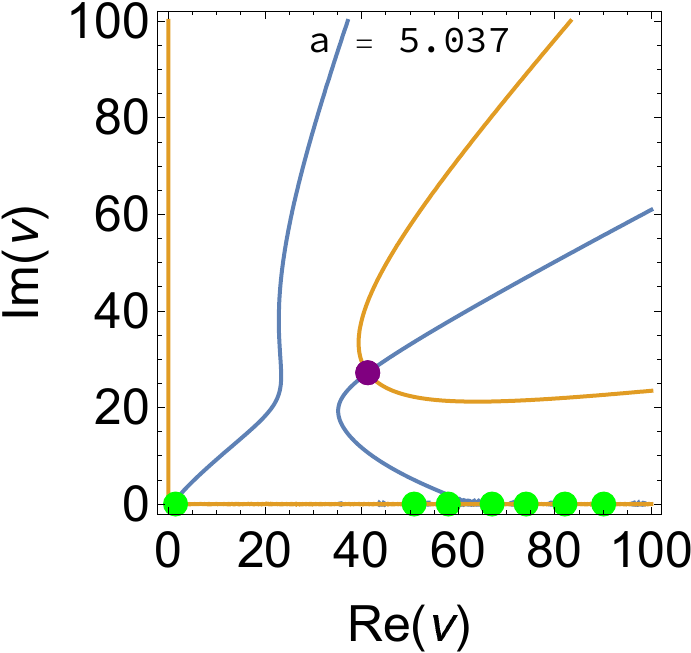}
 \caption{\it ${\tilde{\b}_\text{eff}} = -3$}
 \end{subfigure}
 \begin{subfigure}{.3\textwidth}
 \includegraphics[width=\textwidth]{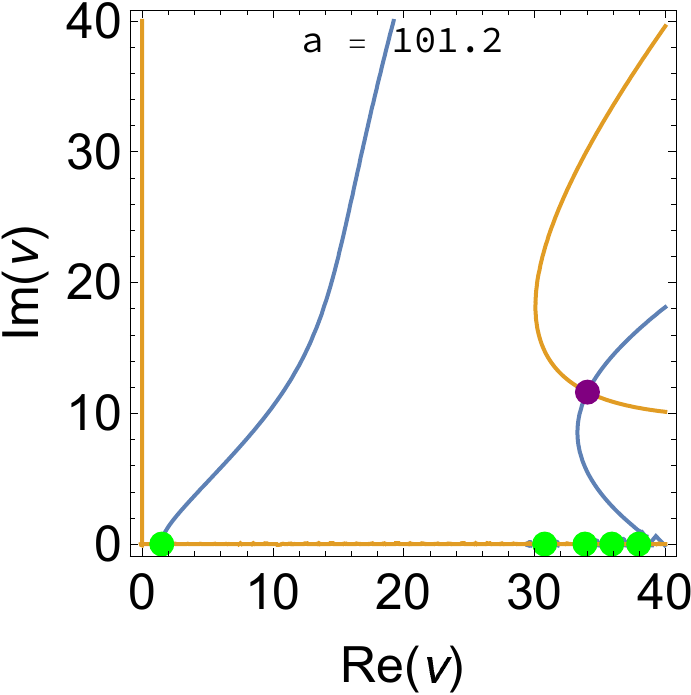}
 \caption{\it ${\tilde{\b}_\text{eff}} = 0$}
 \end{subfigure}
 \begin{subfigure}{.3\textwidth}
 \includegraphics[width=\textwidth]{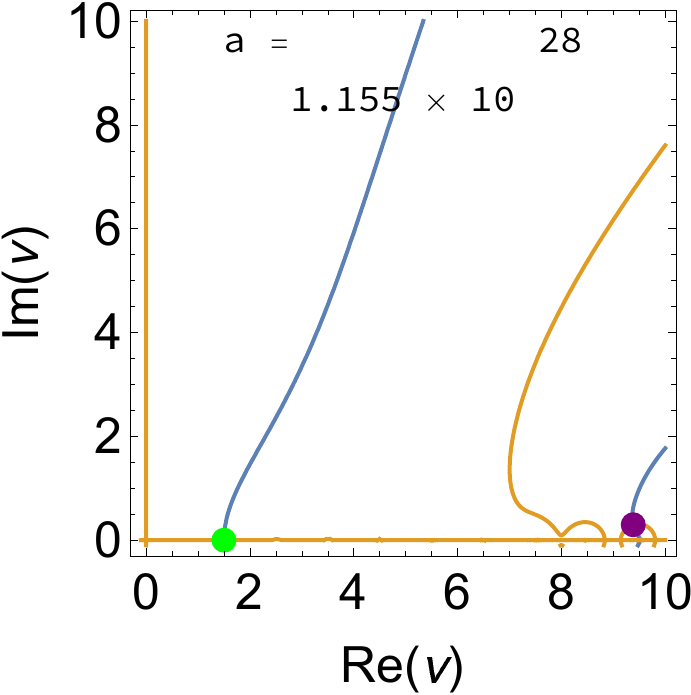}
 \caption{\it ${\tilde{\b}_\text{eff}} = 60$}
 \end{subfigure}
 \begin{subfigure}{.3\textwidth}
 \includegraphics[width=\textwidth]{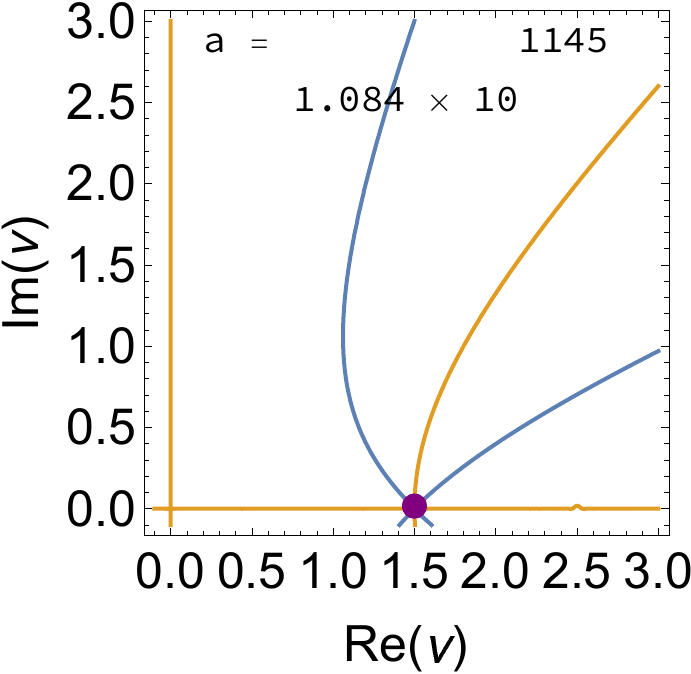}
 \caption{\it ${\tilde{\b}_\text{eff}} = 2631.92$}
 \end{subfigure}
 \begin{subfigure}{.3\textwidth}
 \includegraphics[width=\textwidth]{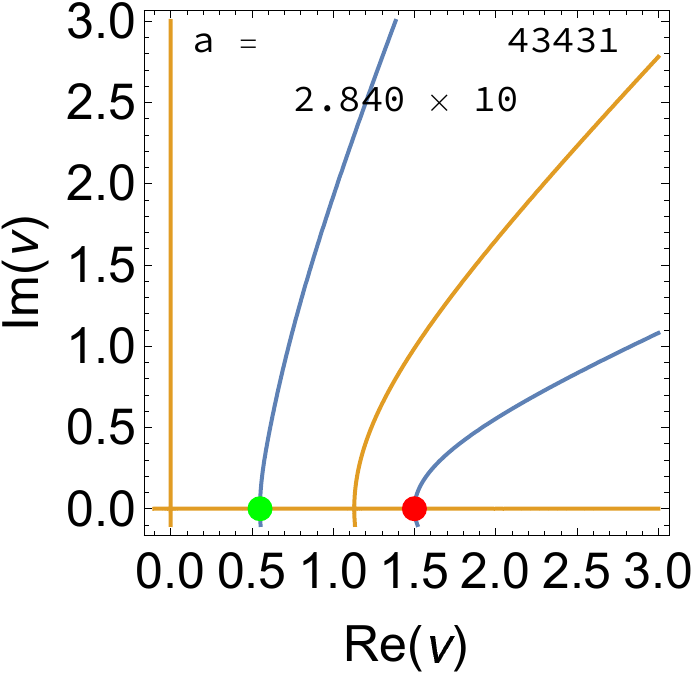}
 \caption{\it ${\tilde{\b}_\text{eff}} = 100000$}
 \end{subfigure}

 \caption{\it AdS, $\tilde{\a} = 1000$, $GN^2\chi^2 = 0.01$.}
 \label{AdS_alpha1000_chi0.01}
 \end{figure}

 \begin{figure}[ht]
 \centering
 \begin{subfigure}{.3\textwidth}
 \includegraphics[width=\textwidth]{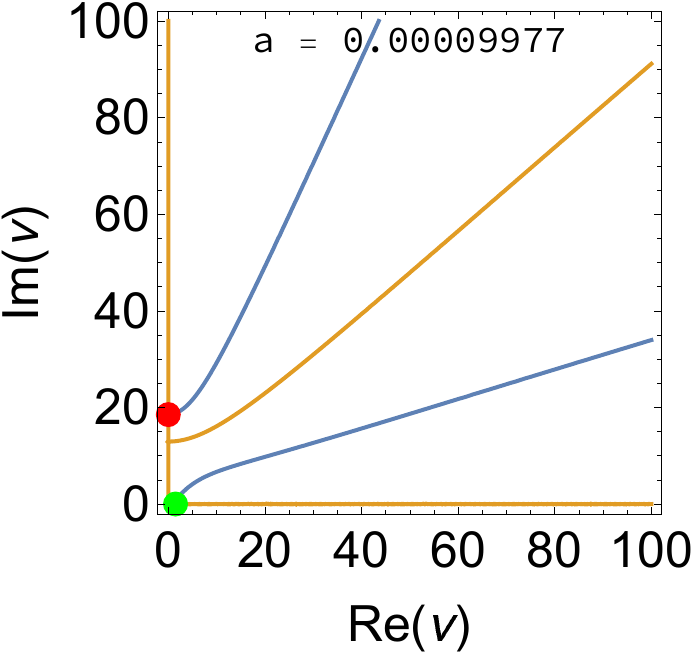}
 \caption{\it ${\tilde{\b}_\text{eff}} = -20$}
 \end{subfigure}
 \begin{subfigure}{.3\textwidth}
 \includegraphics[width=\textwidth]{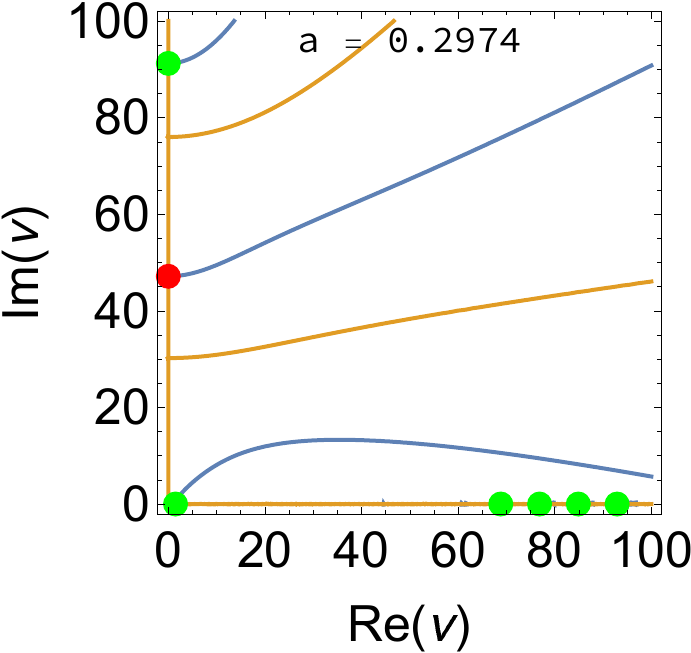}
 \caption{\it ${\tilde{\b}_\text{eff}} = -12$}
 \end{subfigure}
 \begin{subfigure}{.3\textwidth}
 \includegraphics[width=\textwidth]{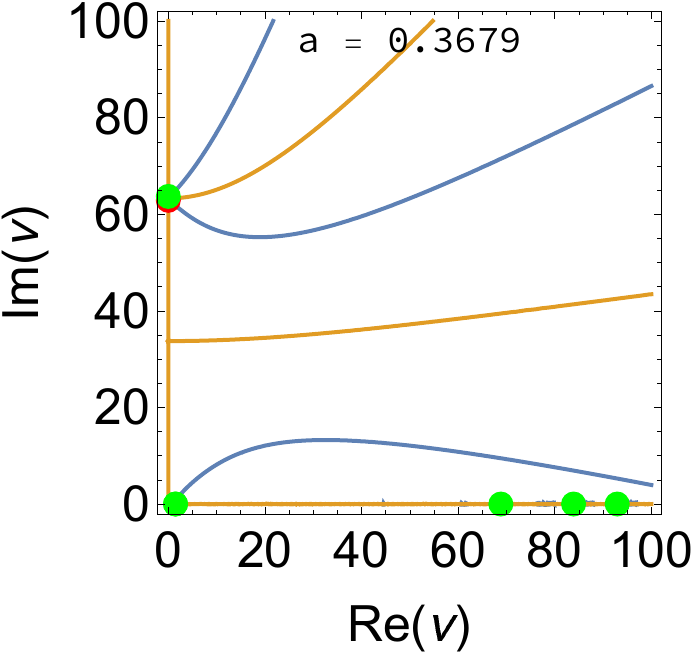}
 \caption{\it ${\tilde{\b}_\text{eff}} = -11.7874$}
 \end{subfigure}
 \begin{subfigure}{.3\textwidth}
 \includegraphics[width=\textwidth]{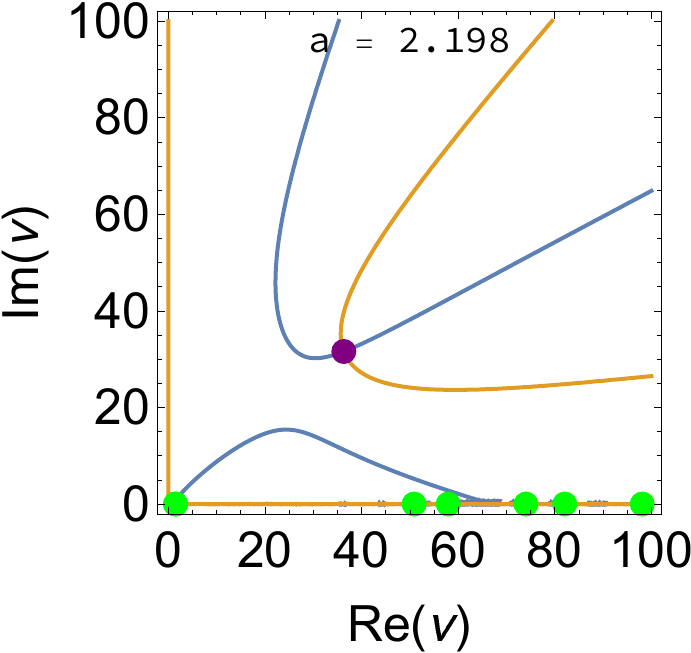}
 \caption{\it ${\tilde{\b}_\text{eff}} = -10$}
 \end{subfigure}
 \begin{subfigure}{.3\textwidth}
 \includegraphics[width=\textwidth]{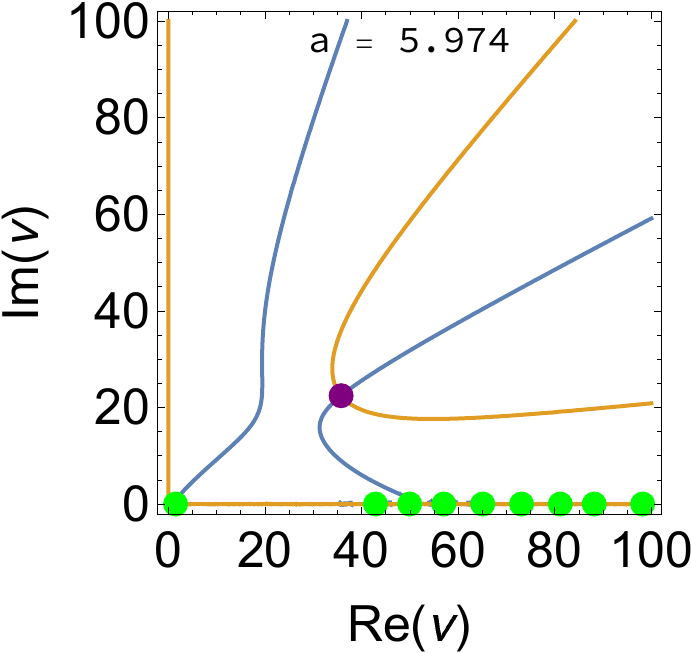}
 \caption{\it ${\tilde{\b}_\text{eff}} = -9$}
 \end{subfigure}
 \begin{subfigure}{.3\textwidth}
 \includegraphics[width=\textwidth]{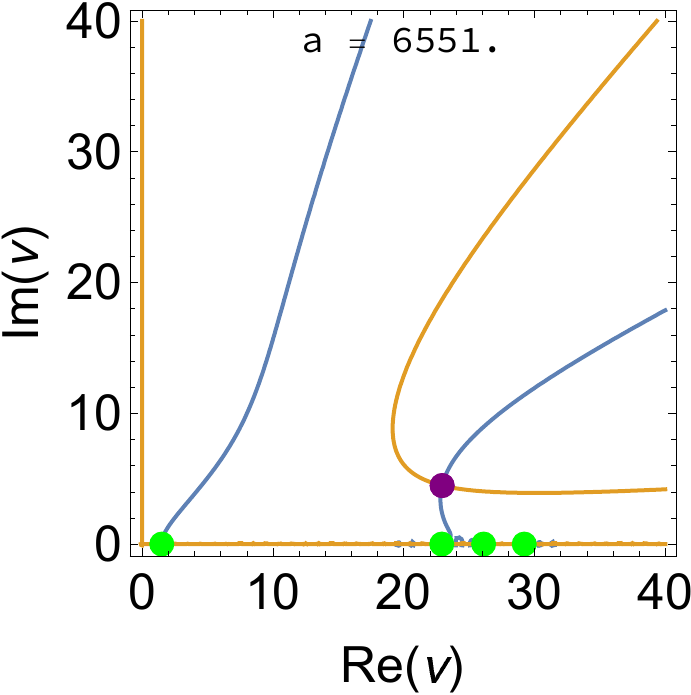}
 \caption{\it ${\tilde{\b}_\text{eff}} = -2$}
 \end{subfigure}
 \begin{subfigure}{.3\textwidth}
 \includegraphics[width=\textwidth]{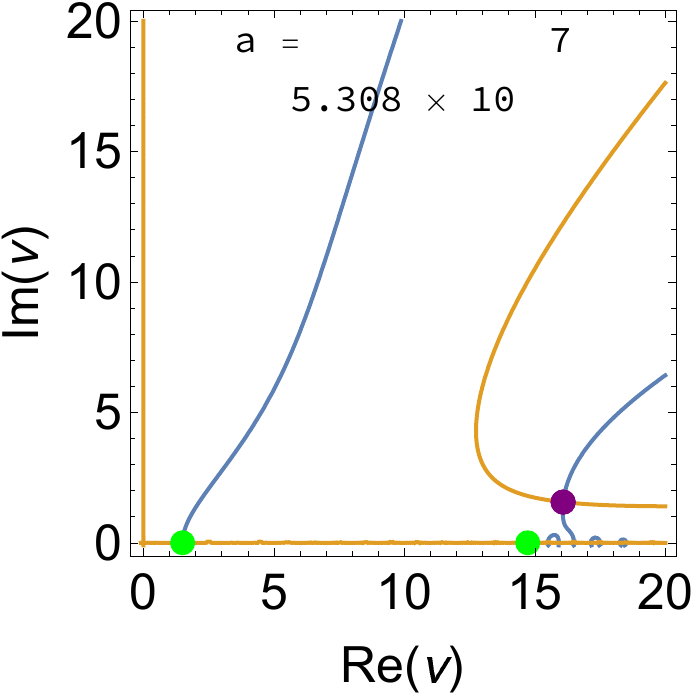}
 \caption{\it ${\tilde{\b}_\text{eff}} = 7$}
 \end{subfigure}
 \begin{subfigure}{.3\textwidth}
 \includegraphics[width=\textwidth]{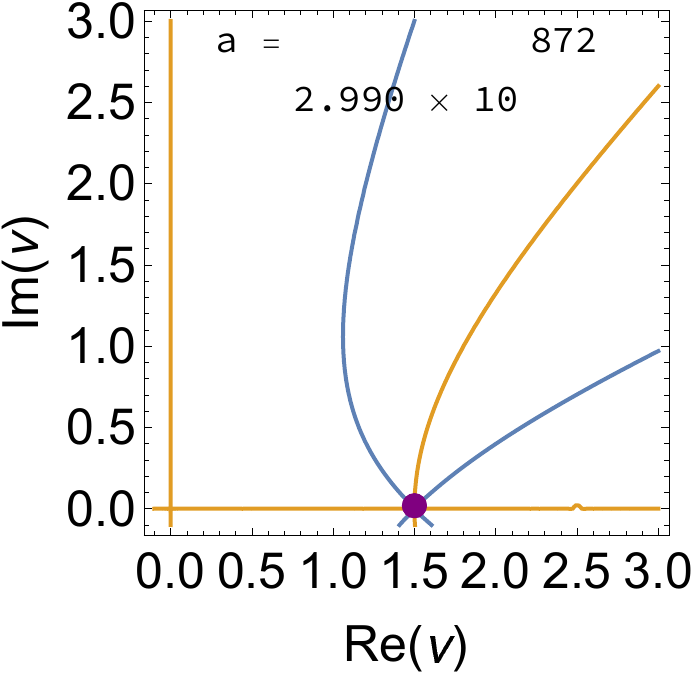}
 \caption{\it ${\tilde{\b}_\text{eff}} = 1998.16$}
 \end{subfigure}
 \begin{subfigure}{.3\textwidth}
 \includegraphics[width=\textwidth]{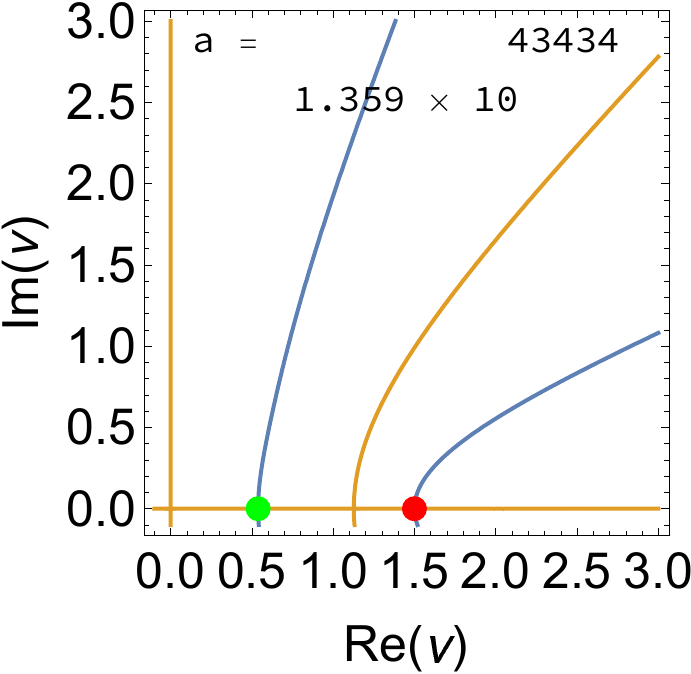}
 \caption{\it ${\tilde{\b}_\text{eff}} = 100000$}
 \end{subfigure}
 \caption{\it AdS, $\tilde{\a} = 1000$, $GN^2\chi^2 = 2 \pi$.}
 \label{AdS_alpha1000_chi2pi}
 \end{figure}

 \begin{figure}[ht]
 \centering
 \begin{subfigure}{.3\textwidth}
 \includegraphics[width=\textwidth]{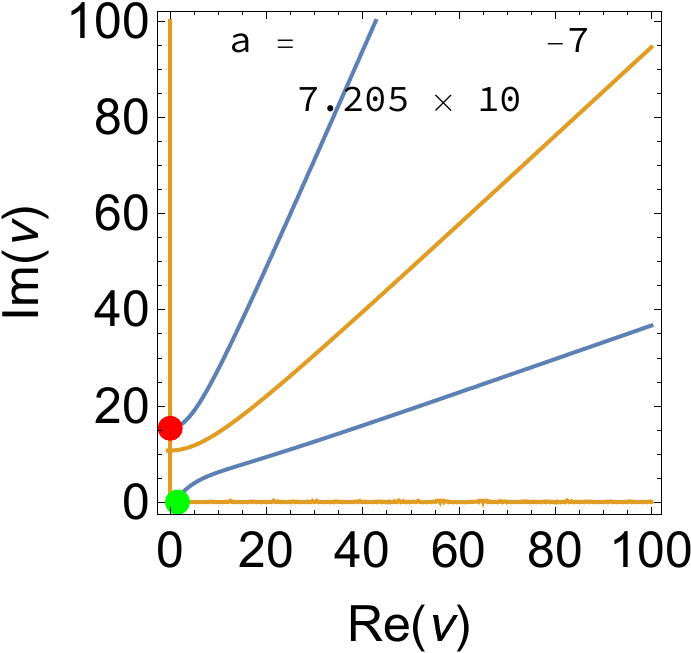}
 \caption{\it ${\tilde{\b}_\text{eff}} = -30$}
 \end{subfigure}
 \begin{subfigure}{.3\textwidth}
 \includegraphics[width=\textwidth]{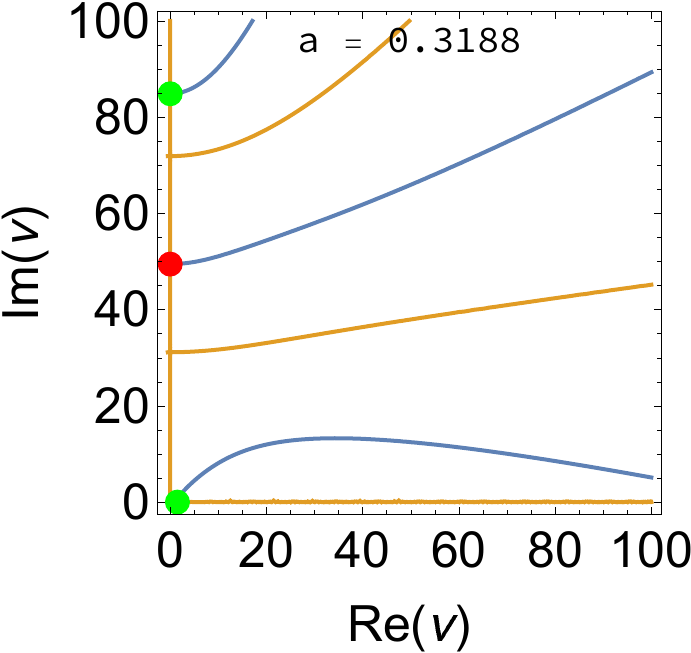}
 \caption{\it ${\tilde{\b}_\text{eff}} = -17$}
 \end{subfigure}
 \begin{subfigure}{.3\textwidth}
 \includegraphics[width=\textwidth]{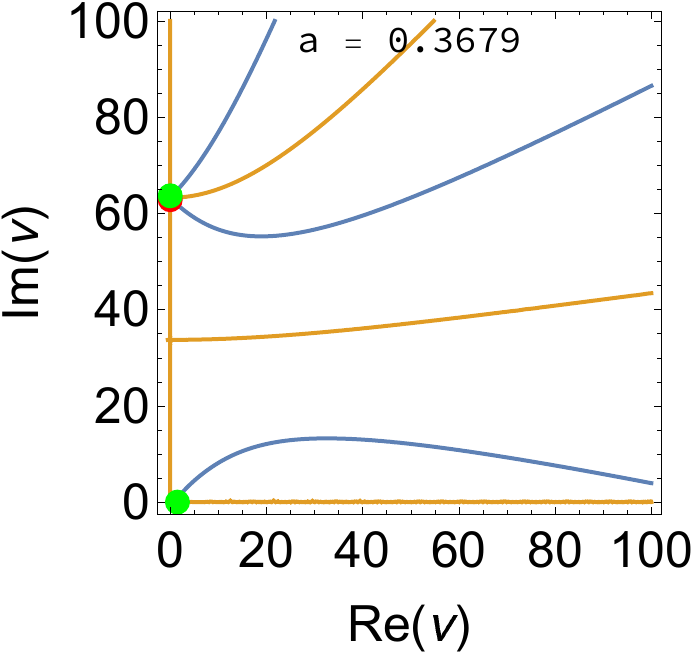}
 \caption{\it ${\tilde{\b}_\text{eff}} = -16.8567$}
 \end{subfigure}
 \begin{subfigure}{.3\textwidth}
 \includegraphics[width=\textwidth]{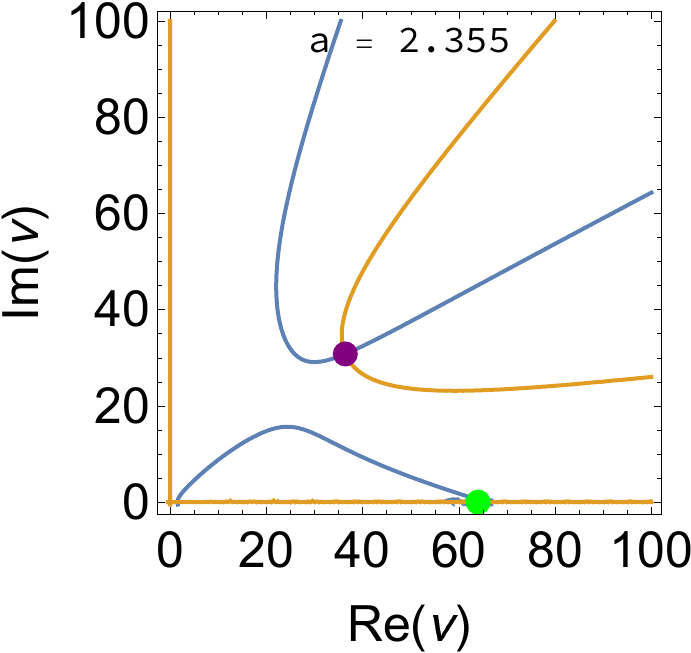}
 \caption{\it ${\tilde{\b}_\text{eff}} = -15$}
 \end{subfigure}
 \begin{subfigure}{.3\textwidth}
 \includegraphics[width=\textwidth]{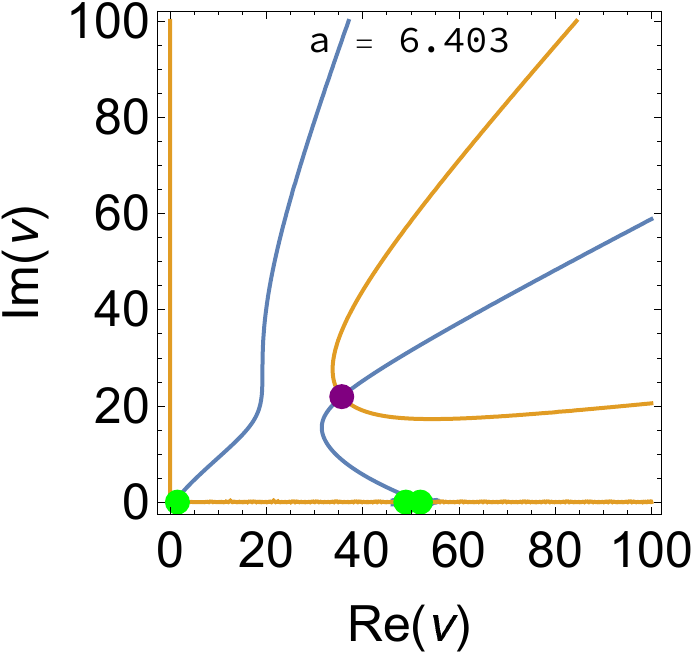}
 \caption{\it ${\tilde{\b}_\text{eff}} = -14$}
 \end{subfigure}
 \begin{subfigure}{.3\textwidth}
 \includegraphics[width=\textwidth]{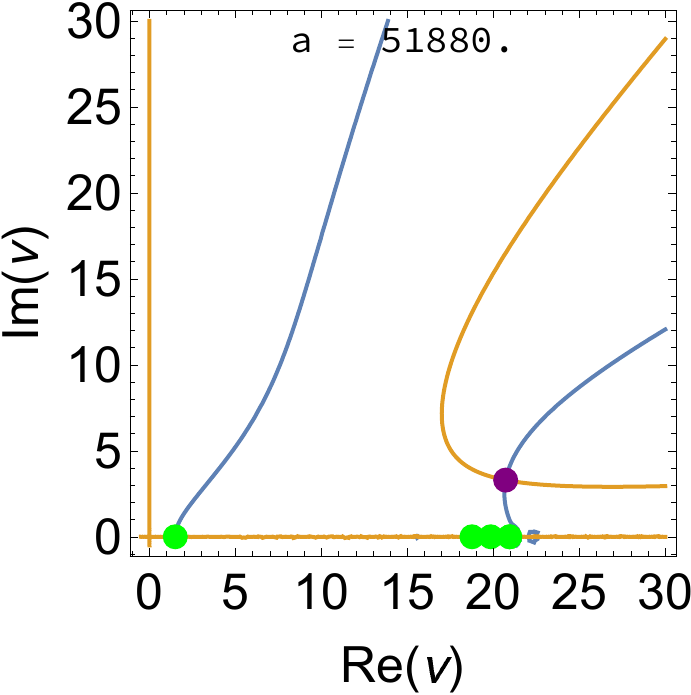}
 \caption{\it ${\tilde{\b}_\text{eff}} = -5$}
 \end{subfigure}
 \begin{subfigure}{.3\textwidth}
 \includegraphics[width=\textwidth]{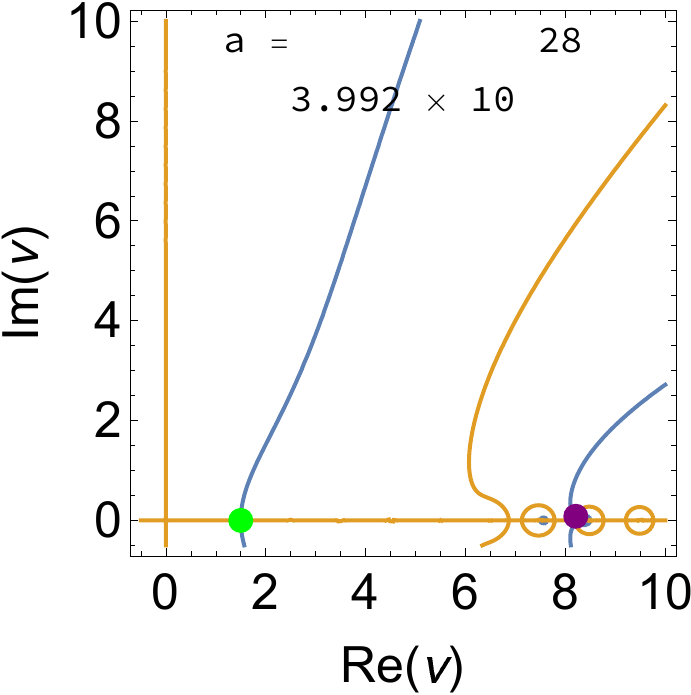}
 \caption{\it ${\tilde{\b}_\text{eff}} = 50$}
 \end{subfigure}
 \begin{subfigure}{.3\textwidth}
 \includegraphics[width=\textwidth]{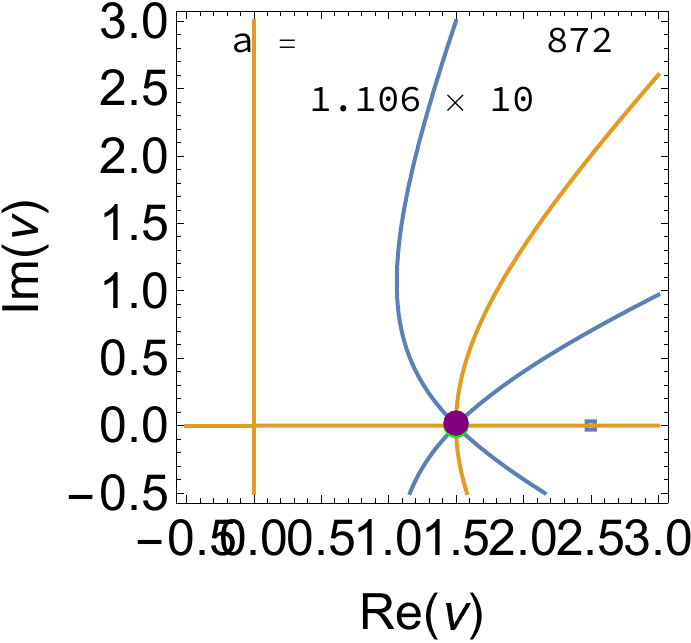}
 \caption{\it ${\tilde{\b}_\text{eff}} = 1992.1$}
 \end{subfigure}
 \begin{subfigure}{.3\textwidth}
 \includegraphics[width=\textwidth]{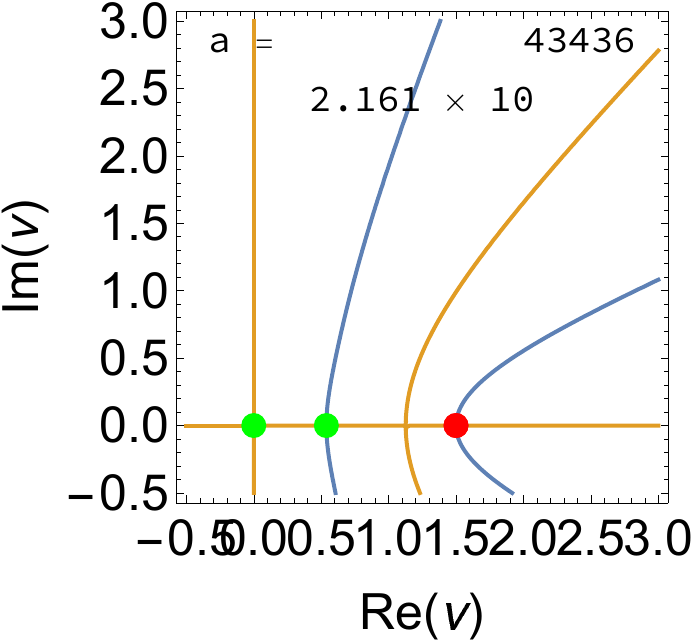}
 \caption{\it ${\tilde{\b}_\text{eff}} = 100000$}
 \end{subfigure}
 \caption{\it AdS, $\tilde{\a} = 1000$, $GN^2\chi^2 = 1000$.}
 \label{AdS_alpha1000_chi1000}
 \end{figure}

 \begin{figure}[ht]
 \centering
 \begin{subfigure}{.3\textwidth}
 \includegraphics[width=\textwidth]{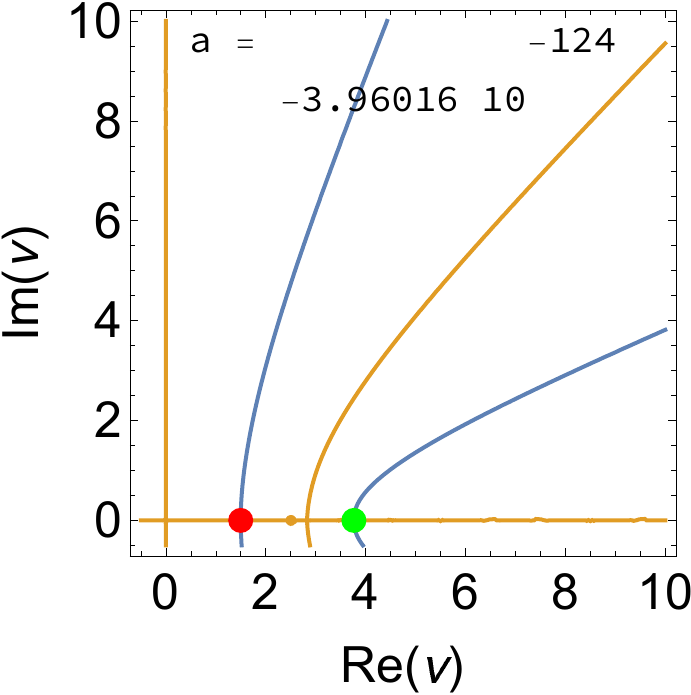}
 \caption{\it ${\tilde{\b}_\text{eff}} = -300$}
 \end{subfigure}
 \begin{subfigure}{.3\textwidth}
 \includegraphics[width=\textwidth]{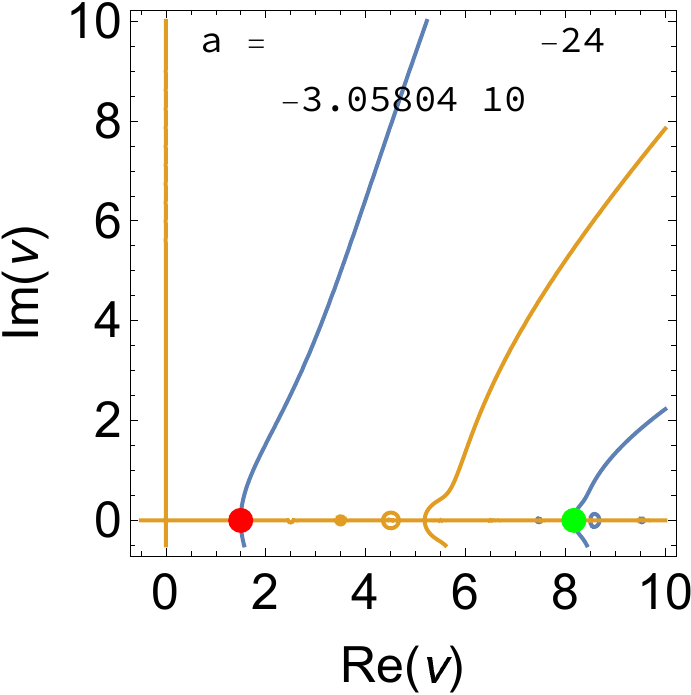}
 \caption{\it ${\tilde{\b}_\text{eff}} = -70$}
 \end{subfigure}
 \begin{subfigure}{.3\textwidth}
 \includegraphics[width=\textwidth]{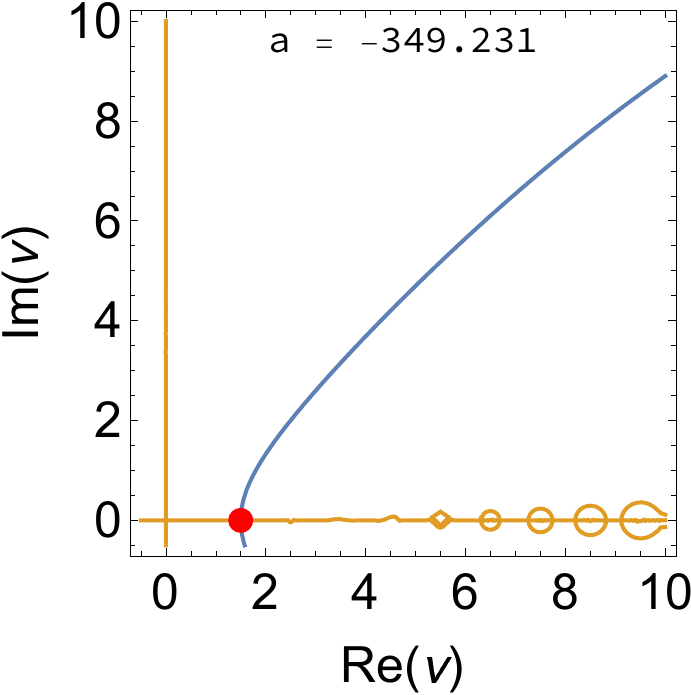}
 \caption{\it ${\tilde{\b}_\text{eff}} = -10$}
 \end{subfigure}
 \begin{subfigure}{.3\textwidth}
 \includegraphics[width=\textwidth]{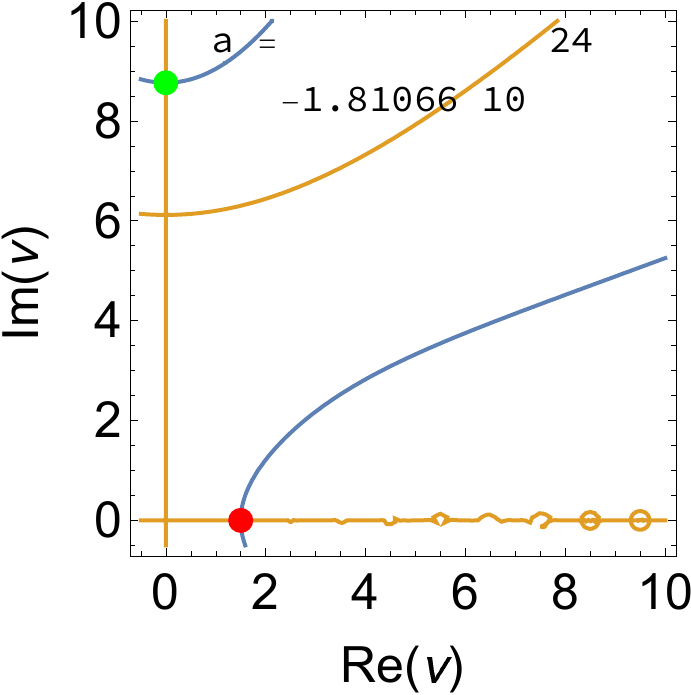}
 \caption{\it ${\tilde{\b}_\text{eff}} = 40$}
 \end{subfigure}
 \begin{subfigure}{.3\textwidth}
 \includegraphics[width=\textwidth]{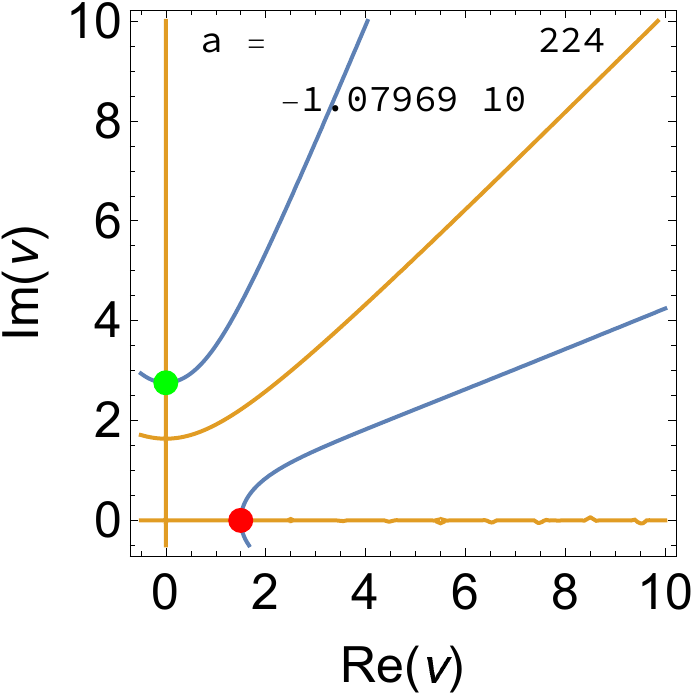}
 \caption{\it ${\tilde{\b}_\text{eff}} = 500$}
 \end{subfigure}
 \begin{subfigure}{.3\textwidth}
 \includegraphics[width=\textwidth]{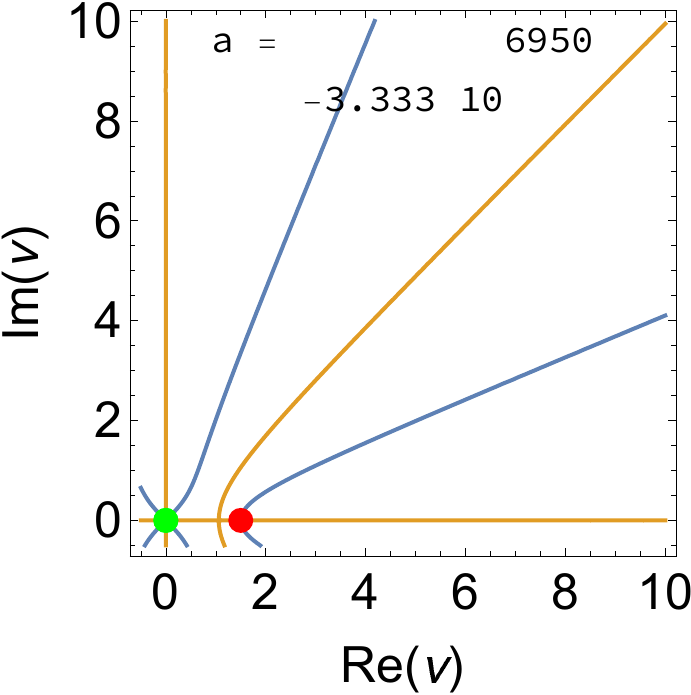}
 \caption{\it ${\tilde{\b}_\text{eff}} = 15988.3$}
 \end{subfigure}
 \begin{subfigure}{.3\textwidth}
 \includegraphics[width=\textwidth]{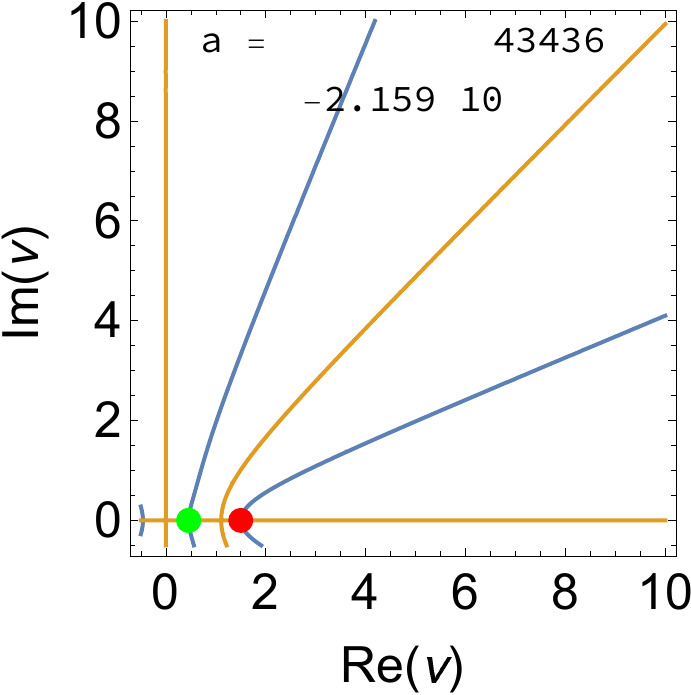}
 \caption{\it ${\tilde{\b}_\text{eff}} = 100000$}
 \end{subfigure}
 \caption{\it AdS, $\tilde{\a} = -1000$, $GN^2\chi^2 = 1000$.}
 \label{AdS_alpha-1000_chi1000}
 \end{figure}
 %
 %

 \begin{figure}[ht]
 \centering
 \begin{subfigure}{.3\textwidth}
 \includegraphics[width=\textwidth]{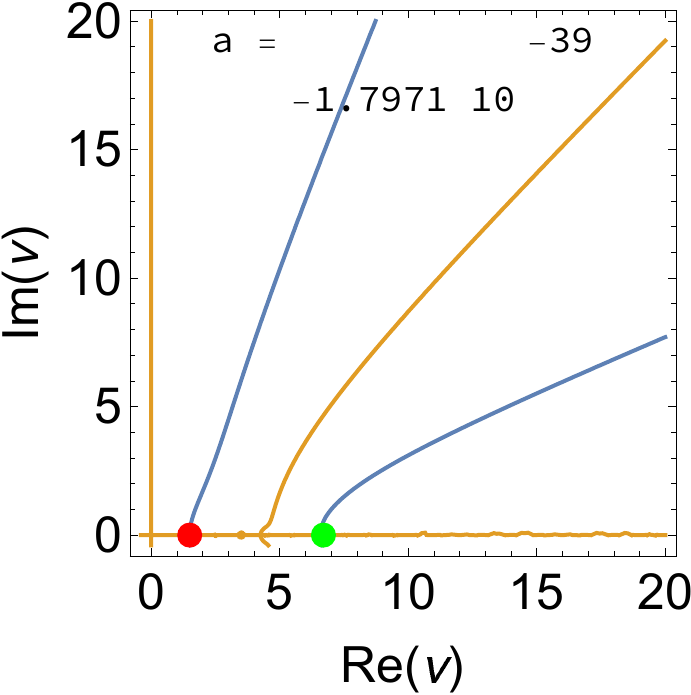}
 \caption{\it ${\tilde{\b}_\text{eff}} = -100$}
 \end{subfigure}
 \begin{subfigure}{.3\textwidth}
 \includegraphics[width=\textwidth]{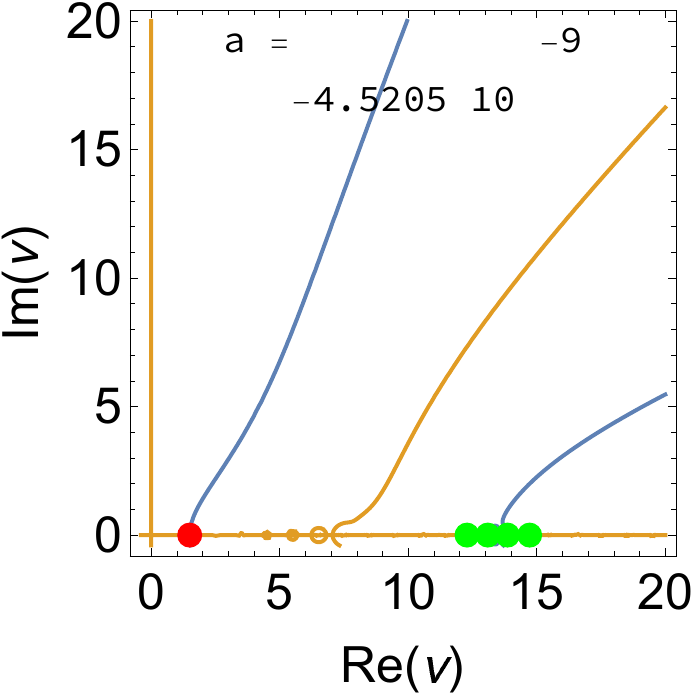}
 \caption{\it ${\tilde{\b}_\text{eff}} = -30$}
 \end{subfigure}
 \begin{subfigure}{.3\textwidth}
 \includegraphics[width=\textwidth]{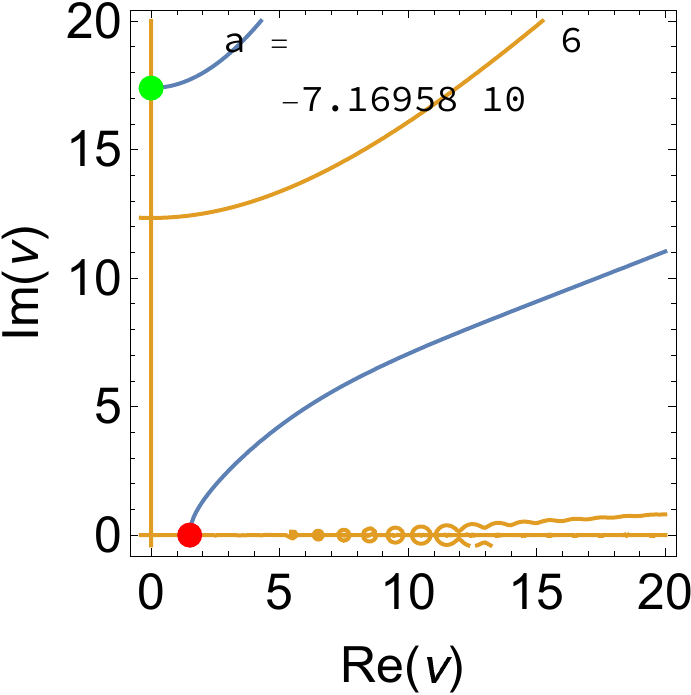}
 \caption{\it ${\tilde{\b}_\text{eff}} = 5$}
 \end{subfigure}
 \begin{subfigure}{.3\textwidth}
 \includegraphics[width=\textwidth]{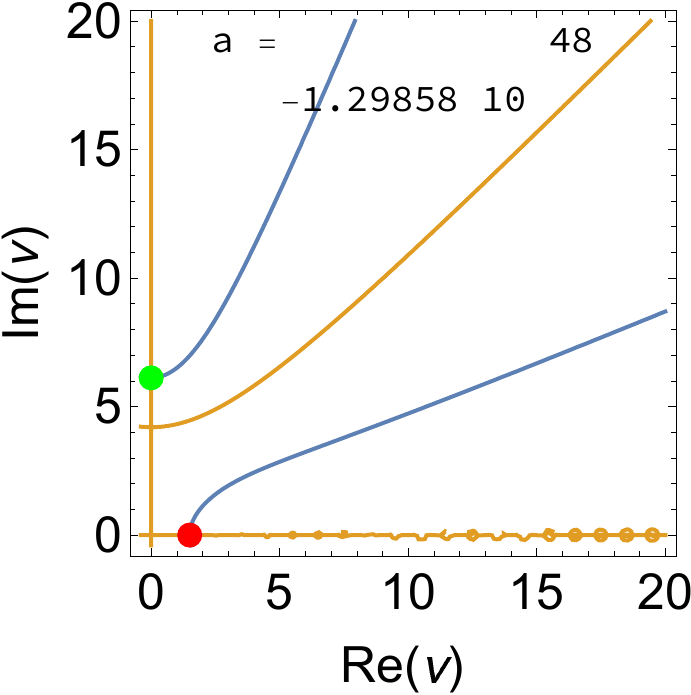}
 \caption{\it ${\tilde{\b}_\text{eff}} = 100$}
 \end{subfigure}
 \begin{subfigure}{.3\textwidth}
 \includegraphics[width=\textwidth]{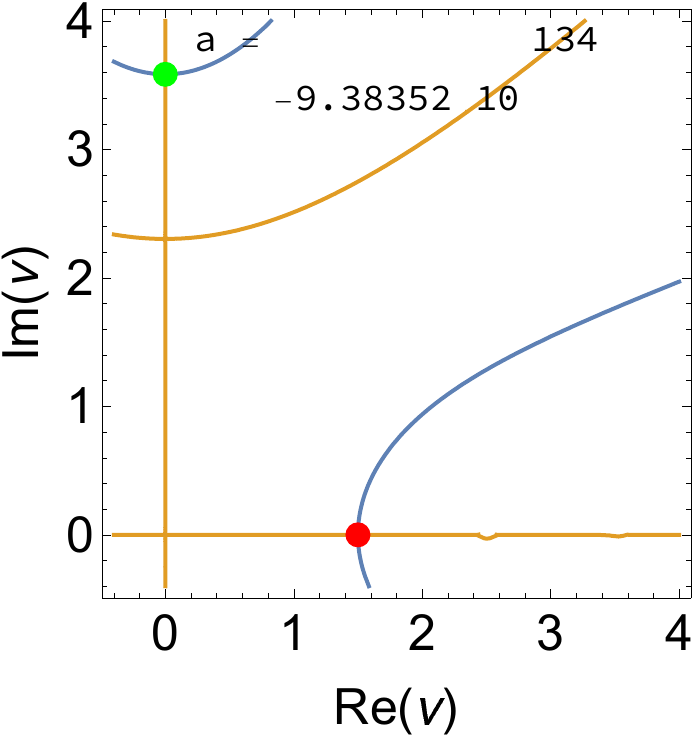}
 \caption{\it ${\tilde{\b}_\text{eff}} = 300$}
 \end{subfigure}
 \begin{subfigure}{.3\textwidth}
 \includegraphics[width=\textwidth]{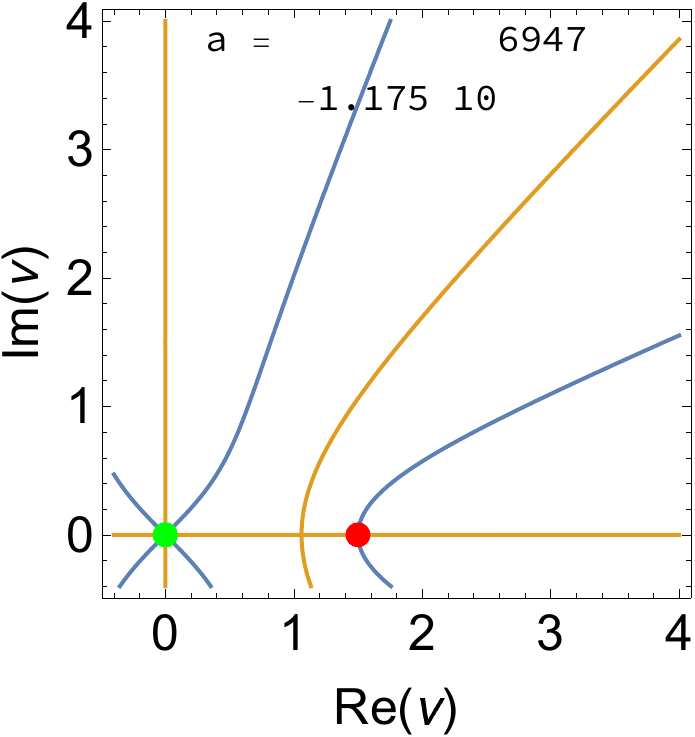}
 \caption{\it ${\tilde{\b}_\text{eff}} = 15985.4$}
 \end{subfigure}
 \begin{subfigure}{.3\textwidth}
 \includegraphics[width=\textwidth]{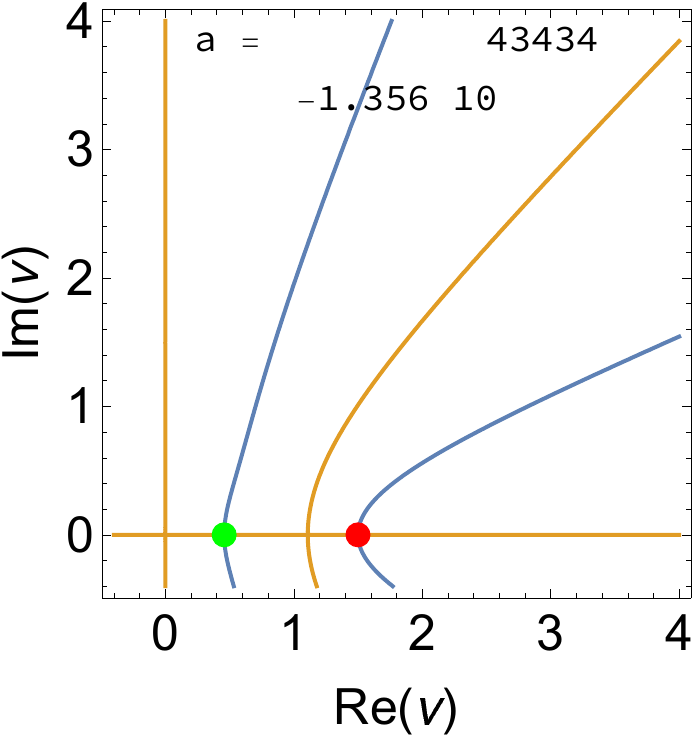}
 \caption{\it ${\tilde{\b}_\text{eff}} = 100000$}
 \end{subfigure}
 \caption{\it AdS, $\tilde{\a} = -1000$, $GN^2\chi^2 = 2\pi$.}
 \label{AdS_alpha-1000_chi2pi}
 \end{figure}
 %
 %

 \begin{figure}[ht]
 \centering
 \begin{subfigure}{.3\textwidth}
 \includegraphics[width=\textwidth]{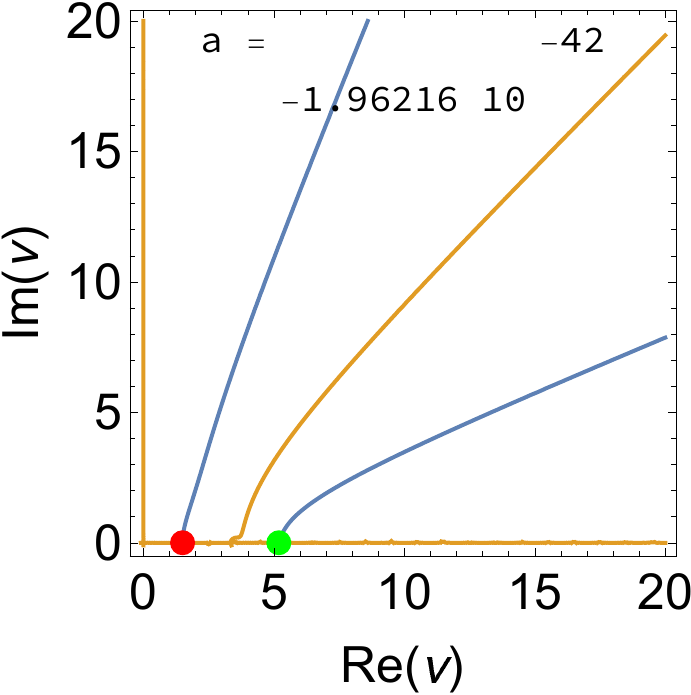}
 \caption{\it ${\tilde{\b}_\text{eff}} = -100$}
 \end{subfigure}
 \begin{subfigure}{.3\textwidth}
 \includegraphics[width=\textwidth]{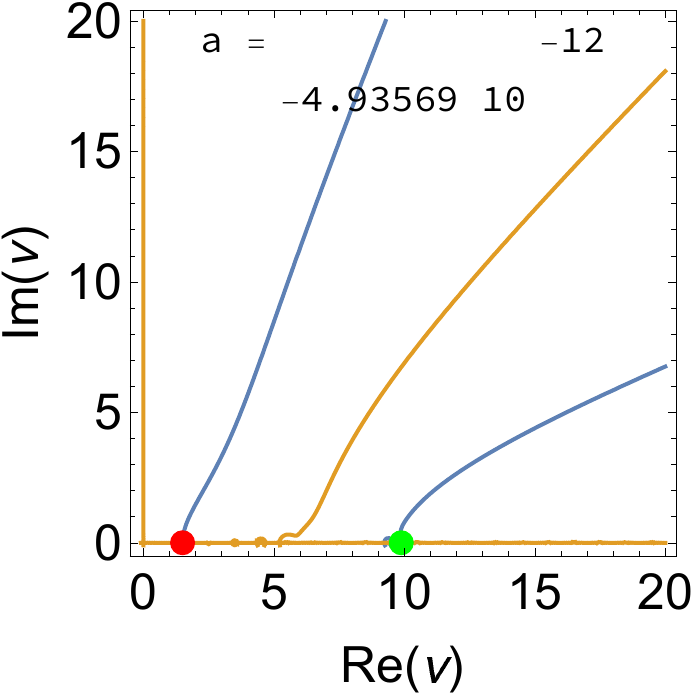}
 \caption{\it ${\tilde{\b}_\text{eff}} = -30$}
 \end{subfigure}
 \begin{subfigure}{.3\textwidth}
 \includegraphics[width=\textwidth]{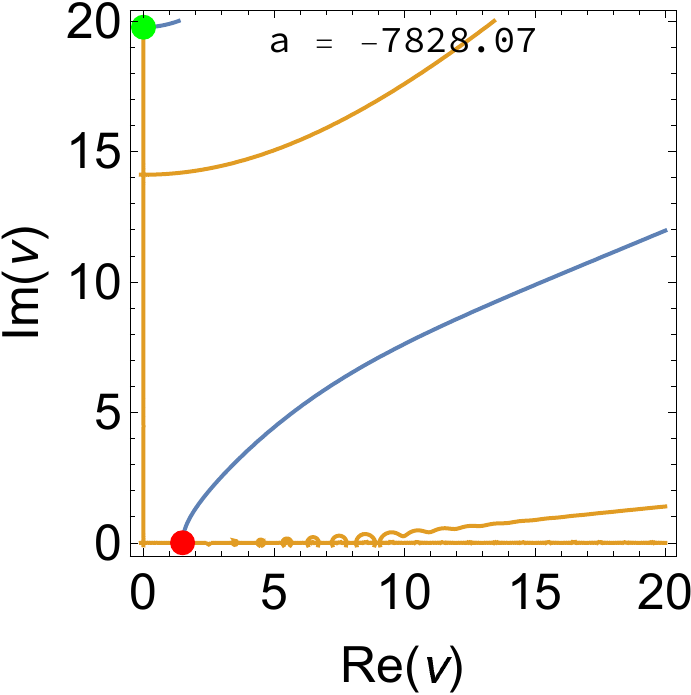}
 \caption{\it ${\tilde{\b}_\text{eff}} = 5$}
 \end{subfigure}
 \begin{subfigure}{.3\textwidth}
 \includegraphics[width=\textwidth]{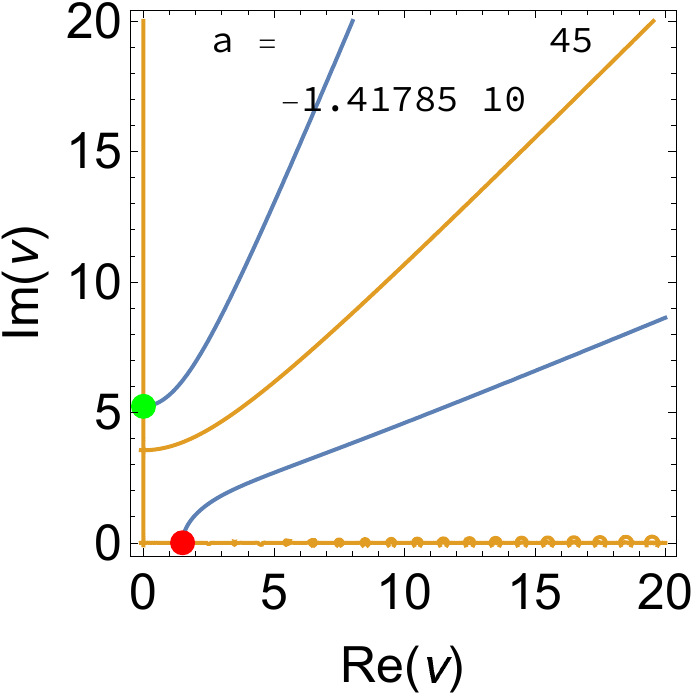}
 \caption{\it ${\tilde{\b}_\text{eff}} = 100$}
 \end{subfigure}
 \begin{subfigure}{.3\textwidth}
 \includegraphics[width=\textwidth]{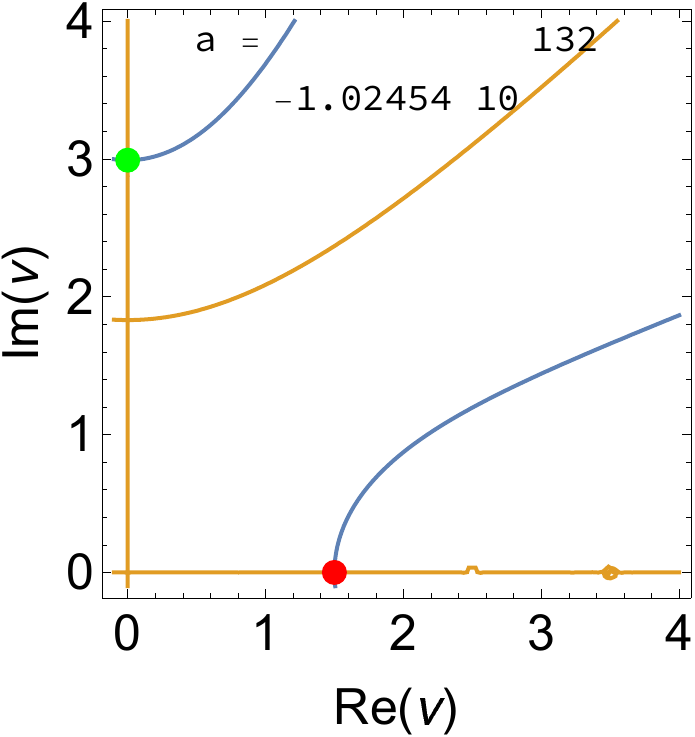}
 \caption{\it ${\tilde{\b}_\text{eff}} = 300$}
 \end{subfigure}
 \begin{subfigure}{.3\textwidth}
 \includegraphics[width=\textwidth]{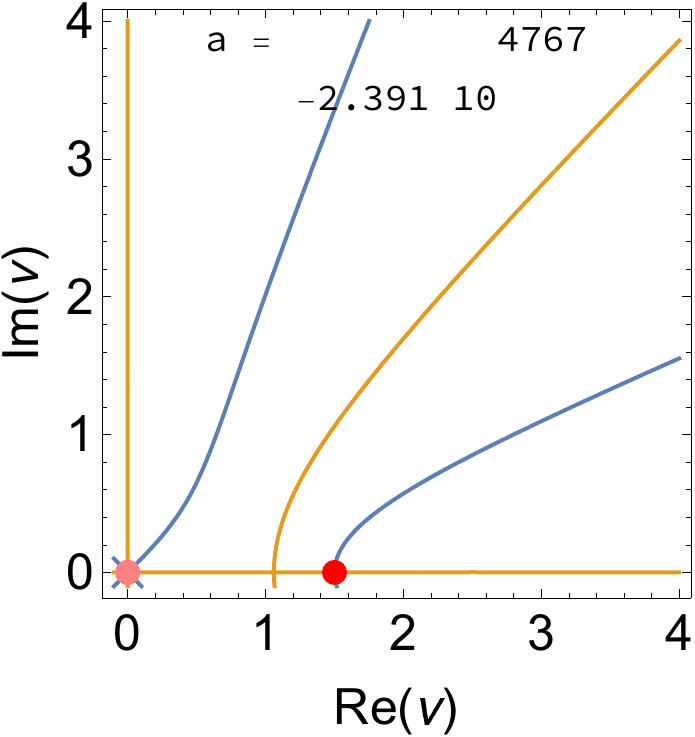}
 \caption{\it ${\tilde{\b}_\text{eff}} = 10973.3$}
 \end{subfigure}
 \begin{subfigure}{.3\textwidth}
 \includegraphics[width=\textwidth]{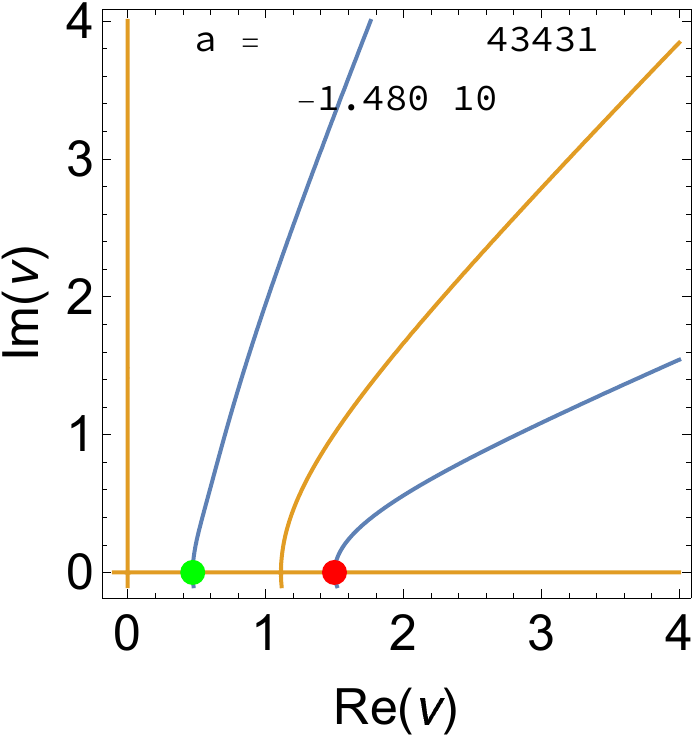}
 \caption{\it ${\tilde{\b}_\text{eff}} = 100000$}
 \end{subfigure}
 \caption{\it AdS, $\tilde{\a} = -1000$, $GN^2\chi^2 = 0.01$.}
 \label{AdS_alpha-1000_chi0.01}
 \end{figure}
 %

 \begin{figure}[ht]
 \centering
 \begin{subfigure}{.3\textwidth}
 \includegraphics[width=\textwidth]{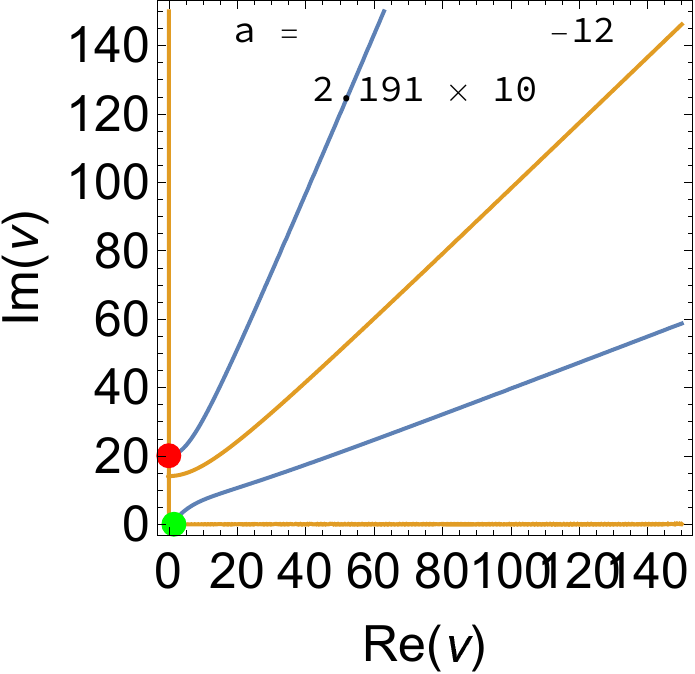}
 \caption{\it ${\tilde{\b}_\text{eff}} = -30$}
 \end{subfigure}
 \begin{subfigure}{.3\textwidth}
 \includegraphics[width=\textwidth]{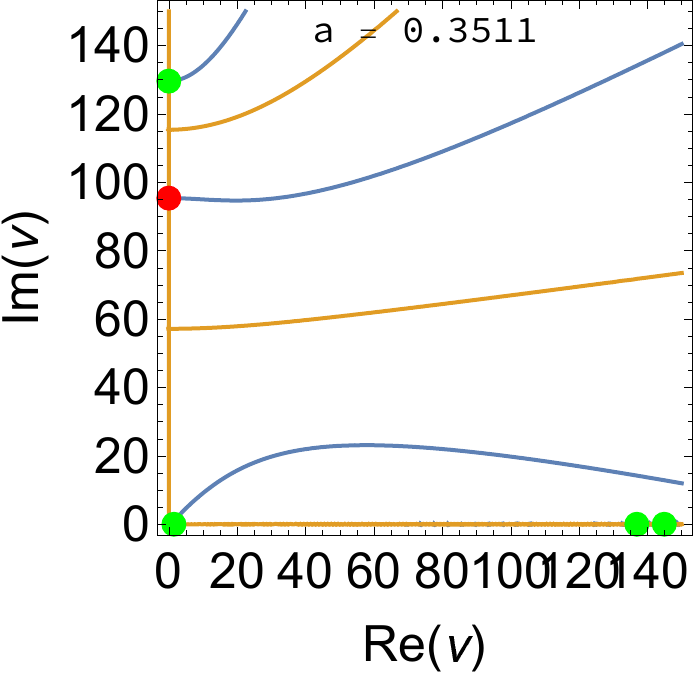}
 \caption{\it ${\tilde{\b}_\text{eff}} = -4.2$}
 \end{subfigure}
 \begin{subfigure}{.3\textwidth}
 \includegraphics[width=\textwidth]{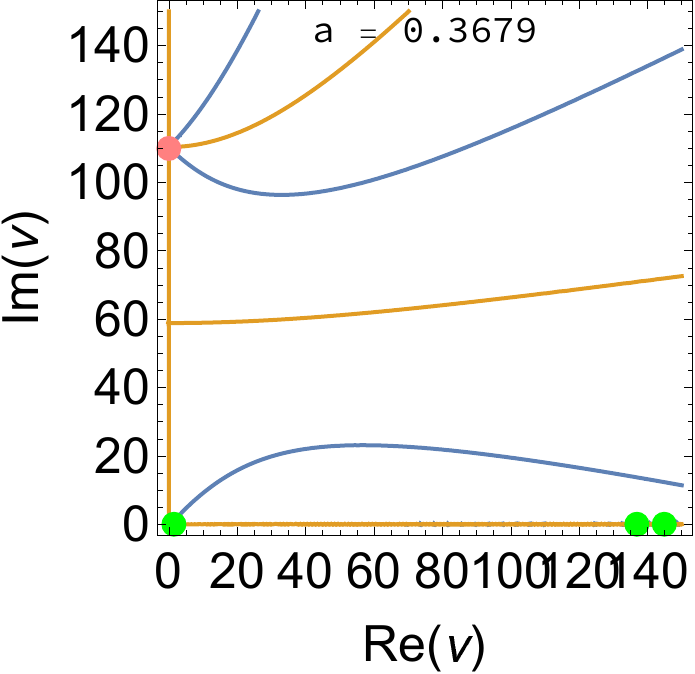}
 \caption{\it ${\tilde{\b}_\text{eff}} = -4.15327$}
 \end{subfigure}
 \begin{subfigure}{.3\textwidth}
 \includegraphics[width=\textwidth]{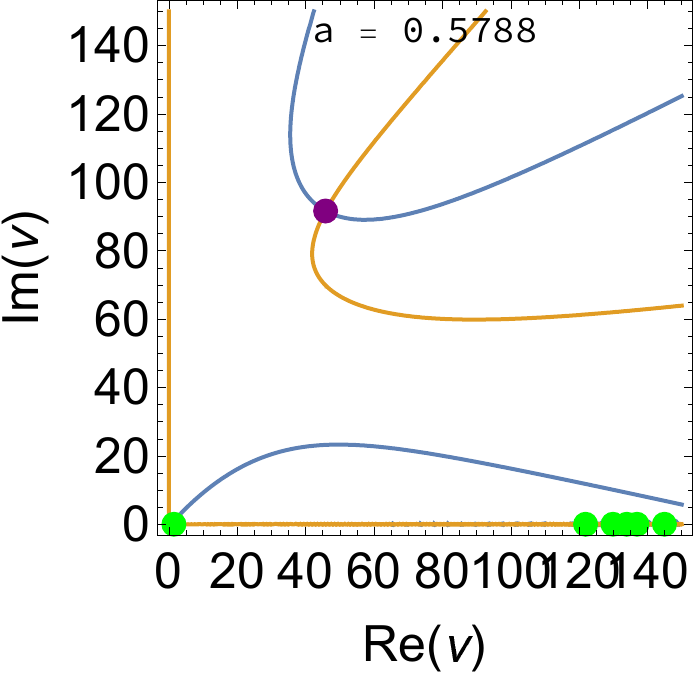}
 \caption{\it ${\tilde{\b}_\text{eff}} = -3.7$}
 \end{subfigure}
 \begin{subfigure}{.3\textwidth}
 \includegraphics[width=\textwidth]{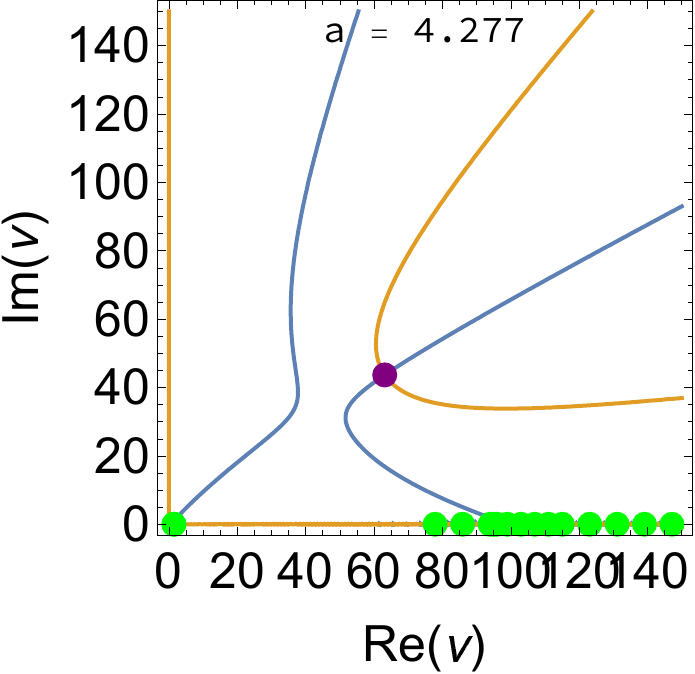}
 \caption{\it ${\tilde{\b}_\text{eff}} = -1.7$}
 \end{subfigure}
 \begin{subfigure}{.3\textwidth}
 \includegraphics[width=\textwidth]{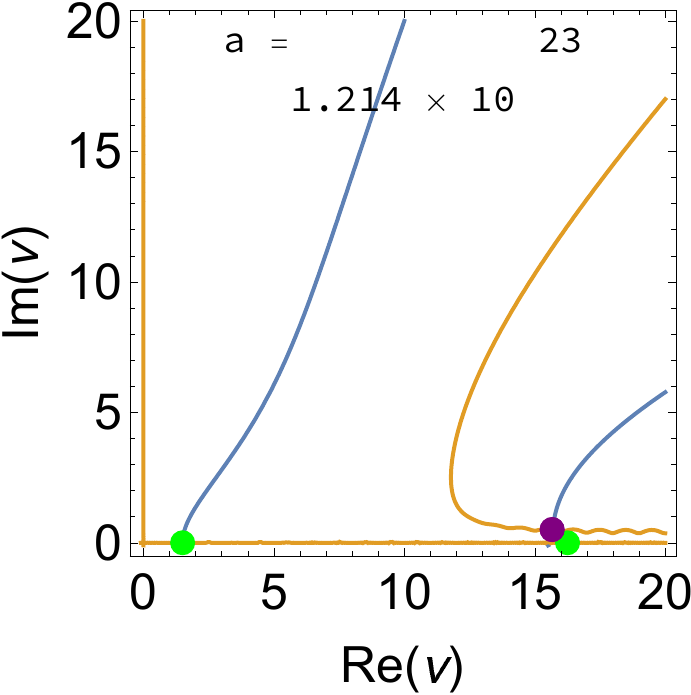}
 \caption{\it ${\tilde{\b}_\text{eff}} = 50$}
 \end{subfigure}
 \begin{subfigure}{.3\textwidth}
 \includegraphics[width=\textwidth]{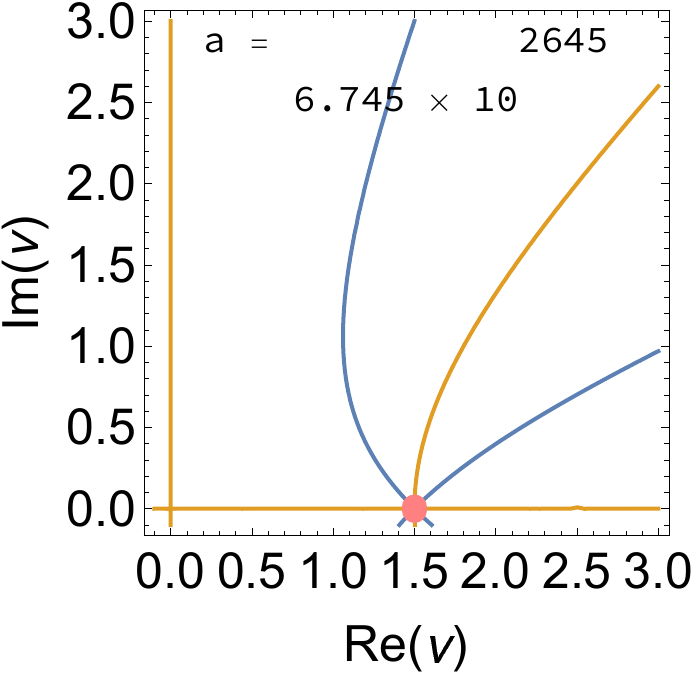}
 \caption{\it ${\tilde{\b}_\text{eff}} = 6089.09$}
 \end{subfigure}
 \begin{subfigure}{.3\textwidth}
 \includegraphics[width=\textwidth]{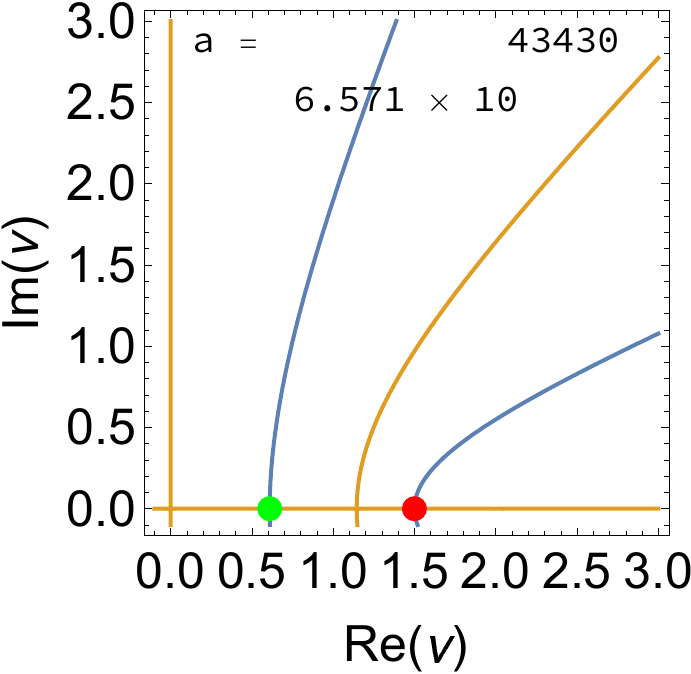}
 \caption{\it ${\tilde{\b}_\text{eff}} = 100000$}
 \end{subfigure}
 \caption{\it AdS, $\tilde{\a} = -100$, $GN^2\chi^2 = 0.001$.}
 \label{AdS_alpha-100_chi0.001}
 \end{figure}

 \begin{figure}[ht]
 \centering
 \begin{subfigure}{.3\textwidth}
 \includegraphics[width=\textwidth]{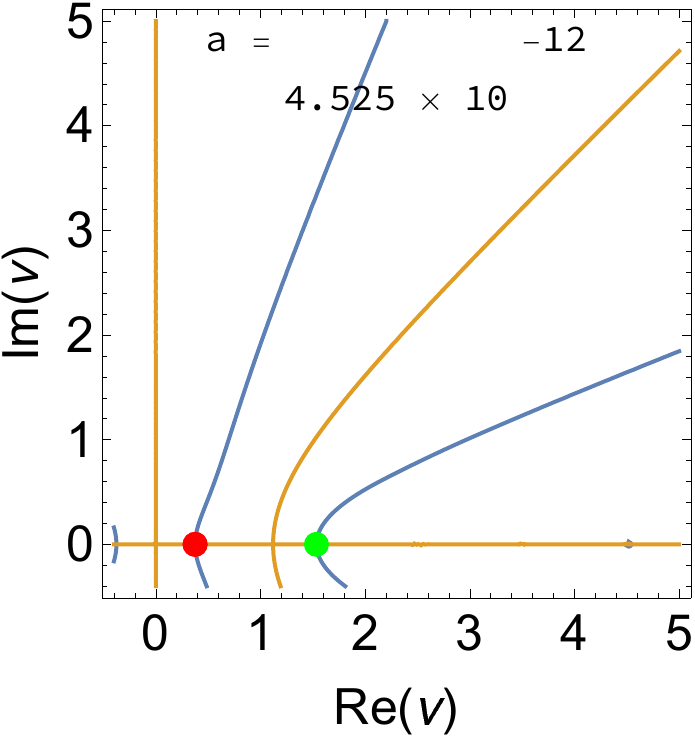}
 \caption{\it ${\tilde{\b}_\text{eff}} = -30$}
 \end{subfigure}
 \begin{subfigure}{.3\textwidth}
 \includegraphics[width=\textwidth]{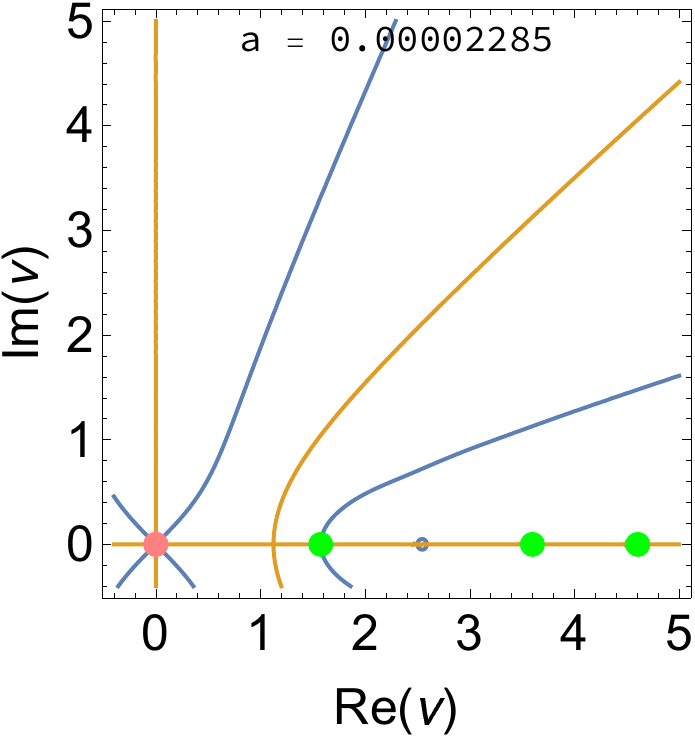}
 \caption{\it ${\tilde{\b}_\text{eff}} = -14.5653 $ (*)}
 \end{subfigure}
 \begin{subfigure}{.3\textwidth}
 \includegraphics[width=\textwidth]{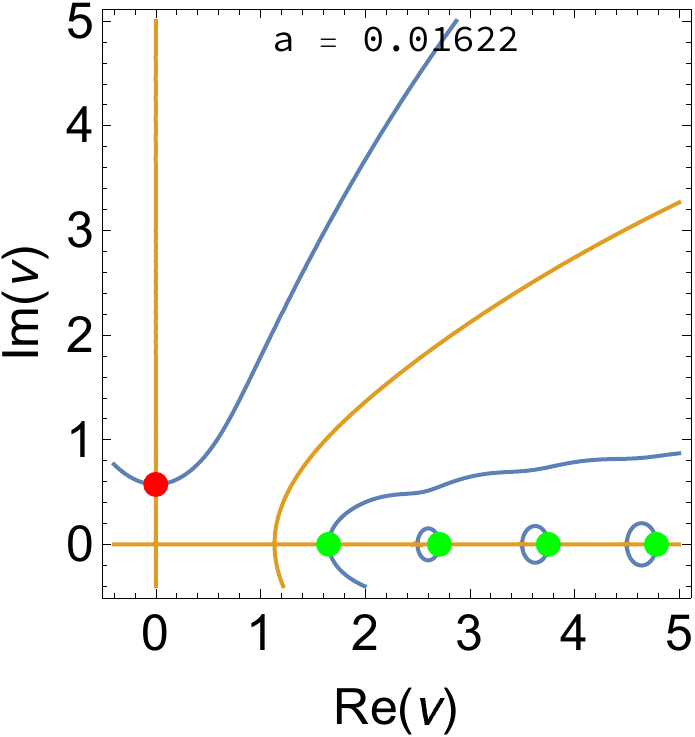}
 \caption{\it ${\tilde{\b}_\text{eff}} = -8$}
 \end{subfigure}
 \begin{subfigure}{.3\textwidth}
 \includegraphics[width=\textwidth]{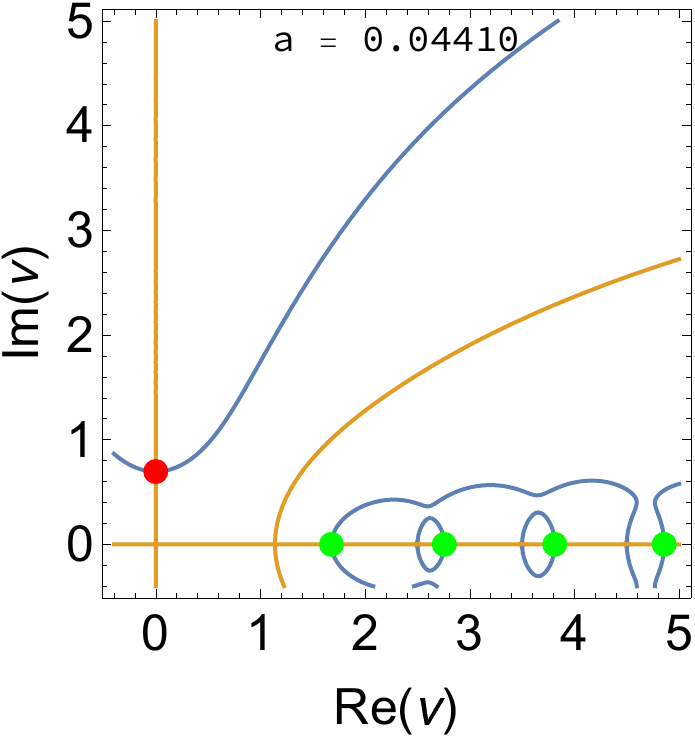}
 \caption{\it ${\tilde{\b}_\text{eff}} = -7$}
 \end{subfigure}
 \begin{subfigure}{.3\textwidth}
 \includegraphics[width=\textwidth]{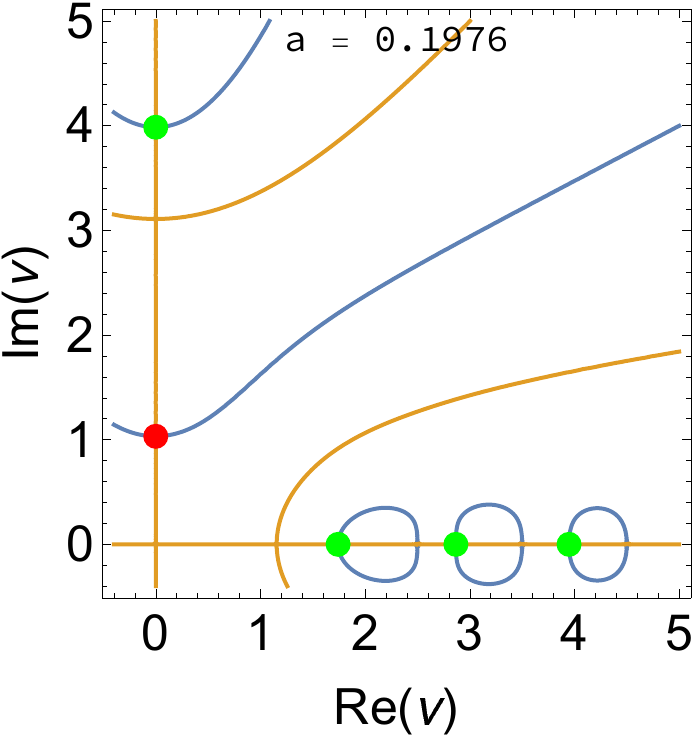}
 \caption{\it ${\tilde{\b}_\text{eff}} = -5.5$}
 \end{subfigure}
 \begin{subfigure}{.3\textwidth}
 \includegraphics[width=\textwidth]{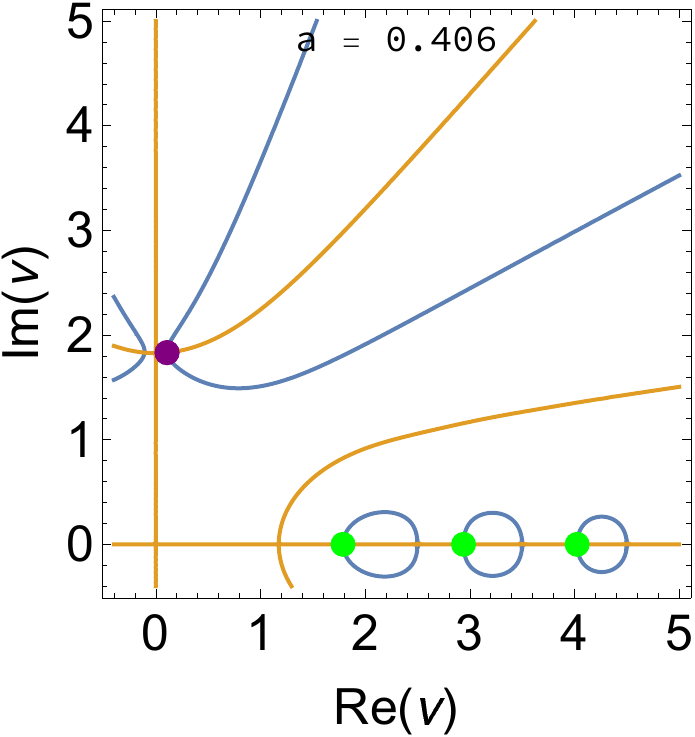}
 \caption{\it ${\tilde{\b}_\text{eff}} = -4.78$}
 \end{subfigure}
 \begin{subfigure}{.3\textwidth}
 \includegraphics[width=\textwidth]{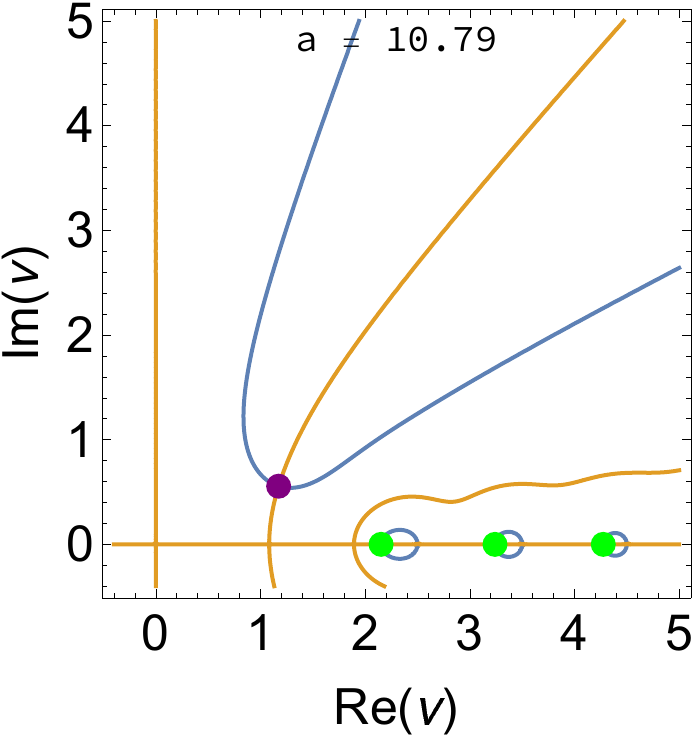}
 \caption{\it ${\tilde{\b}_\text{eff}} = -1.5$}
 \end{subfigure}
 \begin{subfigure}{.3\textwidth}
 \includegraphics[width=\textwidth]{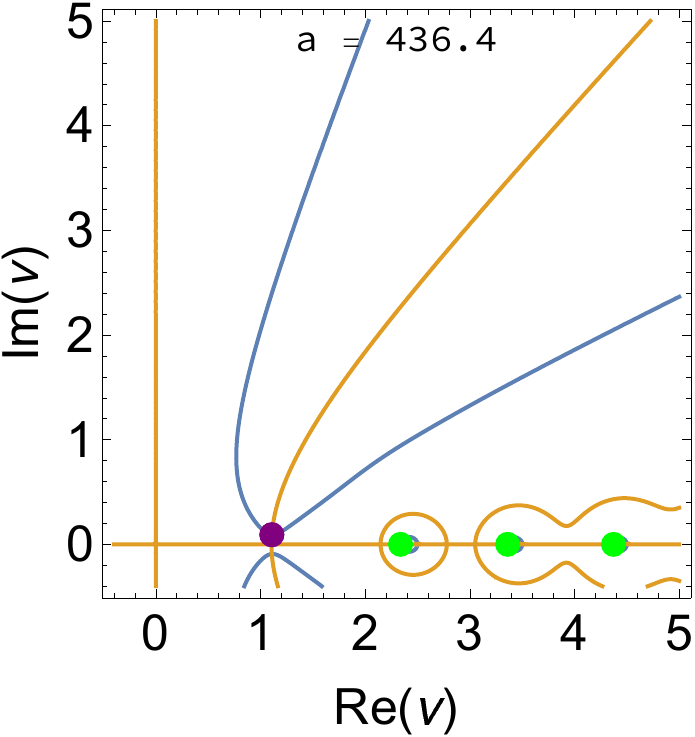}
 \caption{\it ${\tilde{\b}_\text{eff}} = 2.2$}
 \end{subfigure}
 \begin{subfigure}{.3\textwidth}
 \includegraphics[width=\textwidth]{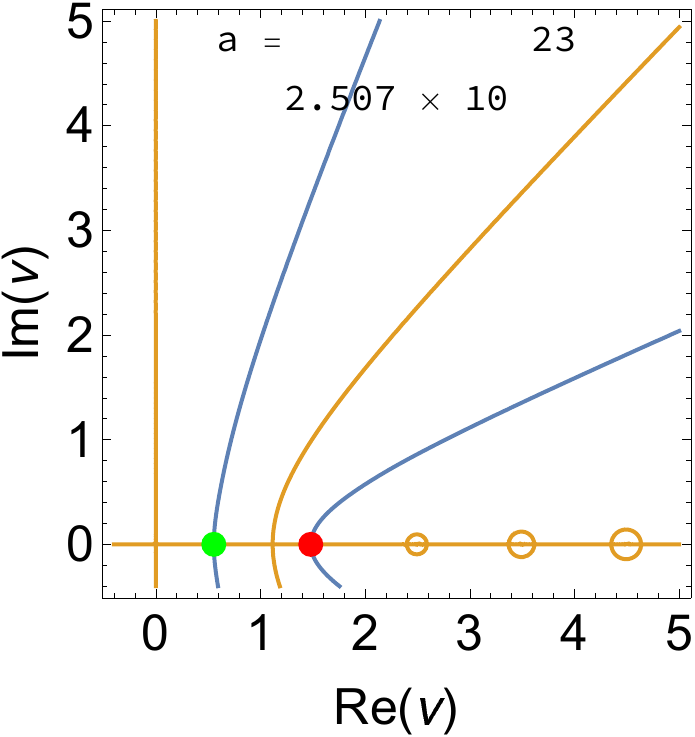}
 \caption{\it ${\tilde{\b}_\text{eff}} = 50$}
 \end{subfigure}

 \caption{\it AdS, $\tilde{\a} = 0$, $GN^2\chi^2 = 2\pi$.}
 \label{AdS_alpha0_chi2pi}
 \end{figure}

 \begin{figure}[h!]
\centering
\begin{subfigure}{.3\textwidth}
\includegraphics[width=\textwidth]{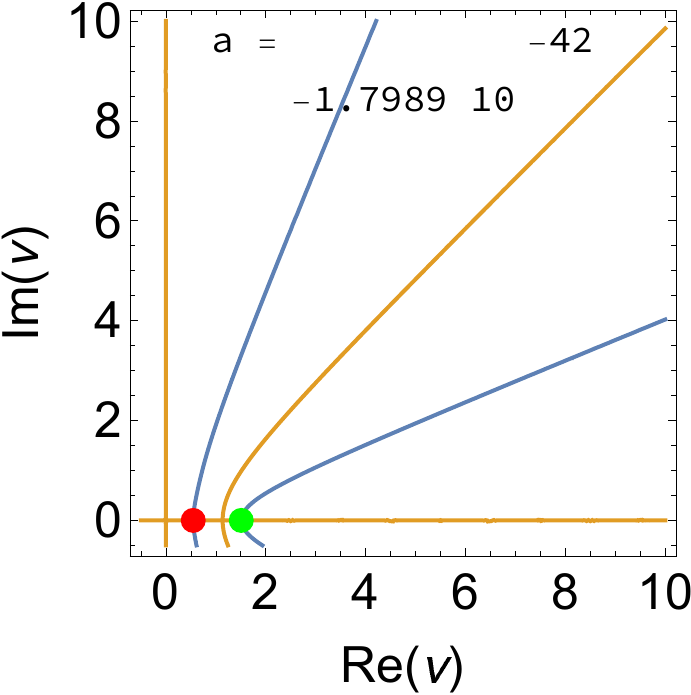}
\caption{\it ${\tilde{\b}_\text{eff}} = -100$}
\end{subfigure}
\begin{subfigure}{.3\textwidth}
\includegraphics[width=\textwidth]{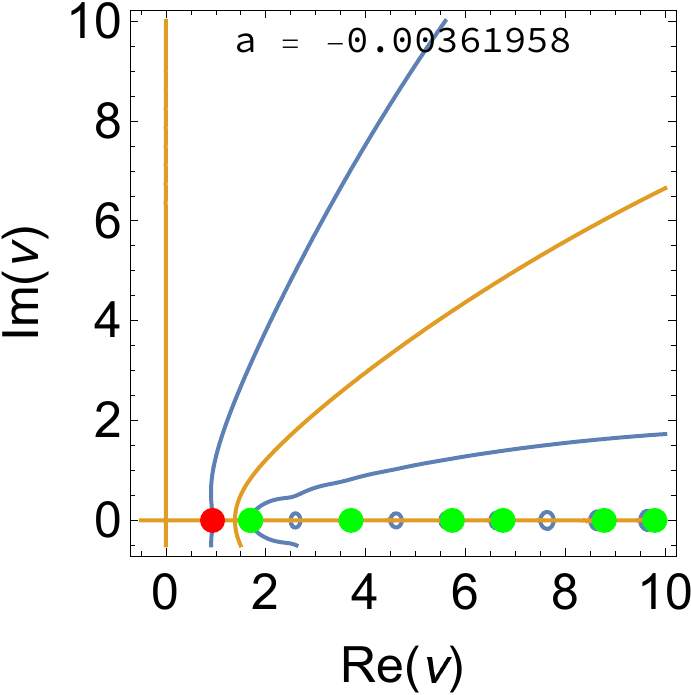}
\caption{\it ${\tilde{\b}_\text{eff}} = -9.5$}
\end{subfigure}
\begin{subfigure}{.3\textwidth}
\includegraphics[width=\textwidth]{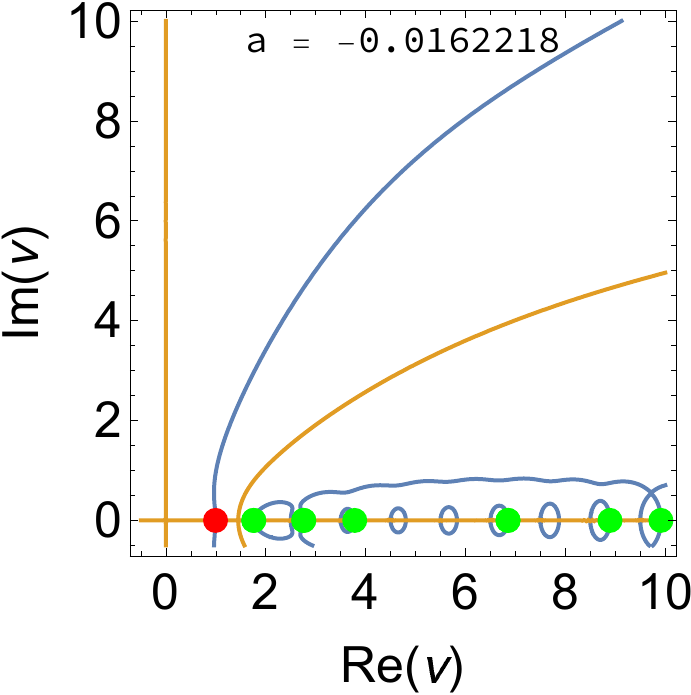}
\caption{\it ${\tilde{\b}_\text{eff}} = -8$}
\end{subfigure}
\begin{subfigure}{.3\textwidth}
\includegraphics[width=\textwidth]{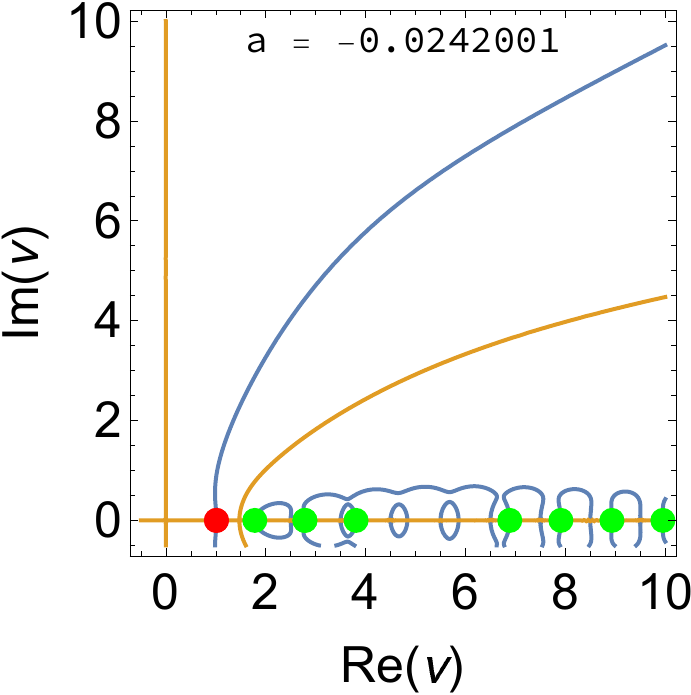}
\caption{\it ${\tilde{\b}_\text{eff}} = -7.6$}
\end{subfigure}
\begin{subfigure}{.3\textwidth}
\includegraphics[width=\textwidth]{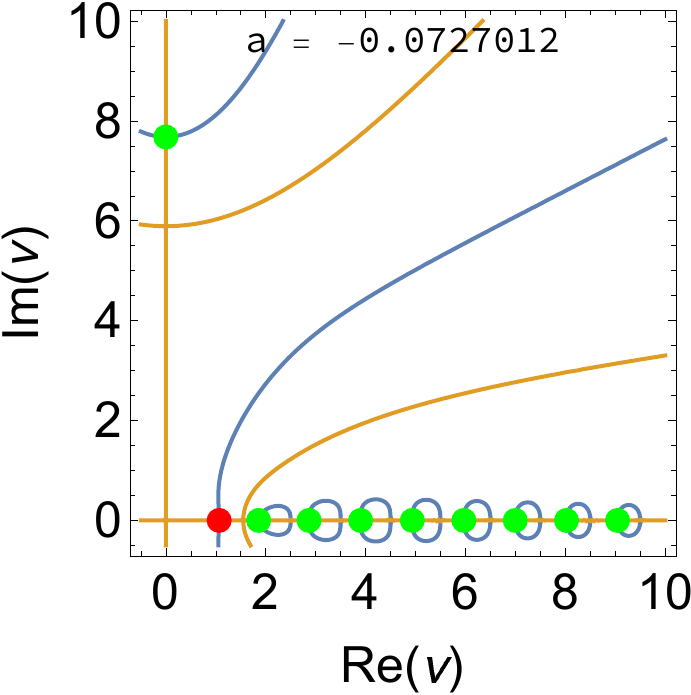}
\caption{\it ${\tilde{\b}_\text{eff}} = -6.5$}
\end{subfigure}
\begin{subfigure}{.3\textwidth}
\includegraphics[width=\textwidth]{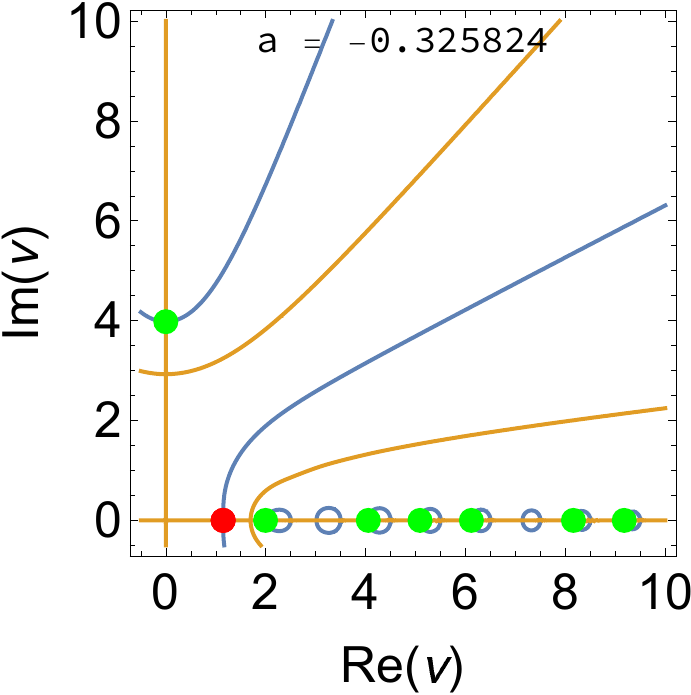}
\caption{\it ${\tilde{\b}_\text{eff}} = -5$}
\end{subfigure}
\begin{subfigure}{.3\textwidth}
\includegraphics[width=\textwidth]{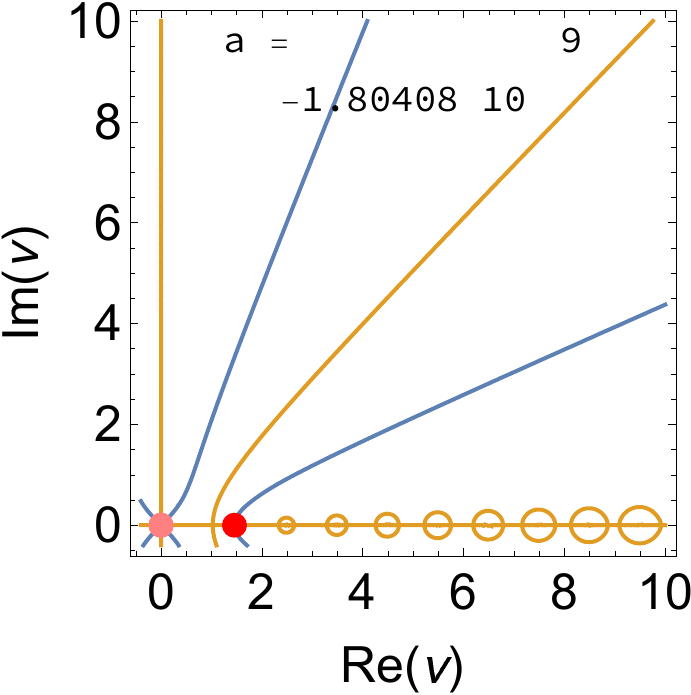}
\caption{\it ${\tilde{\b}_\text{eff}}={\pi\b_\text{BF}\over N^2} = 17.4347$}
\end{subfigure}
\begin{subfigure}{.3\textwidth}
\includegraphics[width=\textwidth]{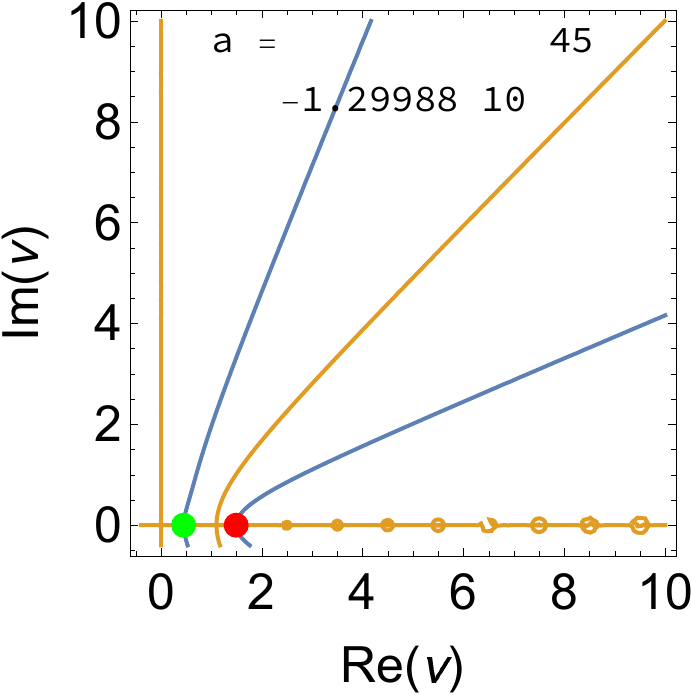}
\caption{\it ${\tilde{\b}_\text{eff}} = 100$}
\end{subfigure}
\caption{\it $\a = -2$, $GN^2\chi^2 = 2\pi$.}
\label{AdS_chi2pi_am2}
\end{figure}
%

 \begin{figure}[ht]
 \centering
 \begin{subfigure}{.3\textwidth}
 \includegraphics[width=\textwidth]{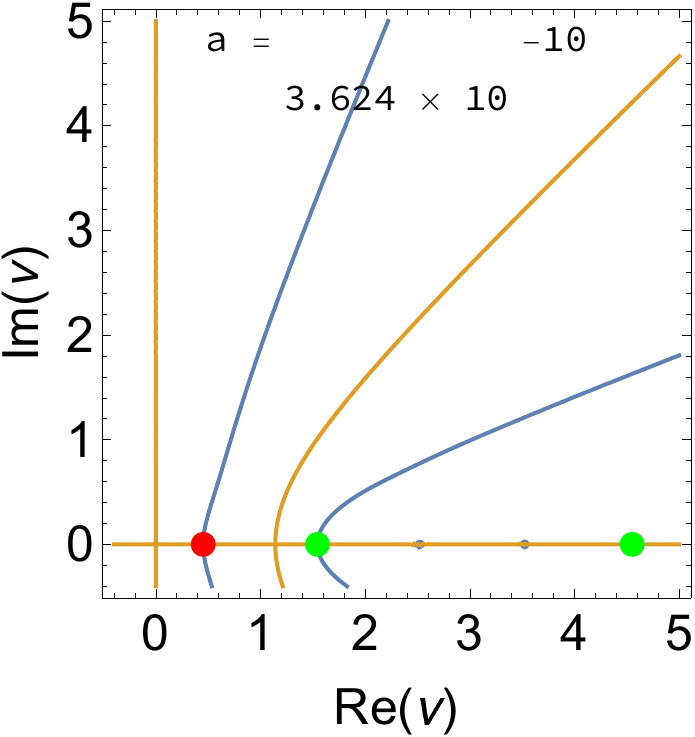}
 \caption{\it ${\tilde{\b}_\text{eff}} = -30$}
 \end{subfigure}
 \begin{subfigure}{.3\textwidth}
 \includegraphics[width=\textwidth]{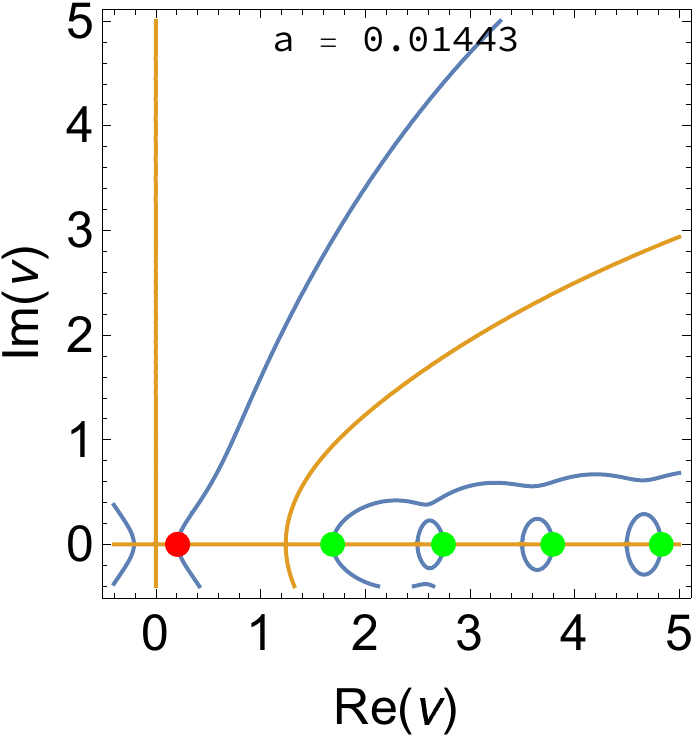}
 \caption{\it ${\tilde{\b}_\text{eff}} = -12.5 $}
 \end{subfigure}
 \begin{subfigure}{.3\textwidth}
 \includegraphics[width=\textwidth]{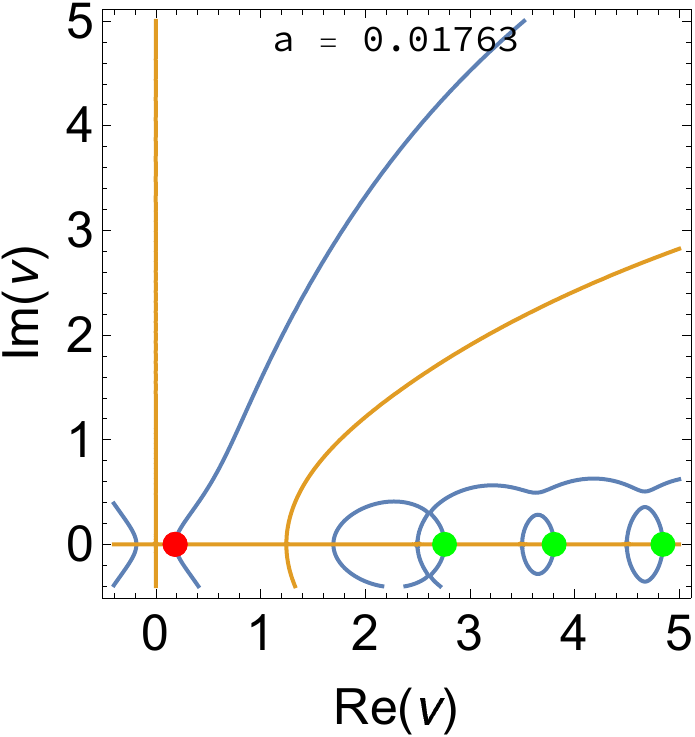}
 \caption{\it ${\tilde{\b}_\text{eff}} = -12.3$}
 \end{subfigure}
 \begin{subfigure}{.3\textwidth}
 \includegraphics[width=\textwidth]{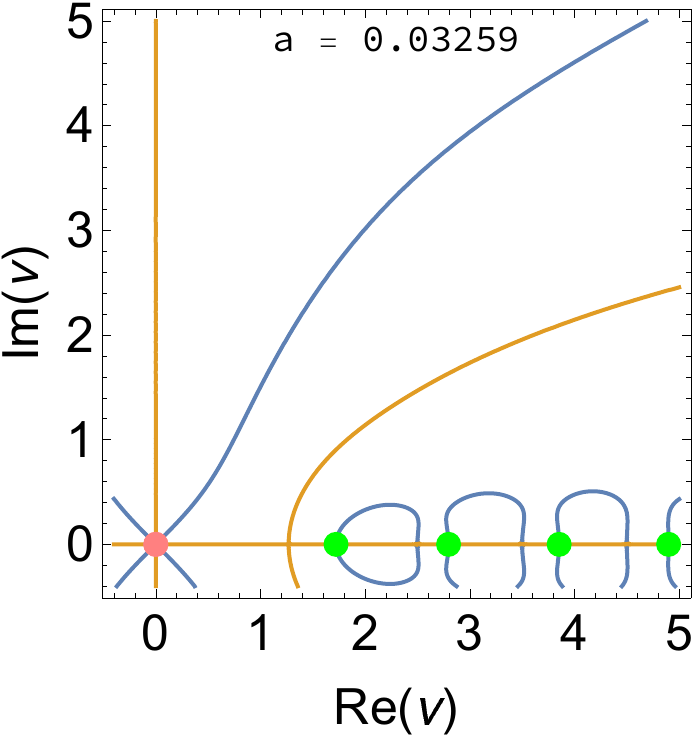}
 \caption{\it ${\tilde{\b}_\text{eff}} = -11.6854 (*)$}
 \end{subfigure}
 \begin{subfigure}{.3\textwidth}
 \includegraphics[width=\textwidth]{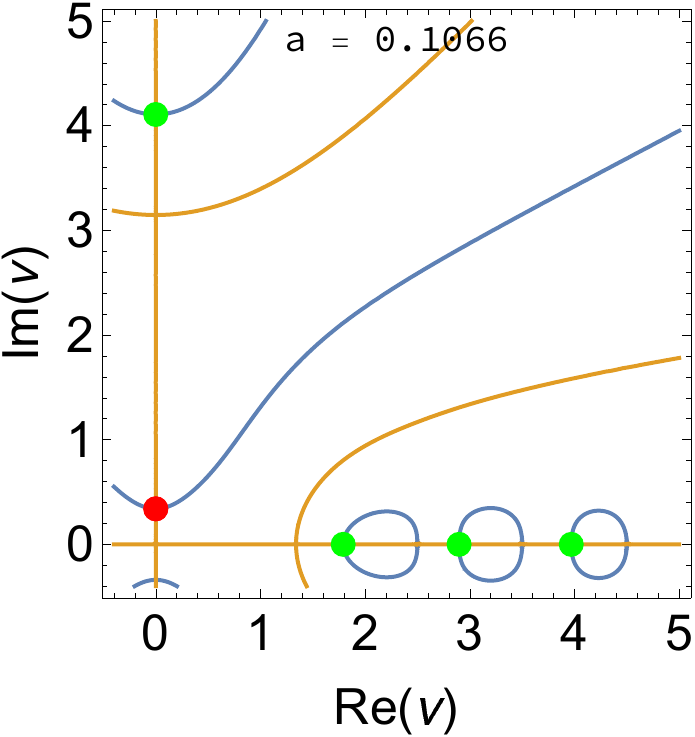}
 \caption{\it ${\tilde{\b}_\text{eff}} = -10.5$}
 \end{subfigure}
 \begin{subfigure}{.3\textwidth}
 \includegraphics[width=\textwidth]{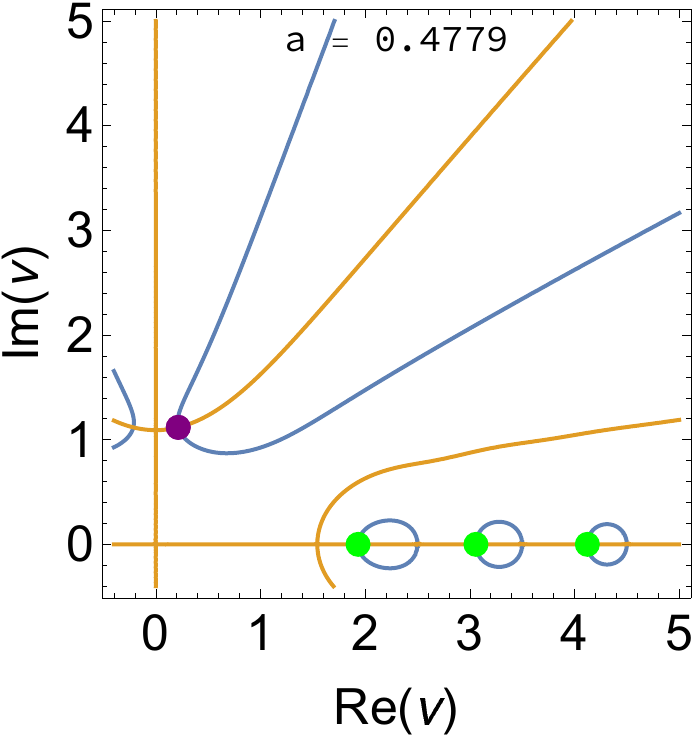}
 \caption{\it ${\tilde{\b}_\text{eff}} = -9$}
 \end{subfigure}
 \begin{subfigure}{.3\textwidth}
 \includegraphics[width=\textwidth]{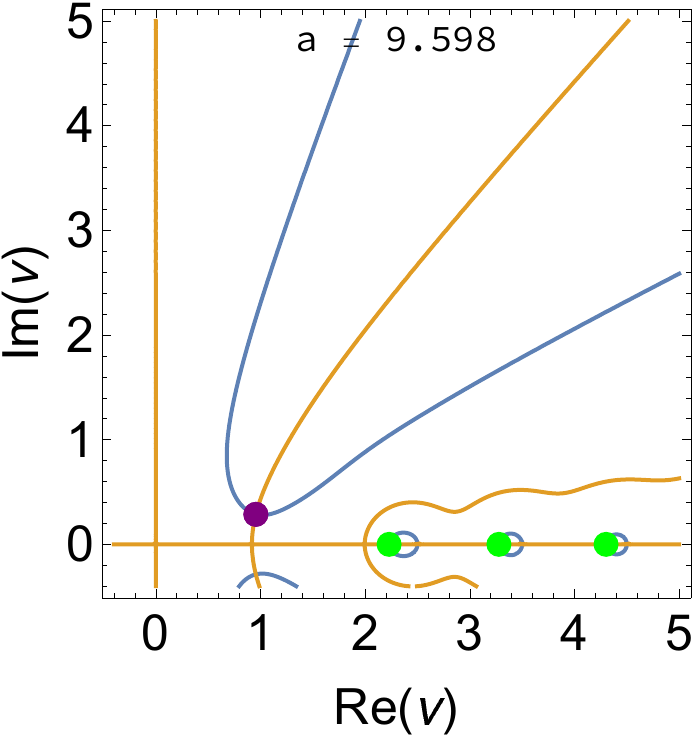}
 \caption{\it ${\tilde{\b}_\text{eff}} = -6$}
 \end{subfigure}
 \begin{subfigure}{.3\textwidth}
 \includegraphics[width=\textwidth]{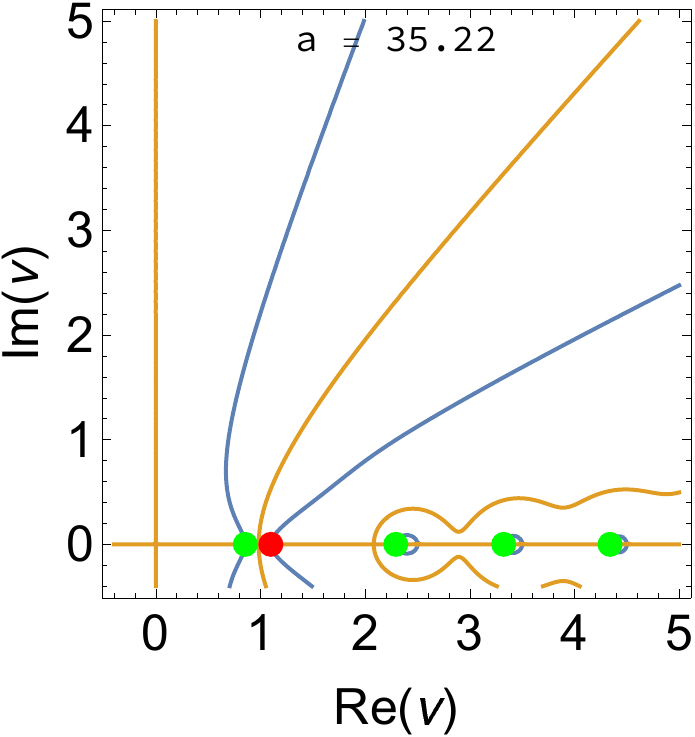}
 \caption{\it ${\tilde{\b}_\text{eff}} = -4.7$}
 \end{subfigure}
 \begin{subfigure}{.3\textwidth}
 \includegraphics[width=\textwidth]{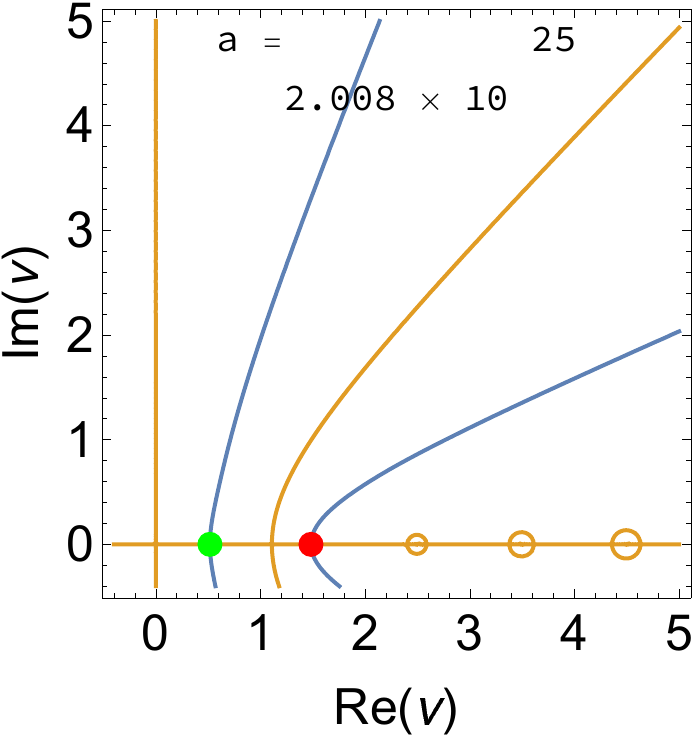}
 \caption{\it ${\tilde{\b}_\text{eff}} = 50$}
 \end{subfigure}

 \caption{\it AdS, $\tilde{\a} = 0$, $GN^2\chi^2 = 1000$.}
 \label{AdS_alpha0_chi1000}
 \end{figure}

 \begin{figure}[ht]
 \centering
 \begin{subfigure}{.3\textwidth}
 \includegraphics[width=\textwidth]{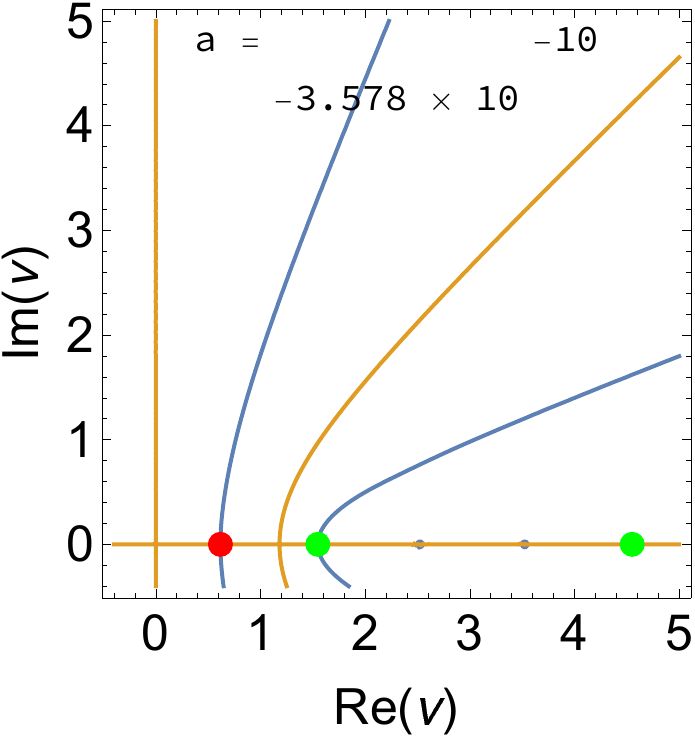}
 \caption{\it ${\tilde{\b}_\text{eff}} = -30$}
 \end{subfigure}
 \begin{subfigure}{.3\textwidth}
 \includegraphics[width=\textwidth]{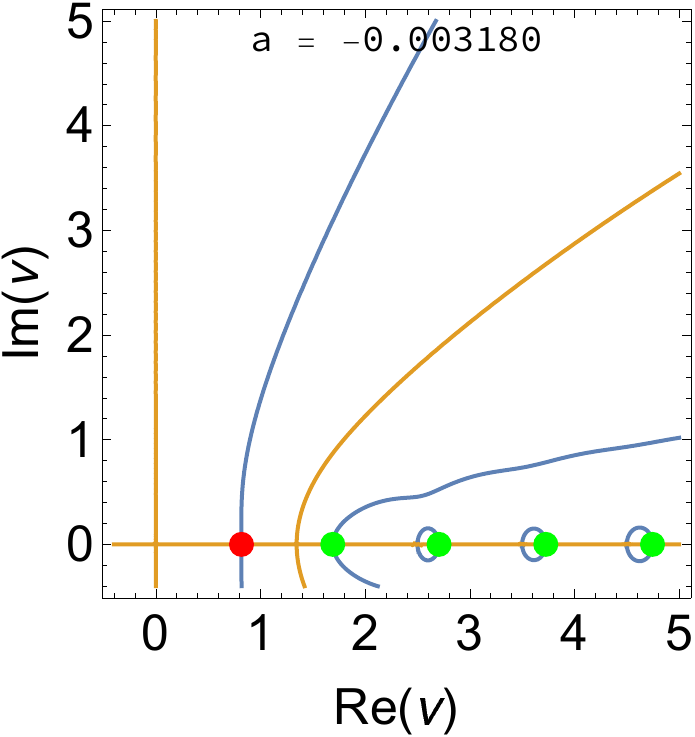}
 \caption{\it ${\tilde{\b}_\text{eff}} = -14 $}
 \end{subfigure}
 \begin{subfigure}{.3\textwidth}
 \includegraphics[width=\textwidth]{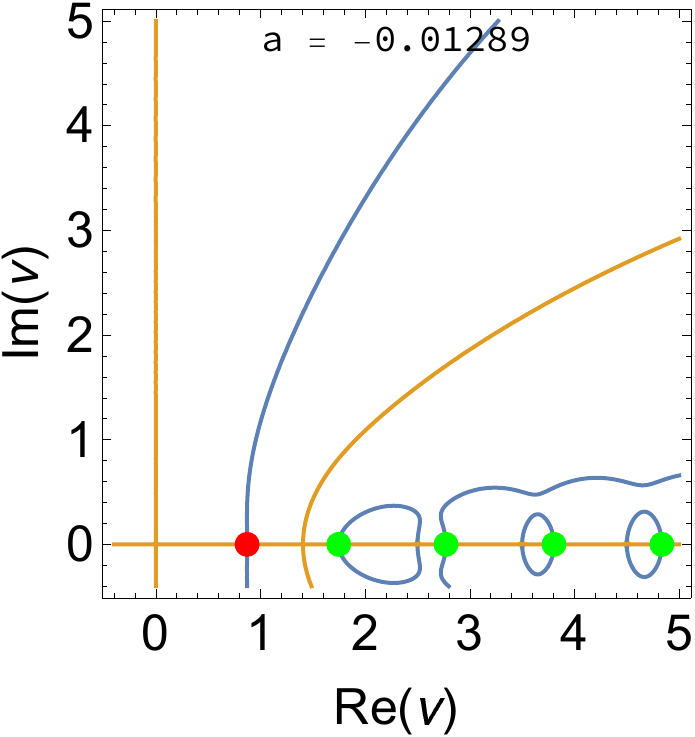}
 \caption{\it ${\tilde{\b}_\text{eff}} = -12.6$}
 \end{subfigure}
 \begin{subfigure}{.3\textwidth}
 \includegraphics[width=\textwidth]{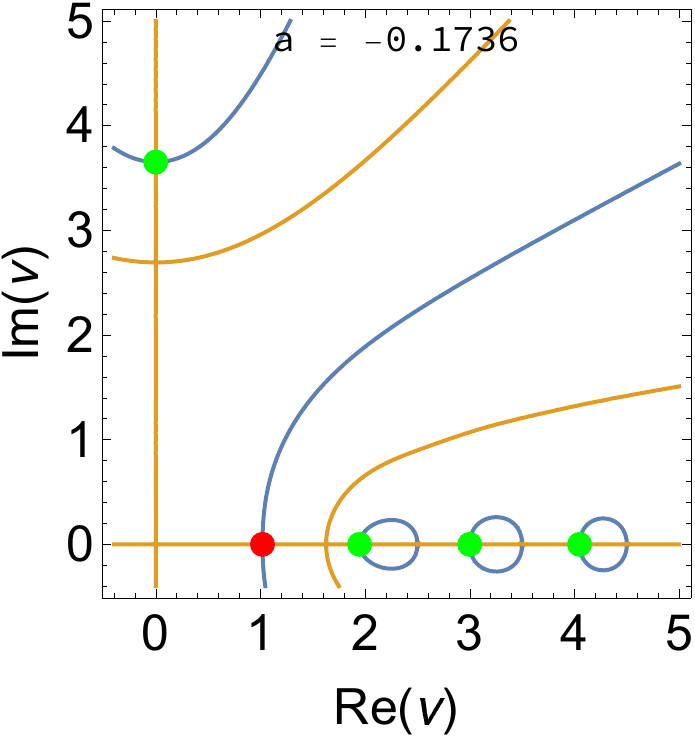}
 \caption{\it ${\tilde{\b}_\text{eff}} = -10$}
 \end{subfigure}
 \begin{subfigure}{.3\textwidth}
 \includegraphics[width=\textwidth]{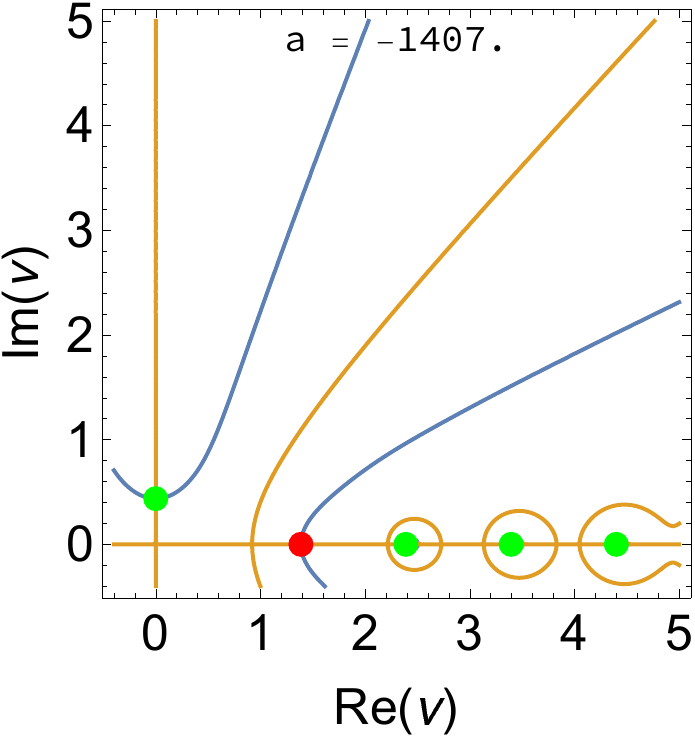}
 \caption{\it ${\tilde{\b}_\text{eff}} = -1$}
 \end{subfigure}
 \begin{subfigure}{.3\textwidth}
 \includegraphics[width=\textwidth]{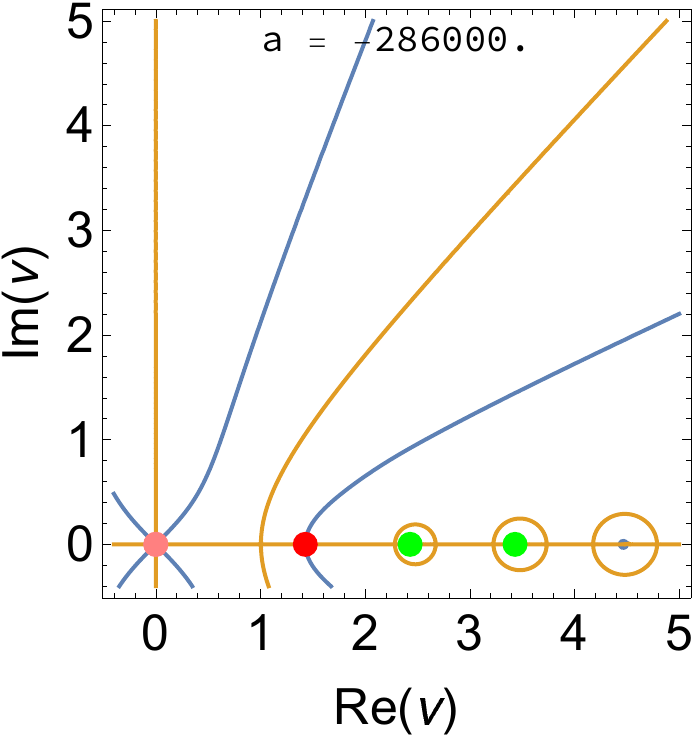}
 \caption{\it ${\tilde{\b}_\text{eff}} = 4.31457$ (*)}
 \end{subfigure}
 \begin{subfigure}{.3\textwidth}
 \includegraphics[width=\textwidth]{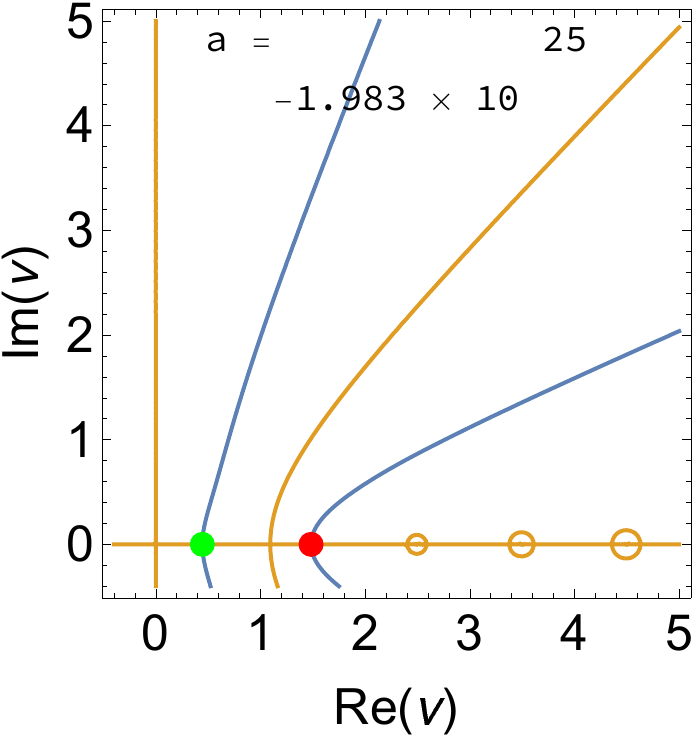}
 \caption{\it ${\tilde{\b}_\text{eff}} = 50$}
 \end{subfigure}

 \caption{\it AdS, $\tilde{\a} = -1$, $GN^2\chi^2 = 1000$.}
 \label{AdS_alpha-1_chi1000}
 \end{figure}

\clearpage


\end{document}